%% file: ReflexivePolygons4.tex
\documentclass[12pt]{article}
\usepackage{jheppub}
\pdfoutput=1

\usepackage{amsmath,bbm,array,amsfonts,graphicx,wrapfig,lscape,float,mathtools,slashbox,multirow,longtable}

\input{pref}

\title{Brane Tilings and Reflexive Polygons}

\author{Amihay Hanany}
\author{and Rak-Kyeong Seong}

\affiliation{
Theoretical Physics Group, The Blackett Laboratory,
Imperial College London, \\
Prince Consort Road, London SW7 2AZ, UK
}

\emailAdd{a.hanany@imperial.ac.uk}
\emailAdd{rak-kyeong.seong@imperial.ac.uk}

\preprint{Imperial/TP/12/AH/01}

\abstract{
Reflexive polygons have attracted great interest both in mathematics and in physics. This paper discusses a new aspect of the existing study in the context of quiver gauge theories. These theories are 4d supersymmetric worldvolume theories of D3 branes with toric Calabi-Yau moduli spaces that are conveniently described with brane tilings. We find all $30$ theories corresponding to the $16$ reflexive polygons, some of the theories being toric (Seiberg) dual to each other. The mesonic generators of the moduli spaces are identified through the Hilbert series. It is shown that the lattice of generators is the dual reflexive polygon of the toric diagram. Thus, the duality forms pairs of quiver gauge theories with the lattice of generators being the toric diagram of the dual and vice versa.
}

\begin{document}

\maketitle

\section{Introduction}

The study of $\mathcal{N}=1$ supersymmetric gauge theories living on D-branes probing singular non-compact Calabi-Yau $3$-folds has been an immensely active and fruitful endeavour in string theory. The matter content of the $4$ dimensional worldvolume theories is encoded in a graph known as the \textbf{quiver} \cite{Douglas:1996sw}.\footnote{For more mathematical reviews on quivers see for example \cite{2007arXiv0710.1898I,2007arXiv0704.0649D}.} An interesting subset of these theories possess mesonic moduli spaces which are toric and are associated to convex lattice polygons. These polygons are known as \textbf{toric diagrams} \cite{1997hep.th...11013L} of the Calabi-Yau singularity.

 In the last two decades, a particular type of polytope caught the attention in string theory in the context of \textbf{mirror symmetry} \cite{Candelas:1989hd,morrison-1993-6,Batyrev:1994hm,Batyrev:1994ju,Batyrev:1997tv,cox1999mirror,mirrorbook}. This polytope is known as a \textbf{reflexive polytope}.

A reflexive polytope is a convex lattice polytope which possesses a single internal lattice point.\footnote{From Latin \textit{reflexus}, Medieval Latin \textit{reflexivus}, meaning to be turned back or reflected.} For a long time, del Pezzo surfaces \cite{Feng:2000mi,Feng:2001xr,Feng:2002fv,Feng:2002zw,Verlinde:2005jr} and more generally Fano varieties \cite{0025-5726-11-3-A03,0025-5726-12-3-A04,ellingsrud1992complex,springerlink:10.1007/BF01170131,springerlink:10.1007/BFb0093585,springerlink:10.1007/BF01160118,0025-5726-19-1-A02,1998math1107B,2007math2890K,2007arXiv0704.0049O,Davey:2011mz} have been associated to a range of reflexive polytopes. 

When Type II superstring theory is compactified on a Calabi-Yau $3$-fold, its worldsheet theory is a $\mathcal{N}=(2,2)$ superconformal field theory. By swapping the Hodge numbers $h_{11}$ and $h_{12}$ associated to the Calabi-Yau 3-fold, one obtains another Calabi-Yau $3$-fold. If one flips the signs of the U(1) R-charges of the left and right moving components of the theory's superalgebra, one obtains another superconformal field theory which is the one compactified on the ``mirror'' of the original Calabi-Yau 3-fold.

Reflexive polytopes have played an important role in studying the relationship between mirror paired Calabi-Yau manifolds and the corresponding superconformal field theories. The reflexive polytopes are used for constructing Calabi-Yau manifolds as hypersurfaces in toric varieties. The underlying property of reflexive polytopes is that they have a polar dual partner which in turn is reflexive and relates to the mirror Calabi-Yau manifold. This property led to a systematic study of mirror paired Calabi-Yau manifolds. The resulting classification \cite{1997CMaPh.185..495K,Kreuzer:1998vb,Kreuzer:2000qv,Kreuzer:2000xy,2008arXiv0802.3376B,2008arXiv0809.4681C} found connections to for instance heterotic string compactifications \cite{1998NuPhB.511..295C,He:2009wi,2010MPLA...25...79H} or to F-theory backgrounds \cite{1997NuPhB.502..613C,1997PhLB..413...63C,Skarke:2000zx,Knapp:2011wk}.

In the following work, reflexive polygons are used to study mesonic moduli spaces of 4d supersymmetric quiver gauge theories dual to Type IIB string theory on $\text{AdS}_5 \times X_{5}$ where $X_5$ is a Sasaki-Einstein $5$-manifold. There are $16$ distinct reflexive polygons and the corresponding theories are worldvolume theories of D$3$-branes probing Calabi-Yau 3-fold singularities. The mesonic moduli spaces are toric Calabi-Yau 3-folds and the reflexive polygons are the corresponding toric diagrams.

The aim of the following work is to identify all $4d$ supersymmetric quiver gauge theories whose moduli space is represented by a reflexive polygon. In order to do so, extensive use is made of \textbf{brane tilings} \cite{Hanany:2005ve,Franco:2005rj}\footnote{For applications of brane tilings see for example \cite{Franco:2005sm,Hanany:2005ss,Hanany:2006nm,Kennaway:2007tq,Yamazaki:2008bt}.} which combine the matter content and the superpotential of the quiver theory on a periodic graph on $\mathbb{T}^2$.

Every consistent brane tiling relates to a consistent quiver gauge theory. Starting from the brane tiling for the orbifold of the form $\mathbb{C}^3/\mathbb{Z}_4 \times \mathbb{Z}_4$ with orbifold action $(1,0,3)(0,1,3)$ \cite{Hanany:2010cx,Davey:2010px,Hanany:2010ne,Davey:2011dd,Hanany:2011iw}, one applies the \textbf{Higgs mechanism} \cite{Feng:2002fv} and uses \textbf{Seiberg duality} \cite{Feng:2000mi,Feng:2001xr,Feng:2002zw,Seiberg:1994pq,Feng:2001bn,2001JHEP...12..001B,Franco:2003ea} on brane tilings in order to find that there exist exactly $30$ quiver gauge theories corresponding to the 16 reflexive polygons. Seiberg duality, also known as toric duality in this context, relates theories with different matter content and superpotential to the same mesonic moduli space.

In order to have a complete classification of the mesonic moduli spaces, the moduli space generators for all $30$ quiver gauge theories are found by computing a generating function known as the \textbf{Hilbert series} \cite{Benvenuti:2006qr,Hanany:2006uc,Feng:2007ur,Butti:2007jv,Forcella:2009vw}. The Hilbert series encodes information about the moduli space generators. They are identified using a method known as \textbf{plethystics} \cite{Hanany:2007zz}. The lattice of generators formed by the mesonic charges is the dual reflexive polygon for the 16 toric diagrams. It is shown that this is the case for all 30 quiver gauge theories.

The complete classification of $4d$ $\mathcal{N}=1$ supersymmetric gauge theories corresponding to the 16 reflexive polygons leads to new observations. The most important observation is that of a new duality which we name \textbf{specular duality}. It relates quiver theories with different mesonic moduli spaces under a swap of external and internal points of the toric diagram. Specular duality partitions the set of $30$ quiver gauge theories in dual pairs and illustrates interesting physics at work. An illustration of this new duality is given at the concluding section, and it is of great interest to explore it further in future work.

The work is structured as follows. In section \sref{s1}, the concepts and motivations behind studying reflexive polygons are reviewed. In addition, the ideas behind brane tilings and the mesonic Hilbert series are reviewed. A key ingredient of the discussion is the lattice of mesonic generators which is reviewed in section \sref{s1}. Sections \sref{sm1} to \sref{sm16} summarize the $30$ quiver gauge theories associated to reflexive polygons, and illustrate the duality between the toric diagram and generator lattices. In section \sref{strees}, the trees illustrating the relationships between toric (Seiberg) dual brane tiling models corresponding to the same reflexive polygon are presented. For the purpose of having a self-contained discussion, appendix \sref{sapp1} reviews the concepts of toric (Seiberg) duality and the Higgs mechanism in the context of brane tiling models. As part of the concluding section, the concept behind specular duality of the $30$ brane tiling models corresponding to reflexive polygons is introduced. 
\\

\section{Background and Motivation \label{s1}}

\subsection{Reflexive Polytopes \label{s1_2}}

\noindent\textbf{Mirror Symmetry.} Reflexive polytopes have been introduced in string theory in the context of mirror symmetry \cite{Candelas:1989hd,morrison-1993-6,Batyrev:1994hm,Batyrev:1994ju,Batyrev:1997tv,cox1999mirror,mirrorbook}. A way to study mirror symmetry is to consider Type II superstring theory compactified on a Calabi-Yau $3$-fold. Its string worldsheet theory is a $\mathcal{N}=(2,2)$ superconformal field theory. It contains a superalgebra with left and right moving components. When one flips the signs of the $U(1)$ R-symmetry charges of the left and right moving components, the Calabi-Yau transitions to a different Calabi-Yau manifold with its Hodge numbers $h_{11}$ and $h_{12}$ being interchanged.

The understanding of mirror symmetry in the context of compactified superstring theory led to a search of mirror paired Calabi-Yau manifolds. Batyrev-Borisov \cite{Batyrev:1994hm,Batyrev:1997tv} laid the foundations for industrialising the search for mirror paired Calabi-Yau manifolds by formulating the construction of Calabi-Yau manifolds as hypersurfaces in toric varieties represented by reflexive polytopes. These reflexive polytopes are on a lattice with the dual polytope and hence corresponding mirror Calabi-Yau manifold being identified by a straightforward geometrical transformation.

Let the following summary review the notion of a reflexive polytope and the concept of its dual:
\begin{itemize}
\item A \textbf{reflexive polytope} is a convex polytope with points in a lattice $\mathbb{Z}^{d}$ and the origin $(0,\dots,0)$ being the unique interior point of the polytope.

\item A \textbf{dual (polar) polytope} exists for every reflexive polytope. The dual of polytope $\Delta$, $\Delta^{\circ}$, is another lattice polytope with points
\beal{es00_20}
\Delta^{\circ}=\{
v^{\circ}\in\mathbb{Z}^d ~|~ \langle v^{\circ},v \rangle \geq -1 ~\forall v\in \Delta
\}
\eea
 The dual of every reflexive polygon is another reflexive polygon. A reflexive polygon can be self-dual, $\Delta=\Delta^{\circ}$.

\item A \textbf{classification of reflexive polytopes} \cite{Kreuzer:1998vb,Kreuzer:2000qv,Kreuzer:2000xy} is available for the dimensions $d\leq 4$ with the number of reflexive polytopes given in \tref{tpolycount}. It is unknown how many exist for higher dimensions.
\end{itemize}

\begin{figure}[H]
\begin{center}
\resizebox{0.9\hsize}{!}{
\includegraphics[trim=0cm 0cm 0cm 0cm,totalheight=18 cm]{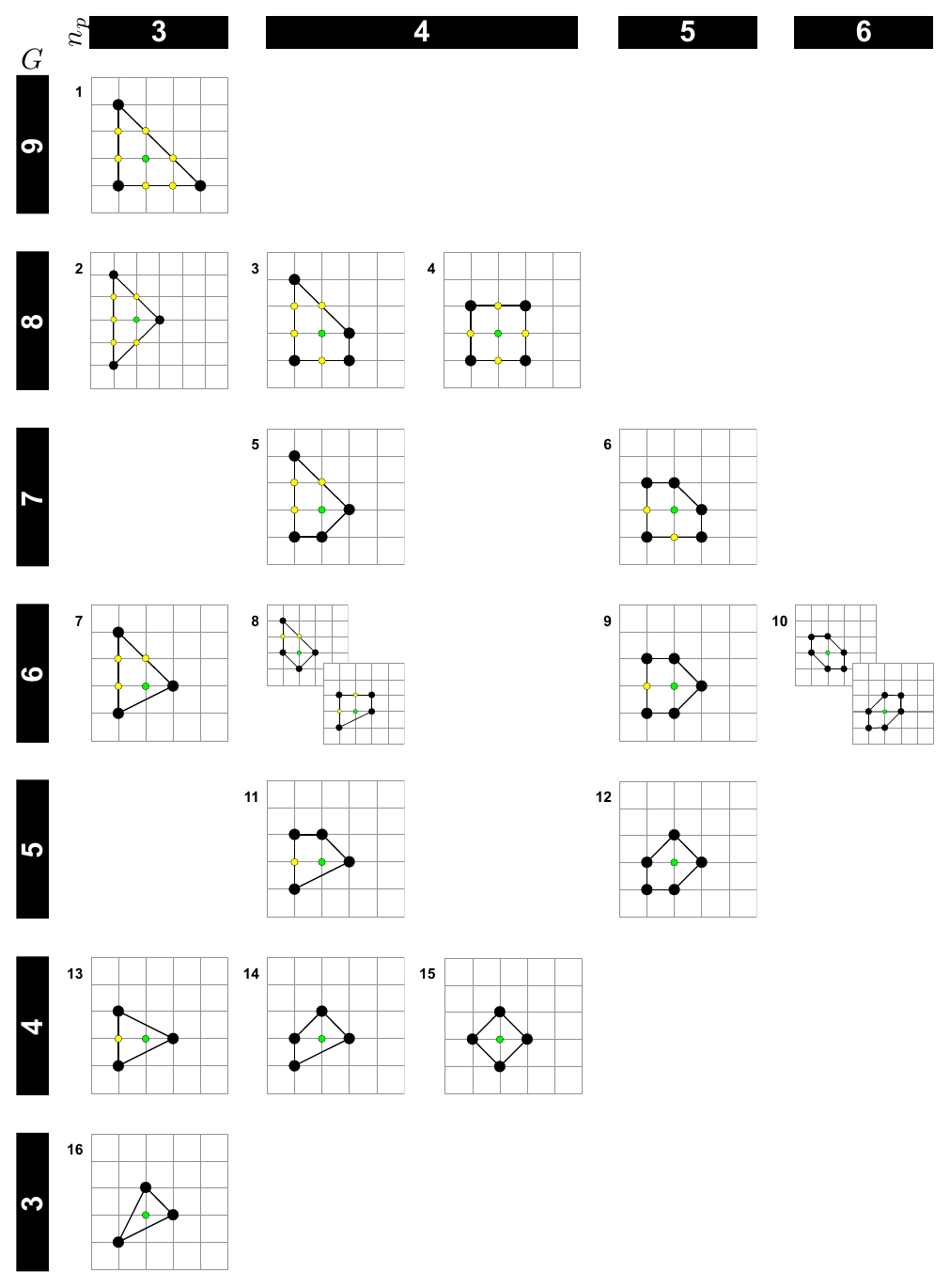}
}
\caption{The $16$ convex polygons which are reflexive. The polygons have been $GL(2,\mathbb{Z})$ adjusted to reflect the duality under (\protect\ref{es00_20}). The green internal points are the origins. $G$ is the area of the polygon with the smallest lattice triangle having normalized area $1$, and $n_G$ is the number of extremal points which are in black. The $4$ polygons with $G=6$ are self-dual. The paired polygons in 8 and 10 are $GL(2,\mathbb{Z})$ equivalent and are each others dual polygon.
\label{f_sumtoric}}
 \end{center}
 \end{figure}

\begin{table}
\centering
\begin{tabular}{|c|c|}
\hline
$d$ & Number of Polytopes\\
\hline\hline
1 & 1 \\
\hline
2 & 16\\
\hline 
3 & 4319\\
\hline
4 & 473800776\\
\hline
\end{tabular}
\caption{Number of reflexive lattice polytopes in dimension $d\leq 4$. The number of polytopes forms a sequence which has the identifier A090045 on OEIS.} \label{tpolycount}
\end{table}

\noindent\textbf{D-branes on Calabi-Yau.} Next to the study of mirror symmetry, reflexive polytopes are playing an interesting role in a different context in string theory. Witten described in 1993 an $\mathcal{N}=(2,2)$ supersymmetric field theory with $U(1)$ gauge groups \cite{Witten:1993yc} in the language of what is today known as gauge linear sigma models (GLSM). He illustrated how the Fayet-Iliopoulos parameter of the $\mathcal{N}=(2,2)$ supersymmetric field theory interpolates between the Landau-Ginzburg and Calabi-Yau phases of the theory. The large parameter limit leads to the space of classical vacua as toric Calabi-Yau spaces determined by the D- and F-terms of the supersymmetric field theory. The formulation of GLSM is going to be used in the context of D-brane gauge theories in this work even though the FI terms will not play a crucial role during the discussion.

Let the focus be on worldvolume theories living on a stack of D3-branes probing Calabi-Yau 3-fold singularities. The gravity dual of these theories is Type IIB string theory on the background $AdS_5\times X_5$ where $X_5$ is a Sasaki-Einstein $5$-manifold. The worldvolume theories are $4d$ $\mathcal{N}=1$ supersymmetric quiver gauge theories whose space of vacua being toric Calabi-Yau $3$-fold are described by lattice polygons on $\mathbb{Z}^2$ known as the toric diagrams. 

A restriction that the toric diagrams are reflexive polygons is introduced for the purpose of the study. A motivation for introducing the restriction is the fact that there are only a finite number $16$ of these reflexive polygons. The natural question to ask, and the question which is fully answered in the following discussion, is which supersymmetric quiver gauge theories exist whose space of vacua correspond to the $16$ reflexive polygons.

There are useful properties of the quiver gauge theories which are considered in this work. These properties provide the essential tools for finding all quiver gauge theories corresponding to reflexive polygons:
 \begin{itemize}
\item \textbf{Brane Tilings (Dimers)} \cite{Hanany:2005ve,Franco:2005rj,Franco:2005sm,Hanany:2005ss,Hanany:2006nm,Kennaway:2007tq,Yamazaki:2008bt} can be used to represent D$3$-brane worldvolume theories whose vacuum moduli space is toric Calabi-Yau. A brane tiling encodes the bifundamental matter content (quiver) and superpotential of the gauge theory. Every consistent brane tiling represents a consistent combination of a quiver and superpotential, and hence a consistent quiver gauge theory.

\item The \textbf{Higgs Mechanism} \cite{Feng:2002fv} in the context of quiver gauge theories has a natural interpretation in terms of the geometrical blow down, i.e. `higgsing', or blow up, i.e. `un-higgsing', of the toric variety corresponding to the gauge theory vacuum moduli space. All $16$ reflexive polygons and the corresponding toric varieties can be related by the geometrical blow downs starting from the abelian orbifold of the form $\mathbb{C}^3/\mathbb{Z}_4\times\mathbb{Z}_4$ with orbifold action $(1,0,3)(0,1,3)$ \cite{Hanany:2010cx,Davey:2010px,Hanany:2010ne,Davey:2011dd,Hanany:2011iw}. For the purpose of a self-contained discussion, the Higgs mechanism in the context of brane tiling theories is reviewed in Appendix \sref{sapp_higgs}.

\item \textbf{Toric (Seiberg) Duality} \cite{Feng:2000mi,Feng:2001xr,Feng:2002zw,Seiberg:1994pq,Feng:2001bn,2001JHEP...12..001B,Franco:2003ea} in the context of quiver gauge theories relates theories with the same vacuum moduli space. In other words, two toric dual theories relate to the same reflexive polygon. Consequently, a single toric variety can be the vacuum moduli space of multiple quiver gauge theories. Such dual quiver gauge theories are known as \textit{toric phases} of the moduli space. More generally, Seiberg duality relates an infinite number of quiver gauge theories by allowing the ranks of gauge groups in the theory to be greater than one. In the following discussion based on brane tilings, only $U(1)$ gauge groups are taken. The search for brane tilings corresponding to the $16$ reflexive polygons uses toric duality in order to identify all toric phases. It turns out that there are $30$ brane tiling theories corresponding to the 16 reflexive polygons. For the purpose of a self-contained discussion, toric (Seiberg) duality in the context of quiver gauge theories and their brane tilings is reviewed in Appendix \sref{sapp_seiberg}.

\end{itemize}

Many of the quiver gauge theories related to reflexive polygons have been studied in the past. A selection of the available literature is given in \tref{tpolylit}. With the following work, a complete classification of all $30$ quiver gauge theories related to reflexive polygons in Witten's language of \textbf{GLSM fields} is provided for the first time. GLSM fields relate the points of the toric diagram with the matter fields of the quiver gauge theory. The F-term and D-term constraint charges on the GLSM fields are used to obtain the \textbf{mesonic Hilbert series}. The mesonic Hilbert series encodes the moduli space \textbf{generators}. 

An intriguing property of theories corresponding to reflexive polygons, which is exemplified in the work below, is as follows: 
\begin{center}
\textit{
The global charges on moduli space generators form a lattice polygon on $\mathbb{Z}^2$ which is reflexive and which is precisely the dual polygon of the toric diagram.}
\end{center}

\begin{figure}[H]
\begin{center}
\resizebox{0.84\hsize}{!}{
\includegraphics[trim=0cm 0cm 0cm 0cm,totalheight=18 cm]{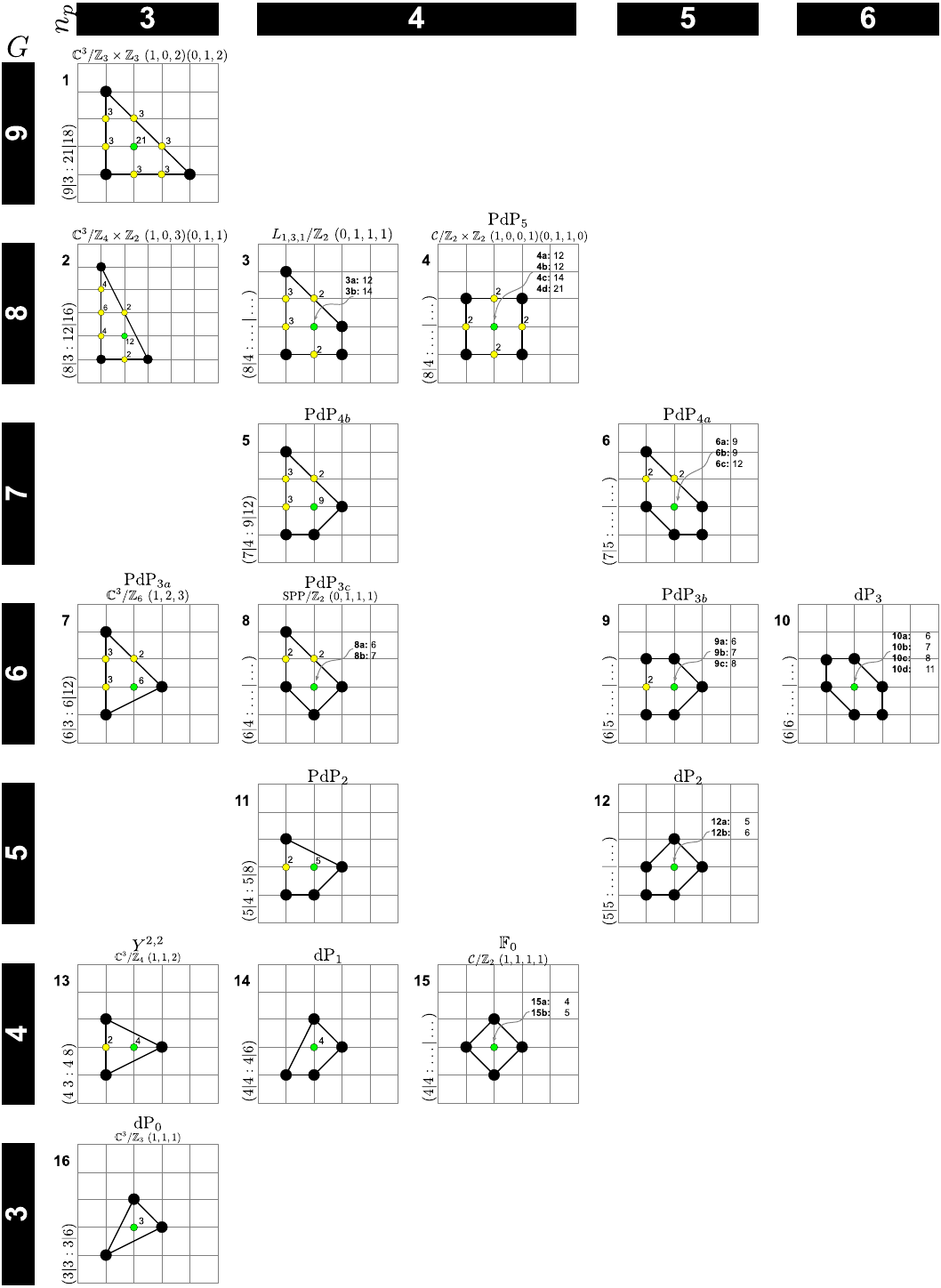}
}
  \caption{The $16$ reflexive polygons as toric diagrams for $30$ brane tilings. The $16$ polygons have been $GL(2,\mathbb{Z})$ transformed to illustrate the blow down from $\mathbb{C}^3/\mathbb{Z}_{4} \times\mathbb{Z}_{4}~(1,0,3)(0,1,3)$ whose toric diagram contains all 16 reflexive polygons. Each polygon is labelled by $(G|n_p:n_i|n_W)$, where $G$ corresponds to the number of $U(n)$ gauge groups, $n_p$ to the number of GLSM fields with non-zero R-charge (number of extremal points in the toric diagram or just the order of the polygon), $n_i$ to the multiplicity of the single interior point of the toric diagram, and $n_W$ to the number of superpotential terms. A reflexive polygon can correspond to multiple quiver gauge theories which are related by toric (Seiberg) duality and distinguished via $n_i$ and $n_W$.
  \label{f_sumtoric2}}
 \end{center}
 \end{figure}

The two sections below provide a review of the physical concepts involved in order to proceed with the complete classification of quiver gauge theories corresponding to reflexive polygons.

\begin{table}
\centering
\resizebox{1\hsize}{!}{
\begin{tabular}{|c|c|c|c|c|c|}
\hline
Model \# & Model Name & $\begin{array}{c} \text{Quiver \& $W$} \\ \text{(Brane Tiling)} \end{array}$ & Toric Data & Mesonic HS & $\begin{array}{c} \text{Generators \&} \\ \text{Generator Lattice} \end{array}$\\
\hline\hline
1 & $\mathbb{C}^3/\mathbb{Z}_3\times\mathbb{Z}_3~(1,0,2)(0,1,2)$
& \cite{Hanany:2005ve,2001JHEP...12..001B}
& 
&
&
\\
2 & $\mathbb{C}^3/\mathbb{Z}_{4}\times\mathbb{Z}_{2}~(1,0,3)(0,1,1)$
& \cite{Hanany:2005ve}
&
&
&
\\
3 & $L_{1,3,1}/\mathbb{Z}_2~(0,1,1,1)$
& \cite{Franco:2005sm,Butti:2005vn}
& \cite{Butti:2005vn}
&
&
\\
4 & $\text{PdP}_5~,~\mathcal{C}/\mathbb{Z}_{2}\times\mathbb{Z}_{2}~(1,0,0,1)(0,1,1,0)$
& \cite{Feng:2002fv,Hanany:2005ve,Franco:2005rj,Forcella:2008bb}
& \cite{Feng:2002fv,Franco:2005rj,Forcella:2008bb}
&
&
\\
5 & $\text{PdP}_{4b}$
&
&
&
&
\\
6 & $\text{PdP}_{4a}$
& \cite{Feng:2002fv,Forcella:2008bb,Butti:2006hc}
& \cite{Feng:2002fv,Forcella:2008bb,Butti:2006hc}
& \cite{Benvenuti:2006qr}
&
\\
7 & $\text{PdP}_{3a}~,~\mathbb{C}^3/\mathbb{Z}_6~(1,2,3)$
& \cite{Hanany:2005ve,2001JHEP...12..001B}
& \cite{Hanany:2005ve}
&
&
\\
8 & $\text{PdP}_{3c}~,~\text{SPP}/\mathbb{Z}_{2}~(0,1,1,1)$
& \cite{Feng:2002fv,2001JHEP...12..001B,Feng:2004uq}
& \cite{Feng:2002fv,Feng:2004uq}
&
&
\\
9 & $\text{PdP}_{3b}$
& \cite{Feng:2002fv,2001JHEP...12..001B,Feng:2004uq}
& \cite{Feng:2002fv,Feng:2004uq}
&
&
\\
10 & $\text{dP}_3$
& \cite{Feng:2002fv,Feng:2002zw,Franco:2005rj,2001JHEP...12..001B,Feng:2004uq,Forcella:2008bb,Forcella:2008ng}
& \cite{Feng:2001xr,Feng:2002fv,Franco:2005rj,Forcella:2008bb,Feng:2004uq,Forcella:2008ng}
& \cite{Benvenuti:2006qr}
&
\\
11 & $\text{PdP}_2$
& \cite{Feng:2002fv,Feng:2004uq}
& \cite{Feng:2002fv,Feng:2004uq}
& 
&
\\
12 & $\text{dP}_2$
& \cite{Feng:2002zw,Franco:2005rj,Forcella:2008bb,Feng:2004uq,Bertolini:2004xf,Forcella:2008ng,Pinansky:2005ex,Davey:2009bp}
& \cite{Feng:2001xr,Franco:2005rj,Forcella:2008bb,Feng:2004uq,Forcella:2008ng,Pinansky:2005ex}
& \cite{Benvenuti:2006qr}
& \cite{Pinansky:2005ex}
\\
13 & $Y^{2,2}~,~\mathbb{C}^3/\mathbb{Z}_{4}~(1,1,2)$
& \cite{Hanany:2005ve,Franco:2005rj}
& \cite{Hanany:2005hq}
& \cite{Benvenuti:2006qr}
& \cite{Benvenuti:2004wx,Benvenuti:2004dy}
\\
14 & $Y^{2,1}~,~\text{dP}_1$
& \cite{Feng:2002zw,Franco:2005rj,Forcella:2008bb,Feng:2004uq,Bertolini:2004xf,Davey:2009bp}
& \cite{Feng:2001xr,Forcella:2008bb,Feng:2004uq,Hanany:2005hq}
& \cite{Benvenuti:2006qr,Butti:2007jv}
& \cite{Benvenuti:2004wx,Benvenuti:2004dy}
\\
15 & $\mathbb{F}_{0}~,~Y^{2,0}~,~\mathcal{C}/\mathbb{Z}_2~(1,1,1,1)$
& \cite{Feng:2001xr,Franco:2005rj,Hanany:2005ve,Davey:2009bp,Feng:2004uq,Forcella:2008bb,Forcella:2008ng,Forcella:2009bv}
& \cite{Feng:2001xr,Forcella:2008bb,Feng:2004uq,Forcella:2008ng,Hanany:2005hq,Forcella:2009bv}
& \cite{Benvenuti:2006qr}
& \cite{Benvenuti:2004wx,Benvenuti:2004dy}
\\
16 & $\text{dP}_0~,~\mathbb{C}^3/\mathbb{Z}_{3}~(1,1,1)$
& \cite{Feng:2002zw,Hanany:2005ve,Hanany:2005ss,Davey:2009bp,Hanany:2010zz}
& \cite{Feng:2001xr,Feng:2002fv,Hanany:2010zz}
& \cite{Benvenuti:2006qr,Butti:2007jv,Hanany:2010zz}
& 
\\
\hline
\end{tabular}
}
\caption{A selection of the literature on quiver gauge theories corresponding to reflexive polygons.} \label{tpolylit}
\end{table}

\subsection{The Brane Tiling and the Forward Algorithm \label{s1_1}}

The worldvolume theory of a stack of $n$ D3-branes probing singular non-compact Calabi-Yau $3$-folds is a $3+1$ dimensional $\mathcal{N}=1$ supersymmetric gauge theory. The corresponding Lagrangian is specified by the theory's gauge groups, matter content and superpotential.

The probed Calabi-Yau $3$-fold is toric, and is the mesonic moduli space of the worldvolume theory.  It is of great interest to associate to each worldvolume theory the corresponding mesonic moduli space. For the purpose of a self-contained discussion, a brief review on the \textbf{forward algorithm} \cite{Feng:2000mi,Gulotta:2008ef} which translates the gauge theory information into toric data is provided below.
\\

\begin{figure}[H]
\begin{center}
\includegraphics[trim=0cm 0cm 0cm 0cm,width=14cm]{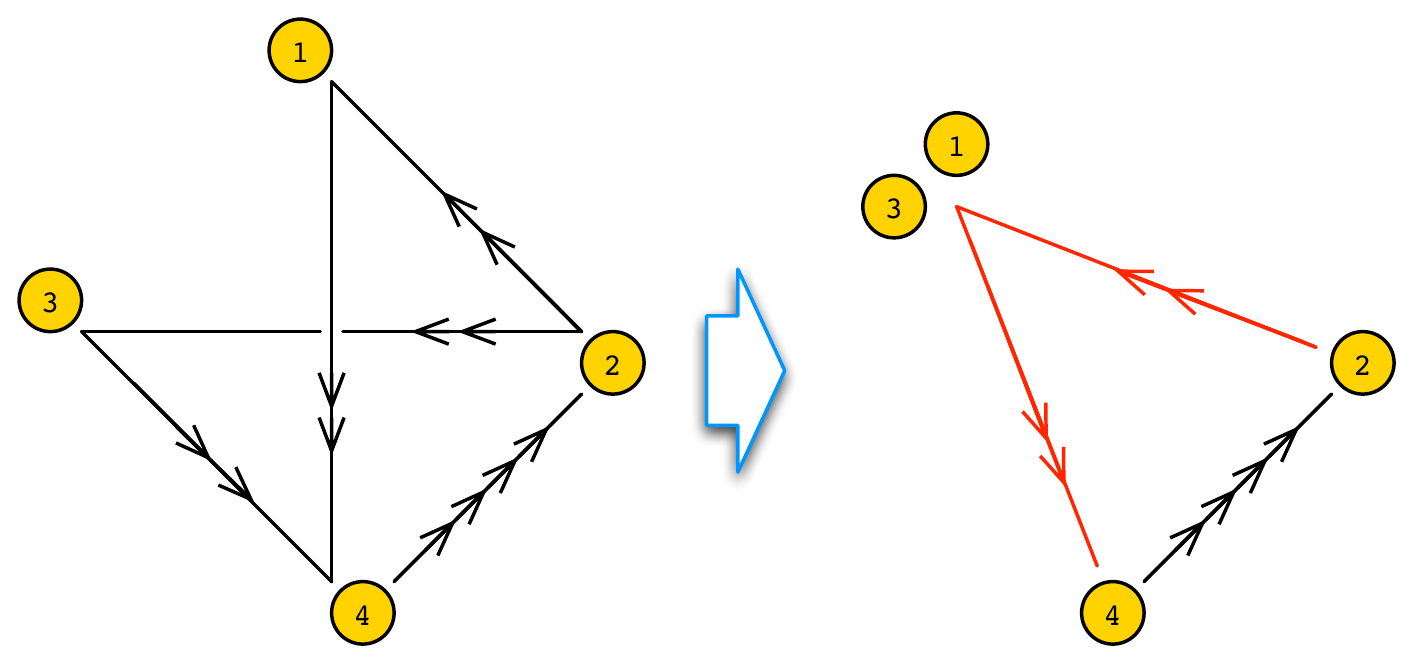}
\caption{The quiver for phase b of the Hirzebruch $\mathbb{F}_0$ model. Vertices $1$ and $3$ share the same incidence information with no matter fields between them. They are combined into a block. All matter fields intersecting the block are colored red and are combined such that a red arrow represents all possible connections from and to all vertices within the block. \label{fblockdia}}
 \end{center}
 \end{figure}

\noindent\textbf{Quiver $\mathcal{Q}$.}  The matter content of the gauge theory is specified by a graph known as the \textbf{quiver} \cite{Douglas:1996sw,2007arXiv0710.1898I,2007arXiv0704.0649D}. It consists of the following components:
\begin{itemize}
	\item \textbf{Vertices} in $\mathcal{Q}$ correspond to $U(n_i)$ gauge groups with $i=1,\dots,G$.
	\item \textbf{Edges} in $\mathcal{Q}$ correspond to the matter fields $X_{ij}$. The matter fields are bifundamental and transform under the fundamental of $U(n_i)$ and antifundamental of $U(n_j)$, imposing a direction on the quiver edges, $i\rightarrow j$. The anomaly cancellation condition for the quiver gauge theory sets  the number of incoming and outgoing edges on a quiver vertex to be equal. Every matter field appears precisely once in a positive and negative term in $W$, with the number of positive and negative terms in $W$ being the same. This is known as the \textbf{toric condition} \cite{Feng:2002zw}.
	\item The \textbf{incidence matrix} $d_{G \times E}$ for $E$ bifundamental matter fields encodes the quiver. Its entry for a gauge group $U(n_i)$ is $-1$ for $X_{ij}$, $+1$ for $X_{ji}$, and $0$ otherwise. The matrix $d_{G \times E}$ has $G-1$ independent rows which can be collected in a new matrix called $\Delta_{(G-1)\times e}$.
\end{itemize}

If two or more quiver vertices share the same intersection number with other quiver vertices and have no matter fields between any two of them, then the quiver vertices can be grouped into a \textbf{block} \cite{1997alg.geom..3027K,Franco:2004rt}. This property is illustrated in the example for phase b of the Hirzebruch $\mathbb{F}_0$ model as shown in \fref{fblockdia}.
\\

\noindent\textbf{Brane Tilings/Dimers:}
The superpotential and the quiver can be combined into a single representation of the supersymmetric gauge theory. The representation is known as a brane tiling or dimer \cite{Hanany:2005ve,Franco:2005rj,Franco:2005sm,2007arXiv0710.1898I}. It is a periodic bipartite graph on $\mathbb{T}^2$ and has the following components:
\begin{itemize}
	\item \textbf{White (resp. black) nodes} correspond to positive (negative) terms in the superpotential. They have a clockwise (anti-clockwise) orientation.
	\item \textbf{Edges} connect to nodes and correspond to the bifundamental fields in the superpotential. Going along the induced orientations around nodes, one can identify the matter fields associated to a specific superpotential term in the correct cyclic order.
	\item \textbf{Faces} correspond to $U(n_i)$ gauge groups. Every edge $X_{ij}$ in the tiling has two neighbouring faces corresponding to $U(n_i)$ and $U(n_j)$. The quiver orientation of the bifundamental field $X_{ij}$ is given by the orientation around the black and white nodes at the two ends of the corresponding tiling edge.
\end{itemize}

The fundamental domain of the $2$-torus on which the brane tiling is drawn is interpreted as a section of the periodic tiling which contains the quiver and superpotential information without repetition. Copying the domain along the fundamental cycles of the torus reproduces the complete brane tiling.
\\

\noindent\textbf{Perfect Matchings/GLSM fields and F-and D-term charges.} A new basis of fields can be defined from the set of bifundamental matter fields. The purpose of the new basis of fields is to describe both F-term and D-term constraints of the supersymmetric gauge theory with a common setting. The new fields are known as gauge linear sigma model fields (GLSM) and are represented as perfect matchings in the brane tiling. They have the following properties:
\begin{itemize}
	\item A \textbf{perfect matching} $p_\alpha$ is a set of bifundamental fields which connect all nodes in the brane tiling uniquely once. The perfect matchings corresponding to \textbf{extremal} (corner) points of the toric diagram have non-zero $U(1)_R$ R-charge. The internal as well as all \textbf{non-extremal} toric points on the perimeter of the toric diagram have zero R-charge. All points on the perimeter are called \textbf{external}, including extremal ones. They can be summarized in a matrix $P_{E\times c}$ where $E$ is the number of matter fields and $c$ the number of perfect matchings.
	\item \textbf{F-terms} are encoded in the perfect matching matrix $P_{E\times c}$. The charges under the F-term constraints are given by the kernel,
	\beal{es00_1c0}
	Q_{F~(c-G-2)\times c} =  \ker{(P_{E \times c})}~~.
	\eea
	\item \textbf{D-terms} are of the form \cite{Witten:1993yc},
\beal{es00_2}
D_i = - e^2 (\sum_{a} d_{ia}|X_a|^2 - \zeta_i)~~,
\eea
where $X_a$ is the matter field corresponding to the $a$-th column of the incidence matrix $d_{G\times E}$, $i$ runs over the $U(n)$ gauge groups in the quiver, $e$ is the gauge coupling, and $\zeta_i$ is the Fayet-Iliopoulos (FI) parameter. The D-terms are encoded via the reduced quiver matrix $\Delta_{(G-1)\times E}$\footnote{Since the sum of rows in $d_{G\times E}$ vanishes, there are $G-1$ independent rows giving the reduced matrix $\Delta_{(G-1)\times E}$.} and are related to the perfect matching matrix as follows,
\beal{es00_3}
\Delta_{(G-1)\times E} = Q_{D~(G-1)\times c}.P^{t}_{c \times E}~~,
\eea
where the $Q_{D~(G-1) \times c}$ matrix is the charge matrix under D-term constraints.
Equivalently, in terms of an interim matrix $\tilde{Q}_{G\times c}$, which maps perfect matchings into their quiver charges, one has the relation
\beal{es00_1c2}
d_{G\times E} = \tilde{Q}_{G \times c}.P^{t}_{c \times E}~~.
\eea
\end{itemize}

Overall, the charge matrices $Q_F$ and $Q_D$ can be concatenated to form a $(c-3) \times c$ matrix,
\beal{es00_5}
Q_t = \left( \ba{c} Q_F \\ Q_D \ea \right)~~.
\eea
The kernel of the charge matrix,
\beal{es00_6}
G_t = \ker{(Q_t)}~~,
\eea
precisely encodes the coordinates of the \textbf{toric diagram} points with columns and hence perfect matchings and GLSM fields corresponding to points of the toric diagram.
\\

\subsection{Hilbert Series and Lattice of Generators \label{s1_3}}

The generating function of mesonic gauge invariant operators (GIOs) is known as the \textbf{mesonic Hilbert series} \cite{Benvenuti:2006qr,Hanany:2006uc,Feng:2007ur,Butti:2007jv,Forcella:2009vw}. The Hilbert series encodes the generators of the associated moduli space. These are essential for a complete classification of the mesonic moduli spaces of brane tilings corresponding to reflexive polygons. The moduli space generators can be extracted from the Hilbert series using a method known as \textbf{plethystics}. These carry charges under the \textbf{mesonic symmetry}. The charges on a $\mathbb{Z}_2$ lattice form a convex polygon which is the dual polygon of the toric diagram.

Let the section below provide a review of the concepts involved.
\\

\noindent\textbf{Mesonic Symmetry.} The mesonic moduli space of a given brane tiling is a non-compact toric Calabi-Yau $3$-fold. The mesonic symmetry of the associated quiver gauge theory takes one of the following forms,
\begin{itemize}
\item $U(1)_{f_1} \times U(1)_{f_2} \times U(1)_R$
\item $SU(2)_{x} \times U(1)_{f} \times U(1)_R$
\item $SU(2)_{x_1} \times SU(2)_{x_2} \times U(1)_R$
\item $SU(3)_{x_1,x_2} \times U(1)_R$~~~,
\end{itemize}
where the lower case indices denote fugacities of the gauge group with the exemption of the R-symmetry group $U(1)_R$. The fugacity associated to the $U(1)_R$ charge is $t$.

The above global symmetries derive from the isometry group of the Calabi-Yau 3-fold. The enhancement of a $U(1)$ flavour to $SU(2)$ or $SU(3)$ is indicated by repeated columns in the total charge matrix $Q_t$.
\\

\noindent\textbf{Mesonic Hilbert Series.} The mesonic moduli space is the space of invariants under F-term charges $Q_F$ and D-term charges $Q_D$. The $c$ GLSM fields corresponding to perfect matchings of the brane tiling form the space $\mathbb{C}^c$ known as the space of perfect matchings. 

\begin{itemize}
\item
The \textbf{Symplectic Quotient}
\beal{es12_1}
\mathcal{M}^{mes}= (\mathbb{C}^c// Q_F) // Q_D ~~.
\eea
is the \textit{mesonic moduli space} of the quiver gauge theory.\footnote{The symplectic quotient $\mathcal{F}^{\flat}=\mathbb{C}^c // Q_F$ is known as the \textit{Master space} \cite{Forcella:2008bb,Forcella:2008eh,Forcella:2008ng,Zaffaroni:2008zz,Forcella:2009bv,Hanany:2010zz} and is the space of invariants including both mesonic and baryonic degrees of freedom.}  The invariants under the symplectic quotient are mesonic GIOs.

\item 
The \textbf{mesonic Hilbert series} is a generating function which counts mesonic GIOs on the moduli space. The mesonic Hilbert series is obtained via the Molien integral formula,
\beal{es12_2}
g_1(y_\alpha; \mathcal{M}^{mes}) =
\prod_{i=1}^{c-3}  \oint_{|z_i|=1} \frac{\ud z_i}{2\pi i z_i} 
\prod_{\alpha=1}^{c} 
\frac{
1
}{
(1-y_\alpha \prod_{j=1}^{c-3} z_j^{(Q_t)_{j\alpha}})
}~~,
\eea
where $c$ is the number of perfect matchings labelled by $\alpha=1,\dots,c$ and $Q_t$ is the total charge matrix in \eref{es00_5}. 
GLSM fields corresponding to extremal perfect matchings $p_\alpha$ carry non-zero R-charges and have fugacities denoted by $y_\alpha=t_\alpha$. For all other GLSM fields $s_m$ with zero R-charges one assigns the fugacity $y_\alpha = y_{s_m}$. The perfect matchings associated to these fields are non-extremal. Certain products of non-extremal perfect matchings such as $s=\prod_m s_m$ are assigned a single fugacity denoted by $y_s$.

\item The \textbf{plethystic logarithm} of the Hilbert series encodes information about the generators of the moduli space and the relations formed by them. It is defined as
\beal{es12_3}
PL[g_1(y_\alpha; \mathcal{M})] = 
\sum_{k=1}^{\infty} \frac{\mu(k)}{k} \log\left[
g_1(y_\alpha^k; \mathcal{M})
\right]~~,
\eea
where $\mu(k)$ is the M\"obius function.
If the expansion of the plethystic logarithm is finite, the moduli space is a \textit{complete intersection} generated by a finite number of generators subject to a finite number of relations. If the expansion is infinite, the moduli space is a non-complete intersection. The first positive terms of the expansion refer to generators of the moduli space.\footnote{The Groebner basis of the set of gauge invariant operators forms the generators of the moduli space.} All higher order terms refer to relations among generators and relations among relations called \textit{syzygies}.
\end{itemize}

\noindent\textbf{R-charges.}\footnote{We review here volume minimisation as a means to calculate R-charges. For alternative methods see for example \cite{Butti:2005vn,Butti:2005ps,Hanany:2011bs}.} The mesonic moduli space is a toric Calabi-Yau cone over a Sasaki-Einstein $5$-manifold whose volume is related under minimization to the $U(1)$ R-charges of the \textit{divisors} of the toric geometry \cite{Martelli:2005tp,Martelli:2006yb,Butti:2006au}. The toric divisors relate to the extremal points of the toric diagram and the corresponding GLSM fields.

The volume of the Sasaki Einstein $5$-manifold $X_5$ is
\beal{es2_4}
\text{Vol}(r_\alpha; X_5) = \frac{8 \pi^3}{27}\lim_{\mu\rightarrow 0} \mu^3 g_{1}(t_\alpha=e^{- \mu  r_\alpha};\mathcal{M}=\mathcal{C}(X_5))~~.
\eea
where $g_1(t_\alpha; \mathcal{M})$ is the mesonic Hilbert series in \eref{es12_2}, $t_\alpha$ is the fugacity for GLSM field $p_\alpha$, and $r_\alpha$ is the corresponding minimization parameter. The Hilbert series related to the divisor $D_\alpha$ and the corresponding GLSM field $p_\alpha$ is obtained through the following modified Molien integral,
\beal{es2_5}
g^{D_{\alpha}}(t_\alpha; \mathcal{M}^{mes})&=&
\prod_{i=1}^{c-3}
\oint_{|z_i|=1} \frac{\ud z_i}{2 \pi i z_i}
~
\left(t_\alpha ~
\prod_{k=1}^{c-3}
z_{k}^{(Q_t)_{k\alpha}}
\right)^{-1}~
g_{1}(t_\alpha,z_i;\mathbb{C}^c)
\nn\\
&=&
\prod_{i=1}^{c-3}
\oint_{|z_i|=1} \frac{\ud z_i}{2 \pi i z_i}
~
\prod_{\beta=1}^{c}
\frac{
\left(t_\alpha ~
\prod_{k=1}^{c-3}
z_{k}^{(Q_t)_{k\alpha}}
\right)^{-1}
}{
1-
t_\beta ~
\prod_{j=1}^{c-3}
z_{j}^{(Q_t)_{j\beta}}
}~~.
\eea
 The associated R-charge is then
\beal{es2_6}
R_{\alpha}
= 
\lim_{\mu\rightarrow 0} \frac{1}{\mu}
\left[
\frac{
g^{D_\alpha}(e^{-\mu r_\alpha}; \mathcal{M}^{mes})
}{
g^{mes}(e^{-\mu r_\alpha};\mathcal{M}^{mes})
}
-1
\right]
~~.
\eea

For superconformality, the superpotential has $R$-charge $2$ which sets the following restriction on the R-charges
\beal{es2_7}
\sum_{\alpha} R_\alpha =2.
\eea
\\

\noindent\textbf{Lattice of Generators.} The lattice of generators is determined by the mesonic charges carried by the generators of the mesonic moduli space. Ignoring the $U(1)_R$ factor, the remaining flavour symmetries have ranks which sum up to $2$. Hence, there are always $2$ fugacities which count flavour charges. The pair of flavour charges carried by each generator is taken as coordinates of a point on the plane. The convex hull of the collection of points corresponding to the collection of moduli space generators forms a convex polygon. This is known as the lattice of generators.

For a non-vanishing convex polygon on $\mathbb{Z}^2$, the flavour charges are subject to the following constraints:
\begin{itemize}
\item The pairs of flavour charges carried by all $n_p$ extremal perfect matchings form a pair of $n_p$-dimensional charge vectors. For a non-trivial choice of flavour charges, the charge vectors are linearly independent.
\item The elements of the $n_p$-dimensional charge vectors sum up to zero.
\item The charges on GLSM fields are scaled such that the charges on mesonic moduli space generators take integer values and the lattice of generators is on $\mathbb{Z}^2$.
\end{itemize}
The lattice of generators subject to the constraints above still exhibits a remaining $GL(2,\mathbb{Z})$ degree of freedom. Moreover, each generator also carries a R-charge which plays the role of a third coordinate for each point in the lattice of generators. In order to remove these remaining degrees of freedom, one makes use of a particular property of generator lattices introduced below.
\\

\noindent\textbf{Duality between Generator Lattices and Toric Diagrams.} \begin{center}
\textit{
The lattice of generators of a brane tiling is \\ 
the dual of the toric diagram.
}
\end{center}
The duality between \textit{reflexive} polygons follows \eref{es00_20}. Hence, for reflexive polygons as toric diagrams, the lattice of generators is another reflexive polygon in $\mathbb{Z}^2$. Accordingly, the remaining $GL(2,\mathbb{Z})$ degree of freedom on the lattice of generators can be removed by making the duality for reflexive polygons exact as defined in \eref{es00_20}. In addition, for reflexive polygons the lattice of generators always lies on $\mathbb{Z}^2$.

When the lattice of generators is considered as a toric diagram of a new brane tiling, the duality between reflexive polygons manifestly relates between two quiver gauge theories with toric moduli spaces. In terms of the number of $U(n)$ gauge groups $G$ and the number of GLSM fields with non-zero R-charge $n_p$, the duality map takes the form
\beal{es2_8}
\text{Model A} &~~\leftrightarrow~~& \text{Model B} \nn\\
G &~~\leftrightarrow~~& 12-G \nn\\
n_p &~~\leftrightarrow~~& n_p
\eea
as illustrated in \fref{f_sumtoric2}.
\\

In the following sections, all $30$ quiver gauge theories with their brane tilings corresponding to the $16$ reflexive polygons are classified. All $30$ quiver gauge theories are obtained by higgsing and toric (Seiberg) dualizing the theory related to the abelian orbifold of the form $\mathbb{C}^3/\mathbb{Z}_{4}\times\mathbb{Z}_4$ with orbifold action $(1,0,3)(0,1,3)$. The details for the \textit{parent} theory for all reflexive polygon theories are given in appendix \sref{s_parent}. 
\\

\clearpage
\section{Model 1: $\mathbb{C}^3/\mathbb{Z}_3\times\mathbb{Z}_3~(1,0,2)(0,1,2)$ \label{sm1}}

\begin{figure}[ht!]
\begin{center}
\includegraphics[trim=0cm 0cm 0cm 0cm,height=4 cm]{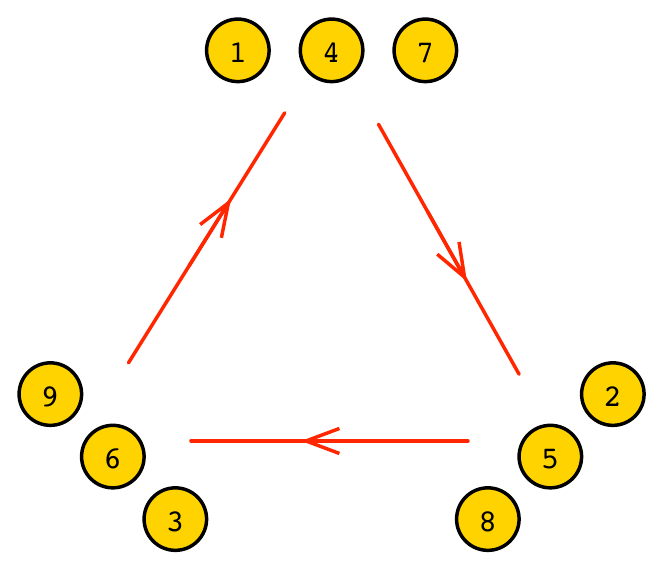}
\includegraphics[width=5.5 cm]{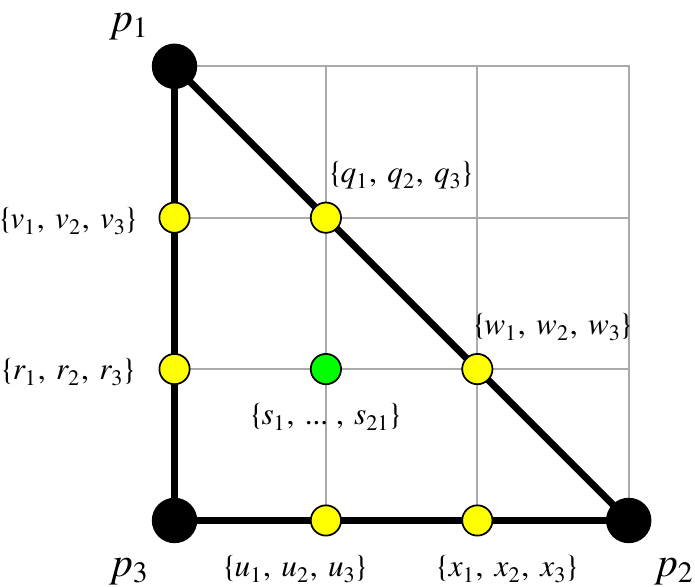}
\includegraphics[width=5 cm]{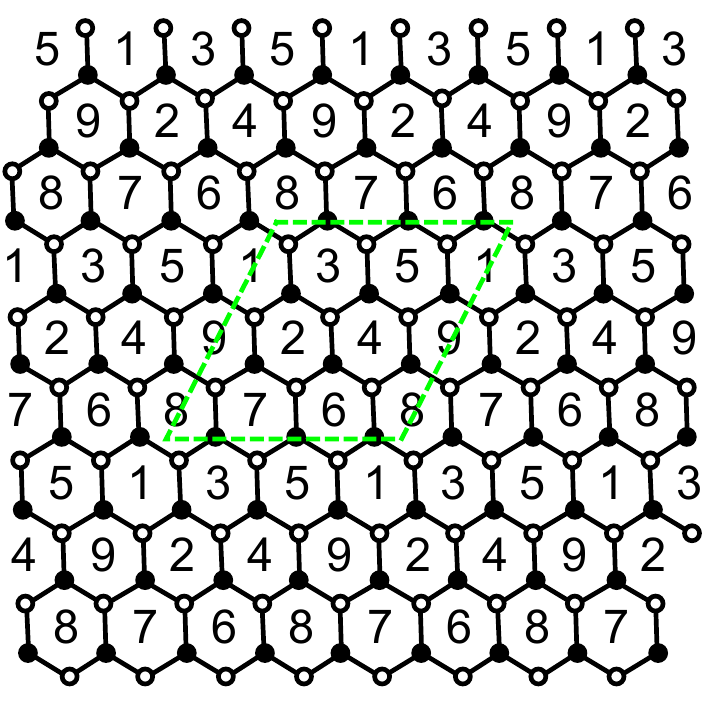}
\caption{The quiver, toric diagram, and brane tiling of Model 1. The red arrows in the quiver indicate all possible connections between blocks of nodes.\label{f1}}
 \end{center}
 \end{figure}
 
\noindent The superpotential is 
\beal{esm1_00}
W&=& 
+ X_{15} X_{56} X_{61}  
+ X_{29} X_{91} X_{12}  
+ X_{31} X_{18} X_{83} 
+ X_{42} X_{23} X_{34}  
+ X_{53} X_{37} X_{75}  
+ X_{67} X_{72} X_{26} 
\nn\\ 
&&
+ X_{78} X_{89} X_{97}  
+ X_{86} X_{64} X_{48}  
+ X_{94} X_{45} X_{59}  
- X_{15} X_{59} X_{91}  
- X_{29} X_{97} X_{72}  
- X_{31} X_{12} X_{23}  
\nn\\
&&
- X_{42} X_{26} X_{64}  
- X_{53} X_{34} X_{45}  
- X_{67} X_{75} X_{56}  
- X_{78} X_{83} X_{37}  
- X_{86} X_{61} X_{18}  
- X_{94} X_{48} X_{89} 
~.
\nn\\
  \eea
 
 \noindent The perfect matching matrix is 
\vspace{0.2cm}

\noindent\makebox[\textwidth]{%
\tiny
$
P=
\left(
\begin{array}{c|ccc|ccc|ccc|ccc|ccc|ccc|ccc|ccccccccccccccccccccc}
 \; & p_1 & p_2 & p_3 & q_1 & q_2 & q_3 & r_1 & r_2 & r_3 & u_1 & u_2 & u_3 & v_1 & v_2 & v_3 & w_1 & w_2 & w_3 & x_1 & x_2 & x_3 & s_1 & s_2 &s_3 &s_4 &s_5 &s_6 &s_7 &s_8 &s_9 &s_{10} &s_{11} &s_{12} &s_{13} &s_{14} &s_{15} &s_{16} &s_{17} &s_{18} &s_{19} &s_{20} &s_{21} \\
 \hline
  X_{89} & 1 & 0 & 0 & 1 & 1 & 0 & 1 & 0 & 0 & 0 & 0 & 0 & 1 & 0 & 1 & 1 & 0 & 0 &
   0 & 0 & 0 & 0 & 0 & 1 & 1 & 0 & 0 & 0 & 0 & 0 & 0 & 1 & 0 & 0 & 0 & 0 & 0 & 0 & 1 &
   1 & 1 & 1 \\
 X_{37} & 1 & 0 & 0 & 1 & 1 & 0 & 0 & 1 & 0 & 0 & 0 & 0 & 0 & 1 & 1 & 1 & 0 & 0 &
   0 & 0 & 0 & 0 & 0 & 0 & 0 & 0 & 0 & 0 & 0 & 0 & 1 & 1 & 1 & 1 & 1 & 1 & 0 & 0 & 1 &
   0 & 0 & 0 \\
 X_{45} & 1 & 0 & 0 & 1 & 1 & 0 & 0 & 0 & 1 & 0 & 0 & 0 & 1 & 1 & 0 & 1 & 0 & 0 &
   0 & 0 & 0 & 0 & 0 & 1 & 1 & 1 & 1 & 1 & 0 & 0 & 1 & 0 & 1 & 0 & 0 & 0 & 0 & 0 & 0 &
   0 & 0 & 0 \\
 X_{64} & 1 & 0 & 0 & 1 & 0 & 1 & 1 & 0 & 0 & 0 & 0 & 0 & 1 & 0 & 1 & 0 & 1 & 0 &
   0 & 0 & 0 & 0 & 0 & 0 & 0 & 0 & 0 & 0 & 1 & 1 & 0 & 1 & 0 & 1 & 1 & 0 & 1 & 0 & 0 &
   1 & 0 & 0 \\
 X_{18} & 1 & 0 & 0 & 1 & 0 & 1 & 0 & 1 & 0 & 0 & 0 & 0 & 0 & 1 & 1 & 0 & 1 & 0 &
   0 & 0 & 0 & 1 & 1 & 0 & 0 & 1 & 1 & 0 & 1 & 0 & 1 & 0 & 0 & 1 & 0 & 0 & 0 & 0 & 0 &
   0 & 0 & 0 \\
 X_{23} & 1 & 0 & 0 & 1 & 0 & 1 & 0 & 0 & 1 & 0 & 0 & 0 & 1 & 1 & 0 & 0 & 1 & 0 &
   0 & 0 & 0 & 1 & 0 & 1 & 0 & 1 & 0 & 0 & 0 & 0 & 0 & 0 & 0 & 0 & 0 & 0 & 1 & 1 & 0 &
   1 & 1 & 0 \\
 X_{72} & 1 & 0 & 0 & 0 & 1 & 1 & 1 & 0 & 0 & 0 & 0 & 0 & 1 & 0 & 1 & 0 & 0 & 1 &
   0 & 0 & 0 & 0 & 1 & 0 & 1 & 0 & 1 & 1 & 1 & 1 & 0 & 0 & 0 & 0 & 0 & 0 & 0 & 0 & 0 &
   0 & 0 & 1 \\
 X_{56} & 1 & 0 & 0 & 0 & 1 & 1 & 0 & 1 & 0 & 0 & 0 & 0 & 0 & 1 & 1 & 0 & 0 & 1 &
   0 & 0 & 0 & 1 & 1 & 0 & 0 & 0 & 0 & 0 & 0 & 0 & 0 & 0 & 0 & 0 & 0 & 1 & 0 & 1 & 1 &
   0 & 1 & 1 \\
 X_{91} & 1 & 0 & 0 & 0 & 1 & 1 & 0 & 0 & 1 & 0 & 0 & 0 & 1 & 1 & 0 & 0 & 0 & 1 &
   0 & 0 & 0 & 0 & 0 & 0 & 0 & 0 & 0 & 1 & 0 & 1 & 0 & 0 & 1 & 0 & 1 & 1 & 1 & 1 & 0 &
   0 & 0 & 0 \\
 X_{29} & 0 & 1 & 0 & 1 & 0 & 0 & 0 & 0 & 0 & 1 & 0 & 0 & 0 & 0 & 0 & 1 & 1 & 0 &
   1 & 0 & 1 & 1 & 0 & 1 & 0 & 1 & 0 & 0 & 0 & 0 & 0 & 1 & 0 & 0 & 0 & 0 & 0 & 0 & 1 &
   1 & 1 & 0 \\
 X_{67} & 0 & 1 & 0 & 1 & 0 & 0 & 0 & 0 & 0 & 0 & 1 & 0 & 0 & 0 & 0 & 1 & 1 & 0 &
   0 & 1 & 1 & 0 & 0 & 0 & 0 & 0 & 0 & 0 & 0 & 0 & 1 & 1 & 1 & 1 & 1 & 0 & 1 & 0 & 0 &
   1 & 0 & 0 \\
 X_{15} & 0 & 1 & 0 & 1 & 0 & 0 & 0 & 0 & 0 & 0 & 0 & 1 & 0 & 0 & 0 & 1 & 1 & 0 &
   1 & 1 & 0 & 0 & 0 & 1 & 1 & 1 & 1 & 0 & 1 & 0 & 1 & 0 & 0 & 1 & 0 & 0 & 0 & 0 & 0 &
   0 & 0 & 0 \\
 X_{31} & 0 & 1 & 0 & 0 & 1 & 0 & 0 & 0 & 0 & 1 & 0 & 0 & 0 & 0 & 0 & 1 & 0 & 1 &
   1 & 0 & 1 & 0 & 0 & 0 & 0 & 0 & 0 & 1 & 0 & 1 & 0 & 1 & 1 & 0 & 1 & 1 & 0 & 0 & 1 &
   0 & 0 & 0 \\
 X_{42} & 0 & 1 & 0 & 0 & 1 & 0 & 0 & 0 & 0 & 0 & 1 & 0 & 0 & 0 & 0 & 1 & 0 & 1 &
   0 & 1 & 1 & 0 & 1 & 0 & 1 & 0 & 1 & 1 & 0 & 0 & 1 & 0 & 1 & 0 & 0 & 0 & 0 & 0 & 0 &
   0 & 0 & 1 \\
 X_{86} & 0 & 1 & 0 & 0 & 1 & 0 & 0 & 0 & 0 & 0 & 0 & 1 & 0 & 0 & 0 & 1 & 0 & 1 &
   1 & 1 & 0 & 0 & 0 & 1 & 1 & 0 & 0 & 0 & 0 & 0 & 0 & 0 & 0 & 0 & 0 & 1 & 0 & 1 & 1 &
   0 & 1 & 1 \\
 X_{78} & 0 & 1 & 0 & 0 & 0 & 1 & 0 & 0 & 0 & 1 & 0 & 0 & 0 & 0 & 0 & 0 & 1 & 1 &
   1 & 0 & 1 & 1 & 1 & 0 & 0 & 1 & 1 & 1 & 1 & 1 & 0 & 0 & 0 & 0 & 0 & 0 & 0 & 0 & 0 &
   0 & 0 & 0 \\
 X_{53} & 0 & 1 & 0 & 0 & 0 & 1 & 0 & 0 & 0 & 0 & 1 & 0 & 0 & 0 & 0 & 0 & 1 & 1 &
   0 & 1 & 1 & 1 & 1 & 0 & 0 & 0 & 0 & 0 & 0 & 0 & 0 & 0 & 0 & 0 & 0 & 0 & 1 & 1 & 0 &
   1 & 1 & 1 \\
 X_{94} & 0 & 1 & 0 & 0 & 0 & 1 & 0 & 0 & 0 & 0 & 0 & 1 & 0 & 0 & 0 & 0 & 1 & 1 &
   1 & 1 & 0 & 0 & 0 & 0 & 0 & 0 & 0 & 0 & 1 & 1 & 0 & 0 & 0 & 1 & 1 & 1 & 1 & 1 & 0 &
   0 & 0 & 0 \\
 X_{59} & 0 & 0 & 1 & 0 & 0 & 0 & 1 & 1 & 0 & 1 & 1 & 0 & 0 & 0 & 1 & 0 & 0 & 0 &
   0 & 0 & 1 & 1 & 1 & 0 & 0 & 0 & 0 & 0 & 0 & 0 & 0 & 1 & 0 & 0 & 0 & 0 & 0 & 0 & 1 &
   1 & 1 & 1 \\
 X_{34} & 0 & 0 & 1 & 0 & 0 & 0 & 1 & 1 & 0 & 1 & 0 & 1 & 0 & 0 & 1 & 0 & 0 & 0 &
   1 & 0 & 0 & 0 & 0 & 0 & 0 & 0 & 0 & 0 & 1 & 1 & 0 & 1 & 0 & 1 & 1 & 1 & 0 & 0 & 1 &
   0 & 0 & 0 \\
 X_{12} & 0 & 0 & 1 & 0 & 0 & 0 & 1 & 1 & 0 & 0 & 1 & 1 & 0 & 0 & 1 & 0 & 0 & 0 &
   0 & 1 & 0 & 0 & 1 & 0 & 1 & 0 & 1 & 0 & 1 & 0 & 1 & 0 & 0 & 1 & 0 & 0 & 0 & 0 & 0 &
   0 & 0 & 1 \\
 X_{61} & 0 & 0 & 1 & 0 & 0 & 0 & 1 & 0 & 1 & 1 & 1 & 0 & 1 & 0 & 0 & 0 & 0 & 0 &
   0 & 0 & 1 & 0 & 0 & 0 & 0 & 0 & 0 & 1 & 0 & 1 & 0 & 1 & 1 & 0 & 1 & 0 & 1 & 0 & 0 &
   1 & 0 & 0 \\
 X_{75} & 0 & 0 & 1 & 0 & 0 & 0 & 1 & 0 & 1 & 1 & 0 & 1 & 1 & 0 & 0 & 0 & 0 & 0 &
   1 & 0 & 0 & 0 & 0 & 1 & 1 & 1 & 1 & 1 & 1 & 1 & 0 & 0 & 0 & 0 & 0 & 0 & 0 & 0 & 0 &
   0 & 0 & 0 \\
 X_{83} & 0 & 0 & 1 & 0 & 0 & 0 & 1 & 0 & 1 & 0 & 1 & 1 & 1 & 0 & 0 & 0 & 0 & 0 &
   0 & 1 & 0 & 0 & 0 & 1 & 1 & 0 & 0 & 0 & 0 & 0 & 0 & 0 & 0 & 0 & 0 & 0 & 1 & 1 & 0 &
   1 & 1 & 1 \\
 X_{48} & 0 & 0 & 1 & 0 & 0 & 0 & 0 & 1 & 1 & 1 & 1 & 0 & 0 & 1 & 0 & 0 & 0 & 0 &
   0 & 0 & 1 & 1 & 1 & 0 & 0 & 1 & 1 & 1 & 0 & 0 & 1 & 0 & 1 & 0 & 0 & 0 & 0 & 0 & 0 &
   0 & 0 & 0 \\
 X_{26} & 0 & 0 & 1 & 0 & 0 & 0 & 0 & 1 & 1 & 1 & 0 & 1 & 0 & 1 & 0 & 0 & 0 & 0 &
   1 & 0 & 0 & 1 & 0 & 1 & 0 & 1 & 0 & 0 & 0 & 0 & 0 & 0 & 0 & 0 & 0 & 1 & 0 & 1 & 1 &
   0 & 1 & 0 \\
 X_{97} & 0 & 0 & 1 & 0 & 0 & 0 & 0 & 1 & 1 & 0 & 1 & 1 & 0 & 1 & 0 & 0 & 0 & 0 &
   0 & 1 & 0 & 0 & 0 & 0 & 0 & 0 & 0 & 0 & 0 & 0 & 1 & 0 & 1 & 1 & 1 & 1 & 1 & 1 & 0 &
   0 & 0 & 0
\end{array}
\right)
$
}
\vspace{1cm}

 \noindent The F-term charge matrix $Q_F=\ker{(P)}$ is

\noindent\makebox[\textwidth]{%
\tiny
$
Q_F=
\left(
\begin{array}{ccc|ccc|ccc|ccc|ccc|ccc|ccc|ccccccccccccccccccccc}
 p_1 & p_2 & p_3 & q_1 & q_2 & q_3 & r_1 & r_2 & r_3 & u_1 & u_2 & u_3 & v_1 & v_2 & v_3 & w_1 & w_2 & w_3 & x_1 & x_2 & x_3 & s_1 & s_2 &s_3 &s_4 &s_5 &s_6 &s_7 &s_8 &s_9 &s_{10} &s_{11} &s_{12} &s_{13} &s_{14} &s_{15} &s_{16} &s_{17} &s_{18} &s_{19} &s_{20} &s_{21} \\
 \hline
 1 & 1 & 0 & 0 & -1 & 0 & 0 & 0 & 0 & 0 & 0 & 0 & 0 & 0 & 0 & 0 & -1 & 0 & 0 & 0 & 0 & 0 & 0 & 0 & 0 & 0 & 0 & 0 & 0 & 0 & 0 & 0 & 0 & 0 & 0 & 0 & 0 & 0 & 0 & 0
   & 0 & 0 \\
 1 & 1 & 0 & -1 & 0 & 0 & 0 & 0 & 0 & 0 & 0 & 0 & 0 & 0 & 0 & 0 & 0 & -1 & 0 & 0 & 0 & 0 & 0 & 0 & 0 & 0 & 0 & 0 & 0 & 0 & 0 & 0 & 0 & 0 & 0 & 0 & 0 & 0 & 0 & 0
   & 0 & 0 \\
 1 & 0 & 1 & 0 & 0 & 0 & 0 & -1 & 0 & 0 & 0 & 0 & -1 & 0 & 0 & 0 & 0 & 0 & 0 & 0 & 0 & 0 & 0 & 0 & 0 & 0 & 0 & 0 & 0 & 0 & 0 & 0 & 0 & 0 & 0 & 0 & 0 & 0 & 0 & 0
   & 0 & 0 \\
 1 & 0 & 1 & 0 & 0 & 0 & -1 & 0 & 0 & 0 & 0 & 0 & 0 & -1 & 0 & 0 & 0 & 0 & 0 & 0 & 0 & 0 & 0 & 0 & 0 & 0 & 0 & 0 & 0 & 0 & 0 & 0 & 0 & 0 & 0 & 0 & 0 & 0 & 0 & 0
   & 0 & 0 \\
 1 & 0 & 0 & -1 & -1 & 0 & 0 & 0 & 0 & 0 & 0 & 0 & 0 & 0 & 0 & 1 & 0 & 0 & 0 & 0 & 0 & 0 & 0 & 0 & 0 & 0 & 0 & 0 & 0 & 0 & 0 & 0 & 0 & 0 & 0 & 0 & 0 & 0 & 0 & 0
   & 0 & 0 \\
 1 & 0 & 0 & -1 & 0 & -1 & 0 & 0 & 0 & 0 & 0 & 0 & 0 & 0 & 0 & 0 & 1 & 0 & 0 & 0 & 0 & 0 & 0 & 0 & 0 & 0 & 0 & 0 & 0 & 0 & 0 & 0 & 0 & 0 & 0 & 0 & 0 & 0 & 0 & 0
   & 0 & 0 \\
 1 & 0 & 0 & -1 & 0 & 0 & 0 & -1 & 0 & 1 & 0 & 0 & 0 & 0 & 0 & 0 & 0 & 0 & 0 & 0 & 0 & 0 & 0 & 0 & 0 & 0 & 0 & 0 & 0 & 0 & 0 & 0 & 0 & 1 & 0 & 0 & 0 & 0 & 0 & 0
   & -1 & 0 \\
 1 & 0 & 0 & 0 & -1 & 0 & -1 & 0 & 0 & 1 & 0 & 0 & 0 & 0 & 0 & 0 & 0 & 0 & 0 & 0 & 0 & 0 & 0 & 0 & 1 & 0 & 0 & 0 & 0 & 0 & 0 & 0 & 0 & 0 & 0 & 0 & 0 & 0 & -1 &
   0 & 0 & 0 \\
 1 & 0 & 0 & 0 & 0 & 0 & -1 & 0 & 0 & 1 & 1 & 0 & 0 & 0 & 0 & 0 & 0 & 0 & 0 & 0 & 0 & -1 & 0 & 0 & 0 & 0 & 0 & 0 & 0 & 0 & 0 & 0 & 0 & 0 & -1 & 0 & 0 & 0 & 0 &
   0 & 0 & 0 \\
 0 & 1 & 1 & 0 & 0 & 0 & 0 & 0 & 0 & 0 & -1 & 0 & 0 & 0 & 0 & 0 & 0 & 0 & -1 & 0 & 0 & 0 & 0 & 0 & 0 & 0 & 0 & 0 & 0 & 0 & 0 & 0 & 0 & 0 & 0 & 0 & 0 & 0 & 0 & 0
   & 0 & 0 \\
 0 & 1 & 1 & 0 & 0 & 0 & 0 & 0 & 0 & -1 & 0 & 0 & 0 & 0 & 0 & 0 & 0 & 0 & 0 & -1 & 0 & 0 & 0 & 0 & 0 & 0 & 0 & 0 & 0 & 0 & 0 & 0 & 0 & 0 & 0 & 0 & 0 & 0 & 0 & 0
   & 0 & 0 \\
 0 & 0 & 1 & 1 & 0 & 0 & -1 & 0 & 0 & 0 & 0 & 0 & 0 & 0 & 0 & 0 & 0 & 0 & 0 & 0 & 0 & -1 & 1 & 0 & 0 & 0 & 0 & 0 & 0 & 0 & 0 & 0 & 0 & -1 & 0 & 0 & 0 & 0 & 0 &
   0 & 0 & 0 \\
 0 & 0 & 1 & 0 & 0 & 0 & -1 & -1 & 0 & 0 & 0 & 0 & 0 & 0 & 1 & 0 & 0 & 0 & 0 & 0 & 0 & 0 & 0 & 0 & 0 & 0 & 0 & 0 & 0 & 0 & 0 & 0 & 0 & 0 & 0 & 0 & 0 & 0 & 0 & 0
   & 0 & 0 \\
 0 & 0 & 1 & 0 & 0 & 0 & -1 & 0 & -1 & 0 & 0 & 0 & 1 & 0 & 0 & 0 & 0 & 0 & 0 & 0 & 0 & 0 & 0 & 0 & 0 & 0 & 0 & 0 & 0 & 0 & 0 & 0 & 0 & 0 & 0 & 0 & 0 & 0 & 0 & 0
   & 0 & 0 \\
 0 & 0 & 1 & 0 & 0 & 0 & 0 & 0 & 0 & -1 & -1 & 0 & 0 & 0 & 0 & 0 & 0 & 0 & 0 & 0 & 1 & 0 & 0 & 0 & 0 & 0 & 0 & 0 & 0 & 0 & 0 & 0 & 0 & 0 & 0 & 0 & 0 & 0 & 0 & 0
   & 0 & 0 \\
 0 & 0 & 1 & 0 & 0 & 0 & 0 & 0 & 0 & -1 & 0 & -1 & 0 & 0 & 0 & 0 & 0 & 0 & 1 & 0 & 0 & 0 & 0 & 0 & 0 & 0 & 0 & 0 & 0 & 0 & 0 & 0 & 0 & 0 & 0 & 0 & 0 & 0 & 0 & 0
   & 0 & 0 \\
 0 & 0 & 1 & 0 & 0 & 0 & -1 & 0 & 0 & 0 & -1 & 0 & 0 & 0 & 0 & 0 & 0 & 0 & 0 & 0 & 0 & -1 & 1 & 0 & 0 & 1 & 0 & 0 & 0 & 0 & 0 & 0 & 0 & 0 & 0 & 0 & 0 & 0 & 0 &
   0 & 0 & 0 \\
 0 & 0 & 1 & 0 & 0 & 0 & 0 & -1 & 0 & 0 & 0 & 0 & 0 & 0 & 0 & 1 & 0 & 0 & 0 & 0 & -1 & 1 & 0 & -1 & 0 & 0 & 0 & 0 & 0 & 0 & 0 & 0 & 0 & 0 & 0 & 0 & 0 & 0 & 0 &
   0 & 0 & 0 \\
 0 & 0 & 1 & 0 & 0 & 0 & 0 & -1 & 0 & -1 & 0 & 0 & 0 & 0 & 0 & 0 & 0 & 0 & 0 & 0 & 0 & 1 & 0 & 0 & 0 & 0 & 0 & 0 & 1 & 0 & 0 & 0 & 0 & 0 & 0 & 0 & 0 & 0 & 0 & 0
   & 0 & -1 \\
 0 & 0 & 1 & 0 & 0 & 0 & 0 & -1 & 0 & 0 & -1 & 0 & 0 & 0 & 0 & 0 & 0 & 0 & 0 & 0 & 0 & 0 & 1 & 0 & 0 & 0 & 0 & 0 & 0 & 0 & 1 & 0 & 0 & 0 & 0 & -1 & 0 & 0 & 0 &
   0 & 0 & 0 \\
 0 & 0 & 0 & 0 & 0 & 1 & 0 & 0 & 0 & 0 & 1 & 0 & 0 & 0 & 0 & 0 & 0 & 0 & 0 & 0 & 0 & 0 & -1 & 0 & 0 & 0 & 0 & 0 & 0 & 0 & 0 & 0 & 0 & 0 & 0 & 0 & -1 & 0 & 0 & 0
   & 0 & 0 \\
 0 & 0 & 0 & 0 & 0 & 1 & 0 & 0 & 0 & 1 & 0 & 0 & 0 & 0 & 0 & 0 & 0 & 0 & 0 & 0 & 0 & -1 & 0 & 0 & 0 & 0 & 0 & 0 & 0 & 0 & 0 & 0 & 0 & 0 & 0 & 0 & 0 & -1 & 0 & 0
   & 0 & 0 \\
 0 & 0 & 0 & 0 & 0 & 0 & 0 & 1 & 0 & 0 & 0 & 0 & 0 & 0 & 0 & 0 & 1 & 0 & 0 & 0 & 0 & -1 & 0 & 0 & 0 & 0 & 0 & 0 & 0 & 0 & -1 & 0 & 0 & 0 & 0 & 0 & 0 & 0 & 0 & 0
   & 0 & 0 \\
 0 & 0 & 0 & 0 & 0 & 0 & 0 & 1 & 0 & 0 & 0 & 0 & 0 & 0 & 0 & 0 & 0 & 1 & 0 & 0 & 0 & 0 & -1 & 0 & 0 & 0 & 0 & 0 & -1 & 0 & 0 & 0 & 0 & 0 & 0 & 0 & 0 & 0 & 0 & 0
   & 0 & 0 \\
 0 & 0 & 0 & 0 & 0 & 0 & 0 & 0 & 0 & 0 & 0 & 0 & 0 & 0 & 0 & 0 & 0 & 0 & 0 & 0 & 0 & 1 & -1 & -1 & 1 & 0 & 0 & 0 & 0 & 0 & 0 & 0 & 0 & 0 & 0 & 0 & 0 & 0 & 0 & 0
   & 0 & 0 \\
 0 & 0 & 0 & 0 & 0 & 0 & 0 & 0 & 0 & 0 & 0 & 0 & 0 & 0 & 0 & 0 & 0 & 0 & 0 & 0 & 0 & 1 & -1 & 0 & 0 & 0 & -1 & 1 & 0 & 0 & 0 & 0 & 0 & 0 & 0 & 0 & 0 & 0 & 0 & 0
   & 0 & 0 \\
 0 & 0 & 0 & 0 & 0 & 0 & 0 & 0 & 0 & 0 & 0 & 0 & 0 & 0 & 0 & 0 & 0 & 0 & 0 & 0 & 0 & 1 & -1 & 0 & 0 & 0 & 0 & 0 & 0 & 0 & 0 & 0 & 0 & 0 & 0 & 0 & 0 & 0 & -1 & 1
   & 0 & 0 \\
 0 & 0 & 0 & 0 & 0 & 0 & 0 & 0 & 0 & 0 & 0 & 0 & 0 & 0 & 0 & 0 & 0 & 0 & 0 & 0 & 0 & 1 & 0 & 0 & 1 & 0 & -1 & 0 & 0 & 0 & 0 & 0 & 0 & 0 & 0 & 0 & 0 & 0 & 0 & -1
   & 0 & 0 \\
 0 & 0 & 0 & 0 & 0 & 0 & 0 & 0 & 0 & 0 & 0 & 0 & 0 & 0 & 0 & 0 & 0 & 0 & 0 & 0 & 0 & 0 & 0 & 0 & 0 & 1 & -1 & 0 & 0 & -1 & 0 & 1 & 0 & 0 & 0 & 0 & 0 & 0 & 0 & 0
   & 0 & 0 \\
 0 & 0 & 0 & 0 & 0 & 1 & 0 & 0 & 0 & 0 & 1 & 0 & 0 & 0 & 0 & 0 & 0 & 0 & 0 & 0 & 0 & 0 & -1 & 0 & 0 & -1 & 0 & 0 & 0 & 1 & 0 & 0 & -1 & 0 & 0 & 0 & 0 & 0 & 0 &
   0 & 0 & 0 \\
 0 & 0 & 0 & 0 & 0 & 0 & 0 & 0 & 0 & 1 & 0 & 0 & 0 & 0 & 0 & 1 & 0 & 0 & 0 & 0 & -1 & 0 & 0 & -1 & 0 & 0 & 1 & 0 & 0 & 0 & 0 & -1 & 0 & 0 & 0 & 0 & 0 & 0 & 0 &
   0 & 0 & 0
\end{array}
\right)
$
}
\vspace{0.5cm}

\noindent The D-term charge matrix is

\noindent\makebox[\textwidth]{%
\tiny
$
Q_D=
\left(
\begin{array}{ccc|ccc|ccc|ccc|ccc|ccc|ccc|ccccccccccccccccccccc}
 p_1 & p_2 & p_3 & q_1 & q_2 & q_3 & r_1 & r_2 & r_3 & u_1 & u_2 & u_3 & v_1 & v_2 & v_3 & w_1 & w_2 & w_3 & x_1 & x_2 & x_3 & s_1 & s_2 &s_3 &s_4 &s_5 &s_6 &s_7 &s_8 &s_9 &s_{10} &s_{11} &s_{12} &s_{13} &s_{14} &s_{15} &s_{16} &s_{17} &s_{18} &s_{19} &s_{20} &s_{21} \\
 \hline
 0 & 0 & 0 & 0 & 0 & 0 & 0 & 0 & 0 & 0 & 0 & 0 & 0 & 0 & 0 & 0 & 0 & 0 & 0 & 0 & 0 & 1
   & -1 & 0 & 0 & 0 & 0 & 0 & 0 & 0 & 0 & 0 & 0 & 0 & 0 & 0 & 0 & 0 & 0 & 0 & 0 & 0 \\
 0 & 0 & 0 & 0 & 0 & 0 & 0 & 0 & 0 & 0 & 0 & 0 & 0 & 0 & 0 & 0 & 0 & 0 & 0 & 0 & 0 & 0
   & 1 & -1 & 0 & 0 & 0 & 0 & 0 & 0 & 0 & 0 & 0 & 0 & 0 & 0 & 0 & 0 & 0 & 0 & 0 & 0 \\
 0 & 0 & 0 & 0 & 0 & 0 & 0 & 0 & 0 & 0 & 0 & 0 & 0 & 0 & 0 & 0 & 0 & 0 & 0 & 0 & 0 & 0
   & 0 & 0 & 1 & -1 & 0 & 0 & 0 & 0 & 0 & 0 & 0 & 0 & 0 & 0 & 0 & 0 & 0 & 0 & 0 & 0 \\
 0 & 0 & 0 & 0 & 0 & 0 & 0 & 0 & 0 & 0 & 0 & 0 & 0 & 0 & 0 & 0 & 0 & 0 & 0 & 0 & 0 & 0
   & 0 & 0 & 0 & 1 & -1 & 0 & 0 & 0 & 0 & 0 & 0 & 0 & 0 & 0 & 0 & 0 & 0 & 0 & 0 & 0 \\
 0 & 0 & 0 & 0 & 0 & 0 & 0 & 0 & 0 & 0 & 0 & 0 & 0 & 0 & 0 & 0 & 0 & 0 & 0 & 0 & 0 & 0
   & 0 & 0 & 0 & 0 & 0 & 1 & -1 & 0 & 0 & 0 & 0 & 0 & 0 & 0 & 0 & 0 & 0 & 0 & 0 & 0 \\
 0 & 0 & 0 & 0 & 0 & 0 & 0 & 0 & 0 & 0 & 0 & 0 & 0 & 0 & 0 & 0 & 0 & 0 & 0 & 0 & 0 & 0
   & 0 & 0 & 0 & 0 & 0 & 0 & 1 & -1 & 0 & 0 & 0 & 0 & 0 & 0 & 0 & 0 & 0 & 0 & 0 & 0 \\
 0 & 0 & 0 & 0 & 0 & 0 & 0 & 0 & 0 & 0 & 0 & 0 & 0 & 0 & 0 & 0 & 0 & 0 & 0 & 0 & 0 & 0
   & 0 & 0 & 0 & 0 & 0 & 0 & 0 & 0 & 1 & -1 & 0 & 0 & 0 & 0 & 0 & 0 & 0 & 0 & 0 & 0 \\
 0 & 0 & 0 & 0 & 0 & 0 & 0 & 0 & 0 & 0 & 0 & 0 & 0 & 0 & 0 & 0 & 0 & 0 & 0 & 0 & 0 & 0
   & 0 & 0 & 0 & 0 & 0 & 0 & 0 & 0 & 0 & 0 & 1 & -1 & 0 & 0 & 0 & 0 & 0 & 0 & 0 & 0
\end{array}
\right)
$
}
\vspace{0.5cm}

The total charge matrix $Q_t$ exhibits no repeated columns. Accordingly, the global symmetry group is $U(1)_{f_1} \times U(1)_{f_2} \times U(1)_R$. Following the discussion on flavour symmetry and R-charges in section \sref{s1_3}, the charges on GLSM fields with non-zero R-charges are chosen as shown in \tref{t1}.

\begin{table}[H]
\centering
\begin{tabular}{|c||c|c|c||l|} 
\hline
\; & $U(1)_{f_1}$ & $U(1)_{f_2}$ & $U(1)_R$ & fugacity \\
\hline
\hline
$p_1$ & 1/3 	&  0 	& $2/3$ &  	$t_1$\\
$p_2$ &-1/3 	& -1/3 	& $2/3$ &  	$t_2$\\
$p_3$ &   0 	&  1/3	& $2/3$ &  	$t_3$\\
\hline
\end{tabular}
\caption{The GLSM fields corresponding to extremal points of the toric diagram with their mesonic charges (Model 1).\label{t1}}
\end{table}

Products of non-extremal perfect matchings are labelled by a single variable as follows,
\beal{esm1_fug2}
&&
q = q_1 q_2 q_3 ~,~
r = r_1 r_2 r_3 ~,~
u = u_1 u_2 u_3 ~,~
v = v_1 v_2 v_3 ~,~
\nn\\
&&
w = w_1 w_2 w_3 ~,~
x = x_1 x_2 x_3 ~,~
s = \prod_{m=1}^{21} s_m ~~.
\eea
The fugacities $t_\alpha$ count extremal perfect matchings corresponding to GLSM fields with non-zero R-charge. The fugacity of the form $y_q$ counts the product of non-extremal perfect matchings $q$ above.

The mesonic Hilbert series of Model 1 is calculated using the Molien integral formula in \eref{es12_2}. It is
 \beal{esm1_1}
&&g_{1}(t_\alpha,y_{q},y_{r},y_{u},y_{v},y_{w},y_{x},y_{s}; \mathcal{M}^{mes}_{1})=
\nn\\
&& \hspace{0.3cm}
\frac{
1 - y_{q}^3 y_{r}^3 y_{u}^3 y_{v}^3 y_{w}^3 y_{x}^3 y_{s}^3 ~ t_1^3 t_2^3 t_3^3 
}{
(1 - y_{q}^2 y_{r} y_{v}^2 y_{w} y_{s} ~ t_{1}^{3}) 
(1 - y_{q} y_{u} y_{w}^2 y_{x}^2 y_{s} ~ t_{2}^{3}) 
(1 - y_{r}^2 y_{u}^2 y_{v} y_{x} y_{s} ~ t_{3}^{3}) 
(1 - y_{q} y_{r} y_{u} y_{v} y_{w} y_{x} y_{s} ~ t_1 t_2 t_3)
}
~~.
\nn\\
\eea
 The plethystic logarithm of the mesonic Hilbert series is
\beal{esm1_3}
&&
PL[g_1(t_\alpha,y_{q},y_{r},y_{u},y_{v},y_{w},y_{x},y_{s};\mathcal{M}_{1}^{mes})]=
y_{q} y_{r} y_{u} y_{v} y_{w} y_{x} y_{s} ~ t_1 t_2 t_3 
+ y_{q}^2 y_{r} y_{v}^2 y_{w} y_{s} ~t_1^3 
\nn\\
&& \hspace{1cm}
+ y_{r}^2 y_{u}^2 y_{v} y_{x} y_{s} ~ t_3^3 
+ y_{q} y_{u} y_{w}^2 y_{x}^2 y_{s} ~ t_2^3 
- y_{q}^3 y_{r}^3 y_{u}^3 y_{v}^3 y_{w}^3 y_{x}^3 y_{s}^3 ~ t_1^3 t_2^3 t_3^3 ~~.
\eea
The finite plethystic logarithm indicates that the mesonic moduli space is a complete intersection.

In terms of the fugacity map 
\beal{esm1_y1}
f_1 = \frac{y_q y_v ~ t_1^2}{y_u y_x ~ t_2 t_3}~,~
f_2 = \frac{y_r y_u ~t_3^2}{y_q y_w ~ t_1 t_2}~,~
t = y_q^{1/3} y_r^{1/3} y_u^{1/3} y_v^{1/3} y_w^{1/3} y_x^{1/3} y_s^{1/3} ~ t_1^{1/3} t_2^{1/3} t_3^{1/3}~~,
\eea
where $f_1$, $f_2$ and $t$ are the fugacities counting the mesonic charges, 
the above plethystic logarithm becomes
\beal{esm3b_3}
PL[g_1(t,f_1,f_2;\mathcal{M}_{1}^{mes})]
&=&
\left(1 + f_1 + f_2 + \frac{1}{f_1 f_2}\right) t^3 - t^9
   \eea
The above plethystics logarithm identifies both the moduli space generators and the mesonic charges carried by them. The generators and the corresponding mesonic charges are summarized in \tref{t1gen}. The generators can be presented on a charge lattice. It is a convex polygon as shown in \tref{t1gen} and is the dual reflexive polygon of the toric diagram of Model 16.

The relation formed among the generators is as follows,
\beal{esm1_3b}
A_1 A_2 A_3 = B^3~~.
\eea
\\

\begin{table}[H]
\resizebox{.95\hsize}{!}{
\begin{minipage}[!b]{0.5\textwidth}
\begin{tabular}{|l|c|c|}
\hline
Generator & $U(1)_{f_1}$ & $U(1)_{f_2}$ 
\\
\hline
\hline
$
A_1=
p_1^3 ~
q^2~
r ~
v^2 ~ 
w ~
s
$
& 1 & 0
\nn\\
$
A_2=
p_2^3 ~
q ~ 
u ~ 
w^2 ~
x^2 ~
s
$
& -1 & -1
\nn\\
$
A_3=
p_3^3 ~ 
r^2 ~
u^2~
v ~ 
x ~
s
$
& 0 & 1
\nn\\
$
B=
p_1 p_2 p_3 ~
q ~
r ~
u ~
v ~
w ~
x ~
s
$
& 0 & 0
\nn\\
\hline
\end{tabular}
\end{minipage}
\hspace{3cm}
\begin{minipage}[!b]{0.3\textwidth}
\includegraphics[width=3.5 cm]{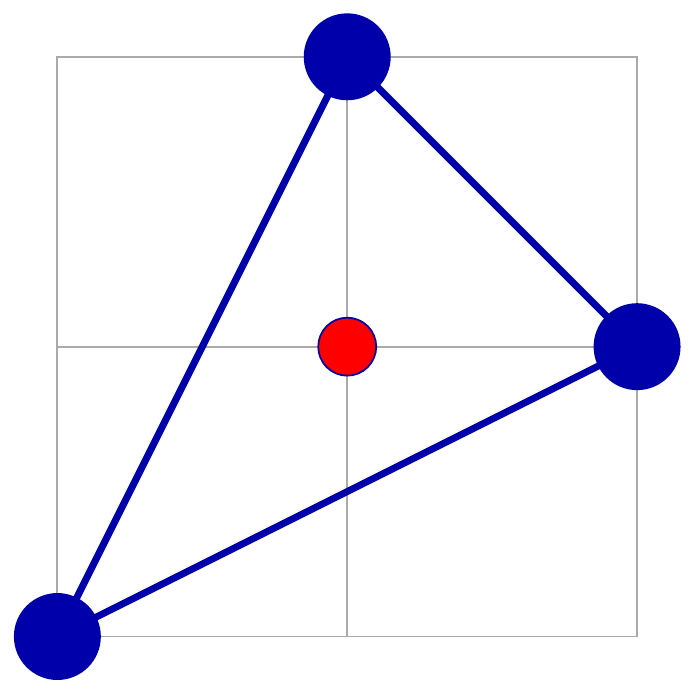}
\end{minipage}
}

\caption{The generators and lattice of generators of the mesonic moduli space of Model 1 in terms of GLSM fields with the corresponding flavor charges.\label{t1gen}} 
\end{table}

\begin{table}[H]
\centering

\resizebox{1\hsize}{!}{

\begin{tabular}{|l|c|c|}
\hline
Generator & $U(1)_{f_1}$ & $U(1)_{f_2}$ 
\\
\hline
\hline
$
X_{18} X_{89} X_{91}=  X_{23} X_{37} X_{72}=  X_{45} X_{56} X_{64}
$
& 1 & 0
\nn\\
$
X_{15} X_{53} X_{31}=  X_{29} X_{94} X_{42}=  X_{67} X_{78} X_{86}
$
& -1 & -1
\nn\\
$
X_{12} X_{26} X_{61}=  X_{34} X_{48} X_{83}=  X_{59} X_{97} X_{75}
$
& 0 & 1
\nn\\
$
X_{12} X_{23} X_{31}=  X_{12} X_{29} X_{91}=  X_{15} X_{56} X_{61}=  X_{15} X_{59} X_{91}=  X_{18} X_{83} X_{31}=  X_{18} X_{86} X_{61}=  X_{23} X_{34} X_{42}=  X_{26} X_{64} X_{42}=  X_{26} X_{67} X_{72}
$
& 0 & 0
\nn\\
$
=X_{29} X_{97} X_{72}=  X_{34} X_{45} X_{53}=  X_{37} X_{75} X_{53}=  X_{37} X_{78} X_{83}=  X_{45} X_{59} X_{94}=  X_{48} X_{86} X_{64}=  X_{48} X_{89} X_{94}=  X_{56} X_{67} X_{75}=  X_{78} X_{89} X_{97}
$
& &
\nn\\
\hline
\end{tabular}
}

\caption{The generators in terms of bifundamental fields (Model 1).\label{t1gen2}} 
\end{table}

With the following fugacity map
\beal{esm1_xx1}
&&
T_1 =
f_1^{1/3} ~ t 
=
y_q^{2/3} y_r^{1/3} y_v^{2/3} y_w^{1/3} y_s^{1/3} t_1
~,~
\nn\\
&&
T_2 =
f_1^{-1/3} f_2^{-1/3} ~t
=
y_q^{1/3} y_u^{1/3} y_w^{2/3} y_x^{2/3} y_s^{1/3} t_2 
~,~
\nn\\
&&
T_3 =
f_2^{1/3} ~t
=
y_r^{2/3} y_u^{2/3} y_v^{1/3} y_x^{1/3} y_s^{1/3} t_3
~,~
\eea
the mesonic Hilbert series becomes
\beal{esm1_xx2}
g_1(T_1,T_2,T_3;\mathcal{M}^{mes}_1) = 
\frac{
1-T_1^3 T_2^3 T_3^3
}{
(1-T_1^3)
(1-T_2^3)
(1-T_3^3)
(1-T_1 T_2 T_3)
}
\eea
with the plethystic logarithm being
\beal{esm1_xx3}
PL[g_1(T_1,T_2,T_3;\mathcal{M}^{mes}_1)]=
T_1 T_2 T_3 + T_1^3 + T_3^3 + T_2^3  - T_1^3 T_2^3 T_3^3
~~.
\eea
The above refinement of the Hilbert series exemplifies the conical structure of the toric Calabi-Yau space.
\\

\section{Model 2: $\mathbb{C}^3/\mathbb{Z}_4\times\mathbb{Z}_2~(1,0,3)(0,1,1)$ \label{sm2}}

\begin{figure}[H]
\begin{center}
\includegraphics[trim=0cm 0cm 0cm 0cm,height=4.5 cm]{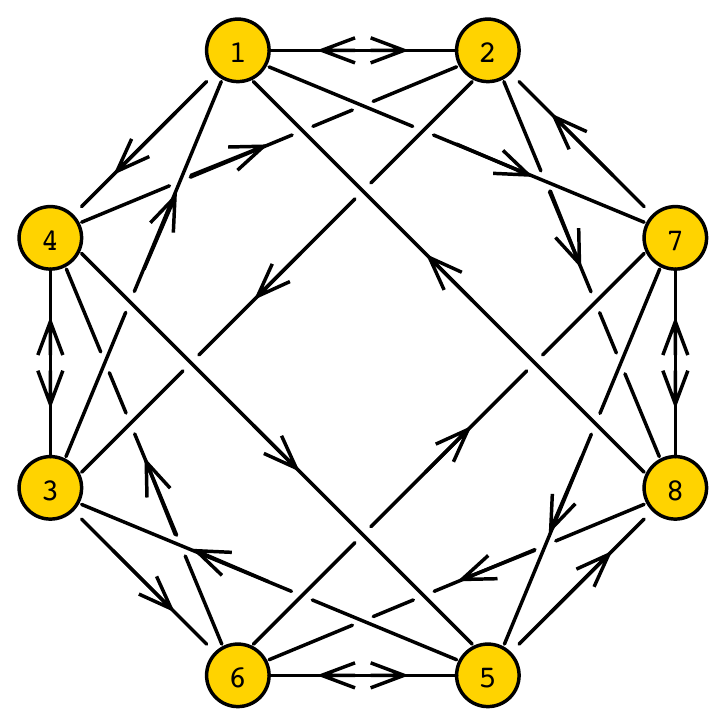}
\includegraphics[width=5 cm]{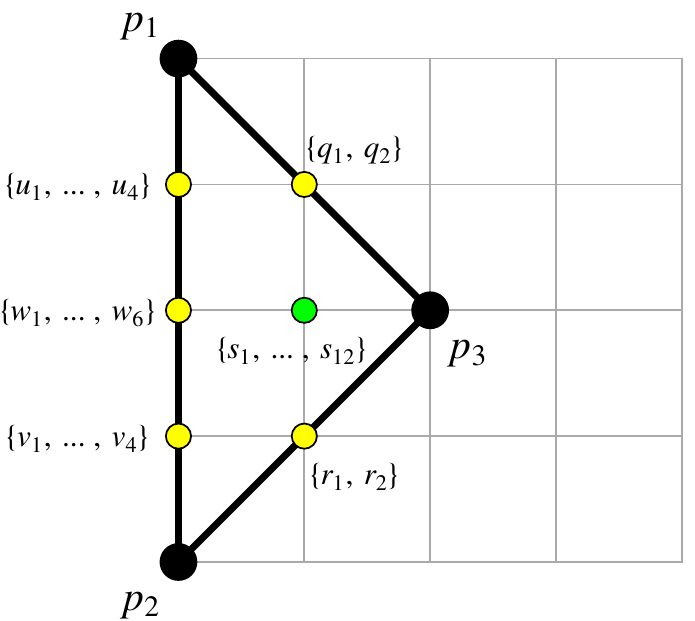}
\includegraphics[width=5 cm]{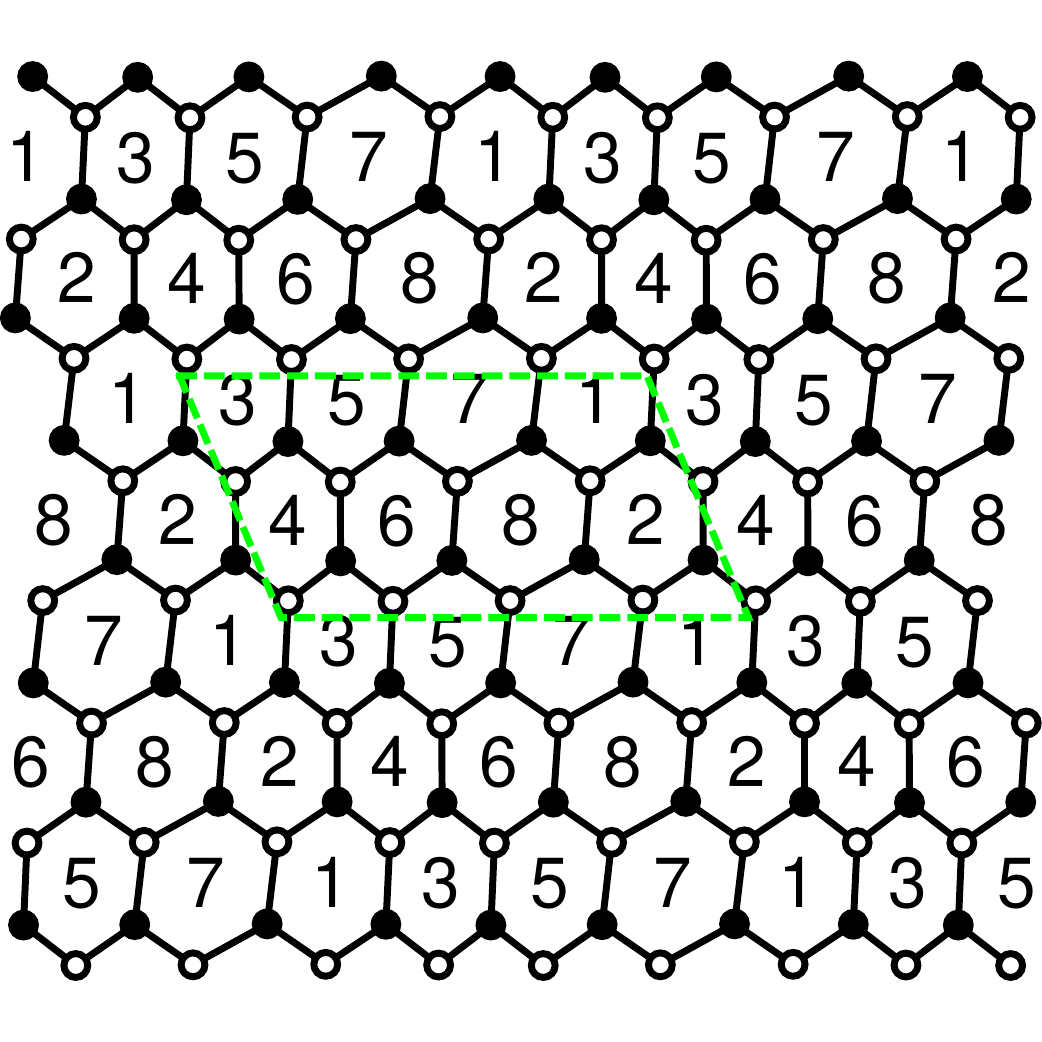}
\caption{The quiver, toric diagram, and brane tiling of Model 2.\label{f2}}
 \end{center}
 \end{figure}
 
 The superpotential is 
\beal{esm2_00}
W&=& 
+ X_{17} X_{72} X_{21}  
+ X_{28} X_{81} X_{12}
+ X_{31} X_{14} X_{43}  
+ X_{42} X_{23} X_{34} 
\nn\\
&& 
+ X_{53} X_{36} X_{65}  
+ X_{64} X_{45} X_{56}  
+ X_{75} X_{58} X_{87}  
+ X_{86} X_{67} X_{78} 
\nn\\
&&
- X_{17} X_{78} X_{81}  
- X_{28} X_{87} X_{72}  
- X_{31} X_{12} X_{23}  
- X_{42} X_{21} X_{14}  
\nn\\
&&
- X_{53} X_{34} X_{45}  
- X_{64} X_{43} X_{36}  
- X_{75} X_{56} X_{67}  
- X_{86} X_{65} X_{58}  
~.
  \eea
 
 \noindent The perfect matching matrix is 
 
\noindent\makebox[\textwidth]{%
\scriptsize
$
P=
\left(
\begin{array}{c|ccc|cc|cc|cccc|cccc|cccccc|cccccccccccc}
 \; & p_1& p_2& p_3& q_1& q_2& r_1& r_2& u_1& u_2& u_3& u_4& v_1& v_2& v_3& v_4& w_1& w_2& w_3& w_4& w_5& w_6& s_1& s_2& s_3& s_4& s_5& s_6& s_7& s_8& s_9& s_{10}& s_{11}& s_{12} \\
 \hline
 X_{67} & 1 & 0 & 0 & 1 & 0 & 0 & 0 & 1 & 1 & 1 & 0 & 1 & 0 & 0 & 0 & 1 & 0 & 1 &
   1 & 0 & 0 & 1 & 0 & 0 & 0 & 1 & 1 & 0 & 0 & 0 & 0 & 0 & 0 \\
 X_{45} & 1 & 0 & 0 & 1 & 0 & 0 & 0 & 1 & 1 & 0 & 1 & 0 & 1 & 0 & 0 & 1 & 0 & 0 &
   0 & 1 & 1 & 1 & 0 & 0 & 0 & 0 & 0 & 1 & 1 & 0 & 0 & 0 & 0 \\
 X_{58} & 1 & 0 & 0 & 0 & 1 & 0 & 0 & 1 & 1 & 1 & 0 & 1 & 0 & 0 & 0 & 1 & 0 & 1 &
   1 & 0 & 0 & 0 & 1 & 0 & 0 & 0 & 0 & 0 & 0 & 1 & 1 & 0 & 0 \\
 X_{36} & 1 & 0 & 0 & 0 & 1 & 0 & 0 & 1 & 1 & 0 & 1 & 0 & 1 & 0 & 0 & 1 & 0 & 0 &
   0 & 1 & 1 & 0 & 1 & 0 & 0 & 0 & 0 & 0 & 0 & 0 & 0 & 1 & 1 \\
 X_{23} & 1 & 0 & 0 & 1 & 0 & 0 & 0 & 1 & 0 & 1 & 1 & 0 & 0 & 1 & 0 & 0 & 1 & 1 &
   0 & 1 & 0 & 0 & 0 & 1 & 0 & 0 & 0 & 1 & 0 & 1 & 0 & 0 & 0 \\
 X_{81} & 1 & 0 & 0 & 1 & 0 & 0 & 0 & 0 & 1 & 1 & 1 & 0 & 0 & 0 & 1 & 0 & 1 & 0 &
   1 & 0 & 1 & 0 & 0 & 1 & 0 & 1 & 0 & 0 & 0 & 0 & 0 & 0 & 1 \\
 X_{14} & 1 & 0 & 0 & 0 & 1 & 0 & 0 & 1 & 0 & 1 & 1 & 0 & 0 & 1 & 0 & 0 & 1 & 1 &
   0 & 1 & 0 & 0 & 0 & 0 & 1 & 0 & 1 & 0 & 0 & 0 & 0 & 1 & 0 \\
 X_{72} & 1 & 0 & 0 & 0 & 1 & 0 & 0 & 0 & 1 & 1 & 1 & 0 & 0 & 0 & 1 & 0 & 1 & 0 &
   1 & 0 & 1 & 0 & 0 & 0 & 1 & 0 & 0 & 0 & 1 & 0 & 1 & 0 & 0 \\
 X_{28} & 0 & 1 & 0 & 0 & 0 & 1 & 0 & 1 & 0 & 0 & 0 & 1 & 1 & 1 & 0 & 1 & 0 & 1 &
   0 & 1 & 0 & 0 & 1 & 0 & 0 & 0 & 0 & 1 & 0 & 1 & 0 & 0 & 0 \\
 X_{31} & 0 & 1 & 0 & 0 & 0 & 1 & 0 & 0 & 1 & 0 & 0 & 1 & 1 & 0 & 1 & 1 & 0 & 0 &
   1 & 0 & 1 & 0 & 1 & 0 & 0 & 1 & 0 & 0 & 0 & 0 & 0 & 0 & 1 \\
 X_{17} & 0 & 1 & 0 & 0 & 0 & 0 & 1 & 1 & 0 & 0 & 0 & 1 & 1 & 1 & 0 & 1 & 0 & 1 &
   0 & 1 & 0 & 1 & 0 & 0 & 0 & 0 & 1 & 0 & 0 & 0 & 0 & 1 & 0 \\
 X_{42} & 0 & 1 & 0 & 0 & 0 & 0 & 1 & 0 & 1 & 0 & 0 & 1 & 1 & 0 & 1 & 1 & 0 & 0 &
   1 & 0 & 1 & 1 & 0 & 0 & 0 & 0 & 0 & 0 & 1 & 0 & 1 & 0 & 0 \\
 X_{64} & 0 & 1 & 0 & 0 & 0 & 1 & 0 & 0 & 0 & 1 & 0 & 1 & 0 & 1 & 1 & 0 & 1 & 1 &
   1 & 0 & 0 & 0 & 0 & 0 & 1 & 1 & 1 & 0 & 0 & 0 & 0 & 0 & 0 \\
 X_{75} & 0 & 1 & 0 & 0 & 0 & 1 & 0 & 0 & 0 & 0 & 1 & 0 & 1 & 1 & 1 & 0 & 1 & 0 &
   0 & 1 & 1 & 0 & 0 & 0 & 1 & 0 & 0 & 1 & 1 & 0 & 0 & 0 & 0 \\
 X_{53} & 0 & 1 & 0 & 0 & 0 & 0 & 1 & 0 & 0 & 1 & 0 & 1 & 0 & 1 & 1 & 0 & 1 & 1 &
   1 & 0 & 0 & 0 & 0 & 1 & 0 & 0 & 0 & 0 & 0 & 1 & 1 & 0 & 0 \\
 X_{86} & 0 & 1 & 0 & 0 & 0 & 0 & 1 & 0 & 0 & 0 & 1 & 0 & 1 & 1 & 1 & 0 & 1 & 0 &
   0 & 1 & 1 & 0 & 0 & 1 & 0 & 0 & 0 & 0 & 0 & 0 & 0 & 1 & 1 \\
 X_{65} & 0 & 0 & 1 & 1 & 0 & 1 & 0 & 0 & 0 & 0 & 0 & 0 & 0 & 0 & 0 & 0 & 0 & 0 &
   0 & 0 & 0 & 1 & 0 & 0 & 1 & 1 & 1 & 1 & 1 & 0 & 0 & 0 & 0 \\
 X_{21} & 0 & 0 & 1 & 1 & 0 & 1 & 0 & 0 & 0 & 0 & 0 & 0 & 0 & 0 & 0 & 0 & 0 & 0 &
   0 & 0 & 0 & 0 & 1 & 1 & 0 & 1 & 0 & 1 & 0 & 1 & 0 & 0 & 1 \\
 X_{43} & 0 & 0 & 1 & 1 & 0 & 0 & 1 & 0 & 0 & 0 & 0 & 0 & 0 & 0 & 0 & 0 & 0 & 0 &
   0 & 0 & 0 & 1 & 0 & 1 & 0 & 0 & 0 & 1 & 1 & 1 & 1 & 0 & 0 \\
 X_{87} & 0 & 0 & 1 & 1 & 0 & 0 & 1 & 0 & 0 & 0 & 0 & 0 & 0 & 0 & 0 & 0 & 0 & 0 &
   0 & 0 & 0 & 1 & 0 & 1 & 0 & 1 & 1 & 0 & 0 & 0 & 0 & 1 & 1 \\
 X_{78} & 0 & 0 & 1 & 0 & 1 & 1 & 0 & 0 & 0 & 0 & 0 & 0 & 0 & 0 & 0 & 0 & 0 & 0 &
   0 & 0 & 0 & 0 & 1 & 0 & 1 & 0 & 0 & 1 & 1 & 1 & 1 & 0 & 0 \\
 X_{34} & 0 & 0 & 1 & 0 & 1 & 1 & 0 & 0 & 0 & 0 & 0 & 0 & 0 & 0 & 0 & 0 & 0 & 0 &
   0 & 0 & 0 & 0 & 1 & 0 & 1 & 1 & 1 & 0 & 0 & 0 & 0 & 1 & 1 \\
 X_{12} & 0 & 0 & 1 & 0 & 1 & 0 & 1 & 0 & 0 & 0 & 0 & 0 & 0 & 0 & 0 & 0 & 0 & 0 &
   0 & 0 & 0 & 1 & 0 & 0 & 1 & 0 & 1 & 0 & 1 & 0 & 1 & 1 & 0 \\
 X_{56} & 0 & 0 & 1 & 0 & 1 & 0 & 1 & 0 & 0 & 0 & 0 & 0 & 0 & 0 & 0 & 0 & 0 & 0 &
   0 & 0 & 0 & 0 & 1 & 1 & 0 & 0 & 0 & 0 & 0 & 1 & 1 & 1 & 1
\end{array}
\right)
$
}
\vspace{0.5cm}

 \noindent The F-term charge matrix $Q_F=\ker{(P)}$ is

\noindent\makebox[\textwidth]{%
\scriptsize
$
Q_F=
\left(
\begin{array}{ccc|cc|cc|cccc|cccc|cccccc|cccccccccccc}
  p_1& p_2& p_3& q_1& q_2& r_1& r_2& u_1& u_2& u_3& u_4& v_1& v_2& v_3& v_4& w_1& w_2& w_3& w_4& w_5& w_6& s_1& s_2& s_3& s_4& s_5& s_6& s_7& s_8& s_9& s_{10}& s_{11}& s_{12} \\
 \hline
1 & 1 & 0 & 0 & 0 & 0 & 0 & -1 & 0 & 0 & 0 & 0 & 0 & 0 & -1 & 0 & 0 & 0 & 0 & 0 & 0 & 0 & 0 & 0 & 0 & 0 & 0 & 0 & 0 & 0 & 0 & 0 & 0 \\
 1 & 1 & 0 & 0 & 0 & 0 & 0 & 0 & -1 & 0 & 0 & 0 & 0 & -1 & 0 & 0 & 0 & 0 & 0 & 0 & 0 & 0 & 0 & 0 & 0 & 0 & 0 & 0 & 0 & 0 & 0 & 0 & 0 \\
 1 & 1 & 0 & 0 & 0 & 0 & 0 & 0 & 0 & -1 & 0 & 0 & -1 & 0 & 0 & 0 & 0 & 0 & 0 & 0 & 0 & 0 & 0 & 0 & 0 & 0 & 0 & 0 & 0 & 0 & 0 & 0 & 0 \\
 1 & 0 & 1 & -1 & -1 & 0 & 0 & 0 & 0 & 0 & 0 & 0 & 0 & 0 & 0 & 0 & 0 & 0 & 0 & 0 & 0 & 0 & 0 & 0 & 0 & 0 & 0 & 0 & 0 & 0 & 0 & 0 & 0 \\
 1 & 0 & 0 & -1 & 0 & 0 & 0 & 0 & -1 & 0 & 0 & 0 & 0 & 0 & 0 & 0 & 0 & 0 & 0 & 0 & 0 & 1 & 0 & 0 & 0 & 0 & 0 & 0 & 0 & 0 & 0 & -1 & 1 \\
 1 & 0 & 0 & -1 & 0 & 0 & 0 & 0 & 0 & -1 & 0 & 0 & 0 & 0 & 0 & 0 & 0 & 0 & 0 & 0 & 0 & 0 & -1 & 0 & 0 & 1 & 0 & 0 & 0 & 1 & 0 & 0 & 0 \\
 1 & 0 & 0 & 0 & 0 & 1 & 0 & -1 & 0 & 0 & 0 & 0 & 0 & 0 & 0 & 0 & 0 & 0 & 0 & 0 & 0 & 0 & 0 & 0 & -1 & -1 & 1 & 0 & 0 & 0 & 0 & 0 & 0 \\
 1 & 0 & 0 & 0 & 0 & 0 & 0 & -1 & 0 & 0 & -1 & 0 & 0 & 0 & 0 & 0 & 0 & 0 & 0 & 1 & 0 & 0 & 0 & 0 & 0 & 0 & 0 & 0 & 0 & 0 & 0 & 0 & 0 \\
 1 & 0 & 0 & 0 & 0 & 0 & 0 & 0 & -1 & 0 & -1 & 0 & 0 & 0 & 0 & 0 & 0 & 0 & 0 & 0 & 1 & 0 & 0 & 0 & 0 & 0 & 0 & 0 & 0 & 0 & 0 & 0 & 0 \\
 1 & 0 & 0 & 0 & 0 & 0 & 0 & 0 & 0 & -1 & -1 & 0 & 0 & 0 & 0 & 0 & 1 & 0 & 0 & 0 & 0 & 0 & 0 & 0 & 0 & 0 & 0 & 0 & 0 & 0 & 0 & 0 & 0 \\
 1 & 0 & 0 & 0 & 0 & 0 & 0 & -1 & -1 & 0 & 0 & 0 & 0 & 0 & 0 & 1 & 0 & 0 & 0 & 0 & 0 & 0 & 0 & 0 & 0 & 0 & 0 & 0 & 0 & 0 & 0 & 0 & 0 \\
 1 & 0 & 0 & 0 & 0 & 0 & 0 & -1 & 0 & -1 & 0 & 0 & 0 & 0 & 0 & 0 & 0 & 1 & 0 & 0 & 0 & 0 & 0 & 0 & 0 & 0 & 0 & 0 & 0 & 0 & 0 & 0 & 0 \\
 1 & 0 & 0 & 0 & 0 & 0 & 0 & 0 & -1 & -1 & 0 & 0 & 0 & 0 & 0 & 0 & 0 & 0 & 1 & 0 & 0 & 0 & 0 & 0 & 0 & 0 & 0 & 0 & 0 & 0 & 0 & 0 & 0 \\
 1 & 0 & 0 & 0 & 0 & 0 & 0 & 0 & 0 & -1 & 0 & 1 & 0 & 0 & 0 & -1 & 0 & 0 & 0 & 0 & 0 & 0 & 0 & 0 & 0 & 0 & 0 & 0 & 0 & 0 & 0 & 0 & 0 \\
 1 & 0 & 0 & 0 & 0 & 0 & 0 & 0 & 0 & 0 & -1 & 0 & 1 & 0 & 0 & -1 & 0 & 0 & 0 & 0 & 0 & 0 & 0 & 0 & 0 & 0 & 0 & 0 & 0 & 0 & 0 & 0 & 0 \\
 0 & 1 & 1 & 0 & 0 & -1 & -1 & 0 & 0 & 0 & 0 & 0 & 0 & 0 & 0 & 0 & 0 & 0 & 0 & 0 & 0 & 0 & 0 & 0 & 0 & 0 & 0 & 0 & 0 & 0 & 0 & 0 & 0 \\
 0 & 0 & 1 & -1 & 0 & -1 & 0 & 0 & 0 & 0 & 0 & 0 & 0 & 0 & 0 & 0 & 0 & 0 & 0 & 1 & 0 & 0 & 0 & 0 & 0 & 1 & 0 & 0 & 0 & 0 & 0 & -1 & 0 \\
 0 & 0 & 1 & 0 & 0 & 0 & 0 & 0 & 0 & 0 & 0 & 0 & 0 & 0 & 0 & 1 & 0 & 0 & 0 & 0 & 0 & -1 & -1 & 0 & 0 & 0 & 0 & 0 & 0 & 0 & 0 & 0 & 0 \\
 0 & 0 & 1 & 0 & 0 & 0 & 0 & 0 & 0 & 0 & 0 & 0 & 0 & 0 & 0 & 0 & 0 & 0 & 1 & 0 & 0 & 0 & 0 & 0 & 0 & -1 & 0 & 0 & 0 & 0 & -1 & 0 & 0 \\
 0 & 0 & 0 & 1 & 0 & 1 & 0 & 0 & 0 & 0 & 0 & 0 & 0 & 0 & 0 & 0 & 0 & 0 & 0 & 0 & 0 & 0 & 0 & 0 & 0 & -1 & 0 & -1 & 0 & 0 & 0 & 0 & 0 \\
 0 & 0 & 0 & 1 & 0 & 0 & 1 & 0 & 0 & 0 & 0 & 0 & 0 & 0 & 0 & 0 & 0 & 0 & 0 & 0 & 0 & -1 & 0 & -1 & 0 & 0 & 0 & 0 & 0 & 0 & 0 & 0 & 0 \\
 0 & 0 & 0 & 0 & 0 & 0 & 0 & 0 & 0 & 0 & 0 & 0 & 0 & 0 & 0 & 0 & 0 & 0 & 0 & 0 & 0 & 1 & 0 & 0 & 1 & 0 & -1 & 0 & -1 & 0 & 0 & 0 & 0 \\
 0 & 0 & 0 & 0 & 0 & 0 & 0 & 0 & 0 & 0 & 0 & 0 & 0 & 0 & 0 & 0 & 0 & 0 & 0 & 0 & 0 & 0 & 0 & 0 & 0 & 1 & -1 & 0 & 0 & 0 & 0 & 1 & -1
\end{array}
\right)
$
}
\vspace{0.5cm}

\noindent The D-term charge matrix is

\noindent\makebox[\textwidth]{%
\scriptsize
$
Q_D=
\left(
\begin{array}{ccc|cc|cc|cccc|cccc|cccccc|cccccccccccc}
  p_1& p_2& p_3& q_1& q_2& r_1& r_2& u_1& u_2& u_3& u_4& v_1& v_2& v_3& v_4& w_1& w_2& w_3& w_4& w_5& w_6& s_1& s_2& s_3& s_4& s_5& s_6& s_7& s_8& s_9& s_{10}& s_{11}& s_{12} \\
 \hline
 0 & 0 & 0 & 0 & 0 & 0 & 0 & 0 & 0 & 0 & 0 & 0 & 0 & 0 & 0 & 0 & 0 & 0 & 0 & 0 & 0 & 0
   & 1 & -1 & 0 & 0 & 0 & 0 & 0 & 0 & 0 & 0 & 0 \\
 0 & 0 & 0 & 0 & 0 & 0 & 0 & 0 & 0 & 0 & 0 & 0 & 0 & 0 & 0 & 0 & 0 & 0 & 0 & 0 & 0 & 0
   & 0 & 1 & -1 & 0 & 0 & 0 & 0 & 0 & 0 & 0 & 0 \\
 0 & 0 & 0 & 0 & 0 & 0 & 0 & 0 & 0 & 0 & 0 & 0 & 0 & 0 & 0 & 0 & 0 & 0 & 0 & 0 & 0 & 0
   & 0 & 0 & 1 & -1 & 0 & 0 & 0 & 0 & 0 & 0 & 0 \\
 0 & 0 & 0 & 0 & 0 & 0 & 0 & 0 & 0 & 0 & 0 & 0 & 0 & 0 & 0 & 0 & 0 & 0 & 0 & 0 & 0 & 0
   & 0 & 0 & 0 & 1 & -1 & 0 & 0 & 0 & 0 & 0 & 0 \\
 0 & 0 & 0 & 0 & 0 & 0 & 0 & 0 & 0 & 0 & 0 & 0 & 0 & 0 & 0 & 0 & 0 & 0 & 0 & 0 & 0 & 0
   & 0 & 0 & 0 & 0 & 1 & -1 & 0 & 0 & 0 & 0 & 0 \\
 0 & 0 & 0 & 0 & 0 & 0 & 0 & 0 & 0 & 0 & 0 & 0 & 0 & 0 & 0 & 0 & 0 & 0 & 0 & 0 & 0 & 0
   & 0 & 0 & 0 & 0 & 0 & 1 & -1 & 0 & 0 & 0 & 0 \\
 0 & 0 & 0 & 0 & 0 & 0 & 0 & 0 & 0 & 0 & 0 & 0 & 0 & 0 & 0 & 0 & 0 & 0 & 0 & 0 & 0 & 0
   & 0 & 0 & 0 & 0 & 0 & 0 & 1 & -1 & 0 & 0 & 0
\end{array}
\right)
$
}
\vspace{0.5cm}

The total charge matrix $Q_t$ does not exhibit repeated columns. Accordingly, the global symmetry is $U(1)_{f_1} \times U(1)_{f_2} \times U(1)_R$. Following the discussion in \sref{s1_3}, the flavour and R-charges on the extremal prefect matchings are found as shown in \tref{t2}.

\begin{table}[H]
\centering
\begin{tabular}{|c||c|c|c||l|} 
\hline
\; & $U(1)_{f_1}$ & $U(1)_{f_2}$ & $U(1)_R$ & fugacity \\
\hline
\hline
$p_1$ & -1/4 	&  1/4 	& $2/3$ &  	$t_1$\\
$p_2$ & -1/4 	& -1/4 	& $2/3$ &  	$t_2$\\
$p_3$ &  1/2 	&  0		& $2/3$ &  	$t_3$\\
\hline
\end{tabular}
\caption{The GLSM fields corresponding to extremal points of the toric diagram with their mesonic charges (Model 2). \label{t2}}
\end{table}

Products of non-extremal perfect matchings are set to be associated with a single variable as follows,
\beal{esm2_fug2}
q= q_1 q_2 ~,~
r= r_1 r_2 ~,~
u= u_1 u_2 u_3 u_4 ~,~
v= v_1 v_2 v_3 v_4 ~,~
w= w_1 w_2 w_3 w_4 w_5 w_6 ~,~
s= \prod_{m=1}^{12} s_m
~.
\nn\\
\eea
The fugacities $t_\alpha$ counts extremal perfect matchings $p_\alpha$ with non-zero R-charge. The fugacity $y_q$ counts the product of non-extremal perfect matchings $q$ above.

The mesonic Hilbert series of Model 2 is calculated using the Molien integral formula in \eref{es12_2}. It is 
 \beal{esm2_1}
 &&
g_{1}(t_\alpha,y_{q},y_{r},y_{u},y_{v},y_{w},y_{s}; \mathcal{M}^{mes}_{2})=
(1 - y_{q}^2 y_{r}^2 y_{u}^4 y_{v}^4 y_{w}^4 y_{s}^2 ~ t_1^4 t_2^4) 
(1 - y_{q}^2 y_{r}^2 y_{u}^2 y_{v}^2 y_{w}^2 y_{s}^2~ t_1^2 t_2^2 t_3^2)
\nn\\
&&
\hspace{1cm}
\times
\frac{
1
}{
(1 - y_{q}^2 y_{u}^3 y_{v} y_{w}^2 y_{s} ~ t_1^4) 
(1 - y_{r}^2 y_{u} y_{v}^3 y_{w}^2 y_{s} ~ t_2^4) 
(1 - y_{q} y_{r} y_{s} ~ t_3^2) 
}
\nn\\
&&
\hspace{1cm}
\times
\frac{1}{
(1 - y_{q} y_{r} y_{u}^2 y_{v}^2 y_{w}^2 y_{s}~ t_1^2 t_2^2) 
(1 - y_{q} y_{r} y_{u} y_{v} y_{w} y_{s} ~ t_1 t_2 t_3)
}~~.
\eea
 The plethystic logarithm of the mesonic Hilbert series is
\beal{esm2_3}
&&
PL[g_1(t_\alpha,y_{q},y_{r},y_{u},y_{v},y_{w},y_{s};\mathcal{M}_{2}^{mes})]=
y_{q} y_{r} y_{s} ~ t_{3}^2
+ y_{q} y_{r} y_{u} y_{v} y_{w} y_{s} ~ t_{1} t_{2} t_{3}
\nn\\
&& \hspace{1cm}
+ y_{q} y_{r} y_{u}^2 y_{v}^2 y_{w}^2 y_{s} ~ t_{1}^2 t_{2}^2
+ y_{q}^2 y_{u}^3 y_{v} y_{w}^2 y_{s} ~ t_{1}^4
+ y_{r}^2 y_{u} y_{v}^3 y_{w}^2 y_{s} ~ t_{2}^4
\nn\\
&& \hspace{1cm}
- y_{q}^2 y_{r}^2 y_{u}^2 y_{v}^2 y_{w}^2 y_{s}^2 ~ t_{1}^2 t_{2}^2 t_{3}^2
- y_{q}^2 y_{r}^2 y_{u}^4 y_{v}^4 y_{w}^4 y_{s}^2 ~ t_{1}^4 t_{2}^4
~~.
\eea
The finite plethystic logarithm indicates that the mesonic moduli space is a complete intersection.

With the fugacity map
\beal{esm2_y1}
f_1 &=&
y_q^{1/3} y_r^{1/3} y_u^{-2/3} y_v^{-2/3} y_w^{-2/3} y_s^{-2/3} t_1^{-2/3} t_2^{-2/3} t_3^{4/3}
~,~
\nn\\
f_2 &=&
y_q y_r^{-1} y_u y_v^{-1} ~ t_1^2 t_2^{-2}
~,~
\nn\\
t &=& y_q^{1/3} y_r^{1/3} y_u^{1/3} y_v^{1/3} y_w^{1/3} y_s^{1/3} t_1^{1/3} t_2^{1/3} t_3^{1/3}~,~
\eea
where $f_1$, $f_2$ and $t$ are the mesonic charge fugacities,
the plethystic logarithm becomes
\beal{esm3b_3}
PL[g_1(t,f_1,f_2;\mathcal{M}_{2}^{mes})]
&=&
f_1 t^2
+t^3
+ \frac{1}{f_1} \left(
1 + f_2 + \frac{1}{f_2}
\right) t^4
- t^6
- \frac{1}{f_{1}^2} t^8
~~.
   \eea
From the above plethystic logarithm, one can identify the moduli space generators as well as their mesonic charges. They are shown in \tref{t2gen}. The charge lattice of generators in \tref{t2gen} is the dual reflexive polygon of the toric diagram of Model 2. The two relations formed by the generators are
\beal{es3b_3b}
A_1 A_3 = A_2^2 ~~,~~
B_1 B_2 = A_3^2
~~.
\eea

\begin{table}[H]
\centering
\resizebox{.95\hsize}{!}{
\begin{minipage}[!b]{0.7\textwidth}
\begin{tabular}{|l|c|c|}
\hline
Generator & $U(1)_{f_1}$ & $U(1)_{f_2}$ 
\\
\hline
\hline
$A_1=
p_3^2~
q~
r~ 
s
$
& 1 & 0
\nn\\
$A_2=
p_1 p_2 p_3 
~q
~r
~u v
~w
~s
$
& 0 & 0
\nn\\
$A_3=
p_1^2 p_2^2~
q~
r~
u^2 v^2~
w^2 ~
s
$
& -1 & 0
\nn\\
$B_1=
p_1^4 ~
q^2 ~
u^3 v~
w^2~
s
$
& -1 & 1
\nn\\
$B_2=
p_2^4~
r^2~
u v^3~
w^2~
s
$
& -1 & -1
\nn\\
\hline
\end{tabular}
\end{minipage}
\hspace{2cm}
\begin{minipage}[!b]{0.25\textwidth}
\includegraphics[width=3.5 cm]{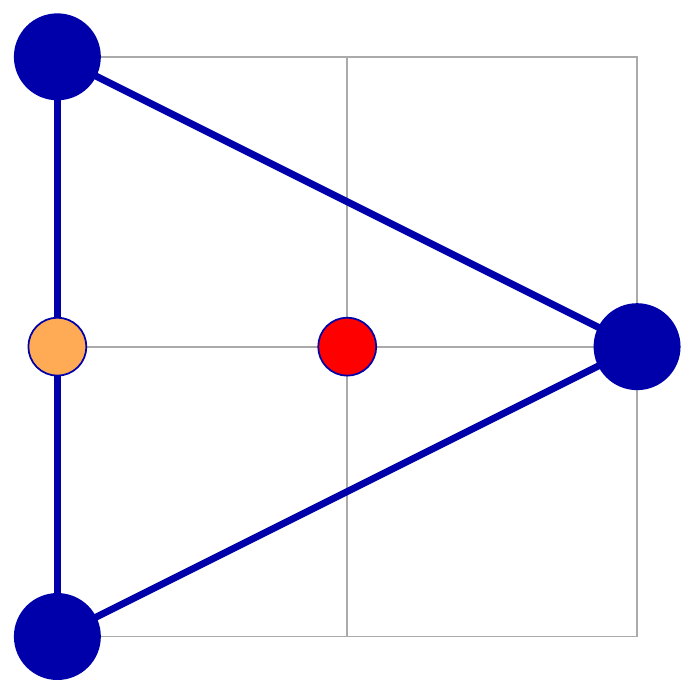}
\end{minipage}

}

\caption{The generators and lattice of generators of the mesonic moduli space of Model 2 in terms of GLSM fields with the corresponding flavor charges.\label{t2gen}} 
\end{table}

\begin{table}[H]
\centering

\resizebox{1\hsize}{!}{
\begin{tabular}{|l|c|c|}
\hline
Generator & $U(1)_{f_1}$ & $U(1)_{f_2}$ 
\\
\hline
\hline
$
X_{12} X_{21}=  X_{34} X_{43}=  X_{56} X_{65}=  X_{78} X_{87}
$
& 1 & 0
\nn\\
$
X_{12} X_{23} X_{31}=  X_{12} X_{28} X_{81}=  X_{14} X_{42} X_{21}=  X_{14} X_{43} X_{31}=  X_{17} X_{72} X_{21}=  X_{17} X_{78} X_{81}=  X_{23} X_{34} X_{42}=  X_{28} X_{87} X_{72}=  X_{34} X_{45} X_{53}
$
& 0 & 0
\nn\\
$=  X_{36} X_{64} X_{43}=  X_{36} X_{65} X_{53}=  X_{45} X_{56} X_{64}=  X_{56} X_{67} X_{75}=  X_{58} X_{86} X_{65}=  X_{58} X_{87} X_{75}=  X_{67} X_{78} X_{86}
$
& &
\nn\\
$
X_{14} X_{42} X_{23} X_{31}=  X_{14} X_{42} X_{28} X_{81}=  X_{14} X_{45} X_{53} X_{31}=  X_{17} X_{72} X_{23} X_{31}=  X_{17} X_{72} X_{28} X_{81}=  X_{17} X_{75} X_{58} X_{81}=  X_{23} X_{36} X_{64} X_{42}
$
& -1 & 0
\nn\\
$
=  X_{28} X_{86} X_{67} X_{72}=  X_{36} X_{64} X_{45} X_{53}=  X_{36} X_{67} X_{75} X_{53}=  X_{45} X_{58} X_{86} X_{64}=  X_{58} X_{86} X_{67} X_{75}
$
& &
\nn\\
$
 X_{14} X_{45} X_{58} X_{81}=  X_{23} X_{36} X_{67} X_{72}
$
& -1 & 1
\nn\\
$
X_{17} X_{75} X_{53} X_{31}=  X_{28} X_{86} X_{64} X_{42}
$
& -1 & -1
\nn\\
\hline
\end{tabular}
}

\caption{The generators in terms of bifundamental fields (Model 2). \label{t2gen2}} 
\end{table}

With the fugacity map
\beal{esm3b_3x}
&&
T_1 = f_1^{-1/4} f_2^{1/4} ~ t
= y_{q}^{1/2} y_{u}^{3/4} y_{v}^{1/4} y_{w}^{1/2} y_{s}^{1/4} ~ t_1
~,~
\nn\\
&&
T_2 = f_1^{-1/4} f_2^{-1/4} ~ t
= y_{r}^{1/2} y_{u}^{1/4} y_{v}^{3/4} y_{w}^{1/2} y_{s}^{1/4} ~ t_2 ~,~
\nn\\
&&
T_3 = f_1^{1/2} ~ t
= y_{q}^{1/2} y_{r}^{1/2} y_{s}^{1/2} ~ t_3~,
\eea   
the mesonic Hilbert series takes the form
\beal{esm3b_3x2}
g_1(T_1,T_2,T_3;\mathcal{M}^{mes}_2)
=
\frac{
(1-T_1^4 T_2^4) (1-T_1^2 T_2^2 T_3^2)
}{
(1- T_1^4)(1-T_2^4)(1-T_3^2)
(1-T_1^2 T_2^2) (1-T_1 T_2 T_3)
}~~,
\eea
with the plethystic logarithm being
\beal{esm3b_3x3}
PL[g_1(T_1,T_2,T_3;\mathcal{M}^{mes}_2)]=
T_3^2 + T_1 T_2 T_3 + T_1^2 T_2^2 + T_1^4 
+ T_2^4 - T_1^2 T_2^2 T_3^2 -T_1^4 T_2^4 ~.
\nn\\ 
\eea
The above refinement of the mesonic Hilbert series emphasises the conical structure of the toric Calabi-Yau space.
\\

\section{Model 3: $L_{1,3,1}/\mathbb{Z}_2~(0,1,1,1)$}
\subsection{Model 3 Phase a}

\begin{figure}[H]
\begin{center}
\includegraphics[trim=0cm 0cm 0cm 0cm,height=4.5 cm]{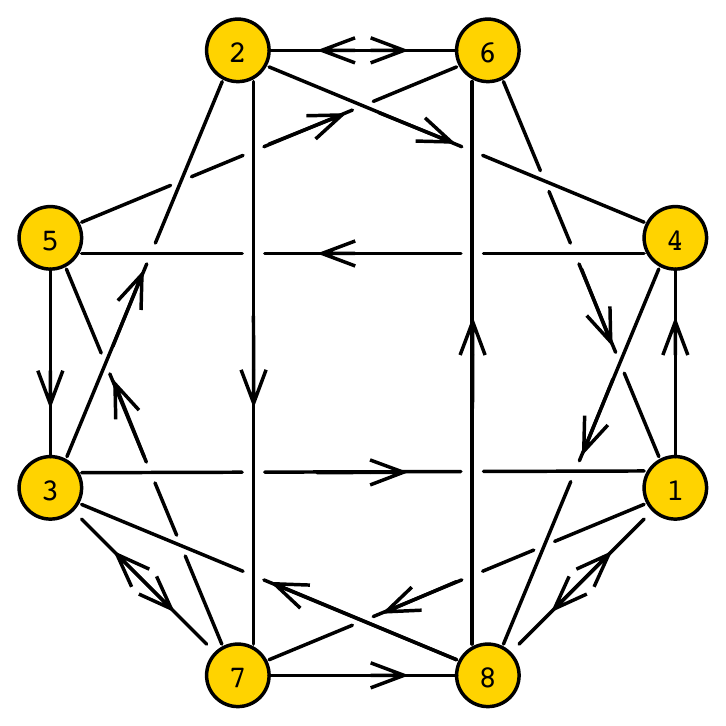}
\includegraphics[width=5 cm]{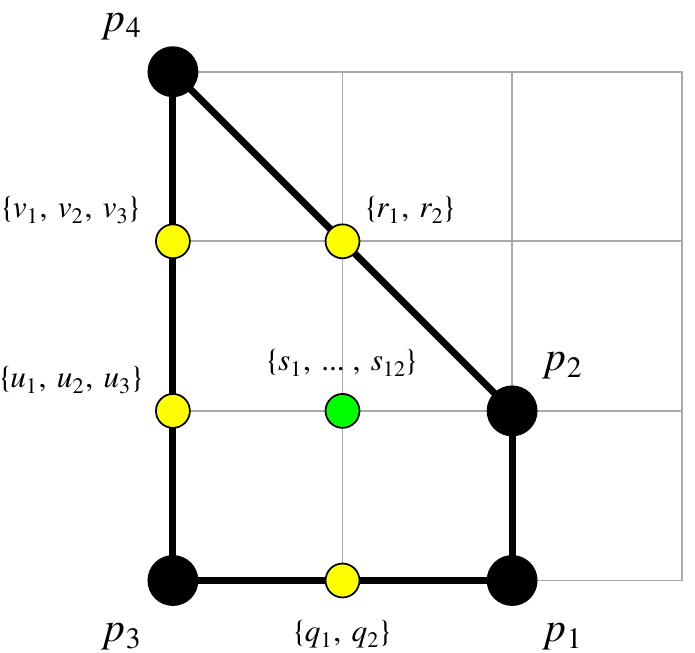}
\includegraphics[width=5 cm]{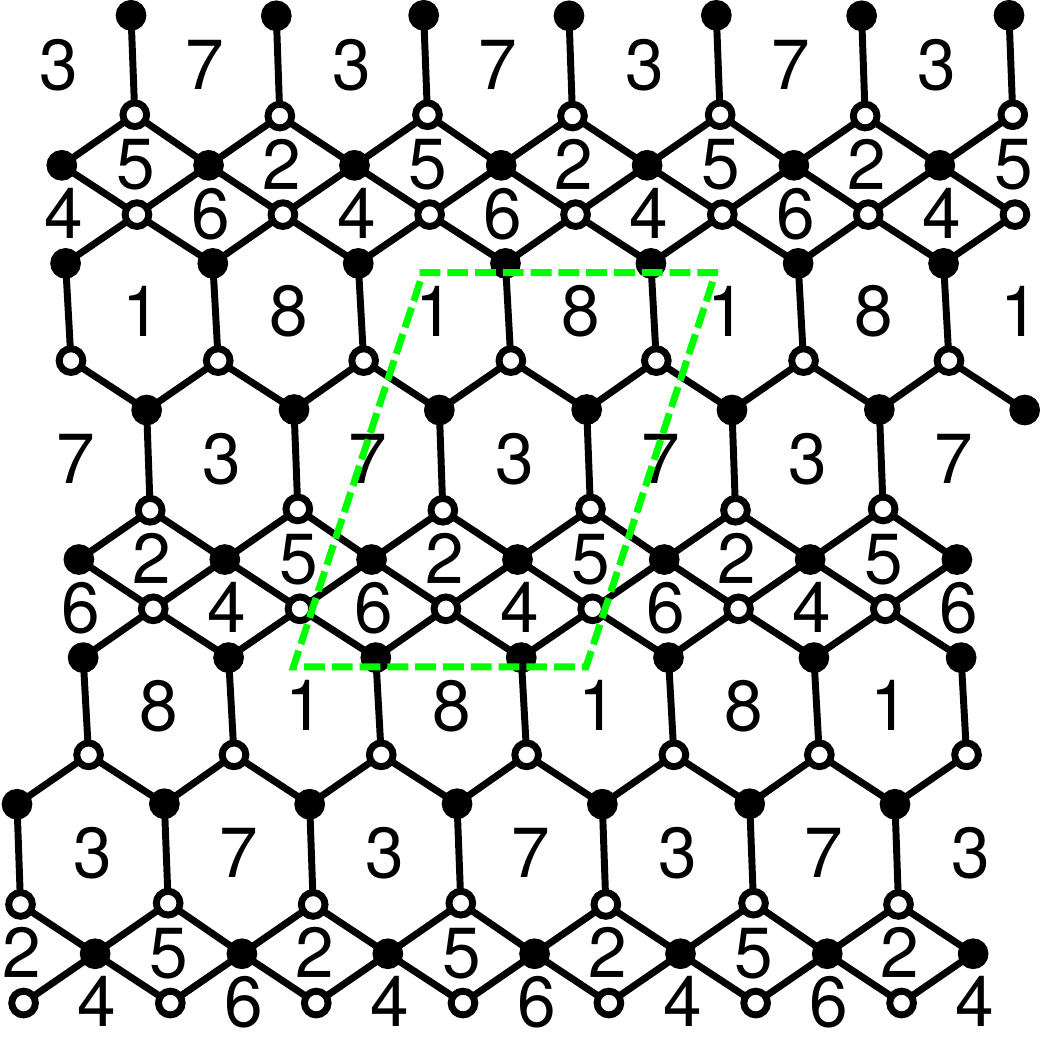}
\caption{The quiver, toric diagram, and brane tiling of Model 3a.\label{f3a}}
 \end{center}
 \end{figure}
 
 \noindent The superpotential is 
\beal{esm3a_00}
W&=& 
+ X_{31} X_{18} X_{83}  
+ X_{32} X_{27} X_{73}  
+ X_{53} X_{37} X_{75}  
+ X_{78} X_{81} X_{17} 
\nn\\
&& 
- X_{14} X_{48} X_{81}  
- X_{31} X_{17} X_{73}   
- X_{78} X_{83} X_{37}  
- X_{86} X_{61} X_{18} 
\nn\\
&&
+ X_{14} X_{45} X_{56} X_{61}  
+ X_{62} X_{24} X_{48} X_{86} 
- X_{32} X_{24} X_{45} X_{53}  
- X_{62} X_{27} X_{75} X_{56} 
~.
  \eea
 
 \noindent The perfect matching matrix is 
 
\noindent\makebox[\textwidth]{%
\scriptsize
$
P=
\left(
\begin{array}{c|cccc|cc|cc|ccc|ccc|cccccccccccc}
 \; &p_1 & p_2& p_3& p_4& q_1& q_2& r_1& r_2& u_1& u_2& u_3& v_1& v_2& v_3& s_1& s_2& s_3& s_4& s_5& s_6& s_7& s_8& s_9& s_{10}& s_{11}& s_{12} \\
 \hline
 X_{81} & 1 & 1 & 0 & 0 & 1 & 0 & 1 & 0 & 0 & 0 & 0 & 0 & 0 & 0 & 1 & 0 & 1 & 0 &
   1 & 1 & 1 & 1 & 0 & 0 & 0 & 0 \\
 X_{73} & 1 & 1 & 0 & 0 & 1 & 0 & 0 & 1 & 0 & 0 & 0 & 0 & 0 & 0 & 1 & 0 & 1 & 0 &
   0 & 0 & 1 & 0 & 1 & 1 & 0 & 1 \\
 X_{37} & 1 & 1 & 0 & 0 & 0 & 1 & 1 & 0 & 0 & 0 & 0 & 0 & 0 & 0 & 0 & 1 & 0 & 1 &
   1 & 1 & 0 & 1 & 0 & 0 & 1 & 0 \\
 X_{18} & 1 & 1 & 0 & 0 & 0 & 1 & 0 & 1 & 0 & 0 & 0 & 0 & 0 & 0 & 0 & 1 & 0 & 1 &
   0 & 0 & 0 & 0 & 1 & 1 & 1 & 1 \\
 X_{24} & 1 & 0 & 0 & 0 & 1 & 0 & 0 & 0 & 0 & 0 & 0 & 0 & 0 & 0 & 0 & 0 & 0 & 0 &
   1 & 0 & 0 & 0 & 0 & 1 & 1 & 0 \\
 X_{56} & 1 & 0 & 0 & 0 & 0 & 1 & 0 & 0 & 0 & 0 & 0 & 0 & 0 & 0 & 1 & 0 & 0 & 0 &
   0 & 0 & 0 & 1 & 1 & 0 & 0 & 0 \\
 X_{45} & 0 & 1 & 0 & 0 & 0 & 0 & 1 & 0 & 0 & 0 & 0 & 0 & 0 & 0 & 0 & 1 & 1 & 0 &
   0 & 0 & 0 & 0 & 0 & 0 & 0 & 1 \\
 X_{62} & 0 & 1 & 0 & 0 & 0 & 0 & 0 & 1 & 0 & 0 & 0 & 0 & 0 & 0 & 0 & 0 & 0 & 1 &
   0 & 1 & 1 & 0 & 0 & 0 & 0 & 0 \\
 X_{83} & 0 & 0 & 1 & 0 & 1 & 0 & 0 & 0 & 1 & 1 & 0 & 0 & 1 & 0 & 1 & 0 & 1 & 0 &
   0 & 0 & 1 & 0 & 0 & 0 & 0 & 0 \\
 X_{61} & 0 & 0 & 1 & 0 & 1 & 0 & 0 & 0 & 1 & 0 & 1 & 0 & 0 & 1 & 0 & 0 & 0 & 0 &
   1 & 1 & 1 & 0 & 0 & 0 & 0 & 0 \\
 X_{17} & 0 & 0 & 1 & 0 & 0 & 1 & 0 & 0 & 1 & 1 & 0 & 0 & 1 & 0 & 0 & 1 & 0 & 1 &
   0 & 0 & 0 & 0 & 0 & 0 & 1 & 0 \\
 X_{48} & 0 & 0 & 1 & 0 & 0 & 1 & 0 & 0 & 1 & 0 & 1 & 0 & 0 & 1 & 0 & 1 & 0 & 0 &
   0 & 0 & 0 & 0 & 1 & 0 & 0 & 1 \\
 X_{75} & 0 & 0 & 1 & 0 & 1 & 0 & 0 & 0 & 0 & 1 & 1 & 1 & 0 & 0 & 0 & 0 & 1 & 0 &
   0 & 0 & 0 & 0 & 0 & 1 & 0 & 1 \\
 X_{32} & 0 & 0 & 1 & 0 & 0 & 1 & 0 & 0 & 0 & 1 & 1 & 1 & 0 & 0 & 0 & 0 & 0 & 1 &
   0 & 1 & 0 & 1 & 0 & 0 & 0 & 0 \\
 X_{27} & 0 & 0 & 0 & 1 & 0 & 0 & 1 & 0 & 1 & 0 & 0 & 0 & 1 & 1 & 0 & 1 & 0 & 0 &
   1 & 0 & 0 & 0 & 0 & 0 & 1 & 0 \\
 X_{86} & 0 & 0 & 0 & 1 & 0 & 0 & 1 & 0 & 0 & 1 & 0 & 1 & 1 & 0 & 1 & 0 & 1 & 0 &
   0 & 0 & 0 & 1 & 0 & 0 & 0 & 0 \\
 X_{53} & 0 & 0 & 0 & 1 & 0 & 0 & 0 & 1 & 1 & 0 & 0 & 0 & 1 & 1 & 1 & 0 & 0 & 0 &
   0 & 0 & 1 & 0 & 1 & 0 & 0 & 0 \\
 X_{14} & 0 & 0 & 0 & 1 & 0 & 0 & 0 & 1 & 0 & 1 & 0 & 1 & 1 & 0 & 0 & 0 & 0 & 1 &
   0 & 0 & 0 & 0 & 0 & 1 & 1 & 0 \\
 X_{31} & 0 & 0 & 0 & 1 & 0 & 0 & 1 & 0 & 0 & 0 & 1 & 1 & 0 & 1 & 0 & 0 & 0 & 0 &
   1 & 1 & 0 & 1 & 0 & 0 & 0 & 0 \\
 X_{78} & 0 & 0 & 0 & 1 & 0 & 0 & 0 & 1 & 0 & 0 & 1 & 1 & 0 & 1 & 0 & 0 & 0 & 0 &
   0 & 0 & 0 & 0 & 1 & 1 & 0 & 1
\end{array}
\right)
$
}
\vspace{0.5cm}

 \noindent The F-term charge matrix $Q_F=\ker{(P)}$ is

\noindent\makebox[\textwidth]{%
\scriptsize
$
Q_F=
\left(
\begin{array}{cccc|cc|cc|ccc|ccc|cccccccccccc}
 p_1 & p_2& p_3& p_4& q_1& q_2& r_1& r_2& u_1& u_2& u_3& v_1& v_2& v_3& s_1& s_2& s_3& s_4& s_5& s_6& s_7& s_8& s_9& s_{10}& s_{11}& s_{12} \\
 \hline
 1 & 0 & 1 & 0 & -1 & -1 & 0 & 0 & 0 & 0 & 0 & 0 & 0 & 0 & 0 & 0 & 0 & 0 & 0 & 0 & 0 & 0 & 0 & 0 & 0 & 0 \\
 0 & 1 & 0 & 1 & 0 & 0 & -1 & -1 & 0 & 0 & 0 & 0 & 0 & 0 & 0 & 0 & 0 & 0 & 0 & 0 & 0 & 0 & 0 & 0 & 0 & 0 \\
  1 & 0 & 0 & 1 & 0 & 0 & -1 & 0 & 0 & 0 & 0 & 0 & 0 & 0 & -1 & 0 & 1 & 0 & 0 & 0 & 0 & 0 & 0 & -1 & 0 & 0 \\
 0 & 1 & 1 & 0 & -1 & 0 & 0 & 0 & 0 & 0 & 0 & 0 & 0 & 0 & 0 & -1 & 0 & 0 & 1 & -1 & 0 & 0 & 0 & 0 & 0 & 0 \\
 1 & 0 & 0 & 0 & 0 & 0 & 0 & 0 & 0 & 0 & 0 & 0 & 1 & 0 & -1 & 0 & 0 & 0 & 0 & 0 & 0 & 0 & 0 & 0 & -1 & 0 \\
 0 & 1 & 0 & 0 & 0 & 0 & 0 & 0 & 1 & 0 & 0 & 0 & 0 & 0 & 0 & -1 & 0 & 0 & 0 & 0 & -1 & 0 & 0 & 0 & 0 & 0 \\
 1 & 0 & 0 & 0 & 0 & 0 & 1 & 0 & 1 & 0 & 0 & 0 & 0 & 0 & -1 & -1 & 0 & 0 & -1 & 0 & 0 & 0 & 0 & 0 & 0 & 0 \\
 0 & 1 & 0 & 0 & 0 & 0 & 0 & 0 & 0 & 1 & 0 & 0 & 0 & 0 & 0 & 0 & -1 & -1 & 0 & 0 & 0 & 0 & 0 & 0 & 0 & 0 \\
 0 & 0 & 1 & 1 & 0 & 0 & 0 & 0 & -1 & 0 & 0 & -1 & 0 & 0 & 0 & 0 & 0 & 0 & 0 & 0 & 0 & 0 & 0 & 0 & 0 & 0 \\
 0 & 0 & 1 & 1 & 0 & 0 & 0 & 0 & 0 & -1 & 0 & 0 & 0 & -1 & 0 & 0 & 0 & 0 & 0 & 0 & 0 & 0 & 0 & 0 & 0 & 0 \\
 0 & 0 & 1 & 0 & 0 & 0 & 0 & 0 & -1 & -1 & 0 & 0 & 1 & 0 & 0 & 0 & 0 & 0 & 0 & 0 & 0 & 0 & 0 & 0 & 0 & 0 \\
 0 & 0 & 1 & 0 & 0 & 0 & 0 & 0 & 0 & -1 & -1 & 1 & 0 & 0 & 0 & 0 & 0 & 0 & 0 & 0 & 0 & 0 & 0 & 0 & 0 & 0 \\
 0 & 0 & 1 & 0 & -1 & 0 & 0 & 0 & 0 & -1 & 0 & 0 & 0 & 0 & 1 & 0 & 0 & 0 & 0 & 0 & 0 & 0 & -1 & 1 & 0 & 0 \\
 0 & 0 & 0 & 0 & 1 & 0 & 1 & 0 & 0 & 0 & 0 & 0 & 0 & 0 & 0 & 0 & -1 & 0 & -1 & 0 & 0 & 0 & 0 & 0 & 0 & 0 \\
 0 & 0 & 0 & 0 & 0 & 1 & 1 & 0 & 0 & 0 & 0 & 0 & 0 & 0 & 0 & -1 & 0 & 0 & 0 & 0 & 0 & -1 & 0 & 0 & 0 & 0 \\
 0 & 0 & 0 & 0 & 0 & 0 & 0 & 0 & 0 & 0 & 0 & 0 & 0 & 0 & 1 & 0 & -1 & 0 & 0 & 0 & 0 & 0 & -1 & 0 & 0 & 1
    \end{array}
\right)
$
}
\vspace{0.5cm}

\noindent The D-term charge matrix is

\noindent\makebox[\textwidth]{%
\scriptsize
$
Q_D=
\left(
\begin{array}{cccc|cc|cc|ccc|ccc|cccccccccccc}
 p_1 & p_2& p_3& p_4& q_1& q_2& r_1& r_2& u_1& u_2& u_3& v_1& v_2& v_3& s_1& s_2& s_3& s_4& s_5& s_6& s_7& s_8& s_9& s_{10}& s_{11}& s_{12} \\
 \hline
 0 & 0 & 0 & 0 & 0 & 0 & 0 & 0 & 0 & 0 & 0 & 0 & 0 & 0 & 0 & 0 & 0 & 0 & 1 & -1 & 0 & 0
   & 0 & 0 & 0 & 0 \\
 0 & 0 & 0 & 0 & 0 & 0 & 0 & 0 & 0 & 0 & 0 & 0 & 0 & 0 & 0 & 0 & 0 & 0 & 0 & 1 & -1 & 0
   & 0 & 0 & 0 & 0 \\
 0 & 0 & 0 & 0 & 0 & 0 & 0 & 0 & 0 & 0 & 0 & 0 & 0 & 0 & 0 & 0 & 0 & 0 & 0 & 0 & 1 & -1
   & 0 & 0 & 0 & 0 \\
 0 & 0 & 0 & 0 & 0 & 0 & 0 & 0 & 0 & 0 & 0 & 0 & 0 & 0 & 0 & 0 & 0 & 0 & 0 & 0 & 0 & 1
   & -1 & 0 & 0 & 0 \\
 0 & 0 & 0 & 0 & 0 & 0 & 0 & 0 & 0 & 0 & 0 & 0 & 0 & 0 & 0 & 0 & 0 & 0 & 0 & 0 & 0 & 0
   & 1 & -1 & 0 & 0 \\
 0 & 0 & 0 & 0 & 0 & 0 & 0 & 0 & 0 & 0 & 0 & 0 & 0 & 0 & 0 & 0 & 0 & 0 & 0 & 0 & 0 & 0
   & 0 & 1 & -1 & 0 \\
 0 & 0 & 0 & 0 & 0 & 0 & 0 & 0 & 0 & 0 & 0 & 0 & 0 & 0 & 0 & 0 & 0 & 0 & 0 & 0 & 0 & 0
   & 0 & 0 & 1 & -1
\end{array}
\right)
$
}
\vspace{0.5cm}

The total charge matrix does not exhibit repeated columns. Accordingly, the global symmetry is $U(1)_{f_1} \times U(1)_{f_2} \times U(1)_R$. Following the discussion in \sref{s1_3}, the mesonic charges on the extremal perfect matchings are found as shown in \tref{t3a}.

\begin{table}[H]
\centering
\begin{tabular}{|c||c|c|c||l|} 
\hline
\; & $U(1)_{f_1}$ & $U(1)_{f_2}$ & $U(1)_R$ & fugacity \\
\hline
\hline
$p_1$ & 1/2 	&  1/2 & $R_1=\frac{1}{6}\left(5-\sqrt{7}\right)$ &  	$t_1$\\
$p_2$ & 0 	& -1/2 & $R_1=\frac{1}{6}\left(5-\sqrt{7}\right)$ &  	$t_2$\\
$p_3$ &-1/2 	& -1/2 & $R_2=\frac{1}{6}\left(1+\sqrt{7}\right)$ &  	$t_3$\\
$p_4$ & 0 	&  1/2 & $R_2=\frac{1}{6}\left(1+\sqrt{7}\right)$ &  	$t_4$\\
\hline
\end{tabular}
\caption{The GLSM fields corresponding to extremal points of the toric diagram with their mesonic charges (Model 3a). The R-charges are obtained using a-maximization.\label{t3a}}
\end{table}

Products of non-extremal perfect matchings are associated to a single variable as follows
\beal{esx3a_1}
q = q_1 q_2 ~,~
r = r_1 r_2 ~,~
u = u_1 u_2 u_3 ~,~
v = v_1 v_2 v_3 ~,~
s = \prod_{m=1}^{12} s_m~~.
\eea
The fugacity $t_\alpha$ counts extremal perfect matchings. The fugacity $y_q$ counts the product of non-extremal perfect matchings $q$ above.

The mesonic Hilbert series of Model 3a is calculated using the Molien integral formula in \eref{es12_2}. It is 
\beal{esm3a_1}
&&g_{1}(t_\alpha,y_{q},y_{r},y_{u},y_{v},y_{s}; \mathcal{M}^{mes}_{3a})=
(1 - y_{q}^2 y_{r}^2 y_{u}^2 y_{v}^2 y_{s}^2 ~ t_1^2 t_2^2 t_3^2 t_4^2) 
(1 - y_{q}^2 y_{r}^2 y_{u}^3 y_{v}^3 y_{s}^2 ~ t_1 t_2 t_3^3 t_4^3)
\nn\\
&&
\hspace{1cm}
\times
\frac{
1
}{
(1 - y_{q} y_{r} y_{s} ~ t_1^2 t_2^2) 
(1 - y_{q}^2 y_{u}^2 y_{v} y_{s} ~ t_1 t_3^3) 
(1 - y_{q} y_{r} y_{u}^2 y_{v}^2 y_{s} ~ t_3^2 t_4^2) 
}
\nn\\
&&
\hspace{1cm}
\times
\frac{1}{
(1 - y_{r}^2 y_{u} y_{v}^2 y_{s} ~ t_2 t_4^3) 
(1 - y_{q} y_{r} y_{u} y_{v} y_{s} ~ t_1 t_2 t_3 t_4)
}
~~.
\nn\\
\eea
 The plethystic logarithm of the mesonic Hilbert series is
\beal{esm3a_2}
&&
PL[g_1(t_\alpha,y_{q},y_{r},y_{u},y_{v},y_{s};\mathcal{M}_{3a}^{mes})]
=
y_{q} y_{r} y_{s} ~ t_{1}^2 t_{2}^2
+ y_{q} y_{r} y_{u} y_{v} y_{s} ~ t_{1} t_{2} t_{3} t_{4}
+ y_{q}^2 y_{u}^2 y_{v} y_{s} ~ t_{1} t_{3}^3
\nn\\
&&
\hspace{1cm}
+ y_{r}^2 y_{u} y_{v}^2 y_{s} ~ t_{2} t_{4}^3
+ y_{q} y_{r} y_{u}^2 y_{v}^2 y_{s} ~ t_{3}^2 t_{4}^2
- y_{q}^2 y_{r}^2 y_{u}^2 y_{v}^2 y_{s}^2 ~ t_{1}^2 t_{2}^2 t_{3}^2 t_{4}^2
- y_{q}^2 y_{r}^2 y_{u}^3 y_{v}^3 y_{s}^2 ~ t_{1} t_{2} t_{3}^3 t_{4}^3
~~.
\nn\\
\eea
The finite plethystic logarithm indicates that the mesonic moduli space is a complete intersection.

Consider the fugacity map
\beal{esm3a_y1}
f_1 &=& \frac{1}{y_u y_v}
~,~
\nn\\
f_2 &=& \frac{
y_r y_v ~ t_2^{1/2} t_4^{3/2}
}{
y_q ~ t_1^{1/2} t_3^{3/2} 
}
~,~
\nn\\
\tilde{t}_1 &=& y_q^{1/4} y_r^{1/4} y_u^{1/4} y_v^{1/4} y_s^{1/4} ~ t_1^{1/2} t_2^{1/2} 
~,~
\nn\\
\tilde{t}_2 &=& y_q^{1/4} y_r^{1/4} y_u^{1/4} y_v^{1/4} y_s^{1/4} ~ t_3^{1/2} t_4^{1/2}~,~
\eea
where $f_1$ and $f_2$ are the flavor fugacities, and $\tilde{t}_1$ and $\tilde{t}_2$ are the fugacities for the R-charges $R_1$ and $R_2$ in \tref{t3a} respectively. Under the above fugacity map, the plethystic logarithm becomes
\beal{esm3a_3}
PL[g_1(t_\alpha,f_1,f_2;\mathcal{M}_{3a}^{mes})]
&=&
f_{1} \tilde{t}_{1}^4
+\tilde{t}_{1}^2 \tilde{t}_{2}^2
+\left(
\frac{1}{f_1 f_2}
+ f_2 
\right)
\tilde{t}_{1} \tilde{t}_{2}^3
+\frac{\tilde{t}_{2}^4}{f_{1}}
-\tilde{t}_{1}^4 \tilde{t}_{2}^4
-\frac{\tilde{t}_{1}^2 \tilde{t}_{2}^6}{f_{1}}
~.\nn\\
   \eea
The above plethystic logarithm indicates both the moduli space generators as well as their mesonic charges. They are summarized in \tref{t3agen}. The generators can be presented on a charge lattice. The convex polygon formed by the generators in \tref{t3agen} is the dual reflexive polygon of the toric diagram of Model 3a. The generators satisfy the following relations
\beal{esm3a_3b}
A_1 A_2 = B^2 ~~,~~
A_2 B = C_1 C_2 ~~.
\eea

\begin{table}[H]
\centering

\resizebox{\hsize}{!}{

\begin{minipage}[!b]{0.6\textwidth}
\begin{tabular}{|l|c|c|}
\hline
Generator & $U(1)_{f_1}$ & $U(1)_{f_2}$ 
\\
\hline
\hline
$
A_1=
p_1^2 p_2^2~
q~
r~
s$
& 1 & 0
\nn\\
$
A_2=
p_3^2 p_4^2~
q~
r~
u^2 v^2~
s
$
& -1 & 0
\nn\\
$
B=
p_1 p_2 p_3 p_4~
q~
r~
u v~
s
$
& 0 & 0
\nn\\
$
C_1=
p_1 p_3^3~
q^2~
u^2 v~
s
$
& -1 & -1
\nn\\
$
C_2=
p_2 p_4^3~
r^2~
u v^2~
s
$
& 0 & 1
\nn\\
\hline
\end{tabular}
\end{minipage}
\hspace{2cm}
\begin{minipage}[!b]{0.25\textwidth}
\includegraphics[width=3.5 cm]{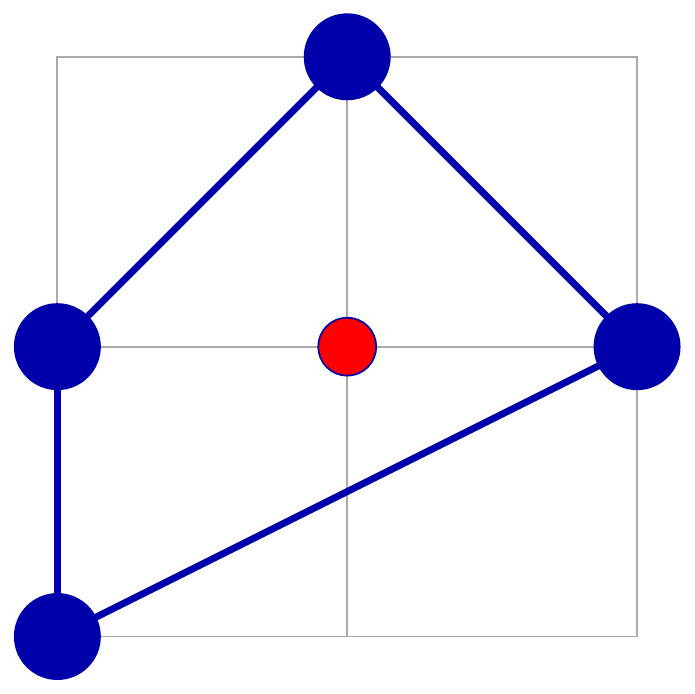}
\end{minipage}

}

\caption{The generators and lattice of generators of the mesonic moduli space of Model 3a in terms of GLSM fields with the corresponding flavor charges.\label{t3agen}} 
\end{table}

\begin{table}[H]
\centering

\resizebox{1\hsize}{!}{
\begin{tabular}{|l|c|c|}
\hline
Generator & $U(1)_{f_1}$ & $U(1)_{f_2}$ 
\\
\hline
\hline
$
 X_{24} X_{45} X_{56} X_{62}=  X_{18} X_{81}=  X_{37} X_{73}
$
& 1 & 0
\nn\\
$
 X_{14} X_{48} X_{83} X_{31}=  X_{14} X_{48} X_{86} X_{61}=  X_{17} X_{75} X_{53} X_{31}=  X_{17} X_{78} X_{83} X_{31}$
 &-1 &0
 \nn\\
$= X_{17} X_{78} X_{86} X_{61}=  X_{27} X_{75} X_{53} X_{32}=  X_{27} X_{78} X_{83} X_{32}
$
& &
\nn\\
$ X_{14} X_{45} X_{56} X_{61}=  X_{24} X_{45} X_{53} X_{32}=  X_{24} X_{48} X_{86} X_{62}=  X_{27} X_{75} X_{56} X_{62}=  X_{14} X_{48} X_{81}$
&0 &0
\nn\\
$=  X_{17} X_{73} X_{31}=  X_{17} X_{78} X_{81}=  X_{18} X_{83} X_{31}=  X_{18} X_{86} X_{61}=  X_{27} X_{73} X_{32}=  X_{37} X_{75} X_{53}=  X_{37} X_{78} X_{83}$
& &
\nn\\
$
 X_{17} X_{75} X_{56} X_{61}=  X_{24} X_{48} X_{83} X_{32}
$
& -1 & -1
\nn\\
$
 X_{14} X_{45} X_{53} X_{31}=  X_{27} X_{78} X_{86} X_{62}
$
& 0 & 1
\nn\\
\hline
\end{tabular}
}

\caption{The generators in terms of bifundamental fields (Model 3a).\label{t3agen}} 
\end{table}

The mesonic Hilbert series and the plethystic logarithm can be re-expressed in terms of the following $3$ fugacities,
\beal{esm3a_x1}
T_1 
= \frac{
f_2
}{
\tilde{t}_1^3 \tilde{t}_2
} 
= \frac{t_4}{y_{q}^2 y_{u} y_{s} ~t_1^2 t_2 t_3^2} 
~,~
T_2 
= \frac{1}{f_1 f_2}~\tilde{t}_1 \tilde{t}_2^3 
= y_{q}^2 y_{u}^2 y_{v} y_{s} ~ t_1 t_3^3~,~
T_3 = f_1 ~ \tilde{t}_1^4 = y_{q} y_{r} y_{s} ~ t_1^2 t_2^2 ~~,
\nn\\
\eea
such that
\beal{esm3a_x2}
g_1(T_1,T_2,T_3;\mathcal{M}^{mes}_{3a})
=
\frac{
(1 - T_1^2 T_2^2 T_3^2) (1 - T_1^3 T_2^3 T_3^2)
}{
(1 - T_3) (1 - T_2) (1 - T_1^2 T_2^2 T_3) (1 - T_1^3 T_2^2 T_3^2) (1 - T_1 T_2 T_3) 
}
\nn\\
\eea
and
\beal{esm3a_x3}
PL[g_1(T_1,T_2,T_3;\mathcal{M}^{mes}_{3a})]&=&
T_3 + T_1 T_2 T_3 + T_2 + T_1^2 T_2^2 T_3 + T_1^3 T_2^2 T_3^2
\nn\\
&&
  - T_1^2 T_2^2 T_3^2
- T_1^3 T_2^3 T_3^2~~.
\eea
The above refinement of the mesonic Hilbert series and the plethystic logarithm illustrates the conical structure of the toric Calalbi-Yau 3-fold.
\\

\subsection{Model 3 Phase b}

\begin{figure}[H]
\begin{center}
\includegraphics[trim=0cm 0cm 0cm 0cm,height=4.5 cm]{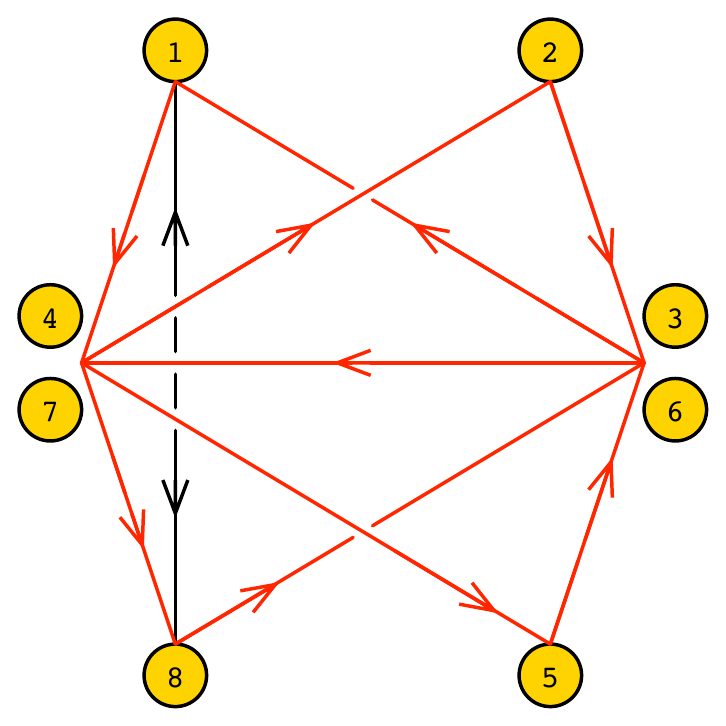}
\includegraphics[width=5 cm]{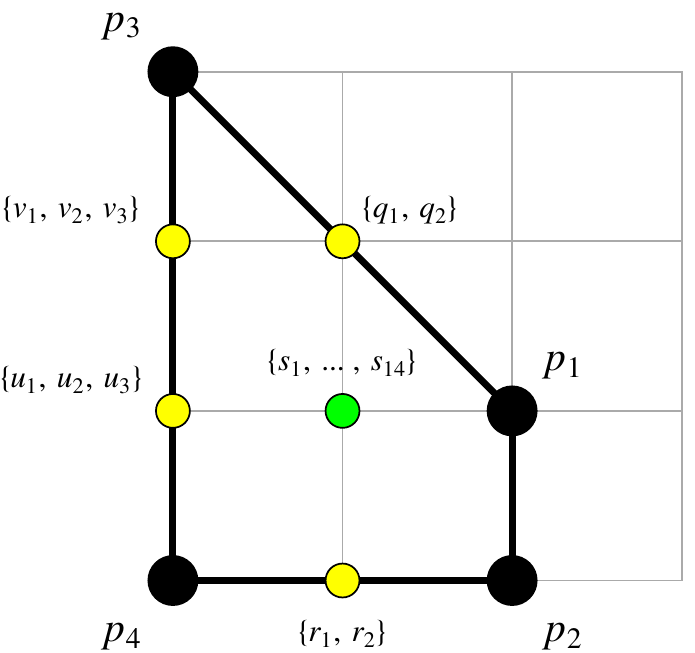}
\includegraphics[width=5 cm]{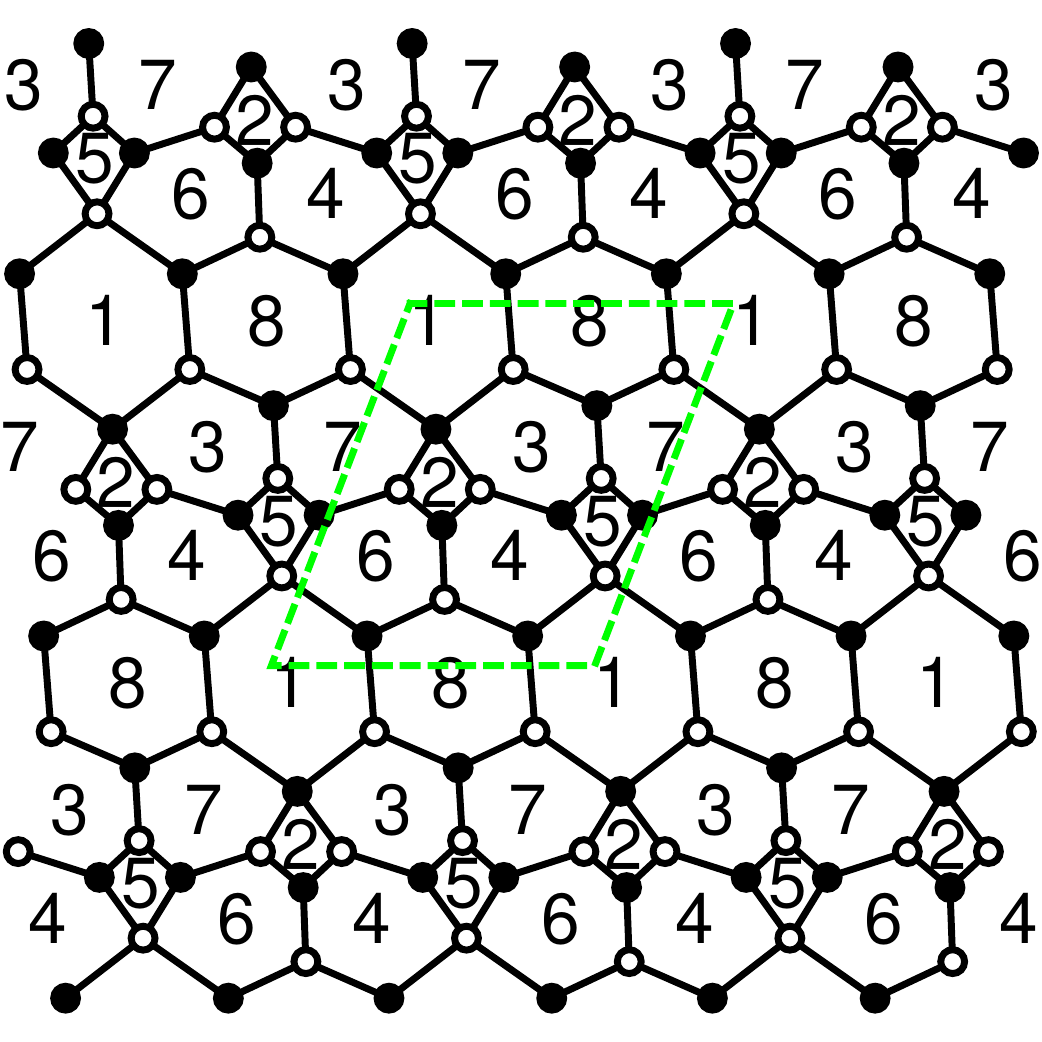}
\caption{The quiver, toric diagram, and brane tiling of Model 3b. The red arrows in the quiver indicate all possible connections between blocks of nodes. \label{f3b}}
 \end{center}
 \end{figure}
 
 \noindent The superpotential is 
\beal{esm3b_00}
W&=& 
+ X_{31} X_{18} X_{83}  
+ X_{42} X_{23} X_{34}  
+ X_{53} X_{37} X_{75}  
+ X_{67} X_{72} X_{26}    
\nn\\
&&
- X_{14} X_{48} X_{81}  
- X_{42} X_{26} X_{64}  
- X_{53} X_{34} X_{45}  
- X_{67} X_{75} X_{56}  
\nn\\
&&
+ X_{78} X_{81} X_{17}  
+ X_{86} X_{64} X_{48}  
+ X_{14} X_{45} X_{56} X_{61}
\nn\\
&&
- X_{78} X_{83} X_{37}  
- X_{86} X_{61} X_{18}  
- X_{17} X_{72} X_{23} X_{31} 
~.
  \eea
 
\noindent The perfect matching matrix is 
 
\noindent\makebox[\textwidth]{%
\scriptsize
$
P=
\left(
\begin{array}{c|cccc|cc|cc|ccc|ccc|cccccccccccccc}
 \; & p_1& p_2& p_3& p_4& q_1& q_2& r_1& r_2& u_1& u_2& u_3& v_1& v_2& v_3& s_1& s_2& s_3& s_4& s_5& s_6& s_7& s_8& s_9& s_{10}& s_{11}& s_{12}& s_{13}& s_{14}\\
 \hline
 X_{37} & 1 & 1 & 0 & 0 & 1 & 0 & 1 & 0 & 0 & 0 & 0 & 0 & 0 & 0 & 1 & 1 & 1 & 1 &
   0 & 0 & 0 & 0 & 0 & 0 & 0 & 0 & 0 & 0 \\
 X_{18} & 1 & 1 & 0 & 0 & 0 & 1 & 1 & 0 & 0 & 0 & 0 & 0 & 0 & 0 & 1 & 0 & 1 & 0 &
   1 & 0 & 1 & 1 & 1 & 1 & 0 & 0 & 0 & 0 \\
 X_{81} & 1 & 1 & 0 & 0 & 1 & 0 & 0 & 1 & 0 & 0 & 0 & 0 & 0 & 0 & 0 & 1 & 0 & 1 &
   0 & 1 & 0 & 0 & 0 & 0 & 1 & 1 & 1 & 1 \\
 X_{64} & 1 & 1 & 0 & 0 & 0 & 1 & 0 & 1 & 0 & 0 & 0 & 0 & 0 & 0 & 1 & 1 & 0 & 0 &
   1 & 1 & 0 & 0 & 0 & 0 & 0 & 0 & 0 & 0 \\
 X_{67} & 1 & 0 & 1 & 0 & 1 & 1 & 0 & 0 & 0 & 0 & 1 & 0 & 1 & 1 & 1 & 1 & 1 & 0 &
   0 & 1 & 0 & 0 & 0 & 0 & 0 & 0 & 0 & 0 \\
 X_{34} & 0 & 1 & 0 & 1 & 0 & 0 & 1 & 1 & 1 & 1 & 0 & 1 & 0 & 0 & 1 & 1 & 0 & 1 &
   1 & 0 & 0 & 0 & 0 & 0 & 0 & 0 & 0 & 0 \\
 X_{45} & 1 & 0 & 0 & 0 & 1 & 0 & 0 & 0 & 0 & 0 & 0 & 0 & 0 & 0 & 0 & 0 & 1 & 0 &
   0 & 0 & 1 & 1 & 0 & 0 & 1 & 1 & 0 & 0 \\
 X_{23} & 1 & 0 & 0 & 0 & 0 & 1 & 0 & 0 & 0 & 0 & 0 & 0 & 0 & 0 & 0 & 0 & 0 & 0 &
   0 & 1 & 1 & 0 & 1 & 0 & 1 & 0 & 1 & 0 \\
 X_{56} & 0 & 1 & 0 & 0 & 0 & 0 & 1 & 0 & 0 & 0 & 0 & 0 & 0 & 0 & 0 & 0 & 0 & 1 &
   0 & 0 & 0 & 0 & 1 & 1 & 0 & 0 & 1 & 1 \\
 X_{72} & 0 & 1 & 0 & 0 & 0 & 0 & 0 & 1 & 0 & 0 & 0 & 0 & 0 & 0 & 0 & 0 & 0 & 0 &
   1 & 0 & 0 & 1 & 0 & 1 & 0 & 1 & 0 & 1 \\
 X_{86} & 0 & 0 & 1 & 0 & 1 & 0 & 0 & 0 & 1 & 0 & 0 & 1 & 1 & 0 & 0 & 0 & 0 & 1 &
   0 & 0 & 0 & 0 & 0 & 0 & 1 & 1 & 1 & 1 \\
 X_{31} & 0 & 0 & 1 & 0 & 1 & 0 & 0 & 0 & 0 & 1 & 0 & 1 & 0 & 1 & 0 & 1 & 0 & 1 &
   0 & 0 & 0 & 0 & 0 & 0 & 0 & 0 & 0 & 0 \\
 X_{14} & 0 & 0 & 1 & 0 & 0 & 1 & 0 & 0 & 1 & 0 & 0 & 1 & 1 & 0 & 1 & 0 & 0 & 0 &
   1 & 0 & 0 & 0 & 0 & 0 & 0 & 0 & 0 & 0 \\
 X_{78} & 0 & 0 & 1 & 0 & 0 & 1 & 0 & 0 & 0 & 1 & 0 & 1 & 0 & 1 & 0 & 0 & 0 & 0 &
   1 & 0 & 1 & 1 & 1 & 1 & 0 & 0 & 0 & 0 \\
 X_{42} & 0 & 0 & 1 & 0 & 1 & 0 & 0 & 0 & 0 & 0 & 1 & 0 & 1 & 1 & 0 & 0 & 1 & 0 &
   0 & 0 & 0 & 1 & 0 & 1 & 0 & 1 & 0 & 1 \\
 X_{53} & 0 & 0 & 1 & 0 & 0 & 1 & 0 & 0 & 0 & 0 & 1 & 0 & 1 & 1 & 0 & 0 & 0 & 0 &
   0 & 1 & 0 & 0 & 1 & 1 & 0 & 0 & 1 & 1 \\
 X_{17} & 0 & 0 & 0 & 1 & 0 & 0 & 1 & 0 & 1 & 0 & 1 & 0 & 1 & 0 & 1 & 0 & 1 & 0 &
   0 & 0 & 0 & 0 & 0 & 0 & 0 & 0 & 0 & 0 \\
 X_{48} & 0 & 0 & 0 & 1 & 0 & 0 & 1 & 0 & 0 & 1 & 1 & 0 & 0 & 1 & 0 & 0 & 1 & 0 &
   0 & 0 & 1 & 1 & 1 & 1 & 0 & 0 & 0 & 0 \\
 X_{83} & 0 & 0 & 0 & 1 & 0 & 0 & 0 & 1 & 1 & 0 & 1 & 0 & 1 & 0 & 0 & 0 & 0 & 0 &
   0 & 1 & 0 & 0 & 0 & 0 & 1 & 1 & 1 & 1 \\
 X_{61} & 0 & 0 & 0 & 1 & 0 & 0 & 0 & 1 & 0 & 1 & 1 & 0 & 0 & 1 & 0 & 1 & 0 & 0 &
   0 & 1 & 0 & 0 & 0 & 0 & 0 & 0 & 0 & 0 \\
 X_{26} & 0 & 0 & 0 & 1 & 0 & 0 & 1 & 0 & 1 & 1 & 0 & 1 & 0 & 0 & 0 & 0 & 0 & 1 &
   0 & 0 & 1 & 0 & 1 & 0 & 1 & 0 & 1 & 0 \\
 X_{75} & 0 & 0 & 0 & 1 & 0 & 0 & 0 & 1 & 1 & 1 & 0 & 1 & 0 & 0 & 0 & 0 & 0 & 0 &
   1 & 0 & 1 & 1 & 0 & 0 & 1 & 1 & 0 & 0
\end{array}
\right)
$
}
\vspace{0.5cm}

 \noindent The F-term charge matrix $Q_F=\ker{(P)}$ is

\noindent\makebox[\textwidth]{%
\scriptsize
$
Q_F=
\left(
\begin{array}{cccc|cc|cc|ccc|ccc|cccccccccccccc}
 p_1& p_2& p_3& p_4& q_1& q_2& r_1& r_2& u_1& u_2& u_3& v_1& v_2& v_3& s_1& s_2& s_3& s_4& s_5& s_6& s_7& s_8& s_9& s_{10}& s_{11}& s_{12}& s_{13}& s_{14}\\
 \hline
 1 & 0 & 1 & 0 & -1 & -1 & 0 & 0 & 0 & 0 & 0 & 0 & 0 & 0 & 0 & 0 & 0 & 0 & 0 & 0 & 0 & 0 & 0 & 0 & 0 & 0 & 0 & 0 \\
 0 & 1 & 0 & 1 & 0 & 0 & -1 & -1 & 0 & 0 & 0 & 0 & 0 & 0 & 0 & 0 & 0 & 0 & 0 & 0 & 0 & 0 & 0 & 0 & 0 & 0 & 0 & 0 \\
 1 & 0 & 0 & 0 & 0 & 0 & 0 & 0 & 1 & 0 & 0 & 0 & 0 & 0 & -1 & 0 & 0 & 0 & 0 & 0 & 0 & 0 & 0 & 0 & -1 & 0 & 0 & 0 \\
 1 & 0 & 0 & 0 & 0 & 0 & 0 & 0 & 0 & 1 & 0 & 0 & 0 & 0 & 0 & -1 & 0 & 0 & 0 & 0 & -1 & 0 & 0 & 0 & 0 & 0 & 0 & 0 \\
 0 & 1 & 0 & 0 & 0 & 0 & 0 & 0 & 0 & 0 & 0 & 0 & 1 & 0 & -1 & 0 & 0 & 0 & 0 & 0 & 0 & 0 & 0 & 0 & 0 & 0 & 0 & -1 \\
 0 & 1 & 0 & 0 & 0 & 0 & 0 & 0 & 0 & 0 & 0 & 0 & 0 & 1 & 0 & -1 & 0 & 0 & 0 & 0 & 0 & 0 & 0 & -1 & 0 & 0 & 0 & 0 \\
 0 & 1 & 0 & 0 & 0 & 0 & -1 & 0 & 0 & 1 & 0 & 0 & 0 & 0 & 1 & -1 & 0 & 0 & -1 & 0 & 0 & 0 & 0 & 0 & 0 & 0 & 0 & 0 \\
 0 & 0 & 1 & 1 & 0 & 0 & 0 & 0 & -1 & 0 & 0 & 0 & 0 & -1 & 0 & 0 & 0 & 0 & 0 & 0 & 0 & 0 & 0 & 0 & 0 & 0 & 0 & 0 \\
 0 & 0 & 1 & 1 & 0 & 0 & 0 & 0 & 0 & -1 & 0 & 0 & -1 & 0 & 0 & 0 & 0 & 0 & 0 & 0 & 0 & 0 & 0 & 0 & 0 & 0 & 0 & 0 \\
 0 & 0 & 0 & 1 & 0 & 0 & 0 & 0 & -1 & 0 & -1 & 0 & 1 & 0 & 0 & 0 & 0 & 0 & 0 & 0 & 0 & 0 & 0 & 0 & 0 & 0 & 0 & 0 \\
 0 & 0 & 0 & 1 & 1 & 0 & 0 & 0 & -1 & 0 & 0 & 0 & 0 & 0 & 1 & -1 & -1 & 0 & 0 & 0 & 0 & 0 & 0 & 0 & 0 & 0 & 0 & 0 \\
 0 & 0 & 0 & 1 & 0 & 0 & 0 & 0 & -1 & -1 & 0 & 1 & 0 & 0 & 0 & 0 & 0 & 0 & 0 & 0 & 0 & 0 & 0 & 0 & 0 & 0 & 0 & 0 \\
 0 & 0 & 0 & 1 & 1 & 0 & 0 & 0 & -1 & 0 & 0 & 0 & 0 & 0 & 0 & -1 & 0 & 0 & 1 & 0 & 0 & -1 & 0 & 0 & 0 & 0 & 0 & 0 \\
 0 & 0 & 0 & 1 & 0 & 1 & 0 & 0 & 0 & -1 & 0 & 0 & 0 & 0 & -1 & 1 & 0 & 0 & 0 & -1 & 0 & 0 & 0 & 0 & 0 & 0 & 0 & 0 \\
 0 & 0 & 0 & 1 & 0 & 0 & -1 & 0 & -1 & 0 & 0 & 0 & 0 & 0 & 1 & -1 & 0 & 1 & 0 & 0 & 0 & 0 & 0 & 0 & 0 & 0 & 0 & 0 \\
 0 & 0 & 0 & 0 & 1 & 0 & 0 & 1 & 0 & 0 & 0 & 0 & 0 & 0 & 0 & -1 & 0 & 0 & 0 & 0 & 0 & 0 & 0 & 0 & 0 & -1 & 0 & 0 \\
 0 & 0 & 0 & 0 & 0 & 1 & 1 & 0 & 0 & 0 & 0 & 0 & 0 & 0 & -1 & 0 & 0 & 0 & 0 & 0 & 0 & 0 & -1 & 0 & 0 & 0 & 0 & 0 \\
 0 & 0 & 0 & 0 & 0 & 0 & 0 & 0 & 0 & 0 & 0 & 0 & 0 & 0 & 0 & 1 & 0 & -1 & 0 & -1 & 0 & 0 & 0 & 0 & 0 & 0 & 1 & 0
\end{array}
\right)
$
}
\vspace{0.5cm}

\noindent The D-term charge matrix is

\noindent\makebox[\textwidth]{%
\scriptsize
$
Q_D=
\left(
\begin{array}{cccc|cc|cc|ccc|ccc|cccccccccccccc}
 p_1& p_2& p_3& p_4& q_1& q_2& r_1& r_2& u_1& u_2& u_3& v_1& v_2& v_3& s_1& s_2& s_3& s_4& s_5& s_6& s_7& s_8& s_9& s_{10}& s_{11}& s_{12}& s_{13}& s_{14}\\
 \hline
 0 & 0 & 0 & 0 & 0 & 0 & 0 & 0 & 0 & 0 & 0 & 0 & 0 & 0 & 0 & 1 & -1 & 0 & 0 & 0 & 0 & 0
   & 0 & 0 & 0 & 0 & 0 & 0 \\
 0 & 0 & 0 & 0 & 0 & 0 & 0 & 0 & 0 & 0 & 0 & 0 & 0 & 0 & 0 & 0 & 1 & -1 & 0 & 0 & 0 & 0
   & 0 & 0 & 0 & 0 & 0 & 0 \\
 0 & 0 & 0 & 0 & 0 & 0 & 0 & 0 & 0 & 0 & 0 & 0 & 0 & 0 & 0 & 0 & 0 & 1 & -1 & 0 & 0 & 0
   & 0 & 0 & 0 & 0 & 0 & 0 \\
 0 & 0 & 0 & 0 & 0 & 0 & 0 & 0 & 0 & 0 & 0 & 0 & 0 & 0 & 0 & 0 & 0 & 0 & 1 & -1 & 0 & 0
   & 0 & 0 & 0 & 0 & 0 & 0 \\
 0 & 0 & 0 & 0 & 0 & 0 & 0 & 0 & 0 & 0 & 0 & 0 & 0 & 0 & 0 & 0 & 0 & 0 & 0 & 1 & -1 & 0
   & 0 & 0 & 0 & 0 & 0 & 0 \\
 0 & 0 & 0 & 0 & 0 & 0 & 0 & 0 & 0 & 0 & 0 & 0 & 0 & 0 & 0 & 0 & 0 & 0 & 0 & 0 & 1 & -1
   & 0 & 0 & 0 & 0 & 0 & 0 \\
 0 & 0 & 0 & 0 & 0 & 0 & 0 & 0 & 0 & 0 & 0 & 0 & 0 & 0 & 0 & 0 & 0 & 0 & 0 & 0 & 0 & 1
   & -1 & 0 & 0 & 0 & 0 & 0
\end{array}
\right)
$
}
\vspace{0.5cm}

The total charge matrix does not exhibit repeated columns. Accordingly, the global symmetry is $U(1)_{f_1} \times U(1)_{f_2} \times U(1)_R$. The mesonic charges on the GLSM fields with non-zero R-charges are the same as for Model 3a and are shown in \tref{t3a}.

Products of non-extremal perfect matchings are expressed in terms of single variables as follows
\beal{esx3b_1}
q = q_1 q_2 ~,~
r = r_1 r_2 ~,~
u = u_1 u_2 u_3 ~,~
v = v_1 v_2 v_3 ~,~
s = \prod_{m=1}^{14} s_m~.
\eea
The fugacity $t_\alpha$ counts GLSM fields corresponding to extremal perfect matchings $p_\alpha$. The fugacity $y_q$ for instance counts the product of non-extremal perfect matchings $q$ shown above.

The refined mesonic Hilbert series and the corresponding plethystic logarithm are found using the Molien integral formula in \eref{es12_2}. The Hilbert series is found to be the same as the one for Model 3a given in \eref{esm3a_1}, \eref{esm3a_2} and \eref{esm3a_3}. Accordingly, the mesonic moduli spaces of Model 3a and 3b are the same, with the corresponding quiver gauge theories being toric (Seiberg) duals.

The generators in terms of all perfect matchings of Model 3b are given in \tref{t3agen} with the corresponding mesonic symmetry charges. The corresponding mesonic generators in terms of quiver fields are given in \tref{t3bgen2}. The mesonic moduli space is a complete intersection, and the generators satisfy the relation in \eref{esm3a_3b}.

\comment{
\begin{table}[h!]
\centering

\resizebox{\hsize}{!}{

\begin{minipage}[!b]{0.6\textwidth}
\begin{tabular}{|l|c|c|}
\hline
Generator & $U(1)_{f_1}$ & $U(1)_{f_2}$ 
\\
\hline
\hline
$A_1=
p_1^2 p_2^2~
q~
r~
s$
& 1 & 0
\nn\\
$A_2=
p_3^2 p_4^2~
q~
r_1 r_2~
\prod_{i=1}^{3} u_i^2 v_i^2~
\prod_{m=1}^{14} s_m$
& -1 & 0
\nn\\
$B=
p_1 p_2 p_3 p_4~
q_1 q_2~
r_1 r_2~
\prod_{i=1}^{3} u_i v_i~
\prod_{m=1}^{14} s_m$
& 0 & 0
\nn\\
$C_1=
p_1 p_3^3~
q_1^2 q_2^2~
\prod_{i=1}^{3} u_i^2 v_i~
\prod_{m=1}^{14} s_m$
& -1 & -1
\nn\\
$C_2=
p_2 p_4^3~
r_1^2 r_2^2~
\prod_{i=1}^{3} u_i v_i^2~
\prod_{m=1}^{14} s_m$
& 0 & 1
\nn\\
\hline
\end{tabular}
\end{minipage}
\hspace{2cm}
\begin{minipage}[!b]{0.25\textwidth}
\includegraphics[width=3.5 cm]{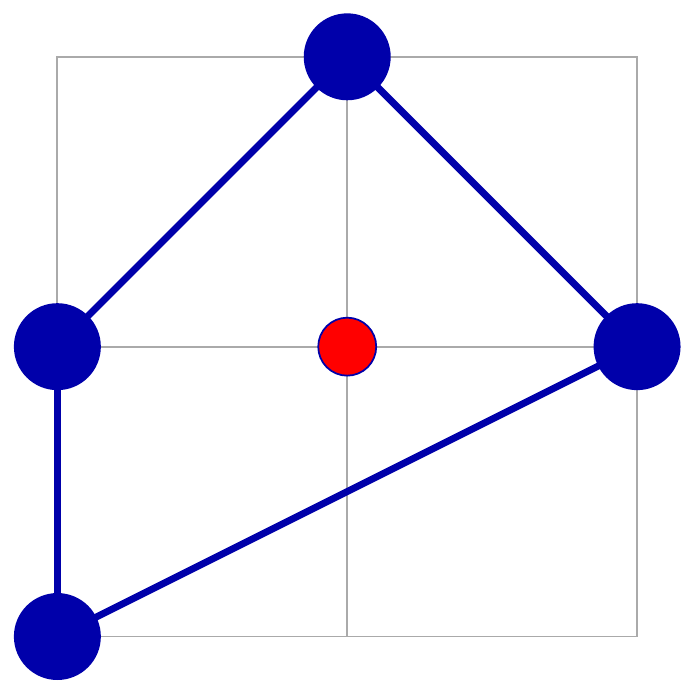}
\end{minipage}

}

\caption{The generators and lattice of generators of the mesonic moduli space of Model 3b in terms of GLSM fields with the corresponding flavor charges.\label{t3bgen}\label{f3bgen}} 
\end{table}
}

\begin{table}[h!]
\centering

\resizebox{\hsize}{!}{
\begin{tabular}{|l|c|c|}
\hline
Generator & $U(1)_{f_1}$ & $U(1)_{f_2}$ 
\\
\hline
\hline
$
X_{18} X_{81}=  X_{23} X_{37} X_{72}=  X_{45} X_{56} X_{64}
$
& 1 & 0
\nn\\
$
X_{14} X_{42} X_{26} X_{61}=  X_{14} X_{48} X_{83} X_{31}=  X_{14} X_{48} X_{86} X_{61}=  X_{17} X_{75} X_{53} X_{31}=  X_{17} X_{78} X_{83} X_{31}=  X_{17} X_{78} X_{86} X_{61}
$
& -1 & 0
\nn\\
$X_{14} X_{45} X_{56} X_{61}=  X_{17} X_{72} X_{23} X_{31}=  X_{14} X_{48} X_{81}=  X_{17} X_{78} X_{81}=  X_{18} X_{83} X_{31}=  X_{18} X_{86} X_{61}=  X_{23} X_{34} X_{42}$
& 0 & 0
\nn\\
$
=  X_{26} X_{64} X_{42}=  X_{26} X_{67} X_{72}=  X_{34} X_{45} X_{53}=  X_{37} X_{75} X_{53}=  X_{37} X_{78} X_{83}=  X_{48} X_{86} X_{64}=  X_{56} X_{67} X_{75}$
& &
\nn\\
$
 X_{34} X_{48} X_{83}=  X_{17} X_{72} X_{26} X_{61}=  X_{17} X_{75} X_{56} X_{61}
$
& -1 & -1
\nn\\
$
 X_{67} X_{78} X_{86}=  X_{14} X_{42} X_{23} X_{31}=  X_{14} X_{45} X_{53} X_{31}
$
& 0 & 1
\nn\\
\hline
\end{tabular}
}

\caption{The generators in terms of bifundamental fields (Model 3b).\label{t3bgen2}\label{f3bgen2}} 
\end{table}

\section{Model 4: $\mathcal{C}/\mathbb{Z}_2\times\mathbb{Z}_2~(1,0,0,1)(0,1,1,0),~\text{PdP}_5$}

\subsection{Model 4 Phase a}

\begin{figure}[H]
\begin{center}
\includegraphics[trim=0cm 0cm 0cm 0cm,width=4.5 cm]{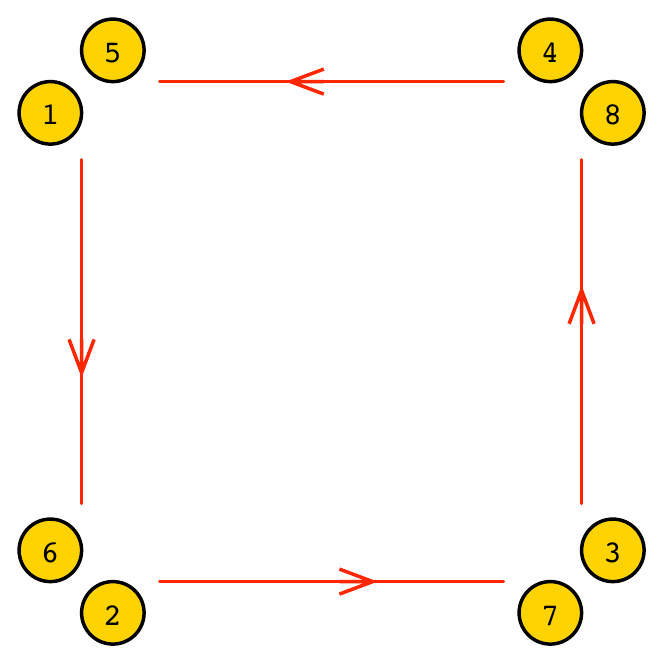}
\includegraphics[width=5 cm]{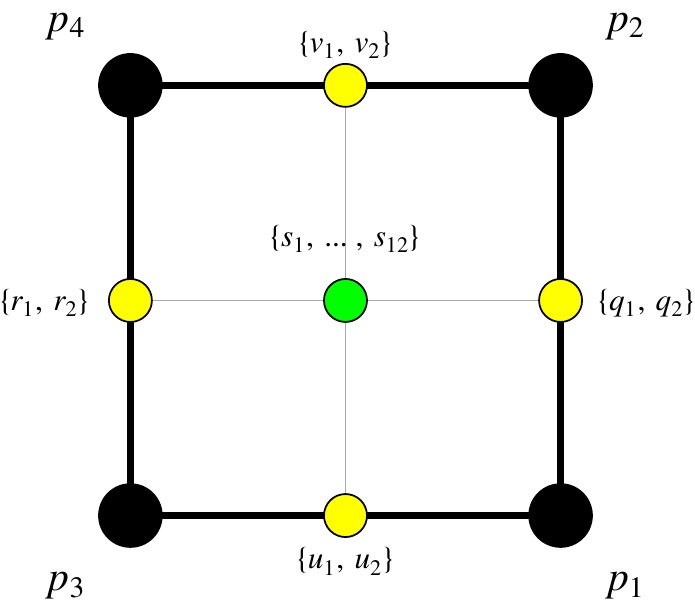}
\includegraphics[width=5 cm]{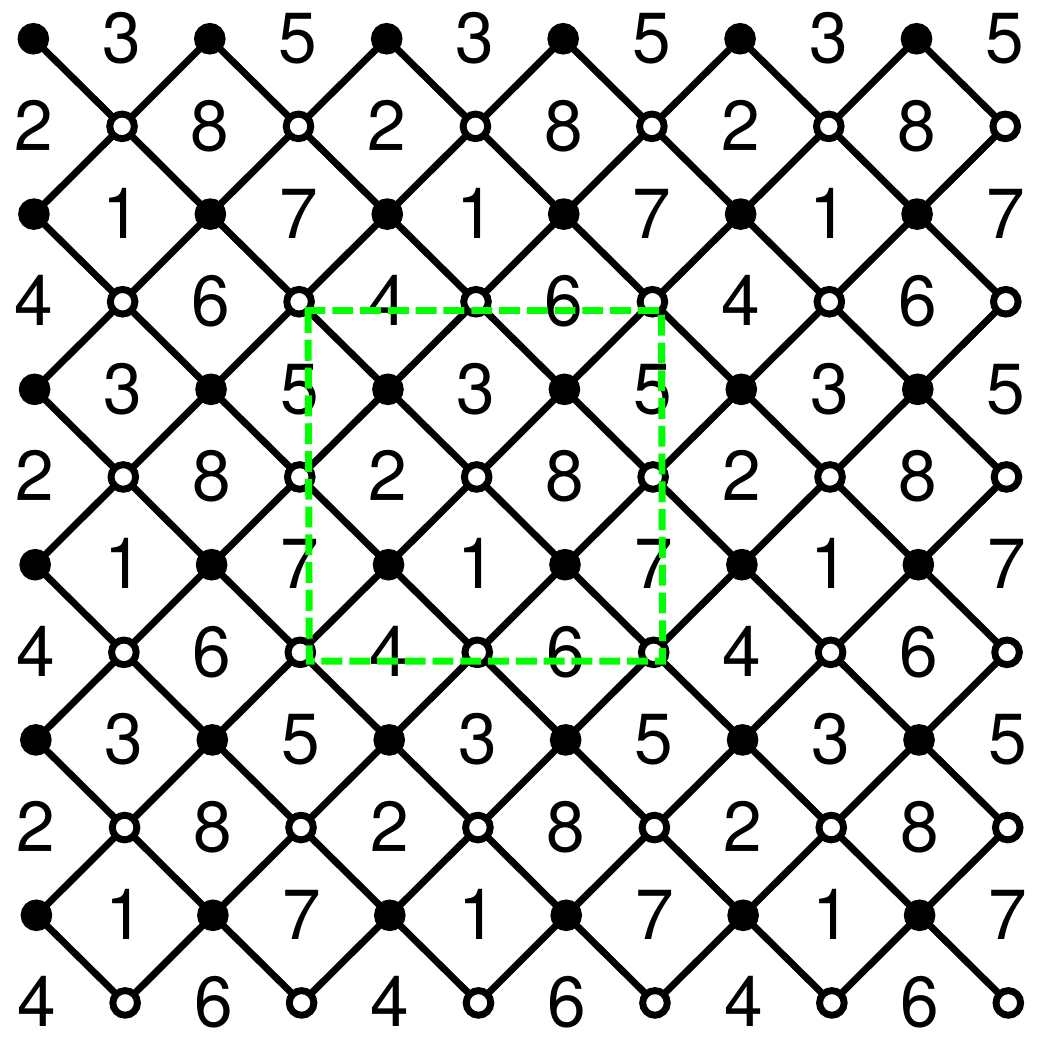}
\caption{The quiver, toric diagram, and brane tiling of Model 4a. The red arrows in the quiver indicate all possible connections between blocks of nodes. \label{f4a}}
 \end{center}
 \end{figure}
 
 \noindent The superpotential is 
\beal{esm4a_00}
W&=&
+ X_{23} X_{38} X_{81} X_{12}    
+ X_{41} X_{16} X_{63} X_{34}  
+ X_{67} X_{74} X_{45} X_{56}  
+ X_{85} X_{52} X_{27} X_{78} 
\nn\\
&&
- X_{27} X_{74} X_{41} X_{12}  
- X_{45} X_{52} X_{23} X_{34}  
- X_{63} X_{38} X_{85} X_{56}
- X_{81} X_{16} X_{67} X_{78}  
\nn\\
  \eea
 
 \noindent The perfect matching matrix is 
 
\noindent\makebox[\textwidth]{%
\footnotesize
$
P=
\left(
\begin{array}{c|cccc|cc|cc|cc|cc|cccccccccccc}
 \; & p_1& p_2& p_3& p_4& q_1& q_2& r_1& r_2& u_1& u_2& v_1& v_2& s_1& s_2& s_3& s_4& s_5& s_6& s_7& s_8& s_9& s_{10}& s_{11}& s_{12} \\
 \hline
 X_{23} & 1 & 0 & 0 & 0 & 1 & 0 & 0 & 0 & 1 & 0 & 0 & 0 & 1 & 1 & 1 & 0 & 0 & 0 &
   0 & 0 & 0 & 0 & 0 & 0 \\
 X_{41} & 1 & 0 & 0 & 0 & 1 & 0 & 0 & 0 & 0 & 1 & 0 & 0 & 0 & 0 & 0 & 1 & 1 & 1 &
   0 & 0 & 0 & 0 & 0 & 0 \\
 X_{85} & 1 & 0 & 0 & 0 & 0 & 1 & 0 & 0 & 1 & 0 & 0 & 0 & 0 & 0 & 0 & 1 & 0 & 0 &
   1 & 1 & 0 & 0 & 0 & 0 \\
 X_{67} & 1 & 0 & 0 & 0 & 0 & 1 & 0 & 0 & 0 & 1 & 0 & 0 & 1 & 0 & 0 & 0 & 0 & 0 &
   0 & 0 & 1 & 1 & 0 & 0 \\
 X_{56} & 0 & 1 & 0 & 0 & 1 & 0 & 0 & 0 & 0 & 0 & 1 & 0 & 0 & 1 & 0 & 0 & 1 & 0 &
   0 & 0 & 0 & 0 & 1 & 0 \\
 X_{78} & 0 & 1 & 0 & 0 & 1 & 0 & 0 & 0 & 0 & 0 & 0 & 1 & 0 & 0 & 1 & 0 & 0 & 1 &
   0 & 0 & 0 & 0 & 0 & 1 \\
 X_{34} & 0 & 1 & 0 & 0 & 0 & 1 & 0 & 0 & 0 & 0 & 1 & 0 & 0 & 0 & 0 & 0 & 0 & 0 &
   1 & 0 & 1 & 0 & 0 & 1 \\
 X_{12} & 0 & 1 & 0 & 0 & 0 & 1 & 0 & 0 & 0 & 0 & 0 & 1 & 0 & 0 & 0 & 0 & 0 & 0 &
   0 & 1 & 0 & 1 & 1 & 0 \\
 X_{74} & 0 & 0 & 1 & 0 & 0 & 0 & 1 & 0 & 1 & 0 & 0 & 0 & 0 & 0 & 1 & 0 & 0 & 0 &
   1 & 0 & 0 & 0 & 0 & 1 \\
 X_{52} & 0 & 0 & 1 & 0 & 0 & 0 & 1 & 0 & 0 & 1 & 0 & 0 & 0 & 0 & 0 & 0 & 1 & 0 &
   0 & 0 & 0 & 1 & 1 & 0 \\
 X_{16} & 0 & 0 & 1 & 0 & 0 & 0 & 0 & 1 & 1 & 0 & 0 & 0 & 0 & 1 & 0 & 0 & 0 & 0 &
   0 & 1 & 0 & 0 & 1 & 0 \\
 X_{38} & 0 & 0 & 1 & 0 & 0 & 0 & 0 & 1 & 0 & 1 & 0 & 0 & 0 & 0 & 0 & 0 & 0 & 1 &
   0 & 0 & 1 & 0 & 0 & 1 \\
 X_{81} & 0 & 0 & 0 & 1 & 0 & 0 & 1 & 0 & 0 & 0 & 1 & 0 & 0 & 0 & 0 & 1 & 1 & 0 &
   1 & 0 & 0 & 0 & 0 & 0 \\
 X_{63} & 0 & 0 & 0 & 1 & 0 & 0 & 1 & 0 & 0 & 0 & 0 & 1 & 1 & 0 & 1 & 0 & 0 & 0 &
   0 & 0 & 0 & 1 & 0 & 0 \\
 X_{27} & 0 & 0 & 0 & 1 & 0 & 0 & 0 & 1 & 0 & 0 & 1 & 0 & 1 & 1 & 0 & 0 & 0 & 0 &
   0 & 0 & 1 & 0 & 0 & 0 \\
 X_{45} & 0 & 0 & 0 & 1 & 0 & 0 & 0 & 1 & 0 & 0 & 0 & 1 & 0 & 0 & 0 & 1 & 0 & 1 &
   0 & 1 & 0 & 0 & 0 & 0
\end{array}
\right)
$
}
\vspace{0.5cm}

 \noindent The F-term charge matrix $Q_F=\ker{(P)}$ is

\noindent\makebox[\textwidth]{%
\footnotesize
$
Q_F=
\left(
\begin{array}{cccc|cc|cc|cc|cc|cccccccccccc}
 p_1& p_2& p_3& p_4& q_1& q_2& r_1& r_2& u_1& u_2& v_1& v_2& s_1& s_2& s_3& s_4& s_5& s_6& s_7& s_8& s_9& s_{10}& s_{11}& s_{12} \\
 \hline
 1 & 1 & 0 & 0 & -1 & -1 & 0 & 0 & 0 & 0 & 0 & 0 & 0 & 0 & 0 & 0 & 0 & 0 & 0 & 0 & 0 & 0 & 0 & 0 \\
 0 & 0 & 1 & 1 & 0 & 0 & -1 & -1 & 0 & 0 & 0 & 0 & 0 & 0 & 0 & 0 & 0 & 0 & 0 & 0 & 0 & 0 & 0 & 0 \\
 1 & 0 & 1 & 0 & 0 & 0 & 0 & 0 & -1 & -1 & 0 & 0 & 0 & 0 & 0 & 0 & 0 & 0 & 0 & 0 & 0 & 0 & 0 & 0 \\
 0 & 1 & 0 & 1 & 0 & 0 & 0 & 0 & 0 & 0 & -1 & -1 & 0 & 0 & 0 & 0 & 0 & 0 & 0 & 0 & 0 & 0 & 0 & 0 \\
 1 & 0 & 0 & 0 & 0 & 0 & 1 & 0 & 0 & 0 & 1 & 0 & -1 & 0 & 0 & 0 & -1 & 0 & -1 & 0 & 0 & 0 & 0 & 0 \\
 1 & 0 & 0 & 0 & 0 & 0 & 0 & 1 & 0 & 0 & 1 & 0 & 0 & -1 & 0 & -1 & 0 & 0 & 0 & 0 & -1 & 0 & 0 & 0 \\
 1 & 0 & 0 & 0 & 0 & 0 & 0 & 1 & 0 & 0 & 1 & 0 & 0 & -1 & 0 & 0 & 0 & -1 & -1 & 0 & -1 & 0 & 0 & 1 \\
 1 & 0 & 0 & 1 & 0 & 0 & 0 & 0 & 0 & 0 & 0 & 0 & -1 & 0 & 0 & -1 & 0 & 0 & 0 & 0 & 0 & 0 & 0 & 0 \\
 0 & 0 & 1 & 1 & 0 & 1 & -1 & 0 & 0 & 0 & 0 & 0 & 0 & 0 & 0 & 0 & 0 & 0 & 0 & -1 & -1 & 0 & 0 & 0 \\
 0 & 0 & 0 & 0 & 1 & 0 & 0 & 1 & 0 & 0 & 0 & 0 & 0 & -1 & 0 & 0 & 0 & -1 & 0 & 0 & 0 & 0 & 0 & 0 \\
 0 & 0 & 0 & 0 & 0 & 1 & 1 & 0 & 0 & 0 & 0 & 0 & 0 & 0 & 0 & 0 & 0 & 0 & -1 & 0 & 0 & -1 & 0 & 0 \\
 0 & 0 & 0 & 0 & 0 & 0 & 0 & 0 & 1 & 0 & 1 & 0 & 0 & -1 & 0 & 0 & 0 & 0 & -1 & 0 & 0 & 0 & 0 & 0 \\
 0 & 0 & 0 & 0 & 0 & 0 & 0 & 0 & 0 & 0 & 0 & 0 & 1 & -1 & 0 & 0 & 0 & 0 & 0 & 0 & 0 & -1 & 1 & 0 \\
 0 & 0 & 0 & 0 & 0 & 0 & 0 & 0 & 0 & 0 & 0 & 0 & 1 & 0 & -1 & 0 & 0 & 0 & 0 & 0 & -1 & 0 & 0 & 1
    \end{array}
\right)
$
}
\vspace{0.5cm}

\noindent The D-term charge matrix is

\noindent\makebox[\textwidth]{%
\footnotesize
$
Q_D=
\left(
\begin{array}{cccc|cc|cc|cc|cc|cccccccccccc}
 p_1& p_2& p_3& p_4& q_1& q_2& r_1& r_2& u_1& u_2& v_1& v_2& s_1& s_2& s_3& s_4& s_5& s_6& s_7& s_8& s_9& s_{10}& s_{11}& s_{12} \\
 \hline
 0 & 0 & 0 & 0 & 0 & 0 & 0 & 0 & 0 & 0 & 0 & 0 & 0 & 0 & 1 & -1 & 0 & 0 & 0 & 0 & 0 & 0
   & 0 & 0 \\
 0 & 0 & 0 & 0 & 0 & 0 & 0 & 0 & 0 & 0 & 0 & 0 & 0 & 0 & 0 & 1 & -1 & 0 & 0 & 0 & 0 & 0
   & 0 & 0 \\
 0 & 0 & 0 & 0 & 0 & 0 & 0 & 0 & 0 & 0 & 0 & 0 & 0 & 0 & 0 & 0 & 1 & -1 & 0 & 0 & 0 & 0
   & 0 & 0 \\
 0 & 0 & 0 & 0 & 0 & 0 & 0 & 0 & 0 & 0 & 0 & 0 & 0 & 0 & 0 & 0 & 0 & 1 & -1 & 0 & 0 & 0
   & 0 & 0 \\
 0 & 0 & 0 & 0 & 0 & 0 & 0 & 0 & 0 & 0 & 0 & 0 & 0 & 0 & 0 & 0 & 0 & 0 & 1 & -1 & 0 & 0
   & 0 & 0 \\
 0 & 0 & 0 & 0 & 0 & 0 & 0 & 0 & 0 & 0 & 0 & 0 & 0 & 0 & 0 & 0 & 0 & 0 & 0 & 1 & -1 & 0
   & 0 & 0 \\
 0 & 0 & 0 & 0 & 0 & 0 & 0 & 0 & 0 & 0 & 0 & 0 & 0 & 0 & 0 & 0 & 0 & 0 & 0 & 0 & 1 & -1
   & 0 & 0
\end{array}
\right)
$
}
\vspace{0.5cm}

The total charge matrix $Q_t$ does not exhibit repeated columns. Accordingly, the global symmetry is $U(1)_{f_1} \times U(1)_{f_2} \times U(1)_R$. The mesonic charges on the extremal perfect matchings are found following the discussion in \sref{s1_3}. They are shown in \tref{t4a}.

\begin{table}[H]
\centering
\begin{tabular}{|c||c|c|c||l|} 
\hline
\; & $U(1)_{f_1}$ & $U(1)_{f_2}$ & $U(1)_R$ & fugacity \\
\hline
\hline
$p_1$ & 1/4 & -1/4 & $1/2$ &  	$t_1$\\
$p_2$ & 1/4 &  1/4 & $1/2$ &  	$t_2$\\
$p_3$ &-1/4 & -1/4 & $1/2$ &  	$t_3$\\
$p_4$ &-1/4 &  1/4 & $1/2$ &  	$t_4$\\
\hline
\end{tabular}
\caption{The GLSM fields corresponding to extremal points of the toric diagram with their mesonic charges (Model 4a).\label{t4a}}
\end{table}

Products of GLSM fields corresponding to non-extremal perfect matchings are called by single variables as follows
\beal{esx4a_1}
q = q_1 q_2 ~,~
r = r_1 r_2 ~,~
u = u_1 u_2 ~,~
v = v_1 v_2 ~,~
s = \prod_{m=1}^{12} s_m~.
\eea
The fugacity $t_\alpha$ counts extremal perfect matchings $p_\alpha$. The fugacity $y_q$ for instance corresponds to the product of non-extremal perfect matchings $q$ shown above.

The refined mesonic Hilbert series of Model 4a is calculated using the Molien integral formula in \eref{es12_2}. It is
 \beal{esm4a_1}
&&g_{1}(t_\alpha,y_{q},y_{r} ,y_{u},y_{v},y_{s}; \mathcal{M}^{mes}_{4a})=(1 - y_{q}^2 y_{r}^2 y_{u}^2 y_{v}^2 y_{s}^2 ~ t_1^2 t_2^2 t_3^2 t_4^2)^2
\nn\\
&&
\hspace{1cm}
\times
\frac{
1
}{
(1 - y_{q}^2 y_{u} y_{v} y_{s} ~ t_1^2 t_2^2) 
(1 - y_{q} y_{r} y_{u}^2 y_{s} ~ t_1^2 t_3^2) 
(1 - y_{q} y_{r} y_{v}^2 y_{s} ~ t_2^2 t_4^2) 
}
\nn\\
&&
\hspace{1cm}
\times
\frac{
1
}{
(1 - y_{r}^2 y_{u} y_{v} y_{s}~ t_3^2 t_4^2)
(1 - y_{q} y_{r} y_{u} y_{v} y_{s} ~ t_1 t_2 t_3 t_4)
}
~~.
\nn\\
\eea
 The plethystic logarithm of the mesonic Hilbert series is
\beal{esm4a_2}
&&
PL[g_1(t_\alpha,y_{q},y_{r} ,y_{u},y_{v},y_{s};\mathcal{M}_{4a}^{mes})]
=
y_{q} y_{r} y_{u} y_{v} y_{s} ~ t_{1} t_{2} t_{3} t_{4}
+ y_{q}^2 y_{u} y_{v} y_{s} ~ t_{1}^2 t_{2}^2
+ y_{r}^2 y_{u} y_{v} y_{s} ~ t_{3}^2 t_{4}^2
\nn\\
&&
\hspace{1cm}
+ y_{q} y_{r} y_{v}^2 y_{s} ~ t_{2}^2 t_{4}^2
+ y_{q} y_{r} y_{u}^2 y_{s} ~ t_{1}^2 t_{3}^2
- 2 ~ y_{q}^2 y_{r}^2 y_{u}^2 y_{v}^2 y_{s}^2 ~ t_{1}^2 t_{2}^2 t_{3}^2 t_{4}^2
~~.
\nn\\
\eea
The finite plethystic logarithm indicates that the mesonic moduli space is a complete intersection.

With the fugacity map
\beal{em4a_y1}
f_1 
=
\frac{y_q~ t_1 t_2}{y_r~ t_3 t_4}
~,~
f_2 
=
\frac{y_v~ t_2 t_4}{y_u~ t_1 t_3}
~,~
t
=
y_q^{1/4} y_r^{1/4} y_u^{1/4} y_v^{1/4} y_s^{1/4} ~ t_1^{1/4} t_2^{1/4} t_3^{1/4} t_4^{1/4}
~,~
\eea
where the fugacities $f_1$, $f_2$ and $t$ count mesonic charges, the Hilbert series becomes
\beal{esm4a_3i}
g_1(t,f_1,f_2;\mathcal{M}_{4a}^{mes})=
\frac{
(1 - t^8)^2
}{
(1 - t^4) (1 - \frac{1}{f_1} t^4) (1 - f_1 t^4) (1 - \frac{1}{f_2} t^4) (1 - f_2 t^4)
}~~.
\eea
The corresponding plethystic logarithm is
\beal{esm4a_3}
PL[g_1(t,f_1,f_2;\mathcal{M}_{4a}^{mes})]
&=&
\left(
1+ f_1 + \frac{1}{f_1} + f_2 + \frac{1}{f_2}
\right)t^4
- 2 t^8~~.
   \eea
The above plethystic logarithm identifies the moduli space generators with their mesonic charges. They are summarized in \tref{t4agen}. The charge lattice of generators in \tref{t4agen} is the dual reflexive polygon of the toric diagram of Model 4a. The generators satisfy the following relations
\beal{esm4a_3b}
A_1 A_2 = B_1 B_2 = C^2 ~~.
\eea

\begin{table}[H]
\centering

\resizebox{\hsize}{!}{

\begin{minipage}[!b]{0.6\textwidth}
\begin{tabular}{|l|c|c|}
\hline
Generator & $U(1)_{f_1}$ & $U(1)_{f_2}$ 
\\
\hline
\hline
$
A_1=
p_1^2 p_3^2~
q~
r~
u^2~
s$
& 0 & -1
\nn\\
$
A_2=
p_2^2 p_4^2~
q~
r~
v^2~
s$
& 0 & 1
\nn\\
$
B_1=
p_1^2 p_2^2~
q^2~
u~
v~
s$
& 1 & 0
\nn\\
$
B_2=
p_3^2 p_4^2~ 
r^2~
u~
v~
s$
& -1 & 0
\nn\\
$
C=
p_1 p_2 p_3 p_4~
q~
r~
u~
v~
s$
& 0 & 0
\nn\\
\hline
\end{tabular}
\end{minipage}
\hspace{2cm}
\begin{minipage}[!b]{0.25\textwidth}
\includegraphics[width=3.5 cm]{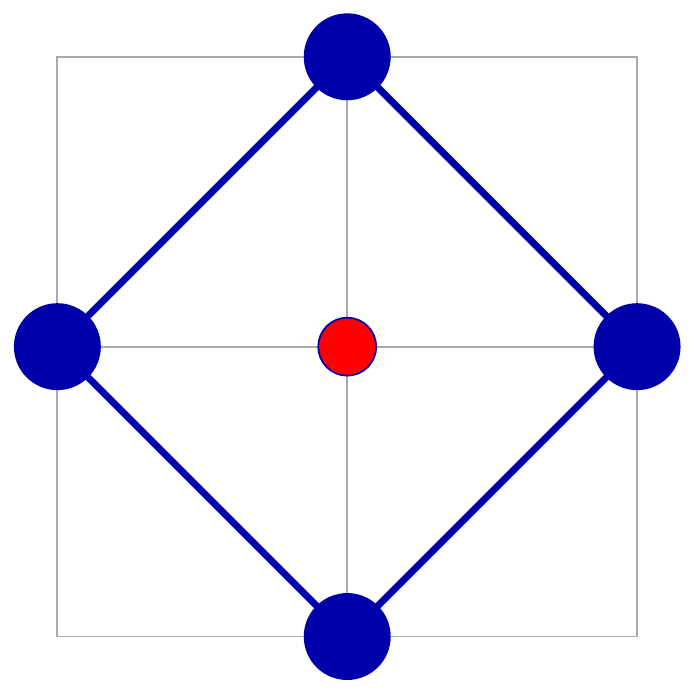}
\end{minipage}

}

\caption{The generators and lattice of generators of the mesonic moduli space of Model 4a in terms of GLSM fields with the corresponding flavor charges.\label{t4agen}\label{f4agen}} 
\end{table}

\begin{table}[H]
\centering

\resizebox{\hsize}{!}{
\begin{tabular}{|l|c|c|}
\hline
Generator & $U(1)_{f_1}$ & $U(1)_{f_2}$ 
\\
\hline
\hline
$ 
X_{16} X_{67} X_{74} X_{41}=  X_{23} X_{38} X_{85} X_{52}
$
& 0 & -1
\nn\\
$
X_{12} X_{23} X_{34} X_{41}=  X_{56} X_{67} X_{78} X_{85}
$
& 1 & 0
\nn\\
$
X_{12} X_{23} X_{38} X_{81}=  X_{12} X_{27} X_{74} X_{41}=  X_{16} X_{63} X_{34} X_{41}=  X_{16} X_{67} X_{78} X_{81}=  X_{23} X_{34} X_{45} X_{52}=  X_{27} X_{78} X_{85} X_{52}=  X_{38} X_{85} X_{56} X_{63}=  X_{45} X_{56} X_{67} X_{74}
$
& 0 & 0
\nn\\
$
X_{16} X_{63} X_{38} X_{81}=  X_{27} X_{74} X_{45} X_{52}
$
& -1 & 0
\nn\\
$
X_{12} X_{27} X_{78} X_{81}=  X_{34} X_{45} X_{56} X_{63}
$
& 0 & 1
\nn\\
\hline
\end{tabular}
}

\caption{The generators in terms of bifundamental fields (Model 4a).\label{t4agen2}\label{f4agen2}} 
\end{table}

The fugacities
\beal{esm4a_x1}
T_1 = \frac{y_r^2 y_u^2 y_s ~ t_1 t_3^3 t_4}{t_2}= \frac{t^4}{f_1 f_2}~,~
T_2 = \frac{y_q ~ t_1 t_2}{y_r ~ t_3 t_4} = f_1~,~
T_3 = \frac{y_v ~ t_2 t_4}{y_u ~ t_1 t_3} = f_2~,
\nn\\
\eea
can be introduced to rewrite the Hilbert series and plethystic logarithm as
\beal{esm4a_x2}
g_1(T_1,T_2,T_3;\mathcal{M}_{4a}^{mes})
=
\frac{(1-T_1^2 T_2^2 T_3^2)^2}{
(1-T_1 T_2 T_3)
(1-T_1 T_3)
(1-T_1 T_2^2 T_3)
(1-T_1 T_2)
(1-T_1 T_2 T_3^2)
}
\nn\\
\eea
and
\beal{esm4a_x3}
PL[g_1(T_1,T_2,T_3;\mathcal{M}_{4a}^{mes})]
=
T_1 T_2 T_3 + T_1 T_2^2 T_3 + T_1 T_3 + T_1 T_2 T_3^2 + T_1 T_2 - T_1^2 T_2^2 T_3^2
\nn\\
\eea
such that powers of the fugacities in the expressions are positive. This illustrates the cone structure of the variety.
\\

\subsection{Model 4 Phase b}

\begin{figure}[H]
\begin{center}
\includegraphics[trim=0cm 0cm 0cm 0cm,height=4.5 cm]{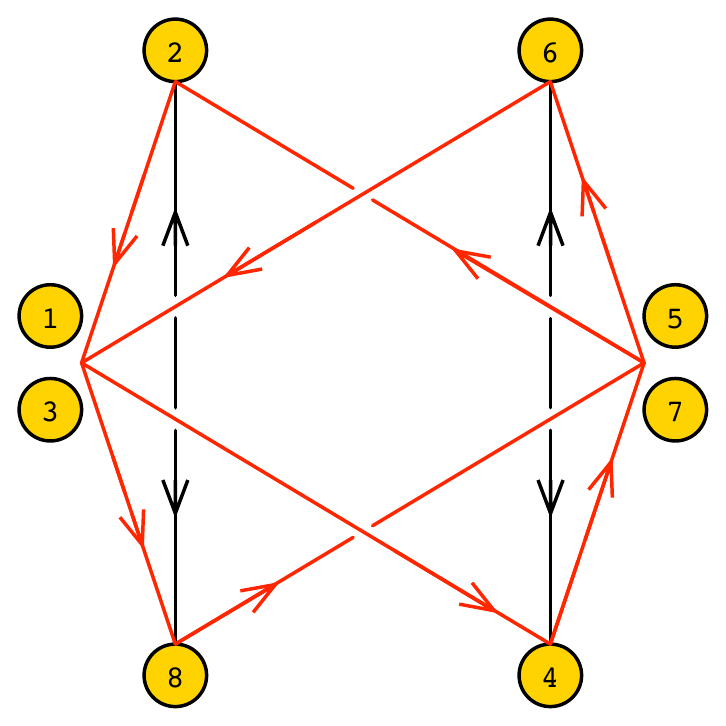}
\includegraphics[width=5 cm]{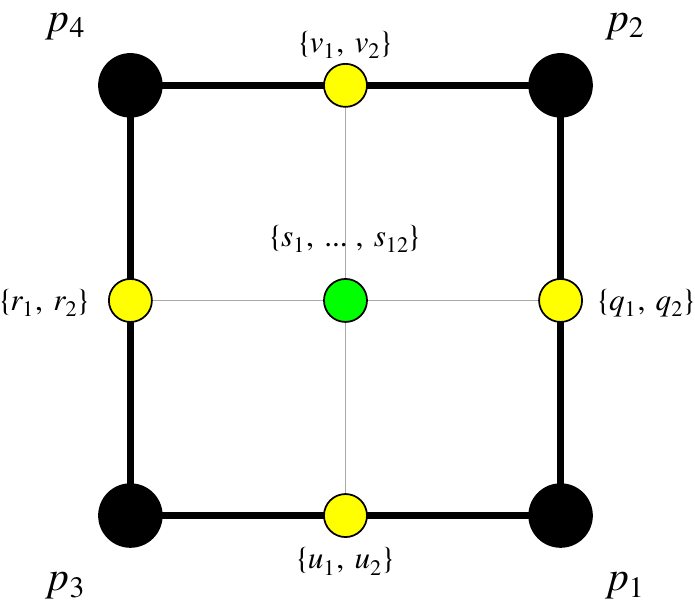}
\includegraphics[width=4.7 cm]{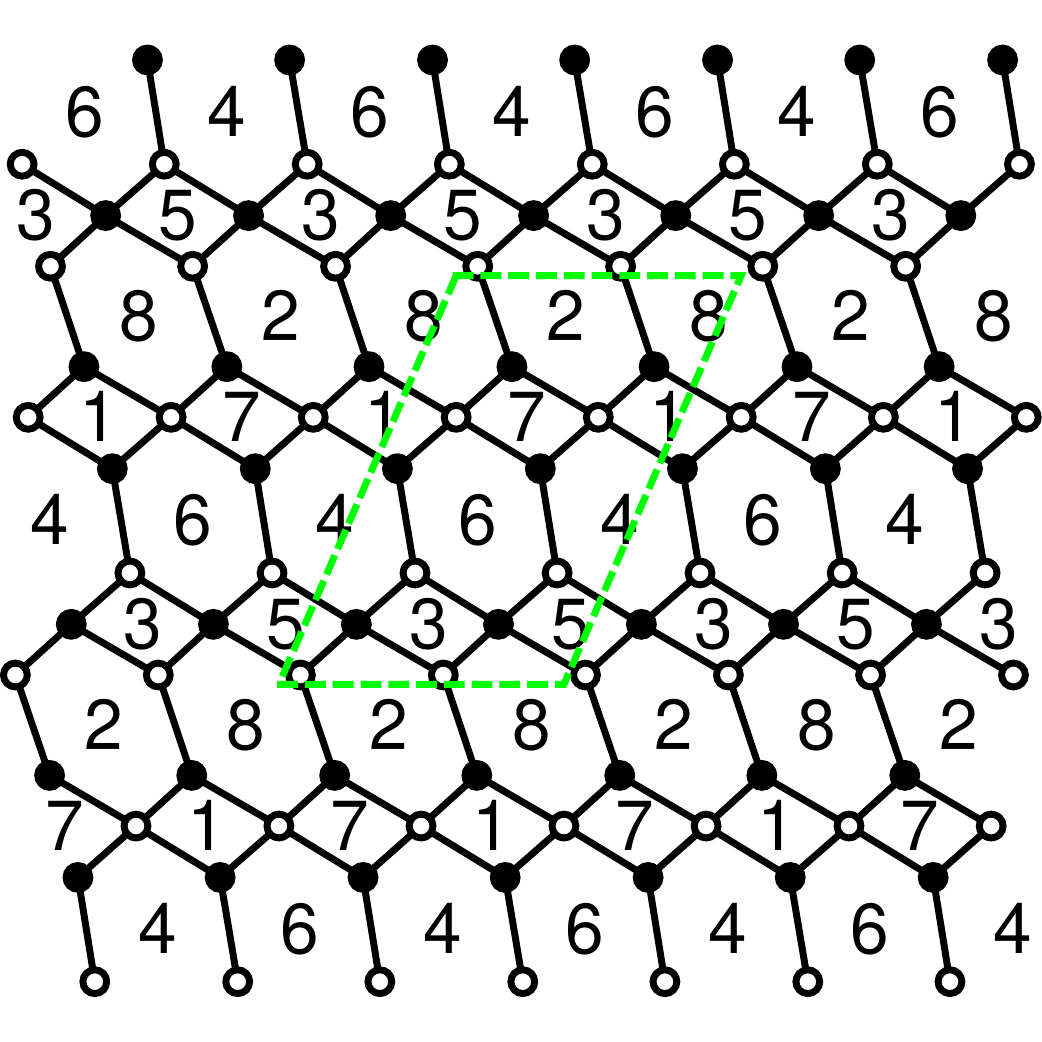}
\caption{The quiver, toric diagram, and brane tiling of Model 4b. The red arrows in the quiver indicate all possible connections between blocks of nodes.\label{f4b}}
 \end{center}
 \end{figure}
 
 \noindent The superpotential is 
\beal{esm4b_00}
W&=&
+ X_{23} X_{38} X_{82} 
+ X_{45} X_{56} X_{64} 
+ X_{63} X_{34} X_{46} 
+ X_{85} X_{52} X_{28} 
+ X_{21} X_{14} X_{47} X_{72} 
+ X_{61} X_{18} X_{87} X_{76} 
\nn\\
&&
- X_{21} X_{18} X_{82} 
- X_{47} X_{76} X_{64} 
- X_{87} X_{72} X_{28} 
- X_{61} X_{14} X_{46} 
- X_{45} X_{52} X_{23} X_{34} 
- X_{63} X_{38} X_{85} X_{56}
  \nn\\
  \eea
 
 \noindent The perfect matching matrix is 
 
\noindent\makebox[\textwidth]{%
\footnotesize
$
P=
\left(
\begin{array}{c|cccc|cc|cc|cc|cc|cccccccccccc}
 \; &p_1& p_2& p_3& p_4& q_1& q_2& 
 & r_2& u_1& u_2& v_1& v_2& s_1& s_2& s_3& s_4& s_5& s_6& s_7& s_8& s_9& s_{10}& s_{11}& s_{12} \\
\hline
 X_{61} & 1 & 0 & 0 & 0 & 1 & 0 & 0 & 0 & 1 & 0 & 0 & 0 & 1 & 0 & 1 & 0 & 1 & 0 &
   0 & 0 & 0 & 0 & 0 & 0 \\
 X_{47} & 1 & 0 & 0 & 0 & 1 & 0 & 0 & 0 & 0 & 1 & 0 & 0 & 0 & 1 & 0 & 1 & 0 & 1 &
   0 & 0 & 0 & 0 & 0 & 0 \\
 X_{34} & 1 & 0 & 0 & 0 & 0 & 1 & 0 & 0 & 1 & 0 & 0 & 0 & 1 & 0 & 0 & 0 & 0 & 0 &
   1 & 0 & 1 & 0 & 0 & 0 \\
 X_{56} & 1 & 0 & 0 & 0 & 0 & 1 & 0 & 0 & 0 & 1 & 0 & 0 & 0 & 1 & 0 & 0 & 0 & 0 &
   0 & 1 & 0 & 1 & 0 & 0 \\
 X_{63} & 0 & 1 & 0 & 0 & 1 & 0 & 0 & 0 & 0 & 0 & 1 & 0 & 0 & 0 & 1 & 0 & 1 & 0 &
   0 & 0 & 0 & 0 & 1 & 0 \\
 X_{45} & 0 & 1 & 0 & 0 & 1 & 0 & 0 & 0 & 0 & 0 & 0 & 1 & 0 & 0 & 0 & 1 & 0 & 1 &
   0 & 0 & 0 & 0 & 0 & 1 \\
 X_{14} & 0 & 1 & 0 & 0 & 0 & 1 & 0 & 0 & 0 & 0 & 1 & 0 & 0 & 0 & 0 & 0 & 0 & 0 &
   1 & 0 & 1 & 0 & 1 & 0 \\
 X_{76} & 0 & 1 & 0 & 0 & 0 & 1 & 0 & 0 & 0 & 0 & 0 & 1 & 0 & 0 & 0 & 0 & 0 & 0 &
   0 & 1 & 0 & 1 & 0 & 1 \\
 X_{85} & 0 & 0 & 1 & 0 & 0 & 0 & 1 & 0 & 1 & 0 & 0 & 0 & 0 & 0 & 0 & 0 & 0 & 1 &
   0 & 0 & 1 & 0 & 0 & 1 \\
 X_{23} & 0 & 0 & 1 & 0 & 0 & 0 & 1 & 0 & 0 & 1 & 0 & 0 & 0 & 0 & 0 & 0 & 1 & 0 &
   0 & 0 & 0 & 1 & 1 & 0 \\
 X_{72} & 0 & 0 & 1 & 0 & 0 & 0 & 0 & 1 & 1 & 0 & 0 & 0 & 0 & 0 & 1 & 0 & 0 & 0 &
   0 & 1 & 0 & 0 & 0 & 1 \\
 X_{18} & 0 & 0 & 1 & 0 & 0 & 0 & 0 & 1 & 0 & 1 & 0 & 0 & 0 & 0 & 0 & 1 & 0 & 0 &
   1 & 0 & 0 & 0 & 1 & 0 \\
 X_{87} & 0 & 0 & 0 & 1 & 0 & 0 & 1 & 0 & 0 & 0 & 1 & 0 & 0 & 1 & 0 & 0 & 0 & 1 &
   0 & 0 & 1 & 0 & 0 & 0 \\
 X_{21} & 0 & 0 & 0 & 1 & 0 & 0 & 1 & 0 & 0 & 0 & 0 & 1 & 1 & 0 & 0 & 0 & 1 & 0 &
   0 & 0 & 0 & 1 & 0 & 0 \\
 X_{52} & 0 & 0 & 0 & 1 & 0 & 0 & 0 & 1 & 0 & 0 & 1 & 0 & 0 & 1 & 1 & 0 & 0 & 0 &
   0 & 1 & 0 & 0 & 0 & 0 \\
 X_{38} & 0 & 0 & 0 & 1 & 0 & 0 & 0 & 1 & 0 & 0 & 0 & 1 & 1 & 0 & 0 & 1 & 0 & 0 &
   1 & 0 & 0 & 0 & 0 & 0 \\
 X_{82} & 1 & 1 & 0 & 0 & 1 & 1 & 0 & 0 & 1 & 0 & 1 & 0 & 0 & 1 & 1 & 0 & 0 & 1 &
   0 & 1 & 1 & 0 & 0 & 1 \\
 X_{28} & 1 & 1 & 0 & 0 & 1 & 1 & 0 & 0 & 0 & 1 & 0 & 1 & 1 & 0 & 0 & 1 & 1 & 0 &
   1 & 0 & 0 & 1 & 1 & 0 \\
 X_{64} & 0 & 0 & 1 & 1 & 0 & 0 & 1 & 1 & 1 & 0 & 1 & 0 & 1 & 0 & 1 & 0 & 1 & 0 &
   1 & 0 & 1 & 0 & 1 & 0 \\
 X_{46} & 0 & 0 & 1 & 1 & 0 & 0 & 1 & 1 & 0 & 1 & 0 & 1 & 0 & 1 & 0 & 1 & 0 & 1 &
   0 & 1 & 0 & 1 & 0 & 1
\end{array}
\right)
$
}
\vspace{0.5cm}

 \noindent The F-term charge matrix $Q_F=\ker{(P)}$ is

\noindent\makebox[\textwidth]{%
\footnotesize
$
Q_F=
\left(
\begin{array}{cccc|cc|cc|cc|cc|cccccccccccc}
p_1& p_2& p_3& p_4& q_1& q_2& r_1& r_2& u_1& u_2& v_1& v_2& s_1& s_2& s_3& s_4& s_5& s_6& s_7& s_8& s_9& s_{10}& s_{11}& s_{12} \\
\hline
 1 & 1 & 0 & 0 & -1 & -1 & 0 & 0 & 0 & 0 & 0 & 0 & 0 & 0 & 0 & 0 & 0 & 0 & 0 & 0 & 0 & 0 & 0 & 0 \\
 0 & 0 & 1 & 1 & 0 & 0 & -1 & -1 & 0 & 0 & 0 & 0 & 0 & 0 & 0 & 0 & 0 & 0 & 0 & 0 & 0 & 0 & 0 & 0 \\
 1 & 0 & 1 & 0 & 0 & 0 & 0 & 0 & -1 & -1 & 0 & 0 & 0 & 0 & 0 & 0 & 0 & 0 & 0 & 0 & 0 & 0 & 0 & 0 \\
 0 & 1 & 0 & 1 & 0 & 0 & 0 & 0 & 0 & 0 & -1 & -1 & 0 & 0 & 0 & 0 & 0 & 0 & 0 & 0 & 0 & 0 & 0 & 0 \\
 1 & 0 & 0 & 1 & 0 & 0 & 0 & 0 & 0 & 0 & 0 & 0 & -1 & -1 & 0 & 0 & 0 & 0 & 0 & 0 & 0 & 0 & 0 & 0 \\
 0 & 0 & 0 & 0 & 1 & 0 & 0 & 1 & 0 & 0 & 0 & 0 & 0 & 0 & -1 & -1 & 0 & 0 & 0 & 0 & 0 & 0 & 0 & 0 \\
 0 & 0 & 0 & 0 & 1 & 0 & 1 & 0 & 0 & 0 & 0 & 0 & 0 & 0 & 0 & 0 & -1 & -1 & 0 & 0 & 0 & 0 & 0 & 0 \\
 0 & 0 & 0 & 0 & 0 & 0 & 0 & 0 & 1 & 0 & 1 & 0 & 0 & 0 & -1 & 0 & 0 & 0 & 0 & 0 & -1 & 0 & 0 & 0 \\
 0 & 0 & 0 & 0 & 0 & 0 & 0 & 0 & 1 & 0 & 0 & 1 & -1 & 0 & 0 & 0 & 0 & 0 & 0 & 0 & 0 & 0 & 0 & -1 \\
 0 & 0 & 1 & 0 & 0 & 0 & -1 & 0 & 0 & 0 & 1 & 0 & 0 & 0 & -1 & 0 & 1 & 0 & 0 & 0 & 0 & 0 & -1 & 0 \\
 0 & 0 & 1 & 0 & 0 & 0 & -1 & 0 & 0 & 0 & 1 & 0 & 1 & 0 & -1 & 0 & 0 & 0 & -1 & 0 & 0 & 0 & 0 & 0 \\
 0 & 0 & 0 & 1 & 0 & 1 & 0 & 0 & 0 & 0 & -1 & 0 & -1 & 0 & 1 & 0 & 0 & 0 & 0 & -1 & 0 & 0 & 0 & 0 \\
 0 & 0 & 0 & 1 & 0 & 0 & -1 & 0 & 1 & 0 & 0 & 0 & -1 & 0 & -1 & 0 & 1 & 0 & 0 & 0 & 0 & 0 & 0 & 0 \\
 0 & 0 & 0 & 0 & 0 & 0 & 0 & 0 & 0 & 0 & 0 & 0 & 0 & 0 & 1 & 0 & -1 & 0 & 0 & -1 & 0 & 1 & 0 & 0
      \end{array}
\right)
$
}
\vspace{0.5cm}

\noindent The D-term charge matrix is

\noindent\makebox[\textwidth]{%
\footnotesize
$
Q_D=
\left(
\begin{array}{cccc|cc|cc|cc|cc|cccccccccccc}
p_1& p_2& p_3& p_4& q_1& q_2& r_1& r_2& u_1& u_2& v_1& v_2& s_1& s_2& s_3& s_4& s_5& s_6& s_7& s_8& s_9& s_{10}& s_{11}& s_{12} \\
\hline
 0 & 0 & 0 & 0 & 0 & 0 & 0 & 0 & 0 & 0 & 0 & 0 & 0 & 0 & 0 & 0 & 1 & -1 & 0 & 0 & 0 & 0
   & 0 & 0 \\
 0 & 0 & 0 & 0 & 0 & 0 & 0 & 0 & 0 & 0 & 0 & 0 & 0 & 0 & 0 & 0 & 0 & 1 & -1 & 0 & 0 & 0
   & 0 & 0 \\
 0 & 0 & 0 & 0 & 0 & 0 & 0 & 0 & 0 & 0 & 0 & 0 & 0 & 0 & 0 & 0 & 0 & 0 & 1 & -1 & 0 & 0
   & 0 & 0 \\
 0 & 0 & 0 & 0 & 0 & 0 & 0 & 0 & 0 & 0 & 0 & 0 & 0 & 0 & 0 & 0 & 0 & 0 & 0 & 1 & -1 & 0
   & 0 & 0 \\
 0 & 0 & 0 & 0 & 0 & 0 & 0 & 0 & 0 & 0 & 0 & 0 & 0 & 0 & 0 & 0 & 0 & 0 & 0 & 0 & 1 & -1
   & 0 & 0 \\
 0 & 0 & 0 & 0 & 0 & 0 & 0 & 0 & 0 & 0 & 0 & 0 & 0 & 0 & 0 & 0 & 0 & 0 & 0 & 0 & 0 & 1
   & -1 & 0 \\
 0 & 0 & 0 & 0 & 0 & 0 & 0 & 0 & 0 & 0 & 0 & 0 & 0 & 0 & 0 & 0 & 0 & 0 & 0 & 0 & 0 & 0
   & 1 & -1
\end{array}
\right)
$
}
\vspace{0.5cm}

The total charge matrix $Q_t$ does not have repeated columns. Accordingly, the global symmetry is $U(1)_{f_1} \times U(1)_{f_2} \times U(1)_R$. This is the same global symmetry as for Model 4a, and the same mesonic charges on extremal perfect matchings are assigned as for Model 4a, as shown in \tref{t4a}.

Let products of non-extremal perfect matchings be associated to a single variable as follows
\beal{esx4b_1}
q = q_1 q_2 ~,~
r = r_1 r_2 ~,~
u = u_1 u_2 ~,~
v = v_1 v_2 ~,~
s = \prod_{m=1}^{12} s_m~.
\eea
The extremal perfect matchings $p_\alpha$ are counted by $t_\alpha$. The fugacity of the form $y_q$ counts the non-extremal perfect matching product $q$ above.

The refined mesonic Hilbert series is calculated using the Molien integral formula in \eref{es12_2}. The Hilbert series and the corresponding plethystic logarithm turn out to be the same as for Model 4a. The mesonic Hilbert series and the refined plethystic logarithms are given in \eref{esm4a_1}, \eref{esm4a_2} and \eref{esm4a_3}. Accordingly, the mesonic moduli spaces of Model 4a and 4b are the same, with the corresponding quiver gauge theories being toric dual.

The generators in terms of perfect matchings of Model 4b are given in \tref{t4agen} with the correspoding mesonic symmetry charges. The corresponding generators in terms of quiver fields are shown in \tref{t4bgen2}. The mesonic moduli space is a complete intersection, with the generators satisfying the relations in \eref{esm4a_3b}.

\comment{
\begin{table}[h!]
\centering

\resizebox{\hsize}{!}{

\begin{minipage}[!b]{0.6\textwidth}
\begin{tabular}{|l|c|c|}
\hline
Generator & $U(1)_{f_1}$ & $U(1)_{f_2}$ 
\\
\hline
\hline
$
A_1=
p_1^2 p_3^2~
q_1 q_2~
r_1 r_2~
u_1^2 u_2^2~
\prod_{m=1}^{12} s_m$
& 0 & -1
\nn\\
$
A_2=
p_2^2 p_4^2~
q_1 q_2~
r_1 r_2~
v_1^2 v_2^2~
\prod_{m=1}^{12} s_m$
& 0 & 1
\nn\\
$
B_1=
p_1^2 p_2^2~
q_1^2 q_2^2~
u_1 u_2~
v_1 v_2~
\prod_{m=1}^{12} s_m$
& 1 & 0
\nn\\
$
B_2=
p_3^2 p_4^2~
r_1^2 r_2^2~
u_1 u_2~
v_1 v_2~
\prod_{m=1}^{12} s_m$
& -1 & 0
\nn\\
$
C=
p_1 p_2 p_3 p_4~
q_1 q_2~

 r_2~
u_1 u_2~
v_1 v_2~
\prod_{m=1}^{12} s_m$
& 0 & 0
\nn\\
\hline
\end{tabular}
\end{minipage}
\hspace{2cm}
\begin{minipage}[!b]{0.2\textwidth}
\includegraphics[width=3.5 cm]{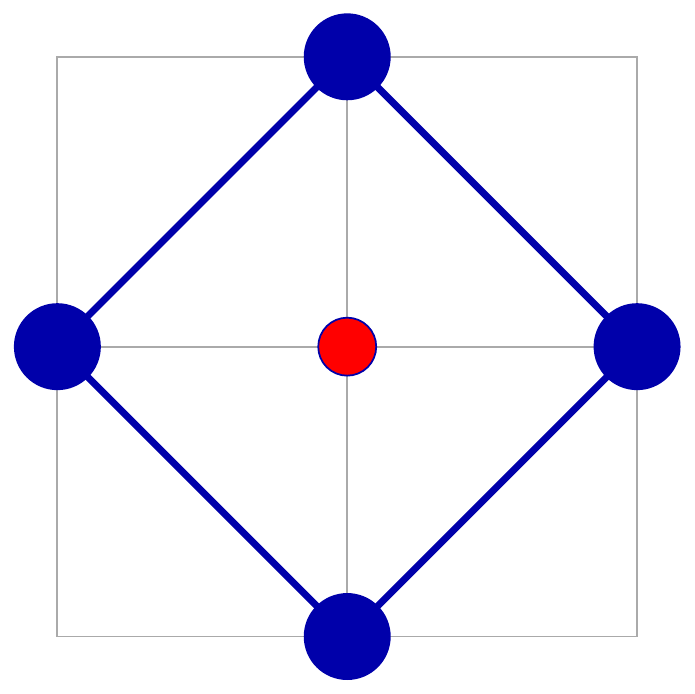}
\end{minipage}

}

\caption{The generators and lattice of generators of the mesonic moduli space of Model 4b in terms of GLSM fields with the corresponding flavor charges.\label{t4bgen}\label{f4bgen}} 
\end{table}
}

\begin{table}[h!]
\centering

\resizebox{\hsize}{!}{

\begin{tabular}{|l|c|c|}
\hline
Generator & $U(1)_{f_1}$ & $U(1)_{f_2}$ 
\\
\hline
\hline
$
X_{56} X_{18} X_{85} X_{61}=  X_{23} X_{34} X_{47} X_{72}
$
& 0 & -1
\nn\\
$
X_{28} X_{82}=  X_{14} X_{45} X_{56} X_{61}=  X_{14} X_{47} X_{76} X_{61}=  X_{34} X_{45} X_{56} X_{63}=  X_{34} X_{47} X_{76} X_{63}
$
& 1 & 0
\nn\\
$
X_{21} X_{14} X_{47} X_{72}=  X_{61} X_{18} X_{87} X_{76}=  X_{23} X_{34} X_{45} X_{52}=  X_{56} X_{38} X_{85} X_{63}=  X_{14} X_{46} X_{61}=  X_{21} X_{18} X_{82}
$
& 0 & 0
\nn\\
$=  X_{23} X_{38} X_{82}=  X_{52} X_{28} X_{85}=  X_{72} X_{28} X_{87}=  X_{34} X_{46} X_{63}=  X_{45} X_{56} X_{64}=  X_{64} X_{47} X_{76}
$
& &
\nn\\
$
X_{46} X_{64}=  X_{21} X_{18} X_{85} X_{52}=  X_{21} X_{18} X_{87} X_{72}=  X_{23} X_{38} X_{85} X_{52}=  X_{23} X_{38} X_{87} X_{72}
$
& -1 & 0
\nn\\
$
 X_{21} X_{14} X_{45} X_{52}=  X_{63} X_{38} X_{87} X_{76}
 $
& 0 & 1
\nn\\
\hline
\end{tabular}

}

\caption{The generators in terms of bifundamental fields (Model 4b).\label{t4bgen2}\label{f4bgen2}} 
\end{table}

\subsection{Model 4 Phase c}

\begin{figure}[H]
\begin{center}
\includegraphics[trim=0cm 0cm 0cm 0cm,height=4.5 cm]{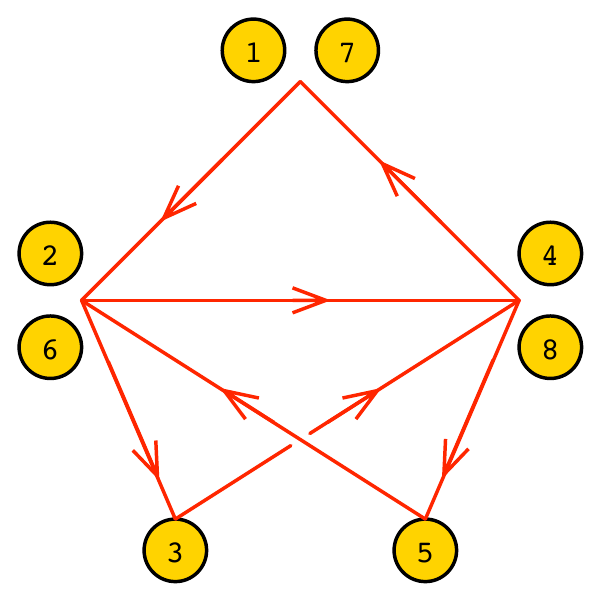}
\includegraphics[width=5 cm]{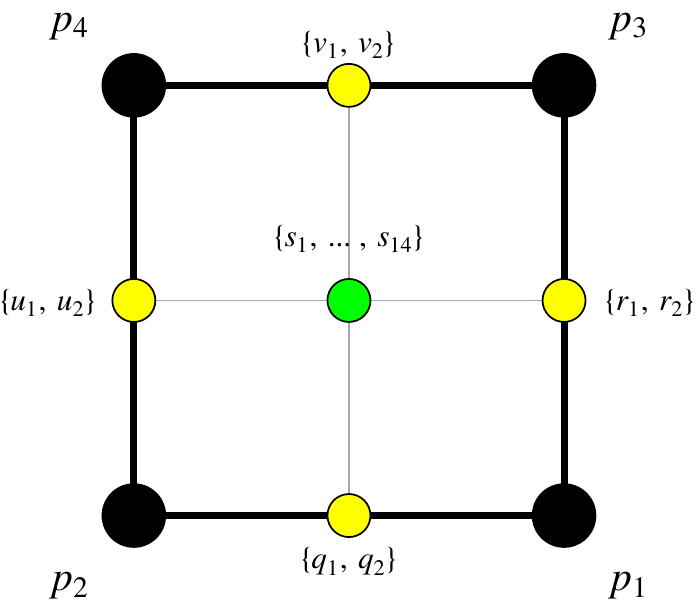}
\includegraphics[width=4.8 cm]{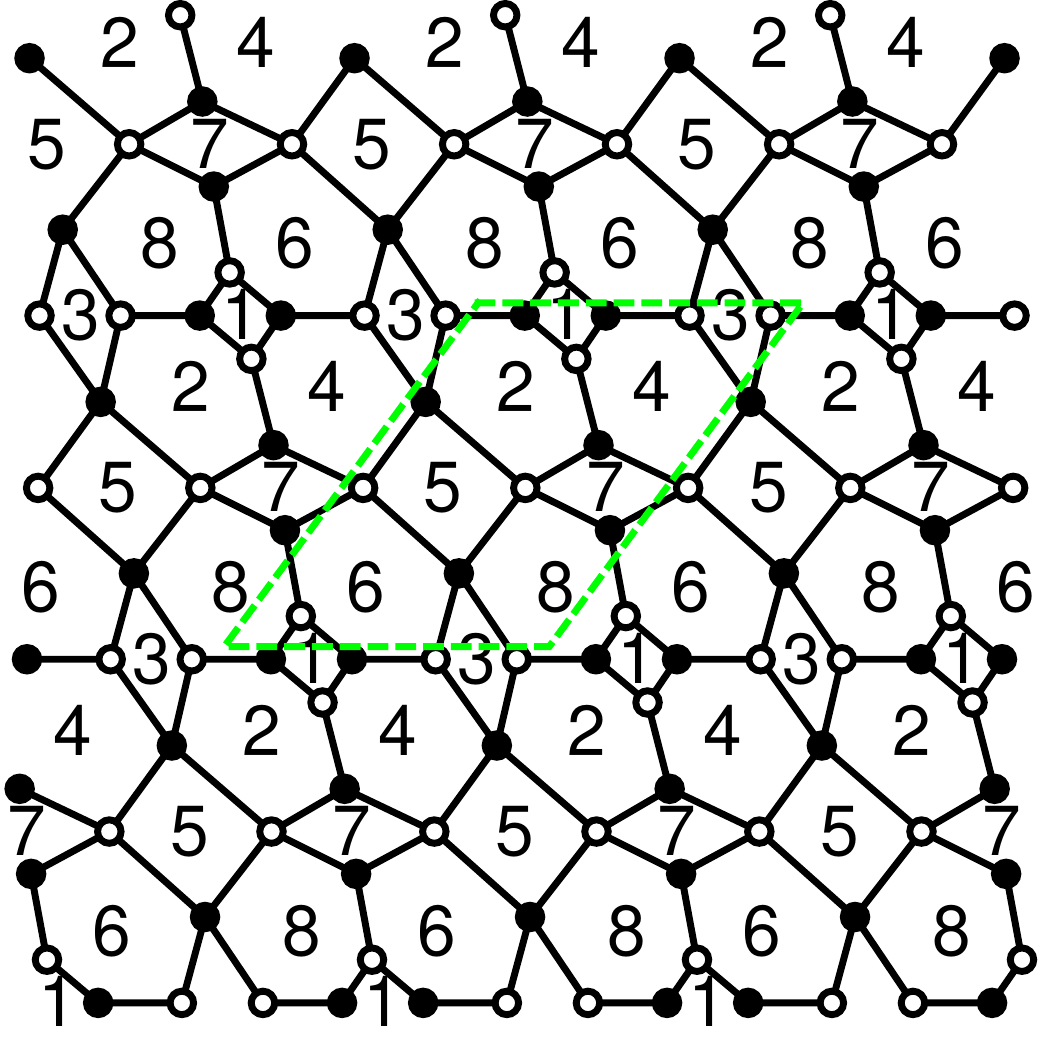}
\caption{The quiver, toric diagram, and brane tiling of Model 4c. The red arrows in the quiver indicate all possible connections between blocks of nodes.\label{f4c}}
 \end{center}
 \end{figure}
 
 \noindent The superpotential is 
\beal{esm4c_00}
W&=&
+ X_{21} X_{14} X_{42}  
+ X_{23} X_{38} X_{82}  
+ X_{61} X_{18} X_{86}  
+ X_{63} X_{34} X_{46}   
+ X_{67} X_{74} X_{45} X_{56}  
+ X_{85} X_{52} X_{27} X_{78} 
\nn\\
&&
- X_{21} X_{18} X_{82}  
- X_{27} X_{74} X_{42}  
- X_{61} X_{14} X_{46}  
- X_{67} X_{78} X_{86}  
- X_{45} X_{52} X_{23} X_{34}  
- X_{63} X_{38} X_{85} X_{56} 
  \nn\\
  \eea
 
 \noindent The perfect matching matrix is 
 
\noindent\makebox[\textwidth]{%
\footnotesize
$
P=
\left(
\begin{array}{c|cccc|cc|cc|cc|cc|cccccccccccccc}
 \; & p_1& p_2& p_3& p_4& q_1& q_2& r_1& r_2& u_1& u_2& v_1& v_2& s_1& s_2& s_3& s_4& s_5& s_6& s_7& s_8& s_9& s_{10}& s_{11}& s_{12}& s_{13}& s_{14} \\
 \hline
 X_{61} & 1 & 0 & 0 & 0 & 1 & 0 & 1 & 0 & 0 & 0 & 0 & 0 & 0 & 0 & 1 & 1 & 1 & 0 &
   1 & 0 & 0 & 0 & 0 & 0 & 1 & 0 \\
 X_{78} & 1 & 0 & 0 & 0 & 1 & 0 & 0 & 1 & 0 & 0 & 0 & 0 & 0 & 0 & 0 & 0 & 1 & 1 &
   0 & 0 & 1 & 0 & 1 & 0 & 1 & 0 \\
 X_{34} & 1 & 0 & 0 & 0 & 0 & 1 & 1 & 0 & 0 & 0 & 0 & 0 & 0 & 0 & 1 & 0 & 0 & 1 &
   0 & 0 & 0 & 1 & 0 & 1 & 1 & 0 \\
 X_{56} & 1 & 0 & 0 & 0 & 0 & 1 & 0 & 1 & 0 & 0 & 0 & 0 & 1 & 0 & 0 & 0 & 0 & 0 &
   0 & 1 & 0 & 0 & 0 & 0 & 0 & 0 \\
 X_{45} & 0 & 1 & 0 & 0 & 1 & 0 & 0 & 0 & 1 & 0 & 0 & 0 & 0 & 1 & 0 & 0 & 0 & 0 &
   0 & 0 & 0 & 0 & 1 & 0 & 0 & 0 \\
 X_{63} & 0 & 1 & 0 & 0 & 1 & 0 & 0 & 0 & 0 & 1 & 0 & 0 & 0 & 0 & 0 & 1 & 1 & 0 &
   1 & 0 & 1 & 0 & 0 & 0 & 0 & 1 \\
 X_{27} & 0 & 1 & 0 & 0 & 0 & 1 & 0 & 0 & 1 & 0 & 0 & 0 & 0 & 0 & 1 & 1 & 0 & 0 &
   0 & 1 & 0 & 1 & 0 & 0 & 0 & 1 \\
 X_{14} & 0 & 1 & 0 & 0 & 0 & 1 & 0 & 0 & 0 & 1 & 0 & 0 & 0 & 0 & 0 & 0 & 0 & 1 &
   0 & 0 & 1 & 1 & 0 & 1 & 0 & 1 \\
 X_{67} & 0 & 0 & 1 & 0 & 0 & 0 & 1 & 0 & 0 & 0 & 1 & 0 & 0 & 0 & 1 & 1 & 0 & 0 &
   1 & 0 & 0 & 1 & 0 & 0 & 0 & 1 \\
 X_{85} & 0 & 0 & 1 & 0 & 0 & 0 & 1 & 0 & 0 & 0 & 0 & 1 & 0 & 1 & 0 & 0 & 0 & 0 &
   0 & 0 & 0 & 0 & 0 & 1 & 0 & 0 \\
 X_{18} & 0 & 0 & 1 & 0 & 0 & 0 & 0 & 1 & 0 & 0 & 1 & 0 & 0 & 0 & 0 & 0 & 0 & 1 &
   0 & 0 & 1 & 1 & 1 & 0 & 0 & 1 \\
 X_{23} & 0 & 0 & 1 & 0 & 0 & 0 & 0 & 1 & 0 & 0 & 0 & 1 & 0 & 0 & 0 & 1 & 1 & 0 &
   0 & 1 & 1 & 0 & 0 & 0 & 0 & 1 \\
 X_{38} & 0 & 0 & 0 & 1 & 0 & 0 & 0 & 0 & 1 & 0 & 1 & 0 & 0 & 0 & 1 & 0 & 0 & 1 &
   0 & 0 & 0 & 1 & 1 & 0 & 1 & 0 \\
 X_{21} & 0 & 0 & 0 & 1 & 0 & 0 & 0 & 0 & 1 & 0 & 0 & 1 & 0 & 0 & 1 & 1 & 1 & 0 &
   0 & 1 & 0 & 0 & 0 & 0 & 1 & 0 \\
 X_{52} & 0 & 0 & 0 & 1 & 0 & 0 & 0 & 0 & 0 & 1 & 1 & 0 & 1 & 0 & 0 & 0 & 0 & 0 &
   1 & 0 & 0 & 0 & 0 & 0 & 0 & 0 \\
 X_{74} & 0 & 0 & 0 & 1 & 0 & 0 & 0 & 0 & 0 & 1 & 0 & 1 & 0 & 0 & 0 & 0 & 1 & 1 &
   0 & 0 & 1 & 0 & 0 & 1 & 1 & 0 \\
 X_{82} & 1 & 1 & 0 & 0 & 1 & 1 & 1 & 0 & 0 & 1 & 0 & 0 & 1 & 1 & 0 & 0 & 0 & 0 &
   1 & 0 & 0 & 0 & 0 & 1 & 0 & 0 \\
 X_{42} & 1 & 0 & 1 & 0 & 1 & 0 & 1 & 1 & 0 & 0 & 1 & 0 & 1 & 1 & 0 & 0 & 0 & 0 &
   1 & 0 & 0 & 0 & 1 & 0 & 0 & 0 \\
 X_{86} & 0 & 1 & 0 & 1 & 0 & 1 & 0 & 0 & 1 & 1 & 0 & 1 & 1 & 1 & 0 & 0 & 0 & 0 &
   0 & 1 & 0 & 0 & 0 & 1 & 0 & 0 \\
 X_{46} & 0 & 0 & 1 & 1 & 0 & 0 & 0 & 1 & 1 & 0 & 1 & 1 & 1 & 1 & 0 & 0 & 0 & 0 &
   0 & 1 & 0 & 0 & 1 & 0 & 0 & 0
\end{array}
\right)
$
}
\vspace{0.5cm}

 \noindent The F-term charge matrix $Q_F=\ker{(P)}$ is

\noindent\makebox[\textwidth]{%
\footnotesize
$
Q_F=
\left(
\begin{array}{cccc|cc|cc|cc|cc|cccccccccccccc}
 p_1& p_2& p_3& p_4& q_1& q_2& r_1& r_2& u_1& u_2& v_1& v_2& s_1& s_2& s_3& s_4& s_5& s_6& s_7& s_8& s_9& s_{10}& s_{11}& s_{12}& s_{13}& s_{14}  \\
\hline
 1 & 1 & 0 & 0 & -1 & -1 & 0 & 0 & 0 & 0 & 0 & 0 & 0 & 0 & 0 & 0 & 0 & 0 & 0 & 0 & 0 & 0 & 0 & 0 & 0 & 0 \\
 1 & 0 & 1 & 0 & 0 & 0 & -1 & -1 & 0 & 0 & 0 & 0 & 0 & 0 & 0 & 0 & 0 & 0 & 0 & 0 & 0 & 0 & 0 & 0 & 0 & 0 \\
 0 & 1 & 0 & 1 & 0 & 0 & 0 & 0 & -1 & -1 & 0 & 0 & 0 & 0 & 0 & 0 & 0 & 0 & 0 & 0 & 0 & 0 & 0 & 0 & 0 & 0 \\
 0 & 0 & 1 & 1 & 0 & 0 & 0 & 0 & 0 & 0 & -1 & -1 & 0 & 0 & 0 & 0 & 0 & 0 & 0 & 0 & 0 & 0 & 0 & 0 & 0 & 0 \\
 1 & 0 & 0 & 1 & 0 & 0 & 0 & 0 & 0 & 0 & 0 & 0 & -1 & 0 & 0 & 0 & 0 & 0 & 0 & 0 & 0 & 0 & 0 & 0 & -1 & 0 \\
 0 & 1 & 1 & 0 & 0 & 0 & 0 & 0 & 0 & 0 & 0 & 0 & 0 & -1 & 0 & 0 & 0 & 0 & 0 & 0 & 0 & 0 & 0 & 0 & 0 & -1 \\
 1 & 0 & 0 & 0 & -1 & 0 & 0 & -1 & 0 & 0 & 0 & 0 & 0 & 0 & 0 & 0 & 0 & -1 & 0 & 0 & 1 & 0 & 1 & 0 & 0 & 0 \\
 1 & 0 & 0 & 0 & -1 & 0 & -1 & 0 & 0 & 0 & 0 & 0 & 0 & 1 & 0 & 1 & 0 & 0 & 0 & -1 & 0 & 0 & 0 & 0 & 0 & 0 \\
 0 & 0 & 1 & 0 & 1 & 0 & -1 & 0 & 0 & 0 & 0 & 0 & 0 & -1 & 0 & 0 & 0 & 0 & 0 & 0 & -1 & 0 & 0 & 1 & 0 & 0 \\
 0 & 0 & 1 & 0 & 0 & 0 & 0 & 0 & 1 & 0 & -1 & 0 & 1 & -1 & 0 & 0 & 0 & 0 & 0 & -1 & 0 & 0 & 0 & 0 & 0 & 0 \\
 0 & 0 & 0 & 0 & 1 & 0 & 0 & 0 & 0 & 0 & 1 & 0 & 0 & 0 & 0 & 0 & 0 & 0 & -1 & 0 & 0 & 0 & -1 & 0 & 0 & 0 \\
 0 & 0 & 0 & 0 & 1 & 0 & 0 & 0 & 0 & 0 & 0 & 1 & 0 & -1 & 0 & 0 & -1 & 0 & 0 & 0 & 0 & 0 & 0 & 0 & 0 & 0 \\
 0 & 0 & 0 & 0 & 0 & 1 & 0 & 0 & 0 & 0 & 1 & 0 & -1 & 0 & 0 & 0 & 0 & 0 & 0 & 0 & 0 & -1 & 0 & 0 & 0 & 0 \\
 0 & 0 & 0 & 0 & 0 & 0 & 1 & 0 & 1 & 0 & 0 & 0 & 0 & -1 & -1 & 0 & 0 & 0 & 0 & 0 & 0 & 0 & 0 & 0 & 0 & 0 \\
 0 & 0 & 0 & 0 & 0 & 0 & 0 & 1 & 0 & 1 & 0 & 0 & -1 & 0 & 0 & 0 & 0 & 0 & 0 & 0 & -1 & 0 & 0 & 0 & 0 & 0 \\
 0 & 0 & 0 & 0 & 0 & 0 & 0 & 0 & 0 & 0 & 0 & 0 & 1 & 0 & 0 & 1 & 0 & 0 & -1 & -1 & 0 & 0 & 0 & 0 & 0 & 0
     \end{array}
\right)
$
}
\vspace{0.5cm}

\noindent The D-term charge matrix is

\noindent\makebox[\textwidth]{%
\footnotesize
$
Q_D=
\left(
\begin{array}{cccc|cc|cc|cc|cc|cccccccccccccc}
 p_1& p_2& p_3& p_4& q_1& q_2& r_1& r_2& u_1& u_2& v_1& v_2& s_1& s_2& s_3& s_4& s_5& s_6& s_7& s_8& s_9& s_{10}& s_{11}& s_{12}& s_{13}& s_{14}  \\
\hline
 0 & 0 & 0 & 0 & 0 & 0 & 0 & 0 & 0 & 0 & 0 & 0 & 0 & 0 & 0 & 0 & 0 & 0 & 1 & -1 & 0 & 0
   & 0 & 0 & 0 & 0 \\
 0 & 0 & 0 & 0 & 0 & 0 & 0 & 0 & 0 & 0 & 0 & 0 & 0 & 0 & 0 & 0 & 0 & 0 & 0 & 1 & -1 & 0
   & 0 & 0 & 0 & 0 \\
 0 & 0 & 0 & 0 & 0 & 0 & 0 & 0 & 0 & 0 & 0 & 0 & 0 & 0 & 0 & 0 & 0 & 0 & 0 & 0 & 1 & -1
   & 0 & 0 & 0 & 0 \\
 0 & 0 & 0 & 0 & 0 & 0 & 0 & 0 & 0 & 0 & 0 & 0 & 0 & 0 & 0 & 0 & 0 & 0 & 0 & 0 & 0 & 1
   & -1 & 0 & 0 & 0 \\
 0 & 0 & 0 & 0 & 0 & 0 & 0 & 0 & 0 & 0 & 0 & 0 & 0 & 0 & 0 & 0 & 0 & 0 & 0 & 0 & 0 & 0
   & 1 & -1 & 0 & 0 \\
 0 & 0 & 0 & 0 & 0 & 0 & 0 & 0 & 0 & 0 & 0 & 0 & 0 & 0 & 0 & 0 & 0 & 0 & 0 & 0 & 0 & 0
   & 0 & 1 & -1 & 0 \\
 0 & 0 & 0 & 0 & 0 & 0 & 0 & 0 & 0 & 0 & 0 & 0 & 0 & 0 & 0 & 0 & 0 & 0 & 0 & 0 & 0 & 0
   & 0 & 0 & 1 & -1
\end{array}
\right)
$
}
\vspace{0.5cm}

The global symmetry is $U(1)_{f_1} \times U(1)_{f_2} \times U(1)_R$. The global symmetry charge assignment on the GLSM fields with non-zero R-charges is the same as for Model 4a and is shown \tref{t4a}.

Products of non-extremal perfect matchings are labelled in terms of single variables as follows
\beal{esx4c_1}
q = q_1 q_2 ~,~
r = r_1 r_2 ~,~
u = u_1 u_2 ~,~
v = v_1 v_2 ~,~
s = \prod_{m=1}^{14} s_m~.
\eea
The fugacity which counts GLSM fields corresponding to extremal perfect matchings $p_\alpha$ is $t_\alpha$. A product non-extremal perfect matchings, for instance $q$, is assigned a fugacity of the form $y_q$.

The mesonic Hilbert series and plethystic logarithm for Model 4c is the same form as for Model 4a. They are given respectively in \eref{esm4a_1}, \eref{esm4a_2} and \eref{esm4a_3}. Accordingly, the mesonic moduli space of Model 4c is the same as for Model 4a. In other words they are toric (Seiberg) duals.

The generators in terms of the perfect matching variables of Model 4c are given in \tref{t4agen} with their mesonic charges. The generators in terms of quiver fields are given in \tref{t4cgen2}. The mesonic moduli space is a complete intersection and the generators satisfy the relations given in \eref{esm4a_3b}.

\comment{
\begin{table}[h!]
\centering

\resizebox{\hsize}{!}{

\begin{minipage}[!b]{0.6\textwidth}
\begin{tabular}{|l|c|c|}
\hline
Generator & $U(1)_{f_1}$ & $U(1)_{f_2}$ 
\\
\hline
\hline
$A_1=
p_1^2 p_3^2~
q_1 q_2~
r_1 r_2~
u_1^2 u_2^2~
\prod_{m=1}^{14} s_m$
& 0 & -1
\nn\\
$A_2=
p_2^2 p_4^2~
q_1 q_2~
r_1 r_2~
v_1^2 v_2^2~
\prod_{m=1}^{14} s_m$
& 0 & 1
\nn\\
$B_1=
p_1^2 p_2^2~
q_1^2 q_2^2~
u_1 u_2~
v_1 v_2~
\prod_{m=1}^{14} s_m$
& 1 & 0
\nn\\
$B_2=
p_3^2 p_4^2~
r_1^2 r_2^2~
u_1 u_2~
v_1 v_2~
\prod_{m=1}^{14} s_m$
& -1 & 0
\nn\\
$C=
p_1 p_2 p_3 p_4~
q_1 q_2~
r_1 r_2~
u_1 u_2~
v_1 v_2~
\prod_{m=1}^{14} s_m$
& 0 & 0
\nn\\
\hline
\end{tabular}
\end{minipage}
\hspace{2cm}
\begin{minipage}[!b]{0.25\textwidth}
\includegraphics[width=3.5 cm]{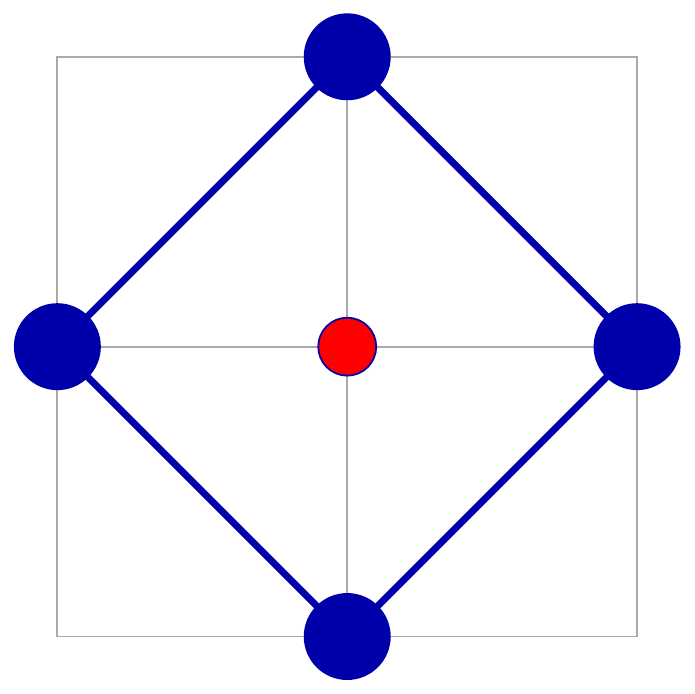}
\end{minipage}
}
\caption{The generators and lattice of generators of the mesonic moduli space of Model 4c in terms of GLSM fields with the corresponding flavor charges.\label{t4cgen}\label{f4cgen}} 
\end{table}
}

\begin{table}[h!]
\centering

\resizebox{\hsize}{!}{

\begin{tabular}{|l|c|c|}
\hline
Generator & $U(1)_{f_1}$ & $U(1)_{f_2}$ 
\\
\hline
\hline
$
X_{27} X_{78} X_{82}=  X_{14} X_{45} X_{56} X_{61}=  X_{34} X_{45} X_{56} X_{63}
$
& 0 & -1
\nn\\
$
X_{23} X_{34} X_{42}=  X_{56} X_{18} X_{85} X_{61}=  X_{56} X_{67} X_{78} X_{85}
$
& 1 & 0
\nn\\
$
X_{23} X_{34} X_{45} X_{52}=  X_{52} X_{27} X_{78} X_{85}=  X_{56} X_{38} X_{85} X_{63}=  X_{45} X_{56} X_{67} X_{74}=  X_{21} X_{14} X_{42}=  X_{14} X_{46} X_{61}
$
& 0 & 0
\nn\\
$
=  X_{21} X_{18} X_{82}=  X_{61} X_{18} X_{86}=  X_{23} X_{38} X_{82}=  X_{42} X_{27} X_{74}=  X_{34} X_{46} X_{63}=  X_{67} X_{78} X_{86}
$
& &
\nn\\
$
X_{63} X_{38} X_{86}=  X_{21} X_{14} X_{45} X_{52}=  X_{45} X_{27} X_{74} X_{52}
$
& -1 & 0
\nn\\
$
X_{46} X_{67} X_{74}=  X_{21} X_{18} X_{85} X_{52}=  X_{23} X_{38} X_{85} X_{52}
$
& 0 & 1
\nn\\
\hline
\end{tabular}
}
\caption{The generators in terms of bifundamental fields (Model 4c).\label{t4cgen2}\label{f4cgen2}} 
\end{table}

\subsection{Model 4 Phase d}

\begin{figure}[H]
\begin{center}
\includegraphics[trim=0cm 0cm 0cm 0cm,height=4.5 cm]{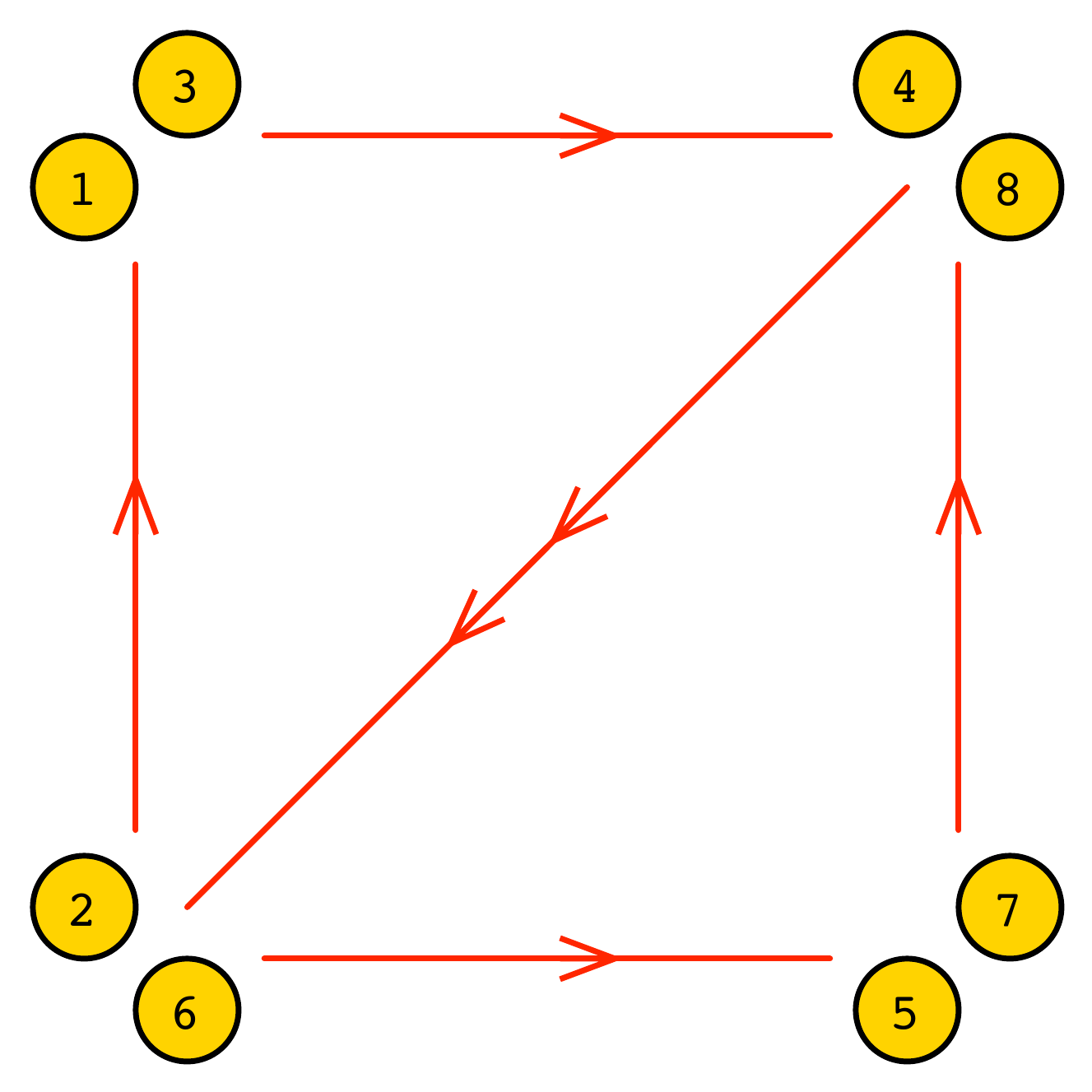}
\includegraphics[width=5 cm]{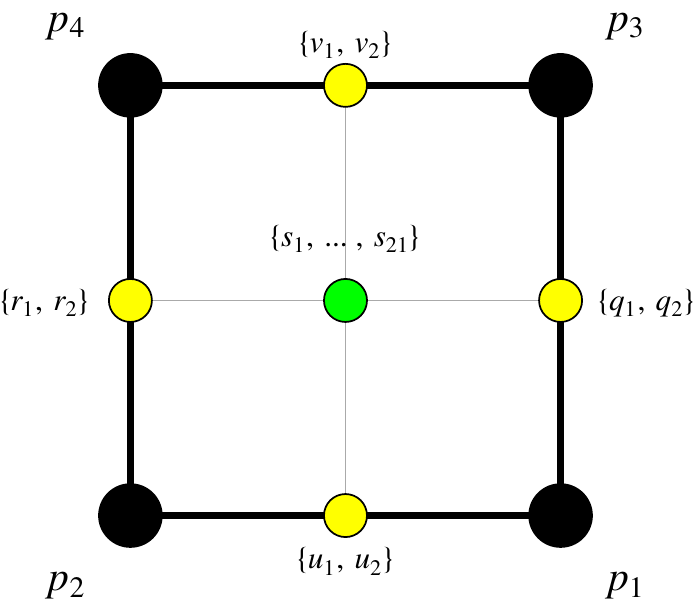}
\includegraphics[width=4.8 cm]{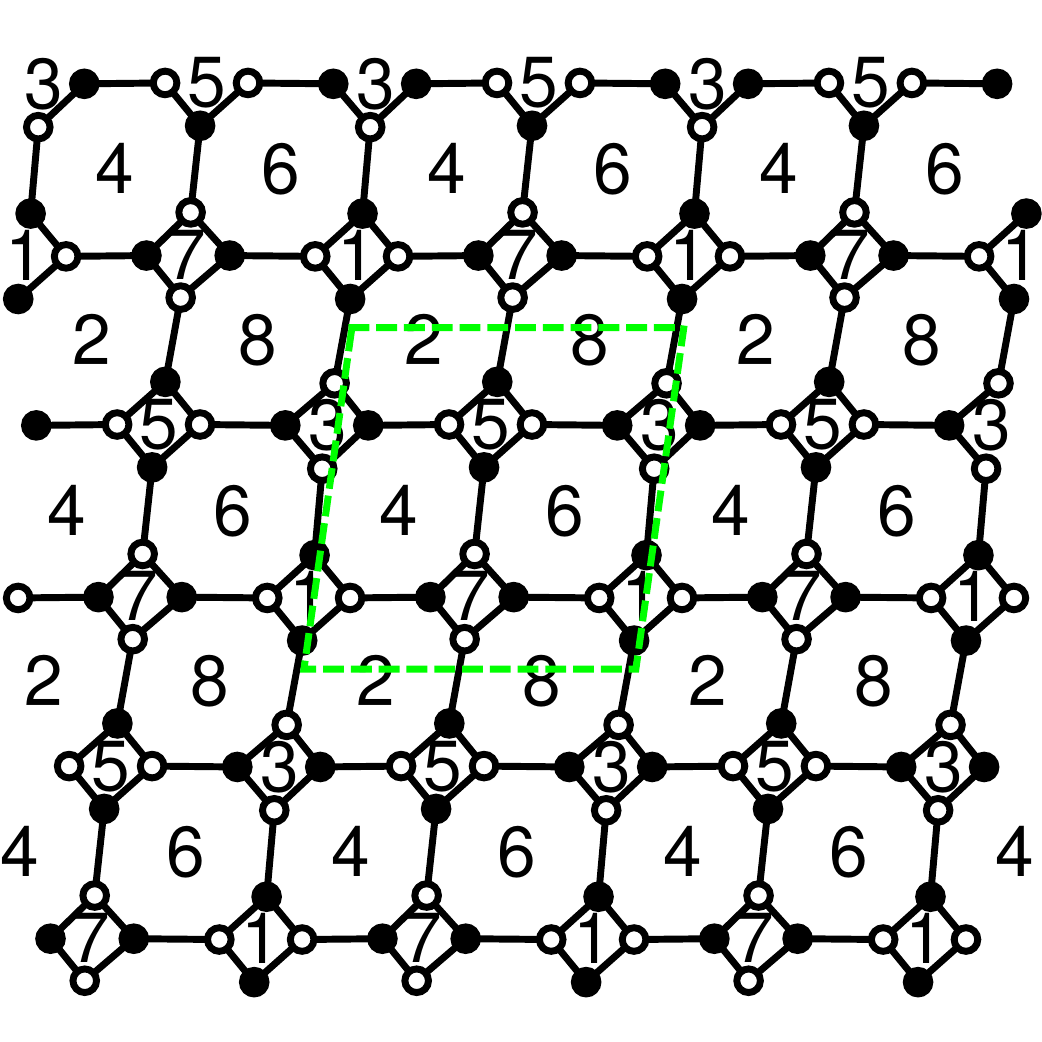}
\caption{The quiver, toric diagram, and brane tiling of Model 4d. The red arrows in the quiver indicate all possible connections between blocks of nodes. \label{f4d}}
 \end{center}
 \end{figure}
 
 \noindent The superpotential is 
\beal{esm4d_00}
W&=&
+ X_{21} X_{14} X_{42}^{1} 
+ X_{23} X_{38} X_{82}^{1} 
+ X_{25} X_{54} X_{42}^{2} 
+ X_{27} X_{78} X_{82}^{2} 
\nn\\
&&
+ X_{61} X_{18} X_{86}^{1} 
+ X_{63} X_{34} X_{46}^{1} 
+ X_{65} X_{58} X_{86}^{2} 
+ X_{67} X_{74} X_{46}^{2}
\nn\\
&&
- X_{21} X_{18} X_{82}^{1} 
- X_{23} X_{34} X_{42}^{2} 
- X_{25} X_{58} X_{82}^{2} 
- X_{27} X_{74} X_{42}^{1} 
\nn\\
&&
- X_{61} X_{14} X_{46}^{1} 
- X_{63} X_{38} X_{86}^{2} 
- X_{65} X_{54} X_{46}^{2} 
- X_{67} X_{78} X_{86}^{1}
  \eea
 
 \noindent The perfect matching matrix is 
 
\noindent\makebox[\textwidth]{%
\scriptsize
$
P=
\left(
\begin{array}{c|cccc|cc|cc|cc|cc|ccccccccccccccccccccc}
 \; & p_1& p_2& p_3& p_4& q_1& q_2& r_1& r_2& u_1& u_2& v_1& v_2& s_1& s_2& s_3& s_4& s_5& s_6& s_7& s_8& s_9& s_{10}& s_{11}& s_{12}& s_{13}& s_{14}& s_{15}& s_{16}& s_{17}& s_{18}& s_{19}& s_{20}& s_{21} \\
 \hline
 X_{42}^{1} & 1 & 1 & 0 & 0 & 1 & 0 & 1 & 0 & 1 & 1 & 0 & 0 & 0 & 0 & 0 & 0 & 0 & 0 &
   0 & 0 & 0 & 0 & 1 & 0 & 1 & 0 & 0 & 0 & 0 & 0 & 0 & 0 & 1 \\
 X_{86} & 1 & 1 & 0 & 0 & 0 & 1 & 0 & 1 & 1 & 1 & 0 & 0 & 0 & 0 & 0 & 0 & 0 & 0 &
   0 & 0 & 0 & 0 & 0 & 1 & 1 & 1 & 0 & 0 & 0 & 0 & 0 & 0 & 0 \\
 X_{46}^{1} & 1 & 0 & 1 & 0 & 1 & 1 & 0 & 0 & 1 & 0 & 1 & 0 & 0 & 0 & 0 & 0 & 0 & 0 &
   0 & 0 & 0 & 0 & 0 & 1 & 1 & 0 & 0 & 0 & 0 & 0 & 0 & 0 & 1 \\
 X_{82} & 1 & 0 & 1 & 0 & 1 & 1 & 0 & 0 & 0 & 1 & 0 & 1 & 0 & 0 & 0 & 0 & 0 & 0 &
   0 & 0 & 0 & 0 & 1 & 0 & 1 & 1 & 0 & 0 & 0 & 0 & 0 & 0 & 0 \\
 X_{58} & 1 & 0 & 0 & 0 & 1 & 0 & 0 & 0 & 1 & 0 & 0 & 0 & 1 & 1 & 1 & 1 & 1 & 0 &
   0 & 0 & 0 & 1 & 0 & 0 & 0 & 0 & 1 & 1 & 0 & 0 & 0 & 0 & 1 \\
 X_{63} & 1 & 0 & 0 & 0 & 1 & 0 & 0 & 0 & 0 & 1 & 0 & 0 & 0 & 0 & 0 & 1 & 1 & 0 &
   0 & 1 & 1 & 0 & 1 & 0 & 0 & 0 & 1 & 1 & 0 & 0 & 1 & 1 & 0 \\
 X_{27} & 1 & 0 & 0 & 0 & 0 & 1 & 0 & 0 & 1 & 0 & 0 & 0 & 1 & 0 & 0 & 0 & 0 & 0 &
   0 & 0 & 0 & 1 & 0 & 1 & 0 & 0 & 1 & 1 & 1 & 1 & 1 & 1 & 0 \\
 X_{14} & 1 & 0 & 0 & 0 & 0 & 1 & 0 & 0 & 0 & 1 & 0 & 0 & 1 & 1 & 0 & 1 & 0 & 1 &
   0 & 1 & 0 & 0 & 0 & 0 & 0 & 1 & 1 & 0 & 1 & 0 & 1 & 0 & 0 \\
 X_{46} & 0 & 1 & 0 & 1 & 0 & 0 & 1 & 1 & 1 & 0 & 1 & 0 & 0 & 0 & 0 & 0 & 0 & 0 &
   0 & 0 & 0 & 0 & 0 & 1 & 1 & 0 & 0 & 0 & 0 & 0 & 0 & 0 & 1 \\
 X_{82}^{1} & 0 & 1 & 0 & 1 & 0 & 0 & 1 & 1 & 0 & 1 & 0 & 1 & 0 & 0 & 0 & 0 & 0 & 0 &
   0 & 0 & 0 & 0 & 1 & 0 & 1 & 1 & 0 & 0 & 0 & 0 & 0 & 0 & 0 \\
 X_{38} & 0 & 1 & 0 & 0 & 0 & 0 & 1 & 0 & 1 & 0 & 0 & 0 & 1 & 1 & 1 & 0 & 0 & 1 &
   1 & 0 & 0 & 1 & 0 & 0 & 0 & 0 & 0 & 0 & 1 & 1 & 0 & 0 & 1 \\
 X_{65} & 0 & 1 & 0 & 0 & 0 & 0 & 1 & 0 & 0 & 1 & 0 & 0 & 0 & 0 & 0 & 0 & 0 & 1 &
   1 & 1 & 1 & 0 & 1 & 0 & 0 & 0 & 0 & 0 & 1 & 1 & 1 & 1 & 0 \\
 X_{21} & 0 & 1 & 0 & 0 & 0 & 0 & 0 & 1 & 1 & 0 & 0 & 0 & 0 & 0 & 1 & 0 & 1 & 0 &
   1 & 0 & 1 & 1 & 0 & 1 & 0 & 0 & 0 & 1 & 0 & 1 & 0 & 1 & 0 \\
 X_{74} & 0 & 1 & 0 & 0 & 0 & 0 & 0 & 1 & 0 & 1 & 0 & 0 & 0 & 1 & 1 & 1 & 1 & 1 &
   1 & 1 & 1 & 0 & 0 & 0 & 0 & 1 & 0 & 0 & 0 & 0 & 0 & 0 & 0 \\
 X_{42} & 0 & 0 & 1 & 1 & 1 & 0 & 1 & 0 & 0 & 0 & 1 & 1 & 0 & 0 & 0 & 0 & 0 & 0 &
   0 & 0 & 0 & 0 & 1 & 0 & 1 & 0 & 0 & 0 & 0 & 0 & 0 & 0 & 1 \\
 X_{86}^{1} & 0 & 0 & 1 & 1 & 0 & 1 & 0 & 1 & 0 & 0 & 1 & 1 & 0 & 0 & 0 & 0 & 0 & 0 &
   0 & 0 & 0 & 0 & 0 & 1 & 1 & 1 & 0 & 0 & 0 & 0 & 0 & 0 & 0 \\
 X_{78} & 0 & 0 & 1 & 0 & 1 & 0 & 0 & 0 & 0 & 0 & 1 & 0 & 0 & 1 & 1 & 1 & 1 & 1 &
   1 & 1 & 1 & 0 & 0 & 0 & 0 & 0 & 0 & 0 & 0 & 0 & 0 & 0 & 1 \\
 X_{61} & 0 & 0 & 1 & 0 & 1 & 0 & 0 & 0 & 0 & 0 & 0 & 1 & 0 & 0 & 1 & 0 & 1 & 0 &
   1 & 0 & 1 & 1 & 1 & 0 & 0 & 0 & 0 & 1 & 0 & 1 & 0 & 1 & 0 \\
 X_{25} & 0 & 0 & 1 & 0 & 0 & 1 & 0 & 0 & 0 & 0 & 1 & 0 & 0 & 0 & 0 & 0 & 0 & 1 &
   1 & 1 & 1 & 0 & 0 & 1 & 0 & 0 & 0 & 0 & 1 & 1 & 1 & 1 & 0 \\
 X_{34} & 0 & 0 & 1 & 0 & 0 & 1 & 0 & 0 & 0 & 0 & 0 & 1 & 1 & 1 & 1 & 0 & 0 & 1 &
   1 & 0 & 0 & 1 & 0 & 0 & 0 & 1 & 0 & 0 & 1 & 1 & 0 & 0 & 0 \\
 X_{18} & 0 & 0 & 0 & 1 & 0 & 0 & 1 & 0 & 0 & 0 & 1 & 0 & 1 & 1 & 0 & 1 & 0 & 1 &
   0 & 1 & 0 & 0 & 0 & 0 & 0 & 0 & 1 & 0 & 1 & 0 & 1 & 0 & 1 \\
 X_{67} & 0 & 0 & 0 & 1 & 0 & 0 & 1 & 0 & 0 & 0 & 0 & 1 & 1 & 0 & 0 & 0 & 0 & 0 &
   0 & 0 & 0 & 1 & 1 & 0 & 0 & 0 & 1 & 1 & 1 & 1 & 1 & 1 & 0 \\
 X_{23} & 0 & 0 & 0 & 1 & 0 & 0 & 0 & 1 & 0 & 0 & 1 & 0 & 0 & 0 & 0 & 1 & 1 & 0 &
   0 & 1 & 1 & 0 & 0 & 1 & 0 & 0 & 1 & 1 & 0 & 0 & 1 & 1 & 0 \\
 X_{54} & 0 & 0 & 0 & 1 & 0 & 0 & 0 & 1 & 0 & 0 & 0 & 1 & 1 & 1 & 1 & 1 & 1 & 0 &
   0 & 0 & 0 & 1 & 0 & 0 & 0 & 1 & 1 & 1 & 0 & 0 & 0 & 0 & 0
\end{array}
\right)
$
}
\vspace{0.5cm}

 \noindent The F-term charge matrix $Q_F=\ker{(P)}$ is

\noindent\makebox[\textwidth]{%
\scriptsize
$
Q_F=
\left(
\begin{array}{cccc|cc|cc|cc|cc|ccccccccccccccccccccc}
  p_1& p_2& p_3& p_4& q_1& q_2& r_1& r_2& u_1& u_2& v_1& v_2& s_1& s_2& s_3& s_4& s_5& s_6& s_7& s_8& s_9& s_{10}& s_{11}& s_{12}& s_{13}& s_{14}& s_{15}& s_{16}& s_{17}& s_{18}& s_{19}& s_{20}& s_{21} \\
 \hline
 1 & 0 & 1 & 0 & -1 & -1 & 0 & 0 & 0 & 0 & 0 & 0 & 0 & 0 & 0 & 0 & 0 &
   0 & 0 & 0 & 0 & 0 & 0 & 0 & 0 & 0 & 0 & 0 & 0 & 0 & 0 & 0 & 0 \\
 0 & 1 & 0 & 1 & 0 & 0 & -1 & -1 & 0 & 0 & 0 & 0 & 0 & 0 & 0 & 0 & 0 &
   0 & 0 & 0 & 0 & 0 & 0 & 0 & 0 & 0 & 0 & 0 & 0 & 0 & 0 & 0 & 0 \\
 1 & 1 & 0 & 0 & 0 & 0 & 0 & 0 & -1 & -1 & 0 & 0 & 0 & 0 & 0 & 0 & 0 &
   0 & 0 & 0 & 0 & 0 & 0 & 0 & 0 & 0 & 0 & 0 & 0 & 0 & 0 & 0 & 0 \\
 0 & 0 & 1 & 1 & 0 & 0 & 0 & 0 & 0 & 0 & -1 & -1 & 0 & 0 & 0 & 0 & 0 &
   0 & 0 & 0 & 0 & 0 & 0 & 0 & 0 & 0 & 0 & 0 & 0 & 0 & 0 & 0 & 0 \\
 0 & 0 & 0 & 1 & 0 & 1 & 0 & 0 & 0 & 0 & -1 & 0 & -1 & 0 & 0 & 0 & 0 &
   0 & 0 & 0 & 0 & 0 & 0 & 0 & -1 & 0 & 0 & 0 & 0 & 0 & 0 & 0 & 1 \\
 0 & 0 & 0 & 0 & 0 & 0 & 0 & 0 & 0 & 0 & 0 & 0 & 1 & -1 & 1 & 0 & 0 & 0
   & 0 & 0 & 0 & -1 & 0 & 0 & 0 & 0 & 0 & 0 & 0 & 0 & 0 & 0 & 0 \\
 0 & 0 & 0 & 0 & 0 & 0 & 0 & 0 & 0 & 0 & 0 & 0 & 0 & 1 & 0 & 0 & 0 & 0
   & 0 & 0 & 0 & 0 & 0 & 0 & 1 & -1 & 0 & 0 & 0 & 0 & 0 & 0 & -1 \\
 0 & 0 & 0 & 0 & 0 & 0 & 0 & 0 & 0 & 0 & 0 & 0 & 0 & 0 & 1 & -1 & 0 & 0
   & 0 & 0 & 0 & -1 & 0 & 0 & 0 & 0 & 1 & 0 & 0 & 0 & 0 & 0 & 0 \\
 0 & 0 & 0 & 0 & 0 & 0 & 0 & 0 & 0 & 0 & 0 & 0 & 0 & 0 & 0 & 1 & -1 &
   -1 & 1 & 0 & 0 & 0 & 0 & 0 & 0 & 0 & 0 & 0 & 0 & 0 & 0 & 0 & 0 \\
 0 & 0 & 0 & 0 & 0 & 0 & 0 & 0 & 0 & 0 & 0 & 0 & 0 & 0 & 0 & 1 & -1 & 0
   & 0 & -1 & 1 & 0 & 0 & 0 & 0 & 0 & 0 & 0 & 0 & 0 & 0 & 0 & 0 \\
 0 & 0 & 0 & 0 & 0 & 0 & 0 & 0 & 0 & 0 & 0 & 0 & 0 & 0 & 0 & 1 & -1 & 0
   & 0 & 0 & 0 & 0 & 0 & 0 & 0 & 0 & -1 & 1 & 0 & 0 & 0 & 0 & 0 \\
 0 & 0 & 0 & 0 & 0 & 0 & 0 & 0 & 0 & 0 & 0 & 0 & 0 & 0 & 0 & 0 & 1 & -1
   & 0 & 0 & 0 & 0 & 0 & 0 & 0 & 0 & 0 & -1 & 1 & 0 & 0 & 0 & 0 \\
 0 & 0 & 0 & 0 & 0 & 0 & 0 & 0 & 0 & 0 & 0 & 0 & 0 & 0 & 0 & 0 & 0 & 1
   & -1 & 0 & 0 & 0 & 0 & 0 & 0 & 0 & 0 & 0 & -1 & 1 & 0 & 0 & 0 \\
 0 & 0 & 0 & 0 & 0 & 0 & 0 & 0 & 0 & 0 & 0 & 0 & 0 & 0 & 0 & 0 & 0 & 0
   & 1 & -1 & 0 & 0 & 0 & 0 & 0 & 0 & 0 & 0 & 0 & -1 & 1 & 0 & 0 \\
 0 & 0 & 0 & 0 & 0 & 0 & 0 & 0 & 0 & 0 & 0 & 0 & 0 & 0 & 0 & 0 & 0 & 0
   & 0 & 1 & -1 & 0 & 0 & 0 & 0 & 0 & 0 & 0 & 0 & 0 & -1 & 1 & 0 \\
 1 & 0 & 0 & 0 & 0 & 0 & 1 & 0 & -1 & 0 & 0 & 0 & 0 & 0 & 0 & -1 & 1 &
   0 & 0 & 0 & 0 & 0 & -1 & 0 & 0 & 0 & 0 & 0 & 0 & 0 & 0 & 0 & 0 \\
 0 & 1 & 0 & 0 & 1 & 0 & 0 & 0 & -1 & 0 & 0 & 0 & 1 & -1 & 0 & 0 & 0 &
   0 & 0 & 0 & 0 & 0 & -1 & 0 & 0 & 0 & 0 & 0 & 0 & 0 & 0 & 0 & 0 \\
 1 & -1 & 0 & 0 & -1 & 0 & 1 & 0 & 0 & 0 & 0 & 0 & -1 & 0 & 1 & 0 & 0 &
   0 & 0 & 0 & 0 & 0 & 0 & 0 & 0 & 0 & 0 & 0 & 0 & 0 & 0 & 0 & 0 \\
 1 & 0 & -1 & 0 & 0 & 0 & 0 & 0 & -1 & 0 & 1 & 0 & 0 & 0 & 0 & 0 & 0 &
   0 & 1 & -1 & 0 & 0 & 0 & 0 & 0 & 0 & 0 & 0 & 0 & 0 & 0 & 0 & 0 \\
 0 & 1 & 0 & 0 & 0 & 0 & -1 & 0 & -1 & 0 & 0 & 0 & 1 & 0 & 0 & 0 & 0 &
   0 & 0 & 0 & 0 & 0 & 0 & 0 & 1 & -1 & 0 & 0 & 0 & 0 & 0 & 0 & 0 \\
 0 & 0 & 1 & 0 & 0 & 0 & 1 & 0 & 0 & 0 & -1 & 0 & 0 & -1 & 0 & 1 & 0 &
   0 & 0 & 0 & 0 & 0 & -1 & 0 & 0 & 0 & 0 & 0 & 0 & 0 & 0 & 0 & 0 \\
 0 & 0 & 0 & 1 & 1 & 0 & 0 & 0 & 0 & 0 & -1 & 0 & 0 & -1 & 0 & 0 & 0 &
   1 & 0 & 0 & 0 & 0 & -1 & 0 & 0 & 0 & 0 & 0 & 0 & 0 & 0 & 0 & 0 \\
 0 & 0 & 0 & 0 & 1 & 0 & 1 & 0 & -1 & 0 & -1 & 0 & 0 & 0 & 0 & 0 & 0 &
   0 & 0 & 0 & 0 & 0 & -1 & 1 & 0 & 0 & 0 & 0 & 0 & 0 & 0 & 0 & 0
\end{array}
\right)
$
}
\vspace{0.5cm}

\noindent The D-term charge matrix is

\noindent\makebox[\textwidth]{%
\scriptsize
$
Q_D=
\left(
\begin{array}{cccc|cc|cc|cc|cc|ccccccccccccccccccccc}
  p_1& p_2& p_3& p_4& q_1& q_2& r_1& r_2& u_1& u_2& v_1& v_2& s_1& s_2& s_3& s_4& s_5& s_6& s_7& s_8& s_9& s_{10}& s_{11}& s_{12}& s_{13}& s_{14}& s_{15}& s_{16}& s_{17}& s_{18}& s_{19}& s_{20}& s_{21} \\
 \hline
 0 & 0 & 0 & 0 & 0 & 0 & 0 & 0 & 0 & 0 & 0 & 0 & 0 & 0 & 0 & 0 & 0 & 0 & 0 & 0 & 1 & -1
   & 0 & 0 & 0 & 0 & 0 & 0 & 0 & 0 & 0 & 0 & 0 \\
 0 & 0 & 0 & 0 & 0 & 0 & 0 & 0 & 0 & 0 & 0 & 0 & 0 & 0 & 0 & 0 & 0 & 0 & 0 & 0 & 0 & 0
   & 1 & -1 & 0 & 0 & 0 & 0 & 0 & 0 & 0 & 0 & 0 \\
 0 & 0 & 0 & 0 & 0 & 0 & 0 & 0 & 0 & 0 & 0 & 0 & 0 & 0 & 0 & 0 & 0 & 0 & 0 & 0 & 0 & 0
   & 0 & 1 & -1 & 0 & 0 & 0 & 0 & 0 & 0 & 0 & 0 \\
 0 & 0 & 0 & 0 & 0 & 0 & 0 & 0 & 0 & 0 & 0 & 0 & 0 & 0 & 0 & 0 & 0 & 0 & 0 & 0 & 0 & 0
   & 0 & 0 & 1 & -1 & 0 & 0 & 0 & 0 & 0 & 0 & 0 \\
 0 & 0 & 0 & 0 & 0 & 0 & 0 & 0 & 0 & 0 & 0 & 0 & 0 & 0 & 0 & 0 & 0 & 0 & 0 & 0 & 0 & 0
   & 0 & 0 & 0 & 1 & -1 & 0 & 0 & 0 & 0 & 0 & 0 \\
 0 & 0 & 0 & 0 & 0 & 0 & 0 & 0 & 0 & 0 & 0 & 0 & 0 & 0 & 0 & 0 & 0 & 0 & 0 & 0 & 0 & 0
   & 0 & 0 & 0 & 0 & 1 & -1 & 0 & 0 & 0 & 0 & 0 \\
 0 & 0 & 0 & 0 & 0 & 0 & 0 & 0 & 0 & 0 & 0 & 0 & 0 & 0 & 0 & 0 & 0 & 0 & 0 & 0 & 0 & 0
   & 0 & 0 & 0 & 0 & 0 & 1 & -1 & 0 & 0 & 0 & 0
\end{array}
\right)
$
}
\vspace{0.5cm}

The global symmetry is $U(1)_{f_1} \times U(1)_{f_2} \times U(1)_R$. The global symmetry charge assignment on perfect matchings with non-zero R-charge is the same as for Model 4a and is shown in \tref{t4a}.

Products of non-extremal perfect matchings are expressed in terms of single variables as follows
\beal{esx4b_1}
q = q_1 q_2 ~,~
r = r_1 r_2 ~,~
u = u_1 u_2 ~,~
v = v_1 v_2 ~,~
s = \prod_{m=1}^{21} s_m~.
\eea
The fugacity which counts extremal perfect matchings is $t_\alpha$. A product of non-extremal perfect matchings such as $q$ is assigned a fugacity of the form $y_q$.

The mesonic Hilbert series and the plethystic logarithm are the same as for Model 4a. The mesonic Hilbert series and the refined plethystic logarithms are given in \eref{esm4a_1}, \eref{esm4a_2} and \eref{esm4a_3} respectively.

The mesonic moduli space generators in terms of perfect matching variables of Model 4d are given in \tref{t4agen}. In terms of quiver fields, the generators with their mesonic charges are shown in \tref{t4dgen2}. The mesonic moduli space is a complete intersection and the generators satisfy the relations in \eref{esm4a_3b}.

\comment{
\begin{table}[h!]
\centering

\resizebox{\hsize}{!}{

\begin{minipage}[!b]{0.6\textwidth}
\begin{tabular}{|l|c|c|}
\hline
Generator & $U(1)_{f_1}$ & $U(1)_{f_2}$ 
\\
\hline
\hline
$A_1=
p_1^2 p_2^2~
q_1 q_2~
r_1 r_2~
u_1^2 u_2^2~
\prod_{m=1}^{21} s_m$
& 1 & 0
\nn\\
$A_2=
p_3^2 p_4^2~ 
q_1 q_2~ 
r_1 r_2~
v_1^2 v_2^2~ 
\prod_{m=1}^{21} s_m$
& -1 & 0
\nn\\
$B_1=
p_1^2 p_3^2~
q_1^2 q_2^2~
u_1 u_2~
v_1 v_2~
\prod_{m=1}^{21} s_m$
& 0 & -1
\nn\\
$B_2=
p_2^2 p_4^2~
r_1^2 r_2^2~
u_1 u_2~
v_1 v_2~
\prod_{m=1}^{21} s_m$
& 0 & 1
\nn\\
$C=
p_1 p_2 p_3 p_4~
q_1 q_2~
r_1 r_2~
u_1 u_2~
v_1 v_2~
\prod_{m=1}^{21} s_m$
& 0 & 0
\nn\\
\hline
\end{tabular}
\end{minipage}
\hspace{2cm}
\begin{minipage}[!b]{0.25\textwidth}
\includegraphics[width=3.5 cm]{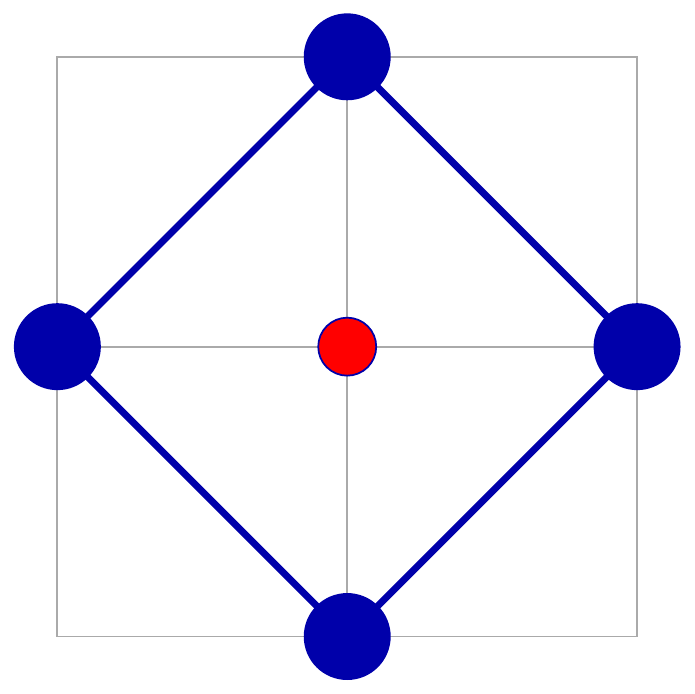}
\end{minipage}
}
\caption{The generators and lattice of generators of the mesonic moduli space of Model 4d in terms of GLSM fields with the corresponding flavor charges.\label{t4dgen}\label{f4dgen}} 
\end{table}
}

\begin{table}[h!]
\centering

\resizebox{\hsize}{!}{

\begin{tabular}{|l|c|c|}
\hline
Generator & $U(1)_{f_1}$ & $U(1)_{f_2}$ 
\\
\hline
\hline
$
X_{21} X_{14} X_ {42}^{2}=  X_ {42}^{2} X_{27} X_{74}=  X_{63} X_{38} X_ {86}^{1}=  X_{65} X_{58} X_ {86}^{1}
$
& 0 & -1
\nn\\
$
 X_{14} X_ {46}^{2} X_{61}=  X_{25} X_{58} X_ {82}^{1}=  X_{27} X_{78} X_ {82}^{1}=  X_{34} X_ {46}^{2} X_{63}
$
& 1 & 0
\nn\\
$
 X_{21} X_{14} X_{42}^{1}=  X_{14} X_{46}^{1} X_{61}=  X_{21} X_{18} X_{82}^{1}=  X_{61} X_{18} X_{86}^{1}=  X_{23} X_{34} X_{42}^{2}=  X_{23} X_{38} X_{82}^{1}
$
& 0 & 0
 \nn\\
 $
 =  X_ {42}^{2} X_{25} X_{54}=  X_{25} X_{58} X_ {82}^{2}=  X_ {42}^{1} X_{27} X_{74}=  X_{27} X_{78} X_ {82}^{2}=  X_{34} X_ {46}^{1} X_{63}=  X_{63} X_{38} X_ {86}^{2}
$
& & 
\nn\\
 $
 =  X_{54} X_ {46}^{2} X_{65}=  X_ {46}^{2} X_{67} X_{74}=  X_{65} X_{58} X_ {86}^{2}=  X_{67} X_{78} X_ {86}^{1}
$
& & 
\nn\\
$
X_{21} X_{18} X_ {82}^{2}=  X_{23} X_{38} X_ {82}^{2}=  X_{54} X_ {46}^{1} X_{65}=  X_ {46}^{1} X_{67} X_{74}
$
& -1 & 0
\nn\\
$
 X_{61} X_{18} X_ {86}^{2}=  X_{23} X_{34} X_ {42}^{1}=  X_ {42}^{1} X_{25} X_{54}=  X_{67} X_{78} X_ {86}^{2}
 $
 & 0 & 1
\nn\\
\hline
\end{tabular}
}
\caption{The generators in terms of bifundamental fields (Model 4d).\label{t4dgen2}\label{f4dgen2}} 
\end{table}

\section{Model 5: $\text{PdP}_{4b}$}

\begin{figure}[H]
\begin{center}
\includegraphics[trim=0cm 0cm 0cm 0cm,width=4.5 cm]{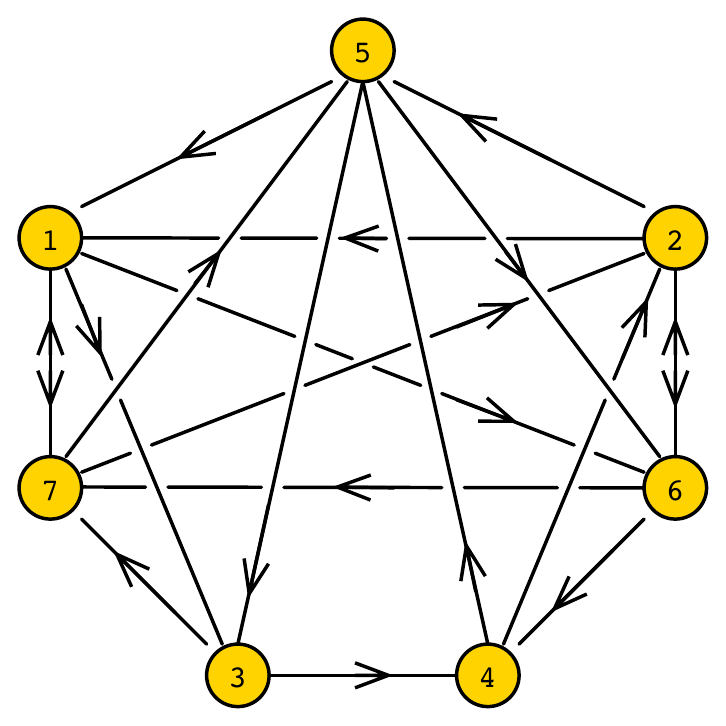}
\includegraphics[width=5 cm]{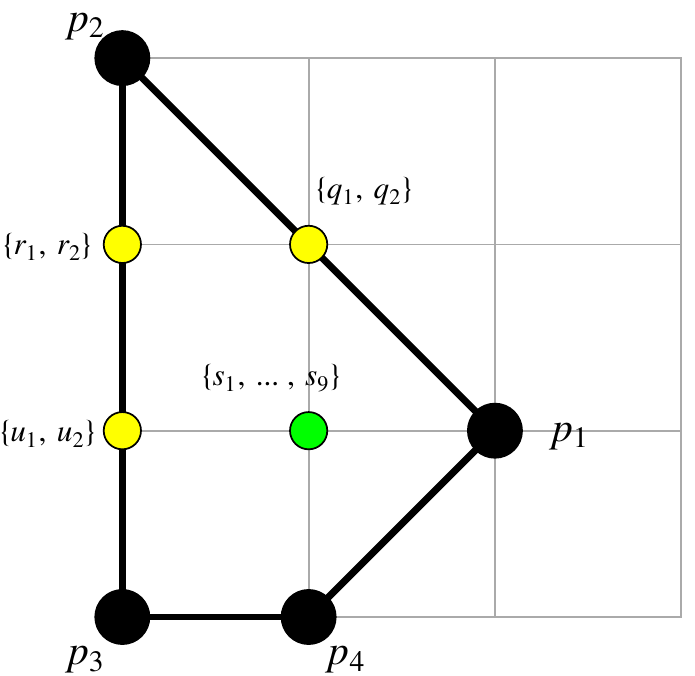}
\includegraphics[width=5 cm]{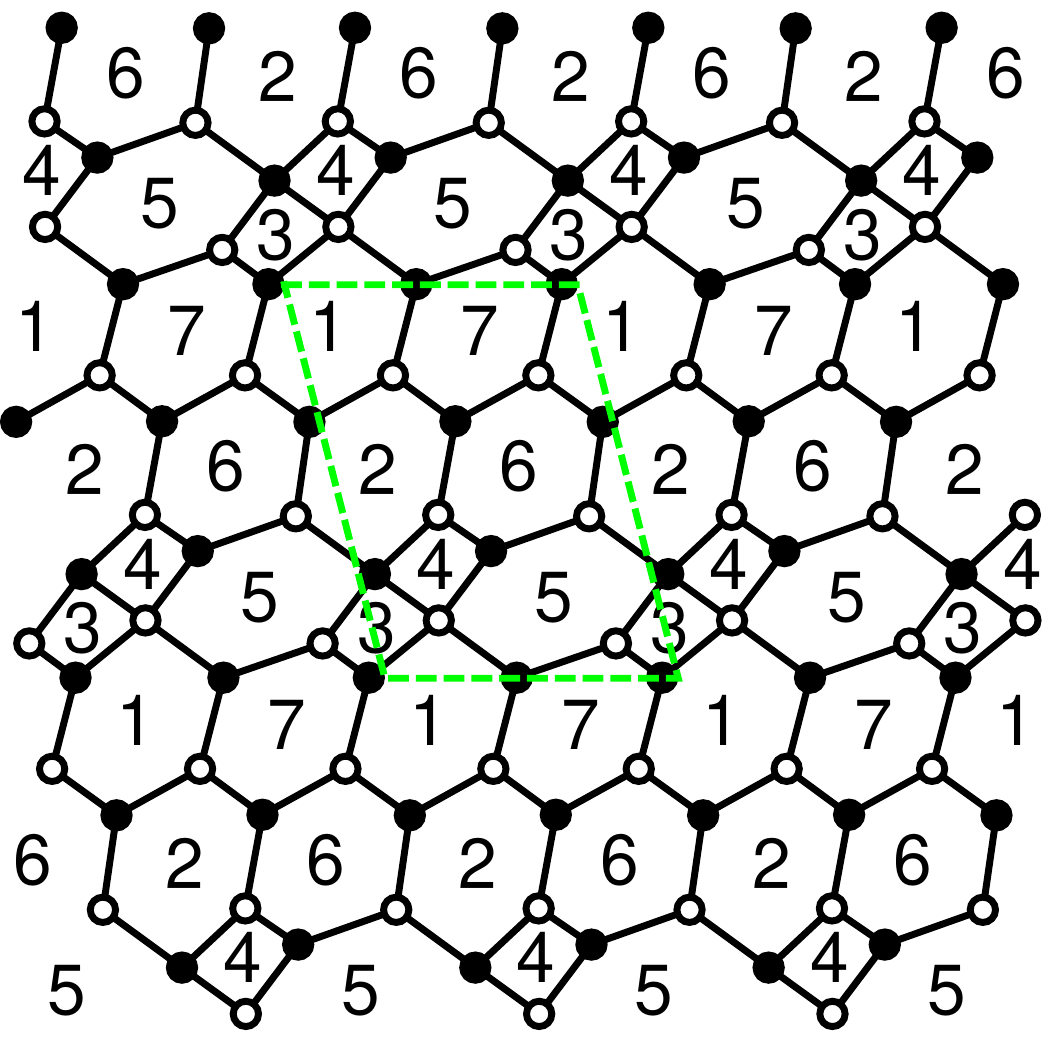}
\caption{The quiver, toric diagram, and brane tiling of Model 5.\label{f5}}
 \end{center}
 \end{figure}
 
\noindent The superpotential is 
\beal{esm5_00}
W&=&
+X_{21} X_{17} X_{72} 
+X_{42} X_{26} X_{64} 
+X_{56} X_{62} X_{25} 
+X_{67} X_{71} X_{16}  
+X_{75} X_{53} X_{37} 
+X_{13} X_{34} X_{45} X_{51} 
\nn\\
&&
-X_{13} X_{37} X_{71} 
-X_{16} X_{62} X_{21} 
-X_{56} X_{64} X_{45} 
-X_{67} X_{72} X_{26} 
-X_{75} X_{51} X_{17}
-X_{25} X_{53} X_{34} X_{42} 
\nn\\
 \eea

\noindent The perfect matching matrix is 

\noindent\makebox[\textwidth]{%
\footnotesize
$
P=
\left(
\begin{array}{c|cccc|cc|ccc|ccc|ccccccccc}
\; & p_{1} & p_{2} & p_{3} & p_{4} & q_1 & q_2 & r_1 & r_2 & r_3 & u_1& u_2 & u_3 & s_1 & s_2 & s_3 & s_4 & s_5 & s_6 & s_7 & s_8 & s_9\\
   \hline
 X_{45} & 1 & 0 & 0 & 0 & 1 & 0 & 0 & 0 & 0 & 0 & 0 & 0 & 1 & 1 & 0 & 0 & 0 & 0 &
   1 & 0 & 0 \\
 X_{53} & 1 & 0 & 0 & 0 & 0 & 1 & 0 & 0 & 0 & 0 & 0 & 0 & 0 & 0 & 1 & 1 & 0 & 0 &
   0 & 1 & 0 \\
 X_{26} & 1 & 0 & 0 & 1 & 1 & 0 & 0 & 0 & 0 & 0 & 0 & 0 & 0 & 0 & 1 & 0 & 1 & 0 &
   1 & 1 & 0 \\
 X_{17} & 1 & 0 & 0 & 0 & 1 & 0 & 0 & 0 & 0 & 0 & 0 & 0 & 1 & 0 & 1 & 1 & 1 & 0 &
   0 & 0 & 1 \\
 X_{62} & 1 & 0 & 0 & 0 & 0 & 1 & 0 & 0 & 0 & 0 & 0 & 0 & 1 & 1 & 0 & 1 & 0 & 1 &
   0 & 0 & 1 \\
 X_{71} & 1 & 0 & 0 & 1 & 0 & 1 & 0 & 0 & 0 & 0 & 0 & 0 & 0 & 1 & 0 & 0 & 0 & 1 &
   1 & 1 & 0 \\
 X_{25} & 0 & 1 & 0 & 0 & 1 & 0 & 1 & 0 & 1 & 0 & 1 & 0 & 0 & 0 & 0 & 0 & 0 & 0 &
   1 & 0 & 0 \\
 X_{75} & 0 & 0 & 1 & 1 & 0 & 0 & 1 & 0 & 0 & 1 & 1 & 0 & 0 & 1 & 0 & 0 & 0 & 1 &
   1 & 0 & 0 \\
 X_{51} & 0 & 1 & 0 & 0 & 0 & 1 & 0 & 1 & 1 & 0 & 0 & 1 & 0 & 0 & 0 & 0 & 0 & 0 &
   0 & 1 & 0 \\
 X_{56} & 0 & 0 & 1 & 1 & 0 & 0 & 0 & 1 & 0 & 1 & 0 & 1 & 0 & 0 & 1 & 0 & 1 & 0 &
   0 & 1 & 0 \\
 X_{37} & 0 & 1 & 0 & 0 & 1 & 0 & 0 & 1 & 1 & 0 & 0 & 1 & 1 & 0 & 0 & 0 & 1 & 0 &
   0 & 0 & 1 \\
 X_{42} & 0 & 0 & 1 & 0 & 0 & 0 & 0 & 1 & 0 & 1 & 0 & 1 & 1 & 1 & 0 & 0 & 0 & 0 &
   0 & 0 & 0 \\
 X_{64} & 0 & 1 & 0 & 0 & 0 & 1 & 1 & 0 & 1 & 0 & 1 & 0 & 0 & 0 & 0 & 1 & 0 & 1 &
   0 & 0 & 1 \\
 X_{13} & 0 & 0 & 1 & 0 & 0 & 0 & 1 & 0 & 0 & 1 & 1 & 0 & 0 & 0 & 1 & 1 & 0 & 0 &
   0 & 0 & 0 \\
 X_{16} & 0 & 1 & 0 & 0 & 1 & 0 & 1 & 1 & 0 & 1 & 0 & 0 & 0 & 0 & 1 & 0 & 1 & 0 &
   0 & 0 & 0 \\
 X_{72} & 0 & 1 & 0 & 0 & 0 & 1 & 1 & 1 & 0 & 1 & 0 & 0 & 0 & 1 & 0 & 0 & 0 & 1 &
   0 & 0 & 0 \\
 X_{21} & 0 & 0 & 1 & 1 & 0 & 0 & 0 & 0 & 1 & 0 & 1 & 1 & 0 & 0 & 0 & 0 & 0 & 0 &
   1 & 1 & 0 \\
 X_{67} & 0 & 0 & 1 & 0 & 0 & 0 & 0 & 0 & 1 & 0 & 1 & 1 & 1 & 0 & 0 & 1 & 0 & 0 &
   0 & 0 & 1 \\
 X_{34} & 0 & 0 & 0 & 1 & 0 & 0 & 0 & 0 & 0 & 0 & 0 & 0 & 0 & 0 & 0 & 0 & 1 & 1 &
   0 & 0 & 1 
   \end{array}
\right)
$
}
\vspace{0.5cm}

\noindent The F-term charge matrix $Q_F=\ker{(P)}$ is

\noindent\makebox[\textwidth]{%
\footnotesize
$
Q_F=
\left(
\begin{array}{cccc|cc|ccc|ccc|ccccccccc}
  p_{1} & p_{2} & p_{3} & p_{4} & q_1 & q_2 & r_1 & r_2 & r_3 & u_1& u_2 & u_3 & s_1 & s_2 & s_3 & s_4 & s_5 & s_6 & s_7 & s_8 & s_9\\
  \hline
  1 & 1 & 0 & 0 & -1 & -1 & 0 & 0 & 0 & 0 & 0 & 0 & 0 & 0 & 0 & 0 & 0 & 0 & 0 & 0 & 0 \\
 1 & 0 & 0 & 0 & 0 & 0 & 0 & 0 & 0 & 1 & 0 & 0 & 0 & -1 & -1 & 0 & 0 & 0 & 0 & 0 & 0 \\
 1 & 0 & 0 & 0 & 0 & 0 & 0 & 0 & 0 & 0 & 0 & 1 & -1 & 0 & 0 & 0 & 0 & 0 & 0 & -1 & 0 \\
  1 & 0 & 0 & 0 & -1 & 0 & 1 & 0 & 0 & 0 & 0 & 0 & 1 & -1 & 0 & -1 & 0 & 0 & 0 & 0 & 0
   \\
 0 & 1 & 0 & 0 & -1 & 0 & 0 & -1 & 0 & 0 & 0 & 0 & 0 & 1 & 0 & 0 & 1 & -1 & 0 & 0 & 0
   \\
 0 & 1 & 0 & 0 & 0 & 0 & -1 & -1 & 0 & 1 & 0 & 0 & 0 & 0 & 0 & 0 & 0 & 0 & 0 & 0 & 0 \\
 0 & 1 & 0 & 0 & 0 & 0 & -1 & 0 & -1 & 0 & 1 & 0 & 0 & 0 & 0 & 0 & 0 & 0 & 0 & 0 & 0 \\
 0 & 1 & 1 & 0 & 0 & 0 & -1 & 0 & 0 & 0 & 0 & -1 & 0 & 0 & 0 & 0 & 0 & 0 & 0 & 0 & 0 \\
 0 & 1 & 1 & 0 & 0 & 0 & 0 & -1 & 0 & 0 & -1 & 0 & 0 & 0 & 0 & 0 & 0 & 0 & 0 & 0 & 0  \\

 0 & 0 & 1 & -1 & 1 & 0 & -1 & 0 & 0 & 0 & 0 & 0 & -1 & 0 & 0 & 0 & 0 & 1 & 0 & 0 & 0
   \\
 0 & 0 & 0 & 0 & 1 & 0 & -1 & 0 & 0 & 0 & 1 & 0 & -1 & 1 & 0 & 0 & 0 & 0 & -1 & 0 & 0\\
 0 & 0 & 0 & 0 & 0 & 0 & 0 & 0 & 0 & 0 & 0 & 0 & 1 & -1 & 0 & 0 & 0 & 1 & 0 & 0 & -1
  \end{array}
\right)
$
}
\vspace{0.5cm}

\noindent The D-term charge matrix is

\noindent\makebox[\textwidth]{%
\footnotesize
$
Q_D=
\left(
\begin{array}{cccc|cc|ccc|ccc|ccccccccc}
  p_{1} & p_{2} & p_{3} & p_{4} & q_1 & q_2 & r_1 & r_2 & r_3 & u_1& u_2 & u_3 & s_1 & s_2 & s_3 & s_4 & s_5 & s_6 & s_7 & s_8 & s_9\\
  \hline
0 & 0 & 0 & 0 & 0 & 0 & 0 & 0 & 0 & 0 & 0 & 0 & 0 & 0 & 1 & -1 & 0 & 0 & 0 & 0 & 0 \\
 0 & 0 & 0 & 0 & 0 & 0 & 0 & 0 & 0 & 0 & 0 & 0 & 0 & 0 & 0 & 1 & -1 & 0 & 0 & 0 & 0 \\
 0 & 0 & 0 & 0 & 0 & 0 & 0 & 0 & 0 & 0 & 0 & 0 & 0 & 0 & 0 & 0 & 1 & -1 & 0 & 0 & 0 \\
 0 & 0 & 0 & 0 & 0 & 0 & 0 & 0 & 0 & 0 & 0 & 0 & 0 & 0 & 0 & 0 & 0 & 1 & -1 & 0 & 0 \\
 0 & 0 & 0 & 0 & 0 & 0 & 0 & 0 & 0 & 0 & 0 & 0 & 0 & 0 & 0 & 0 & 0 & 0 & 1 & -1 & 0 \\
 0 & 0 & 0 & 0 & 0 & 0 & 0 & 0 & 0 & 0 & 0 & 0 & 0 & 0 & 0 & 0 & 0 & 0 & 0 & 1 & -1
  \end{array}
\right)
$
}
\vspace{0.5cm}

The total charge matrix $Q_t$ does not have repeated columns. Accordingly, the global symmetry is $U(1)_{f_1} \times U(1)_{f_2} \times U(1)_{R}$. Following the discussion in \sref{s1_3}, the flavour and R-charges on GLSM fields corresponding to extremal points in the toric diagram in \fref{f5} are found. They are shown in \tref{t5}.

\begin{table}[H]
\centering
\begin{tabular}{|c||c|c|c||l|}
\hline
\; & $U(1)_{f_1}$ & $U(1)_{f_2}$ & $U(1)_R$ & fugacity \\
\hline \hline
$p_1$ &  0	&  -1/2	& $R_1\simeq 0.577$ & $t_1$ \\
$p_2$ &  0 	&   1/2	& $R_2\simeq 0.640$ & $t_2$ \\
$p_3$ & -1   &  -1 	& $R_3\simeq 0.539$ & $t_3$ \\
$p_4$ &  1   &   1 	& $R_4\simeq 0.243$ & $t_4$ \\
\hline
\end{tabular}
\caption{The GLSM fields corresponding to extremal points of the toric diagram with their mesonic charges (Model 5).
 \label{t5}}
\end{table}

\noindent\textit{Fine-tuning R-charges.} The exact R-charges can be expressed in terms of roots of the following polynomials
\beal{es5_p1}
0 &=& 75 + 110 x - 684 x^2 + 162 x^3 + 81 x^4  \nn\\
0 &=& -1124565 + 2218649 x_0 - 1141683 x_0^2 - 16497 x_0^3
\nn\\
&&
+ (746100 - 259716 x_0 + 4428 x_0^2 - 64476 x_0^3) y
\nn\\
&&
+ (775170 + 520182 x_0 - 390258 x_0^2 - 70470 x_0^3) y^2
\nn\\
&&
+ (14580 + 100764 x_0 + 164268 x_0^2 + 26244 x_0^3) y^3 
\nn\\
&&
+ (-110565 - 26487 x_0 - 19683 x_0^2 - 6561 x_0^3) y^4 
\nn\\
&&
+ 38880 y^5~~,
\eea
where the roots satisfy the bounds $0\leq 1-x_0 \leq \frac{2}{3}$ and $0\leq 1-y_0 \leq \frac{2}{3}$. The exact R-charges are
\beal{es5_p2}
R_1 &=&
\frac{1}{8989575077760} 
(
-443015521905 + 10382230129225 x_0 - 1861588105479 x_0^2
\nn\\
&& - 
 1223569555569 x_0^3 + 788576007420 y_0 + 7322446656900 x_0 y_0 - 
 1514870485020 x_0^2 y_0 
\nn\\
&&
 - 803839472100 x_0^3 y_0 + 105890430210 y_0^2 - 
 45532791090 x_0 y_0^2 + 616773772782 x_0^2 y_0^2 
 \nn\\
 &&
 + 
 132554296962 x_0^3 y_0^2 - 87638359380 y_0^3 - 829308203820 x_0 y_0^3 + 
 57898633140 x_0^2 y_0^3 
 \nn\\
 &&
 + 57715867980 x_0^3 y_0^3 + 9044838615 y_0^4 + 
 354606896385 x_0 y_0^4 - 66414222351 x_0^2 y_0^4 
 \nn\\
 &&
 - 37556288361 x_0^3 y_0^4
)
\nn\\
R_2 &=& y_0 ~,~ R_3 = x_0 ~,
\eea
\beal{es5_p2b}
R_4 &=&
\frac{1}{27630249136420257145191668008550400}
(443015521905 - 10382230129225 x_0 
\nn\\
&&
+ 1861588105479 x_0^2 + 
     1223569555569 x_0^3 - 788576007420 y_0 - 7322446656900 x_0 y_0 
\nn\\
&&
+ 1514870485020 x_0^2 y_0 + 803839472100 x_0^3 y_0 - 
     105890430210 y_0^2 + 45532791090 x_0 y_0^2 
\nn\\
&&
- 616773772782 x_0^2 y_0^2 
- 132554296962 x_0^3 y_0^2 + 
     87638359380 y_0^3 + 829308203820 x_0 y_0^3 
\nn\\
&&
- 57898633140 x_0^2 y_0^3 - 57715867980 x_0^3 y_0^3 - 
     9044838615 y_0^4 - 354606896385 x_0 y_0^4 
\nn\\
&&
+ 66414222351 x_0^2 y_0^4 + 
     37556288361 x_0^3 y_0^4)~
(3435680922231398676675 - 
\nn\\
&&
     10875934309383304858731 x_0 + 2208889158465224949597 x_0^2 
\nn\\
&&
+ 1149691223996073074763 x_0^3 + 1308961575315964402860 y_0 
\nn\\
&&
- 
     5303703543601718636316 x_0 y_0 + 1007391627507047358708 x_0^2 y_0 
\nn\\
&&
     + 
     577767803346582055164 x_0^3 y_0 - 41445446612526178750 y_0^2 
     \nn\\
     &&
     + 
     324345443167855962702 x_0 y_0^2 - 
     267480237660960501378 x_0^2 y_0^2 
     \nn\\
     &&
     - 
     83757129586072681230 x_0^3 y_0^2 - 143402222077829778740 y_0^3 
     \nn\\
     &&
     + 
     581897049297268121604 x_0 y_0^3 - 73669737309435993132 x_0^2 y_0^3 
     \nn\\
     &&
     - 
     53860834564699887396 x_0^3 y_0^3 + 46554904501591527955 y_0^4 
     \nn\\
     &&
     - 
     286145797904951411547 x_0 y_0^4 + 58286941395335651277 x_0^2 y_0^4 
     \nn\\
     &&
     + 
     31675092179803827579 x_0^3 y_0^4)~~.
\eea

Products of non-extremal perfect matchings are expressed in terms of single variables as follows
\beal{esx5_1}
q = q_1 q_2 ~,~
r = r_1 r_2 ~,~
u = u_1 u_2 ~,~
s = \prod_{m=1}^{9} s_m~.
\eea
The fugacity which counts extremal perfect matchings is $t_\alpha$. The fugacity of the form $y_q$ counts the product of non-extremal perfect matchings $q$.

The mesonic Hilbert series of Model 5 is found using the Molien integral formula in \eref{es12_2}. It is 
\beal{esm5_1}
&&
g_{1}(t_\alpha,y_{q},y_{r},y_{u},y_{s}; \mathcal{M}^{mes}_5)= 
(1 
+ y_{q} y_{r} y_{u} y_{s} ~ t_1 t_2 t_3 t_4 
+ y_{q} y_{r}^2 y_{u}^2 y_{s} ~ t_2^2 t_3^2 t_4 
- y_{q}^3 y_{r}^3 y_{u}^2 y_{s}^2 ~ t_1^2 t_2^4 t_3 t_4 
\nn\\
&&
\hspace{1cm}
- y_{q}^3 y_{r}^4 y_{u}^3 y_{s}^2 ~ t_1 t_2^5 t_3^2 t_4 
- y_{q}^4 y_{r}^5 y_{u}^4 y_{s}^3 ~ t_1^2 t_2^6 t_3^3 t_4^2
)
\nn\\
&&
\hspace{1cm}
\times
\frac{
1
}{
(1 - y_{q}^2 y_{r}^2 y_{u} y_{s} ~ t_1 t_2^3) 
(1 - y_{q}^2 y_{r}^3 y_{u}^2 y_{s} ~ t_2^4 t_3) 
(1 - y_{q} y_{s} ~ t_1^2 t_4) 
(1 - y_{r} y_{u}^2 y_{s} ~ t_3^3 t_4^2)
}~~.
\eea
The plethystic logarithm of the mesonic Hilbert series is
\beal{esm5_3}
&&
PL[g_1(t_\alpha,y_{q},y_{r},y_{u},y_{s};\mathcal{M}_5^{mes})]=
y_{q} y_{r} y_{u} y_{s} ~ t_1 t_2 t_3 t_4
+ y_{q} y_{s} ~ t_1^2 t_4 
+ y_{q}^2 y_{r}^2 y_{u} y_{s} ~ t_1 t_2^3 
\nn\\
&&
\hspace{1cm}
+ y_{r} y_{u}^2 y_{s} ~ t_3^3 t_4^2
+ y_{q} y_{r}^2 y_{u}^2 y_{s} ~ t_2^2 t_3^2 t_4 
+ y_{q}^2 y_{r}^3 y_{u}^2 y_{s} ~ t_2^4 t_3 
- y_{q}^2 y_{r}^2 y_{u}^2 y_{s}^2 ~ t_1^2 t_2^2 t_3^2 t_4^2
\nn\\
&&
\hspace{1cm}
- y_{q}^3 y_{r}^3 y_{u}^2 y_{s}^2 ~ t_1^2 t_2^4 t_3 t_4 
- y_{q}^2 y_{r}^3 y_{u}^3 y_{s}^2 ~ t_1 t_2^3 t_3^3 t_4^2
- y_{q}^3 y_{r}^4 y_{u}^3 y_{s}^2 ~ t_1 t_2^5 t_3^2 t_4 
- y_{q}^2 y_{r}^4 y_{u}^4 y_{s}^2 ~ t_2^4 t_3^4 t_4^2 
\nn\\
&&
\hspace{1cm}
+ y_{q}^4 y_{r}^4 y_{u}^3 y_{s}^3 ~ t_1^3 t_2^5 t_3^2 t_4^2 
+ \dots~.
\eea

Consider the following fugacity map
\beal{esm5_y1}
&& f_1=
\frac{1}{y_u y_r}
~,~
f_2=
\frac{1}{y_u y_s}
~,~
\nn\\
&&
\tilde{t}_1=
y_q^{1/2} y_r^{1/2} y_u^{1/2} y_s^{1/2} t_1
~,~
\tilde{t}_2=
y_q^{1/2} y_r^{1/2} y_u^{1/2} y_s^{1/2} t_2
~,~
\nn\\
&&
\tilde{t}_3=
t_3
~,~
\tilde{t}_4=
t_4
~,~
\eea
where $f_1$ and $f_2$ are the fugacities for the flavor charges, and $\tilde{t}_i$ is the fugacity for the R-charge $R_i$ in table \tref{t5}.
In terms of the fugacity map above, the plethystic logarithm becomes
\beal{esm5_3b}
&&
PL[g_1(\tilde{t}_\alpha,f_1,f_2;\mathcal{M}_5^{mes})]=
\tilde{t}_{1} \tilde{t}_{2} \tilde{t}_{3} \tilde{t}_{4}
+ f_1 \tilde{t}_{1}^2 \tilde{t}_{4}
+ f_2 \tilde{t}_{1} \tilde{t}_{2}^3
+ \frac{1}{f_{1} f_{2}} \tilde{t}_{3}^3 \tilde{t}_{4}^2
+ \frac{1}{f_1} \tilde{t}_{2}^2 \tilde{t}_{3}^2 \tilde{t}_{4}
+\frac{f_{2}}{f_{1}} \tilde{t}_{2}^4 \tilde{t}_{3}
\nn\\
&&
\hspace{1cm}
-\tilde{t}_{1}^2 \tilde{t}_{2}^2 \tilde{t}_{3}^2 \tilde{t}_{4}^2
-f_2 \tilde{t}_{1}^2 \tilde{t}_{2}^4 \tilde{t}_{3} \tilde{t}_{4}
-\frac{1}{f_1} \tilde{t}_{1} \tilde{t}_{2}^3 \tilde{t}_{3}^3 \tilde{t}_{4}^2
+\dots~.
\eea
The above plethystic logarithm exhibits the moduli space generators with their mesonic charges. 

The generators can be presented as points on a $\mathbb{Z}^{2}$ with the $U(1)_{f_1}\times U(1)_{f_2}$ charges giving the lattice coordinates. The convex polygon formed by the generators on the lattice in \tref{t5gen} is the dual reflexive polygon of the toric diagram of Model 5. \\

\begin{table}[H]
\centering
\resizebox{\hsize}{!}{
\begin{minipage}[!b]{0.6\textwidth}
\begin{tabular}{|l|c|c|}
\hline
Generator & $U(1)_{f_1}$ & $U(1)_{f_2}$ 
\\
\hline
\hline
$
p_{1}^2 p_{4}~
q~
s$
& 1 & 0\\
$p_{1} p_{2} p_{3} p_{4}~
q~
r~
u~
s$
& 0 & 0
\\
$p_{1} p_{2}^3~
q^2~
r^2~
u~
s$
& 0 & 1
\\
$p_{3}^3 p_{4}^2~ 
r~
u^2~
s$
& -1 & -1
\\
$p_{2}^2 p_{3}^2 p_{4}~
q~
r^2~
u^2~
s$
& -1 & 0
\\
$p_{2}^4 p_{3}~
q^2~
r^3~
u^2~
s$
& -1 & 1
\nn\\
\hline
\end{tabular}
\end{minipage}
\hspace{2cm}
\begin{minipage}[!b]{0.25\textwidth}
\includegraphics[width=3.5 cm]{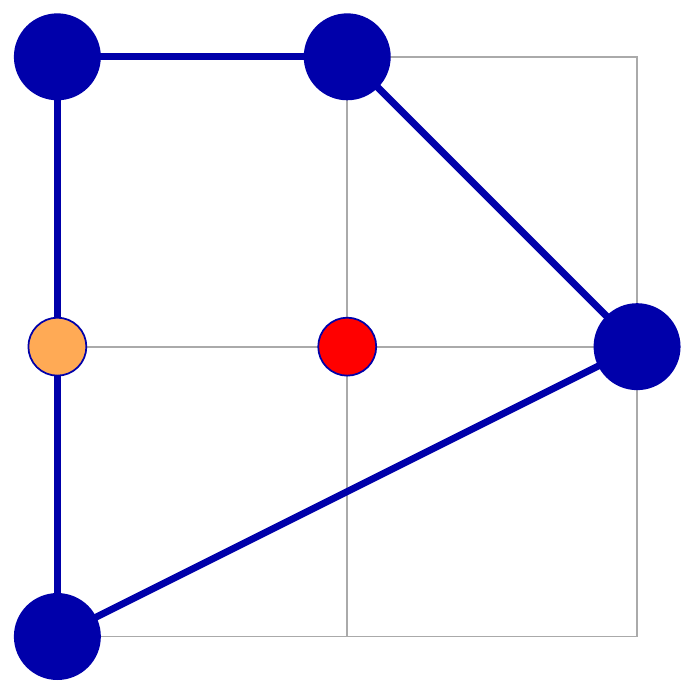}
\end{minipage}
}
\caption{The generators and lattice of generators of the mesonic moduli space of Model 5 in terms of GLSM fields with the corresponding flavor charges.\label{t5gen}\label{f5gen}} 
\end{table}

\begin{table}[H]
\centering

\resizebox{\hsize}{!}{

\begin{tabular}{|l|c|c|}
\hline
Generator & $U(1)_{f_1}$ & $U(1)_{f_2}$ 
\\
\hline
\hline
$
X_{34} X_{45} X_{53}=  X_{17} X_{71}=  X_{26} X_{62}
$
   & 1 & 0
   \\
$
 X_{13} X_{34} X_{45} X_{51}=  X_{25} X_{53} X_{34} X_{42}=  X_{13} X_{37} X_{71}=  X_{16} X_{62} X_{21}=  X_{16} X_{67} X_{71}=  X_{17} X_{72} X_{21}
 $
 & 0 & 0
 \nn\\
 $
 =  X_{17} X_{75} X_{51}=  X_{25} X_{56} X_{62}=  X_{26} X_{64} X_{42}=  X_{26} X_{67} X_{72}=  X_{37} X_{75} X_{53}=  X_{45} X_{56} X_{64}
$
   &   & 
   \\
   $
    X_{16} X_{62} X_{25} X_{51}=  X_{16} X_{64} X_{45} X_{51}=  X_{17} X_{72} X_{25} X_{51}=  X_{25} X_{53} X_{37} X_{72}
   $
   & 0 & 1
   \\
   $
   X_{56} X_{67} X_{75}=  X_{13} X_{34} X_{42} X_{21}
   $
   & -1 & -1
   \\
   $
    X_{13} X_{34} X_{42} X_{25} X_{51}=  X_{13} X_{37} X_{72} X_{21}=  X_{13} X_{37} X_{75} X_{51}=  X_{16} X_{64} X_{42} X_{21}
    $
    & -1 & 0
    \nn\\
    $
    = X_{16} X_{67} X_{72} X_{21}=  X_{16} X_{67} X_{75} X_{51}=  X_{25} X_{56} X_{64} X_{42}=  X_{25} X_{56} X_{67} X_{72}
   $
   & &
   \\
   $
    X_{13} X_{37} X_{72} X_{25} X_{51}=  X_{16} X_{64} X_{42} X_{25} X_{51}=  X_{16} X_{67} X_{72} X_{25} X_{51}
   $
   & -1 & 1
\nn\\
\hline
\end{tabular}
}
\caption{The generators in terms of bifundamental fields (Model 5).\label{t5gen2}\label{f5gen2}} 
\end{table}
   
The Hilbert series and the plethystic logarithm can be re-expressed in terms of just $3$ fugacities
\beal{esm5_x1}
T_1 &=&
\frac{\tilde{t}_3}{f_1 f_2 ~ \tilde{t}_1^2 \tilde{t}_2^2}
=
 \frac{t_3}{y_{q}^2 y_{r} y_{s} ~ t_1^2 t_2^2}~,~
 \nn\\
T_2 &=& 
f_2 ~ \tilde{t}_1 \tilde{t}_2^3
=
y_{q}^2 y_{r}^2 y_{u} y_{s} ~ t_1 t_2^3~,~
\nn\\
T_3 
&=&
f_1 ~ \tilde{t}_1^2 \tilde{t}_4
= y_{q} y_{s} ~ t_1^2 t_4~,~
\eea   
such that
\beal{esm5_x2}
g_1(T_1,T_2,T_3;\mathcal{M}^{mes}_5)
=
\frac{
1 
+ T_1 T_2 T_3 
+ T_1^2 T_2^2 T_3
- T_1 T_2^2 T_3  - T_1^2 T_2^3 T_3 - T_1^3 T_2^4 T_3^2
 }{
 (1 - T_2) (1 - T_1 T_2^2) (1 - T_3) (1 - T_1^3 T_2^2 T_3^2)
 }
\eea   
and
\beal{esm5_x3}
&&
PL[g_1(T_1,T_2,T_3;\mathcal{M}^{mes}_5)]=
T_1 T_2 T_3 
+ T_3  
+ T_2 
+ T_1^3 T_2^2 T_3^2 
+ T_1 T_2^2 
+ T_1^2 T_2^2 T_3 
- T_1 T_2^2 T_3 
\nn\\
&&
- T_1^2 T_2^2 T_3^2  - T_1^2 T_2^3 T_3 - T_1^3 T_2^3 T_3^2 - T_1^4 T_2^4 T_3^2 + T_1^2 T_2^3 T_3^2  + T_1^3 T_2^4 T_3^2 + T_1^4 T_2^4 T_3^3 + T_1^4 T_2^5 T_3^2 
\nn\\
&&
+ T_1^5 T_2^5 T_3^3 - T_1^3 T_2^4 T_3^3 
\dots
~~.
\eea   
The above mesonic Hilbert series and plethystic logarithm illustrates the conical structure of the toric Calalbi-Yau 3-fold.
\\

\section{Model 6: $\text{PdP}_{4a}$}
\subsection{Model 6 Phase a}

\begin{figure}[H]
\begin{center}
\includegraphics[trim=0cm 0cm 0cm 0cm,width=4.5 cm]{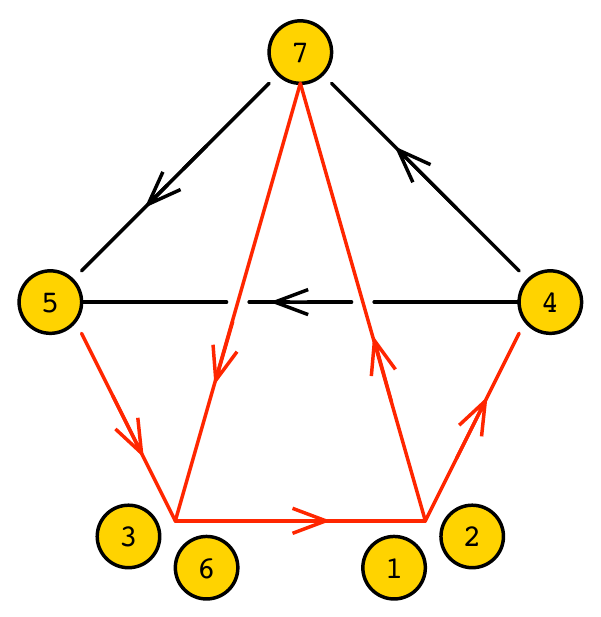}
\includegraphics[width=5 cm]{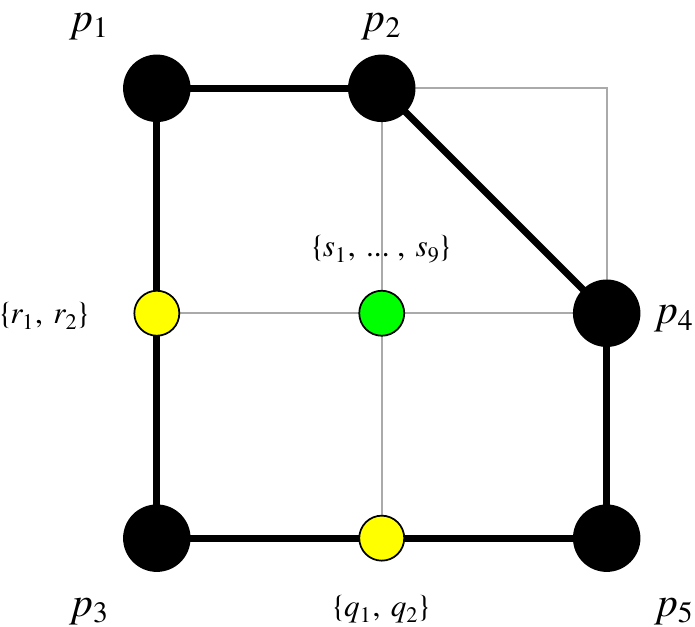}
\includegraphics[width=5 cm]{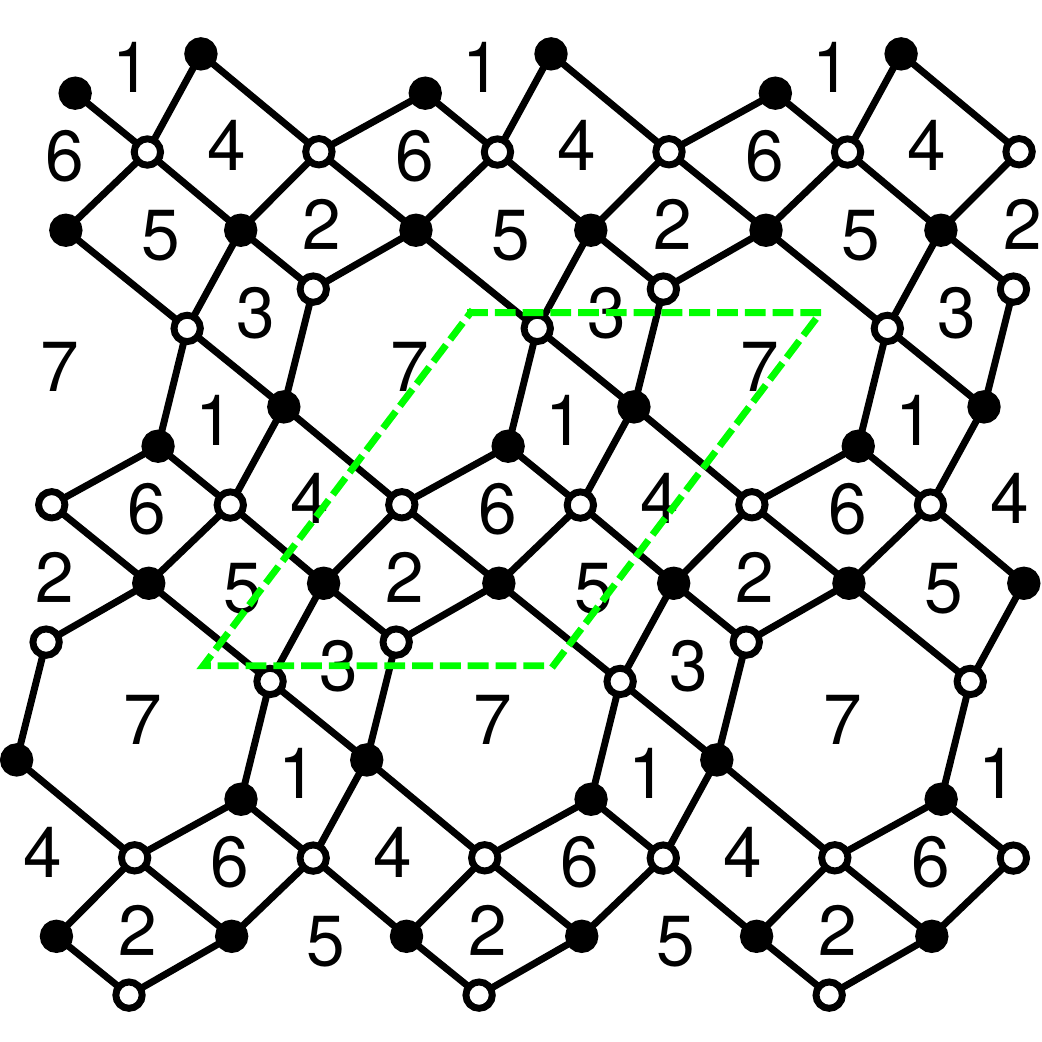}
\caption{The quiver, toric diagram and brane tiling of Model 6a. The red arrows in the quiver indicate all possible connections between blocks of nodes.\label{f6a}}
 \end{center}
 \end{figure}
 
 \noindent The superpotential is 
\beal{esm6a_00}
W&=&
+ X_{32} X_{27} X_{73}  
+ X_{14} X_{45} X_{56} X_{61}  
+ X_{31} X_{17} X_{75} X_{53}  
+ X_{62} X_{24} X_{47} X_{76}  
\nn\\
&&
- X_{76} X_{61} X_{17}  
- X_{31} X_{14} X_{47} X_{73}  
- X_{32} X_{24} X_{45} X_{53}  
- X_{62} X_{27} X_{75} X_{56}   
\nn\\
  \eea
 
 \noindent The perfect matching matrix is 
 
\noindent\makebox[\textwidth]{%
\footnotesize
$
P=
\left(
\begin{array}{c|ccccc|cc|cc|ccccccccc}
 \; & p_1 & p_2& p_3& p_4& p_5& q_1& q_2& r_1& r_2& s_1& s_2& s_3& s_4& s_5& s_6& s_7& s_8& s_9  \\
 \hline
 X_{17} & 1 & 1 & 0 & 0 & 0 & 0 & 0 & 1 & 0 & 1 & 0 & 1 & 0 & 1 & 0 & 0 & 0 & 0 \\
 X_{73} & 1 & 1 & 0 & 0 & 0 & 0 & 0 & 0 & 1 & 0 & 1 & 0 & 1 & 0 & 1 & 0 & 0 & 0 \\
 X_{56} & 1 & 0 & 0 & 0 & 0 & 0 & 0 & 1 & 0 & 0 & 1 & 0 & 0 & 0 & 0 & 1 & 0 & 0 \\
 X_{24} & 1 & 0 & 0 & 0 & 0 & 0 & 0 & 0 & 1 & 1 & 0 & 0 & 0 & 0 & 0 & 0 & 1 & 0 \\
 X_{45} & 0 & 1 & 0 & 1 & 0 & 0 & 0 & 0 & 0 & 0 & 0 & 1 & 0 & 0 & 1 & 0 & 0 & 0 \\
 X_{62} & 0 & 1 & 0 & 0 & 0 & 0 & 0 & 0 & 0 & 0 & 0 & 0 & 1 & 1 & 0 & 0 & 0 & 1 \\
 X_{32} & 0 & 0 & 1 & 0 & 0 & 1 & 0 & 1 & 0 & 0 & 0 & 0 & 0 & 1 & 0 & 1 & 0 & 1 \\
 X_{75} & 0 & 0 & 1 & 0 & 0 & 1 & 0 & 0 & 1 & 0 & 0 & 0 & 0 & 0 & 1 & 0 & 0 & 0 \\
 X_{47} & 0 & 0 & 1 & 0 & 0 & 0 & 1 & 1 & 0 & 0 & 0 & 1 & 0 & 0 & 0 & 0 & 0 & 0 \\
 X_{61} & 0 & 0 & 1 & 0 & 0 & 0 & 1 & 0 & 1 & 0 & 0 & 0 & 1 & 0 & 0 & 0 & 1 & 1 \\
 X_{76} & 0 & 0 & 0 & 1 & 1 & 1 & 0 & 0 & 0 & 0 & 1 & 0 & 0 & 0 & 1 & 1 & 0 & 0 \\
 X_{27} & 0 & 0 & 0 & 1 & 1 & 0 & 1 & 0 & 0 & 1 & 0 & 1 & 0 & 0 & 0 & 0 & 1 & 0 \\
 X_{31} & 0 & 0 & 0 & 1 & 0 & 0 & 0 & 0 & 0 & 0 & 0 & 0 & 0 & 0 & 0 & 1 & 1 & 1 \\
 X_{14} & 0 & 0 & 0 & 0 & 1 & 1 & 0 & 0 & 0 & 1 & 0 & 0 & 0 & 1 & 0 & 0 & 0 & 0 \\
 X_{53} & 0 & 0 & 0 & 0 & 1 & 0 & 1 & 0 & 0 & 0 & 1 & 0 & 1 & 0 & 0 & 0 & 0 & 0
\end{array}
\right)
$
}
\vspace{0.5cm}

 \noindent The F-term charge matrix $Q_F=\ker{(P)}$ is

\noindent\makebox[\textwidth]{%
\footnotesize
$
Q_F=
\left(
\begin{array}{ccccc|cc|cc|ccccccccc}
p_1 & p_2& p_3& p_4& p_5& q_1& q_2& r_1& r_2& s_1& s_2& s_3& s_4& s_5& s_6& s_7& s_8& s_9  \\
\hline 
0 & 0 & 1 & 0 & 1 & -1 & -1 & 0 & 0 & 0 & 0 & 0 & 0 & 0 & 0 & 0 &
   0 & 0 \\
 1 & 0 & 1 & 0 & 0 & 0 & 0 & -1 & -1 & 0 & 0 & 0 & 0 & 0 & 0 & 0 &
   0 & 0 \\
 1 & 0 & 0 & 0 & 1 & 0 & 0 & 0 & 0 & -1 & -1 & 0 & 0 & 0 & 0 & 0 &
   0 & 0 \\
 0 & 1 & 0 & 0 & 0 & 0 & 1 & 0 & 0 & 0 & 0 & -1 & -1 & 0 & 0 & 0 &
   0 & 0 \\
 0 & 1 & 0 & 0 & 0 & 1 & 0 & 0 & 0 & 0 & 0 & 0 & 0 & -1 & -1 & 0 &
   0 & 0 \\
 0 & 0 & 0 & 1 & 0 & 0 & 0 & 1 & 0 & 0 & 0 & -1 & 0 & 0 & 0 & -1 &
   0 & 0 \\
 0 & 0 & 0 & 1 & 0 & 0 & 0 & 0 & 1 & 0 & 0 & 0 & 0 & 0 & -1 & 0 &
   -1 & 0 \\
 0 & 0 & 0 & 0 & 1 & 0 & -1 & 1 & 0 & -1 & 0 & 0 & 0 & 0 & 0 & -1 &
   1 & 0 \\
 0 & 0 & 0 & 0 & 0 & 0 & 0 & 0 & 0 & 1 & 0 & 0 & 0 & -1 & 0 & 0 &
   -1 & 1
\end{array}
\right)
$
}
\vspace{0.5cm}

\noindent The D-term charge matrix is

\noindent\makebox[\textwidth]{%
\footnotesize
$
Q_D=
\left(
\begin{array}{ccccc|cc|cc|ccccccccc}
p_1 & p_2& p_3& p_4& p_5& q_1& q_2& r_1& r_2& s_1& s_2& s_3& s_4& s_5& s_6& s_7& s_8& s_9  \\ 
\hline
 0 & 0 & 0 & 0 & 0 & 0 & 0 & 0 & 0 & 0 & 0 & 1 & -1 & 0 & 0 & 0 & 0 & 0 \\
 0 & 0 & 0 & 0 & 0 & 0 & 0 & 0 & 0 & 0 & 0 & 0 & 1 & -1 & 0 & 0 & 0 & 0 \\
 0 & 0 & 0 & 0 & 0 & 0 & 0 & 0 & 0 & 0 & 0 & 0 & 0 & 1 & -1 & 0 & 0 & 0 \\
 0 & 0 & 0 & 0 & 0 & 0 & 0 & 0 & 0 & 0 & 0 & 0 & 0 & 0 & 1 & -1 & 0 & 0 \\
 0 & 0 & 0 & 0 & 0 & 0 & 0 & 0 & 0 & 0 & 0 & 0 & 0 & 0 & 0 & 1 & -1 & 0 \\
 0 & 0 & 0 & 0 & 0 & 0 & 0 & 0 & 0 & 0 & 0 & 0 & 0 & 0 & 0 & 0 & 1 & -1
\end{array}
\right)
$
}
\vspace{0.5cm}

The total charge matrix $Q_t$ does not exhibit repeated columns. Accordingly, the global symmetry is $U(1)_{f_1} \times U(1)_{f_2} \times U(1)_R$. The mesonic charges on the GLSM fields corresponding to extremal points in the toric diagram in \fref{f6a} are found following the discussion in \sref{s1_3}. They are presented in \tref{t6a}.

\begin{table}[H]
\centering
\begin{tabular}{|c||c|c|c||l|} 
\hline
\; & $U(1)_{f_1}$ & $U(1)_{f_2}$ & $U(1)_R$ & fugacity \\
\hline
\hline
$p_1$ &-1 & 0 & $R_1\simeq 0.427$ &  	$t_1$\\
$p_2$ & 1 & 0 & $R_2\simeq 0.298$ &  	$t_2$\\
$p_3$ & 0 & 0 & $R_3\simeq 0.550$ &  	$t_3$\\
$p_4$ & 0 & 1 & $R_2\simeq 0.298$ &  	$t_4$\\
$p_5$ & 0 &-1 & $R_1\simeq 0.427$ &  	$t_5$\\
\hline
\end{tabular}
\caption{The GLSM fields corresponding to extremal points of the toric diagram with their mesonic charges (Model 6a).\label{t6a}}
\end{table}

\noindent\textit{Fine-tuning R-charges.} The exact R-charges on extremal perfect matchings can be expressed in terms of a root $x_0$ of the following polynomial
\beal{es6a_p1}
0=289 - 695 x + 331 x^2 + 3 x^3~~,
\eea
where the root of interest lies in the range $0\leq 1-x_0 \leq \frac{2}{3}$. The exact R-charges are
\beal{es6a_p2}
R_1 &=& R_5 = x_0~~, \nn\\
R_2 &=& R_4 = {\scriptscriptstyle	\frac{1}{2497416307960655824746468547906174933430973669888}(1791039188638478428147683691212722044339352504896 - 
}
\nn\\
&&
{\scriptscriptstyle	
   14898979385812450997203995618175138834683612621776 x_0 + 
   9465606277116561007612744735839203666371878276840 x_0^2 
   }
   \nn\\
   &&
   {\scriptscriptstyle	
   + 
   81716323060687762935758761257370794928088890023074 x_0^3 - 
   106622759169801872631350808556548913284672579964562 x_0^4 
   }
   \nn\\
   &&
   {\scriptscriptstyle	
   - 
   22312936155603381509800509872608673629726066365173 x_0^5 + 
   47625288680151873547605102674953720401814301943043 x_0^6 
   }
   \nn\\
   &&
   {\scriptscriptstyle	
   + 
   17436573584263377204018474073188553946245197817747 x_0^7 - 
   10640233660391309102082256624734477840137858566189 x_0^8 
   }
   \nn\\
   &&
   {\scriptscriptstyle	
   - 
   5762098668974680244859599181817775913551620378815 x_0^9 + 
   420178930354717433094049925945927510179738217313 x_0^{10} 
   }
   \nn\\
   &&
   {\scriptscriptstyle	
   + 
   721282505298136032927398268634974111953118024491 x_0^{11} + 
   84691631710249529644695474904666891867205565263 x_0^{12} 
   }
   \nn\\
   &&
   {\scriptscriptstyle	
   - 
   28845127177680312829862811387042101533046922792 x_0^{13} - 
   5936715130045788144646704656470430250253226360 x_0^{14} 
   }
   \nn\\
   &&
   {\scriptscriptstyle	
   - 
   98568203174737761263257326460337456059549812 x_0^{15} - 
   427836112588315949366063712216265071084900 x_0^{16})
   }
   \nn\\
   R_3 &=& 
{\scriptscriptstyle	
\frac{1}{162164293596963665649085313948683843212137836604660555443821244188609125275748366817763000746246144}
\times}
\nn\\
&&
{\scriptscriptstyle	
(1169229461732080766319602708065371848435839320818952726286766174485578754720869791380548487029993472 
}
\nn\\
&&
{\scriptscriptstyle	
+ 
   211180778264971290234686689177114661495550847435083609777692608446996489161070763569563200559556608 x_0 
}
\nn\\
&&
{\scriptscriptstyle	
-    8045911260354654893884448259742088551904830575685775809252492449742
813094597380760696064423664722176 x_0^2 
}
\nn\\
&&
{\scriptscriptstyle	
+ 
7868186882915851426335876977581680670251639520854407669554513212398555158000171156489937456815968256 x_0^3 
}
\nn\\
&&
{\scriptscriptstyle	
+ 
   1061412415136716326837022119308869488382612389978875078709377550354824411184572440342496757041597952 x_0^4 }
\nn\\
&&
{\scriptscriptstyle	
- 
   1653502269547432808110213130155065398558657253926330204747817424734038646912023554904414840355605600 x_0^5 }
\nn\\
&&
{\scriptscriptstyle	
- 
   1803409805355686010966266040602399537481777012614017830538582946961232414356541894961178034998651796 x_0^6 }
\nn\\
&&
{\scriptscriptstyle	- 
   549776367467559089730992163878433891954155708884076666297519890732983478315466620106823873137240968 x_0^7 }
\nn\\
&&
{\scriptscriptstyle	
+ 
   1567205800812219625317948680985038429143438706488862950374641790454745258466005289304610895198165728 x_0^8 }
\nn\\
&&
{\scriptscriptstyle	
+ 
   1433721411232234278937225795709815998152998730166082929889466098261318411272932929131404259129653584 x_0^9 }
\nn\\
&&
{\scriptscriptstyle	- 
   613688233093161903664079322747531650516395529165734417290427408319218066807931662878404186231703821 x_0^{10} }
\nn\\
&&
{\scriptscriptstyle	- 
   1113293590933793106422270537761639133335738086439537494201648209333162655868499870321712814024965074 x_0^{11} }
\nn\\
&&
{\scriptscriptstyle	- 
   102041918652529018684594920735103376517462333159418315892949204114090196647595956807850428412457223 x_0^{12} }
\nn\\
&&
{\scriptscriptstyle	
+    423971220164725630883036801237262772103566877143219798793826532397912386224511438398003376083572668 x_0^{13} }
\nn\\
&&
{\scriptscriptstyle	+ 
   180759001526368976093293859900166369755100685781123847882792925416562642901424926786767271598815811 x_0^{14} }
\nn\\
&&
{\scriptscriptstyle	- 
   64076409612708878884915082831557118415463407072251976303703677310275213068268096657416079746613630 x_0^{15} }
\nn\\
&&
{\scriptscriptstyle	
- 
65515048191365797148208738907166511172835001443254598513046452678884061405276488997002820753820879 x_0^{16} }
\nn\\
&&
{\scriptscriptstyle	- 
   6673543248212741805371881957906917086875901203329952658459597394917113521671659599449171717221560 x_0^{17} } \nn
   \eea
   \beal{es6a_p3}
   &&
{\scriptscriptstyle	+ 
9783618126417420629286524671582244856923708960297834037315293570385351437452828996816592454899857 x_0^{18} }
\nn\\
&&
{\scriptscriptstyle	+ 
   3743596998189704676218096923916451542387351120245899948167098322376252076440477648997681642932578 x_0^{19} }
\nn\\
&&
{\scriptscriptstyle	- 
   275998133977857656048993198548594390031696954517741623737712596072996328801012600935299966017093 x_0^{20} }
\nn\\
&&
{\scriptscriptstyle	- 
   476041152324864443368732013757192469363702044100009981148537231549870724895965447800279556079204 x_0^{21} }
\nn\\
&&
{\scriptscriptstyle	- 
   85609276841164659611375420767097192313538344215051215501287679764566381328323514407504142650419 x_0^{22} }
\nn\\
&&
{\scriptscriptstyle	+ 
   17367562182813808407040196634409802339840610442753700821338207976254354309961105906728375495974 x_0^{23} }
\nn\\
&&
{\scriptscriptstyle	+ 
   8815437949275542972852271440501158360572534817622944767660802051044839059890817853038120935475 x_0^{24} }
\nn\\
&&
{\scriptscriptstyle	+ 
   810859117231117720381035609644014422426938987804828817976536807039578657743651484402841788080 x_0^{25} }
\nn\\
&&
{\scriptscriptstyle
- 
   192053072909652328210545003570080037621773138610979153812374936807238481083663630535339645040 x_0^{26} }
\nn\\
&&
{\scriptscriptstyle- 
   53654746591696330685568418173933234993477414863583111739501098102715138908233779767156870480 x_0^{27} }
\nn\\
&&
{\scriptscriptstyle- 
   4633797214013132583423895629091032185087243889634863057878937498434947801893349846356567080 x_0^{28} }
\nn\\
&&
{\scriptscriptstyle- 
   125288849075771386136313950769094507337581594854187196969684084483533817892821528939996160 x_0^{29} }
\nn\\
&&
{\scriptscriptstyle- 
   1502297452596476410349719722105724798487349802028494174267727244065661237915976256430480 x_0^{30} }
\nn\\
&&
{\scriptscriptstyle- 
   8418891003214045205392116768323041884281772276495435205984021439684373541279712292000 x_0^{31} }
\nn\\
&&
{\scriptscriptstyle- 
   18079841511425240505298612186248088798565454098873210645653293047869238161800450000 x_0^{32})
}~~.
\eea

Products of non-extremal perfect matchings are expressed in terms of single variables as follows
\beal{esx6_1}
q = q_1 q_2 ~,~
r = r_1 r_2 ~,~
s = \prod_{m=1}^{9} s_m~. 
\eea
Extremal perfect matchings are counted by the fugacity $t_\alpha$. The fugacity $y_q$ is assigned to the product of non-extremal perfect matchings $q$ above.

The refined mesonic Hilbert series of Model 6a is 
 \beal{esm6a_1}
&&g_{1}(t_\alpha,y_{q},y_{r},y_{s}; \mathcal{M}^{mes}_{6a})=
(1 
+ y_{q} y_{r} y_{s} ~ t_1 t_2 t_3 t_4 t_5 
- y_{q}^2 y_{r}^3 y_{s}^2 ~ t_1^3 t_2^2 t_3^3 t_4 t_5 
- y_{q}^3 y_{r}^3 y_{s}^2 ~ t_1^2 t_2 t_3^4 t_4 t_5^2 
\nn\\
&&
\hspace{0.5cm}
- y_{q}^2 y_{r}^2 y_{s}^2 ~ t_1^2 t_2^2 t_3^2 t_4^2 t_5^2 
- y_{q}^3 y_{r}^2 y_{s}^2 ~ t_1 t_2 t_3^3 t_4^2 t_5^3 
+ y_{q}^4 y_{r}^4 y_{s}^3 ~ t_1^3 t_2^2 t_3^5 t_4^2 t_5^3 
+ y_{q}^5 y_{r}^5 y_{s}^4 ~ t_1^4 t_2^3 t_3^6 t_4^3 t_5^4)
\nn\\
&&
\hspace{0.5cm}
\times
\frac{
1
}{
(1 - y_{q} y_{r}^2 y_{s}~ t_1^2 t_2 t_3^2) 
(1 - y_{r} y_{s} ~ t_1^2 t_2^2 t_4) 
(1 - y_{q}^2 y_{r}^2 y_{s} ~ t_1 t_3^3 t_5) 
}
\nn\\
&&
\hspace{0.5cm}
\times
\frac{1}
{
(1 - y_{q}^2 y_{r} y_{s} ~ t_3^2 t_4 t_5^2) 
(1 - y_{q} y_{s} ~ t_2 t_4^2 t_5^2)}
~~.
\nn\\
\eea
 The plethystic logarithm of the mesonic Hilbert series is
\beal{esm6a_3}
&&
PL[g_1(t_\alpha,y_{q},y_{r},y_{s};\mathcal{M}_{6a}^{mes})]=
y_{q} y_{s} ~ t_2 t_4^2 t_5^2 
+ y_{r} y_{s} ~ t_1^2 t_2^2 t_4 
+ y_{q} y_{r} y_{s} ~ t_1 t_2 t_3 t_4 t_5 
 \nn\\
 &&
  \hspace{1cm}
+ y_{q} y_{r}^2 y_{s} ~ t_1^2 t_2 t_3^2 
+ y_{q}^2 y_{r} y_{s} ~ t_3^2 t_4 t_5^2 
+ y_{q}^2 y_{r}^2 y_{s} ~ t_1 t_3^3 t_5 
- 2 ~ y_{q}^2 y_{r}^2 y_{s}^2 ~ t_1^2 t_2^2 t_3^2 t_4^2 t_5^2 
\nn\\
&&
\hspace{1cm}
- y_{q}^3 y_{r}^3 y_{s}^2 ~ t_1^2 t_2 t_3^4 t_4 t_5^2 
+ \dots~.
\eea

Consider the following fugacity map
\beal{esm6a_y1}
f_1 =
\frac{1}{y_r ~ t_1^2 t_2^2 t_4}
~,~
f_2 =
\frac{1}{y_q ~ t_2 t_4^2 t_5^2}
~,~
\tilde{t}_1 =
y_q^{1/2} y_r^{1/2} y_s^{1/2} ~ t_1 t_5
~,~
\tilde{t}_2 =
t_2 t_4
~,~
\tilde{t}_3 =
\frac{t_3}{t_1 t_2 t_4 t_5}
~,~
\nn\\
\eea
where $f_1$ and $f_2$ are the flavour charge fugacities, and $\tilde{t}_i$ is the fugacity for the R-charge $R_i$ in \tref{t6a}.

In terms of the fugacity map above, the plethystic logarithm becomes
\beal{esm6a_3}
&&
PL[g_1(\tilde{t}_\alpha,f_1,f_2;\mathcal{M}_{6a}^{mes})]=
\left(f_1+f_2\right) \tilde{t}_{1}^2 \tilde{t}_{2}^3
+\tilde{t}_{1}^2 \tilde{t}_{2}^2 \tilde{t}_{3}
+\left(\frac{1}{f_{1}} +\frac{1}{f_{2}}\right) \tilde{t}_{1}^2 \tilde{t}_{2} \tilde{t}_{3}^2
+\frac{1}{f_{1} f_{2}} \tilde{t}_{1}^2 \tilde{t}_{3}^3
   \nn\\
   &&
   \hspace{1cm}
-2 \tilde{t}_{1}^4 \tilde{t}_{2}^4 \tilde{t}_{3}^2 
-\frac{1}{f_{1} f_{2}} \tilde{t}_{1}^4 \tilde{t}_{2}^2 \tilde{t}_{3}^4
    +\dots~.
   \eea
The above plethystic logarithm exhibits the moduli space generators with the corresponding mesonic charges. They are summarized in \tref{t6agen}. The generators can be presented on a charge lattice. The convex polygon formed by the generators in \tref{t6agen} is the dual reflexive polygon of the toric diagram of Model 6a.\\

\begin{table}[H]
\centering
\resizebox{\hsize}{!}{
\begin{minipage}[!b]{0.6\textwidth}
\begin{tabular}{|l|c|c|}
\hline
Generator & $U(1)_{f_1}$ & $U(1)_{f_2}$ 
\\
\hline
\hline
$p_2 p_4^2 p_5^2 ~ q ~ s$ 
& 1 & 0
\nn\\
$p_1^2 p_2^2 p_4 ~ r ~ s$
& 0 & 1
\nn\\
$p_1 p_2 p_3 p_4 p_5 ~ q ~ r ~ s$
& 0 & 0
\nn\\
$p_3^2 p_4 p_5^2 ~ q^2 ~ r ~ s$
& 0 & -1
\nn\\
$p_1^2 p_2 p_3^2 ~ q ~ r^2 ~ s$
& -1 & 0
\nn\\ 
$p_1 p_3^3 p_5 ~ q^2 ~ r^2 ~ s$
& -1 & -1
\nn\\
   \hline
\end{tabular}
\end{minipage}
\hspace{1cm}
\begin{minipage}[!b]{0.3\textwidth}
\includegraphics[width=4 cm]{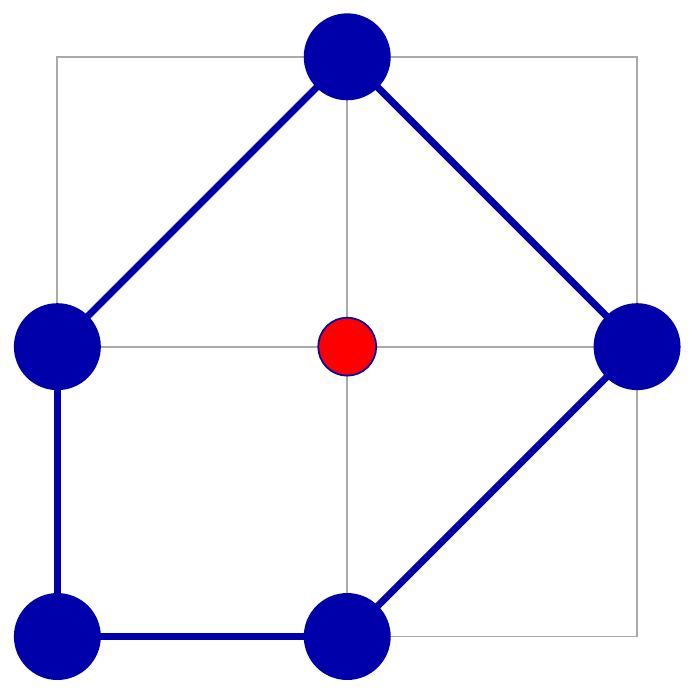}
\end{minipage}
}
\caption{The generators and lattice of generators of the mesonic moduli space of Model 6a in terms of GLSM fields with the corresponding flavor charges.\label{t6agen}\label{f6agen}} 
\end{table}

\begin{table}[H]
\centering

\resizebox{\hsize}{!}{
\begin{tabular}{|l|c|c|}
\hline
Generator & $U(1)_{f_1}$ & $U(1)_{f_2}$ 
\\
\hline
\hline
$
X_{27} X_{76} X_{62}=  X_{14} X_{45} X_{53} X_{31}
$ 
& 1 & 0
\nn\\
$
 X_{17} X_{73} X_{31}=  X_{24} X_{45} X_{56} X_{62}
$
& 0 & 1
\nn\\
$
 X_{17} X_{76} X_{61}=  X_{27} X_{73} X_{32}=  X_{14} X_{47} X_{73} X_{31}=  X_{14} X_{45} X_{56} X_{61}
 $
 & 0 & 0
 \nn\\
 $
 =  X_{17} X_{75} X_{53} X_{31}=  X_{24} X_{45} X_{53} X_{32}=  X_{24} X_{47} X_{76} X_{62}=  X_{27} X_{75} X_{56} X_{62}
$
& & 
\nn\\
$
X_{14} X_{47} X_{75} X_{53} X_{31}=  X_{14} X_{47} X_{76} X_{61}=  X_{27} X_{75} X_{53} X_{32}
$
& 0 & -1
\nn\\
$
 X_{24} X_{47} X_{75} X_{56} X_{62}=  X_{17} X_{75} X_{56} X_{61}=  X_{24} X_{47} X_{73} X_{32}
$
& -1 & 0
\nn\\ 
$
 X_{14} X_{47} X_{75} X_{56} X_{61}=  X_{24} X_{47} X_{75} X_{53} X_{32}
$
& -1 & -1
\nn\\
   \hline
\end{tabular}
}
\caption{The generators in terms of bifundamental fields (Model 6a).\label{t6agen2}\label{f6agen2}} 
\end{table}

The mesonic Hilbert series and plethystic logarithm can be re-expressed in terms of just $3$ fugacities
\beal{esm6a_x1}   
T_1 =
\frac{f_1}{f_2 ~ \tilde{t}_1^2 \tilde{t}_2^2 \tilde{t}_3}
= \frac{t_5}{y_{r}^2 y_{s} ~ t_1^3 t_2^2 t_3}
~,~
T_2 = 
\frac{\tilde{t}_1^2 \tilde{t}_2 \tilde{t}_3^2}{f_1}
=
y_{q} y_{r}^2 y_{s} ~ t_1^2 t_2 t_3^2~,~
T_3 = 
f_2 ~ \tilde{t}_1^2 \tilde{t}_2^3
=
y_{r} y_{s} ~ t_1^2 t_2^2 t_4~,~
\nn\\
\eea
such that   
\beal{esm6a_x2}
&&
g_1(T_1,T_2,T_3;\mathcal{M}^{mes}_{6a})=
\nn\\
&&
\hspace{1cm}
\frac{
1 + T_1 T_2 T_3 - T_1 T_2^2 T_3 - T_1^2 T_2^3 T_3 - T_1^2 T_2^2 T_3^2 - T_1^3 T_2^3 T_3^2 + T_1^3 T_2^4 T_3^2 + T_1^4 T_2^5 T_3^3
}{
(1 - T_2) (1 - T_3) (1 - T_1 T_2^2) (1 - T_1^2 T_2^2 T_3) (1 - T_1^2 T_2 T_3^2)
}
\nn\\
\eea
and
\beal{esm6a_x3}
&&
PL[g_1(T_1,T_2,T_3;\mathcal{M}^{mes}_{6a})]=
T_1^2 T_2 T_3^2 
+ T_3 
+ T_1 T_2 T_3 
+ T_2 
+ T_1^2 T_2^2 T_3 
+ T_1 T_2^2 
\nn\\
&&
\hspace{1cm}
- 2 T_1^2 T_2^2 T_3^2
- T_1^2 T_2^3 T_3 
+ \dots~~.   
\eea
The Hilbert series and plethystic logarithm above illustrate the conical structure of the toric Calabi-Yau 3-fold.   
\\

\subsection{Model 6 Phase b}

\begin{figure}[H]
\begin{center}
\includegraphics[trim=0cm 0cm 0cm 0cm,width=4.5 cm]{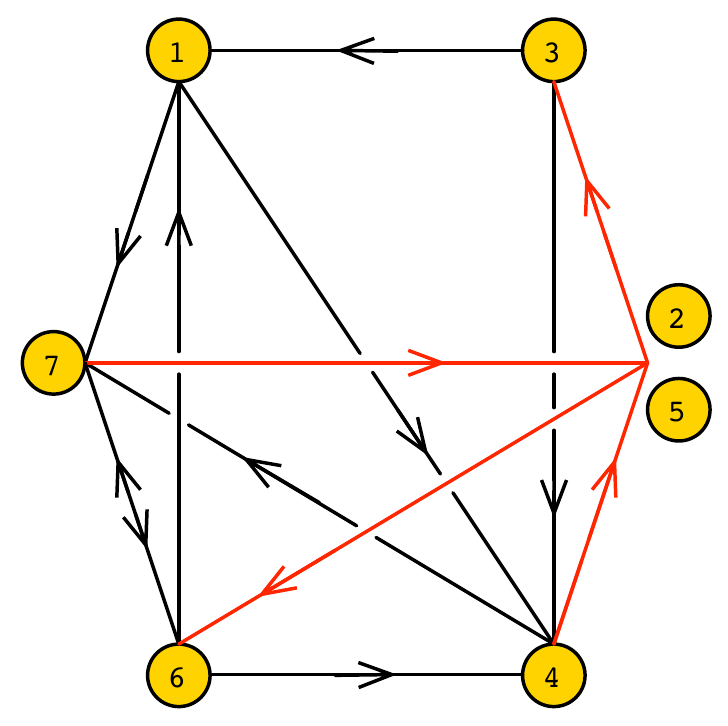}
\includegraphics[width=5 cm]{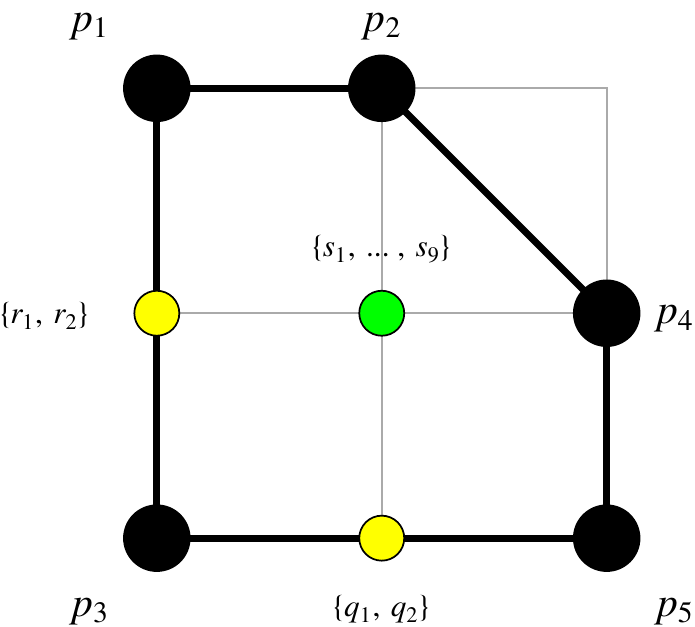}
\includegraphics[width=5 cm]{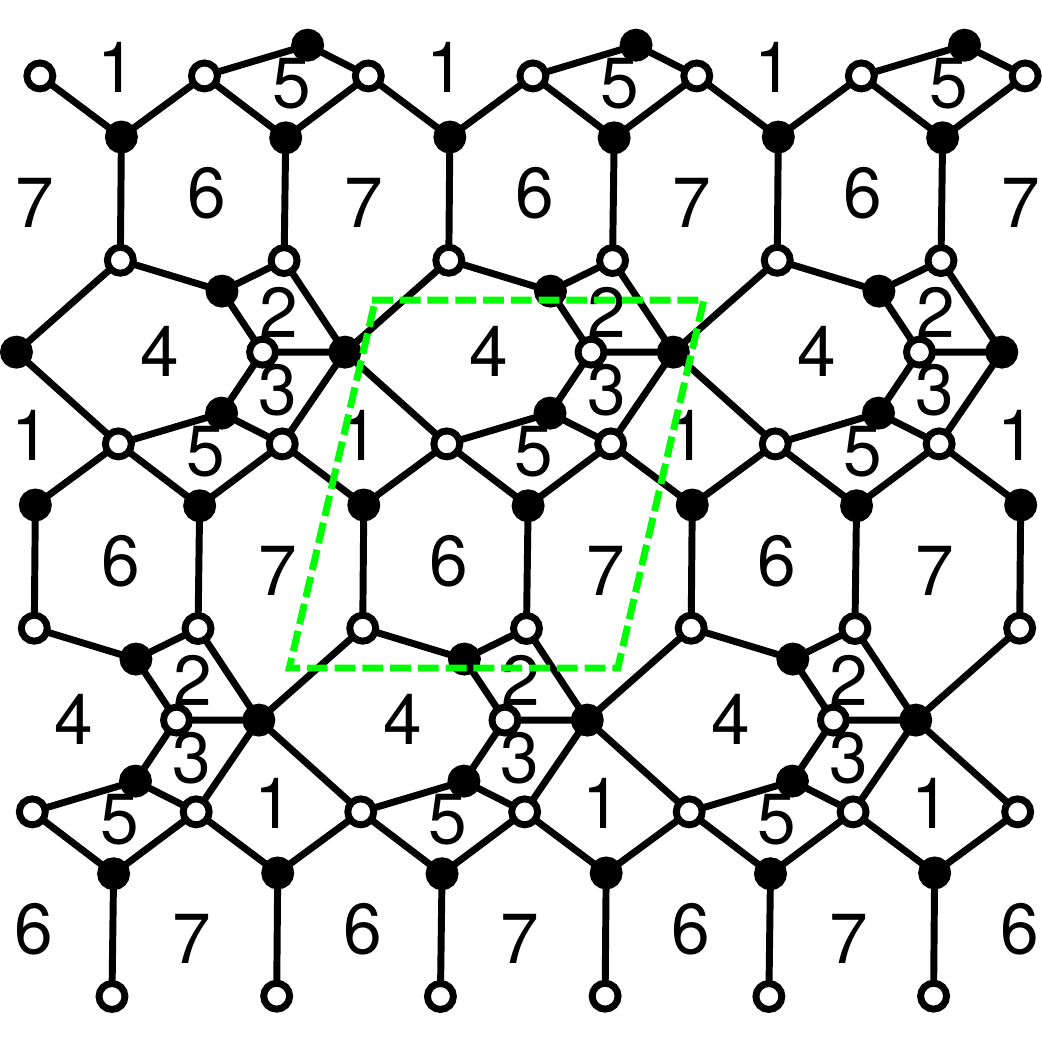}
\caption{The quiver, toric diagram, and brane tiling of Model 6b. The red arrows in the quiver indicate all possible connections between blocks of nodes.\label{f6b}}
 \end{center}
 \end{figure}
 
 \noindent The superpotential is 
\beal{esm6b_00}
W&=&
+ X_{42} X_{23} X_{34}  
+ X_{67} X_{72} X_{26}  
+ X_{76} X_{64} X_{47}  
+ X_{14} X_{45} X_{56} X_{61}  
+ X_{31} X_{17} X_{75} X_{53}  
\nn\\
&&
- X_{67} X_{75} X_{56}  
- X_{76} X_{61} X_{17}  
- X_{42} X_{26} X_{64}  
- X_{53} X_{34} X_{45}  
- X_{14} X_{47} X_{72} X_{23} X_{31} 
\nn\\
  \eea
 
 \noindent The perfect matching matrix is 
 
\noindent\makebox[\textwidth]{%
\footnotesize
$
P=
\left(
\begin{array}{c|ccccc|cc|cc|ccccccccc}
 \; & p_1& p_2& p_3& p_4& p_5& q_1& q_2& r_1& r_2& s_1& s_2& s_3& s_4& s_5& s_6& s_7& s_8& s_9 \\
 \hline
 X_{67} & 1 & 1 & 0 & 1 & 0 & 0 & 0 & 1 & 0 & 1 & 0 & 0 & 0 & 1 & 1 & 1 & 0 & 0
   \\
 X_{76} & 1 & 1 & 0 & 0 & 0 & 0 & 0 & 0 & 1 & 0 & 1 & 1 & 1 & 0 & 0 & 0 & 1 & 1
   \\
 X_{42} & 1 & 1 & 0 & 0 & 0 & 0 & 0 & 1 & 0 & 1 & 1 & 0 & 1 & 0 & 0 & 0 & 0 & 0
   \\
 X_{14} & 1 & 0 & 0 & 0 & 0 & 0 & 0 & 0 & 1 & 0 & 0 & 0 & 0 & 1 & 0 & 0 & 0 & 0
   \\
 X_{53} & 1 & 0 & 0 & 0 & 0 & 0 & 0 & 1 & 0 & 0 & 1 & 0 & 0 & 0 & 1 & 0 & 1 & 0
   \\
 X_{31} & 0 & 1 & 0 & 0 & 0 & 0 & 0 & 0 & 0 & 0 & 0 & 0 & 0 & 0 & 0 & 1 & 0 & 1
   \\
 X_{45} & 0 & 1 & 0 & 1 & 0 & 0 & 0 & 0 & 0 & 1 & 0 & 1 & 1 & 0 & 0 & 0 & 0 & 0
   \\
 X_{34} & 0 & 0 & 1 & 0 & 1 & 1 & 1 & 0 & 1 & 0 & 0 & 0 & 0 & 1 & 0 & 1 & 0 & 1
   \\
 X_{17} & 0 & 0 & 0 & 1 & 1 & 1 & 0 & 0 & 0 & 1 & 0 & 0 & 0 & 1 & 0 & 0 & 0 & 0
   \\
 X_{64} & 0 & 0 & 0 & 1 & 1 & 0 & 1 & 0 & 0 & 0 & 0 & 0 & 0 & 1 & 1 & 1 & 0 & 0
   \\
 X_{72} & 0 & 0 & 0 & 0 & 1 & 0 & 1 & 0 & 0 & 0 & 1 & 0 & 1 & 0 & 0 & 0 & 0 & 0
   \\
 X_{23} & 0 & 0 & 0 & 1 & 0 & 0 & 0 & 0 & 0 & 0 & 0 & 1 & 0 & 0 & 1 & 0 & 1 & 0
   \\
 X_{56} & 0 & 0 & 0 & 0 & 1 & 1 & 0 & 0 & 0 & 0 & 1 & 0 & 0 & 0 & 0 & 0 & 1 & 1
   \\
 X_{26} & 0 & 0 & 1 & 0 & 0 & 1 & 0 & 0 & 1 & 0 & 0 & 1 & 0 & 0 & 0 & 0 & 1 & 1
   \\
 X_{47} & 0 & 0 & 1 & 0 & 0 & 1 & 0 & 1 & 0 & 1 & 0 & 0 & 0 & 0 & 0 & 0 & 0 & 0
   \\
 X_{75} & 0 & 0 & 1 & 0 & 0 & 0 & 1 & 0 & 1 & 0 & 0 & 1 & 1 & 0 & 0 & 0 & 0 & 0
   \\
 X_{61} & 0 & 0 & 1 & 0 & 0 & 0 & 1 & 1 & 0 & 0 & 0 & 0 & 0 & 0 & 1 & 1 & 0 & 0
\end{array}
\right)
$
}
\vspace{0.5cm}

 \noindent The F-term charge matrix $Q_F=\ker{(P)}$ is

\noindent\makebox[\textwidth]{%
\footnotesize
$
Q_F=
\left(
\begin{array}{ccccc|cc|cc|ccccccccc}
p_1& p_2& p_3& p_4& p_5& q_1& q_2& r_1& r_2& s_1& s_2& s_3& s_4& s_5& s_6& s_7& s_8& s_9 \\
 \hline
 1 & 0 & 1 & 0 & 0 & 0 & 0 & -1 & -1 & 0 & 0 & 0 & 0 & 0 & 0 & 0 & 0 & 0 \\
 1 & 0 & 0 & 0 & 1 & 0 & 0 & 0 & 0 & 0 & -1 & 0 & 0 & -1 & 0 & 0 & 0 & 0 \\
 0 & 1 & 0 & 0 & 0 & 1 & 0 & 0 & 0 & -1 & 0 & 0 & 0 & 0 & 0 & 0 & 0 & -1 \\
 0 & 1 & 0 & 0 & 0 & 0 & 1 & 0 & 0 & 0 & 0 & 0 & -1 & 0 & 0 & -1 & 0 & 0 \\
  0 & 0 & 1 & 1 & 0 & -1 & 0 & 0 & 0 & 0 & 1 & 0 & -1 & 0 & -1 & 0 & 0 & 0 \\
 0 & 0 & 1 & 0 & 1 & -1 & -1 & 0 & 0 & 0 & 0 & 0 & 0 & 0 & 0 & 0 & 0 & 0 \\
 0 & 0 & 0 & 1 & 0 & 0 & 0 & 1 & 0 & -1 & 0 & 0 & 0 & 0 & -1 & 0 & 0 & 0 \\
 0 & 0 & 0 & 1 & -1 & 1 & 0 & 0 & 0 & -1 & 1 & 0 & 0 & 0 & 0 & 0 & -1 & 0 \\
 0 & 0 & 0 & 0 & 0 & 0 & 0 & 0 & 0 & 0 & 1 & 1 & -1 & 0 & 0 & 0 & -1 & 0
\end{array}
\right)
$
}
\vspace{0.5cm}

\noindent The D-term charge matrix is

\noindent\makebox[\textwidth]{%
\footnotesize
$
Q_D=
\left(
\begin{array}{ccccc|cc|cc|ccccccccc}
p_1 & p_2& p_3& p_4& p_5& q_1& q_2& r_1& r_2& s_1& s_2& s_3& s_4& s_5& s_6& s_7& s_8& s_9  \\ 
\hline
 0 & 0 & 0 & 0 & 0 & 0 & 0 & 0 & 0 & 1 & -1 & 0 & 0 & 0 & 0 & 0 & 0 & 0 \\
 0 & 0 & 0 & 0 & 0 & 0 & 0 & 0 & 0 & 0 & 1 & -1 & 0 & 0 & 0 & 0 & 0 & 0 \\
 0 & 0 & 0 & 0 & 0 & 0 & 0 & 0 & 0 & 0 & 0 & 1 & -1 & 0 & 0 & 0 & 0 & 0 \\
 0 & 0 & 0 & 0 & 0 & 0 & 0 & 0 & 0 & 0 & 0 & 0 & 1 & -1 & 0 & 0 & 0 & 0 \\
 0 & 0 & 0 & 0 & 0 & 0 & 0 & 0 & 0 & 0 & 0 & 0 & 0 & 1 & -1 & 0 & 0 & 0 \\
 0 & 0 & 0 & 0 & 0 & 0 & 0 & 0 & 0 & 0 & 0 & 0 & 0 & 0 & 1 & -1 & 0 & 0
\end{array}
\right)
$
}
\vspace{0.5cm}

The global symmetry of Model 6b has the form $U(1)_{f_1}\times U(1)_{f_2} \times U(1)_R$. The charges under the global symmetry on the extremal perfect matchings $p_\alpha$ are the same as for Model 6a. They are shown in \tref{t6a}.

Product of non-extremal perfect matchings are expressed in terms of single variables as follows
\beal{esx6b_1}
q = q_1 q_2 ~,~
r = r_1 r_2 ~,~
s = \prod_{m=1}^{9} s_m ~.
\eea
The fugacity counting extremal perfect matchings $p_\alpha$ is $t_\alpha$. The fugacity $y_q$ counts the product of non-extremal perfect matchings $q$.

The refined mesonic Hilbert series of Model 6b is identical to the mesonic Hilbert series for Model 6a. The mesonic Hilbert series and the corresponding plethystic logarithm is shown in \eref{esm6a_1} and \eref{esm6a_3} respectively. The mesonic Hilbert series for Model 6a and 6b are identical and are not complete intersections.

The generators in terms of perfect matchings of Model 6b are shown in \tref{t6agen}. The charge lattice of generators forms a reflexive polygon which is the dual of the toric diagram. The generators in terms of quiver fields of Model 6b are shown in \tref{t6bgen2}.

\comment{
\begin{table}[h!]
\centering
\resizebox{\hsize}{!}{
\begin{minipage}[!b]{0.6\textwidth}
\begin{tabular}{|l|c|c|}
\hline
Generator & $U(1)_{f_1}$ & $U(1)_{f_2}$ 
\\
\hline
\hline
$p_2 p_4^2 p_5^2 ~ q_1 q_2 ~ \prod_{m=1}^{9} s_m$ 
& 1 & 0
\nn\\
$p_1^2 p_2^2 p_4 ~ r_1 r_2 ~ \prod_{m=1}^{9} s_m$
& 0 & 1
\nn\\
$p_1 p_2 p_3 p_4 p_5 ~ q_1 q_2 ~ r_1 r_2 ~\prod_{m=1}^{9} s_m$
& 0 & 0
\nn\\
$p_3^2 p_4 p_5^2 ~ q_1^2 q_2^2 ~ r_1 r_2 ~ \prod_{m=1}^{9} s_m$
& 0 & -1
\nn\\
$p_1^2 p_2 p_3^2 ~ q_1 q_2 ~ r_1^2 r_2^2 ~\prod_{m=1}^{9} s_m$
& -1 & 0
\nn\\ 
$p_1 p_3^3 p_5 ~ q_1^2 q_2^2 ~ r_1^2 r_2^2 ~\prod_{m=1}^{9} s_m$
& -1 & -1
\nn\\
   \hline
\end{tabular}
\end{minipage}
\hspace{1cm}
\begin{minipage}[!b]{0.3\textwidth}
\includegraphics[width=4 cm]{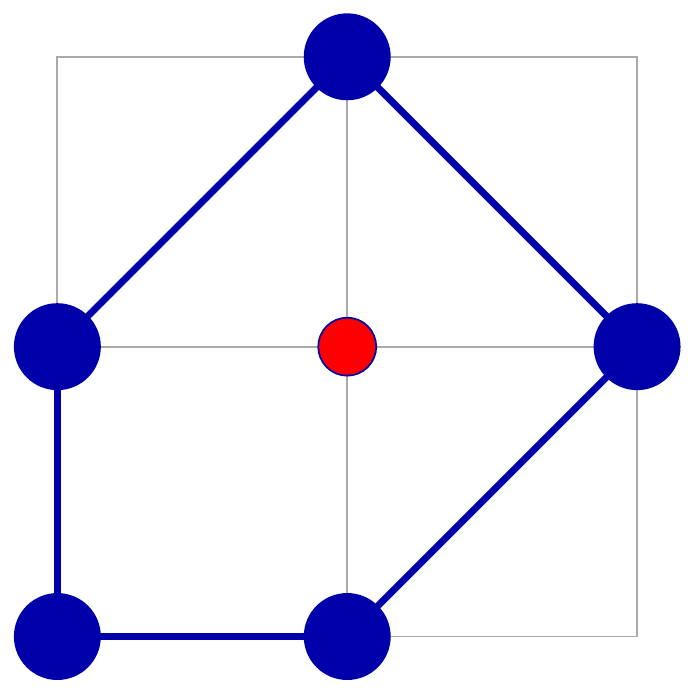}
\end{minipage}
}
\caption{The generators and lattice of generators of the mesonic moduli space of Model 6b in terms of GLSM fields with the corresponding flavor charges.\label{t6bgen}\label{f6bgen}} 
\end{table}
}

\begin{table}[h!]
\centering
\resizebox{\hsize}{!}{
\begin{tabular}{|l|c|c|}
\hline
Generator & $U(1)_{f_1}$ & $U(1)_{f_2}$ 
\\
\hline
\hline
$
X_{45} X_{56} X_{64}=  X_{17} X_{72} X_{23} X_{31}
$
& 0 & 1
\nn\\
$ 
X_{67} X_{76}=  X_{14} X_{42} X_{23} X_{31}=  X_{14} X_{45} X_{53} X_{31}
$
& 1 & 0
\nn\\
$
X_{14} X_{47} X_{72} X_{23} X_{31}=  X_{14} X_{45} X_{56} X_{61}=  X_{17} X_{75} X_{53} X_{31}=  X_{17} X_{76} X_{61}=  X_{23} X_{34} X_{42}
$
& 0 & 0
\nn\\
$
=  X_{26} X_{64} X_{42}=  X_{26} X_{67} X_{72}=  X_{34} X_{45} X_{53}=  X_{47} X_{76} X_{64}=  X_{56} X_{67} X_{75}$
& & 
\nn\\
$
 X_{17} X_{72} X_{26} X_{61}=  X_{17} X_{75} X_{56} X_{61}=  X_{23} X_{34} X_{47} X_{72}=  X_{26} X_{64} X_{47} X_{72}=  X_{47} X_{75} X_{56} X_{64}
$
& -1 & 0
\nn\\
$X_{14} X_{47} X_{75} X_{53} X_{31}=  X_{14} X_{42} X_{26} X_{61}=  X_{14} X_{47} X_{76} X_{61}$
& 0 & -1
\nn\\
$ X_{34} X_{47} X_{75} X_{53}=  X_{14} X_{47} X_{72} X_{26} X_{61}=  X_{14} X_{47} X_{75} X_{56} X_{61}$
& -1 & -1
\nn\\
   \hline
\end{tabular}
}
\caption{The generators in terms of bifundamental fields (Model 6b).\label{t6bgen2}\label{f6bgen2}} 
\end{table}

\subsection{Model 6 Phase c}

\begin{figure}[H]
\begin{center}
\includegraphics[trim=0cm 0cm 0cm 0cm,width=4.5 cm]{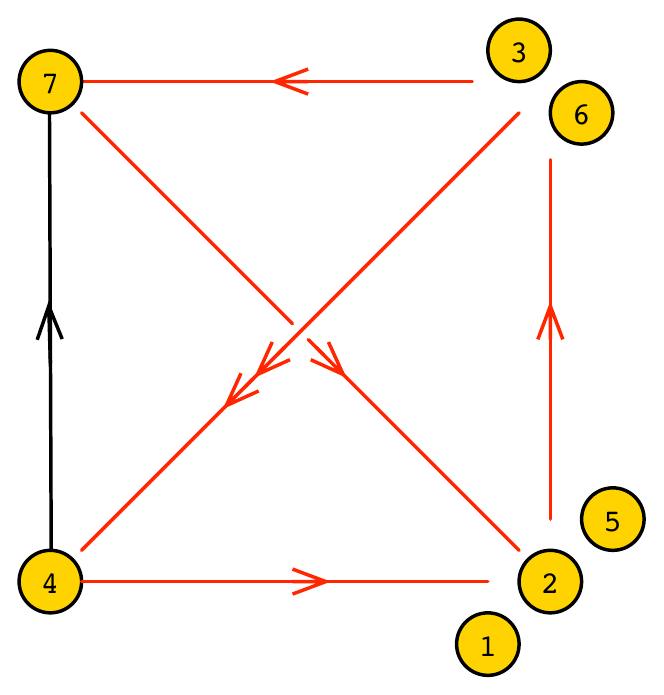}
\includegraphics[width=5 cm]{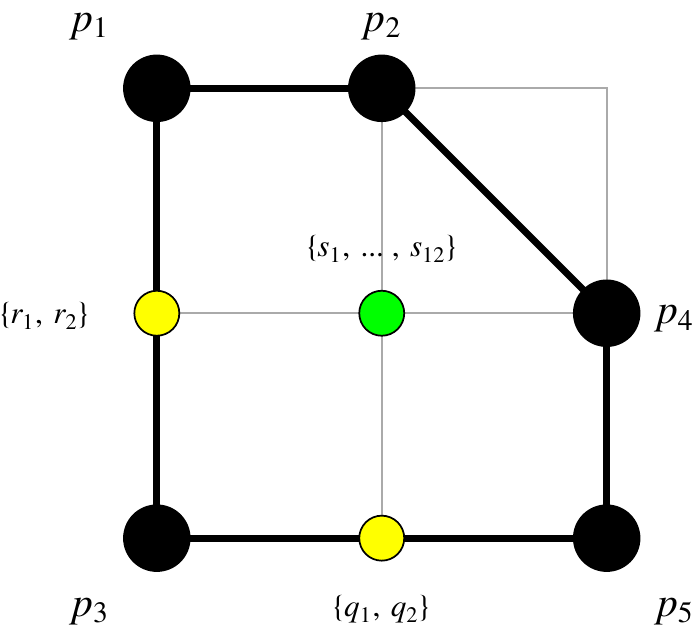}
\includegraphics[width=5 cm]{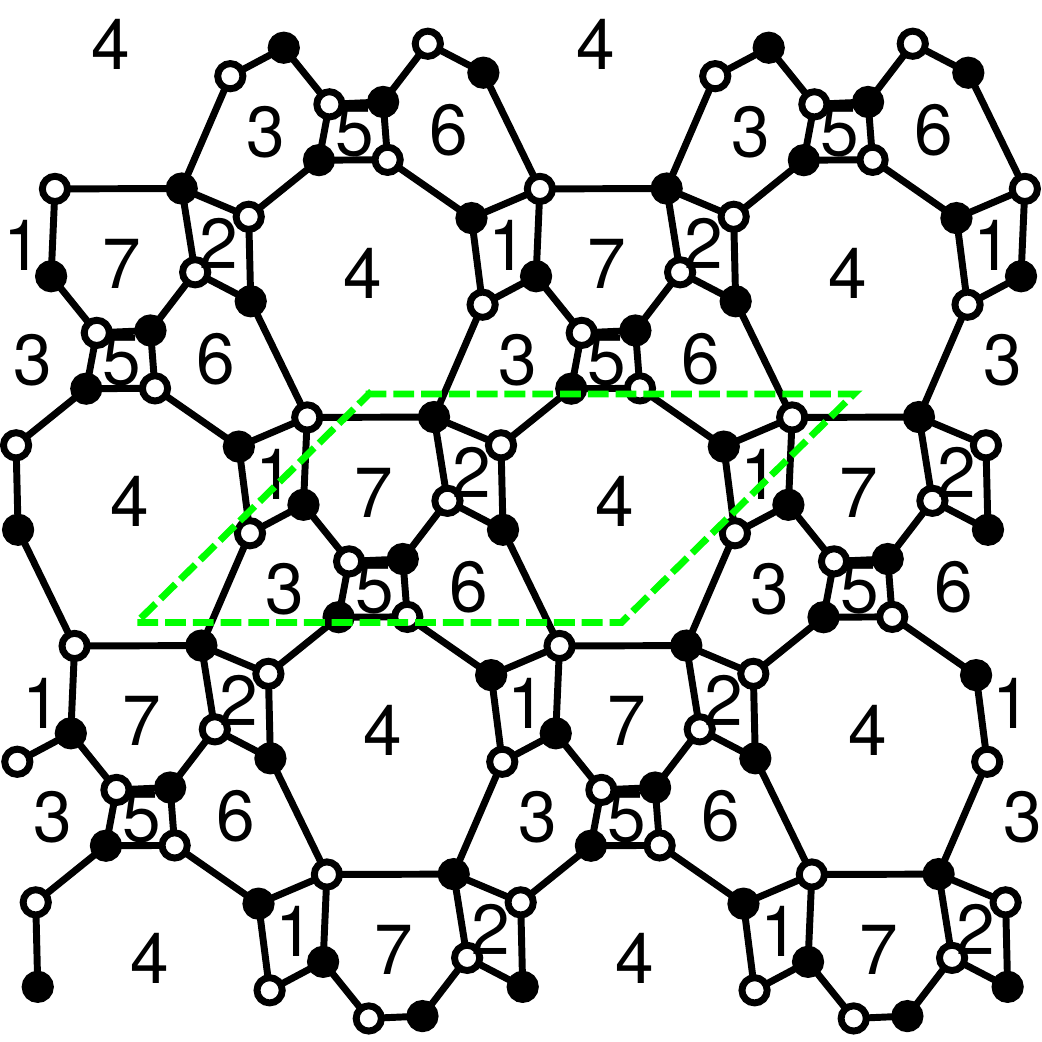}
\caption{The quiver, toric diagram, and brane tiling of Model 6c. The red arrows in the quiver indicate all possible connections between blocks of nodes.\label{f6c}}
 \end{center}
 \end{figure}
 
\noindent The superpotential is 
\beal{esm6c_00}
W&=&
+ X_{41} X_{13} X_{34}^{2} 
+ X_{42} X_{23} X_{34}^{1} 
+ X_{45} X_{56} X_{64}^{2} 
+ X_{67} X_{72} X_{26} 
+ X_{75} X_{53} X_{37} 
+ X_{47} X_{71} X_{16} X_{64}^{1}
\nn\\
&&
- X_{41} X_{16} X_{64}^{2} 
- X_{42} X_{26} X_{64}^{1} 
- X_{45} X_{53} X_{34}^{1}  
- X_{67} X_{75} X_{56} 
- X_{71} X_{13} X_{37} 
- X_{47} X_{72} X_{23} X_{34}^{2} 
\nn\\
  \eea
 
 \noindent The perfect matching matrix is 
 
\noindent\makebox[\textwidth]{%
\footnotesize
$
P=
\left(
\begin{array}{c|ccccc|cc|cc|cccccccccccc}
 \; & p_1& p_2& p_3& p_4& p_5& q_1& q_2& r_1& r_2& s_1& s_2& s_3& s_4& s_5& s_6& s_7& s_8& s_9& s_{10}& s_{11}& s_{12} \\
 \hline
 X_{37} & 1 & 1 & 0 & 1 & 0 & 0 & 0 & 0 & 1 & 1 & 1 & 0 & 1 & 0 & 0 & 0 & 0 & 0 & 0 & 0 & 0 \\
 X_{64}^{1} & 1 & 1 & 0 & 0 & 0 & 0 & 0 & 1 & 0 & 1 & 0 & 1 & 0 & 0 & 0 & 0 & 0 & 0 & 0 & 0 & 0 \\
 X_{41} & 1 & 1 & 0 & 0 & 0 & 0 & 0 & 0 & 1 & 0 & 1 & 0 & 0 & 1 & 0 & 1 & 0 & 1 & 0 & 1 & 0 \\
 X_{34}^{1} & 1 & 0 & 1 & 0 & 0 & 1 & 0 & 1 & 1 & 1 & 0 & 0 & 1 & 0 & 0 & 0 & 0 & 0 & 0 & 0 & 0 \\
 X_{72} & 1 & 0 & 0 & 0 & 0 & 0 & 0 & 1 & 0 & 0 & 0 & 0 & 0 & 1 & 1 & 0 & 0 & 1 & 1 & 0 & 0 \\
 X_{56} & 1 & 0 & 0 & 0 & 0 & 0 & 0 & 0 & 1 & 0 & 0 & 0 & 1 & 0 & 0 & 0 & 0 & 1 & 1 & 1 & 1 \\
 X_{67} & 0 & 1 & 0 & 1 & 1 & 0 & 1 & 0 & 0 & 1 & 1 & 1 & 0 & 0 & 0 & 0 & 0 & 0 & 0 & 0 & 0 \\
 X_{45} & 0 & 1 & 0 & 1 & 0 & 0 & 0 & 0 & 0 & 0 & 1 & 0 & 0 & 1 & 1 & 1 & 1 & 0 & 0 & 0 & 0 \\
 X_{23} & 0 & 1 & 0 & 0 & 0 & 0 & 0 & 0 & 0 & 0 & 0 & 1 & 0 & 0 & 0 & 1 & 1 & 0 & 0 & 1 & 1 \\
 X_{64}^{2} & 0 & 0 & 1 & 0 & 1 & 1 & 1 & 1 & 0 & 1 & 0 & 1 & 0 & 0 & 0 & 0 & 0 & 0 & 0 & 0 & 0 \\
 X_{75} & 0 & 0 & 1 & 0 & 0 & 1 & 0 & 1 & 0 & 0 & 0 & 0 & 0 & 1 & 1 & 1 & 1 & 0 & 0 & 0 & 0 \\
 X_{26} & 0 & 0 & 1 & 0 & 0 & 1 & 0 & 0 & 1 & 0 & 0 & 0 & 1 & 0 & 0 & 1 & 1 & 0 & 0 & 1 & 1 \\
 X_{13} & 0 & 0 & 1 & 0 & 0 & 0 & 1 & 1 & 0 & 0 & 0 & 1 & 0 & 0 & 1 & 0 & 1 & 0 & 1 & 0 & 1 \\
 X_{47} & 0 & 0 & 1 & 0 & 0 & 0 & 1 & 0 & 1 & 0 & 1 & 0 & 0 & 0 & 0 & 0 & 0 & 0 & 0 & 0 & 0 \\
 X_{34}^{2} & 0 & 0 & 0 & 1 & 1 & 1 & 0 & 0 & 0 & 1 & 0 & 0 & 1 & 0 & 0 & 0 & 0 & 0 & 0 & 0 & 0 \\
 X_{42} & 0 & 0 & 0 & 1 & 1 & 0 & 1 & 0 & 0 & 0 & 1 & 0 & 0 & 1 & 1 & 0 & 0 & 1 & 1 & 0 & 0 \\
 X_{16} & 0 & 0 & 0 & 1 & 0 & 0 & 0 & 0 & 0 & 0 & 0 & 0 & 1 & 0 & 1 & 0 & 1 & 0 & 1 & 0 & 1 \\
 X_{71} & 0 & 0 & 0 & 0 & 1 & 1 & 0 & 0 & 0 & 0 & 0 & 0 & 0 & 1 & 0 & 1 & 0 & 1 & 0 & 1 & 0 \\
 X_{53} & 0 & 0 & 0 & 0 & 1 & 0 & 1 & 0 & 0 & 0 & 0 & 1 & 0 & 0 & 0 & 0 & 0 & 1 & 1 & 1 & 1
\end{array}
\right)
$
}
\vspace{0.5cm}

 \noindent The F-term charge matrix $Q_F=\ker{(P)}$ is

\noindent\makebox[\textwidth]{%
\footnotesize
$
Q_F=
\left(
\begin{array}{ccccc|cc|cc|cccccccccccc}
p_1& p_2& p_3& p_4& p_5& q_1& q_2& r_1& r_2& s_1& s_2& s_3& s_4& s_5& s_6& s_7& s_8& s_9& s_{10}& s_{11}& s_{12} \\
\hline
 1 & 0 & 0 & 1 & 0 & 0 & 1 & 0 & 0 & -1 & -1 & 0 & 0 & 0 & 0 & 0 & 0 & 0 & -1 & 0 & 0 \\
 1 & 0 & 0 & 0 & 1 & 0 & 0 & 0 & 0 & -1 & 0 & 0 & 0 & 0 & 0 & 0 & 0 & -1 & 0 & 0 & 0 \\
 0 & 1 & 0 & 0 & 0 & 1 & 0 & 0 & 0 & -1 & 0 & 0 & 0 & 0 & 0 & -1 & 0 & 0 & 0 & 0 & 0 \\
 0 & 1 & 0 & 0 & 0 & 0 & 1 & 0 & 0 & 0 & -1 & -1 & 0 & 0 & 0 & 0 & 0 & 0 & 0 & 0 & 0 \\
 0 & 0 & 1 & 0 & 0 & -1 & 0 & -1 & 0 & 1 & -1 & 0 & 0 & 1 & 0 & 0 & 0 & 0 & 0 & 0 & 0 \\
 0 & 0 & 0 & 1 & 0 & 0 & 0 & 1 & 0 & -1 & 0 & 0 & 0 & 0 & 0 & 0 & -1 & 0 & -1 & 0 & 1 \\
 0 & 0 & 0 & 1 & 0 & 0 & 0 & 1 & 0 & -1 & 0 & 0 & 0 & 0 & 0 & -1 & 0 & 0 & -1 & 1 & 0 \\
 0 & 0 & 0 & 1 & 0 & 0 & 0 & 0 & 1 & 0 & -1 & 0 & -1 & 0 & 0 & 0 & 0 & 0 & 0 & 0 & 0 \\
 0 & 0 & 0 & 0 & 1 & -1 & 0 & 0 & 1 & 0 & -1 & 0 & 0 & 1 & 0 & 0 & 0 & -1 & 0 & 0 & 0 \\
 0 & 0 & 0 & 0 & 1 & 0 & -1 & 1 & 0 & -1 & 1 & 0 & 0 & -1 & 0 & 0 & 0 & 0 & 0 & 0 & 0 \\
 0 & 0 & 0 & 0 & 0 & 0 & 0 & 1 & -1 & -1 & 1 & 0 & 1 & 0 & -1 & 0 & 0 & 0 & 0 & 0 & 0 \\
 0 & 0 & 0 & 0 & 0 & 0 & 0 & 0 & 0 & 0 & 0 & 0 & 0 & -1 & 0 & 1 & 0 & 0 & 1 & 0 & -1
\end{array}
\right)
$
}
\vspace{0.5cm}

\noindent The D-term charge matrix is

\noindent\makebox[\textwidth]{%
\footnotesize
$
Q_D=
\left(
\begin{array}{ccccc|cc|cc|cccccccccccc}
p_1& p_2& p_3& p_4& p_5& q_1& q_2& r_1& r_2& s_1& s_2& s_3& s_4& s_5& s_6& s_7& s_8& s_9& s_{10}& s_{11}& s_{12} \\
\hline
 0 & 0 & 0 & 0 & 0 & 0 & 0 & 0 & 0 & 1 & -1 & 0 & 0 & 0 & 0 & 0 & 0 & 0 & 0 & 0 & 0 \\
 0 & 0 & 0 & 0 & 0 & 0 & 0 & 0 & 0 & 0 & 1 & -1 & 0 & 0 & 0 & 0 & 0 & 0 & 0 & 0 & 0 \\
 0 & 0 & 0 & 0 & 0 & 0 & 0 & 0 & 0 & 0 & 0 & 1 & -1 & 0 & 0 & 0 & 0 & 0 & 0 & 0 & 0 \\
 0 & 0 & 0 & 0 & 0 & 0 & 0 & 0 & 0 & 0 & 0 & 0 & 1 & -1 & 0 & 0 & 0 & 0 & 0 & 0 & 0 \\
 0 & 0 & 0 & 0 & 0 & 0 & 0 & 0 & 0 & 0 & 0 & 0 & 0 & 1 & -1 & 0 & 0 & 0 & 0 & 0 & 0 \\
 0 & 0 & 0 & 0 & 0 & 0 & 0 & 0 & 0 & 0 & 0 & 0 & 0 & 0 & 1 & -1 & 0 & 0 & 0 & 0 & 0
\end{array}
\right)
$
}
\vspace{0.5cm}

The global symmetry of Model 6c is $U(1)_{f_1}\times U(1)_{f_2} \times U(1)_R$. The global symmetry is the same as for Model 6a and 6b. The charges on the extremal perfect matchings are shown in \tref{t6a}.

Products of non-extremal perfect matchings are chosen to be associated to a single variable as shown below
\beal{esx6c}
q = q_1 q_2 ~,~
r = r_1 r_2 ~,~
s = \prod_{m=1}^{12} s_m ~.
\eea
Extremal perfect matchings are counted by the fugacity $t_\alpha$. Products of non-extremal perfect matchings such as $q$ are counted by fugacities of the form $y_q$.

The refined mesonic Hilbert series of Model 6c computed using the Molien integral formula is identical to the mesonic Hilbert series of Model 6a and 6b in \eref{esm6a_1}. Accordingly, the plethystic logarithm are identical as well and hence the mesonic moduli space is a non-complete intersection.

The moduli space generators in terms of perfect matchings of Model 6c are shown in \tref{t6agen}. The lattice of generators is a reflexive polygon and is the dual of the toric diagram. The generators in terms of quiver fields of Model 6c are shown in \tref{t6cgen2}.

\comment{
\begin{table}[h!]
\centering
\resizebox{\hsize}{!}{
\begin{minipage}[!b]{0.6\textwidth}
\begin{tabular}{|l|c|c|}
\hline
Generator & $U(1)_{f_1}$ & $U(1)_{f_2}$ 
\\
\hline
\hline
$p_2 p_4^2 p_5^2 ~ q_1 q_2 ~ \prod_{m=1}^{12} s_m$ 
& 1 & 0
\nn\\
$p_1^2 p_2^2 p_4 ~ r_1 r_2 ~ \prod_{m=1}^{12} s_m$
& 0 & 1
\nn\\
$p_1 p_2 p_3 p_4 p_5 ~ q_1 q_2 ~ r_1 r_2 ~\prod_{m=1}^{12} s_m$
& 0 & 0
\nn\\
$p_3^2 p_4 p_5^2 ~ q_1^2 q_2^2 ~ r_1 r_2 ~ \prod_{m=1}^{12} s_m$
& 0 & -1
\nn\\
$p_1^2 p_2 p_3^2 ~ q_1 q_2 ~ r_1^2 r_2^2 ~\prod_{m=1}^{12} s_m$
& -1 & 0
\nn\\ 
$p_1 p_3^3 p_5 ~ q_1^2 q_2^2 ~ r_1^2 r_2^2 ~\prod_{m=1}^{12} s_m$
& -1 & -1
\nn\\
   \hline
\end{tabular}
\end{minipage}
\centering
\begin{minipage}[!b]{0.2\textwidth}
\includegraphics[width=4 cm]{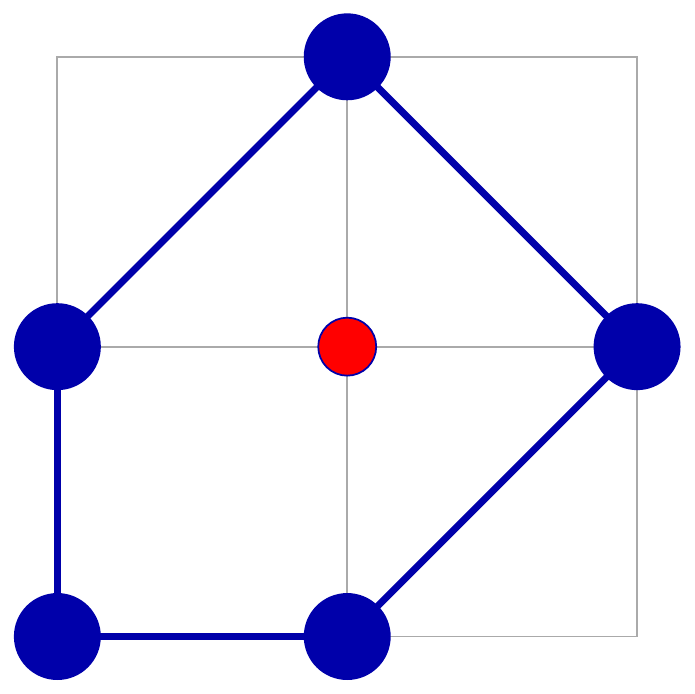}
\end{minipage}
}
\caption{The generators and lattice of generators of the mesonic moduli space of Model 6c in terms of GLSM fields with the corresponding flavor charges.\label{t6cgen}\label{f6cgen}} 
\end{table}
}

\begin{table}[h!]
\centering
\resizebox{\hsize}{!}{
\begin{tabular}{|l|c|c|}
\hline
Generator & $U(1)_{f_1}$ & $U(1)_{f_2}$ 
\\
\hline
\hline
$X_{16} X_{67} X_{71}=  X_{23} X_{34}^{2} X_{42}=  X_{34}^{2} X_{45} X_{53}$
& 1 & 0 
\nn\\
$X_{41} X_{16} X_ {64}^{1}=  X_{23} X_{37} X_{72}=  X_{45} X_{56} X_{64}^{1}$
& 0 & 1
\nn\\
$ X_{47} X_{71} X_{16} X_{64}^{1}=  X_{23} X_{34}^{2} X_{47} X_{72}=  X_{13} X_{34}^{2} X_{41}=  X_{13} X_{37} X_{71}=  X_{41} X_{16} X_{64}^{2}=  X_{23} X_{34}^{1} X_{42}$
& 0 & 0
\nn\\
$
=  X_{42} X_{26} X_{64}^{1}=  X_{26} X_{67} X_{72}=  X_{34}^{1} X_{45} X_{53}=  X_{53} X_{37} X_{75}=  X_{45} X_{56} X_{64}^{2}=  X_{56} X_{67} X_{75}$
&  & 
\nn\\
$X_{42} X_{26} X_ {64}^{2}=  X_{13} X_ {34}^{2} X_{47} X_{71}=  X_{47} X_{71} X_{16} X_ {64}^{2}=  X_ {34}^{2} X_{47} X_{75} X_{53}$
&0 & -1
\nn\\
$ X_{13} X_{34}^{1} X_{41}=  X_{23} X_{34}^{1} X_{47} X_{72}=  X_{47} X_{72} X_{26} X_{64}^{1}=  X_{56} X_{47} X_{75} X_{64}^{1}$
& -1 & 0
\nn\\ 
$X_{13} X_ {34}^{1} X_{47} X_{71}=  X_{47} X_{72} X_{26} X_ {64}^{2}=  X_ {34}^{1} X_{47} X_{75} X_{53}=  X_{56} X_{47} X_{75} X_ {64}^{2}$
&-1 & -1
\nn\\
   \hline
\end{tabular}
}
\caption{The generators in terms of bifundamental fields (Model 6c).\label{t6cgen2}\label{f6cgen2}} 
\end{table}

\section{Model 7: $\mathbb{C}^3/\mathbb{Z}_6~(1,2,3)$, $\text{PdP}_{3a}$}

\begin{figure}[H]
\begin{center}
\includegraphics[trim=0cm 0cm 0cm 0cm,width=5 cm]{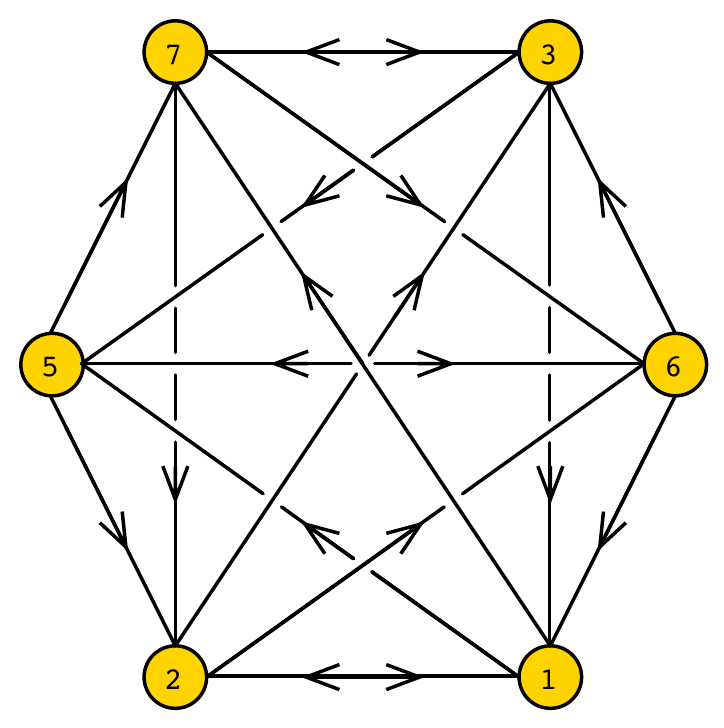}
\includegraphics[width=5 cm]{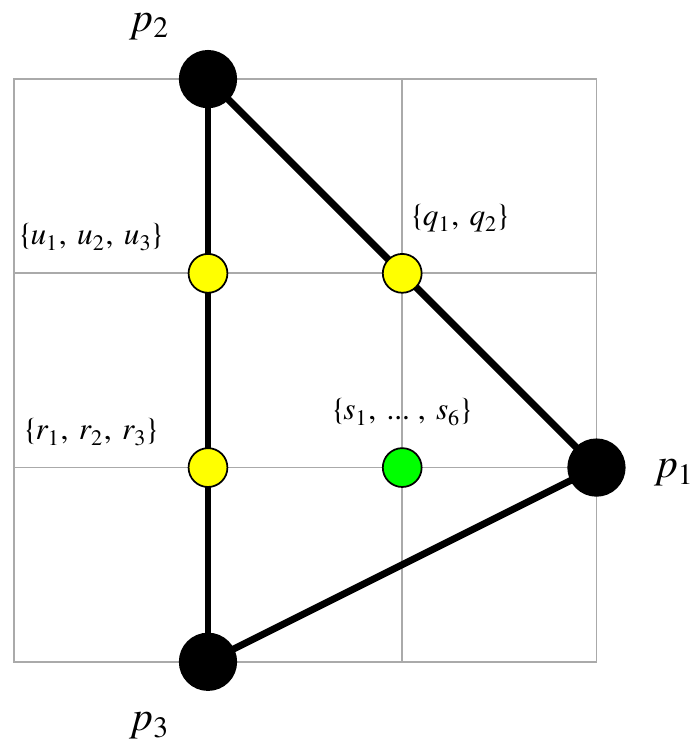}
\includegraphics[width=5 cm]{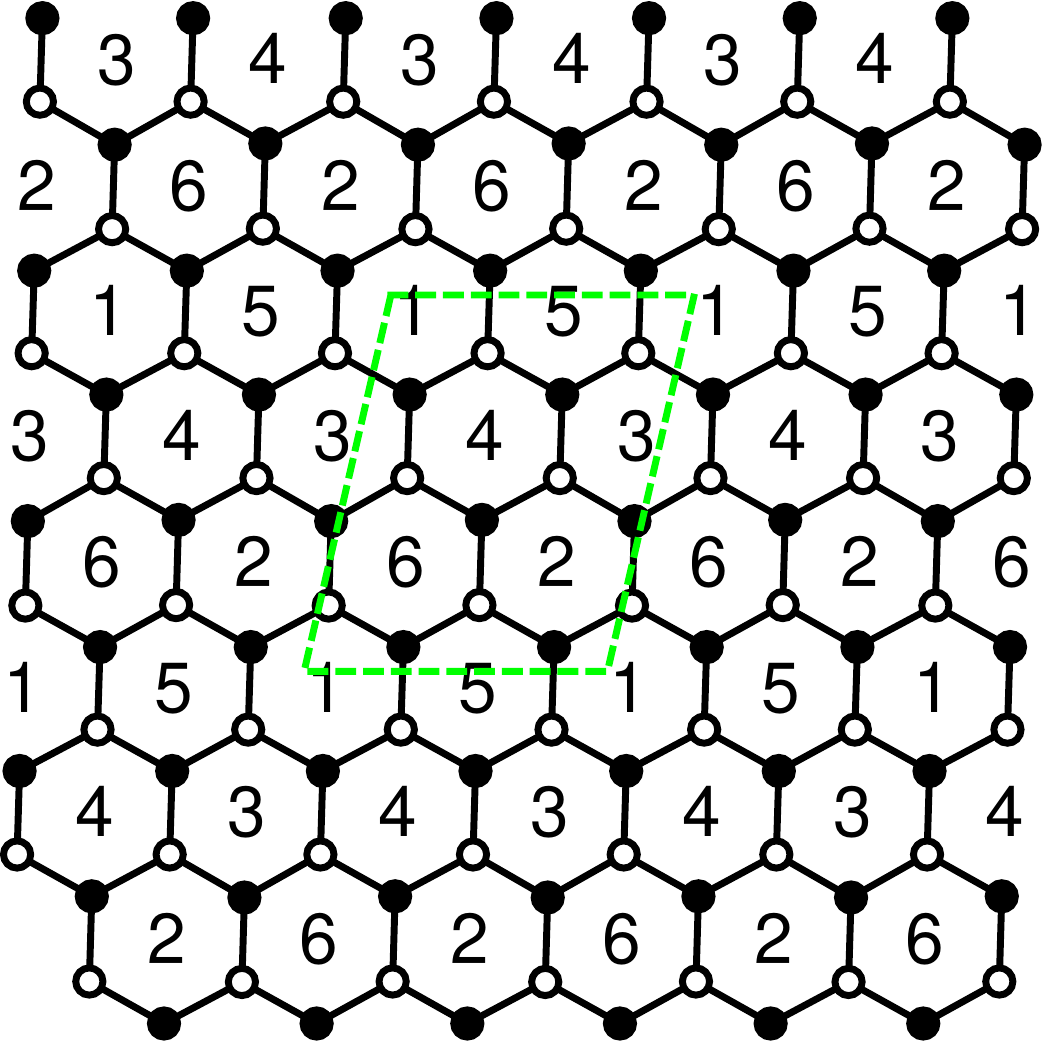}
\caption{The quiver, toric diagram, and brane tiling of Model 7.\label{f7}}
 \end{center}
 \end{figure}
 
 \noindent The superpotential is 
\beal{esm7_00}
W&=&
+ X_{12} X_{26} X_{61}  
+ X_{63} X_{34} X_{46} 
+ X_{24} X_{43} X_{32}  
+ X_{35} X_{51} X_{13}  
+ X_{41} X_{15} X_{54}  
+ X_{56} X_{62} X_{25}  
\nn\\
&&
- X_{12} X_{25} X_{51} 
- X_{63} X_{32} X_{26}  
- X_{24} X_{46} X_{62} 
- X_{35} X_{54} X_{43}
- X_{41} X_{13} X_{34}   
- X_{56} X_{61} X_{15}  
\nn\\
  \eea
 
 \noindent The perfect matching matrix is 
 
\noindent\makebox[\textwidth]{%
\footnotesize
$
P=
\left(
\begin{array}{c|ccc|cc|ccc|ccc|cccccc}
 \; & p_1 & p_2& p_3& q_1& q_2& r_1& r_2& r_3& u_1& u_2& u_3& s_1& s_2& s_3& s_4& s_5& s_6 \\
 \hline
 X_{26} & 1 & 0 & 0 & 1 & 0 & 0 & 0 & 0 & 0 & 0 & 0 & 1 & 0 & 0 & 1 & 1 & 0 \\
 X_{62} & 1 & 0 & 0 & 0 & 1 & 0 & 0 & 0 & 0 & 0 & 0 & 0 & 1 & 1 & 0 & 0 & 1 \\
 X_{15} & 1 & 0 & 0 & 1 & 0 & 0 & 0 & 0 & 0 & 0 & 0 & 0 & 1 & 1 & 0 & 1 & 0 \\
 X_{51} & 1 & 0 & 0 & 0 & 1 & 0 & 0 & 0 & 0 & 0 & 0 & 1 & 0 & 0 & 1 & 0 & 1 \\
 X_{43} & 1 & 0 & 0 & 1 & 0 & 0 & 0 & 0 & 0 & 0 & 0 & 1 & 0 & 1 & 0 & 0 & 1 \\
 X_{34} & 1 & 0 & 0 & 0 & 1 & 0 & 0 & 0 & 0 & 0 & 0 & 0 & 1 & 0 & 1 & 1 & 0 \\
 X_{46} & 0 & 1 & 0 & 1 & 0 & 1 & 0 & 0 & 1 & 1 & 0 & 1 & 0 & 0 & 0 & 0 & 0 \\
 X_{32} & 0 & 1 & 0 & 0 & 1 & 1 & 0 & 0 & 1 & 1 & 0 & 0 & 1 & 0 & 0 & 0 & 0 \\
 X_{13} & 0 & 1 & 0 & 1 & 0 & 0 & 1 & 0 & 1 & 0 & 1 & 0 & 0 & 1 & 0 & 0 & 0 \\
 X_{54} & 0 & 1 & 0 & 0 & 1 & 0 & 1 & 0 & 1 & 0 & 1 & 0 & 0 & 0 & 1 & 0 & 0 \\
 X_{25} & 0 & 1 & 0 & 1 & 0 & 0 & 0 & 1 & 0 & 1 & 1 & 0 & 0 & 0 & 0 & 1 & 0 \\
 X_{61} & 0 & 1 & 0 & 0 & 1 & 0 & 0 & 1 & 0 & 1 & 1 & 0 & 0 & 0 & 0 & 0 & 1 \\
 X_{56} & 0 & 0 & 1 & 0 & 0 & 1 & 1 & 0 & 1 & 0 & 0 & 1 & 0 & 0 & 1 & 0 & 0 \\
 X_{12} & 0 & 0 & 1 & 0 & 0 & 1 & 1 & 0 & 1 & 0 & 0 & 0 & 1 & 1 & 0 & 0 & 0 \\
 X_{41} & 0 & 0 & 1 & 0 & 0 & 1 & 0 & 1 & 0 & 1 & 0 & 1 & 0 & 0 & 0 & 0 & 1 \\
 X_{35} & 0 & 0 & 1 & 0 & 0 & 1 & 0 & 1 & 0 & 1 & 0 & 0 & 1 & 0 & 0 & 1 & 0 \\
 X_{24} & 0 & 0 & 1 & 0 & 0 & 0 & 1 & 1 & 0 & 0 & 1 & 0 & 0 & 0 & 1 & 1 & 0 \\
 X_{63} & 0 & 0 & 1 & 0 & 0 & 0 & 1 & 1 & 0 & 0 & 1 & 0 & 0 & 1 & 0 & 0 & 1
\end{array}
\right)
$
}
\vspace{0.5cm}

 \noindent The F-term charge matrix $Q_F=\ker{(P)}$ is

\noindent\makebox[\textwidth]{%
\footnotesize
$
Q_F=
\left(
\begin{array}{ccc|cc|ccc|ccc|cccccc}
p_1 & p_2& p_3& q_1& q_2& r_1& r_2& r_3& u_1& u_2& u_3& s_1& s_2& s_3& s_4& s_5& s_6 \\ 
\hline
 1 & 1 & 0 & -1 & -1 & 0 & 0 & 0 & 0 & 0 & 0 & 0 & 0 & 0 & 0 & 0 &
   0 \\
 1 & 0 & 0 & 0 & 0 & 1 & 0 & 0 & 0 & 0 & 0 & -1 & -1 & 0 & 0 & 0 &
   0 \\
 1 & 0 & 0 & 0 & 0 & 0 & 1 & 0 & 0 & 0 & 0 & 0 & 0 & -1 & -1 & 0 &
   0 \\
 1 & 0 & 0 & 0 & 0 & 0 & 0 & 1 & 0 & 0 & 0 & 0 & 0 & 0 & 0 & -1 &
   -1 \\
    1 & 0 & 0 & 0 & -1 & 0 & 0 & 0 & 1 & 0 & 0 & -1 & 0 & -1 & 0 & 0 &
   1 \\
 0 & 1 & 1 & 0 & 0 & -1 & 0 & 0 & 0 & 0 & -1 & 0 & 0 & 0 & 0 & 0 &
   0 \\
 0 & 1 & 1 & 0 & 0 & 0 & -1 & 0 & 0 & -1 & 0 & 0 & 0 & 0 & 0 & 0 &
   0 \\
 0 & 1 & 1 & 0 & 0 & 0 & 0 & -1 & -1 & 0 & 0 & 0 & 0 & 0 & 0 & 0 &
   0 \\
 0 & 0 & 1 & 0 & 0 & -1 & -1 & 0 & 1 & 0 & 0 & 0 & 0 & 0 & 0 & 0 &
   0
\end{array}
\right)
$
}
\vspace{0.5cm}

\noindent The D-term charge matrix is

\noindent\makebox[\textwidth]{%
\footnotesize
$
Q_D=
\left(
\begin{array}{ccc|cc|ccc|ccc|cccccc}
p_1 & p_2& p_3& q_1& q_2& r_1& r_2& r_3& u_1& u_2& u_3& s_1& s_2& s_3& s_4& s_5& s_6 \\ 
\hline
 0 & 0 & 0 & 0 & 0 & 0 & 0 & 0 & 0 & 0 & 0 & 1 & -1 & 0 & 0 & 0 & 0 \\
 0 & 0 & 0 & 0 & 0 & 0 & 0 & 0 & 0 & 0 & 0 & 0 & 1 & -1 & 0 & 0 & 0 \\
 0 & 0 & 0 & 0 & 0 & 0 & 0 & 0 & 0 & 0 & 0 & 0 & 0 & 1 & -1 & 0 & 0 \\
 0 & 0 & 0 & 0 & 0 & 0 & 0 & 0 & 0 & 0 & 0 & 0 & 0 & 0 & 1 & -1 & 0 \\
 0 & 0 & 0 & 0 & 0 & 0 & 0 & 0 & 0 & 0 & 0 & 0 & 0 & 0 & 0 & 1 & -1
\end{array}
\right)
$
}
\vspace{0.5cm}

The total charge matrix $Q_t$ does not exhibit repeated columns. Accordingly, the global symmetry is $U(1)_{f_1} \times U(1)_{f_2} \times U(1)_R$. The flavour and R-charges on the GLSM fields corresponding to extremal points in the toric diagram in \fref{f7} are found as shown in \tref{t7} following the discussion in \sref{s1_3}.

\begin{table}[H]
\centering
\begin{tabular}{|c||c|c|c||l|} 
\hline
\; & $U(1)_{f_1}$ & $U(1)_{f_2}$ & $U(1)_R$ & fugacity \\
\hline
\hline
$p_1$ & 1/2 & 0   & $2/3$ &  	$t_1$\\
$p_2$ &-1/6 & 1/3 & $2/3$ &  	$t_2$\\
$p_3$ &-1/3 &-1/3 & $2/3$ &  	$t_3$\\
\hline
\end{tabular}
\caption{The GLSM fields corresponding to extremal points of the toric diagram with their mesonic charges (Model 7). \label{t7}}
\end{table}

Products of non-extremal perfect matchings are expressed in terms of single variables as follows
\beal{esx7_1}
q = q_1 q_2 ~,~
r = r_1 r_2 r_3 ~,~
u = u_1 u_2 u_3 ~,~
s = \prod_{m=1}^{6} s_m~.
\eea
Extremal perfect matchings are counted by the fugacity $t_\alpha$. Products of non-extremal perfect matchings such as $q$ are counted by fugacities of the form $y_q$.

The mesonic Hilbert series of Model 7 is 
 \beal{esm7_1}
&&g_{1}(t_\alpha,y_{q},y_{r},y_{u},y_{s}; \mathcal{M}^{mes}_{7})=
\nn\\
&&
\hspace{0.5cm}
\frac{
1 
+ y_{q}^2 y_{r} y_{u}^2 y_{s} ~ t_1 t_2^3 
+ y_{q} y_{r} y_{u} y_{s} ~ t_1 t_2 t_3 
+ y_{q}^2 y_{r}^2 y_{u}^3 y_{s} ~ t_2^4 t_3 
+ y_{q} y_{r}^2 y_{u}^2 y_{s} ~ t_2^2 t_3^2 
+ y_{q}^3 y_{r}^3 y_{u}^4 y_{s}^2 ~ t_1 t_2^5 t_3^2
}{
(1 - y_{q} y_{s} ~ t_1^2) 
(1 - y_{q}^3 y_{r}^2 y_{u}^4 y_{s} ~ t_2^6) 
(1 - y_{r}^2 y_{u} y_{s} ~ t_3^3)
}~~.
\nn\\
\eea
 The plethystic logarithm of the mesonic Hilbert series is
\beal{esm7_3}
&&
PL[g_1(t_\alpha,y_{q},y_{r},y_{u},y_{s};\mathcal{M}_{7}^{mes})]=
y_{q} y_{s} ~ t_1^2 
+ y_{q} y_{r} y_{u} y_{s} ~ t_1 t_2 t_3 
+ y_{r}^2 y_{u} y_{s} ~ t_3^3
\nn\\
&&
\hspace{1cm}
+ y_{q} y_{r}^2 y_{u}^2 y_{s} ~ t_2^2 t_3^2 
+ y_{q}^2 y_{r} y_{u}^2 y_{s} ~ t_1 t_2^3 
+ y_{q}^2 y_{r}^2 y_{u}^3 y_{s} ~ t_2^4 t_3 
- y_{q}^2 y_{r}^2 y_{u}^2 y_{s}^2 ~ t_1^2 t_2^2 t_3^2 
\nn\\
&&
\hspace{1cm}
+ y_{q}^3 y_{r}^2 y_{u}^4 y_{s} ~ t_2^6 
- y_{q}^2 y_{r}^3 y_{u}^3 y_{s}^2 ~ t_1 t_2^3 t_3^3
- y_{q}^3 y_{r}^2 y_{u}^3 y_s^2 ~ t_1^2 t_2^4 t_3
+ \dots~.
\eea

With the following fugacity map
\beal{esm7_y1}
f_1 &=& 
y_q^{1/3} y_r^{-2/3} y_u^{-2/3} ys^{1/3} ~ t_1^{4/3} t_2^{-2/3} t_3^{-2/3}
~,~
\nn\\
f_2 &=& 
y_q^{2/3} y_r^{-1/3} y_u^{2/3} y_s^{-1/3} ~ t_1^{-1/3} t_2^{5/3} t_3^{-4/3}
~,~
\nn\\
t &=&
y_q^{1/3} y_r^{1/3} y_u^{1/3} y_s^{1/3} t_1^{1/3} t_2^{1/3} t_3^{1/3}
~,~
\eea
where the fugacities $f_1$, $f_2$ and $t$ count the mesonic symmetry charges. Under the fugacity map above, the above plethystic logarithm becomes
\beal{esm7_3}
&&
PL[g_1(t,f_1,f_2;\mathcal{M}_{7}^{mes})]=
f_1 t^2 
+ \left(1  + \frac{1}{f_1 f_2} \right)t^3
+ \left(\frac{1}{f_1} + f_2 \right)t^4
+ \frac{f_2}{f_1} t^5
- t^6 
+ \frac{f_2^2}{f_1} t^6
 \nn\\
 &&
 \hspace{1cm}
 - \left(\frac{1}{f_1} + f_2 \right) t^7
    +\dots~.
   \nn\\
   \eea
The plethystic logarithm above exhibits the moduli space generators with their mesonic charges. They are summarized in \tref{t7gen}. The mesonic generators can be presented on a charge lattice. The convex polygon formed by the generators in \tref{t7gen} is the dual reflexive polygon of the toric diagram of Model 7. For the case of Model 7, the toric diagram is self-dual, and the charge lattice of the generators forms again the toric diagram of Model 7.\\

\begin{table}[H]
\centering
\resizebox{\hsize}{!}{
\begin{minipage}[!b]{0.6\textwidth}
\begin{tabular}{|l|c|c|}
\hline
Generator & $U(1)_{f_1}$ & $U(1)_{f_2}$ 
\\
\hline
\hline
$p_{1}^2 ~ q ~ s$
 & 1 & 0 
\nn\\
$p_{1} p_{2} p_{3} ~ q ~ r ~ u ~ s$
 & 0 & 0 
\nn\\
$p_{1} p_{2}^3 ~ q^2 ~ r ~ u^2 ~ s$
 & 0 & 1
\nn\\
$p_{3}^3 ~ r^2 ~ u ~ s$
 & -1 & -1 
\nn\\
$p_{2}^2 p_{3}^2 ~ q ~ r^2 ~ u^2 ~ s$
 & -1 & 0 
\nn\\
$p_{2}^4 p_{3} ~ q^2 ~ r^2 ~ u^3 ~ s$
 &-1 & 1 
\nn\\
$p_{2}^6 ~ q^3 ~ r^2 ~ u^4 ~ s$ 
& -1 & 2 
   \nn\\
   \hline
\end{tabular}
\end{minipage}
\hspace{2cm}
\begin{minipage}[!b]{0.2\textwidth}
\includegraphics[width=3.5 cm]{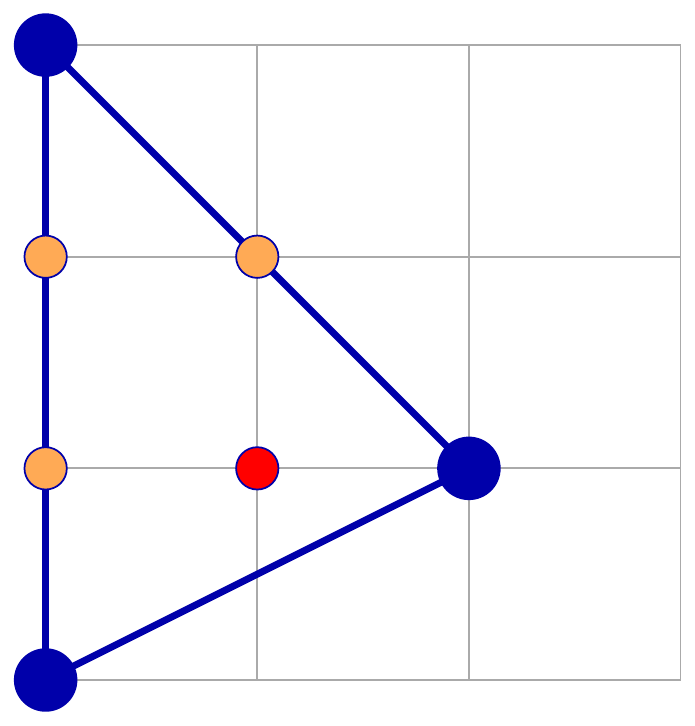}
\end{minipage}
}
\caption{The generators and lattice of generators of the mesonic moduli space of Model 7 in terms of GLSM fields with the corresponding flavor charges.\label{t7gen}\label{f7gen}} 
\end{table}

\begin{table}[H]
\centering
\resizebox{\hsize}{!}{
\begin{tabular}{|l|c|c|}
\hline
Generator & $U(1)_{f_1}$ & $U(1)_{f_2}$ 
\\
\hline
\hline
$
 X_{15} X_{51}=  X_{26} X_{62}=  X_{34} X_{43}
 $
 & 1 & 0 
\nn\\
$
X_{12} X_{25} X_{51}=  X_{12} X_{26} X_{61}=  X_{13} X_{34} X_{41}=  X_{13} X_{35} X_{51}=  X_{15} X_{54} X_{41}=  X_{15} X_{56} X_{61}
$
& 0 & 0
\nn\\
$
=  X_{24} X_{43} X_{32}=  X_{24} X_{46} X_{62}=  X_{25} X_{56} X_{62}=  X_{26} X_{63} X_{32}=  X_{34} X_{46} X_{63}=  X_{35} X_{54} X_{43}
$
 &  &  
\nn\\
$
 X_{13} X_{32} X_{25} X_{51}=  X_{13} X_{32} X_{26} X_{61}=  X_{13} X_{34} X_{46} X_{61}=  X_{15} X_{54} X_{46} X_{61}=  X_{25} X_{54} X_{43} X_{32}=  X_{25} X_{54} X_{46} X_{62}
$
 & 0 & 1
\nn\\
$X_{12} X_{24} X_{41}=  X_{35} X_{56} X_{63}$
 & -1 & -1 
\nn\\
$
 X_{12} X_{24} X_{46} X_{61}=  X_{12} X_{25} X_{54} X_{41}=  X_{12} X_{25} X_{56} X_{61}=  X_{13} X_{32} X_{24} X_{41}=  X_{13} X_{35} X_{54} X_{41}
 $
 & -1 & 0
 \nn\\
 $
 =  X_{13} X_{35} X_{56} X_{61}=  X_{24} X_{46} X_{63} X_{32}=  X_{25} X_{56} X_{63} X_{32}=  X_{35} X_{54} X_{46} X_{63}
$
 & &  
\nn\\
$
 X_{12} X_{25} X_{54} X_{46} X_{61}=  X_{13} X_{32} X_{24} X_{46} X_{61}=  X_{13} X_{32} X_{25} X_{54} X_{41}
 $
 &-1 & 1
 \nn\\
 $
 =  X_{13} X_{32} X_{25} X_{56} X_{61}=  X_{13} X_{35} X_{54} X_{46} X_{61}=  X_{25} X_{54} X_{46} X_{63} X_{32}
$
 & &  
\nn\\
$ X_{13} X_{32} X_{25} X_{54} X_{46} X_{61}$ 
& -1 & 2 
   \nn\\
   \hline
\end{tabular}
}
\caption{The generators in terms of bifundamental fields (Model 7).\label{t7gen2}\label{f7gen2}} 
\end{table}

With the fugacity map
\beal{esm7_3x1}
T_1 &=& f_1^{1/2} ~ t
= y_{q}^{1/2} y_{s}^{1/2} ~ t_1 ~,~
\nn\\
T_2 &=& \frac{f_2^{1/3}~ t}{f_1^{1/6}}
= y_{q}^{1/2} y_{r}^{1/3} y_{u}^{2/3} y_{s}^{1/6} ~ t_2 ~,~
\nn\\
T_3 &=& \frac{t}{f_1^{1/3} f_2^{1/3}}
= y_{r}^{2/3} y_{u}^{1/3} y_{s}^{1/3} ~ t_3
\eea   
the mesonic Hilbert series becomes
\beal{esm7_3x2}
g_{1}(T_1,T_2,T_3;\mathcal{M}^{mes}_{7})
=
\frac{
1+T_1 T_2^3 + T_1 T_2 T_3 + T_2^4 T_3 + T_2^2 T_3^2 + T_1 T_2^5 T_3^2
}{
(1-T_1^2)(1-T_2^6) (1-T_3^2)
}
\eea
with the plethystic logarithm being
\beal{esm6_3x3}
&&
PL[g_{1}(T_1,T_2,T_3;\mathcal{M}^{mes}_{7})]
=
T_1^2 
+ T_1 T_2 T_3 
+ T_3^3
+ T_2 T_3
+ T_1 T_2^3 
\nn\\
&&
\hspace{1cm}
+ T_2^4 T_3
- T_1^2 T_2^2 T_3^2
+ T_2^6
- T_1 T_2^3 T_3^3
- T_1^2 T_2^4 T_3
+ \dots
\eea
The above Hilbert series and plethystic logarithm illustrate the conical structure of the toric Calabi-Yau 3-fold.
\\

\section{Model 8: $\text{SPP}/\mathbb{Z}_{2}~(0,1,1,1),~\text{PdP}_{3c}$}
\subsection{Model 8 Phase a}

\begin{figure}[H]
\begin{center}
\includegraphics[trim=0cm 0cm 0cm 0cm,width=4.5 cm]{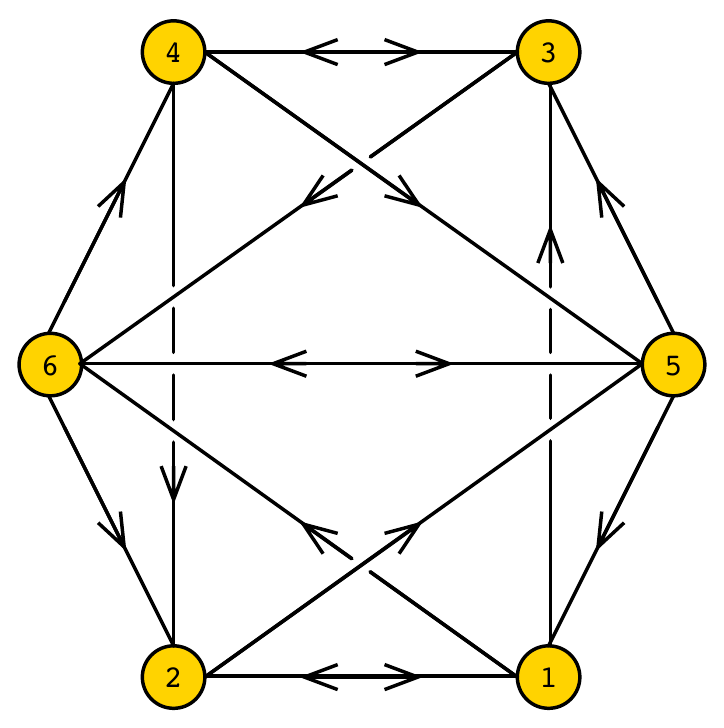}
\includegraphics[width=5 cm]{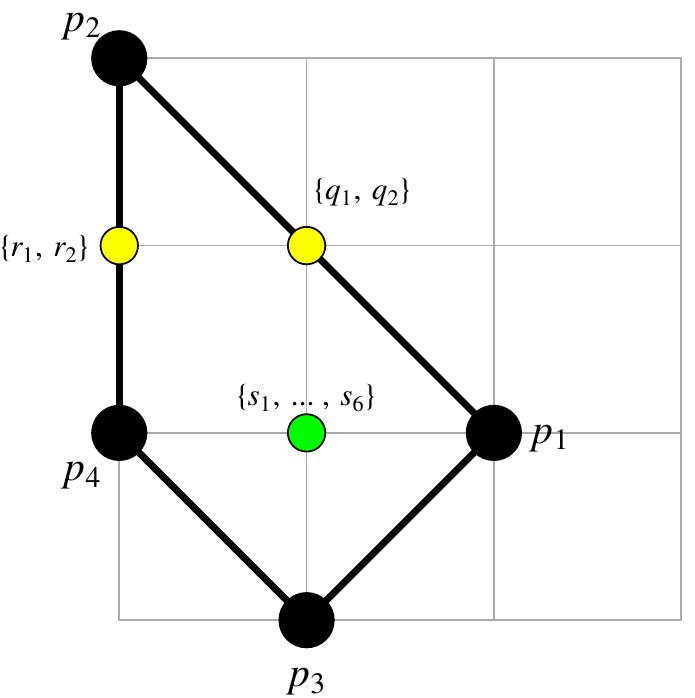}
\includegraphics[width=5 cm]{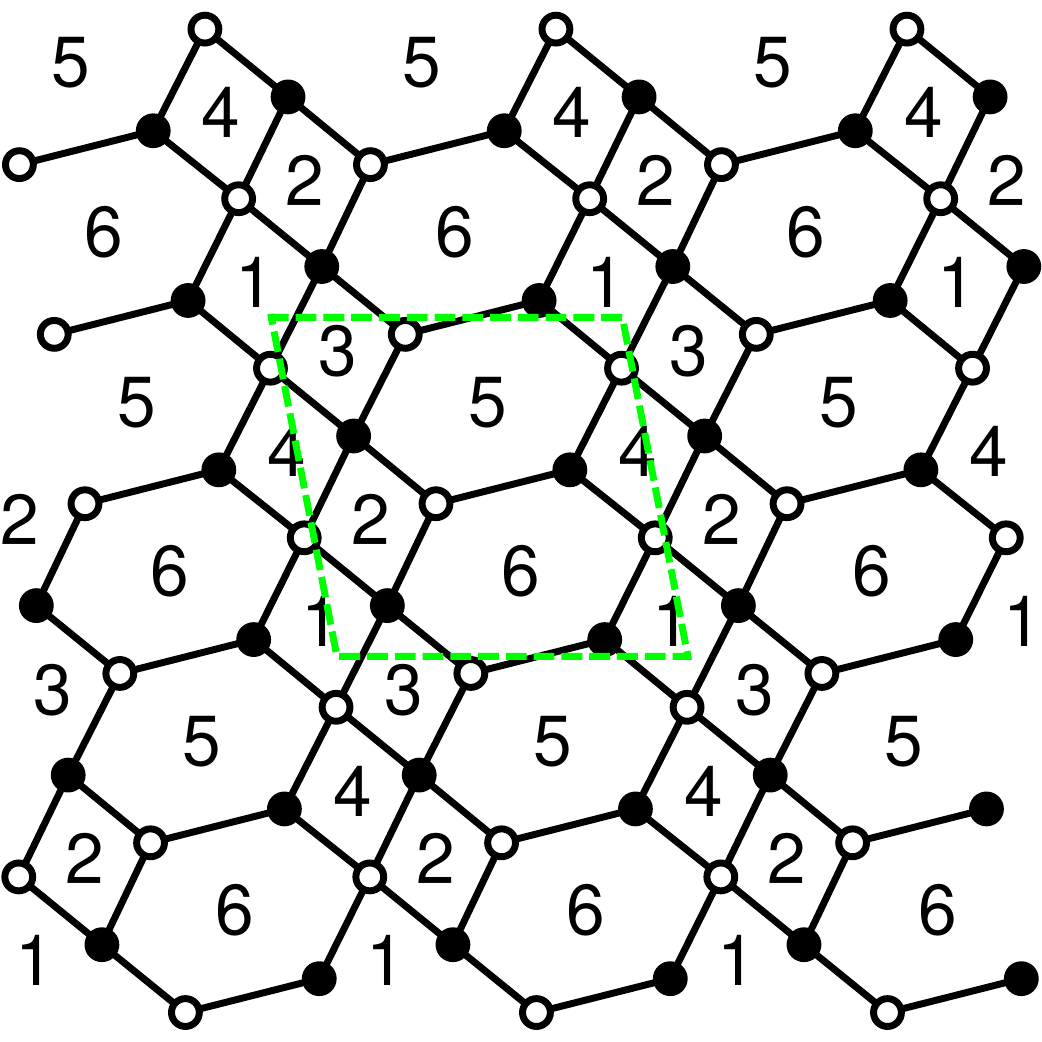}
\caption{The quiver, toric diagram, and brane tiling of Model 8a.\label{f8a}}
 \end{center}
 \end{figure}
 
 \noindent The superpotential is 
\beal{esm8a_00}
W&=&
+ X_{56} X_{62} X_{25}  
+ X_{65} X_{53} X_{36}  
+ X_{13} X_{34} X_{45} X_{51}  
+ X_{21} X_{16} X_{64} X_{42}  
\nn\\
&&
- X_{56} X_{64} X_{45}  
- X_{65} X_{51} X_{16}  
- X_{13} X_{36} X_{62} X_{21}  
- X_{25} X_{53} X_{34} X_{42} 
  \eea
 
 \noindent The perfect matching matrix is 
 
\noindent\makebox[\textwidth]{%
\footnotesize
$
P=
\left(
\begin{array}{c|cccc|cc|cc|cccccc}
 \; & p_{1} & p_{2} & p_{3} & p_{4} & q_{1} & q_{2} &
   r_{1} & r_{2} & s_{1} & s_{2} & s_{3} & s_{4} &
   s_{5} & s_{6} \\
   \hline
 X_{16} & 1 & 0 & 0 & 0 & 1 & 0 & 0 & 0 & 0 & 1 & 1 & 0 & 0 & 0 \\
 X_{45} & 1 & 0 & 0 & 0 & 1 & 0 & 0 & 0 & 0 & 0 & 0 & 0 & 1 & 1 \\
 X_{62} & 1 & 0 & 0 & 0 & 0 & 1 & 0 & 0 & 1 & 0 & 0 & 0 & 0 & 1 \\
 X_{53} & 1 & 0 & 0 & 0 & 0 & 1 & 0 & 0 & 0 & 0 & 1 & 1 & 0 & 0 \\
 X_{36} & 0 & 1 & 0 & 0 & 1 & 0 & 1 & 0 & 0 & 1 & 0 & 0 & 0 & 0 \\
 X_{25} & 0 & 1 & 0 & 0 & 1 & 0 & 0 & 1 & 0 & 0 & 0 & 0 & 1 & 0 \\
 X_{51} & 0 & 1 & 0 & 0 & 0 & 1 & 1 & 0 & 0 & 0 & 0 & 1 & 0 & 0 \\
 X_{64} & 0 & 1 & 0 & 0 & 0 & 1 & 0 & 1 & 1 & 0 & 0 & 0 & 0 & 0 \\
 X_{56} & 0 & 0 & 1 & 1 & 0 & 0 & 1 & 0 & 0 & 1 & 1 & 1 & 0 & 0 \\
 X_{65} & 0 & 0 & 1 & 1 & 0 & 0 & 0 & 1 & 1 & 0 & 0 & 0 & 1 & 1 \\
 X_{34} & 0 & 0 & 1 & 0 & 0 & 0 & 0 & 0 & 1 & 1 & 0 & 0 & 0 & 0 \\
 X_{21} & 0 & 0 & 1 & 0 & 0 & 0 & 0 & 0 & 0 & 0 & 0 & 1 & 1 & 0 \\
 X_{42} & 0 & 0 & 0 & 1 & 0 & 0 & 1 & 0 & 0 & 0 & 0 & 0 & 0 & 1 \\
 X_{13} & 0 & 0 & 0 & 1 & 0 & 0 & 0 & 1 & 0 & 0 & 1 & 0 & 0 & 0
\end{array}
\right)
$
}
\vspace{0.5cm}

 \noindent The F-term charge matrix $Q_F=\ker{(P)}$ is
 
\noindent\makebox[\textwidth]{%
\footnotesize
$
Q_F=
\left(
\begin{array}{cccc|cc|cc|cccccc}
  p_{1} & p_{2} & p_{3} & p_{4} & q_{1} & q_{2} &
   r_{1} & r_{2} & s_{1} & s_{2} & s_{3} & s_{4} &
   s_{5} & s_{6} \\
   \hline
 1 & 1 & 0 & 0 & -1 & -1 & 0 & 0 & 0 & 0 & 0 & 0 & 0 & 0 \\
 1 & 0 & 0 & 0 & -1 & 0 & 0 & 1 & -1 & 1 & -1 & 0 & 0 & 0 \\
 0 & 1 & 0 & 0 & -1 & 0 & -1 & 0 & -1 & 1 & 0 & 0 & 0 & 1 \\
 0 & 1 & 0 & 1 & 0 & 0 & -1 & -1 & 0 & 0 & 0 & 0 & 0 & 0 \\
 0 & 0 & 1 & 0 & 1 & 0 & 0 & 0 & 0 & -1 & 0 & 0 & -1 & 0 \\
 0 & 0 & 1 & 0 & 0 & 1 & 0 & 0 & -1 & 0 & 0 & -1 & 0 & 0
\end{array}
\right)
$
}
\vspace{0.5cm}

\noindent The D-term charge matrix is

\noindent\makebox[\textwidth]{%
\footnotesize
$
Q_D=
\left(
\begin{array}{cccc|cc|cc|cccccc}
  p_{1} & p_{2} & p_{3} & p_{4} & q_{1} & q_{2} &
   r_{1} & r_{2} & s_{1} & s_{2} & s_{3} & s_{4} &
   s_{5} & s_{6} \\
   \hline
 0 & 0 & 0 & 0 & 0 & 0 & 0 & 0 & 1 & -1 & 0 & 0 & 0 & 0 \\
 0 & 0 & 0 & 0 & 0 & 0 & 0 & 0 & 0 & 1 & -1 & 0 & 0 & 0 \\
 0 & 0 & 0 & 0 & 0 & 0 & 0 & 0 & 0 & 0 & 1 & -1 & 0 & 0 \\
 0 & 0 & 0 & 0 & 0 & 0 & 0 & 0 & 0 & 0 & 0 & 1 & -1 & 0 \\
 0 & 0 & 0 & 0 & 0 & 0 & 0 & 0 & 0 & 0 & 0 & 0 & 1 & -1
\end{array}
\right)
$
}
\vspace{0.5cm}

The total charge matrix $Q_t$ does not have repeated columns. Accordingly, the global symmetry is $U(1)_{f_1} \times U(1)_{f_2} \times U(1)_R$. The mesonic charges on the GLSM fields corresponding to extremal points in the toric diagram in \fref{f8a} are presented in \tref{t8a}. The charges have been found using the constraints discussed in \sref{s1_3}.

\begin{table}[H]
\centering
\begin{tabular}{|c||c|c|c||l|} 
\hline
\; & $U(1)_{f_1}$ & $U(1)_{f_2}$ & $U(1)_R$ & fugacity \\
\hline
\hline
$p_1$ & 1 & 0 		& $R_1=1/\sqrt{3}$ 		&  	$t_1$\\ 
$p_2$ &-1/2 & 1/2 	& $R_1=1/\sqrt{3}$ 		&  	$t_2$\\
$p_3$ &-1 & 0 		& $R_2=1-1/\sqrt{3}$ 	&  	$t_3$\\
$p_4$ & 1/2 & -1/2 	& $R_2=1-1/\sqrt{3}$ 	&  	$t_4$\\
\hline
\end{tabular}
\caption{The GLSM fields corresponding to extremal points of the toric diagram with their mesonic charges (Model 8a). The R-charges are obtained using a-maximization.\label{t8a}}
\end{table}

Products of non-extremal perfect matchings are labelled in terms of single variables as follows
\beal{es8a_x1}
q = q_1 q_2 ~,~
r = r_1 r_2 ~,~
s = \prod_{m=1}^{6} s_m~.
\eea
The fugacity which counts extremal perfect matchings $p_\alpha$ is $t_\alpha$. A product of non-extremal perfect matchings such as $q$ above is associated to the fugacity of the form $y_q$.

The mesonic Hilbert series of Model 8a is calculated using the Molien integral formula in \eref{es12_2}. It is
 \beal{esm8a_1}
&&g_{1}(t_\alpha,y_{q},y_{r},y_{s}; \mathcal{M}^{mes}_{8a})= 
(1 
+ y_{q}^2 y_{r}^2 y_{s}~ t_1 t_2^3 t_4 
+ y_{q} y_{r} y_{s} ~ t_1 t_2 t_3 t_4 
- y_{q}^3 y_{r}^2 y_{s}^2 ~ t_1^3 t_2^3 t_3 t_4
\nn\\
&&
\hspace{1cm} 
+ y_{q} y_{r}^2 y_{s} ~ t_2^2 t_3 t_4^2 
- y_{q}^3 y_{r}^3 y_{s}^2 ~ t_1^2 t_2^4 t_3 t_4^2 
- y_{q}^2 y_{r}^2 y_{s}^2 ~ t_1^2 t_2^2 t_3^2 t_4^2 
- y_{q}^4 y_{r}^4 y_{s}^3 ~ t_1^3 t_2^5 t_3^2 t_4^3
)
\nn\\
&&
\hspace{1cm}
\times
\frac{
1
 }{
(1 - y_{q}^2 y_{r} y_{s}~ t_1^2 t_2^2) 
(1 - y_{q}y_{s}~ t_1^2 t_3) 
(1 - y_{q}^2 y_{r}^3 y_{s}~ t_2^4 t_4^2) 
(1 - y_{r} y_{s} ~ t_3^2 t_4^2)
}~~.
\eea
 The plethystic logarithm of the mesonic Hilbert series is
\beal{esm8a_3}
&&
PL[g_1(t_\alpha,y_{q},y_{r},y_{s};\mathcal{M}_{8a}^{mes})]=
y_{q}^2 y_{r} y_{s} ~t_1^2 t_2^2 
+ y_{q} y_{s}~t_1^2 t_3 
+ y_{q}^2 y_{r}^2 y_{s} ~t_1 t_2^3 t_4 
+ y_{q} y_{r} y_{s} ~t_1 t_2 t_3 t_4 
\nn\\
&&
\hspace{1cm}
+ y_{q}^2 y_{r}^3 y_{s} ~t_2^4 t_4^2 
- y_{q}^3 y_{r}^2 y_{s}^2 ~t_1^3 t_2^3 t_3 t_4 
- y_{q}^4 y_{r}^4 y_{s}^2 ~t_1^2 t_2^6 t_4^2 
 + y_{q} y_{r}^2 y_{s} ~t_2^2 t_3 t_4^2 
 - 2 ~y_{q}^3 y_{r}^3 y_{s}^2~t_1^2 t_2^4 t_3 t_4^2 
 \nn\\
 &&
 \hspace{1cm}
+ \dots~.
\eea

Consider the following fugacity map
\beal{esm8a_y1}
f_1 = 
\frac{
t_1 t_3^{1/2}
}{
y_r ~ t_2 t_4^{1/2} 
}
~,~
f_2 =
\frac{
t_2 t_4^{1/2}
}{
y_s ~ t_1 t_3^{1/2}
}
~,~
\tilde{t}_1 =
y_q^{1/2} y_r^{1/2} y_s^{1/2} ~ t_1^{1/2} t_2^{1/2}
~,~
\tilde{t}_2 =
t_3^{1/2} t_4^{1/2}
~,~
\eea
where the fugacities $f_1$ and $f_2$ count flavour charges, and the fugacities $\tilde{t}_1$ and $\tilde{t}_2$ count R-charges $R_1$ and $R_2$ in \tref{t8a} respectively. Under the fugacity map above, the plethystic logarithm becomes
\beal{esm8a_3}
&&
PL[g_1(\tilde{t}_\alpha,f_1,f_2;\mathcal{M}_{8a}^{mes})]=
f_1 f_2 \tilde{t}_1^4 
+ f_1 \tilde{t}_1^2 \tilde{t}_2 
+ f_2 \tilde{t}_1^4 \tilde{t}_2 
+ \tilde{t}_1^2 \tilde{t}_2^2 
+ \frac{f_2}{f_1} \tilde{t}_1^4 \tilde{t}_2^2 
- f_1 f_2 \tilde{t}_1^6 \tilde{t}_2^2
\nn\\
&&
\hspace{1cm}
- f_2^2 \tilde{t}_1^8 \tilde{t}_2^2
+ \frac{1}{f_1} \tilde{t}_1^2 \tilde{t}_2^3 
- 2 f_2 \tilde{t}_1^6 \tilde{t}_2^3 
\dots~.
\eea
The above plethystic logarithm exhibits the moduli space generators with their corresponding mesonic charges. They are summarized in \tref{f8agen}. The generators can be presented on a charge lattice. The convex polygon formed by the generators in \tref{f8agen} is the dual reflexive polygon of the toric diagram of Model 8a. For the case of Model 8a, the toric diagram is self-dual, and the charge lattice of the generators forms again the toric diagram of Model 8a.\\

\begin{table}[H]
\centering
\resizebox{\hsize}{!}{
\begin{minipage}[!b]{0.6\textwidth}
\begin{tabular}{|l|c|c|}
\hline
Generator & $U(1)_{f_1}$ & $U(1)_{f_2}$ 
\\
\hline
\hline
$p_{1}^2 p_{3} ~ q ~ s$
   & 1&0
   \nn\\
$p_{3}^2 p_{4}^2 ~ r ~
s$
   & -1&-1
   \nn\\
$p_{1} p_{2} p_{3} p_{4} ~ q ~ r ~ s$
   & 0&0
   \nn\\
$p_{1}^2 p_{2}^2 ~ q^2 ~ r ~ s$
   & 1&1
   \nn\\
$p_{2}^2 p_{3} p_{4}^2 ~ q~
   r^2 ~ s$
   & -1&0
   \nn\\
$p_{1} p_{2}^3 p_{4} ~ q^2 ~ r^2 ~ s$
   & 0&1
   \nn\\
$p_{2}^4 p_{4}^2 ~ q^2 ~ r^3 ~ s$
   & -1&1
   \nn\\
   \hline
\end{tabular}
\end{minipage}
\hspace{1cm}
\begin{minipage}[!b]{0.3\textwidth}
\includegraphics[width=4 cm]{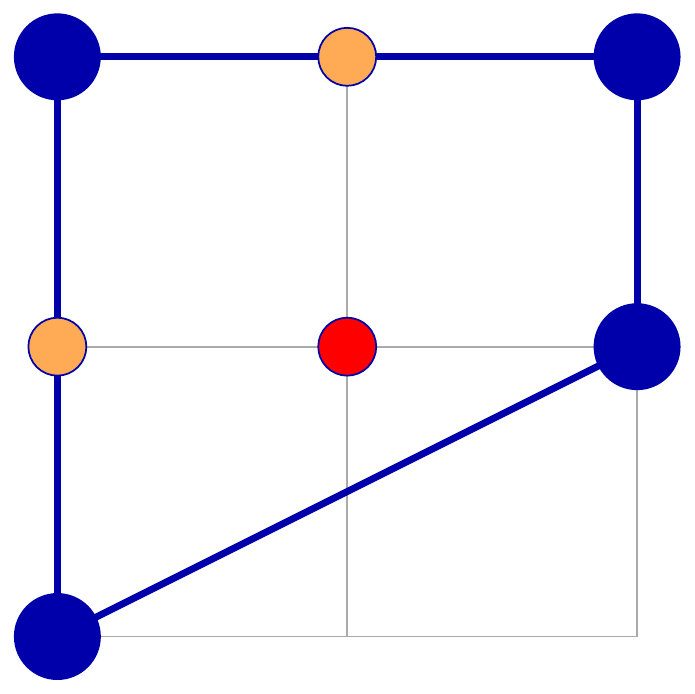}
\end{minipage}
}
\caption{The generators and lattice of generators of the mesonic moduli space of Model 8a in terms of GLSM fields with the corresponding flavor charges.\label{f8agen}} 
\end{table}

\begin{table}[H]
\centering
\resizebox{\hsize}{!}{
\begin{tabular}{|l|c|c|}
\hline
Generator & $U(1)_{f_1}$ & $U(1)_{f_2}$ 
\\
\hline
\hline
$
X_{16} X_{62} X_{21}=  X_{34} X_{45} X_{53}
$
   & 1&0
   \nn\\
$
X_{56} X_{65}=  X_{13} X_{34} X_{42} X_{21}
$
   & -1&-1
   \nn\\
$
 X_{16} X_{65} X_{51}=  X_{25} X_{56} X_{62}=  X_{36} X_{65} X_{53}=  X_{45} X_{56} X_{64}
 $
 & 0 & 0
 \nn\\
 $
 =  X_{13} X_{36} X_{62} X_{21}=  X_{13} X_{34} X_{45} X_{51}=  X_{16} X_{64} X_{42} X_{21}=  X_{25} X_{53} X_{34} X_{42}
$
   & &
   \nn\\
$X_{16} X_{62} X_{25} X_{51}=  X_{16} X_{64} X_{45} X_{51}=  X_{25} X_{53} X_{36} X_{62}=  X_{36} X_{64} X_{45} X_{53}$
   & 1&1
   \nn\\
$
X_{13} X_{36} X_{65} X_{51}=  X_{25} X_{56} X_{64} X_{42}=  X_{13} X_{36} X_{64} X_{42} X_{21}=  X_{13} X_{34} X_{42} X_{25} X_{51}
$
   & -1&0
   \nn\\
$
X_{13} X_{36} X_{62} X_{25} X_{51}=  X_{13} X_{36} X_{64} X_{45} X_{51}=  X_{16} X_{64} X_{42} X_{25} X_{51}=  X_{25} X_{53} X_{36} X_{64} X_{42}
$
   & 0&1
   \nn\\
$X_{13} X_{36} X_{64} X_{42} X_{25} X_{51}$
   & -1&1
   \nn\\
   \hline
\end{tabular}
}
\caption{The generators in terms of bifundamental fields (Model 8a).\label{f8agen2}} 
\end{table}

The mesonic Hilbert series and the plethystic logarithm can be re-expressed in terms of just $3$ fugacities
\beal{esm8a_x1}
T_1 =
\frac{
\tilde{t}_2
}{
f_1^2 f_2 ~ \tilde{t}_1^4
} 
= \frac{t_4}{y_{q}^2 y_{s} ~ t_1^3 t_2}~,~
T_2 = f_1 f_2 ~ \tilde{t}_1^4
= y_{q}^2 y_{r} y_{s}~ t_1^2 t_2^2~,~
T_3 = f_1 ~ \tilde{t}_1^2 \tilde{t}_2
= y_{q} y_{s}~ t_1^2 t_3~,
\nn\\
\eea
such that
\beal{esm8a_x2}
&&
g_1(T_1,T_2,T_3;\mathcal{M}^{mes}_{8a})=
\nn\\
&&
\hspace{1cm}
\frac{
1 + T_1 T_2^2 + T_1 T_2 T_3 - T_1 T_2^2 T_3 + T_1^2 T_2^2 T_3 - T_1^2 T_2^3 T_3 - T_1^2 T_2^2 T_3^2 - T_1^3 T_2^4 T_3^2
}{
(1 - T_2) (1 - T_3) (1 - T_1^2 T_2^3) (1 - T_1^2 T_2 T_3^2)
}
\nn\\
\eea   
and
\beal{esm8a_x3}
&&
PL[g_1(T_1,T_2,T_3;\mathcal{M}^{mes}_{8a})]=
T_2 
+ T_3 
+ T_1 T_2^2 
+ T_1 T_2 T_3 
+ T_1^2 T_2^3
- T_1 T_2^2 T_3  
- T_1^2 T_2^4   
\nn\\
&&
 + T_1^2 T_2^2 T_3 
 - 2 T_1^2 T_2^3 T_3 
+\dots
~~.
\eea
The above Hilbert series and plethystic logarithm in terms of just three fugacities with positive powers illustrate the conical structure of the toric Calabi-Yau 3-fold.
\\

\subsection{Model 8 Phase b}

\begin{figure}[H]
\begin{center}
\includegraphics[trim=0cm 0cm 0cm 0cm,width=4.5 cm]{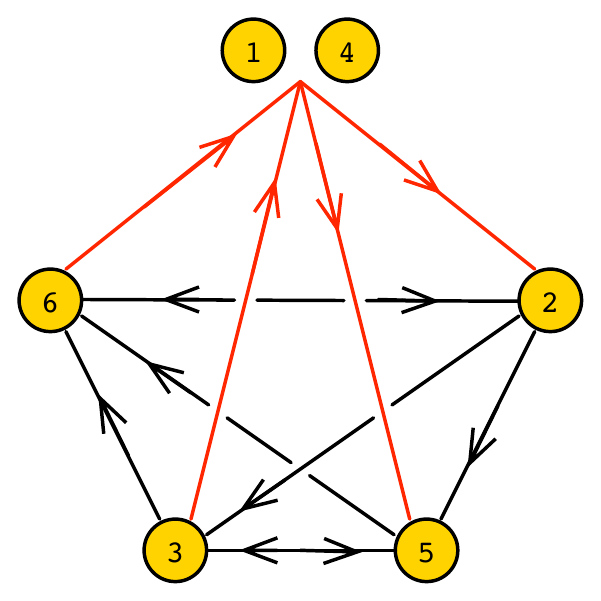}
\includegraphics[width=5 cm]{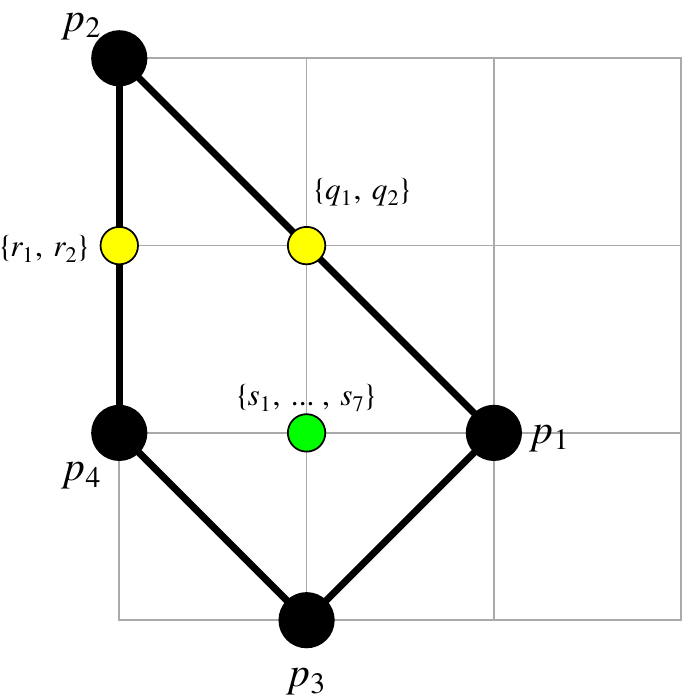}
\includegraphics[width=5 cm]{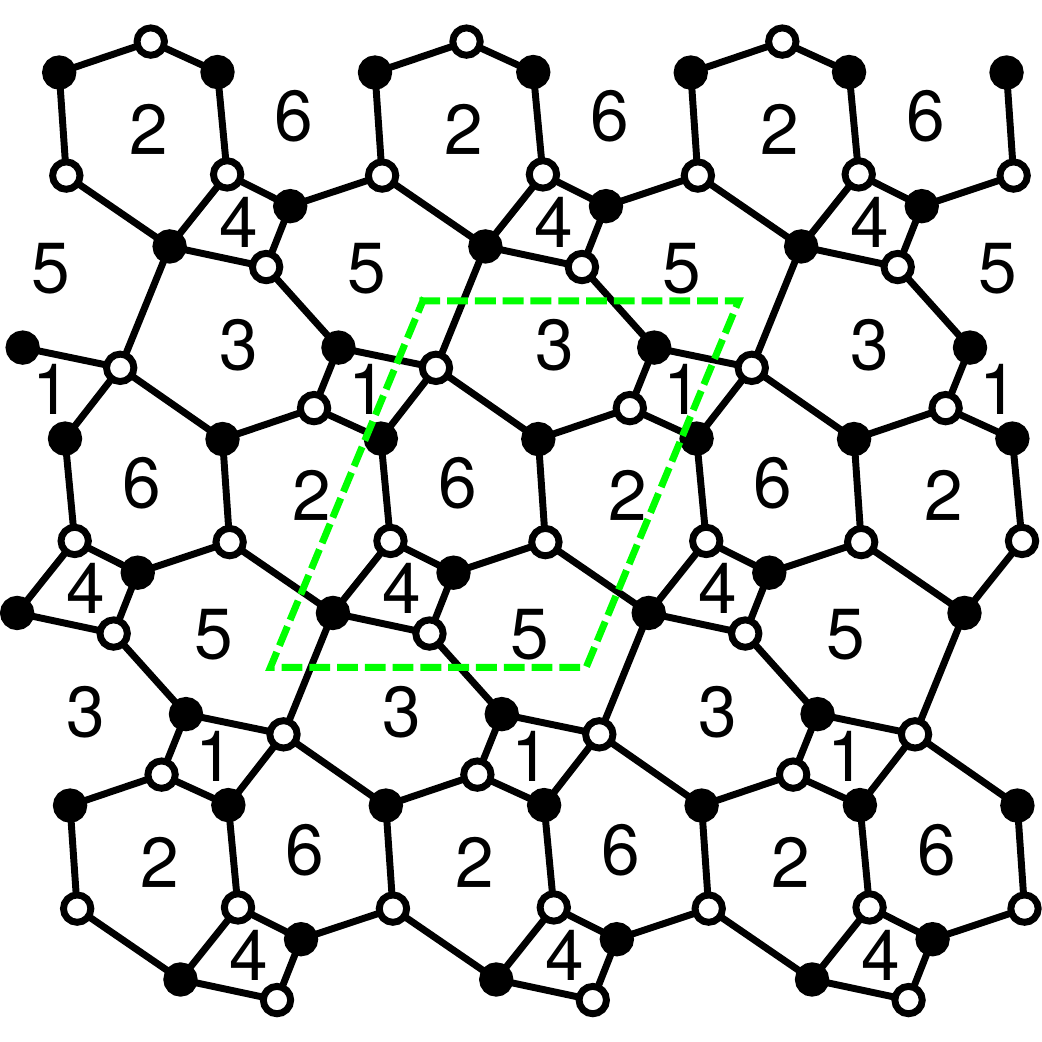}
\caption{The quiver, toric diagram, and brane tiling of Model 8b. The red arrows in the quiver indicate all possible connections between blocks of nodes.\label{f8b}}
 \end{center}
 \end{figure}
 
 \noindent The superpotential is 
\beal{esm8b_00}
W&=&
+ X_{31} X_{12} X_{23}  
+ X_{56} X_{62} X_{25}  
+ X_{64} X_{42} X_{26}  
+ X_{61} X_{15} X_{53}^{1} X_{36} 
+ X_{34} X_{45} X_{53}^{2} 
\nn\\
&&
- X_{31} X_{15} X_{53}^{2} 
- X_{36} X_{62} X_{23}  
- X_{56} X_{64} X_{45}  
- X_{61} X_{12} X_{26}  
- X_{25} X_{53}^{1} X_{34} X_{42}  
~~.
\nn\\
  \eea
 
 \noindent The perfect matching matrix is 
 
\noindent\makebox[\textwidth]{%
\footnotesize
$
P=
\left(
\begin{array}{c|cccc|cc|cc|ccccccc}
 \; & p_{1} & p_{2} & p_{3} & p_{4} & q_{1} & q_{2} &
   r_{1} & r_{2} & s_{1} & s_{2} & s_{3} & s_{4} &
   s_{5} & s_{6} & s_{7} \\
   \hline
   X_{560} & 1 & 1 & 0 & 0 & 0 & 0 & 1 & 0 & 0 &
   0 & 1 & 1 & 0 & 0 & 0 \\
 X_{23} & 1 & 1 & 0 & 0 & 0 & 0 & 0 & 1 & 1 &
   0 & 0 & 1 & 0 & 0 & 0 \\
 X_{26} & 1 & 0 & 1 & 0 & 0 & 1 & 0 & 0 & 1 &
   0 & 1 & 1 & 0 & 0 & 0 \\
 X_{15} & 1 & 0 & 0 & 0 & 0 & 0 & 0 & 0 & 1 &
   0 & 0 & 0 & 1 & 0 & 1 \\
 X_{34} & 1 & 0 & 0 & 0 & 0 & 0 & 0 & 0 & 0 &
   1 & 1 & 0 & 0 & 0 & 1 \\
 X_{53}^{2} & 0 & 1 & 0 & 1 & 1 & 0 & 1 & 1 & 0 &
   0 & 0 & 1 & 0 & 0 & 0 \\
 X_{42} & 0 & 1 & 0 & 0 & 0 & 0 & 1 & 0 & 0 &
   0 & 0 & 0 & 1 & 1 & 0 \\
 X_{61} & 0 & 1 & 0 & 0 & 0 & 0 & 0 & 1 & 0 &
   1 & 0 & 0 & 0 & 1 & 0 \\
 X_{62} & 0 & 0 & 1 & 0 & 1 & 0 & 0 & 0 & 0 &
   1 & 0 & 0 & 1 & 1 & 1 \\
 X_{53}^{1} & 0 & 0 & 1 & 0 & 1 & 0 & 0 & 0 & 0 &
   0 & 0 & 1 & 0 & 0 & 0 \\
 X_{45} & 0 & 0 & 1 & 0 & 0 & 1 & 0 & 0 & 1 &
   0 & 0 & 0 & 1 & 1 & 0 \\
 X_{31} & 0 & 0 & 1 & 0 & 0 & 1 & 0 & 0 & 0 &
   1 & 1 & 0 & 0 & 1 & 0 \\
 X_{12} & 0 & 0 & 0 & 1 & 1 & 0 & 1 & 0 & 0 &
   0 & 0 & 0 & 1 & 0 & 1 \\
 X_{64} & 0 & 0 & 0 & 1 & 1 & 0 & 0 & 1 & 0 &
   1 & 0 & 0 & 0 & 0 & 1 \\
 X_{36} & 0 & 0 & 0 & 1 & 0 & 1 & 1 & 0 & 0 &
   0 & 1 & 0 & 0 & 0 & 0 \\
 X_{25} & 0 & 0 & 0 & 1 & 0 & 1 & 0 & 1 & 1 &
   0 & 0 & 0 & 0 & 0 & 0
\end{array}
\right)
$
}
\vspace{0.5cm}

 \noindent The F-term charge matrix $Q_F=\ker{(P)}$ is

\noindent\makebox[\textwidth]{%
\footnotesize
$
Q_F=
\left(
\begin{array}{cccc|cc|cc|ccccccc}
 p_{1} & p_{2} & p_{3} & p_{4} & q_{1} & q_{2} &
   r_{1} & r_{2} & s_{1} & s_{2} & s_{3} & s_{4} &
   s_{5} & s_{6} & s_{7} \\
   \hline
 1 & 1 & 1 & 1 & 0 & 0 & -1 & 0 & 0 & -1 & 0 & 0 &
   -1 & -1 & 0 \\
 1 & 1 & 0 & 0 & -1 & -1 & 0 & 0 & 0 & 0 & 0 & 0 &
   0 & 0 & 0 \\
 0 & 1 & 1 & 0 & 0 & 0 & -1 & 0 & 1 & 0 & -1 & 0 &
   0 & -1 & 0 \\
    0 & 0 & 1 & 0 & 1 & 0 & 0 & 0 & 0 & 0 & 0 & 0 & 0
   & -1 & -1 \\
 0 & 0 & 0 & 1 & 1 & 0 & -1 & 0 & 0 & -1 & 1 & 0 &
   0 & -1 & 0 \\
 0 & 0 & 0 & 1 & 1 & 0 & -1 & 0 & 1 & 0 & 0 & -1 &
   0 & -1 & 0 \\
 0 & 0 & 0 & 1 & 1 & 0 & 0 & -1 & -1 & 0 & 1 & 0 &
   0 & 0 & -1
\end{array}
\right)
$
}
\vspace{0.5cm}

\noindent The D-term charge matrix is

\noindent\makebox[\textwidth]{%
\footnotesize
$
Q_D=
\left(
\begin{array}{cccc|cc|cc|ccccccc}
 p_{1} & p_{2} & p_{3} & p_{4} & q_{1} & q_{2} &
   r_{1} & r_{2} & s_{1} & s_{2} & s_{3} & s_{4} &
   s_{5} & s_{6} & s_{7} \\
   \hline
 0 & 0 & 0 & 0 & 0 & 0 & 0 & 0 & 0 & 1 & -1 & 0 &
   0 & 0 & 0 \\
 0 & 0 & 0 & 0 & 0 & 0 & 0 & 0 & 0 & 0 & 1 & -1 &
   0 & 0 & 0 \\
 0 & 0 & 0 & 0 & 0 & 0 & 0 & 0 & 0 & 0 & 0 & 1 &
   -1 & 0 & 0 \\
 0 & 0 & 0 & 0 & 0 & 0 & 0 & 0 & 0 & 0 & 0 & 0 & 1
   & -1 & 0 \\
 0 & 0 & 0 & 0 & 0 & 0 & 0 & 0 & 0 & 0 & 0 & 0 & 0
   & 1 & -1
\end{array}
\right)
$
}
\vspace{0.5cm}

The total charge matrix $Q_t$ does not have repeated columns. Accordingly, the global symmetry is $U(1)_{f_1} \times U(1)_{f_2} \times U(1)_R$. The flavour and R-charges on the GLSM fields corresponding to extremal points in the toric diagram are the same as in Model 8a, and are given in \tref{t8a}.

Products of non-extremal perfect matchings are expressed as
\beal{esm8a_x1}
q = q_1 q_2 ~,~
r = r_1 r_2 ~,~
s = \prod_{m=1}^{7} s_m ~.
\eea
The extremal perfect matchings are counted by $t_\alpha$. Products of non-extremal perfect matchings such as $q$ are associated to a fugacity of the form $y_q$.

The mesonic Hilbert series and the plethystic logarithm are identical to the ones for Model 8a and are given in \eref{esm8a_1} and \eref{esm8a_3} respectively. As a result, the mesonic moduli spaces for Models 8a and 8b are the same.

The generators of the mesonic moduli space in terms of all perfect matchings of Model 8b are shown in \tref{f8agen}. In terms of Model 8b quiver fields, the generators are shown in \tref{t8bgen2}. From the plethystic logarithm in \eref{esm8a_3} one observes that the mesonic moduli space is not a complete intersection.

\comment{
\begin{table}[h!]
\centering
\resizebox{\hsize}{!}{
\begin{minipage}[!b]{0.6\textwidth}
\begin{tabular}{|l|c|c|}
\hline
Generator & $U(1)_{f_1}$ & $U(1)_{f_2}$ 
\\
\hline
\hline
$p_{1} p_{3}^2 ~ q_{1} q_{2} ~ \prod_{m=1}^{7} s_m$
&1 &0
\nn\\
$p_{1}^2 p_{2}^2 ~ r_{1} r_{2} ~ \prod_{m=1}^{7} s_m$
   & -1&-1
\nn\\
$p_{1} p_{2} p_{3} p_{4} ~ q_{1} q_{2} ~ r_{1} r_{2} ~ \prod_{m=1}^{7} s_m$
   & 0&0
\nn\\
$p_{3}^2 p_{4}^2 ~ q_{1}^2 q_{2}^2 ~ r_{1} r_{2} ~ \prod_{m=1}^{7} s_m$
   & 1&1
\nn\\
$p_{1} p_{2}^2 p_{4}^2 ~ q_{1} q_{2} ~ r_{1}^2 r_{2}^2 ~ \prod_{m=1}^{7} s_m$
   & -1&0
\nn\\
$p_{2} p_{3} p_{4}^3 ~ q_{1}^2 q_{2}^2 ~ r_{1}^2 r_{2}^2 ~ \prod_{m=1}^{7} s_m$
   & 0&1
\nn\\
$p_{2}^2 p_{4}^4 ~ q_{1}^2 q_{2}^2 ~ r_{1}^3 r_{2}^3 ~ \prod_{m=1}^{7} s_m$
   & -1&1
\nn\\
   \hline
\end{tabular}
\end{minipage}
\hspace{1cm}
\begin{minipage}[!b]{0.3\textwidth}
\includegraphics[width=4 cm]{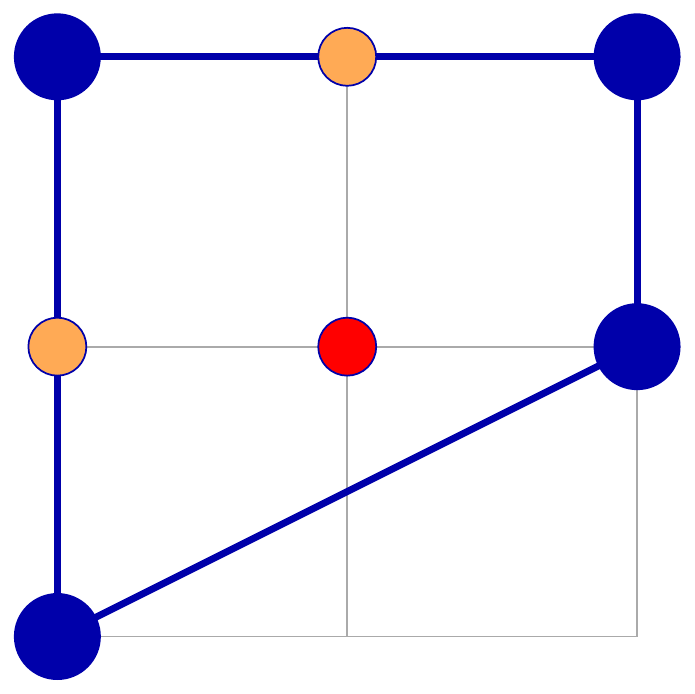}
\end{minipage}
}
\caption{The generators and lattice of generators of the mesonic moduli space of Model 8b in terms of GLSM fields with the corresponding flavor charges.\label{t8bgen}\label{f8bgen}} 
\end{table}
}

\begin{table}[h!]
\centering
\resizebox{\hsize}{!}{
\begin{tabular}{|l|c|c|}
\hline
Generator & $U(1)_{f_1}$ & $U(1)_{f_2}$ 
\\
\hline
\hline
$ X_{26} X_{62}=  X_{15} X_{53}^{1} X_{31}=  X_{34} X_{45} X_{53}^{1}$
&1 &0
\nn\\
$X_{15} X_{56} X_{61}=  X_{23} X_{34} X_{42}$
   & -1&-1
\nn\\
$ X_{15} X_{53}^{1} X_{36} X_{61}=  X_{25} X_{53}^{1} X_{34} X_{42}=  X_{12} X_{23} X_{31}=  X_{12} X_{26} X_{61}=  X_{15} X_{53}^{2} X_{31}$
&0&0
\nn\\
$=  X_{23} X_{36} X_{62}=  X_{25} X_{56} X_{62}=  X_{26} X_{64} X_{42}=  X_{34} X_{45} X_{53}^{2}=  X_{45} X_{56} X_{64}$
   & &
\nn\\
$ X_{12} X_{25} X_{53}^{1} X_{31}=  X_{25} X_{53}^{1} X_{36} X_{62}=  X_{36} X_{64} X_{45} X_{53}^{1}$
   & 1&1
\nn\\
$X_{12} X_{23} X_{36} X_{61}=  X_{12} X_{25} X_{56} X_{61}=  X_{15} X_ {53}^{2} X_{36} X_{61}$
& -1 & 0
\nn\\
$
=  X_{23} X_{36} X_{64} X_{42}=  X_{25} X_ {53}^{2} X_{34} X_{42}=  X_{25} X_{56} X_{64} X_{42}$
   & &
\nn\\
$X_{12} X_{25} X_ {53}^{1} X_{36} X_{61}=  X_{25} X_ {53}^{1} X_{36} X_{64} X_{42}=  X_{12} X_{25} X_ {53}^{2} X_{31}=  X_{25} X_ {53}^{2} X_{36} X_{62}=  X_{36} X_{64} X_{45} X_ {53}^{2}$
   & 0& 1
\nn\\
$ X_{12} X_{25} X_{53}^{2} X_{36} X_{61}=  X_{25} X_{53}^{2} X_{36} X_{64} X_{42}$
   & -1&1
\nn\\
   \hline
\end{tabular}
}
\caption{The generators in terms of bifundamental fields (Model 8b).\label{t8bgen2}\label{f8bgen2}} 
\end{table}

\section{Model 9: $\text{PdP}_{3b}$}
\subsection{Model 9 Phase a}

\begin{figure}[H]
\begin{center}
\includegraphics[trim=0cm 0cm 0cm 0cm,width=4.5 cm]{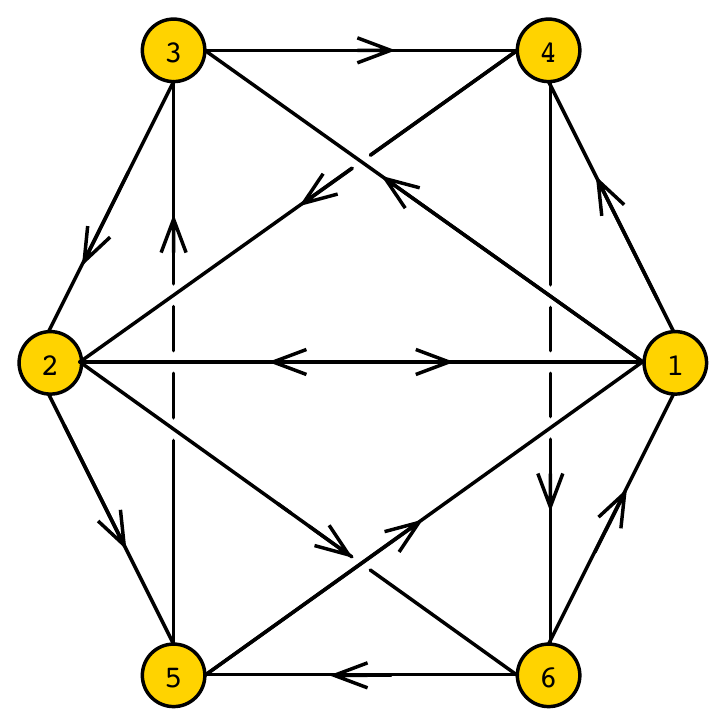}
\includegraphics[width=5 cm]{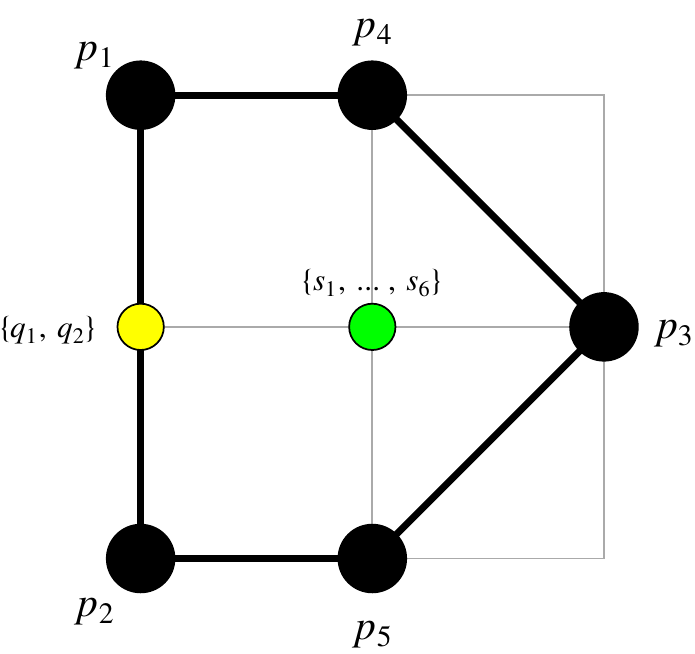}
\includegraphics[width=5 cm]{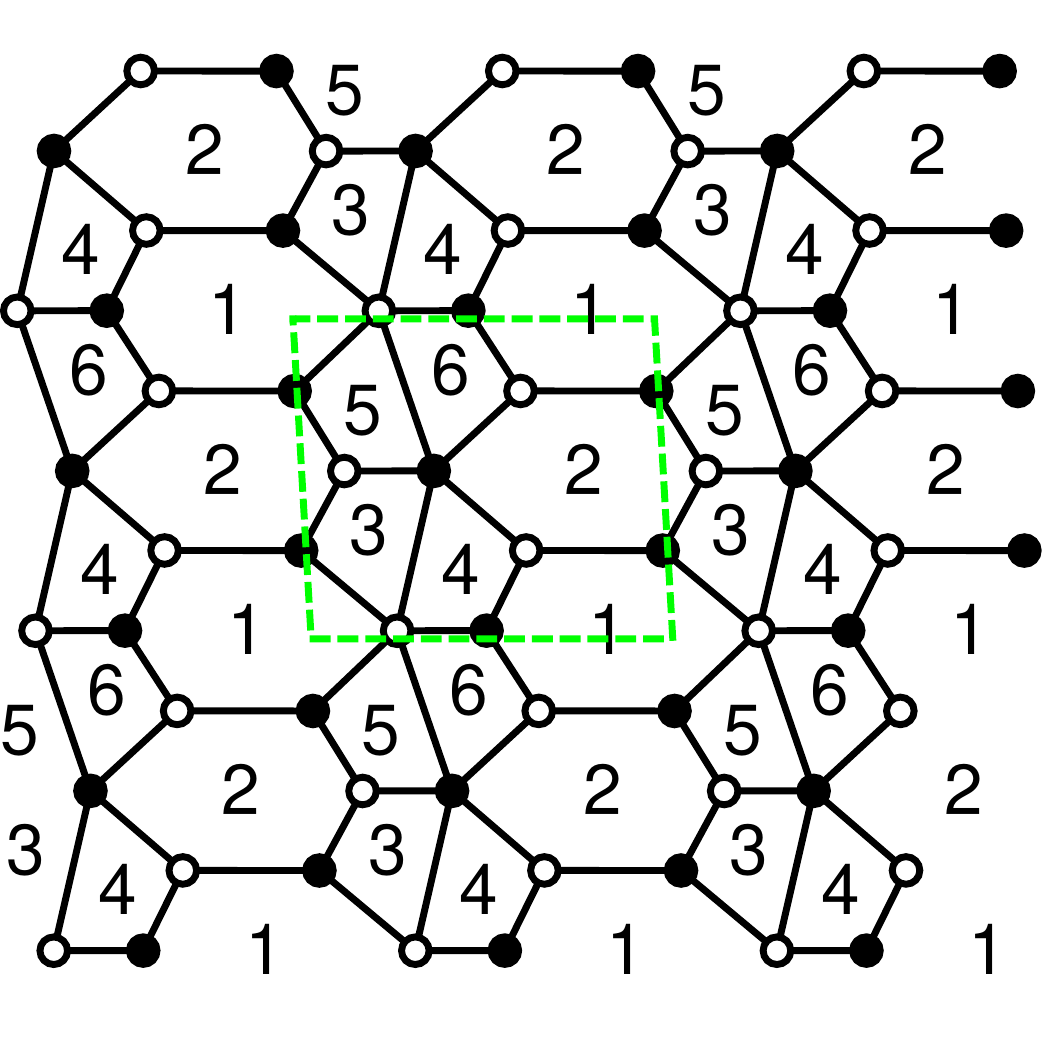}
\caption{The quiver, toric diagram, and brane tiling of Model 9a.\label{f9a}}
 \end{center}
 \end{figure}
 
 \noindent The superpotential is 
\beal{esm9a_00}
W&=&
+X_{12} X_{26} X_{61} 
+X_{25} X_{53} X_{32}  
+X_{42} X_{21} X_{14}  
+X_{13} X_{34} X_{46} X_{65} X_{51} 
\nn\\
&&
-X_{13} X_{32} X_{21}  
-X_{25} X_{51} X_{12}  
-X_{46} X_{61} X_{14}  
-X_{26} X_{65} X_{53} X_{34} X_{42}  
\eea
 
 \noindent The perfect matching matrix is 
 
\noindent\makebox[\textwidth]{%
\footnotesize
$
P=
\left(
\begin{array}{c|ccccc|cc|cccccc}
 \; & p_{1} & p_{2} & p_{3} & p_{4} &
   p_{5} & q_{1} & q_{2} & s_{1} &
   s_{2} & s_{3} & s_{4} & s_{5} &
   s_{6} \\
   \hline
 X_{26} & 1 & 0 & 0 & 0 & 0 & 1 & 0 & 1 & 0 & 0
   & 0 & 0 & 0 \\
 X_{51} & 1 & 0 & 0 & 0 & 0 & 0 & 1 & 0 & 1 & 0
   & 0 & 0 & 0 \\
 X_{13} & 0 & 1 & 0 & 0 & 0 & 1 & 0 & 0 & 0 & 1
   & 0 & 0 & 0 \\
 X_{42} & 0 & 1 & 0 & 0 & 0 & 0 & 1 & 0 & 0 & 0
   & 1 & 0 & 0 \\
 X_{46} & 0 & 0 & 1 & 0 & 0 & 0 & 0 & 1 & 0 & 0
   & 1 & 0 & 0 \\
 X_{53} & 0 & 0 & 1 & 0 & 0 & 0 & 0 & 0 & 1 & 1
   & 0 & 0 & 0 \\
 X_{14} & 1 & 0 & 0 & 1 & 0 & 1 & 0 & 0 & 0 & 1
   & 0 & 1 & 0 \\
 X_{32} & 1 & 0 & 0 & 1 & 0 & 0 & 1 & 0 & 0 & 0
   & 1 & 1 & 0 \\
 X_{25} & 0 & 1 & 0 & 0 & 1 & 1 & 0 & 1 & 0 & 0
   & 0 & 0 & 1 \\
 X_{61} & 0 & 1 & 0 & 0 & 1 & 0 & 1 & 0 & 1 & 0
   & 0 & 0 & 1 \\
 X_{12} & 0 & 0 & 1 & 1 & 0 & 0 & 0 & 0 & 0 & 1
   & 1 & 1 & 0 \\
 X_{21} & 0 & 0 & 1 & 0 & 1 & 0 & 0 & 1 & 1 & 0
   & 0 & 0 & 1 \\
 X_{65} & 0 & 0 & 0 & 1 & 0 & 0 & 0 & 0 & 0 & 0
   & 0 & 0 & 1 \\
 X_{34} & 0 & 0 & 0 & 0 & 1 & 0 & 0 & 0 & 0 & 0
   & 0 & 1 & 0
\end{array}
\right)
$
}
\vspace{0.5cm}

 \noindent The F-term charge matrix $Q_F=\ker{(P)}$ is

\noindent\makebox[\textwidth]{%
\footnotesize
$
Q_F=
\left(
\begin{array}{ccccc|cc|cccccc}
p_{1} & p_{2} & p_{3} & p_{4} &
   p_{5} & q_{1} & q_{2} & s_{1} &
   s_{2} & s_{3} & s_{4} & s_{5} &
   s_{6} \\
   \hline
 1 & 1 & 0 & 0 & 0 & -1 & -1 & 0 & 0 & 0 & 0 & 0 & 0
   \\
 0 & 0 & 0 & 1 & 1 & 0 & 0 & 0 & 0 & 0 & 0 & -1 & -1
 \\
 1 & 0 & 1 & 0 & 1 & 0 & 0 & -1 & -1 & 0 & 0 & -1 & 0
   \\
 0 & 0 & 1 & 0 & 0 & 1 & 0 & -1 & 0 & -1 & 0 & 0 & 0
   \\
 0 & 0 & 1 & 0 & 0 & 0 & 1 & 0 & -1 & 0 & -1 & 0 & 0   
\end{array}
\right)
$
}
\vspace{0.5cm}

\noindent The D-term charge matrix is

\noindent\makebox[\textwidth]{%
\footnotesize
$
Q_D=
\left(
\begin{array}{ccccc|cc|cccccc}
p_{1} & p_{2} & p_{3} & p_{4} &
   p_{5} & q_{1} & q_{2} & s_{1} &
   s_{2} & s_{3} & s_{4} & s_{5} &
   s_{6} \\
   \hline
 0 & 0 & 0 & 0 & 0 & 0 & 0 & 1 & -1 & 0 & 0 & 0 & 0
   \\
 0 & 0 & 0 & 0 & 0 & 0 & 0 & 0 & 1 & -1 & 0 & 0 & 0
   \\
 0 & 0 & 0 & 0 & 0 & 0 & 0 & 0 & 0 & 1 & -1 & 0 & 0
   \\
 0 & 0 & 0 & 0 & 0 & 0 & 0 & 0 & 0 & 0 & 1 & -1 & 0
   \\
 0 & 0 & 0 & 0 & 0 & 0 & 0 & 0 & 0 & 0 & 0 & 1 & -1
\end{array}
\right)
$
}
\vspace{0.5cm}

The total charge matrix does not exhibit repeated columns. Accordingly, the global symmetry is $U(1)_{f_1} \times U(1)_{f_2} \times U(1)_R$. Following the discussion in \sref{s1_3}, the mesonic charges on extremal perfect matchings are found. They are shown in \tref{t9a}.

\begin{table}[H]
\centering
\begin{tabular}{|c||c|c|c||l|} 
\hline
\; & $U(1)_{f_1}$ & $U(1)_{f_2}$ & $U(1)_R$ & fugacity \\
\hline
\hline
$p_1$ & -2/5 & 1/2 & $R_1=2 \left(-2 + \sqrt{5}\right)$ 	&  $t_1$\\
$p_2$ & -1/5 &-1/2 & $R_1=2 \left(-2 + \sqrt{5}\right)$ 	&  $t_2$\\
$p_3$ &  2/5 & 0   & $R_1=2 \left(-2 + \sqrt{5}\right)$ 	&  $t_3$\\
$p_4$ &  1/5 & 0 & $R_2=7 - 3 \sqrt{5}$ 	&  $t_4$\\ 
$p_5$ & 0 & 0 & $R_2=7 - 3 \sqrt{5}$ 		&  $t_5$\\ 
\hline
\end{tabular}
\caption{The GLSM fields corresponding to extremal points of the toric diagram with their mesonic charges (Model 9a). The R-charges are obtained using a-maximization.
\label{t9a}}
\end{table}

Products of non-extremal perfect matchings are expressed as 
\beal{esm9a_x1}
q = q_1 q_2 ~,~ s=\prod_{m=1}^{6} s_m ~.
\eea
Extremal perfect matchings are counted by $t_\alpha$. Products of non-extremal perfect matchings such as $q$ are counted by a fugacity of the form $y_q$.

The mesonic Hilbert series of Model 9a is found using the Molien integral formula in \eref{s1_3}. It is
 \beal{esm9a_1}
 &&
g_{1}(t_\alpha,y_{q},y_{s}; \mathcal{M}^{mes}_{9a})=
\nn\\
&& \hspace{0.7cm}
\frac{
P(t_\alpha)
}{
(1 - y_{q}^2 y_{s} ~ t_1^3 t_2 t_4^2) 
(1 - y_{q} y_{s} ~ t_1^2 t_3 t_4^2) 
(1 - y_{s} ~ t_3^2 t_4 t_5) 
(1 - y_{q}^2 y_{s} ~ t_1 t_2^3 t_5^2) 
(1 - y_{q} y_{s} ~ t_2^2 t_3 t_5^2)
}~~.
\nn\\
\eea
The numerator is given by the polynomial
\beal{esm9a2_1}
P(t_\alpha)&=&
1 
+ y_{q}^2 y_{s} ~ t_1^2 t_2^2 t_4 t_5 
+ y_{q} y_{s} ~ t_1 t_2 t_3 t_4 t_5 
- y_{q}^3 y_{s}^2 ~ t_1^4 t_2^2 t_3 t_4^3 t_5 
- y_{q}^2 y_{s}^2 ~ t_1^3 t_2 t_3^2 t_4^3 t_5 
\nn\\
&&  
- y_{q}^3 y_{s}^2 ~ t_1^3 t_2^3 t_3 t_4^2 t_5^2
- y_{q}^2 y_{s}^2 ~ t_1^2 t_2^2 t_3^2 t_4^2 t_5^2
- y_{q}^3 y_{s}^2 ~ t_1^2 t_2^4 t_3 t_4 t_5^3 
- y_{q}^2 y_{s}^2 ~ t_1 t_2^3 t_3^2 t_4 t_5^3 
\nn\\
&&
+ y_{q}^4 y_{s}^3 ~ t_1^4 t_2^4 t_3^2 t_4^3 t_5^3 
+ y_{q}^3 y_{s}^3 ~ t_1^3 t_2^3 t_3^3 t_4^3 t_5^3 
+ y_{q}^5 y_{s}^4 ~ t_1^5 t_2^5 t_3^3 t_4^4 t_5^4
~~.
\eea
 The plethystic logarithm of the mesonic Hilbert series is
\beal{esm9a_3}
&&
PL[g_1(t_\alpha,y_{q},y_{s};\mathcal{M}_{9a}^{mes})]=
y_{s} ~ t_3^2 t_4 t_5 
+ y_{q} y_{s} ~ t_1 t_2 t_3 t_4 t_5 
+ y_{q} y_{s} ~ t_1^2 t_3 t_4^2 
+ y_{q} y_{s} ~t_2^2 t_3 t_5^2 
\nn\\
&&
\hspace{1cm}
+ y_{q}^2 y_{s} ~ t_1^2 t_2^2 t_4 t_5 
+ y_{q}^2 y_{s}~ t_1 t_2^3 t_5^2 
 + y_{q}^2 y_{s} ~ t_1^3 t_2 t_4^2 
- 2 ~y_{q}^2 y_{s}^2 ~ t_1^2 t_2^2 t_3^2 t_4^2 t_5^2
-  y_{q}^2 y_{s}^2 ~ t_1^3 t_2 t_3^2 t_4^3 t_5
\nn\\
&&
\hspace{1cm}
- y_{q}^2 y_{s}^2 ~ t_1 t_2^3 t_3^2 t_4 t_5^3
+ \dots~.
\eea

Consider the following fugacity map
\beal{esm9a_y1}
f_1 = 
y_q^{-2/3} y_s^{1/3}~ t_1^{-2/3} t_2^{2/3} t_3^{4/3}
~,~
f_2 =
\frac{t_1 t_4}{t_2 t_5}
~,~
\tilde{t}_1 =
y_q^{1/3} y_s^{1/3} ~t_1^{1/3} t_2^{1/3} t_3^{1/3}
~,~
\tilde{t}_2 =
t_4^{1/2} t_5^{1/2}
~,~
\nn\\
\eea
where the fugacities $f_1$ and $f_2$ count flavour charges, and the fugacities $\tilde{t}_1$ and $\tilde{t}_2$ count the R-charges $R_1$ and $R_2$ in \tref{t9a} respectively. Under the fugacity map above, the plethystic logarithm becomes
\beal{esm9a_3}
&&
PL[g_1(\tilde{t}_\alpha,f_1,f_2;\mathcal{M}_{9a}^{mes})]=
f_1 \tilde{t}_1^2 \tilde{t}_2^2
+ \left(
1  
+ f_2
+ \frac{1}{f_2} 
\right) \tilde{t}_1^3 \tilde{t}_2^2
+ \left(
\frac{1}{f_1}
+ \frac{1}{f_1 f_2} 
+ \frac{f_2}{f_1}
\right) \tilde{t}_1^4 \tilde{t}_2^2 
 \nn\\
 &&
 \hspace{1cm}
 - \left(
 2 
 + f_2
 + \frac{1}{f_2} 
 \right) \tilde{t}_1^6 \tilde{t}_2^4
 +\dots ~~.
   \eea
This plethystic logarithm exhibits the moduli space generators with their mesonic charges. They are summarized in \tref{t9agen}. The generators can be presented on a charge lattice. The convex polygon formed by the generators in \tref{t9agen} is the dual reflexive polygon of the toric diagram of Model 9a. For the case of Model 9a, the toric diagram is self-dual, and the charge lattice of the generators forms again the toric diagram of Model 9a.\\

\begin{table}[H]
\centering
\resizebox{\hsize}{!}{
\begin{minipage}[!b]{0.6\textwidth}
\begin{tabular}{|l|c|c|}
\hline
Generator & $U(1)_{f_1}$ & $U(1)_{f_2}$ 
\\
\hline
\hline
$p_{3}^2 p_{4} p_{5} ~ s$
   & 1 & 0
   \nn\\
$p_{1}^2 p_{3} p_{4}^2 ~ q ~s$
   & 0& 1
   \nn\\
$p_{1} p_{2}
   p_{3} p_{4} p_{5} ~q~
   s$
   & 0&0
   \nn\\
$p_{2}^2 p_{3} p_{5}^2 ~
   q ~ s$
   & 0&-1
   \nn\\
$p_{1}^3
   p_{2} p_{4}^2 ~ q^2
  ~ s$
   & -1&1
   \nn\\
$p_{1}^2 p_{2}^2 p_{4}
   p_{5} ~ q^2 ~ s$
   & -1&0
   \nn\\
$p_{1} p_{2}^3 p_{5}^2
  ~ q^2 ~ s$
   & -1&-1
   \nn\\
   \hline
\end{tabular}
\end{minipage}
\hspace{1cm}
\begin{minipage}[!b]{0.3\textwidth}
\includegraphics[width=4 cm]{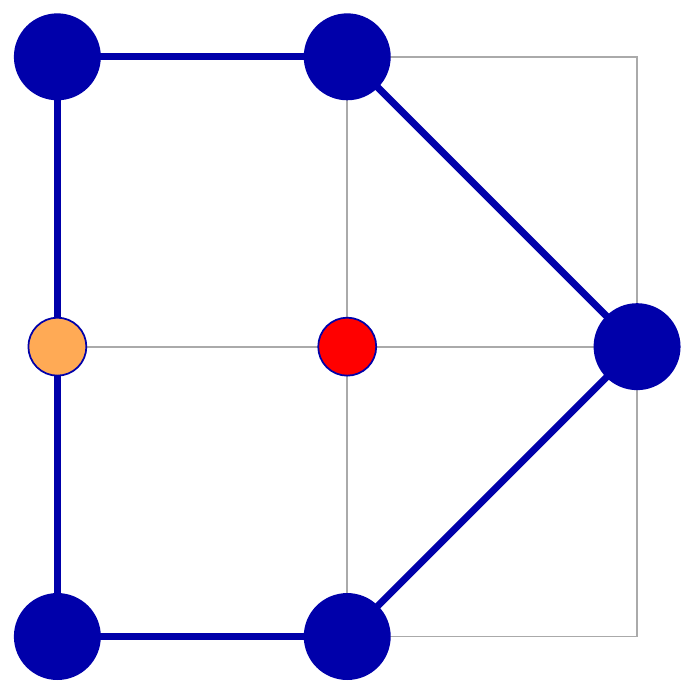}
\end{minipage}
}
\caption{The generators and lattice of generators of the mesonic moduli space of Model 9a in terms of GLSM fields with the corresponding flavor charges.\label{t9agen}\label{f9agen}} 
\end{table}

\begin{table}[H]
\centering
\resizebox{\hsize}{!}{
\begin{tabular}{|l|c|c|}
\hline
Generator & $U(1)_{f_1}$ & $U(1)_{f_2}$ 
\\
\hline
\hline
$X_{12} X_{21}=  X_{34} X_{46} X_{65} X_{53}$
   & 1 & 0
   \nn\\
$ X_{12} X_{26} X_{65} X_{51}=  X_{14} X_{46} X_{65} X_{51}=  X_{26} X_{65} X_{53} X_{32}$
   & 0& 1
   \nn\\
$X_{13} X_{34} X_{46} X_{65} X_{51}=  X_{26} X_{65} X_{53} X_{34} X_{42}=  X_{12} X_{25} X_{51}=  X_{12} X_{26} X_{61}
$
& 0 & 0
\nn\\
$=  X_{13} X_{32} X_{21}=  X_{14} X_{42} X_{21}=  X_{14} X_{46} X_{61}=  X_{25} X_{53} X_{32}$
   & &
   \nn\\
$ X_{13} X_{34} X_{42} X_{21}=  X_{13} X_{34} X_{46} X_{61}=  X_{25} X_{53} X_{34} X_{42}$
   & 0 &-1
   \nn\\
$ X_{13} X_{32} X_{26} X_{65} X_{51}=  X_{14} X_{42} X_{26} X_{65} X_{51}$
   & -1&1
   \nn\\
$X_{13} X_{34} X_{42} X_{26} X_{65} X_{51}=  X_{13} X_{32} X_{25} X_{51}=  X_{13} X_{32} X_{26} X_{61}=  X_{14} X_{42} X_{25} X_{51}=  X_{14} X_{42} X_{26} X_{61}$
   & -1&0
   \nn\\
$ X_{13} X_{34} X_{42} X_{25} X_{51}=  X_{13} X_{34} X_{42} X_{26} X_{61}$
   & -1&-1
   \nn\\
   \hline
\end{tabular}
}
\caption{The generators in terms of bifundamental fields (Model 9a).\label{t9agen2}\label{f9agen2}} 
\end{table}

The mesonic Hilbert series and the plethystic logarithm can be re-expressed in terms of $3$ fugacities
\beal{esm9a_x1}
T_1 = \frac{t_5}{y_{q}^2 y_{s}~ t_1^4 t_4^3}~,~
T_2 = y_{q}^2 y_{s} ~ t_1^3 t_2 t_4^2~,~
T_3 = y_{q} y_{s} ~ t_1^2 t_3 t_4^2~,
\eea
such that
\beal{esm9a_x1}
&&
g_1(T_1,T_2,T_3;\mathcal{M}^{mes}_{9a})=
\nn\\
&&
(1 + T_1 T_2^2 + T_1 T_2 T_3 - T_1 T_2^2 T_3 - T_1 T_2 T_3^2
- T_1^2 T_2^3 T_3 - T_1^2 T_2^2 T_3^2 - T_1^3 T_2^4 T_3   - T_1^3 T_2^3 T_3^2
 \nn\\
&&
\hspace{0.2cm}
+ T_1^3 T_2^4 T_3^2 + T_1^3 T_2^3 T_3^3 + T_1^4 T_2^5 T_3^3)\times
\frac{
1
}{
(1 - T_2) (1 - T_3) (1 - T_1^2 T_2^3) (1 - T_1 T_3^2) (1 - T_1^2 T_2^2 T_3)
}
\nn\\
\eea
and
\beal{esm9a_x1}
&&
PL[g_1(T_1,T_2,T_3;\mathcal{M}^{mes}_{9a})]=
T_1 T_3^2 
+ T_1 T_2 T_3 
+ T_3 
+ T_1^2 T_2^2 T_3
+ T_1 T_2^2 
+ T_1^2 T_2^3 
+ T_2 
\nn\\
&&
\hspace{1cm}
- 2 T_1^2 T_2^2 T_3^2 
- T_1 T_2 T_3^2
- T_1^3 T_2^3 T_3^2
+\dots
~~.
\eea
The above Hilbert series and plethystic logarithm illustrate the conical structure of the toric Calabi-Yau 3-fold.
\\

\subsection{Model 9 Phase b} 

\begin{figure}[H]
\begin{center}
\includegraphics[trim=0cm 0cm 0cm 0cm,width=4.5 cm]{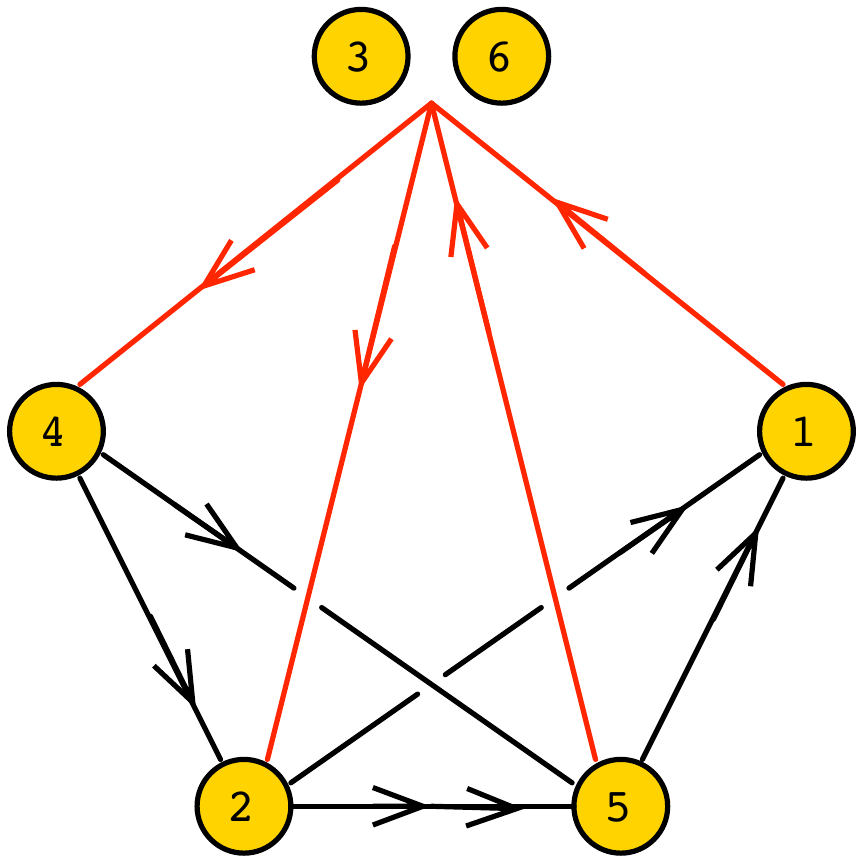}
\includegraphics[width=5 cm]{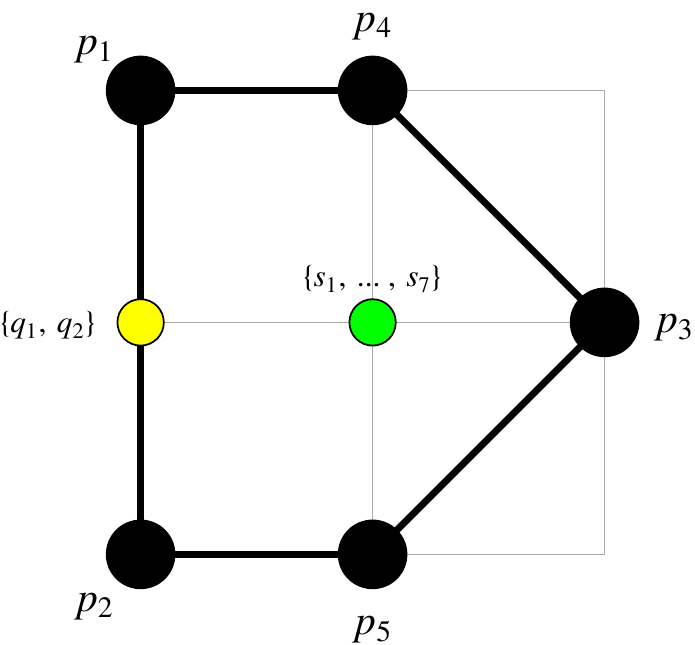}
\includegraphics[width=4.8 cm]{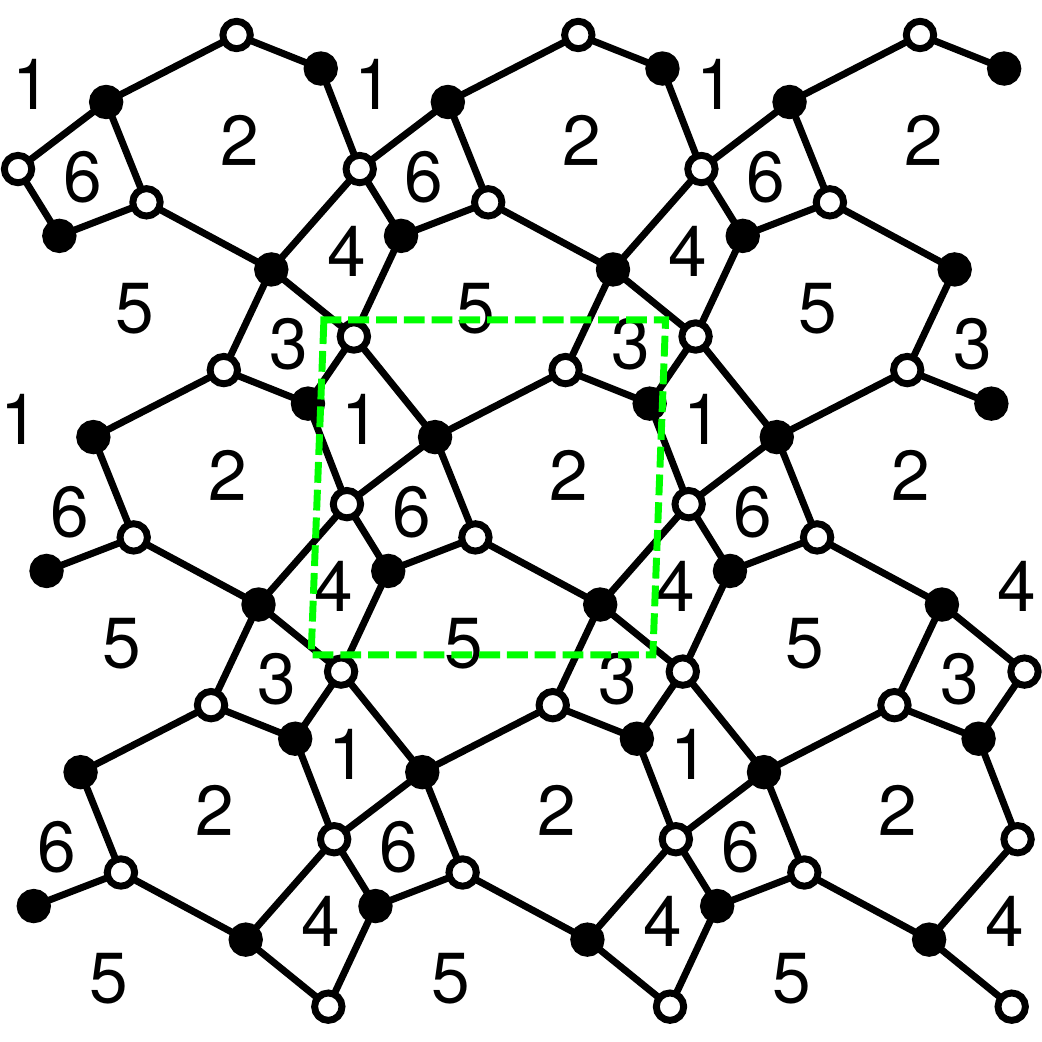}
\caption{The quiver, toric diagram, and brane tiling of Model 9b. The red arrows in the quiver indicate all possible connections between blocks of nodes.}
  \label{f9b}
 \end{center}
 \end{figure}
 
\noindent The superpotential is 
\beal{esm9b_00}
W&=&
+ X_{25}^{2} X_{53} X_{32}  
 + X_{56} X_{62} X_{25}^{1} 
  + X_{13} X_{34} X_{45} X_{51}  
 + X_{21} X_{16} X_{64} X_{42}  
 \nn\\
 &&
-X_{13} X_{32} X_{21}  
 - X_{56} X_{64} X_{45}  
 - X_{16} X_{62} X_{25}^{2} X_{51}  
 - X_{25}^{1} X_{53} X_{34} X_{42} 
  \eea
 
 \noindent The perfect matching matrix is 
 
\noindent\makebox[\textwidth]{%
\footnotesize
$
P=
\left(
\begin{array}{c|ccccc|cc|ccccccc}
 0 & p_{1} & p_{2} & p_{3} & p_{4} &
   p_{5} & q_{1} & q_{2} & s_{1} &
   s_{2} & s_{3} & s_{4} & s_{5} &
   s_{6} & s_{7} \\
   \hline
X_{32} & 1 & 0 & 0 & 1 & 0 & 1 & 0 & 0 & 1 & 0 & 1 & 0 & 1 & 0 \\
 X_{25}^{1} & 1 & 0 & 0 & 1 & 0 & 0 & 1 & 1 & 0 & 0 & 0 & 0 & 0 & 0 \\
 X_{51} & 1 & 0 & 0 & 0 & 0 & 1 & 0 & 0 & 0 & 0 & 0 & 0 & 0 & 1 \\
 X_{64} & 1 & 0 & 0 & 0 & 0 & 0 & 1 & 0 & 0 & 1 & 0 & 0 & 1 & 0 \\
 X_{56} & 0 & 1 & 0 & 0 & 1 & 1 & 0 & 0 & 1 & 0 & 0 & 1 & 0 & 1 \\
 X_{25}^{2} & 0 & 1 & 0 & 0 & 1 & 0 & 1 & 1 & 0 & 0 & 0 & 0 & 0 & 0 \\
 X_{42} & 0 & 1 & 0 & 0 & 0 & 1 & 0 & 0 & 0 & 0 & 1 & 0 & 0 & 0 \\
 X_{13} & 0 & 1 & 0 & 0 & 0 & 0 & 1 & 0 & 0 & 1 & 0 & 1 & 0 & 0 \\
 X_{45} & 0 & 0 & 1 & 1 & 0 & 0 & 0 & 1 & 0 & 0 & 1 & 0 & 0 & 0 \\
 X_{21} & 0 & 0 & 1 & 0 & 1 & 0 & 0 & 1 & 0 & 0 & 0 & 0 & 0 & 1 \\
 X_{62} & 0 & 0 & 1 & 0 & 0 & 0 & 0 & 0 & 0 & 1 & 1 & 0 & 1 & 0 \\
 X_{53} & 0 & 0 & 1 & 0 & 0 & 0 & 0 & 0 & 0 & 1 & 0 & 1 & 0 & 1 \\
 X_{16} & 0 & 0 & 0 & 1 & 0 & 0 & 0 & 0 & 1 & 0 & 0 & 1 & 0 & 0 \\
 X_{34} & 0 & 0 & 0 & 0 & 1 & 0 & 0 & 0 & 1 & 0 & 0 & 0 & 1 & 0
\end{array}
\right)
$
}
\vspace{0.5cm}

 \noindent The F-term charge matrix $Q_F=\ker{(P)}$ is

\noindent\makebox[\textwidth]{%
\footnotesize
$
Q_F=
\left(
\begin{array}{ccccc|cc|ccccccc}
 p_{1} & p_{2} & p_{3} & p_{4} &
   p_{5} & q_{1} & q_{2} & s_{1} &
   s_{2} & s_{3} & s_{4} & s_{5} &
   s_{6} & s_{7} \\
   \hline
 1 & 1 & 0 & 0 & 0 & -1 & -1 & 0 & 0 & 0 & 0 & 0 & 0 & 0 \\
 0 & 0 & 0 & 1 & 1 & 0 & 0 & -1 & -1 & 0 & 0 & 0 & 0 & 0 \\
 1 & 0 & 0 & 0 & 1 & -1 & 0 & -1 & 0 & 0 & 1 & 0 & -1 & 0 \\
 1 & 0 & 0 & 0 & 1 & 0 & -1 & 0 & 0 & 1 & 0 & 0 & -1 & -1\\
 0 & 1 & 1 & 1 & 0 & 0 & 0 & -1 & 0 & 0 & -1 & -1 & 0 & 0 \\
 0 & 0 & 1 & 0 & 0 & 1 & 0 & 0 & 0 & 0 & -1 & 0 & 0 & -1 
 \end{array}
\right)
$
}
\vspace{0.5cm}

\noindent The D-term charge matrix is

\noindent\makebox[\textwidth]{%
\footnotesize
$
Q_D=
\left(
\begin{array}{ccccc|cc|ccccccc}
 p_{1} & p_{2} & p_{3} & p_{4} &
   p_{5} & q_{1} & q_{2} & s_{1} &
   s_{2} & s_{3} & s_{4} & s_{5} &
   s_{6} & s_{7} \\
   \hline 0 & 0 & 0 & 0 & 0 & 0 & 0 & 1 & -1 & 0 & 0 & 0 & 0 &
   0 \\
 0 & 0 & 0 & 0 & 0 & 0 & 0 & 0 & 1 & -1 & 0 & 0 & 0 &
   0 \\
 0 & 0 & 0 & 0 & 0 & 0 & 0 & 0 & 0 & 1 & -1 & 0 & 0 &
   0 \\
 0 & 0 & 0 & 0 & 0 & 0 & 0 & 0 & 0 & 0 & 1 & -1 & 0 &
   0 \\
 0 & 0 & 0 & 0 & 0 & 0 & 0 & 0 & 0 & 0 & 0 & 0 & 1 &
   -1
\end{array}
\right)
$
}
\vspace{0.5cm}

The total charge matrix $Q_t$ does not have repeated columns. Accordingly, the global symmetry group for the Model 9b theory is $U(1)_{f_1}\times U(1)_{f_2} \times U(1)_R$. The flavour and R-charges on the extremal perfect matchings $p_\alpha$ are the same as for Model 9a, and are summarised in \tref{t9a}. They are found following the discussion in \sref{s1_3}.

Products of non-extremal perfect matchings are expressed as
\beal{esm9b_x1}
q = q_1 q_2 ~,~
s = \prod_{m=1}^{7} s_m~.
\eea
The fugacity counting extremal perfect matchings $p_\alpha$ is $t_\alpha$. The fugacity $y_q$ counts the product of non-extremal perfect matchings $q$ above.

The mesonic Hilbert series for Model 9b is identical to the one for Model 9a. The mesonic Hilbert series is shown in \eref{esm9a_1}. The corresponding plethystic logarithm in \eref{esm9a_3} indicates that the mesonic moduli space is not a complete intersection. As a summary, both Model 9a and 9b mesonic moduli spaces are identical.

The generators of the mesonic moduli space in terms of the perfect matching fields of Model 9b are presented in \tref{f9agen}. The charge lattice of mesonic generators forms a convex polygon which is another reflexive polygon precisely being the dual of the toric diagram. The generators of the mesonic moduli space in terms of quiver fields of Model 9b are shown in \tref{f9bgen2}.

\comment{
\begin{table}[h!]
\centering
\resizebox{\hsize}{!}{
\begin{minipage}[!b]{0.6\textwidth}
\begin{tabular}{|l|c|c|}
\hline
Generator & $U(1)_{f_1}$ & $U(1)_{f_2}$ 
\\
\hline
\hline
$p_{3}^2 p_{4} p_{5} ~ \prod_{m=1}^{7} s_m$
   & 1 & 0
   \nn\\
$p_{1}^2 p_{3} p_{4}^2 ~ q_{1} q_{2} ~\prod_{m=1}^{7} s_m$
   & 0& 1
   \nn\\
$p_{1} p_{2}
   p_{3} p_{4} p_{5} ~q_{1} q_{2}~
   \prod_{m=1}^{7} s_m$
   & 0&0
   \nn\\
$p_{2}^2 p_{3} p_{5}^2 ~
   q_{1} q_{2} ~ \prod_{m=1}^{7} s_m$
   & 0&-1
   \nn\\
$p_{1}^3
   p_{2} p_{4}^2 ~q_{1}^2 q_{2}^2
  ~ \prod_{m=1}^{7} s_m$
   & -1&1
   \nn\\
$p_{1}^2 p_{2}^2 p_{4}
   p_{5} ~ q_{1}^2 q_{2}^2 ~ \prod_{m=1}^{7} s_m$
   & -1&0
   \nn\\
$p_{1} p_{2}^3 p_{5}^2
  ~ q_{1}^2 q_{2}^2 ~ \prod_{m=1}^{7} s_m$
   & -1&-1
   \nn\\
\hline
\end{tabular}
\end{minipage}
\hspace{1cm}
\begin{minipage}[!b]{0.3\textwidth}
\includegraphics[width=4 cm]{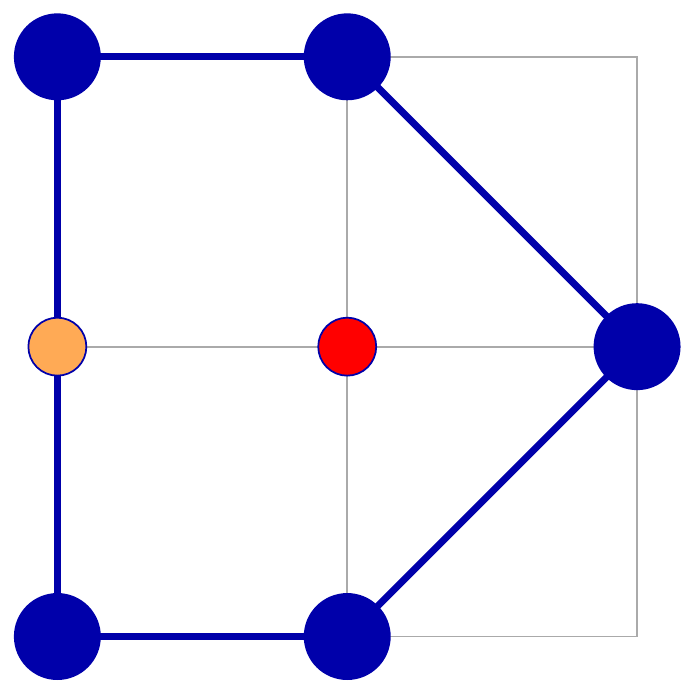}
\end{minipage}
}
\caption{The generators and lattice of generators of the mesonic moduli space of Model 9b in terms of GLSM fields with the corresponding flavor charges.\label{t9bgen}\label{f9bgen}} 
\end{table}
}

\begin{table}[h!]
\centering
\resizebox{\hsize}{!}{
\begin{tabular}{|l|c|c|}
\hline
Generator & $U(1)_{f_1}$ & $U(1)_{f_2}$ 
\\
\hline
\hline
$X_{16} X_{62} X_{21}=  X_{34} X_{45} X_{53}$
   & 1&0
   \nn\\
$X_ {25}^{1} X_{53} X_{32}=  X_{16} X_{62} X_ {25}^{1} X_{51}=  X_{16} X_{64} X_{45} X_{51}$
   & 0&1
   \nn\\
$X_{13} X_{32} X_{21}=  X_ {25}^{1} X_{56} X_{62}=  X_ {25}^{2} X_{53} X_{32}=  X_{45} X_{56} X_{64}$
& 0 & 0
\nn\\
$
=  X_{13} X_{34} X_{45} X_{51}=  X_{16} X_{64} X_{42} X_{21}=  X_{16} X_{62} X_ {25}^{2} X_{51}=  X_ {25}^{1} X_{53} X_{34} X_{42}$
   & &
   \nn\\
$X_ {25}^{2} X_{56} X_{62}=  X_{13} X_{34} X_{42} X_{21}=  X_ {25}^{2} X_{53} X_{34} X_{42}$
   & 0&-1
   \nn\\
$X_{13} X_{32} X_ {25}^{1} X_{51}=  X_{16} X_{64} X_{42} X_ {25}^{1} X_{51}$
   & -1&1
   \nn\\
$ X_{13} X_{32} X_{25}^{2} X_{51}=  X_{25}^{1} X_{56} X_{64} X_{42}=  X_{13} X_{34} X_{42} X_{25}^{1} X_{51}=  X_{16} X_{64} X_{42} X_{25}^{2} X_{51}$
   & -1&0
   \nn\\
$ X_{25}^{2} X_{56} X_{64} X_{42}=  X_{13} X_{34} X_{42} X_{25}^{2} X_{51}$
   & -1&-1
   \nn\\
\hline
\end{tabular}
}
\caption{The generators in terms of bifundamental fields (Model 9b).\label{t9bgen2}\label{f9bgen2}} 
\end{table}

\subsection{Model 9 Phase c}

\begin{figure}[H]
\begin{center}
\includegraphics[trim=0cm 0cm 0cm 0cm,width=4.5 cm]{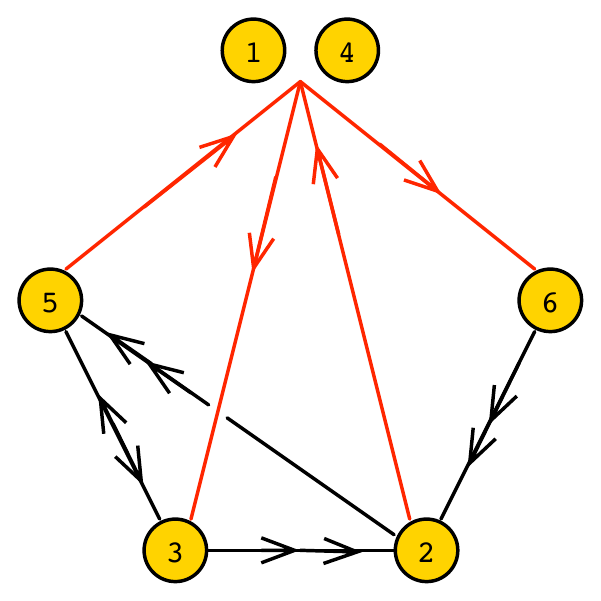}
\includegraphics[width=5 cm]{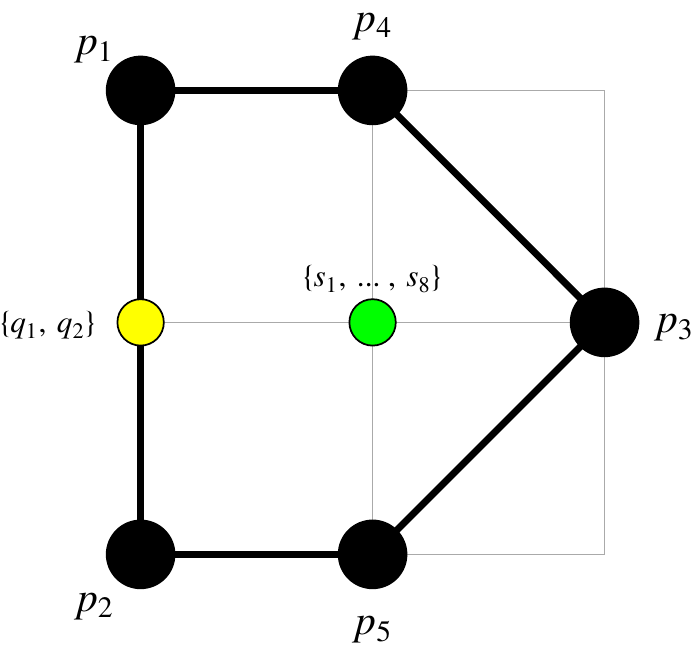}
\includegraphics[width=5 cm]{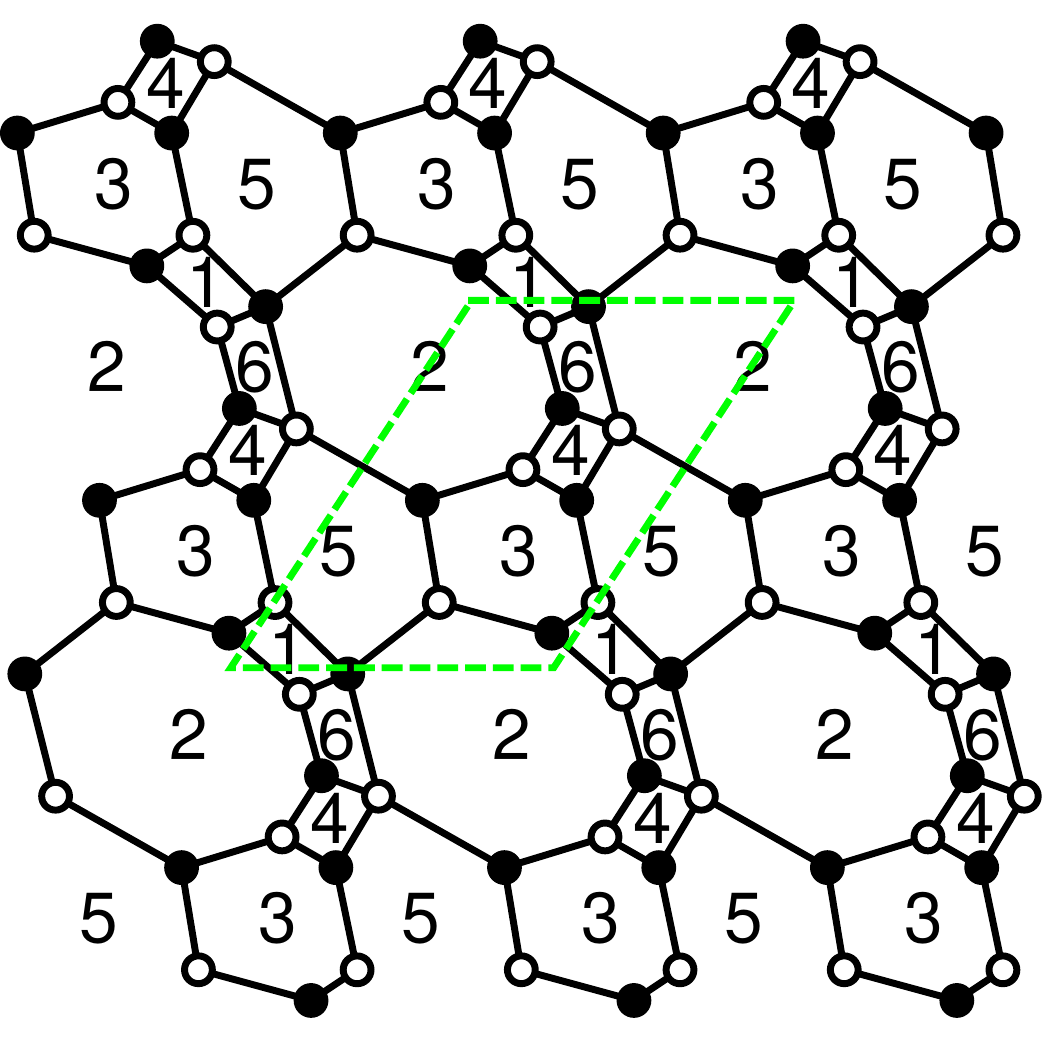}
\caption{The quiver, toric diagram, and brane tiling of Model 9c. The red arrows in the quiver indicate all possible connections between blocks of nodes.}
  \label{f9c}
 \end{center}
 \end{figure}
 
 \noindent The superpotential is 
\beal{esm9c_00}
W&=&
+ X_{21}X_{16}X_{62}^{2} 
+ X_{24}X_{43}X_{32}^{2} 
+ X_{25}^{2} X_{53}X_{32}^{1} 
+ X_{51}X_{13}X_{35} 
+  X_{54}X_{46}X_{62}^{1} X_{25}^{1}
\nn\\
&&
-X_{13} X_{32}^{1} X_{21} 
- X_{24}X_{46}X_{62}^{2} 
- X_{25}^{1} X_{53}X_{32}^{2} 
- X_{54}X_{43} X_{35} 
- X_{16}X_{62}^{1} X_{25}^{2} X_{51} 
\nn\\
  \eea
 
 \noindent The perfect matching matrix is 
 
\noindent\makebox[\textwidth]{%
\footnotesize
$
P=
\left(
\begin{array}{c|ccccc|cc|cccccccc}
 \; & p_{1} & p_{2} & p_{3} & p_{4} & p_{5} & q_{1} &
   q_{2} & s_{1} & s_{2} & s_{3} & s_{4} & s_{5} &
   s_{6} & s_{7} & s_{8} \\
   \hline
 X_{25}^{1} & 1 & 0 & 0 & 1 & 0 & 1 & 0 & 0 & 1 & 0 & 0 & 0 & 0 & 0 & 0 \\
 X_{32}^{1} & 1 & 0 & 0 & 1 & 0 & 0 & 1 & 1 & 0 & 0 & 0 & 0 & 0 & 0 & 1 \\
 X_{25}^{2} & 0 & 1 & 0 & 0 & 1 & 1 & 0 & 0 & 1 & 0 & 0 & 0 & 0 & 0 & 0 \\
 X_{32}^{2} & 0 & 1 & 0 & 0 & 1 & 0 & 1 & 1 & 0 & 0 & 0 & 0 & 0 & 0 & 1 \\
 X_{43} & 1 & 0 & 0 & 0 & 0 & 1 & 0 & 0 & 0 & 1 & 1 & 0 & 1 & 0 & 0 \\
 X_{51} & 1 & 0 & 0 & 0 & 0 & 0 & 1 & 0 & 0 & 0 & 0 & 0 & 1 & 1 & 0 \\
 X_{13} & 0 & 1 & 0 & 0 & 0 & 1 & 0 & 0 & 0 & 1 & 1 & 1 & 0 & 0 & 0 \\
 X_{54} & 0 & 1 & 0 & 0 & 0 & 0 & 1 & 0 & 0 & 0 & 0 & 1 & 0 & 1 & 0 \\
 X_{53} & 0 & 0 & 1 & 0 & 0 & 0 & 0 & 0 & 0 & 1 & 1 & 1 & 1 & 1 & 0 \\
 X_{62}^{1} & 0 & 0 & 1 & 0 & 0 & 0 & 0 & 0 & 0 & 0 & 1 & 0 & 0 & 0 & 1 \\
 X_{62}^{2} & 1 & 1 & 0 & 0 & 0 & 1 & 1 & 0 & 0 & 0 & 1 & 0 & 0 & 0 & 1 \\
 X_{24} & 0 & 0 & 1 & 1 & 0 & 0 & 0 & 0 & 1 & 0 & 0 & 1 & 0 & 1 & 0 \\
 X_{21} & 0 & 0 & 1 & 0 & 1 & 0 & 0 & 0 & 1 & 0 & 0 & 0 & 1 & 1 & 0 \\
 X_{16} & 0 & 0 & 0 & 1 & 0 & 0 & 0 & 1 & 0 & 1 & 0 & 1 & 0 & 0 & 0 \\
 X_{46} & 0 & 0 & 0 & 0 & 1 & 0 & 0 & 1 & 0 & 1 & 0 & 0 & 1 & 0 & 0 \\
 X_{35} & 0 & 0 & 1 & 1 & 1 & 0 & 0 & 1 & 1 & 0 & 0 & 0 & 0 & 0 & 1
\end{array}
\right)
$
}
\vspace{0.5cm}

 \noindent The F-term charge matrix $Q_F=\ker{(P)}$ is

\noindent\makebox[\textwidth]{%
\footnotesize
$
Q_F=
\left(
\begin{array}{ccccc|cc|cccccccc}
 p_{1} & p_{2} & p_{3} & p_{4} & p_{5} & q_{1} &
   q_{2} & s_{1} & s_{2} & s_{3} & s_{4} & s_{5} &
   s_{6} & s_{7} & s_{8} \\
   \hline
 1 & 1 & 0 & 0 & 0 & -1 & -1 & 0 & 0 & 0 & 0 & 0 & 0 & 0 & 0 \\
 0 & 0 & 0 & 1 & 1 & 0 & 0 & -1 & -1 & 0 & 0 & 0 & 0 & 0 & 0 \\
  1 & 0 & 0 & 0 & 1 & -1 & 0 & -1 & 0 & 0 & 0 & 1 & 0 & -1 & 0 \\
 1 & 0 & 0 & 0 & 1 & -1 & 0 & -1 & 0 & 1 & 0 & 0 & -1 & 0 & 0 \\
 0 & 1 & 0 & 1 & 0 & -1 & 0 & -1 & 0 & 1 & 0 & -1 & 0 & 0 & 0 \\
 0 & 0 & 1 & 0 & 0 & 1 & 0 & 0 & -1 & 0 & -1 & 0 & 0 & 0 & 0 \\
 0 & 0 & 0 & 0 & 0 & 0 & 0 & -1 & 0 & 1 & -1 & 0 & 0 & 0 & 1
\end{array}
\right)
$
}
\vspace{0.5cm}

\noindent The D-term charge matrix is

\noindent\makebox[\textwidth]{%
\footnotesize
$
Q_D=
\left(
\begin{array}{ccccc|cc|cccccccc}
 p_{1} & p_{2} & p_{3} & p_{4} & p_{5} & q_{1} &
   q_{2} & s_{1} & s_{2} & s_{3} & s_{4} & s_{5} &
   s_{6} & s_{7} & s_{8} \\
   \hline
 0 & 0 & 0 & 0 & 0 & 0 & 0 & 0 & 1 & -1 & 0 & 0 & 0 & 0 & 0 \\
 0 & 0 & 0 & 0 & 0 & 0 & 0 & 0 & 0 & 1 & -1 & 0 & 0 & 0 & 0 \\
 0 & 0 & 0 & 0 & 0 & 0 & 0 & 0 & 0 & 0 & 1 & -1 & 0 & 0 & 0 \\
 0 & 0 & 0 & 0 & 0 & 0 & 0 & 0 & 0 & 0 & 0 & 1 & -1 & 0 & 0 \\
 0 & 0 & 0 & 0 & 0 & 0 & 0 & 0 & 0 & 0 & 0 & 0 & 0 & 1 & -1
\end{array}
\right)
$
}
\vspace{0.5cm}

The total charge matrix $Q_t$ does not have repeated columns. Accordingly, the global symmetry of Model 9c is the same as for Model 9a and 9b above and takes the form $U(1)_{f_1} \times U(1)_{f_2} \times U(1)_R$. The mesonic charges on the extremal perfect matchings are summarised in \tref{t9a}.

The following products of non-extremal perfect matchings are assigned single variables
\beal{esm9c_x1}
q = q_1 q_2 ~,~
s = \prod_{m=1}^{8} s_m~.
\eea
The extremal perfect matchings are counted by the fugacity $t_\alpha$. Products of non-extremal perfect matchings such as $q$ above are associated to fugacities of the form $y_q$.

The mesonic Hilbert series is identical to the mesonic Hilbert series of Model 9a and 9b. The mesonic Hilbert series is given in \eref{esm9a_1} with the corresponding plethystic logarithm in \eref{esm9a_3}. The mesonic Hilbert series of Models 9a, 9b and 9c are identical and are not complete intersections.

The generators of the mesonic moduli space in terms of Model 9c GLSM fields are shown in \tref{t9agen}. The mesonic charges of the generators correspond to lattice coordinates of points which form a reflexive polygon being the dual of the toric diagram. The generators in terms of quiver fields of Model 9c are shown in \tref{t9cgen2}.

\comment{
\begin{table}[h!]
\centering
\resizebox{\hsize}{!}{
\begin{minipage}[!b]{0.6\textwidth}
\begin{tabular}{|l|c|c|}
\hline
Generator & $U(1)_{f_1}$ & $U(1)_{f_2}$ 
\\
\hline
\hline
$p_{1} p_{2} p_{5}^2 ~ \prod_{m=1}^{8} s_m$
   & 1&0
   \nn\\
$p_{1}^2 p_{3}^2 p_{5} ~
   q_{1} q_{2} ~ \prod_{m=1}^{8} s_m$
   & 0&1
   \nn\\
$p_{1} p_{2} p_{3} p_{4} p_{5} ~ q_{1}
   q_{2} ~ \prod_{m=1}^{8} s_m$
   & 0&0
   \nn\\
$p_{2}^2 p_{4}^2 p_{5} ~ q_{1} q_{2} ~ \prod_{m=1}^{8} s_m$
   & 0&-1
   \nn\\
$p_{1}^2 p_{3}^3 p_{4} ~ q_{1}^2 q_{2}^2 ~ \prod_{m=1}^{8} s_m$
   & -1&1
   \nn\\
$p_{1}
   p_{2} p_{3}^2 p_{4}^2 ~ q_{1}^2 q_{2}^2 ~ \prod_{m=1}^{8} s_m$
   & -1&0
   \nn\\
$p_{2}^2
   p_{3} p_{4}^3 ~ q_{1}^2 q_{2}^2 ~ \prod_{m=1}^{8} s_m$
   & -1&-1
   \nn\\
   \hline
\end{tabular}
\end{minipage}
\hspace{1cm}
\begin{minipage}[!b]{0.3\textwidth}
\includegraphics[width=4 cm]{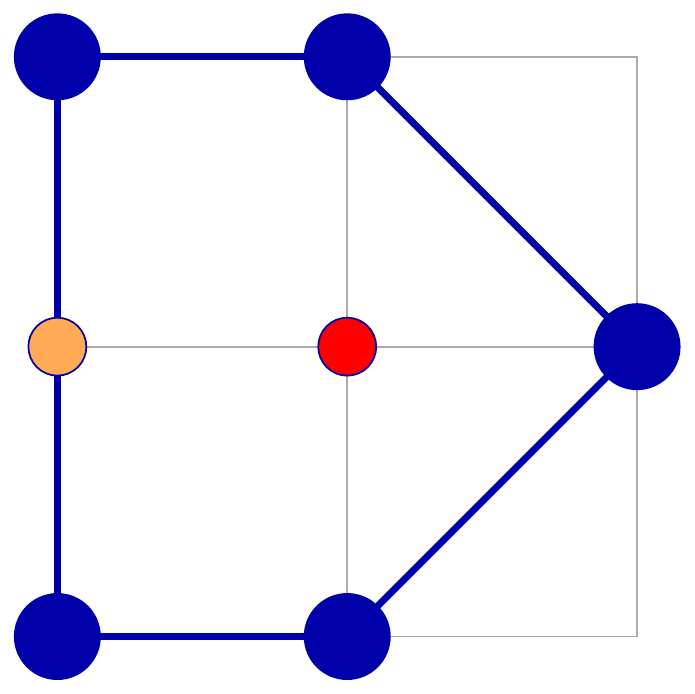}
\end{minipage}
}
\caption{The generators and lattice of generators of the mesonic moduli space of Model 9c in terms of GLSM fields with the corresponding flavor charges. \label{t9cgen} \label{f9cgen}}
\end{table}
}

\begin{table}[h!]
\centering
\resizebox{\hsize}{!}{
\begin{tabular}{|l|c|c|}
\hline
Generator & $U(1)_{f_1}$ & $U(1)_{f_2}$ 
\\
\hline
\hline
$X_{35} X_{53}=  X_{16} X_ {62}^{1} X_{21}=  X_{24} X_{46} X_ {62}^{1}$
   & 1&0
   \nn\\
$X_{16} X_ {62}^{1} X_ {25}^{1} X_{51}=  X_{24} X_{43} X_ {32}^{1}=  X_ {25}^{1} X_{53} X_ {32}^{1}$
   & 0&1
   \nn\\
$ X_{16} X_{62}^{1} X_{25}^{2} X_{51}=  X_{25}^{1} X_{54} X_{46} X_{62}^{1}=  X_{13} X_{32}^{1} X_{21}=  X_{13} X_{35} X_{51}= 
$
& 0 & 0
\nn\\
$
 X_{16} X_{62}^{2} X_{21}=  X_{24} X_{43} X_{32}^{2}=  X_{24} X_{46} X_{62}^{2}=  X_{25}^{1} X_{53} X_{32}^{2}=  X_{25}^{2} X_{53} X_{32}^{1}=  X_{35} X_{54} X_{43}$
   & &
   \nn\\
$ X_{25}^{2} X_{54} X_{46} X_{62}^{1}=  X_{13} X_{32}^{2} X_{21}=  X_{25}^{2} X_{53} X_{32}^{2}$
   & 0&-1
   \nn\\
$X_{13} X_ {32}^{1} X_ {25}^{1} X_{51}=  X_{16} X_ {62}^{2} X_{25}^{1} X_{51}=  X_ {25}^{1} X_{54} X_{43} X_ {32}^{1}$
   &-1&1
   \nn\\
$ X_{13} X_{32}^{2} X_{25}^{1} X_{51}=  X_{13} X_{32}^{1} X_{25}^{2} X_{51}=  X_{16} X_{62}^{2} X_{25}^{2} X_{51}=  X_{25}^{1} X_{54} X_{43} X_{32}^{2}=  X_{25}^{1} X_{54} X_{46} X_{62}^{2}=  X_{25}^{2} X_{54} X_{43} X_{32}^{1}$
   &-1&0
   \nn\\
$ X_{13} X_{32}^{2} X_{25}^{2} X_{51}=  X_{25}^{2} X_{54} X_{43} X_{32}^{2}=  X_{25}^{2} X_{54} X_{46} X_{62}^{2}$
   & -1&-1
   \nn\\
   \hline
\end{tabular}
}
\caption{The generators in terms of bifundamental fields (Model 9c). \label{t9cgen2} \label{f9cgen2}}
\end{table}

\section{Model 10: $\text{dP}_3$}
\subsection{Model 10 Phase a}

\begin{figure}[H]
\begin{center}
\includegraphics[trim=0cm 0cm 0cm 0cm,width=4 cm]{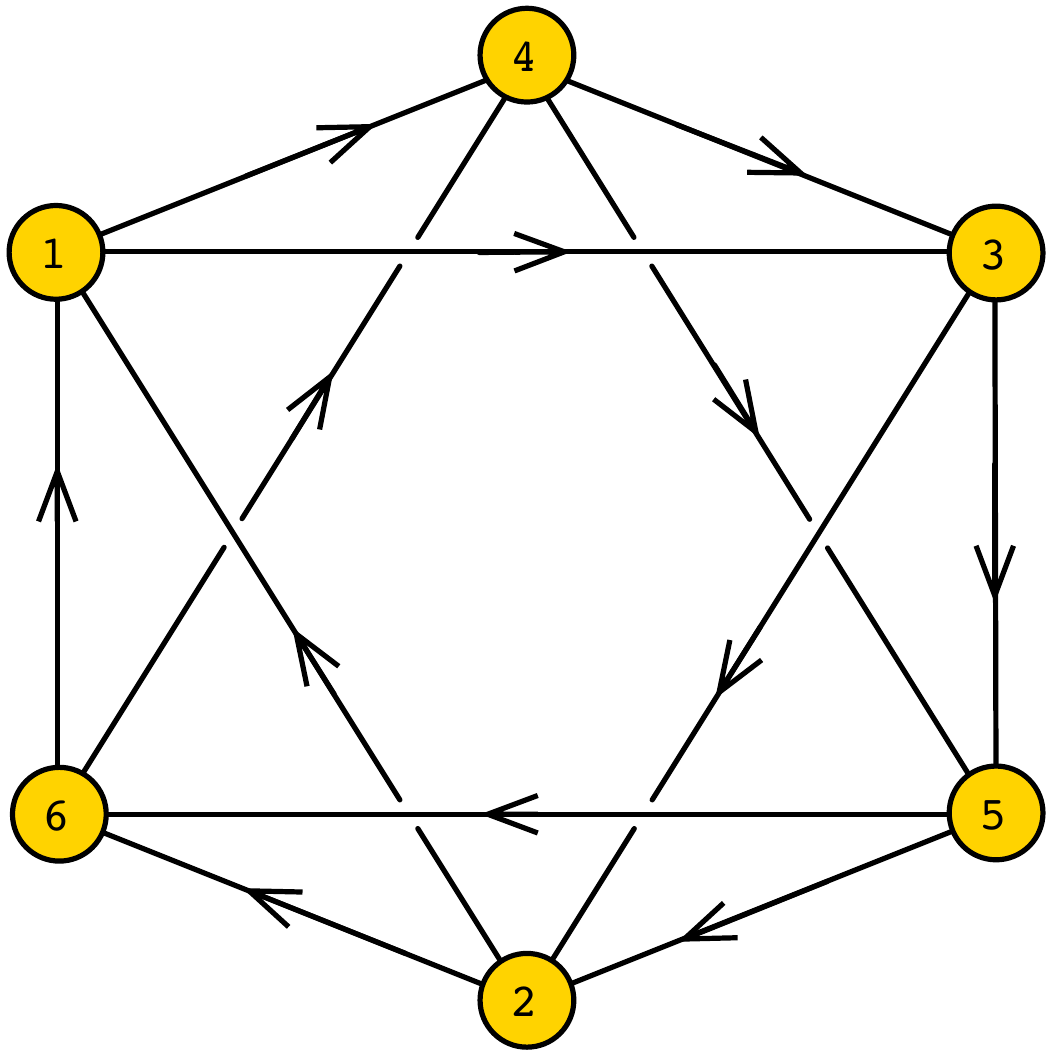}
\includegraphics[width=5 cm]{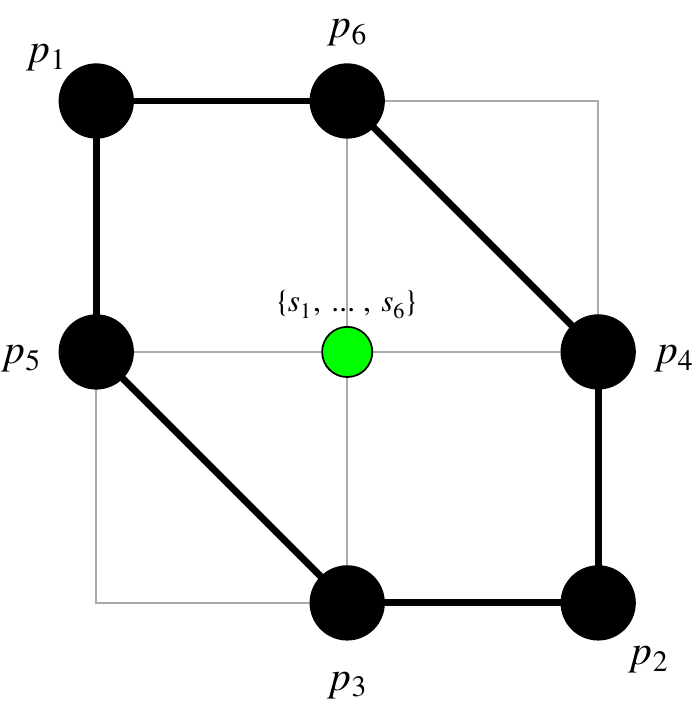}
\includegraphics[width=5 cm]{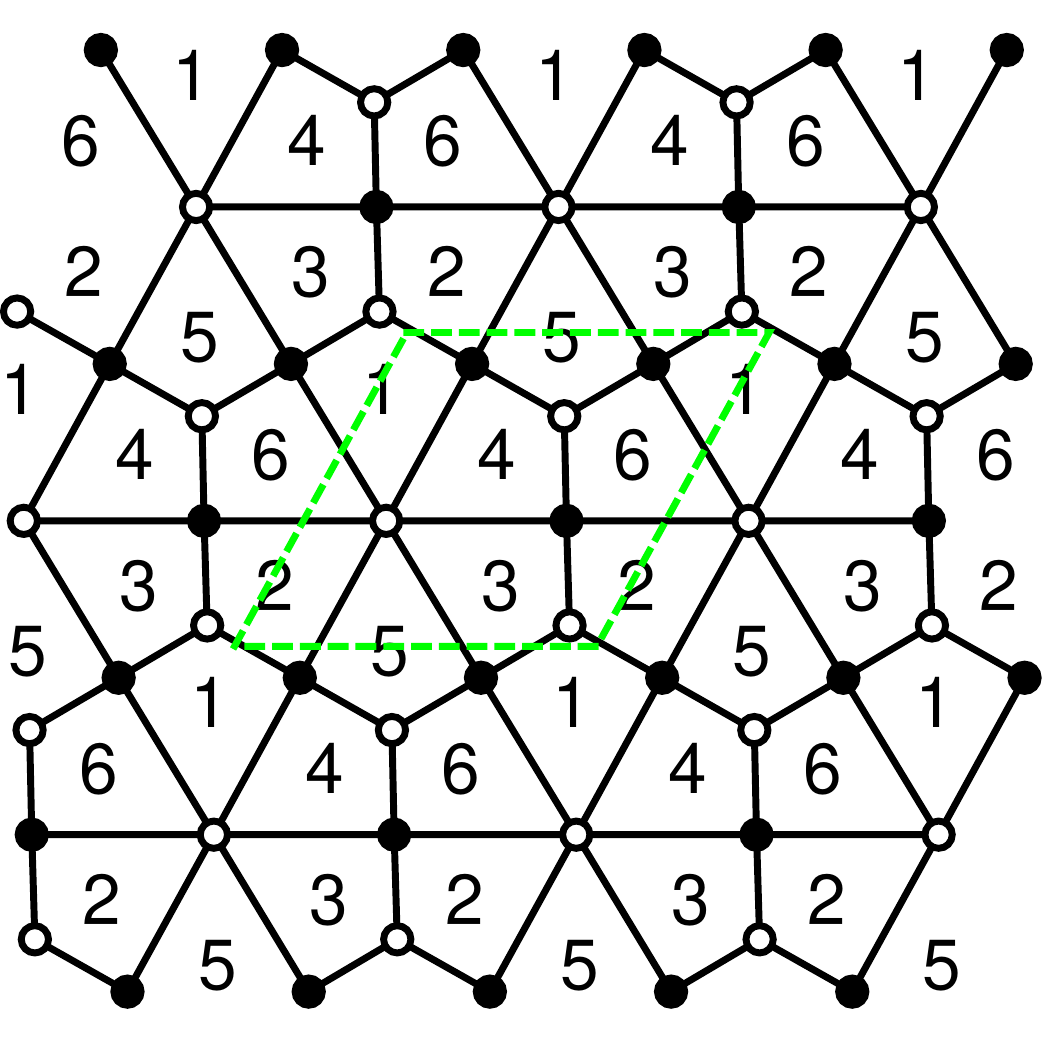}
\caption{The quiver, toric diagram, and brane tiling of Model 10a.}
  \label{f10a}
 \end{center}
 \end{figure}
 
 \noindent The superpotential is 
\beal{esm10a_00}
W&=&
+ X_{13} X_{32} X_{21}  
+ X_{56} X_{64} X_{45}  
+ X_{43} X_{35} X_{52} X_{26} X_{61} X_{14} 
\nn\\
&&
- X_{13} X_{35} X_{56} X_{61}  
- X_{14} X_{45} X_{52} X_{21}  
- X_{26} X_{64} X_{43} X_{32}  
  \eea
 
 \noindent The perfect matching matrix is 
 
\noindent\makebox[\textwidth]{%
\footnotesize
$
P=
\left(
\begin{array}{c|cccccc|cccccc}
 \; & p_{1} & p_{2} & p_{3} & p_{4} & p_{5} & p_{6} & s_{1}
   & s_{2} & s_{3} & s_{4} & s_{5} & s_{6} \\
   \hline
 X_{45} & 1 & 0 & 0 & 0 & 1 & 0 & 1 & 0 & 0 & 0 & 1 & 0 \\
 X_{13} & 1 & 0 & 0 & 0 & 0 & 1 & 1 & 0 & 0 & 0 & 0 & 1 \\
 X_{56} & 0 & 1 & 1 & 0 & 0 & 0 & 0 & 1 & 1 & 0 & 0 & 0 \\
 X_{21} & 0 & 1 & 0 & 1 & 0 & 0 & 0 & 1 & 0 & 1 & 0 & 0 \\
 X_{32} & 0 & 0 & 1 & 0 & 1 & 0 & 0 & 0 & 1 & 0 & 1 & 0 \\
 X_{64} & 0 & 0 & 0 & 1 & 0 & 1 & 0 & 0 & 0 & 1 & 0 & 1 \\
 X_{26} & 1 & 0 & 0 & 0 & 0 & 0 & 0 & 1 & 0 & 0 & 0 & 0 \\
 X_{43} & 0 & 1 & 0 & 0 & 0 & 0 & 1 & 0 & 0 & 0 & 0 & 0 \\
 X_{14} & 0 & 0 & 1 & 0 & 0 & 0 & 0 & 0 & 0 & 0 & 0 & 1 \\
 X_{35} & 0 & 0 & 0 & 1 & 0 & 0 & 0 & 0 & 0 & 0 & 1 & 0 \\
 X_{61} & 0 & 0 & 0 & 0 & 1 & 0 & 0 & 0 & 0 & 1 & 0 & 0 \\
 X_{52} & 0 & 0 & 0 & 0 & 0 & 1 & 0 & 0 & 1 & 0 & 0 & 0
\end{array}
\right)
$
}
\vspace{0.5cm}

 \noindent The F-term charge matrix $Q_F=\ker{(P)}$ is

\noindent\makebox[\textwidth]{%
\footnotesize
$
Q_F=
\left(
\begin{array}{cccccc|cccccc}
 p_{1} & p_{2} & p_{3} & p_{4} & p_{5} & p_{6} & s_{1}
   & s_{2} & s_{3} & s_{4} & s_{5} & s_{6} \\
   \hline
    1 & 1 & 0 & 0 & 0 & 0 & -1 & -1 & 0 & 0 & 0 & 0 \\
     0 & 0 & 0 & 1 & 1 & 0 & 0 & 0 & 0 & -1 & -1 & 0 \\
 0 & 1 & 0 & 0 & 1 & 1 & -1 & 0 & -1 & -1 & 0 & 0 \\
     0 & 0 & 1 & 0 & 0 & 1 & 0 & 0 & -1 & 0 & 0 & -1 
\end{array}
\right)
$
}
\vspace{0.5cm}

\noindent The D-term charge matrix is

\noindent\makebox[\textwidth]{%
\footnotesize
$
Q_D=
\left(
\begin{array}{cccccc|cccccc}
 p_{1} & p_{2} & p_{3} & p_{4} & p_{5} & p_{6} & s_{1}
   & s_{2} & s_{3} & s_{4} & s_{5} & s_{6} \\
   \hline
 0 & 0 & 0 & 0 & 0 & 0 & 1 & -1 & 0 & 0 & 0 & 0 \\
 0 & 0 & 0 & 0 & 0 & 0 & 0 & 1 & -1 & 0 & 0 & 0 \\
 0 & 0 & 0 & 0 & 0 & 0 & 0 & 0 & 1 & -1 & 0 & 0 \\
 0 & 0 & 0 & 0 & 0 & 0 & 0 & 0 & 0 & 1 & -1 & 0 \\
 0 & 0 & 0 & 0 & 0 & 0 & 0 & 0 & 0 & 0 & 1 & -1
\end{array}
\right)
$
}
\vspace{0.5cm}

The total charge matrix $Q_t$ does not exhibit repeated columns. Accordingly, the global symmetry is $U(1)_{f_1} \times U(1)_{f_2} \times U(1)_R$. The mesonic charges on the GLSM fields corresponding to extremal points in the toric diagram in \fref{f10a} are found following the discussion in \sref{s1_3}. They are presented in \tref{t10a}.

\begin{table}[H]
\centering
\begin{tabular}{|c||c|c|c||l|} 
\hline
\; & $U(1)_{f_1}$ & $U(1)_{f_2}$ & $U(1)_R$ & fugacity \\
\hline
\hline
$p_1$ &-1 & 0 & 1/3 &  $t_1$\\
$p_2$ &-1 & 1 & 1/3 &  $t_2$\\
$p_3$ & 1 & 0 & 1/3 &  $t_3$\\
$p_4$ & 1 &-1 & 1/3 &  $t_4$\\ 
$p_5$ & 0 & 0 & 1/3 &  $t_5$\\ 
$p_6$ & 0 & 0 & 1/3 &  $t_6$\\ 
\hline
\end{tabular}
\caption{The GLSM fields corresponding to extremal points of the toric diagram with their mesonic charges (Model 10a).\label{t10a}}
\end{table}

The product of all internal perfect matchings is labelled as follows
\beal{esm10a_x1}
s = \prod_{m=1}^6 s_m~.
\eea
The fugacity counting extremal perfect matchings is $t_\alpha$. The product of internal perfect matchings is associated to the fugacity $y_s$.

The refined mesonic Hilbert series of Model 10a is found using the Molien integral formula in \eref{es12_2}. It is
 \beal{esm10a_1}
&&g_{1}(t_\alpha,y_{s}; \mathcal{M}^{mes}_{10a})=
\frac{
P(t_\alpha)
}{
(1 - y_{s} ~ t_2^2 t_3^2 t_4 t_5) 
(1 - y_{s} ~ t_1 t_2 t_3^2 t_5^2) 
(1 - y_{s} ~ t_2^2 t_3 t_4^2 t_6) 
}
\nn\\
&&
\hspace{1cm}
\times
\frac{1}{
(1 - y_{s} ~ t_1^2 t_3 t_5^2 t_6) 
(1 - y_{s} ~ t_1 t_2 t_4^2 t_6^2) 
(1 - y_{s} ~ t_1^2 t_4 t_5 t_6^2)
}~~.
\nn\\
\eea
The numerator is given by the polynomial
\beal{esm10a2_1}
P(t_\alpha)&=&
1 
+ y_{s} ~ t_1 t_2 t_3 t_4 t_5 t_6 
- y_{s}^2 ~ t_1 t_2^3 t_3^3 t_4^2 t_5^2 t_6 
- y_{s}^2 ~ t_1^2 t_2^2 t_3^3 t_4 t_5^3 t_6 
- y_{s}^2 ~ t_1 t_2^3 t_3^2 t_4^3 t_5 t_6^2 
\nn\\
&&
- 2~y_{s}^2 ~ t_1^2 t_2^2 t_3^2 t_4^2 t_5^2 t_6^2 
- y_{s}^2 ~t_1^3 t_2 t_3^2 t_4 t_5^3 t_6^2 
+ y_{s}^3 ~t_1^2 t_2^4 t_3^4 t_4^3 t_5^3 t_6^2 
+ y_{s}^3 ~t_1^3 t_2^3 t_3^4 t_4^2 t_5^4 t_6^2 
\nn\\
&&
- y_{s}^2 ~t_1^2 t_2^2 t_3 t_4^3 t_5 t_6^3 
- y_{s}^2 ~t_1^3 t_2 t_3 t_4^2 t_5^2 t_6^3 
+ y_{s}^3 ~t_1^2 t_2^4 t_3^3 t_4^4 t_5^2 t_6^3 
+ 2~y_{s}^3 ~ t_1^3 t_2^3 t_3^3 t_4^3 t_5^3 t_6^3 
\nn\\
&&
+ y_{s}^3 ~t_1^4 t_2^2 t_3^3 t_4^2 t_5^4 t_6^3 
+ y_{s}^3 ~t_1^3 t_2^3 t_3^2 t_4^4 t_5^2 t_6^4 
+ y_{s}^3 ~t_1^4 t_2^2 t_3^2 t_4^3 t_5^3 t_6^4 
- y_{s}^4 ~t_1^4 t_2^4 t_3^4 t_4^4 t_5^4 t_6^4 
\nn\\
&&
- y_{s}^5 ~t_1^5 t_2^5 t_3^5 t_4^5 t_5^5 t_6^5
~~.
\eea
 The plethystic logarithm of the mesonic Hilbert series is
\beal{esm10a_3}
&&
PL[g_1(t_\alpha,y_{s};\mathcal{M}_{10a}^{mes})]=
y_{s} ~t_1 t_2 t_3 t_4 t_5 t_6 
+ y_{s} ~t_1^2 t_3 t_5^2 t_6 
+ y_{s} ~t_2^2 t_3 t_4^2 t_6 
+ y_{s} ~t_1 t_2 t_4^2 t_6^2 
+ y_{s} ~t_1 t_2 t_3^2 t_5^2 
\nn\\
&&
\hspace{1cm}
+ y_{s} ~t_1^2 t_4 t_5 t_6^2 
+ y_{s} ~t_2^2 t_3^2 t_4 t_5 
- 3 ~y_{s}^2~t_1^2 t_2^2 t_3^2 t_4^2 t_5^2 t_6^2 
- y_{s}^2 ~t_1^3 t_2 t_3^2 t_4 t_5^3 t_6^2 
- y_{s}^2 ~t_1 t_2^3 t_3^2 t_4^3 t_5 t_6^2 
 \nn\\
 &&
  \hspace{1cm}
- y_{s}^2 ~t_1^2 t_2^2 t_3 t_4^3 t_5 t_6^3 
- y_{s}^2 ~t_1^2 t_2^2 t_3^3 t_4 t_5^3 t_6 
- y_{s}^2 ~t_1^3 t_2 t_3 t_4^2 t_5^2 t_6^3  
- y_{s}^2 ~t_1 t_2^3 t_3^3 t_4^2 t_5^2 t_6 
 +\dots~.
  \nn\\
\eea

\begin{figure}[H]
\centering
\resizebox{\hsize}{!}{
\begin{minipage}[!b]{0.6\textwidth}
\begin{tabular}{|l|c|c|}
\hline
Generator & $U(1)_{f_1}$ & $U(1)_{f_2}$ 
\\
\hline
\hline
$p_{2}^2 p_{3}^2 p_{4} p_{5} ~ s$
& 1 & 1
\\
$p_{1} p_{2} p_{3}^2 p_{5}^2 ~ s$
& 0 & 1
\\
$p_{2}^2 p_{3} p_{4}^2 p_{6} ~ s$
& 1 & 0
\\
$p_{1} p_{2} p_{3} p_{4} p_{5} p_{6} ~ s$
& 0 & 0
\\
$p_{1}^2 p_{3} p_{5}^2 p_{6} ~ s$
& -1 & 0
\\
$p_{1} p_{2} p_{4}^2 p_{6}^2 ~ s$
& 0 & -1
\\
$p_{1}^2 p_{4} p_{5} p_{6}^2 ~ s$
& -1 & -1 
   \\
   \hline
\end{tabular}
\end{minipage}
\hspace{1cm}
\begin{minipage}[!b]{0.3\textwidth}
\includegraphics[width=4 cm]{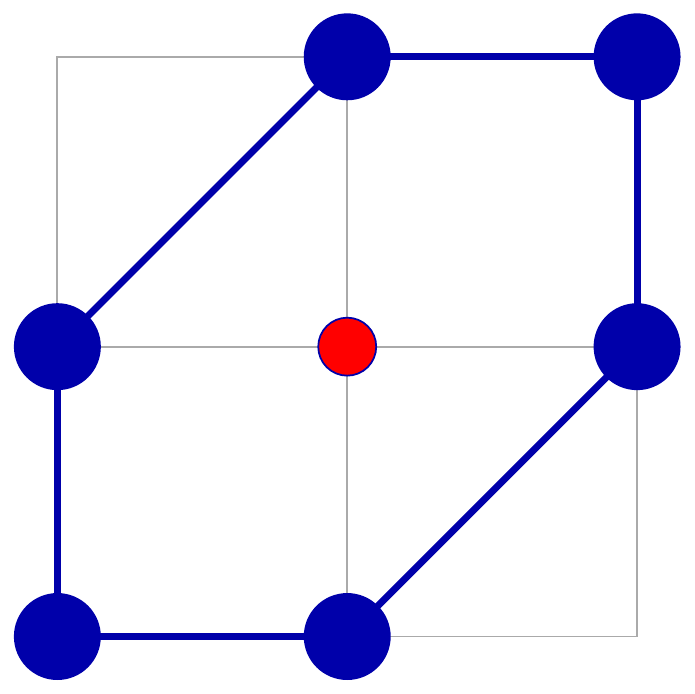}
\end{minipage}
}
\caption{The generators and lattice of generators of the mesonic moduli space of Model 10a in terms of GLSM fields with the corresponding flavor charges.\label{t10agen}\label{f10agen}} 
\end{figure}

Under the following fugacity map
\beal{esm10a_y1}
f_1 = \frac{t_2 t_4}{t_1 t_5}
~,~
f_2 = \frac{t_3 t_5}{t_4 t_6}
~,~
t = y_s^{1/6} ~ t_1^{1/6} t_2^{1/6} t_3^{1/6} t_4^{1/6} t_5^{1/6} t_6^{1/6}
~,~
\eea
where $f_1$, $f_2$ and $t$ are the mesonic charge fugacities, the mesonic Hilbert series and the plethystic logarithm are expressed as
\beal{esm10a_3i}
&&
g_1(t,f_1,f_2;\mathcal{M}_{10a}^{mes})=
\bigg(1 + t^6 
- \Big(2 + \frac{1}{f_1} + f_1 + \frac{1}{f_2} + \frac{1}{f_1 f_2} + f_2 + f_1 f_2
\Big) ~t^{12} 
\nn\\
&&
\hspace{0.5cm}
+ \Big(2 + \frac{1}{f_1} + f_1 + \frac{1}{f_2} + \frac{1}{f_1 f_2} + f_2 + f_1 f_2
\Big)~t^{18} 
- t^{24} - t^{30}
\bigg)
\times
\nn\\
&&
\hspace{0.5cm}
\frac{
1
}{
\left(1 - \frac{1}{f_1} t^6\right) 
(1 - f_1 t^6) 
\left(1 - \frac{1}{f_2} t^6\right) 
\left(1 - \frac{1}{f_1 f_2} t^6\right) 
(1 - f_2 t^6) 
(1 - f_1 f_2 t^6)
}
\eea
and
\beal{esm10a_3}
&&
PL[g_1(t,f_1,f_2;\mathcal{M}_{10a}^{mes})]=
\Big(
1
+ \frac{1}{f_1}
+ f_1
+ \frac{1}{f_2}
+ f_2
+ \frac{1}{f_1 f_2}
+ f_1 f_2
\Big) t^6
- \Big(
3
+\frac{1}{f_{1}}
+f_{1}
+\frac{1}{f_{2}}
   \nn\\
   &&
   \hspace{0.5cm}
+f_{2}
+\frac{1}{f_1 f_2}
+f_1 f_2
\Big) t^{12}
+2\Big(
2
+\frac{1}{f_{1}}
+ f_{1}
+\frac{1}{f_{2}}
+ f_{2}
+\frac{1}{f_{1} f_2}
+ f_1 f_2
\Big) t^{18}
   +\dots~.
   \eea
The above plethystic logarithm exhibits both the moduli space generators and the corresponding mesonic charges. They are summarized in \tref{t10agen}. The generators can be presented on a charge lattice. The convex polygon formed by the generators in \tref{t10agen} is the dual reflexive polygon of the toric diagram of Model 10a.\\

\begin{figure}[H]
\centering
\resizebox{\hsize}{!}{
\begin{tabular}{|l|c|c|}
\hline
Generator & $U(1)_{f_1}$ & $U(1)_{f_2}$ 
\\
\hline
\hline
$X_{14} X_{43} X_{32} X_{21}=  X_{14} X_{43} X_{35} X_{56} X_{61}$
& 1 & 1
\\
$X_{14} X_{45} X_{56} X_{61}=  X_{14} X_{43} X_{32} X_{26} X_{61}$
& 0 & 1
\\
$X_{35} X_{56} X_{64} X_{43}=  X_{14} X_{43} X_{35} X_{52} X_{21}$
& 1 & 0
\\
$X_{14} X_{43} X_{35} X_{52} X_{26} X_{61}=  X_{13} X_{32} X_{21}=  X_{45} X_{56} X_{64}=  X_{13} X_{35} X_{56} X_{61}=  X_{14} X_{45} X_{52} X_{21}=  X_{26} X_{64} X_{43} X_{32}$
& 0 & 0
\\
$X_{13} X_{32} X_{26} X_{61}=  X_{14} X_{45} X_{52} X_{26} X_{61}$
& -1 & 0
\\
$X_{13} X_{35} X_{52} X_{21}=  X_{26} X_{64} X_{43} X_{35} X_{52}$
& 0 & -1
\\
$X_{26} X_{64} X_{45} X_{52}=  X_{13} X_{35} X_{52} X_{26} X_{61}$
& -1 & -1 
   \\
   \hline
\end{tabular}
}
\caption{The generators in terms of bifundamental fields (Model 10a).\label{t10agen2}\label{f10agen2}} 
\end{figure}

Under the following fugacity map
\beal{esm10a_x1}
T_1 = \frac{t^6}{f_1 f_2} = y_{s} ~ t_1^2 t_4 t_5 t_6^2 ~,~
T_2 = f_1 = \frac{t_2 t_4}{t_1 t_5}~,~
T_3 = f_2 = \frac{t_3 t_5}{t_4 t_6}~,
\eea
the mesonic Hilbert series and the plethystic logarithm can be rewritten as
\beal{esm10_x2}
&&
g_1(T_1,T_2,T_3;\mathcal{M}_{10a}^{mes})
=
\big(
1 
+ T_1 T_2 T_3
- (
2 T_1^2 T_2^2 T_3^2
+ T_1^2 T_2 T_3^2
+ T_1^2 T_2^3 T_3^2
+ T_1^2 T_2^2 T_3
\nn\\
&&
\hspace{0.5cm}
+ T_1^2 T_2 T_3
+ T_1^2 T_2^2 T_3^3
+ T_1^2 T_2^3 T_3^3
)
+
(
2 T_1^3 T_2^3 T_3^3
+ T_1^3 T_2^2 T_3^3
+ T_1^3 T_2^4 T_3^3
+ T_1^3 T_2^3 T_3^2
\nn\\
&&
\hspace{0.5cm}
+ T_1^3 T_2^2 T_3^2
+ T_1^3 T_2^3 T_3^4
+ T_1^3 T_2^4 T_3^4
)
- T_1^4 T_2^4 T_3^4
- T_1^5 T_2^5 T_3^5
\big)
\times
\nn\\
&&
\hspace{0.5cm}
\frac{
1
}{
(1- T_1 T_3)
(1- T_1 T_2^2 T_3)
(1- T_1 T_2)
(1- T_1)
(1- T_1 T_2 T_3^2)
(1- T_1 T_2^2 T_3^2)
}
\eea
and
\beal{esm10_x3}
&&
PL[g_1(t,f_1,f_2;\mathcal{M}_{10a}^{mes})]=
T_1 T_2 T_3
+ T_1 T_3
+ T_1 T_2^2 T_3
+ T_1 T_2
+ T_1 T_2 T_3^2
+ T_1
+ T_1 T_2^2 T_3^2
\nn\\
&&
\hspace{0.5cm}
- (
3 T_1^2 T_2^2 T_3^2
+ T_1^2 T_2 T_3^2
+ T_1^2 T_2^3 T_3^2
+ T_1^2 T_2^2 T_3
+ T_1^2 T_2^2 T_3^3
+ T_1^2 T_2 T_3
+ T_1^2 T_2^3 T_3^3
)
\nn\\
&&
\hspace{0.5cm}
+
4 T_1^3 T_2^3 T_3^3
+ T_1^3 T_2^2 T_3^3
+ T_1^3 T_2^4 T_3^3
+ T_1^3 T_2^3 T_3^2
+ T_1^3 T_2^3 T_3^4
+ T_1^3 T_2^2 T_3^2
+ T_1^3 T_2^4 T_3^4
+\dots
\nn\\
\eea
such that the powers of the fugacities are all positive indicating the cone structure of the variety.
\\

\subsection{Model 10 Phase b}

\begin{figure}[H]
\begin{center}
\includegraphics[trim=0cm 0cm 0cm 0cm,width=4.5 cm]{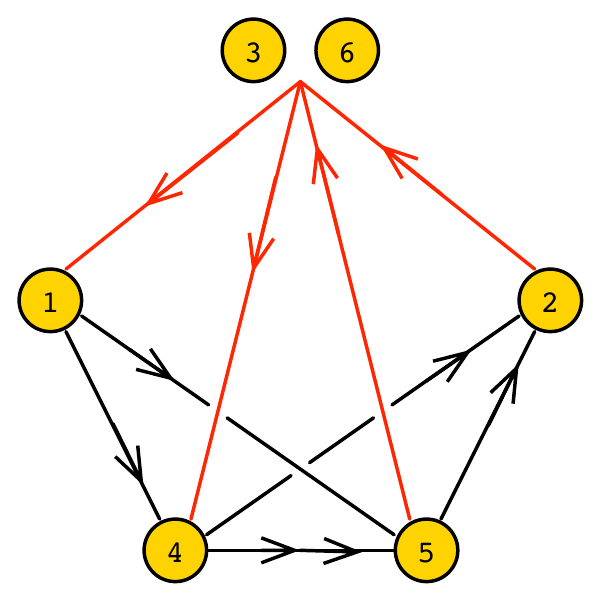}
\includegraphics[width=5 cm]{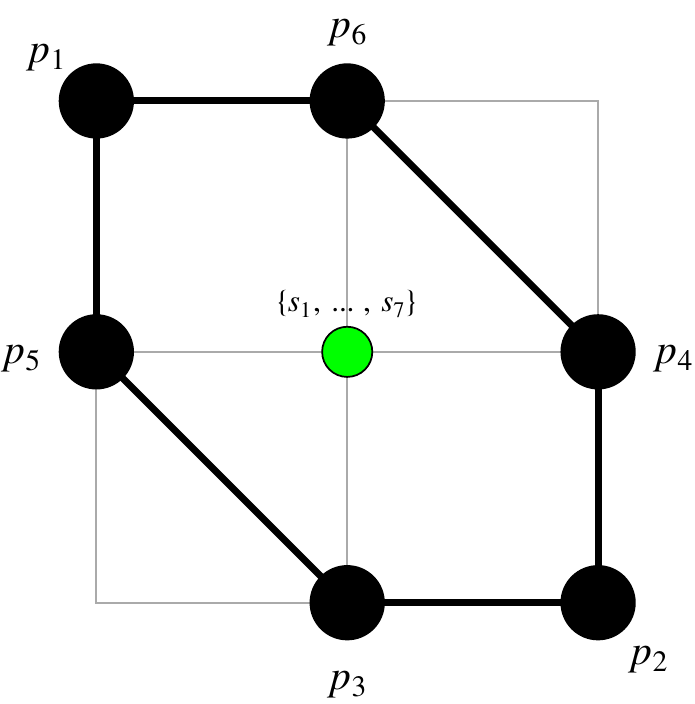}
\includegraphics[width=5 cm]{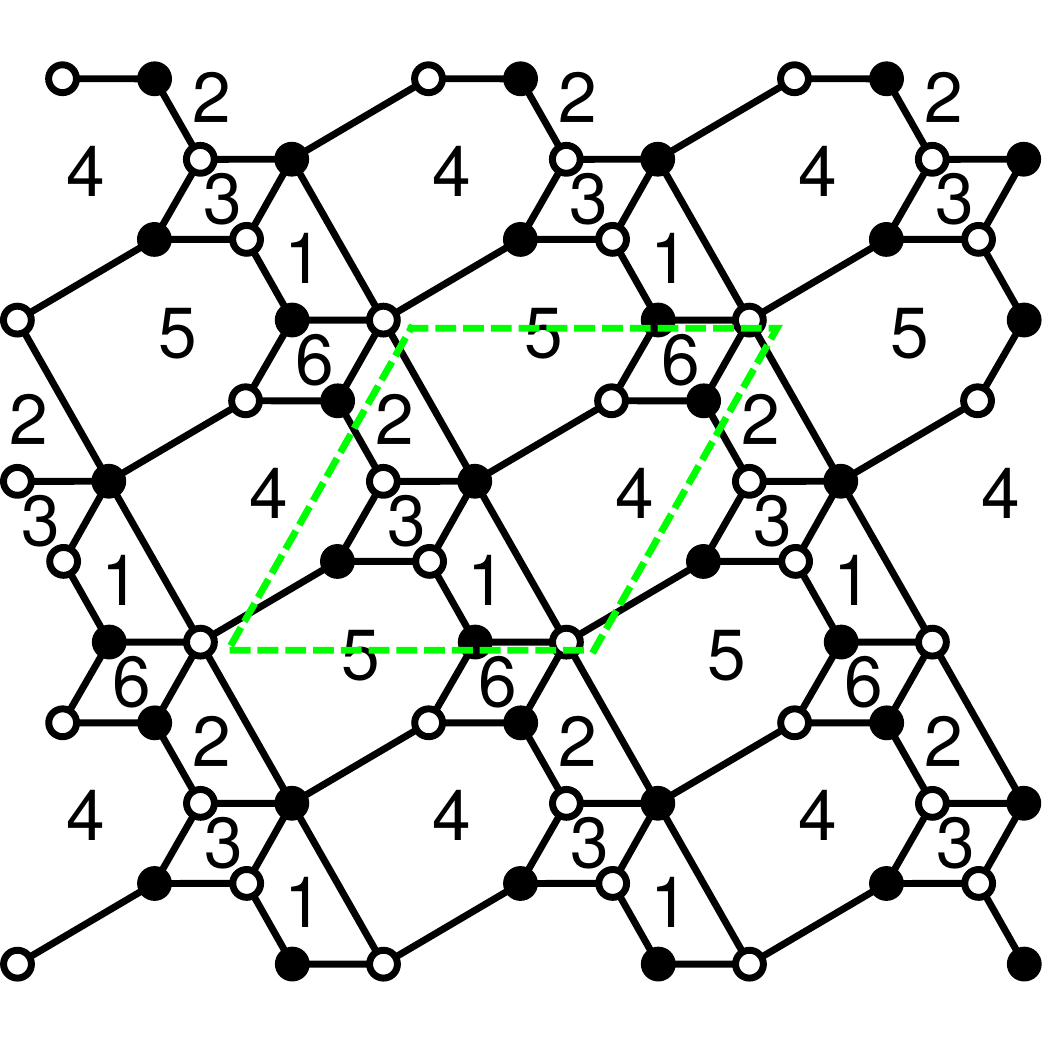}
\caption{The quiver, toric diagram and brane tiling of Model 10b. The red arrows in the quiver indicate all possible connections between blocks of nodes.}
  \label{f10b}
 \end{center}
 \end{figure}
 
 \noindent The superpotential is 
\beal{esm10b_00}
W&=&
+ X_{31} X_{15} X_{53}  
+ X_{42} X_{23} X_{34}  
+ X_{56} X_{64} X_{45}^{2} 
+ X_{52} X_{26} X_{61} X_{14} X_{45}^{1}
\nn\\
&&
- X_{42} X_{26} X_{64}  
- X_{53} X_{34} X_{45}^{1} 
- X_{56} X_{61} X_{15}  
- X_{14} X_{45}^{2} X_{52} X_{23} X_{31}  
  \eea
 
 \noindent The perfect matching matrix is 
 
\noindent\makebox[\textwidth]{%
\footnotesize
$
P=
\left(
\begin{array}{c|cccccc|ccccccc}
 \; & p_{1} & p_{2} & p_{3} & p_{4} & p_{5} & p_{6} & s_{1}
   & s_{2} & s_{3} & s_{4} & s_{5} & s_{6} & s_{7} \\
   \hline
 X_{45}^{2} & 1 & 0 & 0 & 0 & 0 & 1 & 0 & 0 & 0 & 0 & 1 & 0 & 0 \\
 X_{15} & 1 & 0 & 1 & 0 & 1 & 0 & 0 & 0 & 0 & 0 & 1 & 1 & 0 \\
 X_{34} & 1 & 0 & 0 & 0 & 1 & 0 & 1 & 0 & 1 & 0 & 0 & 1 & 0 \\
 X_{26} & 1 & 0 & 0 & 0 & 0 & 0 & 1 & 0 & 0 & 1 & 0 & 0 & 0 \\
 X_{42} & 0 & 1 & 0 & 1 & 0 & 1 & 0 & 0 & 0 & 0 & 1 & 0 & 1 \\
 X_{56} & 0 & 1 & 0 & 1 & 0 & 0 & 1 & 0 & 0 & 1 & 0 & 0 & 1 \\
 X_{45}^{1} & 0 & 1 & 1 & 0 & 0 & 0 & 0 & 0 & 0 & 0 & 1 & 0 & 0 \\
 X_{31} & 0 & 1 & 0 & 0 & 0 & 0 & 1 & 0 & 1 & 0 & 0 & 0 & 0 \\
 X_{64} & 0 & 0 & 1 & 0 & 1 & 0 & 0 & 1 & 1 & 0 & 0 & 1 & 0 \\
 X_{23} & 0 & 0 & 1 & 0 & 0 & 0 & 0 & 1 & 0 & 1 & 0 & 0 & 0 \\
 X_{53} & 0 & 0 & 0 & 1 & 0 & 1 & 0 & 1 & 0 & 1 & 0 & 0 & 1 \\
 X_{14} & 0 & 0 & 0 & 1 & 0 & 0 & 0 & 0 & 0 & 0 & 0 & 1 & 0 \\
 X_{52} & 0 & 0 & 0 & 0 & 1 & 0 & 0 & 0 & 0 & 0 & 0 & 0 & 1 \\
 X_{61} & 0 & 0 & 0 & 0 & 0 & 1 & 0 & 1 & 1 & 0 & 0 & 0 & 0 
 \end{array}
\right)
$
}
\vspace{0.5cm}

 \noindent The F-term charge matrix $Q_F=\ker{(P)}$ is

\noindent\makebox[\textwidth]{%
\footnotesize
$
Q_F=
\left(
\begin{array}{cccccc|ccccccc}
 p_{1} & p_{2} & p_{3} & p_{4} & p_{5} & p_{6} & s_{1}
   & s_{2} & s_{3} & s_{4} & s_{5} & s_{6} & s_{7} \\
   \hline
  1 & 1 & 0 & 0 & 0 & 0 & -1 & 0 & 0 & 0 & -1 & 0 & 0 \\
 1 & 0 & 1 & 0 & -1 & 0 & 0 & 0 & 0 & -1 & -1 & 0 & 1 \\
 1 & 0 & 0 & 1 & 0 & -1 & -1 & 0 & 1 & 0 & 0 & -1 & 0 \\
 0 & 0 & 0 & 1 & 1 & 0 & 0 & 0 & 0 & 0 & 0 & -1 & -1 \\
 0 & 0 & 0 & 0 & 0 & 0 & 1 & 1 & -1 & -1 & 0 & 0 & 0
\end{array}
\right)
$
}
\vspace{0.5cm}

\noindent The D-term charge matrix is

\noindent\makebox[\textwidth]{%
\footnotesize
$
Q_D=
\left(
\begin{array}{cccccc|ccccccc}
 p_{1} & p_{2} & p_{3} & p_{4} & p_{5} & p_{6} & s_{1}
   & s_{2} & s_{3} & s_{4} & s_{5} & s_{6} & s_{7} \\
   \hline
 0 & 0 & 0 & 0 & 0 & 0 & 0 & 1 & -1 & 0 & 0 & 0 & 0 \\
 0 & 0 & 0 & 0 & 0 & 0 & 0 & 0 & 1 & -1 & 0 & 0 & 0 \\
 0 & 0 & 0 & 0 & 0 & 0 & 0 & 0 & 0 & 1 & -1 & 0 & 0 \\
 0 & 0 & 0 & 0 & 0 & 0 & 0 & 0 & 0 & 0 & 1 & -1 & 0 \\
 0 & 0 & 0 & 0 & 0 & 0 & 0 & 0 & 0 & 0 & 0 & 1 & -1
\end{array}
\right)
$
}
\vspace{0.5cm}

The total charge matrix $Q_t$ does not exhibit repeated columns. Accordingly, the global symmetry of Model 10b is identical to the one for Model 10a, $U(1)_{f_1} \times U(1)_{f_2} \times U(1)_R$. The flavour and R-charges on the extremal perfect matchings are found following the discussion in \sref{s1_3}. They are identical to Model 10a, and are shown in \tref{t10a}.

The product of all internal perfect matchings is given by the variable
\beal{esm10b_x1}
s= \prod_{m=1}^{7} s_m~.
\eea
The fugacity for extremal perfect matchings $p_\alpha$ is $t_\alpha$ and the fugacity for the above product of internal perfect matchings is $y_{s}$.

The mesonic Hilbert series of Model 10a and 10b are identical. They are called phases of the same toric moduli space. The Hilbert series is found in \eref{esm10a2_1} with the plethystic logarithm in \eref{esm10a_3}. The moduli space is not a complete intersection. 

The generators of the mesonic moduli space in terms of the perfect matchings of Model 10b are shown in \tref{t10agen}. The generators in terms of quiver fields of Model 10b are shown in \tref{t10agen2}. The charge lattice of generators is the dual reflexive polygon of the toric diagram of Model 10b.

\comment{
\begin{table}[H]
\centering
\resizebox{\hsize}{!}{
\begin{minipage}[!b]{0.6\textwidth}
\begin{tabular}{|l|c|c|}
\hline
Generator & $U(1)_{f_1}$ & $U(1)_{f_2}$ 
\\
\hline
\hline
$p_{2}^2 p_{3}^2 p_{4} p_{5} ~ s$
& 1 & 1
\\
$p_{1} p_{2} p_{3}^2 p_{5}^2 ~ s$
& 0 & 1
\\
$p_{2}^2 p_{3} p_{4}^2 p_{6} ~ s$
& 1 & 0
\\
$p_{1} p_{2} p_{3} p_{4} p_{5} p_{6} ~ s$
& 0 & 0
\\
$p_{1}^2 p_{3} p_{5}^2 p_{6} ~ s$
& -1 & 0
\\
$p_{1} p_{2} p_{4}^2 p_{6}^2 ~ s$
& 0 & -1
\\
$p_{1}^2 p_{4} p_{5} p_{6}^2 ~ s$
& -1 & -1 
   \nn\\
   \hline
\end{tabular}
\end{minipage}
\hspace{1cm}
\begin{minipage}[!b]{0.3\textwidth}
\includegraphics[width=4 cm]{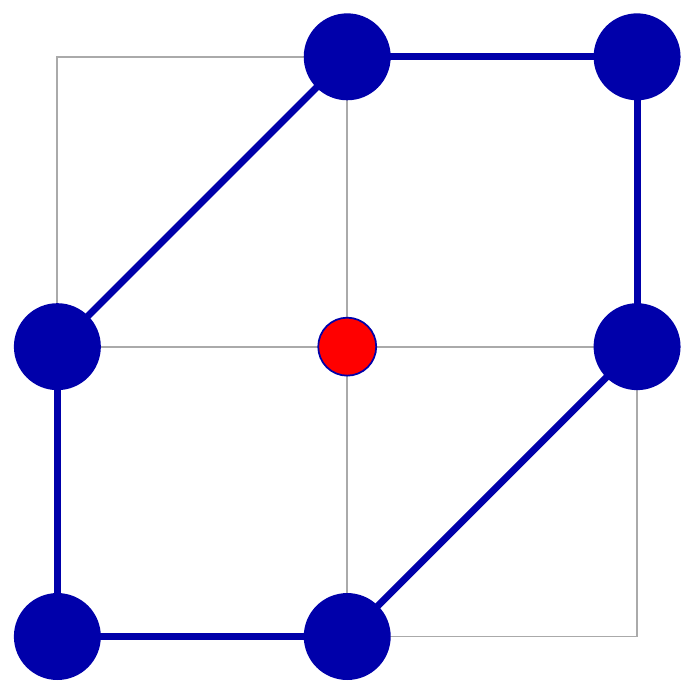}
\end{minipage}
}
\caption{The generators and lattice of generators of the mesonic moduli space of Model 10b in terms of GLSM fields with the corresponding flavor charges. \label{t10bgen}\label{f10bgen}} 
\end{table}
}

\begin{table}[h!]
\centering
\resizebox{\hsize}{!}{
\begin{tabular}{|l|c|c|}
\hline
Generator & $U(1)_{f_1}$ & $U(1)_{f_2}$ 
\\
\hline
\hline
$X_{15} X_{52} X_{23} X_{31}=  X_{23} X_{34} X_ {45}^{1} X_{52}=  X_{26} X_{64} X_ {45}^{1} X_{52}$
   & 1 & 1
   \nn\\
$ X_{15} X_{52} X_{26} X_{61}=  X_{23} X_{34} X_{45}^{2} X_{52}=  X_{26} X_{64} X_{45}^{2} X_{52}$
   & 0 & 1
   \nn\\
$ X_{45}^{1} X_{56} X_{64}=  X_{14} X_{45}^{1} X_{52} X_{23} X_{31}$
   & 1 & 0
   \nn\\
$X_{14} X_ {45}^{2} X_{52} X_{23} X_{31}=  X_{14} X_ {45}^{1} X_{52} X_{26} X_{61}=  X_{15} X_{53} X_{31}=  X_{15} X_{56} X_{61}=  X_{23} X_{34} X_{42}=  X_{26} X_{64} X_{42}=  X_{34} X_ {45}^{1} X_{53}=  X_ {45}^{2} X_{56} X_{64}$
   & 0 & 0
   \nn\\
$X_{34} X_{45}^{2} X_{53}=  X_{14} X_{45}^{2} X_{52} X_{26} X_{61}$
   & -1 & 0
   \nn\\
$X_{14} X_{42} X_{23} X_{31}=  X_{14} X_ {45}^{1} X_{53} X_{31}=  X_{14} X_ {45}^{1} X_{56} X_{61}$
   & 0 & -1
   \nn\\
$ X_{14} X_{42} X_{26} X_{61}=  X_{14} X_{45}^{2} X_{53} X_{31}=  X_{14} X_{45}^{2} X_{56} X_{61}$
   & -1 & -1
   \nn\\
\hline
\end{tabular}
}
\caption{The generators in terms of bifundamental fields (Model 10b).\label{f10bgen2}} 
\end{table}

\subsection{Model 10 Phase c}

\begin{figure}[H]
\begin{center}
\includegraphics[trim=0cm 0cm 0cm 0cm,width=4.5 cm]{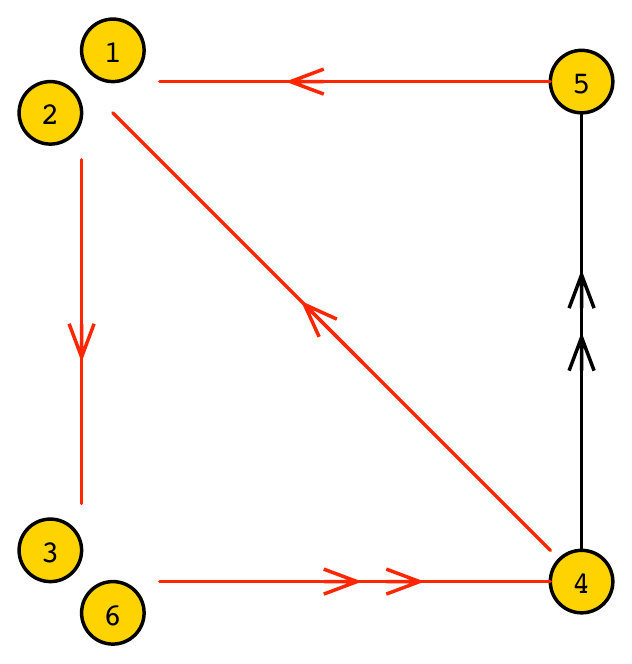}
\includegraphics[width=5 cm]{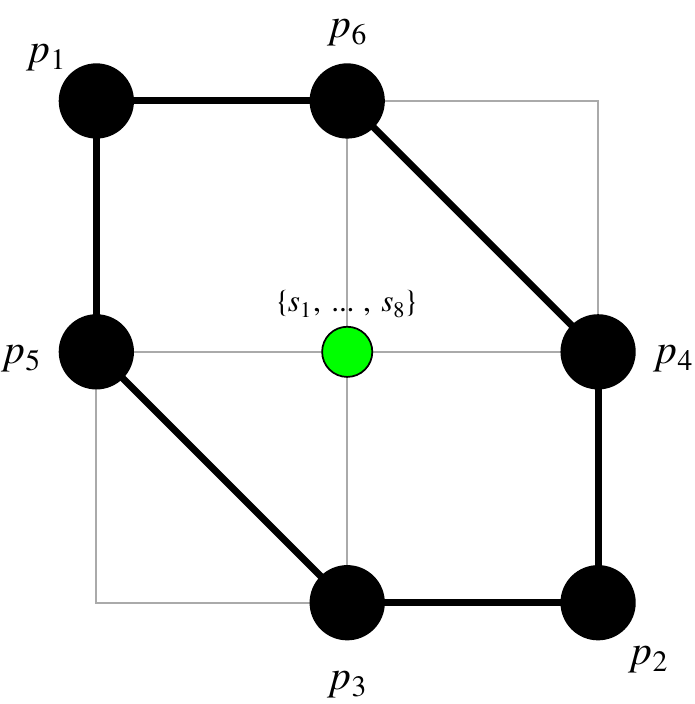}
\includegraphics[width=4.8 cm]{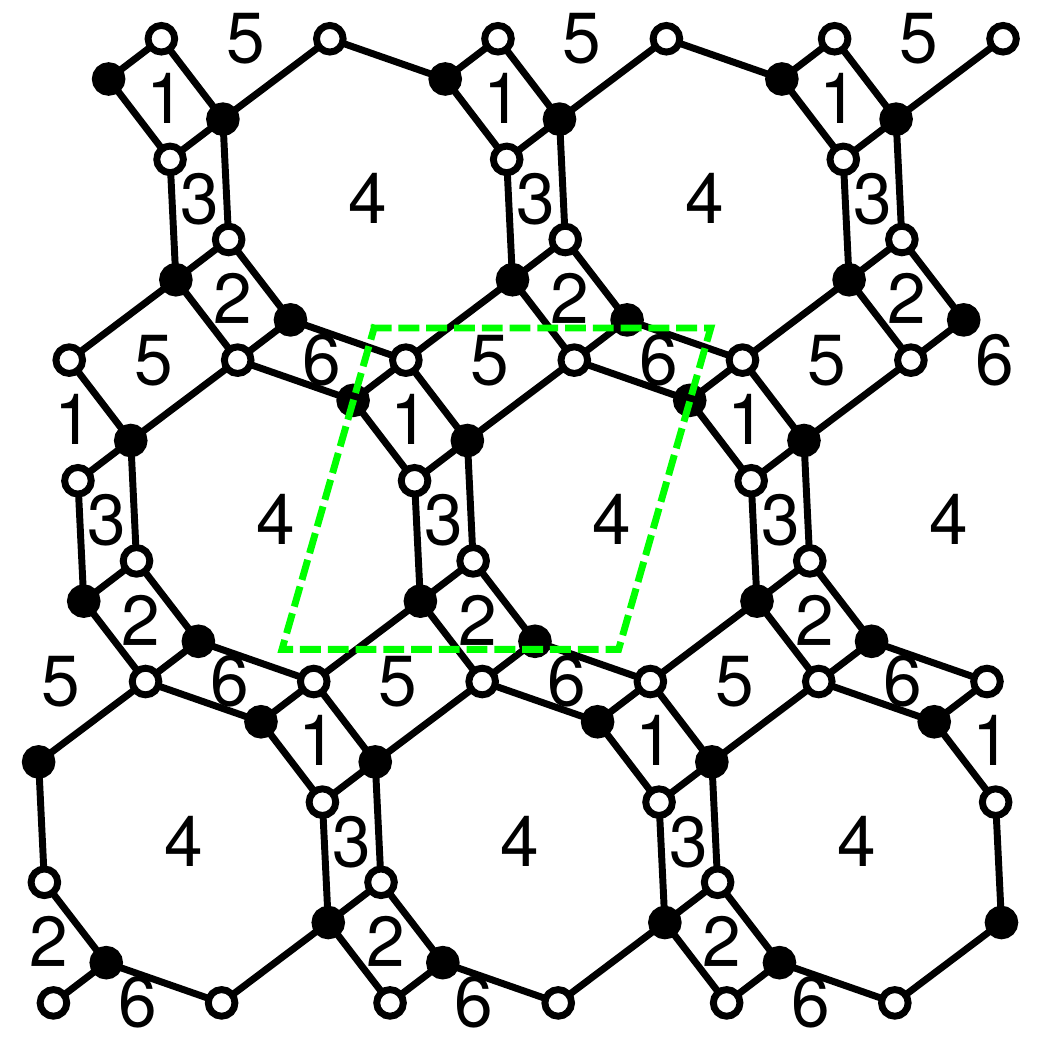}
\caption{The quiver, toric diagram, and brane tiling of Model 10c. The red arrows in the quiver indicate all possible connections between blocks of nodes.}
  \label{f10c}
 \end{center}
 \end{figure}
 
 \noindent The superpotential is 
\beal{esm10c_00}
W&=&
+ X_{41} X_{13} X_{34}^{2} 
+ X_{42} X_{23} X_{34}^{1} 
+ X_{45}^{1} X_{52} X_{26} X_{64}^{2} 
+ X_{51} X_{16} X_{64}^{1} X_{45}^{2}
\nn\\
&&
- X_{41} X_{16} X_{64}^{2} 
- X_{42} X_{26} X_{64}^{1} 
- X_{45}^{2} X_{52} X_{23} X_{34}^{2} 
- X_{51} X_{13} X_{34}^{1} X_{45}^{1} 
  \eea
 
 \noindent The perfect matching matrix is 
 
\noindent\makebox[\textwidth]{%
\footnotesize
$
P=
\left(
\begin{array}{c|cccccc|cccccccc}
 \; & p_{1} & p_{2} & p_{3} & p_{4} & p_{5} & p_{6} & s_{1}
   & s_{2} & s_{3} & s_{4} & s_{5} & s_{6} & s_{7} & s_{8}
   \\
   \hline
 X_{42} & 1 & 0 & 0 & 0 & 1 & 1 & 0 & 0 & 1 & 0 & 1 & 1 & 0 &
   0 \\
 X_{34}^{2} & 1 & 0 & 0 & 0 & 1 & 0 & 0 & 1 & 0 & 0 & 0 & 0 & 0 &
   1 \\
 X_{64}^{2} & 1 & 0 & 0 & 0 & 0 & 1 & 0 & 0 & 0 & 0 & 0 & 0 & 1 &
   1 \\
 X_{51} & 1 & 0 & 0 & 0 & 0 & 0 & 0 & 0 & 0 & 1 & 0 & 1 & 0 &
   0 \\
 X_{41} & 0 & 1 & 1 & 1 & 0 & 0 & 0 & 0 & 1 & 1 & 0 & 1 & 0 &
   0 \\
 X_{64}^{1} & 0 & 1 & 1 & 0 & 0 & 0 & 0 & 0 & 0 & 0 & 0 & 0 & 1 &
   1 \\
 X_{34}^{1} & 0 & 1 & 0 & 1 & 0 & 0 & 0 & 1 & 0 & 0 & 0 & 0 & 0 &
   1 \\
 X_{52} & 0 & 1 & 0 & 0 & 0 & 0 & 0 & 0 & 0 & 0 & 1 & 1 & 0 &
   0 \\
 X_{45}^{1} & 0 & 0 & 1 & 0 & 1 & 0 & 0 & 0 & 1 & 0 & 0 & 0 & 0 &
   0 \\
 X_{23} & 0 & 0 & 1 & 0 & 0 & 0 & 1 & 0 & 0 & 1 & 0 & 0 & 1 &
   0 \\
 X_{45}^{2} & 0 & 0 & 0 & 1 & 0 & 1 & 0 & 0 & 1 & 0 & 0 & 0 & 0 &
   0 \\
 X_{26} & 0 & 0 & 0 & 1 & 0 & 0 & 1 & 1 & 0 & 1 & 0 & 0 & 0 &
   0 \\
 X_{16} & 0 & 0 & 0 & 0 & 1 & 0 & 1 & 1 & 0 & 0 & 1 & 0 & 0 &
   0 \\
 X_{13} & 0 & 0 & 0 & 0 & 0 & 1 & 1 & 0 & 0 & 0 & 1 & 0 & 1 &
   0
    \end{array}
\right)
$
}
\vspace{0.5cm}

 \noindent The F-term charge matrix $Q_F=\ker{(P)}$ is

\noindent\makebox[\textwidth]{%
\footnotesize
$
Q_F=
\left(
\begin{array}{cccccc|cccccccc}
p_{1} & p_{2} & p_{3} & p_{4} & p_{5} & p_{6} & s_{1}
   & s_{2} & s_{3} & s_{4} & s_{5} & s_{6} & s_{7} & s_{8}
   \\
   \hline
 1 & 1 & 0 & 0 & 0 & 0 & 0 & 0 & 0 & 0 & 0 & -1 & 0 & -1 \\
  1 & 0 & 0 & 1 & 0 & -1 & 0 & -1 & 0 & 0 & 1 & -1 & 0 & 0 \\
   0 & 0 & 1 & 0 & 0 & 1 & 0 & 0 & -1 & 0 & 0 & 0 & -1 & 0 \\
 0 & 0 & 0 & 1 & 1 & 0 & 0 & -1 & -1 & 0 & 0 & 0 & 0 & 0 \\
 0 & 0 & 0 & 0 & 0 & 0 & 1 & -1 & 0 & 0 & 0 & 0 & -1 & 1 \\
 0 & 0 & 0 & 0 & 0 & 0 & 1 & 0 & 0 & -1 & -1 & 1 & 0 & 0
 \end{array}
\right)
$
}
\vspace{0.5cm}

\noindent The D-term charge matrix is

\noindent\makebox[\textwidth]{%
\footnotesize
$
Q_D=
\left(
\begin{array}{cccccc|cccccccc}
p_{1} & p_{2} & p_{3} & p_{4} & p_{5} & p_{6} & s_{1}
   & s_{2} & s_{3} & s_{4} & s_{5} & s_{6} & s_{7} & s_{8}
   \\
   \hline
 0 & 0 & 0 & 0 & 0 & 0 & 0 & 0 & 1 & -1 & 0 & 0 & 0 & 0 \\
 0 & 0 & 0 & 0 & 0 & 0 & 0 & 0 & 0 & 1 & -1 & 0 & 0 & 0 \\
 0 & 0 & 0 & 0 & 0 & 0 & 0 & 0 & 0 & 0 & 1 & -1 & 0 & 0 \\
 0 & 0 & 0 & 0 & 0 & 0 & 0 & 0 & 0 & 0 & 0 & 1 & -1 & 0 \\
 0 & 0 & 0 & 0 & 0 & 0 & 0 & 0 & 0 & 0 & 0 & 0 & 1 & -1
\end{array}
\right)
$
}
\vspace{0.5cm}

The global symmetry for Model 10c is identical to the global symmetries of Model 10a and Model 10b, $U(1)_{f_1} \times U(1)_{f_2} \times U(1)_R$. The mesonic charges on the extremal perfect matchings with non-zero R-charge are shown in \tref{t10a}.

The product of all internal perfect matchings is expressed as
\beal{esm10c_x1}
s = \prod_{m=1}^{8} s_m ~.
\eea
The fugacity $t_\alpha$ counts extremal perfect matchings and the fugacity $y_{s}$ counts the above product of internal perfect matchings.

The mesonic Hilbert series is identical to the Hilbert series for Models 10a and 10b in \eref{esm10a_1}. 

The moduli space generators in terms of all perfect matchings of Model 10c are shown in \tref{t10agen}, with the corresponding lattice of generators being the dual reflexive polygon of the toric diagram. The generators in terms of quiver fields of Model 10c are shown in \tref{t10cgen2}.

\comment{
\begin{table}[h!]
\centering
\resizebox{\hsize}{!}{
\begin{minipage}[!b]{0.6\textwidth}
\begin{tabular}{|l|c|c|}
\hline
Generator & $U(1)_{f_1}$ & $U(1)_{f_2}$ 
\\
\hline
\hline
$p_{2}^2 p_{3}^2 p_{4} p_{5} ~ \prod_{m=1}^{8} s_m$
& 1 & 1
\\
$p_{1} p_{2} p_{3}^2 p_{5}^2 ~ \prod_{m=1}^{8} s_m$
& 0 & 1
\\
$p_{2}^2 p_{3} p_{4}^2 p_{6} ~ \prod_{m=1}^{8} s_m$
& 1 & 0
\\
$p_{1} p_{2} p_{3} p_{4} p_{5} p_{6} ~ \prod_{m=1}^{8} s_m$
& 0 & 0
\\
$p_{1}^2 p_{3} p_{5}^2 p_{6} ~ \prod_{m=1}^{8} s_m$
& -1 & 0
\\
$p_{1} p_{2} p_{4}^2 p_{6}^2 ~ \prod_{m=1}^{8} s_m$
& 0 & -1
\\
$p_{1}^2 p_{4} p_{5} p_{6}^2 ~ \prod_{m=1}^{8} s_m$
& -1 & -1 
   \nn\\
\hline
\end{tabular}
\end{minipage}
\hspace{1cm}
\begin{minipage}[!b]{0.3\textwidth}
\includegraphics[width=4 cm]{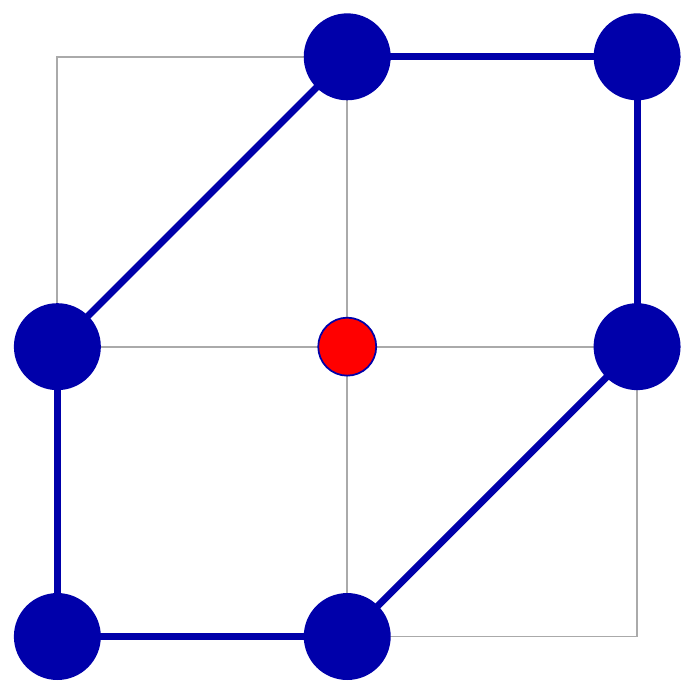}
\end{minipage}
}
\caption{The generators and lattice of generators of the mesonic moduli space of Model 10c in terms of GLSM fields with the corresponding flavor charges.\label{t10cgen}\label{f10cgen}} 
\end{table}
}

\begin{table}[h!]
\centering
\resizebox{\hsize}{!}{
\begin{tabular}{|l|c|c|}
\hline
Generator & $U(1)_{f_1}$ & $U(1)_{f_2}$ 
\\
\hline
\hline
$ X_{16} X_{64}^{1} X_{41}=  X_{23} X_{34}^{1} X_{45}^{1} X_{52}=  X_{26} X_{64}^{1} X_{45}^{1} X_{52}$
   & 1 & 1
   \nn\\
$ X_{13} X_{34}^{1} X_{41}=  X_{23} X_{34}^{1} X_{45}^{2} X_{52}=  X_{26} X_{64}^{1} X_{45}^{2} X_{52}$
   & 0 & 1
   \nn\\
$X_{16} X_{64}^{1} X_{45}^{1} X_{51}=  X_{23} X_{34}^{2} X_{45}^{1} X_{52}$
   & 1 & 0
   \nn\\
$X_{13} X_ {34}^{2} X_{41}=  X_{16} X_ {64}^{2} X_{41}=  X_{23} X_{34}^{1} X_{42}=  X_{26} X_ {64}^{1} X_{42}
$
& 0 & 0
\nn\\
$
=  X_{13} X_ {34}^{1} X_{45}^{1} X_{51}=  X_{16} X_ {64}^{1} X_ {45}^{2} X_{51}=  X_{23} X_{34}^{2} X_ {45}^{2} X_{52}=  X_{26} X_ {64}^{2} X_ {45}^{1} X_{52}$
   & & 
   \nn\\
$X_{13} X_ {34}^{1} X_ {45}^{2} X_{51}=  X_{26} X_ {64}^{2} X_{45}^{2} X_{52}$
   & -1 & 0
   \nn\\
$X_{23} X_ {34}^{2} X_{42}=  X_{13} X_ {34}^{2} X_ {45}^{1} X_{51}= X_{16} X_ {64}^{2} X_ {45}^{1} X_{51}$
   & 0 & -1
   \nn\\
$X_{26} X_ {64}^{2} X_{42}=  X_{13} X_ {34}^{2} X_ {45}^{2} X_{51}=  X_{16} X_ {64}^{2} X_ {45}^{2} X_{51}$
   & -1 & -1
   \nn\\
\hline
\end{tabular}
}
\caption{The generators in terms of bifundamental fields (Model 10c).\label{t10cgen2}\label{f10cgen2}} 
\end{table}

\subsection{Model 10 Phase d}

\begin{figure}[H]
\begin{center}
\includegraphics[trim=0cm 0cm 0cm 0cm,width=4.5 cm]{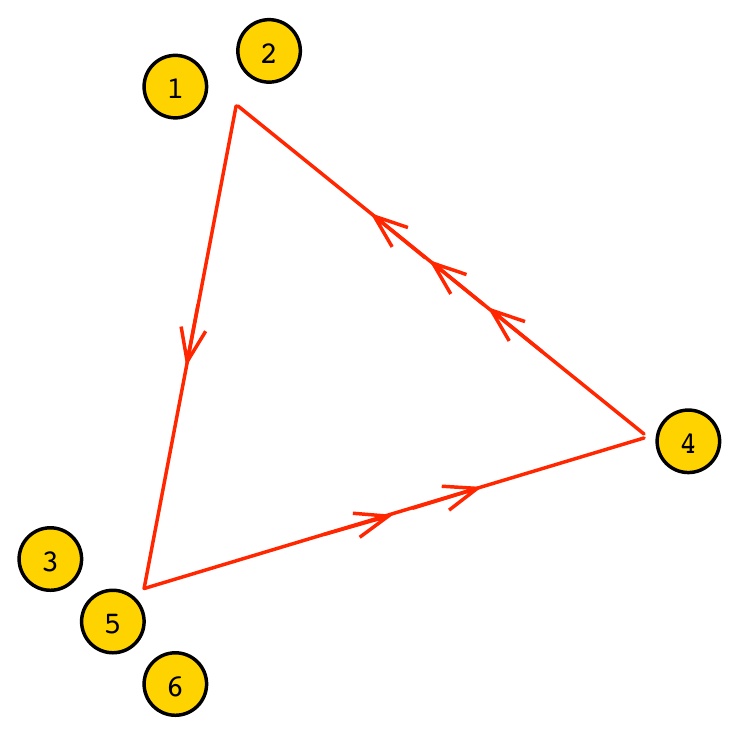}
\includegraphics[width=5 cm]{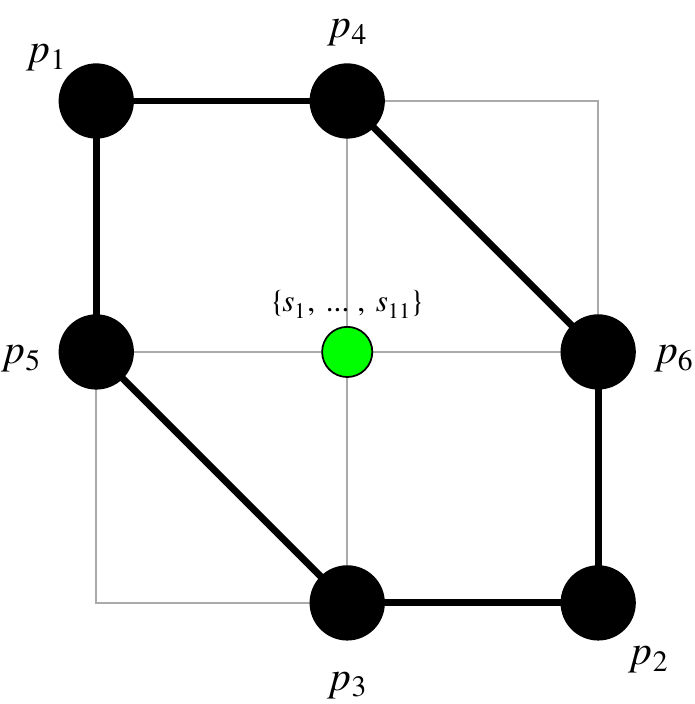}
\includegraphics[width=4.8 cm]{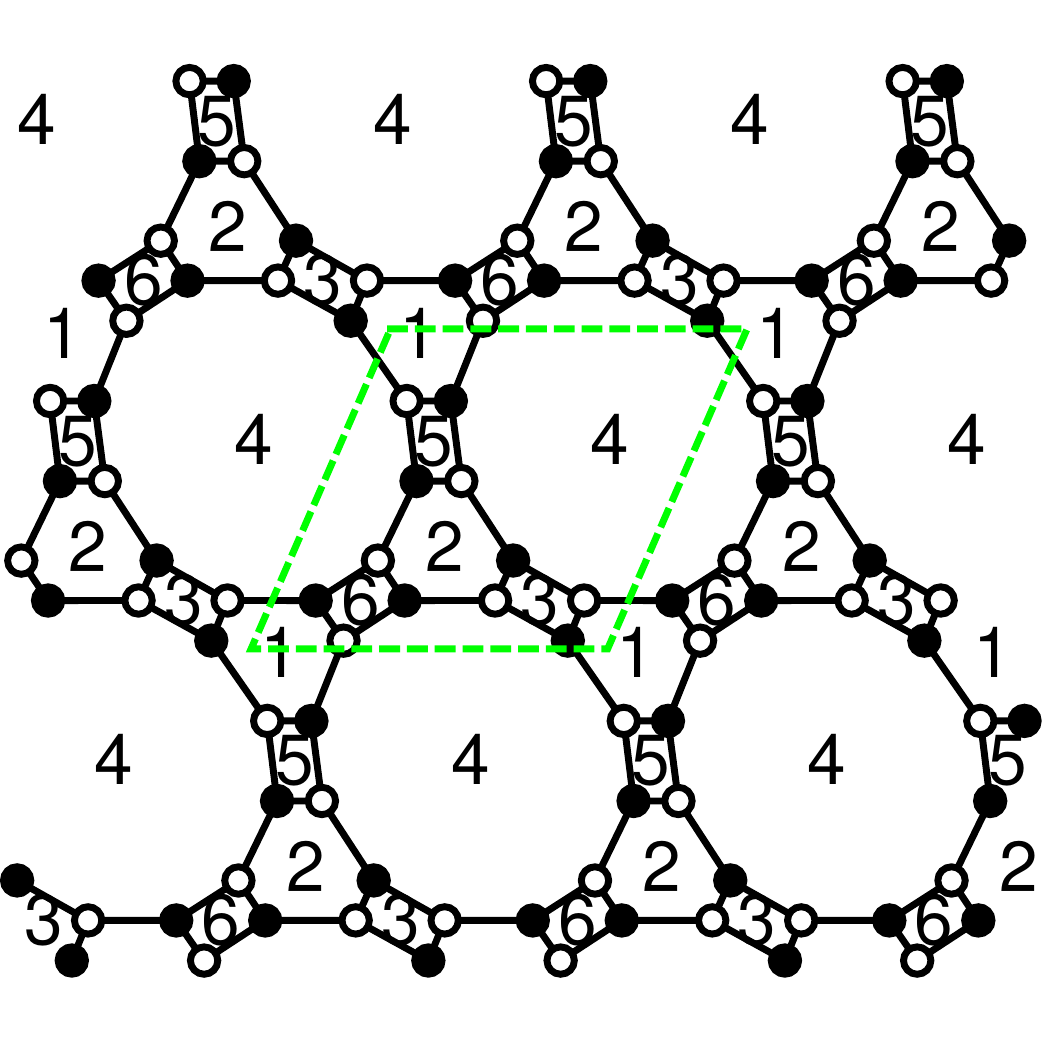}
\caption{The quiver, toric diagram, and brane tiling of Model 10d. The red arrows in the quiver indicate all possible connections between blocks of nodes.}
  \label{f10d}
 \end{center}
 \end{figure}
 
 \noindent The superpotential is 
\beal{esm10d_00}
W&=&
+  X_{15} X_{54}^{1} X_{41}^{2} 
+ X_{25} X_{54}^{2} X_{42}^{2} 
+ X_{26} X_{64}^{2} X_{42}^{3} 
+ X_{41}^{1} X_{13} X_{34}^{2} 
+ X_{16} X_{64}^{1} X_{41}^{3}
+ X_{42}^{1} X_{23} X_{34}^{1} 
\nn\\
&&
- X_{15} X_{54}^{2} X_{41}^{3}  
- X_{13} X_{34}^{1} X_{41}^{2} 
- X_{23} X_{34}^{2} X_{42}^{2} 
- X_{25} X_{54}^{1} X_{42}^{3} 
- X_{41}^{1} X_{16} X_{64}^{2} 
- X_{42}^{1} X_{26} X_{64}^{1}
~~.
\nn\\
  \eea
 
 \noindent The perfect matching matrix is 
 
\noindent\makebox[\textwidth]{%
\footnotesize
$
P=
\left(
\begin{array}{c|cccccc|ccccccccccc}
 \; & p_{1} & p_{2} & p_{3} & p_{4} & p_{5} & p_{6} & s_{1}
   & s_{2} & s_{3} & s_{4} & s_{5} & s_{6} & s_{7} &
   s_{8} & s_{9} & s_{10} & s_{11} \\
   \hline
 X_{42}^{2} & 1 & 0 & 1 & 0 & 1 & 0 & 0 & 0 & 0 & 0 & 0 & 1 & 1 & 0 & 0 &
   0 & 0 \\
 X_{42}^{3} & 1 & 0 & 0 & 1 & 0 & 1 & 0 & 0 & 0 & 0 & 0 & 1 & 1 & 0 & 0 &
   0 & 0 \\
 X_{41}^{1} & 1 & 0 & 0 & 0 & 1 & 1 & 0 & 0 & 0 & 0 & 1 & 0 & 1 & 0 & 0 &
   0 & 0 \\
 X_{34} & 1 & 0 & 0 & 0 & 1 & 0 & 0 & 0 & 1 & 1 & 0 & 0 & 0 & 0 & 1 &
   0 & 1 \\
 X_{64} & 1 & 0 & 0 & 0 & 0 & 1 & 0 & 1 & 1 & 0 & 0 & 0 & 0 & 0 & 0 &
   1 & 1 \\
 X_{15} & 1 & 0 & 0 & 0 & 0 & 0 & 1 & 0 & 0 & 1 & 0 & 1 & 0 & 0 & 0 &
   1 & 1 \\
 X_{41}^{3} & 0 & 1 & 1 & 0 & 1 & 0 & 0 & 0 & 0 & 0 & 1 & 0 & 1 & 0 & 0 &
   0 & 0 \\
 X_{64}^{2} & 0 & 1 & 1 & 0 & 0 & 0 & 0 & 1 & 1 & 0 & 0 & 0 & 0 & 0 & 0 &
   1 & 1 \\
 X_{42}^{1} & 0 & 1 & 1 & 1 & 0 & 0 & 0 & 0 & 0 & 0 & 0 & 1 & 1 & 0 & 0 &
   0 & 0 \\
 X_{41}^{2} & 0 & 1 & 0 & 1 & 0 & 1 & 0 & 0 & 0 & 0 & 1 & 0 & 1 & 0 & 0 &
   0 & 0 \\
 X_{34}^{2} & 0 & 1 & 0 & 1 & 0 & 0 & 0 & 0 & 1 & 1 & 0 & 0 & 0 & 0 & 1 &
   0 & 1 \\
 X_{25} & 0 & 1 & 0 & 0 & 0 & 0 & 1 & 0 & 0 & 1 & 1 & 0 & 0 & 0 & 0 &
   1 & 1 \\
 X_{54} & 0 & 0 & 1 & 0 & 1 & 0 & 0 & 1 & 1 & 0 & 0 & 0 & 0 & 1 & 1 &
   0 & 0 \\
 X_{13} & 0 & 0 & 1 & 0 & 0 & 0 & 1 & 1 & 0 & 0 & 0 & 1 & 0 & 1 & 0 &
   1 & 0 \\
 X_{54}^{2} & 0 & 0 & 0 & 1 & 0 & 1 & 0 & 1 & 1 & 0 & 0 & 0 & 0 & 1 & 1 &
   0 & 0 \\
 X_{16} & 0 & 0 & 0 & 1 & 0 & 0 & 1 & 0 & 0 & 1 & 0 & 1 & 0 & 1 & 1 &
   0 & 0 \\
 X_{26} & 0 & 0 & 0 & 0 & 1 & 0 & 1 & 0 & 0 & 1 & 1 & 0 & 0 & 1 & 1 &
   0 & 0 \\
 X_{23} & 0 & 0 & 0 & 0 & 0 & 1 & 1 & 1 & 0 & 0 & 1 & 0 & 0 & 1 & 0 &
   1 & 0
\end{array}
\right)
$
}
\vspace{0.5cm}

 \noindent The F-term charge matrix $Q_F=\ker{(P)}$ is

\noindent\makebox[\textwidth]{%
\footnotesize
$
Q_F=
\left(
\begin{array}{cccccc|ccccccccccc}
 p_{1} & p_{2} & p_{3} & p_{4} & p_{5} & p_{6} & s_{1}
   & s_{2} & s_{3} & s_{4} & s_{5} & s_{6} & s_{7} &
   s_{8} & s_{9} & s_{10} & s_{11} \\
   \hline
 1 & 1 & 0 & -1 & -1 & 0 & 0 & 0 & 0 & 0 & 0 & 0 & 0 & 1 & 0 & -1 & 0 \\
 1 & 0 & 0 & 1 & 0 & -1 & 0 & 0 & 0 & 0 & 0 & -1 & 0 & 1 & -1 & 0 & 0 \\
 1 & 0 & 0 & 1 & 0 & -1 & 0 & 0 & 0 & 0 & 0 & -1 & 0 & 0 & 0 & 1 & -1 \\
 0 & 1 & 0 & -1 & 0 & 1 & 0 & 0 & 0 & 1 & -1 & 0 & 0 & 0 & 0 & 0 & -1 \\
 0 & 1 & 0 & -1 & 0 & 1 & 0 & 0 & 0 & 0 & 0 & 1 & -1 & 0 & 0 & -1 & 0 \\
  0 & 0 & 1 & 0 & 0 & 1 & 0 & -1 & 0 & 0 & 0 & 0 & -1 & 0 & 0 & 0 & 0 \\
 0 & 0 & 0 & 0 & 0 & 0 & 1 & 0 & 1 & 0 & 0 & 0 & 0 & 0 & -1 & -1 & 0 \\
 0 & 0 & 0 & 0 & 0 & 0 & 0 & 1 & 0 & 1 & 0 & 0 & 0 & 0 & -1 & -1 & 0 \\
 0 & 0 & 0 & 0 & 0 & 0 & 0 & 1 & -1 & 0 & 0 & 0 & 0 & 0 & 0 & -1 & 1
 \end{array}
\right)
$
}
\vspace{0.5cm}

\noindent The D-term charge matrix is

\noindent\makebox[\textwidth]{%
\footnotesize
$
Q_D=
\left(
\begin{array}{cccccc|ccccccccccc}
 p_{1} & p_{2} & p_{3} & p_{4} & p_{5} & p_{6} & s_{1}
   & s_{2} & s_{3} & s_{4} & s_{5} & s_{6} & s_{7} &
   s_{8} & s_{9} & s_{10} & s_{11} \\
   \hline
 0 & 0 & 0 & 0 & 0 & 0 & 0 & 0 & 0 & 0 & 1 & -1 & 0 & 0 & 0 & 0 & 0 \\
 0 & 0 & 0 & 0 & 0 & 0 & 0 & 0 & 0 & 0 & 0 & 1 & -1 & 0 & 0 & 0 & 0 \\
 0 & 0 & 0 & 0 & 0 & 0 & 0 & 0 & 0 & 0 & 0 & 0 & 1 & -1 & 0 & 0 & 0 \\
 0 & 0 & 0 & 0 & 0 & 0 & 0 & 0 & 0 & 0 & 0 & 0 & 0 & 1 & -1 & 0 & 0 \\
 0 & 0 & 0 & 0 & 0 & 0 & 0 & 0 & 0 & 0 & 0 & 0 & 0 & 0 & 1 & -1 & 0
 \end{array}
\right)
$
}
\vspace{0.5cm}

The symmetry $U(1)_{f_1}\times U(1)_{f_2} \times U(1)_R$ of Model 10d is identical to Models 10a to 10c discussed above. The symmetry charges on the extremal perfect matchings with non-zero R-charges are shown in \tref{t10a}.

The product of all internal perfect matchings is
\beal{esm10d_x1}
s = \prod_{m=1}^{11} s_m~.
\eea
The fugacity $y_{s}$ counts the above product of internal perfect matchings whereas the fugacity $t_\alpha$ counts the external perfect matchings $p_\alpha$.

The mesonic Hilbert series of Model 10d is identical to Models 10a, 10b and 10c. This indicates that the mesonic moduli spaces are identical, and given the corresponding plethystic logarithm in \eref{esm10a_3}, the mesonic moduli spaces are not complete intersections.

The moduli space generators in terms of all perfect matchings of Model 10d are shown in \tref{t10agen} with the corresponding charge lattice of generators forming a reflexive polygon which is the dual polygon of the toric diagram. The generators in terms of quiver fields of Model 10d are shown in \tref{t10dgen2}.

\comment{
\begin{figure}[H]
\centering
\resizebox{\hsize}{!}{
\begin{minipage}[!b]{0.6\textwidth}
\begin{tabular}{|l|c|c|}
\hline
Generator & $U(1)_{f_1}$ & $U(1)_{f_2}$ 
\\
\hline
\hline
$p_{2}^2 p_{3}^2 p_{4} p_{5} ~ \prod_{m=1}^{11} s_m$
& 1 & 1
\\
$p_{1} p_{2} p_{3}^2 p_{5}^2 ~ \prod_{m=1}^{11} s_m$
& 0 & 1
\\
$p_{2}^2 p_{3} p_{4}^2 p_{6} ~ \prod_{m=1}^{11} s_m$
& 1 & 0
\\
$p_{1} p_{2} p_{3} p_{4} p_{5} p_{6} ~ \prod_{m=1}^{11} s_m$
& 0 & 0
\\
$p_{1}^2 p_{3} p_{5}^2 p_{6} ~ \prod_{m=1}^{11} s_m$
& -1 & 0
\\
$p_{1} p_{2} p_{4}^2 p_{6}^2 ~ \prod_{m=1}^{11} s_m$
& 0 & -1
\\
$p_{1}^2 p_{4} p_{5} p_{6}^2 ~ \prod_{m=1}^{11} s_m$
& -1 & -1 
   \nn\\
   \hline
\end{tabular}
\end{minipage}
\hspace{1cm}
\begin{minipage}[!b]{0.3\textwidth}
\includegraphics[width=4 cm]{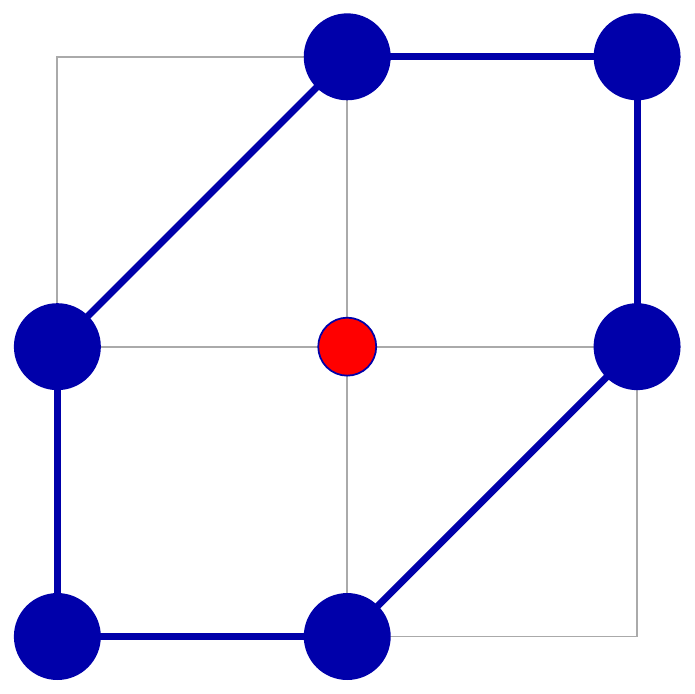}
\end{minipage}
}
\caption{The generators and lattice of generators of the mesonic moduli space of Model 10d in terms of GLSM fields with the corresponding flavor charges.\label{t10dgen}\label{f10dgen}} 
\end{figure}
}

\begin{figure}[H]
\centering
\resizebox{\hsize}{!}{
\begin{tabular}{|l|c|c|}
\hline
Generator & $U(1)_{f_1}$ & $U(1)_{f_2}$ 
\\
\hline
\hline
$X_{13} X_{34}^{2} X_{41}^{3}=  X_{41}^{3} X_{16} X_{64}^{2}=  X_{42}^{1} X_{25} X_{54}^{1}=  X_{42}^{1} X_{26} X_{64}^{2}$   
& 1 & 1
   \\
$X_{13} X_{34}^{1} X_{41}^{3}=  X_{41}^{3} X_{15} X_{54}^{1}=  X_{42}^{2} X_{25} X_{54}^{1}=  X_{42}^{2} X_{26} X_{64}^{2}$
   & 0 & 1
   \\
$X_{13} X_{34}^{2} X_{41}^{2}=  X_{41}^{2} X_{16} X_{64}^{2}=  X_{23} X_{34}^{2} X_{42}^{1}=  X_{42}^{1} X_{25} X_{54}^{2}$
   & 1 & 0
   \\
$X_{13} X_{34}^{1} X_{41}^{2}=  X_{13} X_{34}^{2} X_{41}^{1}=  X_{41}^{2} X_{15} X_{54}^{1}=  X_{41}^{3} X_{15} X_{54}^{2}=  X_{41}^{1} X_{16} X_{64}^{2}=  X_{41}^{3} X_{16} X_{64}^{1}=  X_{23} X_{34}^{1} X_{42}^{1}$
& 0 & 0
\\
$= X_{23} X_{34}^{2} X_{42}^{2}=  X_{42}^{2} X_{25} X_{54}^{2}=  X_{42}^{3} X_{25} X_{54}^{1}=  X_{42}^{1} X_{26} X_{64}^{1}=  X_{42}^{3} X_{26} X_{64}^{2}$
   & & 
   \\
$X_{13} X_{34}^{1} X_{41}^{1}=  X_{41}^{1} X_{15} X_{54}^{1}=  X_{23} X_{34}^{1} X_{42}^{2}=  X_{42}^{2} X_{26} X_{64}^{1}$
   & -1 & 0
   \\
$X_{41}^{2} X_{15} X_{54}^{2}=  X_{41}^{2} X_{16} X_{64}^{1}=  X_{23} X_{34}^{2} X_{42}^{3}=  X_{42}^{3} X_{25} X_{54}^{2}$
   & 0 & -1
   \\
$X_{41}^{1} X_{15} X_{54}^{2}=  X_{41}^{1} X_{16} X_{64}^{1}=  X_{23} X_{34}^{1} X_{42}^{3}=  X_{42}^{3} X_{26} X_{64}^{1}$
   & -1 & -1
   \\
   \hline
\end{tabular}
}
\caption{The generators in terms of bifundamental fields (Model 10d). \label{t10dgen2}\label{f10dgen2}} 
\end{figure}

\section{Model 11: $\text{PdP}_2$}

\begin{figure}[H]
\begin{center}
\includegraphics[trim=0cm 0cm 0cm 0cm,width=4.5 cm]{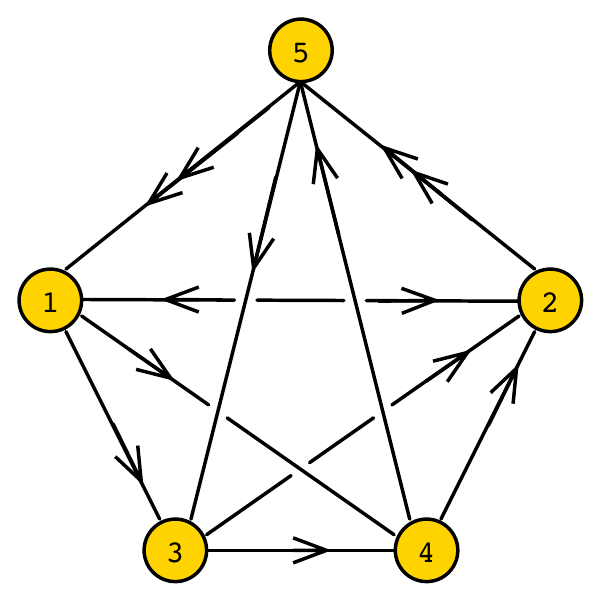}
\includegraphics[width=5 cm]{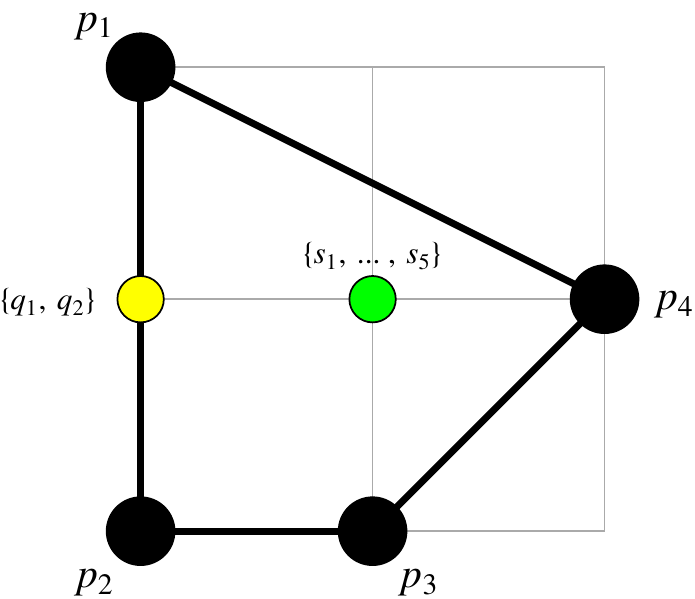}
\includegraphics[width=5 cm]{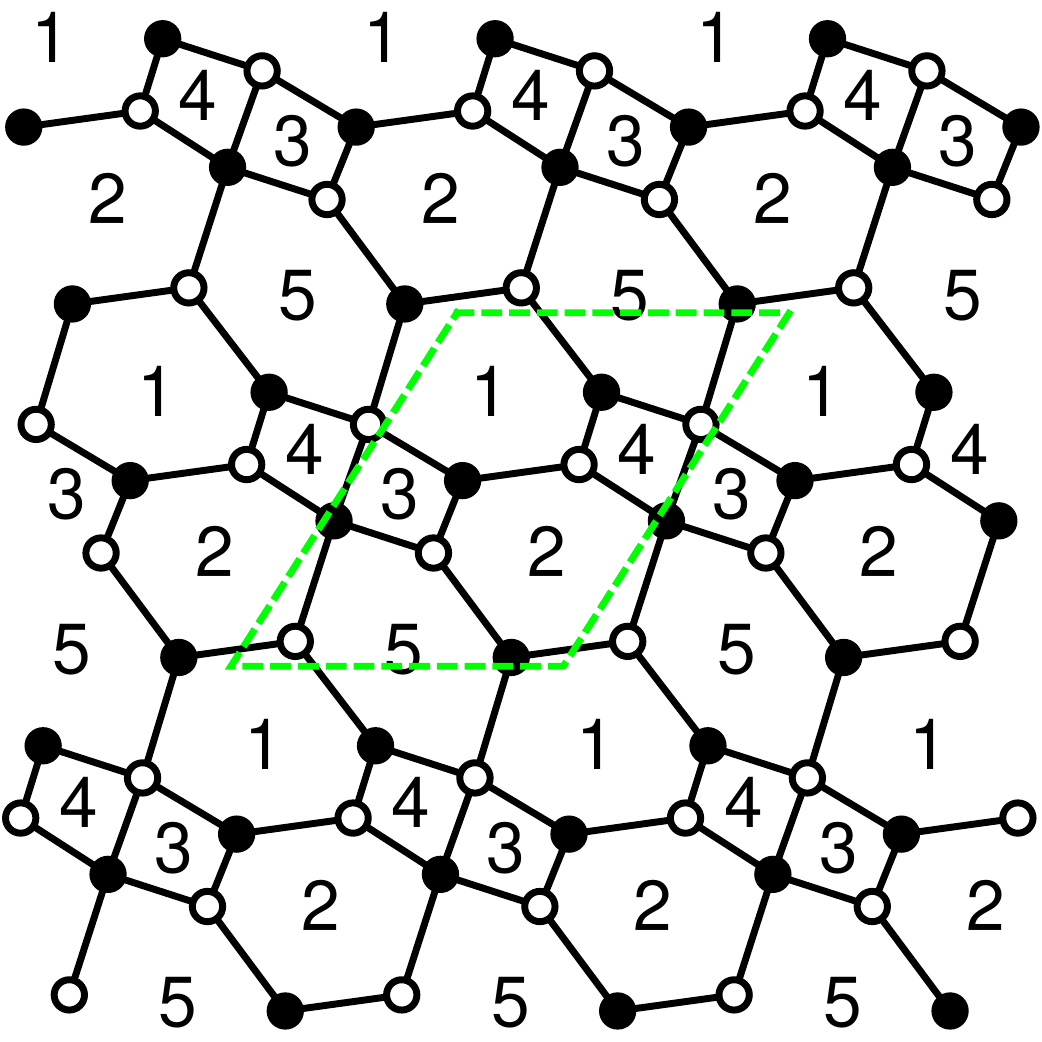}
\caption{The quiver, toric diagram, and brane tiling of Model 11.}
  \label{f11}
 \end{center}
 \end{figure}
 
 \noindent The superpotential is 
\beal{esm11_00}
W&=&
 +X_{21}X_{14} X_{42} 
 +X_{53} X_{32} X_{25}^{2} 
 +X_{51}^{2} X_{12} X_{25}^{1} 
 +X_{13} X_{34} X_{45}X_{51}^{1} 
 \nn\\
 && 
 -X_{13}X_{32} X_{21} 
 -X_{14} X_{45} X_{51}^{2} 
 -X_{51}^{1} X_{12} X_{25}^{2} 
 -X_{53} X_{34} X_{42} X_{25}^{1} 
  \eea
 
 \noindent The perfect matching matrix is 
 
\noindent\makebox[\textwidth]{%
\footnotesize
$
P=
\left(
\begin{array}{c|cccc|cc|ccccc}
 \; & p_1 & p_2 & p_3 & p_4 & q_1 & q_2 & s_1
   & s_2 & s_3 & s_4 & s_5 \\
   \hline
 X_{14} & 1 & 0 & 0 & 0 & 1 & 0 & 1 & 0 & 0 & 1 & 0 \\
 X_{32} & 1 & 0 & 0 & 0 & 0 & 1 & 1 & 0 & 0 & 0 & 1 \\
 X_{25}^{1} & 1 & 0 & 0 & 0 & 1 & 0 & 0 & 1 & 0 & 0 & 0 \\
 X_{25}^{2} & 0 & 1 & 1 & 0 & 1 & 0 & 0 & 1 & 0 & 0 & 0 \\
 X_{51}^{1} & 1 & 0 & 0 & 0 & 0 & 1 & 0 & 0 & 1 & 0 & 0 \\
 X_{51}^{2} & 0 & 1 & 1 & 0 & 0 & 1 & 0 & 0 & 1 & 0 & 0 \\
 X_{13} & 0 & 1 & 0 & 0 & 1 & 0 & 0 & 0 & 0 & 1 & 0 \\
 X_{42} & 0 & 1 & 0 & 0 & 0 & 1 & 0 & 0 & 0 & 0 & 1 \\
 X_{21} & 0 & 0 & 1 & 1 & 0 & 0 & 0 & 1 & 1 & 0 & 0 \\
 X_{12} & 0 & 0 & 0 & 1 & 0 & 0 & 1 & 0 & 0 & 1 & 1 \\
 X_{34} & 0 & 0 & 1 & 0 & 0 & 0 & 1 & 0 & 0 & 0 & 0 \\
 X_{45} & 0 & 0 & 0 & 1 & 0 & 0 & 0 & 1 & 0 & 0 & 1 \\
 X_{53} & 0 & 0 & 0 & 1 & 0 & 0 & 0 & 0 & 1 & 1 & 0
\end{array}
\right)
$
}
\vspace{0.5cm}

 \noindent The F-term charge matrix $Q_F=\ker{(P)}$ is

\noindent\makebox[\textwidth]{%
\footnotesize
$
Q_F=
\left(
\begin{array}{cccc|cc|ccccc}
p_1 & p_2 & p_3 & p_4 & q_1 & q_2 & s_1
   & s_2 & s_3 & s_4 & s_5 \\
   \hline
 1 & 1 & 0 & 0 & -1 & -1 & 0 & 0 & 0 & 0 & 0 \\
  1 & 1 & 0 & 1 & -1 & 0 & 0 & 0 & -1 & 0 & -1 \\
 0 & 1 & -1 & 0 & -1 & 0 & 1 & 1 & 0 & 0 & -1\\
 0 & 0 & 0 & 1 & 1 & 0 & 0 & -1 & 0 & -1 & 0 
\end{array}
\right)
$
}
\vspace{0.5cm}

\noindent The D-term charge matrix is

\noindent\makebox[\textwidth]{%
\footnotesize
$
Q_D=
\left(
\begin{array}{cccc|cc|ccccc}
p_1 & p_2 & p_3 & p_4 & q_1 & q_2 & s_1
   & s_2 & s_3 & s_4 & s_5 \\
   \hline
 0 & 0 & 0 & 0 & 0 & 0 & 1 & -1 & 0 & 0 & 0 \\
 0 & 0 & 0 & 0 & 0 & 0 & 0 & 1 & -1 & 0 & 0 \\
 0 & 0 & 0 & 0 & 0 & 0 & 0 & 0 & 1 & -1 & 0 \\
 0 & 0 & 0 & 0 & 0 & 0 & 0 & 0 & 0 & 1 & -1
  \end{array}
\right)
$
}
\vspace{0.5cm}

The total charge matrix $Q_t$ does not exhibit repeated columns. Accordingly, the global symmetry is $U(1)_{f_1} \times U(1)_{f_2} \times U(1)_R$. The flavour and R-charges on the GLSM fields corresponding to extremal points in the toric diagram in \fref{f11} are found following the discussion in \sref{s1_3}. They are presented in \tref{t11}.

\begin{table}[H]
\centering
\begin{tabular}{|c||c|c|c||l|} 
\hline
\; & $U(1)_{f_1}$ & $U(1)_{f_2}$ & $U(1)_R$ & fugacity \\
\hline
\hline
$p_1$ &-1/4 	&-1/3 	& $R_1 \simeq 0.622$ &  $t_1$\\
$p_2$ &-1/4 	& 0	 	& $R_2 \simeq 0.502$ &  $t_2$\\
$p_3$ & 0 	& 2/3	& $R_3 \simeq 0.306$ &  $t_3$\\ 
$p_4$ & 1/2 	&-1/3 	& $R_4 \simeq 0.570$ &  $t_4$\\
\hline
\end{tabular}
\caption{The GLSM fields corresponding to extremal points of the toric diagram with their mesonic charges (Model 11).\label{t11}}
\end{table} 

\noindent\textit{Fine-tuning R-charges.} The exact R-charges are expressed in terms of the root $x_0$ in the range $0\leq 1-x_0 \leq \frac{2}{3}$ of the polynomial
\beal{es11_p1}
27 - 42 x - 68 x^2 + 42 x^3 + 9 x^4 = 0,
\eea
where
\beal{es11_p2}
R_1 &=&
1 + \frac{1}{144} 
\left(
-63 + 250 x_0 - 422 x_0^2 - 384 x_0^3 + 261 x_0^4 + 54 x_0^5 \right)
\nn\\
R_2 &=&
1 + \frac{1}{72} \left(
-189 + 281 x_0 + 257 x_0^2 - 177 x_0^3 - 
    36 x_0^4\right)
\nn\\
R_3 &=&
1 + \frac{1}{288} \left(
333 - 1351 x_0 - 294 x_0^2 + 1450 x_0^3 -  327 x_0^4 - 99 x_0^5
\right)
\nn\\
R_4 &=&
1- x_0
~~.
\eea

Products of non-extremal perfect matchings are assigned the following variables
\beal{esm11_x1}
q = q_1 q_2 ~,~
s = \prod_{m=1}^5 s_m~.
\eea
The fugacities $y_{q}$ and $y_{s}$ count respectively the above products of internal perfect matchings. The fugacity $t_\alpha$ counts all other extremal perfect matchings $p_\alpha$.

The mesonic Hilbert series of Model 11 is found using the Molien integral formula in \eref{es12_2}. It is
 \beal{esm5_1}
&&g_{1}(t_\alpha,y_{q},y_{s}; \mathcal{M}^{mes}_{11})= 
(
1 
+ y_{q} y_{s} ~ t_1 t_2 t_3 t_4 
+ y_{q}^2 y_{s} ~ t_1 t_2^3 t_3^2 
+ y_{q}^2 y_{s} ~ t_1^2 t_2^2 t_3 
- y_{q}^2 y_{s}^2 ~t_1^2 t_2^2 t_3^2 t_4^2 
\nn\\
&&
\hspace{1cm}
- y_{q}^2 y_{s}^2 ~t_1^3 t_2 t_3 t_4^2 
- y_{q}^3 y_{s}^2 ~t_1^3 t_2^3 t_3^2 t_4 
- y_{q}^3 y_{s}^2 ~t_1^4 t_2^2 t_3 t_4 
- y_{q}^3 y_{s}^3 ~t_1^4 t_2^4 t_3^3 t_4^2
+ y_{q} y_{s} ~t_2^2 t_3^2 t_4 
)
\nn\\
&&
\hspace{1cm}
\times
\frac{
1
}{
(1 - y_{q}^2 y_{s} ~ t_1^3 t_2) 
(1 - y_{q}^2 y_{s} ~ t_2^4 t_3^3) 
(1 - y_{q} y_{s} ~ t_1^2 t_4) 
(1 - y_{s} ~ t_3 t_4^2)
}
~~.
 \eea
 The plethystic logarithm of the mesonic Hilbert series is
\beal{esm11_3}
&&
PL[g_1(t_\alpha,y_{q},y_{s};\mathcal{M}_{11}^{mes})]=
y_{q} y_{s} ~t_1^2 t_4 
+ y_{s} ~t_3 t_4^2
+ y_{q}^2 y_{s} ~t_1^3 t_2 
+ y_{q} y_{s} ~t_1 t_2 t_3 t_4
+ y_{q}^2 y_{s} ~t_1^2 t_2^2 t_3 
\nn\\
&&
\hspace{0.5cm}
+ y_{q} y_{s} ~t_2^2 t_3^2 t_4
+ y_{q}^2 y_{s} ~t_1 t_2^3 t_3^2 
+ y_{q}^2 y_{s} ~t_2^4 t_3^3 
- y_{q}^2 y_{s}^2 ~t_1^3 t_2 t_3 t_4^2 
- y_{q}^3 y_{s}^2 ~t_1^4 t_2^2 t_3 t_4 
- 2~ y_{q}^2 y_{s}^2 ~t_1^2 t_2^2 t_3^2 t_4^2
\nn\\
&&
\hspace{0.5cm}
+ \dots~.
\eea

Consider the following fugacity map
\beal{esm11_y1}
&& f_1 = y_q^{-3/4} y_s^{1/4}
~,~
f_2 = y_q^{-1/4} y_s^{-1/4}
~,~
\nn\\
&&
\tilde{t}_1 =
y_q^{1/4} y_s^{1/4} ~ t_1
~,~
\tilde{t}_2 = 
y_q^{1/4} y_s^{1/4} ~ t_2
~,~
\tilde{t}_3 = 
y_q^{1/4} y_s^{1/4} ~ t_3
~,~
\tilde{t}_4 = 
y_q^{1/4} y_s^{1/4} ~ t_4
~,~
\eea
where the fugacities $f_1$ and $f_2$ count flavour charges, and the fugacity $\tilde{t}_i$ counts the R-charge $R_i$ in \tref{t11}.

Under the fugacity map above, the plethystic logarithm becomes
\beal{esm11_3}
&&
PL[g_1(\tilde{t}_\alpha,f_1,f_2;\mathcal{M}_{11}^{mes})]=
\frac{1}{f_2} \tilde{t}_1^2 \tilde{t}_4 + f_1 \tilde{t}_3 \tilde{t}_4^2
+ \frac{1}{f_1 f_2} \tilde{t}_1^3 \tilde{t}_2 + \tilde{t}_1 \tilde{t}_2 \tilde{t}_3 \tilde{t}_4
+ \frac{1}{f_1} \tilde{t}_1^2 \tilde{t}_2^2 \tilde{t}_3 
\nn\\
&&
\hspace{0.5cm}
+ f_2 \tilde{t}_2^2 \tilde{t}_3^2 \tilde{t}_4 
+ \frac{f_2}{f_1} \tilde{t}_1 \tilde{t}_2^3 \tilde{t}_3^2 
+ \frac{f_2^2}{f_1} \tilde{t}_2^4 \tilde{t}_3^3 - \frac{1}{f_2}\tilde{t}_1^3 \tilde{t}_2 \tilde{t}_3 \tilde{t}_4^2
- \frac{1}{f_1 f_2} \tilde{t}_1^4 \tilde{t}_2^2 \tilde{t}_3 \tilde{t}_4 - 2 \tilde{t}_1^2 \tilde{t}_2^2 \tilde{t}_3^2 \tilde{t}_4^2
   +\dots~.
   \nn\\
   \eea
The plethsytic logarithm above exhibits the moduli space generators with the corresponding mesonic charges. They are summarized in \tref{t11gen}. The generators can be presented on a charge lattice. The convex polygon formed by the generators in \tref{t11gen} is the dual reflexive polygon of the toric diagram of Model 11.
\\

\begin{table}[H]
\centering
\begin{minipage}[!b]{0.5\textwidth}
\begin{tabular}{|l|c|c|}
\hline
Generator & $U(1)_{f_1}$ & $U(1)_{f_2}$ 
\\
\hline
\hline
$p_{3} p_{4}^2 ~ s$
   & 1 & 0
   \\
$p_{1}^2 p_{4} ~ q ~ s$
   & 0 & -1
   \\
$p_{1} p_{2} p_{3} p_{4} ~ q ~ s$
   & 0 & 0
   \\
$p_{2}^{2}
   p_{3}^2 p_{4} ~ q ~ s$
   & 0 & 1
   \\
   $p_{1}^3 p_{2} ~ q^2 ~ s$
   & -1 & -1
   \\
   $p_{1}^2 p_{2}^2 p_{3}
   ~ q^2 ~ s$
   & -1 & 0
   \\
   $p_{1} p_{2}^3 p_{3}^2 ~ q^2 ~ s$
   & -1 & 1
   \\
   $p_{2}^4 p_{3}^3 ~ q^2 ~ s$
   & -1 & 2
   \\
\hline
\end{tabular}
\end{minipage}
\hspace{1cm}
\begin{minipage}[!b]{0.3\textwidth}
\includegraphics[width=4 cm]{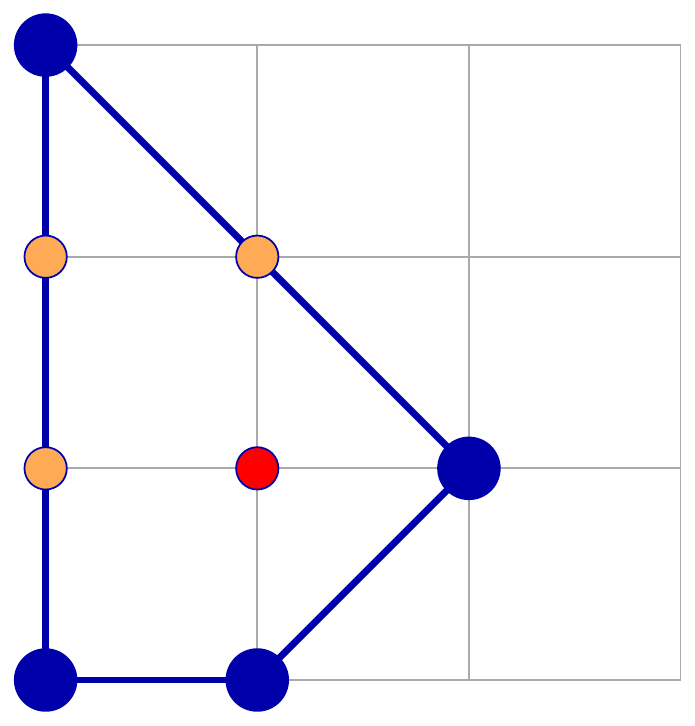}
\end{minipage}
\caption{The generators and lattice of generators of the mesonic moduli space of Model 11 in terms of GLSM fields with the corresponding flavor charges.\label{t11gen}\label{f11gen}} 
\end{table}

\begin{table}[H]
\centering
\resizebox{\hsize}{!}{
\begin{tabular}{|l|c|c|}
\hline
Generator & $U(1)_{f_1}$ & $U(1)_{f_2}$ 
\\
\hline
\hline
$X_{12} X_{21}=  X_{34} X_{45} X_{53}$
   & 1 & 0
   \\
$X_{12} X_{25}^{1} X_{51}^{1}=  X_{14} X_{45} X_{51}^{1}=  X_{32} X_{25}^{1} X_{53}$
   & 0 & -1
   \\
$X_{13} X_{34} X_{45} X_ {51}^{1}=  X_{34} X_ {25}^{1} X_{53} X_{42}=  X_{12} X_ {25}^{1} X_ {51}^{2}=  X_{12} X_ {25}^{2} X_ {51}^{1}$
& 0 & 0
\nn\\
$= X_{21} X_{13} X_{32}=  X_{21} X_{14} X_{42}=  X_{14} X_{45} X_{51}^{2}=  X_{32} X_ {25}^{2} X_{53}$
   & & 
   \\
$X_{12} X_{25}^{2} X_{51}^{2}=  X_{21} X_{13} X_{34} X_{42}=  X_{13} X_{34} X_{45} X_{51}^{2}=  X_{34} X_{25}^{2} X_{53} X_{42}$
   & 0 & 1
   \\
   $ X_{25}^{1} X_{51}^{1} X_{13} X_{32}=  X_{25}^{1} X_{51}^{1} X_{14} X_{42}$
   & -1 & -1
   \\
   $X_{25}^{1} X_ {51}^{1} X_{13} X_{34} X_{42}=  X_ {25}^{1} X_{51}^{2} X_{13} X_{32}=  X_ {25}^{2} X_ {51}^{1} X_{13} X_{32}=  X_{25}^{1} X_ {51}^{2} X_{14} X_{42}=  X_ {25}^{2} X_ {51}^{1} X_{14} X_{42}$
   & -1 & 0
   \\
   $X_ {25}^{2} X_ {51}^{2} X_{13} X_{32}=  X_ {25}^{2} X_ {51}^{2} X_{14} X_{42}=  X_ {25}^{1} X_ {51}^{2} X_{13} X_{34} X_{42}=  X_{25}^{2} X_ {51}^{1} X_{13} X_{34} X_{42}$
   & -1 & 1
   \\
   $X_{25}^{2} X_{51}^{2} X_{13} X_{34} X_{42}$
   & -1 & 2
   \\
\hline
\end{tabular}
}
\caption{The generators in terms of bifundamental fields (Model 11).\label{t11gen2}\label{f11gen2}} 
\end{table}

The mesonic Hilbert series and the plethystic logarithm can be re-expressed in terms of just $3$ fugacities
\beal{esm11_x1}
T_1 = \frac{f_2 ~ \tilde{t}_2}{f_1 ~ \tilde{t}_1 \tilde{t}_4^2}
= \frac{t_2}{y_{s} ~t_1 t_4^2}~,~
T_2 = \frac{1}{f_2} ~ \tilde{t}_1^2 \tilde{t}_4
= y_{q} y_{s}~ t_1^2 t_4 ~,~
T_3 = f_1 ~ \tilde{t}_3 \tilde{t}_4^2
= y_{s} ~t_3 t_4^2~,
\eea
such that
\beal{esm11_x2}
&&
g_1(T_1,T_2,T_3;\mathcal{M}^{mes}_{11})=
\nn\\
&&
\hspace{0.5cm}
(1 + T_1 T_2 T_3 + T_1^3 T_2^2 T_3^2 + T_1^2 T_2^2 T_3- T_1^2 T_2^2 T_3^2 
- T_1 T_2^2 T_3      - T_1^3 T_2^3 T_3^2  - T_1^2 T_2^3 T_3    \nn\\
   &&
   \hspace{0.5cm}
- T_1^4 T_2^4 T_3^3 + T_1^2 T_2 T_3^2 
)
    \times
\frac{
1
   }{
   (1 - T_1 T_2^2) (1 - T_1^4 T_2^2 T_3^3) (1 - T_2)  (1 - T_3) 
     }
\nn\\
\eea   
and
\beal{esm11_x3}
&&
PL[g_1(T_1,T_2,T_3;\mathcal{M}^{mes}_{11})]=
T_2 
+ T_3  
+ T_1 T_2^2 
+ T_1 T_2 T_3 
+ T_1^2 T_2^2 T_3 
+ T_1^2 T_2 T_3^2
 \nn\\
 &&
 \hspace{0.5cm}
 + T_1^3 T_2^2 T_3^2 
 + T_1^4 T_2^2 T_3^3 
 - T_1^2 T_2^3 T_3  
 - T_1 T_2^2 T_3  
 + 2 T_1^2 T_2^2 T_3^2 
 + \dots
~~.
\eea
The powers of the fugacities in the Hilbert series and plethystic logarithm above are all positive. This illustrates the conical structure of the toric Calabi-Yau 3-fold. 
\\

\section{Model 12: $\text{dP}_2$}
\subsection{Model 12 Phase a}

\begin{figure}[H]
\begin{center}
\includegraphics[trim=0cm 0cm 0cm 0cm,width=4.5 cm]{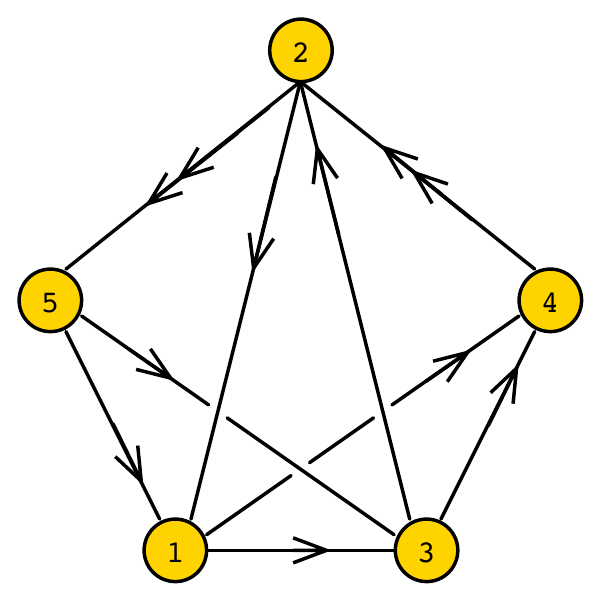}
\includegraphics[width=5 cm]{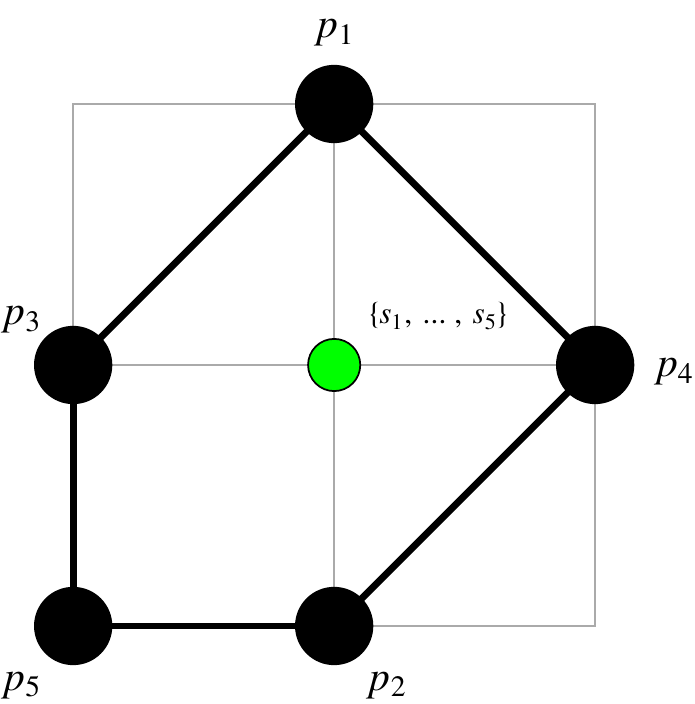}
\includegraphics[width=5 cm]{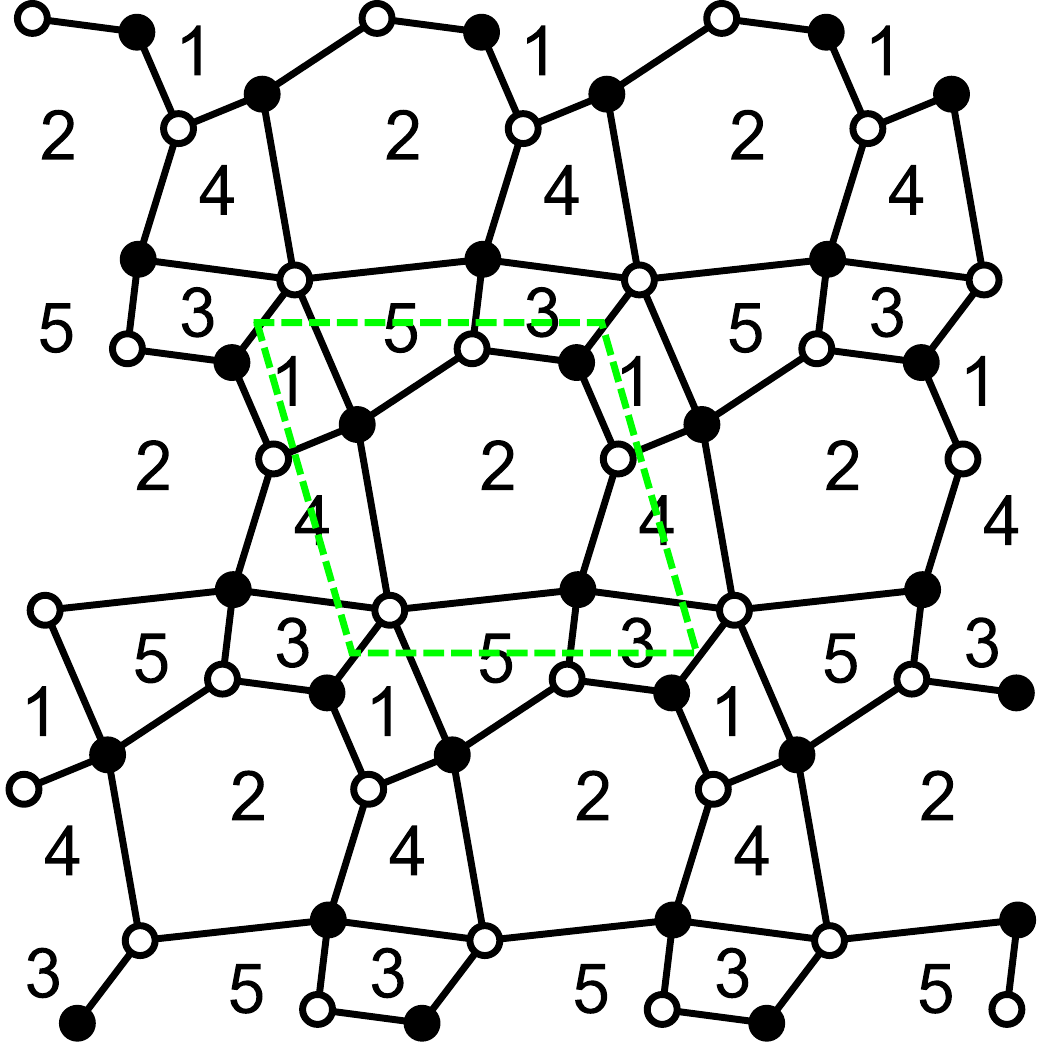}
\caption{The quiver, toric diagram, and brane tiling of Model 12a.}
  \label{f12a}
 \end{center}
 \end{figure}
 
 \noindent The superpotential is 
\beal{esm12a_00}
W&=&
+X_{21} X_{14} X_{42}^{1} 
+X_{25}^{2} X_{53} X_{32} 
+X_{42}^{2} X_{25}^{1} X_{51} X_{13} X_{34}
\nn\\
&&
-X_{13} X_{32} X_{21} 
-X_{14} X_{42}^{2} X_{25}^{2} X_{51} 
-X_{25}^{1} X_{53} X_{34} X_{42}^{1} 
~~.
\eea
 
\noindent The perfect matching matrix is  

\noindent\makebox[\textwidth]{%
\footnotesize
$
P=
\left(
\begin{array}{c|ccccc|ccccc}
 \; & p_{1} & p_{2} & p_{3} & p_{4} & p_{5} & s_{1} &
   s_{2} & s_{3} & s_{4} & s_{5} \\
   \hline
 X_{14}			& 1 & 0 & 0 & 0 & 0 & 1 & 0 & 0 & 0 & 1 \\
 X_{34} 		& 0 & 1 & 0 & 0 & 0 & 1 & 0 & 0 & 0 & 0 \\
 X_{25}^{1} & 1 & 0 & 0 & 0 & 0 & 0 & 1 & 0 & 0 & 0 \\
 X_{25}^{2} & 0 & 1 & 0 & 0 & 1 & 0 & 1 & 0 & 0 & 0 \\
 X_{42}^{1} & 0 & 0 & 1 & 0 & 1 & 0 & 0 & 1 & 0 & 0 \\
 X_{42}^{2} & 0 & 0 & 0 & 1 & 0 & 0 & 0 & 1 & 0 & 0 \\
 X_{32} 		& 1 & 0 & 1 & 0 & 0 & 1 & 0 & 1 & 0 & 0 \\
 X_{21} 		& 0 & 1 & 0 & 1 & 0 & 0 & 1 & 0 & 1 & 0 \\
 X_{51} 		& 0 & 0 & 1 & 0 & 0 & 0 & 0 & 0 & 1 & 0 \\
 X_{53} 		& 0 & 0 & 0 & 1 & 0 & 0 & 0 & 0 & 1 & 1 \\
 X_{13} 		& 0 & 0 & 0 & 0 & 1 & 0 & 0 & 0 & 0 & 1
\end{array}
\right)
$
}
\vspace{0.5cm}

 \noindent The F-term charge matrix $Q_F=\ker{(P)}$ is

\noindent\makebox[\textwidth]{%
\footnotesize
$
Q_F=
\left(
\begin{array}{ccccc|ccccc}
 p_{1} & p_{2} & p_{3} & p_{4} & p_{5} & s_{1} &
   s_{2} & s_{3} & s_{4} & s_{5} \\
   \hline
 1 & 1 & 0 & 0 & 0 & -1 & -1 & 0 & 0 & 0 \\
 0 & 0 & 1 & 1 & 0 & 0 & 0 & -1 & -1 & 0 \\
 0 & 1 & 0 & -1 & -1 & -1 & 0 & 1 & 0 & 1
\end{array}
\right)
$
}
\vspace{0.5cm}

\noindent The D-term charge matrix is

\noindent\makebox[\textwidth]{%
\footnotesize
$
Q_D=
\left(
\begin{array}{ccccc|ccccc}
 p_{1} & p_{2} & p_{3} & p_{4} & p_{5} & s_{1} &
   s_{2} & s_{3} & s_{4} & s_{5} \\
   \hline
 0 & 0 & 0 & 0 & 0 & 1 & -1 & 0 & 0 & 0 \\
 0 & 0 & 0 & 0 & 0 & 0 & 1 & -1 & 0 & 0 \\
 0 & 0 & 0 & 0 & 0 & 0 & 0 & 1 & -1 & 0 \\
 0 & 0 & 0 & 0 & 0 & 0 & 0 & 0 & 1 & -1
   \end{array}
\right)
$
}
\vspace{0.5cm}

The total charge matrix $Q_t$ does not exhibit repeated columns. Accordingly, the global symmetry is $U(1)_{f_1} \times U(1)_{f_2} \times U(1)_R$. The mesonic charges on the extremal perfect matchings are found following the discussion in \sref{s1_3}. They are presented in \tref{t12a}.

\begin{table}[H]
\centering
\begin{tabular}{|c||c|c|c||l|} 
\hline
\; & $U(1)_{f_1}$ & $U(1)_{f_2}$ & $U(1)_R$ & fugacity \\
\hline
\hline
$p_1$ & 1/2 & 0 & $R_1=\frac{1}{16} \left(-21 + 5 \sqrt{33}\right)$ 	&  $t_1$\\
$p_2$ &-1/2 & 0 & $R_2=\frac{3}{16} \left(19 - 3 \sqrt{33}\right)$ 	&  $t_2$\\
$p_3$ & 0 &-1/2 &  $R_2=\frac{3}{16} \left(19 - 3 \sqrt{33}\right)$  	&  $t_3$\\
$p_4$ & 0 & 1/2 &$R_1=\frac{1}{16} \left(-21 + 5 \sqrt{33}\right)$ 	&  $t_4$\\ 
$p_5$ & 0 & 0 & $R_3=\frac{1}{2} \left(-5 + \sqrt{33}\right)$ 	&  $t_5$\\ 
\hline
\end{tabular}
\caption{The GLSM fields corresponding to extremal points of the toric diagram with their mesonic charges (Model 12a). The R-charges are obtained using a-maximization \cite{Bertolini:2004xf}.\label{t12a}}
\end{table}

The product of all internal perfect matchings is
\beal{esm12a_x1}
s = \prod_{m=1}^{5} s_m ~.
\eea
The above product is counted by the fugacity $y_{s}$. The extremal perfect matchings $p_\alpha$ are counted by $t_\alpha$.

The mesonic Hilbert series of Model 12a is calculated using the Molien integral formula in \eref{es12_2}. It is 
 \beal{esm12a_1}
g_{1}(t_\alpha,y_{s}; \mathcal{M}^{mes}_{12a})
=
\frac{
P(t_\alpha)
 }{
(1 - y_{s} ~t_1^2 t_3 t_4) (1 - y_{s} ~t_1 t_2 t_4^2) (1 - y_{s} ~t_1^2 t_3^2 t_5) 
(1 - y_{s} ~t_2^2 t_4^2 t_5) (1 - y_{s} ~t_2^2 t_3^2 t_5^3)
}~~,
\nn\\
 \eea
 where the numerator is the polynomial
 \beal{esm12a_1b}
 P(t_\alpha)&=&
 1 
+ y_{s} ~t_1 t_2 t_3 t_4 t_5 
- y_{s}^2 ~t_1^3 t_2 t_3^2 t_4^2 t_5 
- y_{s}^2 ~t_1^2 t_2^2 t_3 t_4^3 t_5 
+ y_{s} ~t_1 t_2 t_3^2 t_5^2 
+ y_{s} ~t_2^2 t_3 t_4 t_5^2 
\nn\\
&&
- y_{s}^2 ~t_1^3 t_2 t_3^3 t_4 t_5^2 
- 2~y_{s}^2 ~ t_1^2 t_2^2 t_3^2 t_4^2 t_5^2 
- y_{s}^2 ~t_1 t_2^3 t_3 t_4^3 t_5^2 
+ y_{s}^3 ~t_1^4 t_2^2 t_3^3 t_4^3 t_5^2 
+ y_{s}^3 ~t_1^3 t_2^3 t_3^2 t_4^4 t_5^2 
\nn\\
&&
- y_{s}^2 ~t_1^2 t_2^2 t_3^3 t_4 t_5^3 
- y_{s}^2 ~t_1 t_2^3 t_3^2 t_4^2 t_5^3 
+ y_{s}^3 ~t_1^3 t_2^3 t_3^3 t_4^3 t_5^3 
+ y_{s}^4 ~t_1^4 t_2^4 t_3^4 t_4^4 t_5^4
~~.
 \eea
The mesonic moduli space of Model 12a is not a complete intersection.
 The plethystic logarithm of the mesonic Hilbert series is
\beal{esm12a_2}
&&
PL[g_1(t_\alpha,y_{s};\mathcal{M}_{12a}^{mes})]=
y_{s} ~t_1^2 t_3 t_4 
+ y_{s} ~t_1 t_2 t_4^2
+ y_{s} ~t_1 t_2 t_3 t_4 t_5 
+ y_{s} ~t_1^2 t_3^2 t_5 
+ y_{s} ~t_2^2 t_4^2 t_5
\nn\\
&&
\hspace{1cm}
+ y_{s} ~t_2^2 t_3 t_4 t_5^2
+ y_{s} ~t_1 t_2 t_3^2 t_5^2 
+ y_{s} ~t_2^2 t_3^2 t_5^3 
- y_{s}^2 ~t_1^3 t_2 t_3^2 t_4^2 t_5 
- y_{s}^2 ~t_1^2 t_2^2 t_3 t_4^3 t_5 
- 3 ~ y_{s}^2 ~t_1^2 t_2^2 t_3^2 t_4^2 t_5^2 
\nn\\
&&
\hspace{1cm}
- y_{s}^2 ~t_1^3 t_2 t_3^3 t_4 t_5^2 
- y_{s}^2 ~t_1 t_2^3 t_3 t_4^3 t_5^2
+ \dots~.
\eea

Consider the following fugacity map
\beal{esm12a_y1}
f_1 = t_3 t_4
~,~
f_2 = \frac{t_2 t_4^2}{t_1}
~,~
\tilde{t}_1 = y_s^{1/4} ~ t_1^{1/2}
~,~
\tilde{t}_2 = y_s^{1/4} ~ t_1^{1/2}
~,~
\tilde{t}_3 = \frac{t_2 t_3 t_4 t_5}{t_1}
~,~
\eea
where $f_1$ and $f_2$ are flavour charge fugacities, and $\tilde{t}_i$ is the fugacity for R-charge $R_i$ in \tref{t12a}. Under the fugacity map above, the above plethystic logarithm becomes
\beal{esm12a_3}
&&
PL[g_1(\tilde{t}_\alpha,f_1,f_2;\mathcal{M}_{12a}^{mes})]=
\left(
f_1  
+ f_2
\right) \tilde{t}_1^3 \tilde{t}_2
+ \left(
1
+ \frac{f_1}{f_2} 
+ \frac{f_2}{f_1}
\right) \tilde{t}_1^2 \tilde{t}_2^2 \tilde{t}_3 
\nn\\
&&
\hspace{1cm}
+ \left(
\frac{1}{f_1} 
+ \frac{1}{f_2}
\right) \tilde{t}_1 \tilde{t}_2^3 \tilde{t}_3^2
+ \frac{1}{f_1 f_2}\tilde{t}_2^4 \tilde{t}_3^3
- \left(
f_1
+ f_2
\right) \tilde{t}_1^5 \tilde{t}_2^3 \tilde{t}_3 
- \left(
3 
- \frac{f_1}{f_2} 
- \frac{f_2}{f_1}
\right) \tilde{t}_1^4 \tilde{t}_2^4 \tilde{t}_3^2 
+\dots~.
   \nn\\
   \eea
The above plethystic logarithm with its refinement exhibits all the moduli space generators with their mesonic charges. 
They are summarized in \tref{t12agen}. The generators can be presented on a charge lattice. The convex polygon formed by the generators in \tref{t12agen} is the dual reflexive polygon of the toric diagram of Model 12a.
\\

\begin{table}[H]
\centering
\resizebox{\hsize}{!}{
\begin{minipage}[!b]{0.5\textwidth}
\begin{tabular}{|l|c|c|}
\hline
Generator & $U(1)_{f_1}$ & $U(1)_{f_2}$ 
\\
\hline
\hline
$p_{1}^2 p_{3} p_{4} ~ s$
& 1 & 0
\nn\\
$p_{1} p_{2} p_{4}^2 ~ s$
   & 0 & 1
   \nn\\
$p_{1}^2 p_{3}^2 p_{5} ~ s$
   & 1 &-1
   \nn\\
$p_{1} p_{2} p_{3} p_{4} p_{5}
  ~ s$
   & 0 & 0
   \nn\\
$p_{2}^2 p_{4}^2 p_{5} ~ s$
& -1 & 1
\nn\\
   $p_{1} p_{2} p_{3}^2 p_{5}^2 ~ s$
   & 0 & -1
   \nn\\
$p_{2}^2 p_{3} p_{4} p_{5}^2 ~ s$
& -1 & 0
\nn\\
$p_{2}^2 p_{3}^2 p_{5}^3 ~ s$
   & -1 & -1
   \\
   \hline
\end{tabular}
\end{minipage}
\hspace{1cm}
\begin{minipage}[!b]{0.3\textwidth}
\includegraphics[width=4 cm]{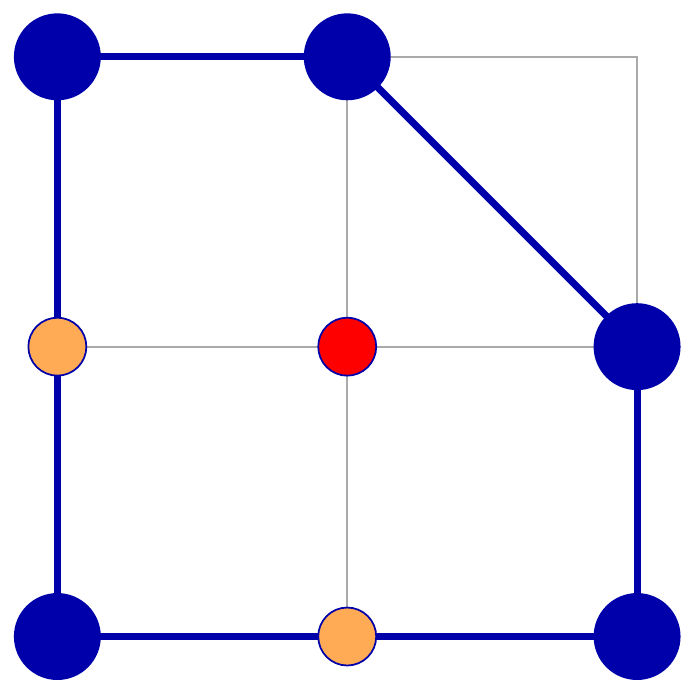}
\end{minipage}
}
\caption{The generators and lattice of generators of the mesonic moduli space of Model 12a in terms of GLSM fields with the corresponding flavor charges.\label{t12agen}\label{f12agen}} 
\end{table}

\begin{table}[H]
\centering
\resizebox{\hsize}{!}{
\begin{tabular}{|l|c|c|}
\hline
Generator & $U(1)_{f_1}$ & $U(1)_{f_2}$ 
\\
\hline
\hline
$X_ {25}^{1} X_{53} X_{32}=  X_{14} X_ {42}^{2} X_ {25}^{1} X_{51}$
& 1 & 0
\nn\\
$X_{14} X_ {42}^{2} X_{21}=  X_ {25}^{1} X_{53} X_{34} X_ {42}^{2}$
   & 0 & 1
   \nn\\
$X_{13} X_{32} X_ {25}^{1} X_{51}=  X_{14} X_ {42}^{1} X_ {25}^{1} X_{51}$
   & 1 & -1
   \nn\\
$ X_{13} X_{34} X_{42}^{2} X_{25}^{1} X_{51}=  X_{14} X_{42}^{2} X_{25}^{2} X_{51}=  X_{25}^{1} X_{53} X_{34} X_{42}^{1}=  X_{13} X_{32} X_{21}=  X_{14} X_{42}^{1} X_{21}=  X_{25}^{2} X_{53} X_{32}$
   & 0 & 0
   \nn\\
$ X_{13} X_{34} X_{42}^{2} X_{21}=  X_{25}^{2} X_{53} X_{34} X_{42}^{2}$
& -1 & 1
\nn\\
   $X_{13} X_{34} X_ {42}^{1} X_ {25}^{1} X_{51}=  X_{13} X_{32} X_{25}^{2} X_{51}=  X_{14} X_ {42}^{1} X_ {25}^{2} X_{51}$
   & 0 & 1
   \nn\\
$X_{13} X_{34} X_ {42}^{2} X_ {25}^{2} X_{51}=  X_{13} X_{34} X_{42}^{1} X_{21}=  X_ {25}^{2} X_{53} X_{34} X_ {42}^{1}$
& -1 & 0
\nn\\
$X_{13} X_{34} X_ {42}^{1} X_ {25}^{2} X_{51}$
   & -1 & -1
   \\
   \hline
\end{tabular}
}
\caption{The generators in terms of bifundamental fields (Model 12a).\label{t12agen2}\label{f12agen2}} 
\end{table}

The mesonic Hilbert series and the plethystic logarithm can be re-expressed in terms of just $3$ fugacities
\beal{esm12a_x1}
T_1 = \frac{\tilde{t}_3}{f_1 f_2 ~ \tilde{t}_1^4}
= \frac{t_5}{y_{s} ~t_1^2 t_4^2} ~,~
T_2 = f_1 ~ \tilde{t}_1^3 \tilde{t}_2
= y_{s} ~t_1^2 t_3 t_4~,~
T_3 = f_2 ~ \tilde{t}_1^3 \tilde{t}_2 
= y_{s} ~t_1 t_2 t_4^2~,
\eea
such that
\beal{esm12a_x2}
&&
g_1(T_1,T_2,T_3;\mathcal{M}^{mes}_{12a})=
\nn\\&&
   \hspace{0.8cm}
(1 + T_1 T_2 T_3 - T_1 T_2^2 T_3 - 
   T_1 T_2 T_3^2
+ T_1^2 T_2^2 T_3 + T_1^2 T_2 T_3^2 - T_1^2 T_2^3 T_3   - 2 T_1^2 T_2^2 T_3^2 
   \nn\\
   &&
   \hspace{1cm}
   - T_1^2 T_2 T_3^3 
   + T_1^2 T_2^3 T_3^2+ T_1^2 T_2^2 T_3^3
    -T_1^3 T_2^3 T_3^2  - 
   T_1^3 T_2^2 T_3^3 + T_1^3 T_2^3 T_3^3 + 
   T_1^4 T_2^4 T_3^4)
   \nn\\
   &&
   \hspace{0.8cm}
   \times
\frac{1}{
(1 - T_2)  (1 - T_3) (1 - T_1 T_2^2) (1 - T_1 T_3^2) (1 - T_1^3 T_2^2 T_3^2)
}
\nn\\
\eea
and
\beal{esm12a_x3}
&&
PL[g_1(T_1,T_2,T_3;\mathcal{M}^{mes}_{12a})]=
T_2  
+ T_3 
+ T_1 T_2 T_3   
+ T_1 T_2^2 
+ T_1 T_3^2 
+ T_1^2 T_2 T_3^2 
+ T_1^2 T_2^2 T_3 
 \nn\\
 &&
 \hspace{1cm}
+ T_1^3 T_2^2 T_3^2 
- T_1 T_2^2 T_3 
- T_1 T_2 T_3^2 
- T_1^2 T_2^3 T_3
- 3 T_1^2 T_2^2 T_3^2   
- T_1^2 T_2^3 T_3 
- T_1^2 T_2 T_3^3 
 \nn\\
 &&
 \hspace{1cm}
  +\dots~~.
\eea
The above Hilbert series and plethystic logarithm illustrate the conical structure of the toric Calabi-Yau 3-fold.
\\

\subsection{Model 12 Phase b}

\begin{figure}[H]
\begin{center}
\includegraphics[trim=0cm 0cm 0cm 0cm,width=4.5 cm]{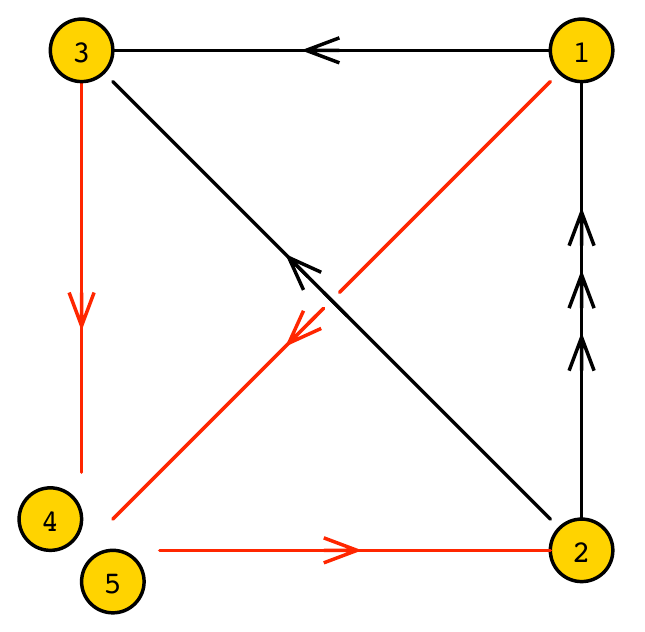}
\includegraphics[width=5 cm]{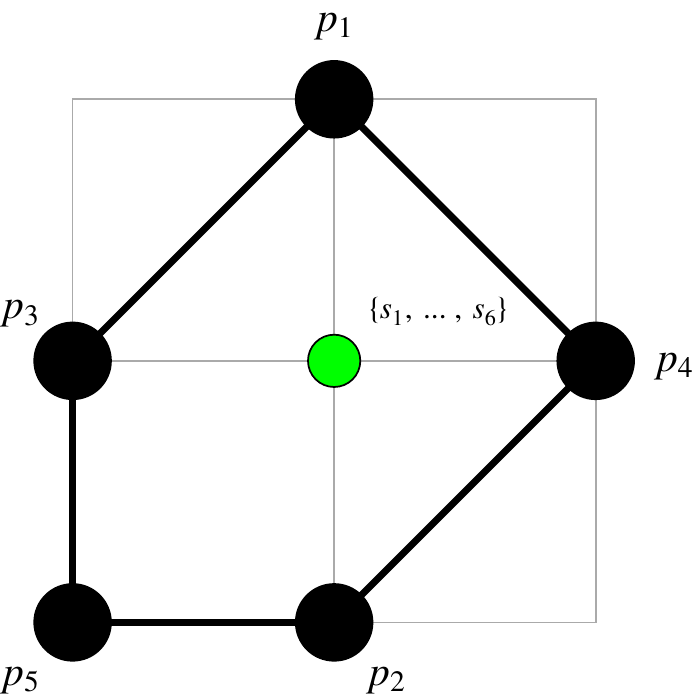}
\includegraphics[width=5 cm]{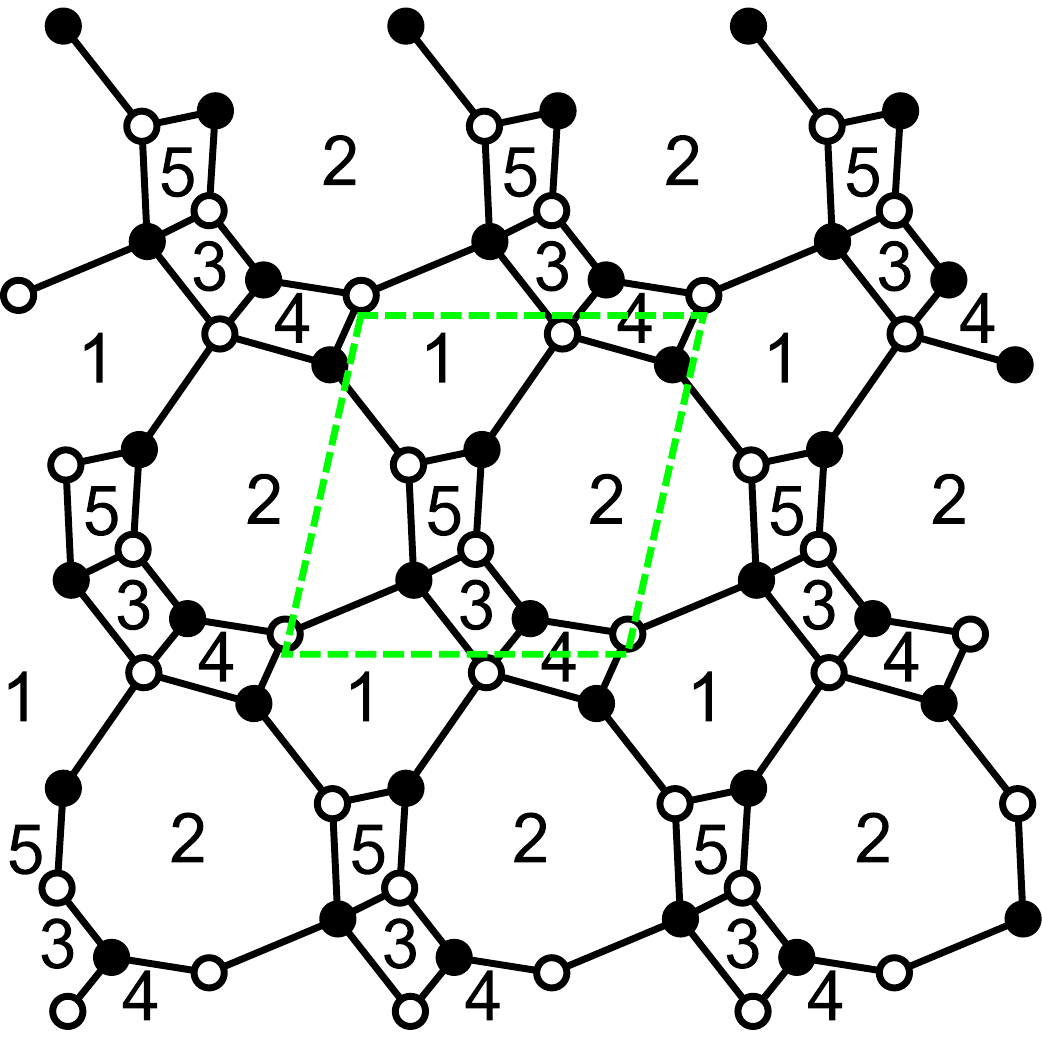}
\caption{The quiver, toric diagram, and brane tiling of Model 12b. The red arrows in the quiver indicate all possible connections between blocks of nodes.}
  \label{f12b}
 \end{center}
 \end{figure}
 
 \noindent The superpotential is 
\beal{esm12b_00}
W&=&
+X_{15} X_{52}^{2} X_{21}^{2} 
+X_{21}^{1} X_{14} X_{42}^{1} 
+X_{35} X_{52}^{1} X_{23} 
+X_{13} X_{34} X_{42}^{2} X_{21}^{3} 
\nn\\
&&
-X_{14} X_{42}^{2} X_{21}^{2} 
-X_{15} X_{52}^{1} X_{21}^{3} 
-X_{34} X_{42}^{1} X_{23} 
-X_{21}^{1} X_{13} X_{35} X_{52}^{2}
~~.
\eea
 
\noindent The perfect matching matrix is 

\noindent\makebox[\textwidth]{%
\footnotesize
$
P=
\left(
\begin{array}{c|ccccc|cccccc}
 \; & p_{1} & p_{2} & p_{3} & p_{4} & p_{5} & s_{1} &
   s_{2} & s_{3} & s_{4} & s_{5} & s_{6} \\
   \hline
 X_{21}^{1} & 1 & 0 & 1 & 0 & 0 & 1 & 0 & 0 & 0 & 0 & 0 \\
 X_{42}^{2} & 1 & 0 & 0 & 0 & 0 & 0 & 1 & 1 & 0 & 0 & 0 \\
 X_{21}^{2} & 0 & 1 & 1 & 0 & 1 & 1 & 0 & 0 & 0 & 0 & 0 \\
 X_{21}^{3} & 0 & 1 & 0 & 1 & 0 & 1 & 0 & 0 & 0 & 0 & 0 \\
 X_{23} & 1 & 0 & 0 & 1 & 0 & 1 & 0 & 0 & 0 & 0 & 1 \\
 X_{42}^{1} & 0 & 1 & 0 & 0 & 1 & 0 & 1 & 1 & 0 & 0 & 0 \\
 X_{52}^{1} & 0 & 0 & 1 & 0 & 1 & 0 & 1 & 0 & 1 & 0 & 0 \\
 X_{52}^{2} & 0 & 0 & 0 & 1 & 0 & 0 & 1 & 0 & 1 & 0 & 0 \\
 X_{15} & 1 & 0 & 0 & 0 & 0 & 0 & 0 & 1 & 0 & 1 & 1 \\
 X_{35} & 0 & 1 & 0 & 0 & 0 & 0 & 0 & 1 & 0 & 1 & 0 \\
 X_{34} & 0 & 0 & 1 & 0 & 0 & 0 & 0 & 0 & 1 & 1 & 0 \\
 X_{14} & 0 & 0 & 0 & 1 & 0 & 0 & 0 & 0 & 1 & 1 & 1 \\
 X_{13} & 0 & 0 & 0 & 0 & 1 & 0 & 0 & 0 & 0 & 0 & 1
\end{array}
\right)
$
}
\vspace{0.5cm}

 \noindent The F-term charge matrix $Q_F=\ker{(P)}$ is

\noindent\makebox[\textwidth]{%
\footnotesize
$
Q_F=
\left(
\begin{array}{ccccc|cccccc}
 p_{1} & p_{2} & p_{3} & p_{4} & p_{5} & s_{1} &
   s_{2} & s_{3} & s_{4} & s_{5} & s_{6} \\
   \hline
 1 & 1 & 0 & 0 & 0 & -1 & 0 & -1 & 0 & 0 & 0 \\
 0 & 0 & 1 & 1 & 0 & -1 & 0 & 0 & -1 & 0 & 0 \\
 0 & 1 & 1 & 0 & -1 & -1 & 0 & 0 & 0 & -1 & 1 \\
 0 & 0 & 0 & 0 & 0 & 0 & 1 & -1 & -1 & 1 & 0
\end{array}
\right)
$
}
\vspace{0.5cm}

\noindent The D-term charge matrix is

\noindent\makebox[\textwidth]{%
\footnotesize
$
Q_D=
\left(
\begin{array}{ccccc|cccccc}
 p_{1} & p_{2} & p_{3} & p_{4} & p_{5} & s_{1} &
   s_{2} & s_{3} & s_{4} & s_{5} & s_{6} \\
   \hline
 0 & 0 & 0 & 0 & 0 & 1 & -1 & 0 & 0 & 0 & 0 \\
 0 & 0 & 0 & 0 & 0 & 0 & 0 & 1 & -1 & 0 & 0 \\
 0 & 0 & 0 & 0 & 0 & 0 & 0 & 0 & 1 & -1 & 0 \\
 0 & 0 & 0 & 0 & 0 & 0 & 0 & 0 & 0 & 1 & -1
   \end{array}
\right)
$
}
\vspace{0.5cm}

The total charge matrix $Q_t$ does not have repeated columns. Accordingly, the global symmetry is $U(1)_{f_1} \times U(1)_{f_2} \times U(1)_R$. The charge assignment on the extremal perfect matchings with non-zero R-charge is the the same as for Model 12a in \tref{t12a}.

The product of all internal perfect matchings is expressed as
\beal{esm12b_x1}
s=\prod_{m=1}^{6} s_m ~.
\eea
The product is counted by the fugacity $y_{s}$. The remaining extremal perfect matchings $p_\alpha$ are counted by the fugacity $t_\alpha$.

The mesonic Hilbert series and the plethystic logarithm of the Hilbert series is the same as for Model 12a. They are shown respectively in \eref{esm12a_1}, \eref{esm12a_2} and \eref{esm12a_3}. Accordingly, the mesonic moduli spaces of Model 12a and 12b are toric duals. 

The moduli space generators in terms of perfect matching variables of Model 12b are shown in \tref{t12agen} with their corresponding mesonic charges. The generators in terms of quiver fields are shown in \tref{t12bgen2}.

\comment{
\begin{table}[H]
\centering
\resizebox{\hsize}{!}{
\begin{minipage}[!b]{0.5\textwidth}
\begin{tabular}{|l|c|c|}
\hline
Generator & $U(1)_{f_1}$ & $U(1)_{f_2}$ 
\\
\hline
\hline
$p_{1}^2 p_{3} p_{4} ~ \prod_{m=1}^{6} s_m$
& 1 & 0
\nn\\
$p_{1} p_{2} p_{4}^2 ~ \prod_{m=1}^{6} s_m$
   & 0 & 1
   \nn\\
$p_{1}^2 p_{3}^2 p_{5} ~ \prod_{m=1}^{6} s_m$
   & 1 &-1
   \nn\\
$p_{1} p_{2} p_{3} p_{4} p_{5}
  ~ \prod_{m=1}^{6} s_m$
   & 0 & 0
   \nn\\
$p_{2}^2 p_{4}^2 p_{5} ~ \prod_{m=1}^{6} s_m$
& -1 & 1
\nn\\
   $p_{1} p_{2} p_{3}^2 p_{5}^2 ~ \prod_{m=1}^{6} s_m$
   & 0 & -1
   \nn\\
$p_{2}^2 p_{3} p_{4} p_{5}^2 ~ \prod_{m=1}^{6} s_m$
& -1 & 0
\nn\\
$p_{2}^2 p_{3}^2 p_{5}^3 ~ \prod_{m=1}^{6} s_m$
   & -1 & -1
   \\
   \hline
\end{tabular}
\end{minipage}
\hspace{1cm}
\begin{minipage}[!b]{0.3\textwidth}
\includegraphics[width=4 cm]{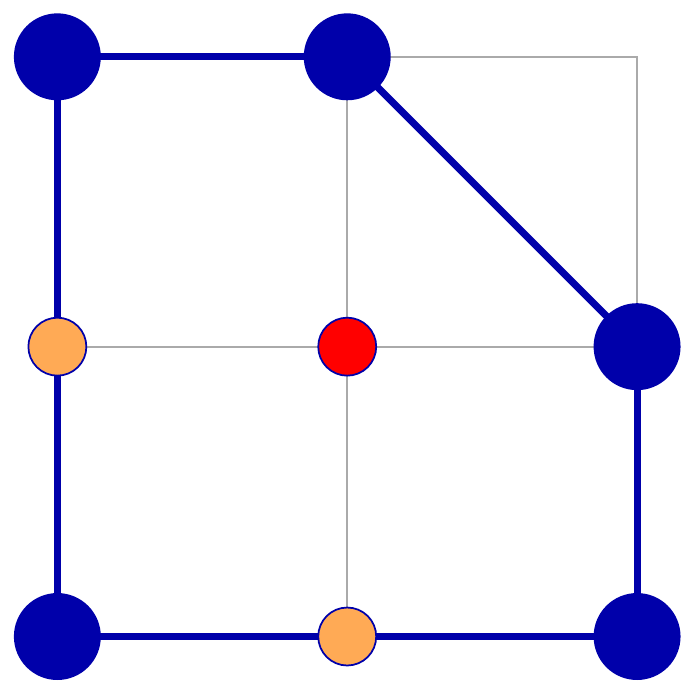}
\end{minipage}
}
\caption{The generators and lattice of generators of the mesonic moduli space of Model 12b in terms of GLSM fields with the corresponding flavor charges. The lattice of generators is precisely the toric diagram of Model 6 (PdP$_4$).\label{t12bgen}\label{f12bgen}} 
\end{table}
}

\begin{table}[H]
\centering
\resizebox{\hsize}{!}{
\begin{tabular}{|l|c|c|}
\hline
Generator & $U(1)_{f_1}$ & $U(1)_{f_2}$ 
\\
\hline
\hline
$X_{14} X_ {42}^{2} X_ {21}^{1}=  X_{15} X_ {52}^{2} X_ {21}^{1}=  X_{23} X_{34} X_ {42}^{2}$
   & 1 & 0
   \\
$ X_{14} X_{42}^{2} X_{21}^{3}=  X_{15} X_{52}^{2} X_{21}^{3}=  X_{23} X_{35} X_{52}^{2}$
   & 0 & 1
   \\
$ X_{15} X_{52}^{1} X_{21}^{1}=  X_{13} X_{34} X_{42}^{2} X_{21}^{1}$
   & 1 & -1
   \\
$X_{13} X_{35} X_ {52}^{2} X_ {21}^{1}=  X_{13} X_{34} X_ {42}^{2} X_{21}^{3}=  X_{14} X_ {42}^{1} X_ {21}^{1}=  X_{14} X_ {42}^{2} X_{21}^{2}=  X_{15} X_ {52}^{2} X_ {21}^{2}=  X_{15} X_ {52}^{1} X_{21}^{3}=  X_{23} X_{34} X_ {42}^{1}=  X_{23} X_{35} X_ {52}^{1}$
   & 0 & 0
   \\
$ X_{14} X_{42}^{1} X_{21}^{3}=  X_{13} X_{35} X_{52}^{2} X_{21}^{3}$
   & -1 & 1
   \\
$ X_{15} X_{52}^{1} X_{21}^{2}=  X_{13} X_{34} X_{42}^{1} X_{21}^{1}=  X_{13} X_{35} X_{52}^{1} X_{21}^{1}=  X_{13} X_{34} X_{42}^{2} X_{21}^{2}$
   & 0 & -1
   \\
$X_{14} X_ {42}^{1} X_ {21}^{2}=  X_{13} X_{35} X_ {52}^{2} X_{21}^{2}=  X_{13} X_{34} X_ {42}^{1} X_ {21}^{3}=  X_{13} X_{35} X_{52}^{1} X_ {21}^{3}$
   & -1 & 0
   \\
   $ X_{13} X_{34} X_{42}^{1} X_{21}^{2}=  X_{13} X_{35} X_{52}^{1} X_{21}^{2}$
   & -1 & -1
   \\
   \hline
\end{tabular}
}
\caption{The generators in terms of bifundamental fields (Model 12b).\label{t12bgen2}\label{f12bgen2}} 
\end{table}

\section{Model 13: $\mathbb{C}^3/\mathbb{Z}_{4,(1,1,2)}$,~$Y^{2,2}$}

\begin{figure}[H]
\begin{center}
\includegraphics[trim=0cm 0cm 0cm 0cm,width=4.5 cm]{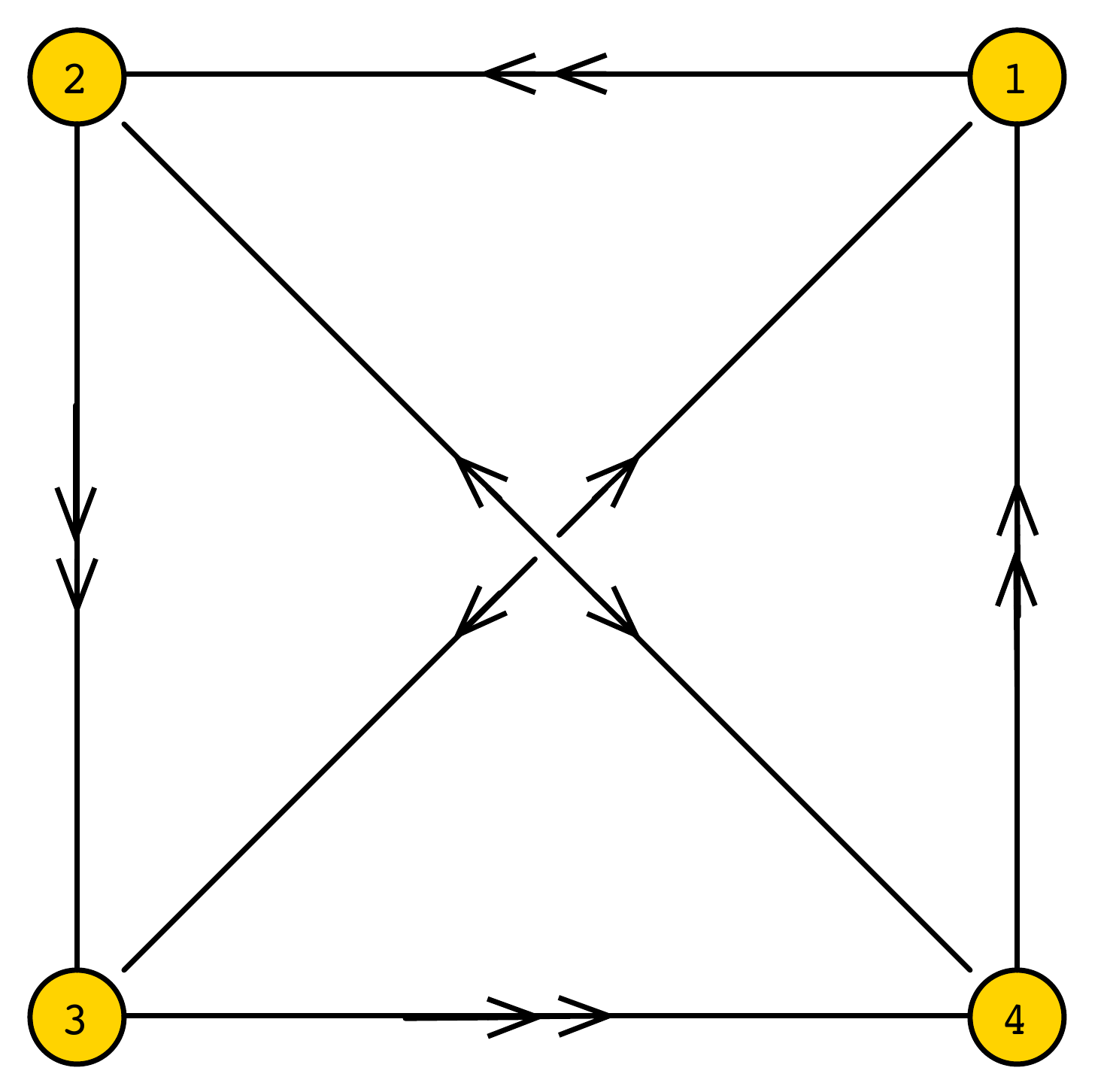}
\includegraphics[width=5 cm]{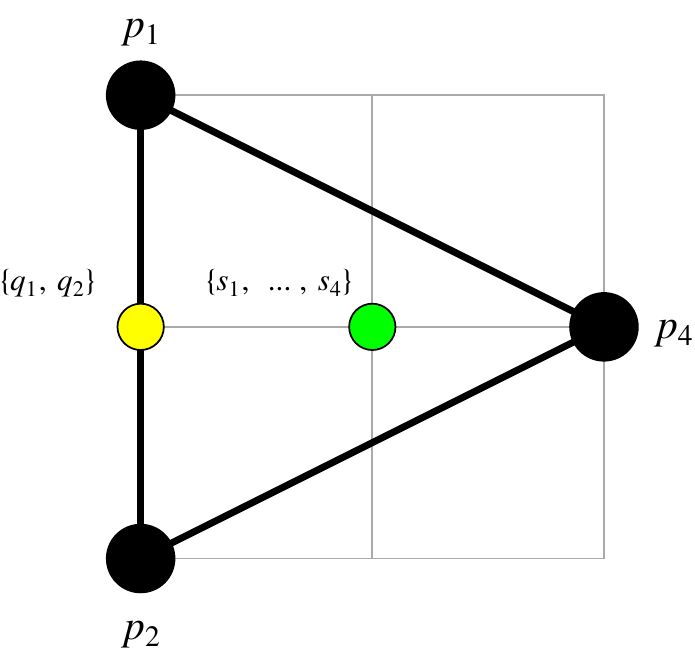}
\includegraphics[width=5 cm]{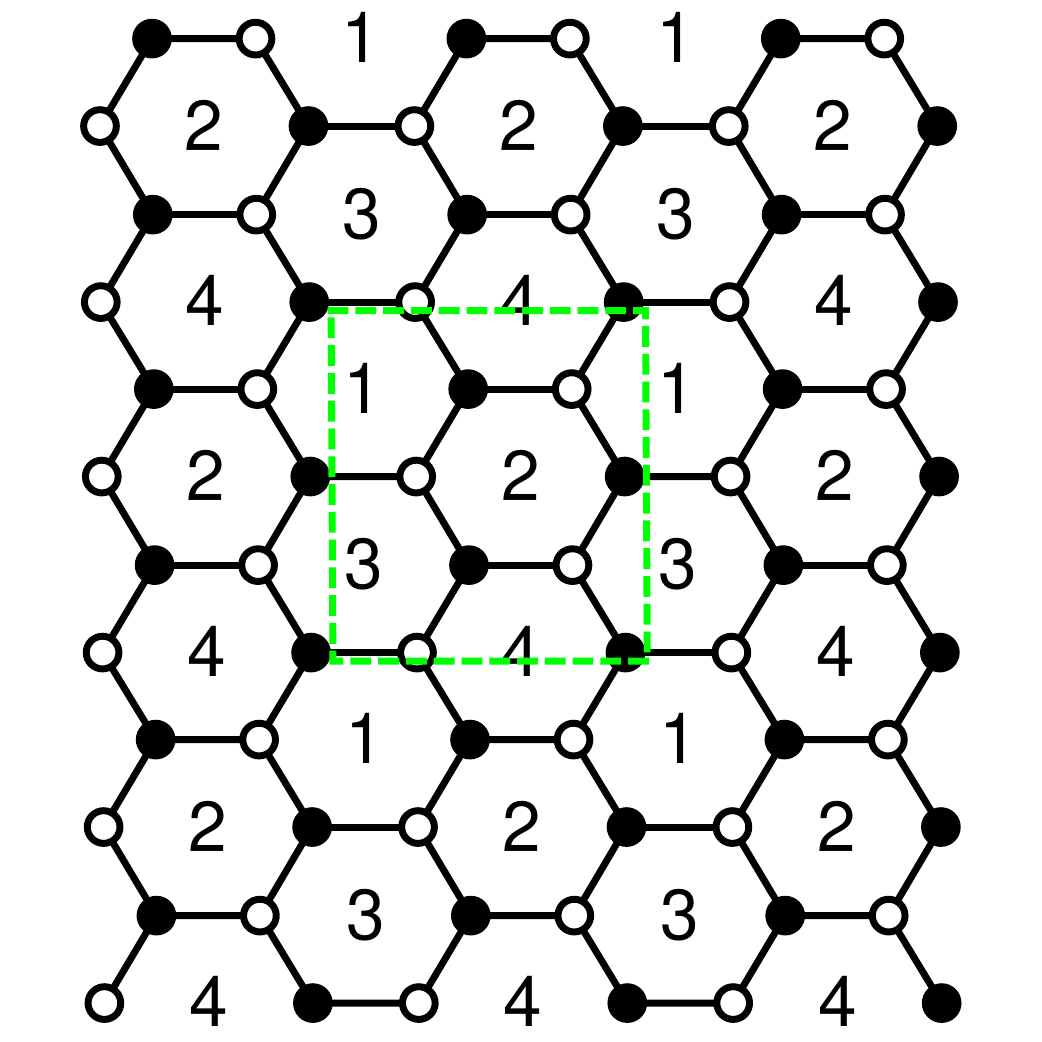}
\caption{The quiver, toric diagram, and brane tiling Model 13.}
  \label{f13}
 \end{center}
 \end{figure}
 
 \noindent The superpotential is 
\beal{esm13_00}
W&=&
+ X_{12} X_{24} X_{41}^{1} 
+ X_{31} X_{12}^{2} X_{23}^{2} 
+ X_{41}^{2} X_{13} X_{34}^{1}
+ X_{34}^{2} X_{42} X_{23}^{1} 
\nn\\
&&
- X_{12} X_{23} X_{31} 
- X_{13} X_{34}^{2} X_{41}^{1} 
- X_{41}^{2} X_{12}^{2} X_{24} 
- X_{34}^{1} X_{42} X_{23}^{2} 
~~.
\eea
 
\noindent The perfect matching matrix is 

\noindent\makebox[\textwidth]{%
\footnotesize
$
P=
\left(
\begin{array}{c|ccc|cc|cccc}
 \; & p_{1} & p_{2} & p_{3} & q_{1} & q_{2} & s_{1} &
   s_{2} & s_{3} & s_{4} \\
   \hline
 X_{34}^{1} & 1 & 0 & 0 & 1 & 0 & 1 & 0 & 0 & 0 \\
 X_{34}^{2} & 0 & 1 & 0 & 1 & 0 & 1 & 0 & 0 & 0 \\
 X_{12}^{2} & 1 & 0 & 0 & 1 & 0 & 0 & 1 & 0 & 0 \\
 X_{12}^{1} & 0 & 1 & 0 & 1 & 0 & 0 & 1 & 0 & 0 \\
 X_{23}^{1} & 1 & 0 & 0 & 0 & 1 & 0 & 0 & 1 & 0 \\
 X_{23}^{2} & 0 & 1 & 0 & 0 & 1 & 0 & 0 & 1 & 0 \\
 X_{41}^{1} & 1 & 0 & 0 & 0 & 1 & 0 & 0 & 0 & 1 \\
 X_{41}^{2} & 0 & 1 & 0 & 0 & 1 & 0 & 0 & 0 & 1 \\
 X_{24} & 0 & 0 & 1 & 0 & 0 & 1 & 0 & 1 & 0 \\
 X_{31} & 0 & 0 & 1 & 0 & 0 & 1 & 0 & 0 & 1 \\
 X_{13} & 0 & 0 & 1 & 0 & 0 & 0 & 1 & 1 & 0 \\
 X_{42} & 0 & 0 & 1 & 0 & 0 & 0 & 1 & 0 & 1
\end{array}
\right)
$
}
\vspace{0.5cm}

 \noindent The F-term charge matrix $Q_F=\ker{(P)}$ is

\noindent\makebox[\textwidth]{%
\footnotesize
$
Q_F=
\left(
\begin{array}{ccc|cc|cccc}
 p_{1} & p_{2} & p_{3} & q_{1} & q_{2} & s_{1} &
   s_{2} & s_{3} & s_{4} \\
   \hline
    1 & 1 & 0 & -1 & -1 & 0 & 0 & 0 & 0 \\
 0 & 0 & 1 & 1 & 0 & -1 & -1 & 0 & 0 \\
 0 & 0 & 1 & 0 & 1 & 0 & 0 & -1 & -1
\end{array}
\right)
$
}
\vspace{0.5cm}

\noindent The D-term charge matrix is

\noindent\makebox[\textwidth]{%
\footnotesize
$
Q_D=
\left(
\begin{array}{ccc|cc|cccc}
p_{1} & p_{2} & p_{3} & q_{1} & q_{2} & s_{1} &
   s_{2} & s_{3} & s_{4} \\
   \hline
 0 & 0 & 0 & 0 & 0 & 1 & -1 & 0 & 0 \\
 0 & 0 & 0 & 0 & 0 & 0 & 1 & -1 & 0 \\
 0 & 0 & 0 & 0 & 0 & 0 & 0 & 1 & -1
   \end{array}
\right)
$
}
\vspace{0.5cm}

The GLSM fields $p_1$ and $p_2$ are equally charged under the F-term and D-term constraints. This is shown by the corresponding columns in the total charge matrix $Q_t$ which are identical. Accordingly, the global symmetry is enhanced from $U(1)^3$ to $SU(2)_{x} \times U(1)_{f} \times U(1)_R$ with $U(1)_R$ being the R-symmetry. The mesonic charges on the GLSM fields corresponding to extremal points in the toric diagram in \fref{f13} are found following the discussion in \sref{s1_3}. They are presented in \tref{t13}.

\begin{table}[H]
\centering
\begin{tabular}{|c||c|c|c||l|} 
\hline
\; & $U(1)_f$ & $SU(2)_x$ & $U(1)_R$ & fugacity \\
\hline
\hline
$p_1$ &-1/4 	& 1/2 	& 2/3 &  $t_1$\\
$p_2$ &-1/4 	&-1/2 	& 2/3 &  $t_2$\\
$p_3$ &1/2 	& 0 		& 2/3 &  $t_3$\\
\hline
\end{tabular}
\caption{The GLSM fields corresponding to extremal points of the toric diagram with their mesonic charges (Model 13).\label{t13}}
\end{table}

Products of non-extremal perfect matchings are expressed as follows
\beal{esm13}
q = q_1 q_2 ~,~
s = \prod_{m=1}^{4} s_m~.
\eea
The fugacities counting the above products are respectively $y_{q}$ and $y_{s}$. The fugacity which counts extremal perfect matchings is $t_\alpha$.

The mesonic Hilbert series of Model 13 is computed using the Molien integral formula in \eref{es12_2}. It is 
 \beal{esm13_1}
 &&
g_{1}(t_\alpha,y_{q},y_{s}; \mathcal{M}^{mes}_{13})
=
\nn\\
&&
\hspace{0.5cm}
\frac{
1 
+ y_{q}^2 y_{s} ~t_1^3 t_2 
+ y_{q}^2 y_{s} ~t_1^2 t_2^2 
+ y_{q}^2 y_{s} ~t_1 t_2^3 
+ y_{q} y_{s} ~t_1^2 t_3 
+ y_{q} y_{s} ~t_1 t_2 t_3 
+ y_{q} y_{s} ~t_2^2 t_3 
+ y_{q}^3 y_{s}^2 ~t_1^3 t_2^3 t_3
 }{
(1 - y_{q}^2 y_{s} ~t_1^4) 
(1 - y_{q}^2 y_{s} ~t_2^4) 
(1 - y_{s} ~t_3^2)
}~~.
\nn\\
 \eea
The mesonic moduli space of Model 13 is not a complete intersection.
 The plethystic logarithm of the mesonic Hilbert series is
\beal{esm13_3}
&&
PL[g_1(t_\alpha,y_{q},y_{s};\mathcal{M}_{13}^{mes})]=
y_{s}~t_{3}^2
+ y_{q}y_{s}~t_{1} t_{2} t_{3}
+ y_{q}y_{s}~t_{1}^2 t_{3} 
+ y_{q}y_{s}~t_{2}^2 t_{3}
+ y_{q}^2 y_{s}~t_{1}^4
\nn\\
&&
\hspace{1cm}
+ y_{q}^2 y_{s}~t_{1}^3 t_{2}
+ y_{q}^2 y_{s}~t_{1}^2 t_{2}^2
+ y_{q}^2 y_{s}~t_{1} t_{2}^3
+ y_{q}^2 y_{s}~t_{2}^4
- 2~y_{q}^2 y_{s}^2~ t_{1}^2 t_{2}^2 t_{3}^2 
+ \dots~.
\eea

Consider the following fugacity map
\beal{esm13}
f = y_q^{-2/3} y_s^{1/3} ~ t_1^{-2/3} t_2^{-2/3} t_3^{4/3}
~,~
\tilde{x}^2 = x = \frac{t_1}{t_2}
~,~
t = y_q^{1/3} y_s^{1/3} ~ t_1^{1/3} t_2^{1/3} t_3^{1/3}
~,~
\eea 
where the fugacities $f$, $x$ and $t$ are mesonic charge fugacities. $x$ is the charge fugacity for the enhanced symmetry $SU(2)_x$. Using the redefinition of this fugacity to $\tilde{x}=\sqrt{x}$ and the fugacities $f$ and $t$, one can rewrite the expansion of the Hilbert series in terms of characters of irreducible representations of $SU(2)$ as follows
\beal{esm13_1b}
g_{1}(t,\tilde{x},f;\mathcal{M}^{mes}_{13})=
\sum_{m=0}^{\infty}\sum_{n=0}^{\infty}
~
\left(
[2m]_{\tilde{x}} f^{n} t^{2n+3m}
+
[4(n+1)+2m]_{\tilde{x}} f^{-(n+1)} t^{4(n+1)+3m}
\right)~~.
\nn\\
\eea
The corresponding plethystic logarithm is
\beal{esm13_3c}
&&
PL[g_1(t,\tilde{x},f;\mathcal{M}_{13}^{mes})]
=
f t^2 
+[2]_{\tilde{x}} t^3 
+ [4]_{\tilde{x}} \frac{1}{f} t^4 
- (1+[4]_{\tilde{x}}) t^6 
- ([2]_{\tilde{x}}+[4]_{\tilde{x}}) \frac{1}{f} t^7 
\nn\\
&&
\hspace{1cm}
- (1+[4]_{\tilde{x}}) \frac{1}{f^2} t^8 
+ ([2]_{\tilde{x}}+[4]_{\tilde{x}}) t^9 
+ (1+2[2]_{\tilde{x}}+2[4]_{\tilde{x}} +[6]_{\tilde{x}}) \frac{1}{f} t^{10} 
+ \dots~.
\nn\\
\eea
In terms of the mesonic charge fugacities $f$, $x$ and $t$, the above plethystic logarithm exhibits the moduli space generators and their mesonic charges. They are summarized in \tref{t13gen}. The flavour charges of generators are integers using $f$ and $x$. They can be presented on a charge lattice. The convex polygon formed by the generators is the dual reflexive polygon of the toric diagram. 

As indicated in \eref{esm13_3c}, the generators fall into irreducible representation of $SU(2)$ with the characters
\beal{esm13_xx5}
f t^2 
+[2]_{\tilde{x}} t^3 
+ [4]_{\tilde{x}} \frac{1}{f} t^4
=
f t^2 
+
\left(
\tilde{x}^2 + 1 + \frac{1}{\tilde{x}^2}
\right) 
t^3
+
\left(
\tilde{x}^4 + \tilde{x}^2 + 1 + \frac{1}{\tilde{x}^2} + \frac{1}{\tilde{x}^4}
\right)\frac{1}{f}
t^4
~~.
\nn\\
\eea
The above three terms correspond to the three columns of points in the lattice of generators in \tref{f13gen}. The generators in terms of quiver fields are shown in \tref{f13gen2}.
\\

\begin{table}[H]
\centering
\resizebox{\hsize}{!}{
\begin{minipage}[!b]{0.6\textwidth}
\begin{tabular}{|l|c|c|}
\hline
Generator & $U(1)_{f}$ & $SU(2)_{x}$ 
\\
\hline
\hline
$p_{3}^2 ~ s$
& 1 & 0
\nn\\
$p_{1}^2 p_{3} ~ q ~ s$
& 0 & 1
\nn\\
$p_{1} p_{2} p_{3} ~ q ~ s$
& 0 & 0
\nn\\
$p_{2}^2 p_{3} ~ q ~ s$
& 0 & -1
\nn\\
$p_{1}^4 ~ q^2 ~ s$
& -1 & 2
\nn\\
$p_{1}^3 p_{2} ~ q^2 ~ s$
& -1 & 1
\nn\\
$p_{1}^2 p_{2}^2 ~ q^2 ~ s$
& -1 & 0
\nn\\
$p_{1} p_{2}^3 ~ q^2 ~ s$
& -1 & -1
\nn\\
$p_{2}^4 ~ q^2 ~ s$
& -1 & -2
   \\
   \hline
\end{tabular}
\end{minipage}
\hspace{1cm}
\begin{minipage}[!b]{0.3\textwidth}
\includegraphics[width=4 cm]{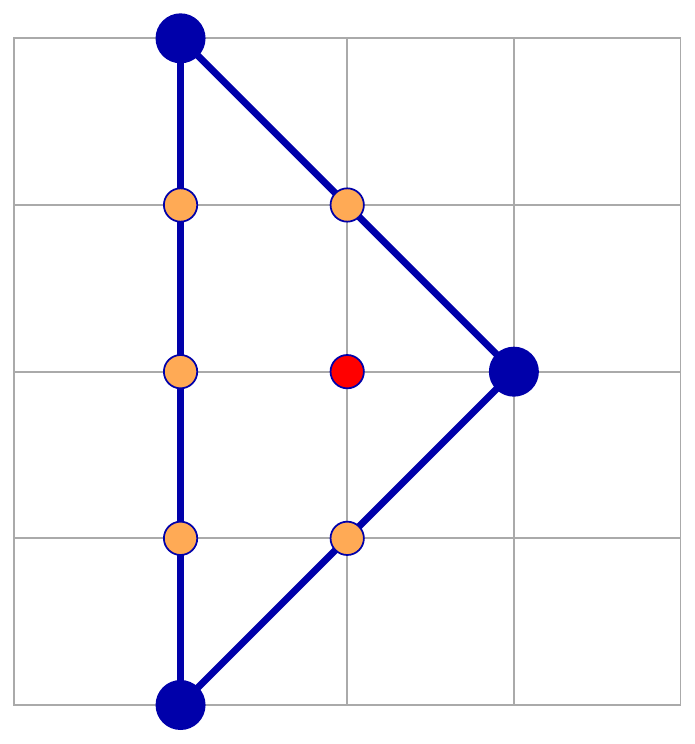}
\end{minipage}
}
\caption{The generators and lattice of generators of the mesonic moduli space of Model 13 in terms of GLSM fields with the corresponding flavor charges. \label{t13gen}\label{f13gen}} 
\end{table}

\begin{table}[H]
\centering
\resizebox{\hsize}{!}{
\begin{tabular}{|l|c|c|}
\hline
Generator & $U(1)_{f}$ & $SU(2)_{x}$ 
\\
\hline
\hline
$X_{13} X_{31}=  X_{24} X_{42}$
& 1 & 0
\nn\\
$X_ {12}^{2} X_ {23}^{1} X_{31}=  X_ {12}^{2} X_{24} X_ {41}^{1}=  X_{13} X_ {34}^{1} X_ {41}^{1}=  X_ {23}^{1} X_ {34}^{1} X_{42}$
& 0 & 1
\nn\\
$X_ {12}^{1} X_ {23}^{1} X_{31}=  X_ {12}^{1} X_{24} X_ {41}^{1}=  X_{12}^{2} X_ {23}^{2} X_{31}=  X_ {12}^{2} X_{24} X_ {41}^{2}=  X_{13} X_ {34}^{1} X_ {41}^{2}=  X_{13} X_ {34}^{2} X_ {41}^{1}=  X_{23}^{1} X_ {34}^{2} X_{42}=  X_ {23}^{2} X_ {34}^{1} X_{42}$
& 0 & 0
\nn\\
$ X_{12}^{1} X_{23}^{2} X_{31}=  X_{12}^{1} X_{24} X_{41}^{2}=  X_{13} X_{34}^{2} X_{41}^{2}=  X_{23}^{2} X_{34}^{2} X_{42}$
& 0 & -1
\nn\\
$ X_{12}^{2} X_{23}^{1} X_{34}^{1} X_{41}^{1}$
& -1 & 2
\nn\\
$X_ {12}^{1} X_ {23}^{1} X_ {34}^{1} X_ {41}^{1}=  X_ {12}^{2} X_{23}^{1} X_ {34}^{1} X_ {41}^{2}=  X_ {12}^{2} X_ {23}^{1} X_{34}^{2} X_ {41}^{1}=  X_ {12}^{2} X_ {23}^{2} X_ {34}^{1} X_ {41}^{1}$
& 1 & -1
\nn\\
$ X_{12}^{1} X_{23}^{1} X_{34}^{1} X_{41}^{2}=  X_{12}^{1} X_{23}^{1} X_{34}^{2} X_{41}^{1}=  X_{12}^{1} X_{23}^{2} X_{34}^{1} X_{41}^{1}=  X_{12}^{2} X_{23}^{1} X_{34}^{2} X_{41}^{2}=  X_{12}^{2} X_{23}^{2} X_{34}^{1} X_{41}^{2}=  X_{12}^{2} X_{23}^{2} X_{34}^{2} X_{41}^{1}$
& -1 & 0
\nn\\
$X_ {12}^{1} X_ {23}^{1} X_ {34}^{2} X_ {41}^{2}=  X_ {12}^{1} X_{23}^{2} X_ {34}^{1} X_ {41}^{2}=  X_ {12}^{1} X_ {23}^{2} X_{34}^{2} X_ {41}^{1}=  X_ {12}^{2} X_ {23}^{2} X_ {34}^{2} X_ {41}^{2}$
& -1 & -1
\nn\\
$X_ {12}^{1} X_ {23}^{2} X_ {34}^{2} X_ {41}^{2}$
& -1 & -2
   \\
   \hline
\end{tabular}
}
\caption{The generators in terms of bifundamental fields (Model 13).\label{t13gen2}\label{f13gen2}} 
\end{table}

With the fugacity map
\beal{esm13_xx1}
T_1 = f^{-1/4} x^{1/2} ~ t
= y_{q}^{1/2} y_{s}^{1/4} t_1 ~,~
T_2 = f^{-1/4} x^{-1/2} ~ t
= y_{q}^{1/2} y_{s}^{1/4} t_2~,~
T_3 = f^{1/2} ~ t
= y_{s}^{1/2} t_3~,
\nn\\
\eea
the mesonic Hilbert series takes the form
\beal{esm13_xx2}
g_1(T_1,T_2,T_3;\mathcal{M}^{mes}_{13})
=
\frac{
1 
+ T_1^3 T_2
+ T_1^2 T_2^2
+ T_1 T_2^3
+ T_1^2 T_3
+ T_1 T_2 T_3
+ T_2^2 T_3
+ T_1^3 T_2^3 T_3
}{
(1-T_1^4)
(1-T_2^4)
(1-T_3^2)
}~~,
\nn\\
\eea
with the plethystic logarithm becoming
\beal{esm13_xx3}
&&
PL[g_1(T_1,T_2,T_3;\mathcal{M}^{mes}_{13})]
=
T_3^2 
+ T_1 T_2 T_3
+ T_1^2 T_3
+ T_2^2 T_3 
+ T_1^4
+ T_1^3 T_2
+ T_1^2 T_2^2
\nn\\
&&
\hspace{0.5cm}
+ T_1 T_2^3
+ T_2^4
- 2 T_1^2 T_2^2 T_3^2 
+\dots ~~.
\eea
The above Hilbert series and plethystic logarithm is written in terms of just three fugacities with positive powers. This illustrates the conical structure of the toric Calabi-Yau 3-fold.
\\

\section{Model 14: $\text{dP}_1$}

\begin{figure}[H]
\begin{center}
\includegraphics[trim=0cm 0cm 0cm 0cm,width=4 cm]{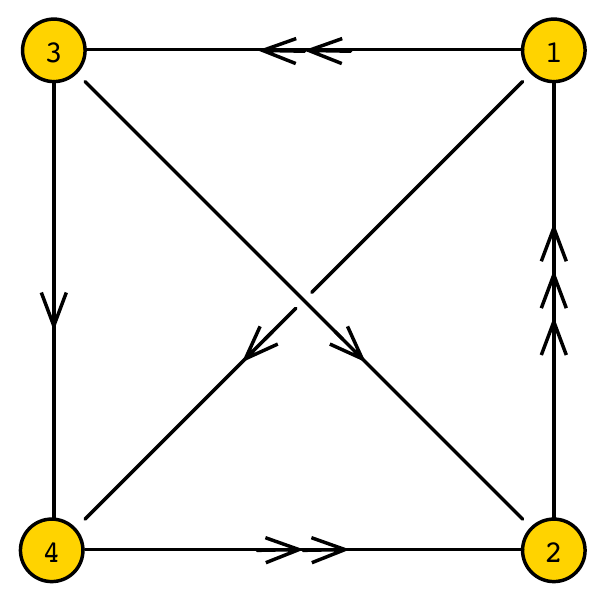}
\includegraphics[width=5 cm]{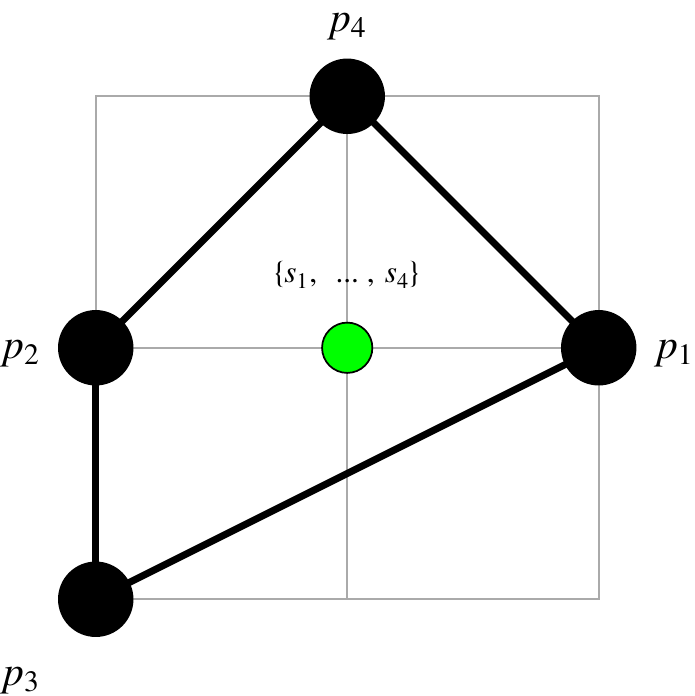}
\includegraphics[width=5 cm]{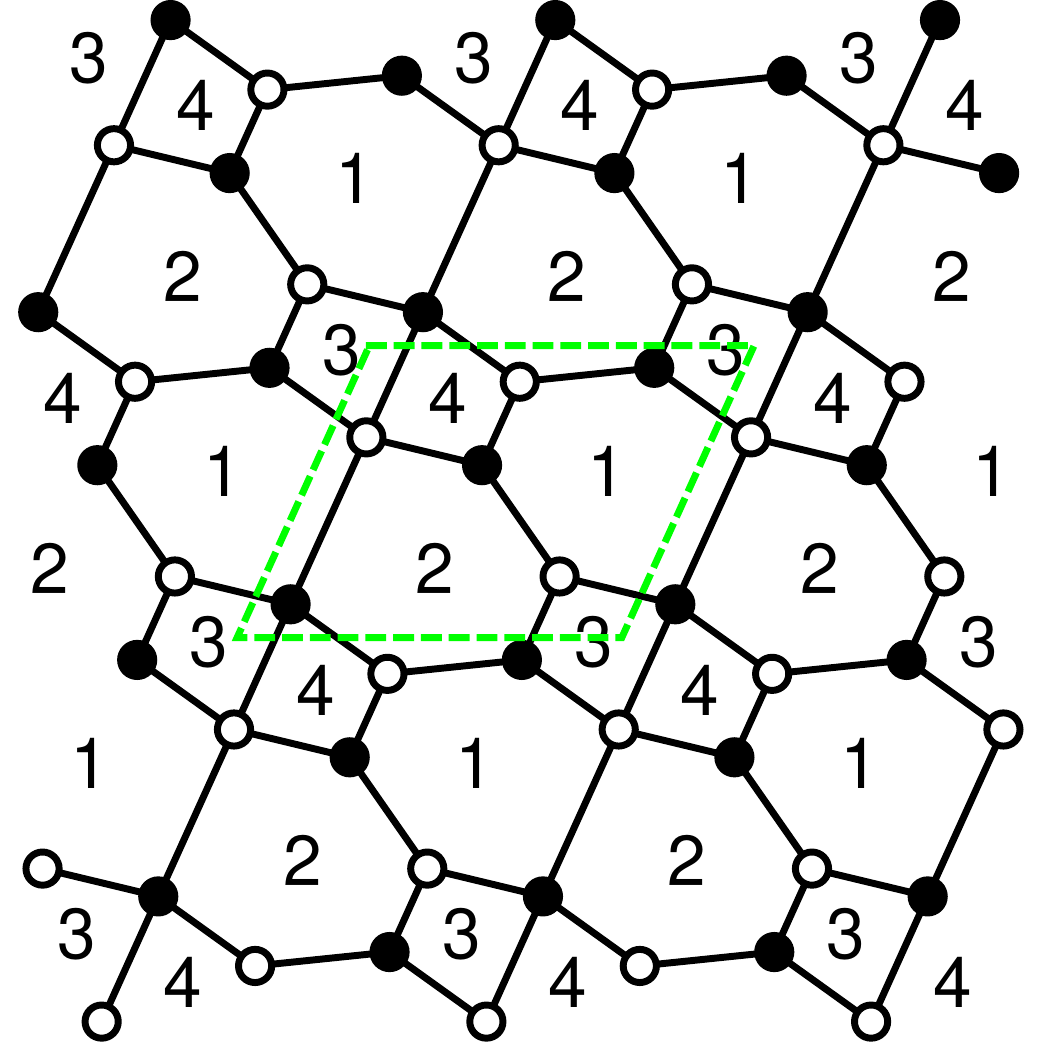}
\caption{The quiver, toric diagram, and brane tiling of Model 14.}
  \label{f14}
 \end{center}
 \end{figure}
 
 \noindent The superpotential is 
\beal{esm14_00}
W&=&
+ X_{21}^{1} X_{14}  X_{42}^{1} 
+ X_{21}^{3} X_{13}^{2} X_{32}  
+ X_{42}^{2} X_{21}^{2} X_{13}^{1} X_{34} 
\nn\\
&&
-X_{13}^{1} X_{32} X_{21}^{1} 
- X_{14} X_{42}^{2} X_{21}^{3} 
- X_{21}^{2} X_{13}^{2} X_{34} X_{42}^{1}
\eea
 
 \noindent The perfect matching matrix is 
 
\noindent\makebox[\textwidth]{%
\footnotesize
$
P=
\left(
\begin{array}{c|cccc|cccc}
 \; & p_1 & p_2 & p_3 & p_4 & s_1 & s_2 & s_3 & s_4 \\
 \hline
 X_{21}^{2} & 1 & 0 & 0 & 0 & 1 & 0 & 0 & 0 \\
 X_{32}     & 1 & 0 & 0 & 0 & 0 & 1 & 0 & 1 \\
 X_{21}^{3} & 0 & 1 & 1 & 0 & 1 & 0 & 0 & 0 \\
 X_{21}^{1} & 0 & 1 & 0 & 1 & 1 & 0 & 0 & 0 \\
 X_{42}^{1} & 0 & 0 & 1 & 0 & 0 & 1 & 0 & 0 \\
 X_{42}^{2} & 0 & 0 & 0 & 1 & 0 & 1 & 0 & 0 \\
 X_{13}^{1} & 0 & 0 & 1 & 0 & 0 & 0 & 1 & 0 \\
 X_{13}^{2} & 0 & 0 & 0 & 1 & 0 & 0 & 1 & 0 \\
 X_{14} & 1 & 0 & 0 & 0 & 0 & 0 & 1 & 1 \\
 X_{34} & 0 & 1 & 0 & 0 & 0 & 0 & 0 & 1
\end{array}
\right)
$
}
\vspace{0.5cm}

 \noindent The F-term charge matrix $Q_F=\ker{(P)}$ is

\noindent\makebox[\textwidth]{%
\footnotesize
$
Q_F=
\left(
\begin{array}{cccc|cccc}
 p_1 & p_2 & p_3 & p_4 & s_1 & s_2 & s_3 & s_4 \\
\hline
 1 & 1 & 0 & 0 & -1 & 0 & 0 & -1 \\
 1 & 0 & 1 & 1 & -1 & -1 & -1 & 0
\end{array}
\right)
$
}
\vspace{0.5cm}

\noindent The D-term charge matrix is

\noindent\makebox[\textwidth]{%
\footnotesize
$
Q_D=
\left(
\begin{array}{cccc|cccc}
 p_1 & p_2 & p_3 & p_4 & s_1 & s_2 & s_3 & s_4 \\
\hline
 0 & 0 & 0 & 0 & 1 & -1 & 0 & 0 \\
 0 & 0 & 0 & 0 & 0 & 1 & -1 & 0 \\
 0 & 0 & 0 & 0 & 0 & 0 & 1 & -1
\end{array}
\right)
$
}
\vspace{0.5cm}

The total charge matrix $Q_t$ does not have repeated columns. Accordingly, the global symmetry is $U(1)_{f_1} \times U(1)_{f_2} \times U(1)_R$. The flavour and R-charges on the GLSM fields corresponding to extremal points in the toric diagram in \fref{f14} are found following the discussion in \sref{s1_3}. They are presented in \tref{t14}.

\begin{table}[ht!]
\centering
\begin{tabular}{|c||c|c|c||l|} 
\hline
\; & $U(1)_{f_1}$ & $U(1)_{f_2}$ & $U(1)_R$ & fugacity \\
\hline
\hline
$p_1$ & 1 & 0 & $R_1 =\sqrt{13}-3$ 		&  $t_1$\\
$p_2$ & 1 & 1 & $R_2 =(5\sqrt{13}-17)/3$ 	&  $t_2$\\
$p_3$ &-1 &-1 & $R_3 =4(4-\sqrt{13})/3$ 	&  $t_3$\\
$p_4$ &-1 & 0 & $R_3 =4(4-\sqrt{13})/3$ 	&  $t_4$\\ 
\hline
\end{tabular}
\caption{The GLSM fields corresponding to extremal points of the toric diagram with their mesonic charges (Model 14). The R-charges are obtained using a-maximization \cite{Bertolini:2004xf}.\label{t14}}
\end{table}

The product of all internal perfect matchings is 
\beal{esm14_x1}
s = \prod_{m=1}^{4} s_m~.
\eea
The fugacity counting the above product is $y_{s}$. The fugacity which counts the remaining extremal perfect matchings $p_\alpha$ is $t_\alpha$.

The mesonic Hilbert series of Model 14 is found using the Molien integral formula in \eref{es12_2}. It is 
 \beal{esm14_1}
g_{1}(t_\alpha,y_{s}; \mathcal{M}^{mes}_{14})= 
\frac{
P(t_\alpha)
}{
(1 - y_{s}~ t_1^2 t_3) (1 - y_{s}~t_2^2 t_3^3) (1 - y_{s}~t_1^2 t_4) (1 - y_{s}~t_2^2 t_4^3)
}~~,
 \eea
where the numerator is given by the polynomial
\beal{esm14_1b}
P(t_\alpha)&=&
1 
+ y_{s}~t_1 t_2 t_3^2 
+ y_{s}~t_1 t_2 t_3 t_4 
- y_{s}^2~t_1^3 t_2 t_3^2 t_4 
+ y_{s}~t_2^2 t_3^2 t_4 
- y_{s}^2~t_1^2 t_2^2 t_3^3 t_4  
\nn\\
&&
+ y_{s}~t_1 t_2 t_4^2 
- y_{s}^2~t_1^3 t_2 t_3 t_4^2
+ y_{s}~t_2^2 t_3 t_4^2 
- y_{s}^2~t_1^2 t_2^2 t_3^2 t_4^2 
- y_{s}^2~t_1^2 t_2^2 t_3 t_4^3 
- y_{s}^3~t_1^3 t_2^3 t_3^3 t_4^3
~~.
\nn\\
\eea
 The plethystic logarithm of the mesonic Hilbert series is
\beal{esm14_3}
&&
PL[g_1(t_\alpha,y_{s};\mathcal{M}_{14}^{mes})]=
y_{s}~t_{1}^2 t_{4}
+y_{s}~t_{1}^2 t_{3}
+y_{s}~t_{1} t_{2} t_{3} t_{4}
+y_{s}~t_{1} t_{2} t_{4}^2
+y_{s}~ t_{1} t_{2} t_{3}^2
\nn\\
&&
+y_{s}~t_{2}^2 t_{3}^2 t_{4}
+y_{s}~ t_{2}^2 t_{3}^3
+y_{s}~t_{2}^2 t_{3} t_{4}^2
+y_{s}~t_{2}^2 t_{4}^3
- y_{s}^2~t_{1}^3 t_{2} t_{3} t_{4}^2 
- y_{s}^2~t_{1}^3 t_{2} t_{3}^2 t_{4}
+ \dots~.
\eea

Consider the following fugacity map
\beal{esm14_y1}
f_1 = t_3^{-1/2} t_4^{1/2}
~,~
f_2 = \frac{t_4}{t_3}
~,~
\tilde{t}_1 =
y_s^{1/2} ~ t_1
~,~
\tilde{t}_2 =
y_s^{1/2} ~ t_2
~,~
\tilde{t}_3 =
t_3^{1/2} t_4^{1/2}
~,~
\eea
where the fugacities $f_1$ and $f_2$ count flavour charges, and the fugacity $\tilde{t}_i$ count the R-charge $R_i$ in \tref{t14}. Accordingly, the plethystic logarithm becomes
\beal{esm14_3}
&&
PL[g_1(\tilde{t}_\alpha,f_1,f_2;\mathcal{M}_{14}^{mes})]
=
\left(f_1 + \frac{f_1}{f_2}\right) \tilde{t}_{1}^2 \tilde{t}_{3}
+ \left(1+f_{2} +\frac{1}{f_{2}}\right) \tilde{t}_{1} \tilde{t}_{2} \tilde{t}_{3}^2
\nn\\
&&
\hspace{1cm}
+ \left(\frac{1}{f_1} + \frac{1}{f_1 f_2} + \frac{f_2}{f_1} + \frac{f_2^2}{f_1}\right) \tilde{t}_{2}^2 \tilde{t}_{3}^3
-\left(f_1 + \frac{f_1}{f_2}\right) \tilde{t}_{1}^3 \tilde{t}_{2} \tilde{t}_{3}^3
\comment{
- \left(
3+ \frac{1}{f_2^2} + 2 \frac{1}{f_2} + 2 f_2 + f_2^2
 \right) \tilde{t}_{1}^2 \tilde{t}_{2}^2 \tilde{t}_{3}^4
\nn\\
&&
-\left(
2 \frac{1}{f_1} + \frac{1}{f_1 f_2} + 2 \frac{f_2}{f_1} + \frac{f_2^2}{f_1}
\right) \tilde{t}_{1} \tilde{t}_{2}^3 \tilde{t}_{3}^5
-\left(
\frac{1}{f_1^2}+\frac{f_2}{f_1^2} + \frac{f_2^2}{f_1^2}
\right) \tilde{t}_{2}^4 \tilde{t}_{3}^6
}
   +\dots~.
   \eea
The first positive terms in the above plethystic logarithm correspond to moduli space generators with the corresponding flavour charge counted by the fugacities $f_1$ and $f_2$. The generators and the corresponding mesonic charges are shown in \tref{t14gen}. The generators can be presented on a charge lattice. The convex polygon formed by the generators in \tref{t14gen} is the dual reflexive polygon of the toric diagram of Model 14.
\\

\begin{table}[H]
\centering
\resizebox{\hsize}{!}{
\begin{minipage}[!b]{0.6\textwidth}
\begin{tabular}{|l|c|c|}
\hline
Generator & $U(1)_{f_1}$ & $U(1)_{f_2}$ 
\\
\hline
\hline
$p_{1}^2 p_{3} ~ s$
& 1 & -1
\nn\\
$p_{1} p_{2} p_{3}^2 ~ s$
& 0 & -1
\nn\\
$p_{2}^2 p_{3}^3 ~ s$
& -1 & -1
\nn\\
$p_{1}^2 p_{4} ~ s$
& 1 & 0
\nn\\
$p_{1} p_{2} p_{3} p_{4} ~ s$
& 0 & 0
\nn\\
$p_{2}^2 p_{3}^2 p_{4} ~ s$
& -1 & 0
\nn\\
$p_{1} p_{2} p_{4}^2 ~ s$
& 0 & 1
\nn\\
$p_{2}^2 p_{3} p_{4}^2 ~ s$
& -1 & 1
\nn\\
$p_{2}^2 p_{4}^3 ~ s$
& -1 & 2
   \\
   \hline
\end{tabular}
\end{minipage}
\hspace{1cm}
\begin{minipage}[!b]{0.3\textwidth}
\includegraphics[width=4 cm]{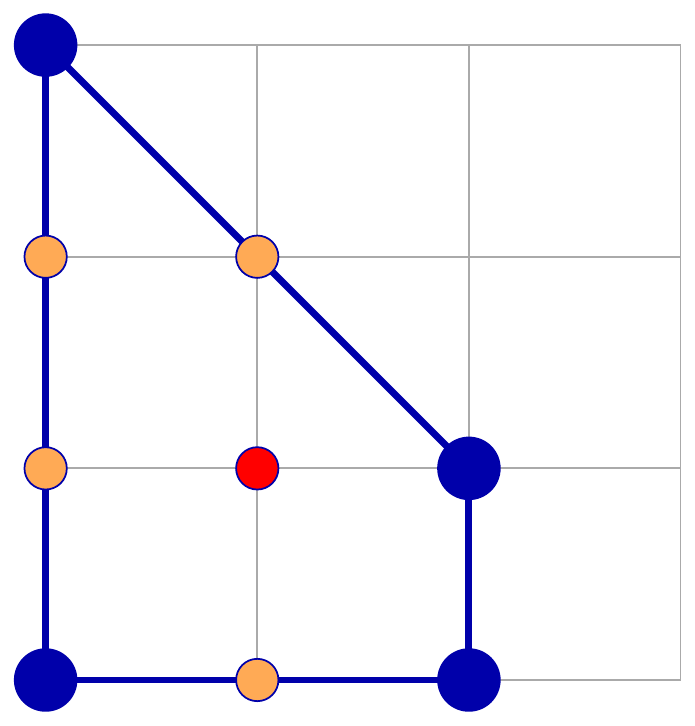}
\end{minipage}
}
\caption{The generators and lattice of generators of the mesonic moduli space of Model 14 in terms of GLSM fields with the corresponding flavor charges. The lattice of generators is the toric diagram of Model 3. \label{t14gen}\label{f14gen}}
\end{table}

\begin{table}[H]
\centering
\resizebox{\hsize}{!}{
\begin{tabular}{|l|c|c|}
\hline
Generator & $U(1)_{f_1}$ & $U(1)_{f_2}$ 
\\
\hline
\hline
$X_ {13}^{1} X_{32} X_ {21}^{2}=  X_{14} X_ {42}^{1} X_ {21}^{2}$
& 1 & -1
\nn\\
$ X_{13}^{1} X_{34} X_{42}^{1} X_{21}^{2}=  X_{13}^{1} X_{32} X_{21}^{3}=  X_{14} X_{42}^{1} X_{21}^{3}$
& 0 & -1
\nn\\
$X_{13}^{1} X_{34} X_{42}^{1} X_{21}^{3}$
& -1 & -1
\nn\\
$X_ {13}^{2} X_{32} X_ {21}^{2}=  X_{14} X_ {42}^{2} X_ {21}^{2}$
& 1 & 0
\nn\\
$X_ {13}^{1} X_{34} X_ {42}^{2} X_ {21}^{2}=  X_ {13}^{2} X_{34} X_{42}^{1} X_ {21}^{2}=  X_ {13}^{1} X_{32} X_ {21}^{1}=  X_ {13}^{2} X_{32} X_ {21}^{3}=  X_{14} X_ {42}^{1} X_ {21}^{1}=  X_{14} X_{42}^{2} X_ {21}^{3}$
& 0 & 0
\nn\\
$X_ {13}^{1} X_{34} X_ {42}^{1} X_ {21}^{1}=  X_ {13}^{1} X_{34} X_{42}^{2} X_ {21}^{3}=  X_ {13}^{2} X_{34} X_ {42}^{1} X_ {21}^{3}$
& -1 & 0
\nn\\
$X_ {13}^{2} X_{34} X_ {42}^{2} X_ {21}^{2}=  X_ {13}^{2} X_{32} X_{21}^{1}=  X_{14} X_ {42}^{2} X_ {21}^{1}$
& 0 & 1
\nn\\
$X_ {13}^{1} X_{34} X_ {42}^{2} X_ {21}^{1}=  X_ {13}^{2} X_{34} X_{42}^{1} X_ {21}^{1}=  X_ {13}^{2} X_{34} X_ {42}^{2} X_ {21}^{3}$
& -1 & 1
\nn\\
$X_ {13}^{2} X_{34} X_ {42}^{2} X_ {21}^{1}$
& -1 & 2
   \\
   \hline
\end{tabular}
}
\caption{The generators in terms of bifundamental fields (Model 14).\label{t14gen2}\label{f14gen2}}
\end{table}

The mesonic Hilbert series and the plethystic logarithm can be re-expressed in terms of just $3$ fugacities
\beal{esm14_x1}
T_1 = \frac{f_2 ~ \tilde{t}_2}{f_1^2 ~ \tilde{t}_1^3}
=\frac{t_2}{y_{s}~t_1^3}~,~
T_2 = \frac{f_1}{f_2} ~ \tilde{t}_1^2 \tilde{t}_3
=y_{s}~ t_1^2 t_3~,~
T_3 = f_1 ~ \tilde{t}_1^2 \tilde{t}_3
=y_{s}~ t_1^2 t_4~,~
\eea
such that
\beal{esm14_x2}
&&
g_1(T_1,T_2,T_3;\mathcal{M}_{14}^{mes})=
\nn\\
&&
\hspace{0.75cm}
(1 + T_1 T_2^2 + T_1 T_2 T_3 - T_1 T_2^2 T_3 + T_1^2 T_2^2 T_3 - 
   T_1^2 T_2^3 T_3 + T_1 T_3^2 - T_1 T_2 T_3^2 + T_1^2 T_2 T_3^2
   \nn\\
   &&
   \hspace{0.8cm}
    - T_1^2 T_2^2 T_3^2 - T_1^2 T_2 T_3^3 - 
   T_1^3 T_2^3 T_3^3)
   \times
   \frac{1}{
 (1 - T_2) (1 - T_1^2 T_2^3) (1 - T_3) (1 - 
     T_1^2 T_3^3)
     }
\nn\\
\eea
and
\beal{esm14_x3}
&&
PL[g_1(T_1,T_2,T_3;\mathcal{M}_{14}^{mes})]=
T_3 
+ T_2 
+ T_1 T_2 T_3 
+ T_1 T_3^2 
+ T_1 T_2^2  
+ T_1^2 T_2^2 T_3
+ T_1^2 T_2^3  
 \nn\\
 &&
 \hspace{1cm}
+ T_1^2 T_2 T_3^2 
+ T_1^2 T_3^3 
- T_1 T_2 T_3^2  
- T_1 T_2^2 T_3 
+\dots
~~.
\eea
The above Hilbert series and plethystic logarithm illustrate the conical structure of the toric Calabi-Yau 3-fold.
\\

\section{Model 15: $\mathcal{C}/\mathbb{Z}_{2}~(1,1,1,1),~\mathbb{F}_0$}
\subsection{Model 15 Phase a}

\begin{figure}[H]
\begin{center}
\includegraphics[trim=0cm 0cm 0cm 0cm, width=4.5 cm]{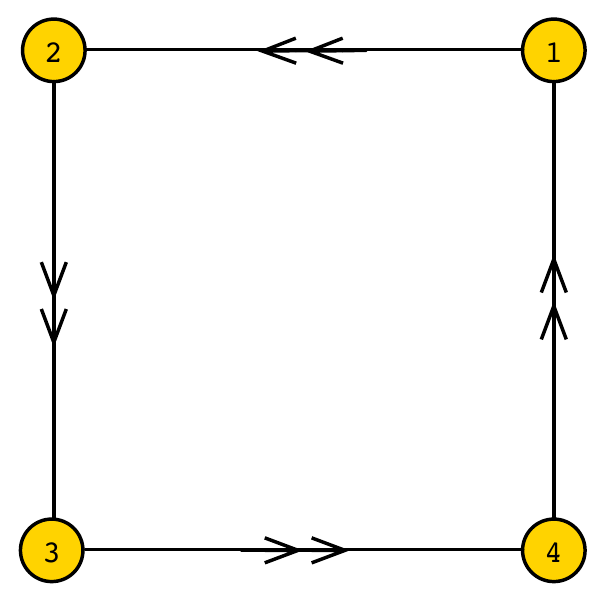}
\includegraphics[width=5 cm]{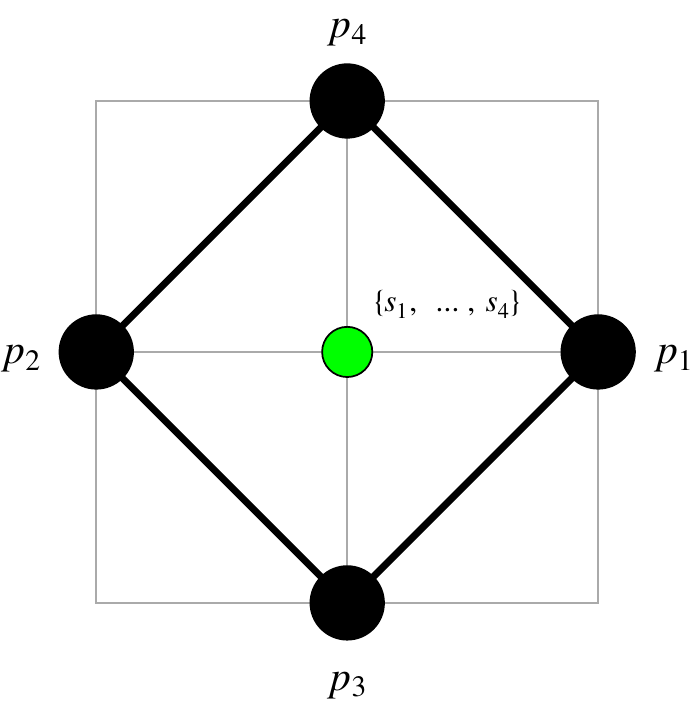}
\includegraphics[width=5 cm]{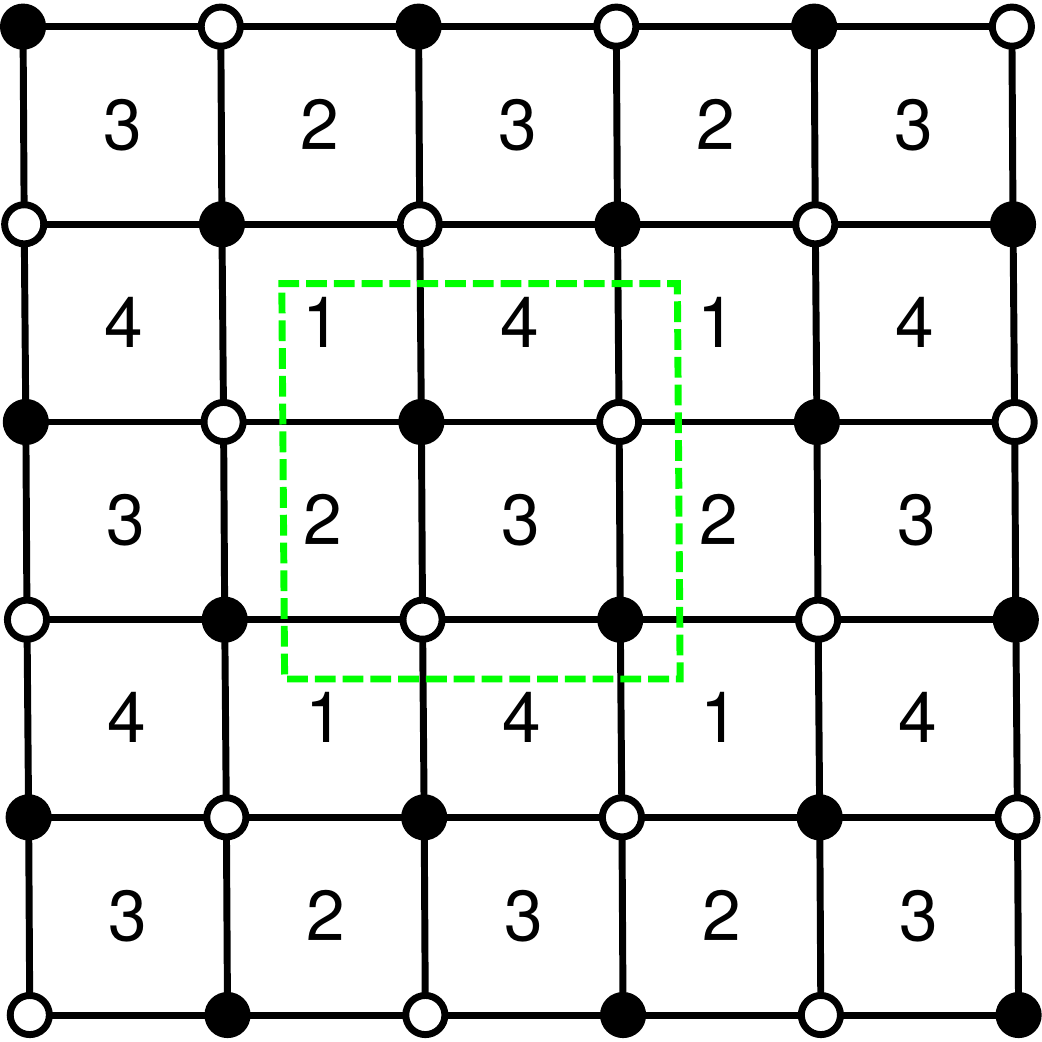}
\caption{The quiver, toric diagram, and brane tiling of Model 15a.}
  \label{f15a}
 \end{center}
 \end{figure}
 
 \noindent The superpotential is 
\beal{esm15a_00}
W&=&
+ X_{12}^{1} X_{23}^{1} X_{34}^{2} X_{41}^{2} 
+ X_{12}^{2} X_{23}^{2} X_{34}^{1} X_{41}^{1} 
- X_{12}^{1} X_{23}^{2} X_{34}^{2} X_{41}^{1} 
- X_{12}^{2} X_{23}^{1} X_{34}^{1} X_{41}^{2} 
~~.
\nn\\
\eea
 
 \noindent The perfect matching matrix is 
 
\noindent\makebox[\textwidth]{%
\footnotesize
$
P=
\left(
\begin{array}{c|cccc|cccc}
 \; & p_1 & p_2 & p_3 & p_4 & s_1 & s_2 & s_3 & s_4 \\
 \hline
 X_{12}^{1} & 1 & 0 & 0 & 0 & 1 & 0 & 0 & 0 \\
 X_{12}^{2} & 0 & 1 & 0 & 0 & 1 & 0 & 0 & 0 \\
 X_{34}^{1} & 1 & 0 & 0 & 0 & 0 & 1 & 0 & 0 \\
 X_{34}^{2} & 0 & 1 & 0 & 0 & 0 & 1 & 0 & 0 \\
 X_{23}^{1} & 0 & 0 & 1 & 0 & 0 & 0 & 1 & 0 \\
 X_{23}^{2} & 0 & 0 & 0 & 1 & 0 & 0 & 1 & 0 \\
 X_{41}^{1} & 0 & 0 & 1 & 0 & 0 & 0 & 0 & 1 \\
 X_{41}^{2} & 0 & 0 & 0 & 1 & 0 & 0 & 0 & 1
\end{array}
\right)
$
}
\vspace{0.5cm}

 \noindent The F-term charge matrix $Q_F=\ker{(P)}$ is

\noindent\makebox[\textwidth]{%
\footnotesize
$
Q_F=
\left(
\begin{array}{cccc|cccc}
 p_1 & p_2 & p_3 & p_4 & s_1 & s_2 & s_3 & s_4 \\
 \hline
 1 & 1 & 0 & 0 & -1 & -1 & 0 & 0 \\
 0 & 0 & 1 & 1 & 0 & 0 & -1 & -1
\end{array}
\right)
$
}
\vspace{0.5cm}

\noindent The D-term charge matrix is

\noindent\makebox[\textwidth]{%
\footnotesize
$
Q_D=
\left(
\begin{array}{cccc|cccc}
 p_1 & p_2 & p_3 & p_4 & s_1 & s_2 & s_3 & s_4 \\
 \hline
 0 & 0 & 0 & 0 & 1 & -1 & 0 & 0 \\
 0 & 0 & 0 & 0 & 0 & 1 & -1 & 0 \\
 0 & 0 & 0 & 0 & 0 & 0 & 1 & -1
\end{array}
\right)
$
}
\vspace{0.5cm}

The pairs of GLSM fields $\{p_1,p_2\}$ and $\{p_3,p_4\}$ have the same charge under the F-term and D-term constraints. This is shown by the identical columns in the total charge matrix $Q_t$. Accordingly, the global symmetry is enhanced from $U(1)^2\times U(1)_R$ to
 $SU(1)_{x_1} \times SU(2)_{x_2} \times U(1)_R$. The mesonic charges on the GLSM fields corresponding to extremal points in the toric diagram in \fref{f15a} are found following the discussion in \sref{s1_3}. They are presented in \tref{t15a}.

\begin{table}[H]
\centering
\begin{tabular}{|c||c|c|c||c|} 
\hline
\; & $SU(2)_{x_1}$ & $SU(2)_{x_2}$ & $U(1)_R$ & fugacity \\
\hline
\hline
$p_1$ & 1/2 & 0 & $1/2$ &  $t_1$\\
$p_2$ &-1/2 & 0 & $1/2$ &  $t_2$\\
$p_3$ & 0 & 1/2 & $1/2$ &  $t_3$\\
$p_4$ & 0 &-1/2 & $1/2$ &  $t_4$\\ 
\hline
\end{tabular}
\caption{The GLSM fields corresponding to extremal points of the toric diagram with their mesonic charges (Model 15a).\label{t15a}}
\end{table}

The product of all internal perfect matchings labelled by
\beal{esm15a_x1}
s = \prod_{m=1}^{4} s_m~.
\eea
The above product is counted by the fugacity $y_{s}$. All remaining extremal perfect matchings $p_\alpha$ are counted by the fugacity $t_\alpha$.

The mesonic Hilbert series of Model 15a is calculated using the Molien integral formula in \eref{es12_2}. It is 
 \beal{esm15a_1}
g_{1}(t_\alpha,y_{s}; \mathcal{M}^{mes}_{15a})= 
\frac{
P(t_\alpha)
}{
(1 - y_{s}~t_1^2 t_3^2) (1 - y_{s}~t_2^2 t_3^2) (1 - y_{s}~t_1^2 t_4^2) (1 - y_{s}~t_2^2 t_4^2)
}~~,
 \eea
where the numerator is given by the polynomial
\beal{esm15a_1b}
P(t_\alpha)&=&
1 
+ y_{s}~ t_1 t_2 t_3^2 
+ y_{s}~t_1^2 t_3 t_4 
+ y_{s}~t_1 t_2 t_3 t_4 
+ y_{s}~t_2^2 t_3 t_4 
- y_{s}^2~t_1^2 t_2^2 t_3^3 t_4 
\nn\\
&&
+ y_{s}~t_1 t_2 t_4^2 
- y_{s}^2~t_1^3 t_2 t_3^2 t_4^2 
- y_{s}^2~t_1^2 t_2^2 t_3^2 t_4^2 
- y_{s}^2~t_1 t_2^3 t_3^2 t_4^2 
- y_{s}^2~t_1^2 t_2^2 t_3 t_4^3 
- y_{s}^3~t_1^3 t_2^3 t_3^3 t_4^3
~~.
\nn\\
\eea
 The plethystic logarithm of the mesonic Hilbert series is
\beal{esm15a_3}
&&
PL[g_1(t_\alpha,y_{s};\mathcal{M}_{15a}^{mes})]=
y_{s}~t_1^2 t_3^2 
+ y_{s}~t_1 t_2 t_3^2 
+ y_{s}~t_2^2 t_3^2 
+ y_{s}~t_1^2 t_3 t_4 
+ y_{s}~t_1 t_2 t_3 t_4 
+ y_{s}~t_2^2 t_3 t_4 
\nn\\
&&
\hspace{0.5cm}
+ y_{s}~t_1^2 t_4^2 
+ y_{s}~t_1 t_2 t_4^2 
+ y_{s}~t_2^2 t_4^2
- y_{s}^2~t_1^2 t_2^2 t_3^4 
- y_{s}^2~t_1^3 t_2 t_3^3 t_4 
- 2 ~y_{s}^2~t_1^2 t_2^2 t_3^3 t_4 
- y_{s}^2~t_1 t_2^3 t_3^3 t_4 
\nn\\
&&
\hspace{0.5cm}
- y_{s}^2~t_1^4 t_3^2 t_4^2 
- 2~y_{s}^2~ t_1^3 t_2 t_3^2 t_4^2 
- 4~y_{s}^2~ t_1^2 t_2^2 t_3^2 t_4^2 
- 2~y_{s}^2~ t_1 t_2^3 t_3^2 t_4^2 
- y_{s}^2~t_2^4 t_3^2 t_4^2 
- y_{s}^2~t_1^3 t_2 t_3 t_4^3 
\nn\\
&&
\hspace{0.5cm}
- 2~y_{s}^2~ t_1^2 t_2^2 t_3 t_4^3 
- y_{s}^2~t_1 t_2^3 t_3 t_4^3 
- y_{s}^2~t_1^2 t_2^2 t_4^4
+ \dots~.
\eea
From the infinite plethystic logarithm one concludes that the moduli space is not a complete intersection.

Consider the following fugacity map
\beal{esm15a}
\tilde{x}_1^2 = x_1 =
\frac{t_1}{t_2}
~,~
\tilde{x}_2^2 = x_2 =
\frac{t_3}{t_4}
~,~
t =
y_s^{1/4} ~ t_1^{1/4} t_2^{1/4} t_3^{1/4} t_4^{1/4}
~,~
\eea
where $x_1$, $x_2$ and $t$ are mesonic charge fugacities. In terms of $\tilde{x}_1$ and $\tilde{x}_2$ both the Hilbert series and the plethystic logarithm can be expressed in terms of characters of irreducible representations of $SU(2)\times SU(2)$. The Taylor expansion of the Hilbert series takes the form
\beal{esm15a2_1b}
g_{1}(t,\tilde{x}_1,\tilde{x}_2; \mathcal{M}^{mes}_{15a})= 
\sum_{n=0}^{\infty}~
[2n;2n]_{\tilde{x}_1,\tilde{x}_2}~t^{4n}
~~.
\eea
The plethystic logarithm in terms of characters of irreducible representations of $SU(2)\times SU(2)$ is
\beal{esm15A_3}
&&
PL[g_1(t,\tilde{x}_1,\tilde{x}_2;\mathcal{M}_{15a}^{mes})]=
[2;2]_{\tilde{x}_1,\tilde{x}_2} t^4
- (
1+[4;0]_{\tilde{x}_1,\tilde{x}_2}+[2;2]_{\tilde{x}_1,\tilde{x}_2}+[0;4]_{\tilde{x}_1,\tilde{x}_2}
) t^8
\nn\\
&&
\hspace{0.5cm}
+ (
[2;0]_{\tilde{x}_1,\tilde{x}_2}
+ [4;0]_{\tilde{x}_1,\tilde{x}_2}
+ [0;2]_{\tilde{x}_1,\tilde{x}_2}
+ 2[2;2]_{\tilde{x}_1,\tilde{x}_2}
+ [4;2]_{\tilde{x}_1,\tilde{x}_2}
+ [0;4]_{\tilde{x}_1,\tilde{x}_2}
+ [2;4]_{\tilde{x}_1,\tilde{x}_2}
) t^{12}
\nn\\
&&
\hspace{0.5cm}
- (
4 [2;0]_{\tilde{x}_1,\tilde{x}_2}
+ [4;0]_{\tilde{x}_1,\tilde{x}_2}
+ [6;0]_{\tilde{x}_1,\tilde{x}_2}
+ 4[0;2]_{\tilde{x}_1,\tilde{x}_2}
+ 5[2;2]_{\tilde{x}_1,\tilde{x}_2}
+ 4[4;2]_{\tilde{x}_1,\tilde{x}_2}
+ [6;2]_{\tilde{x}_1,\tilde{x}_2}
\nn\\
&&
\hspace{0.5cm}
+ [0;4]_{\tilde{x}_1,\tilde{x}_2}
+ 4[2;4]_{\tilde{x}_1,\tilde{x}_2}
+ [4;4]_{\tilde{x}_1,\tilde{x}_2}
+ [0;6]_{\tilde{x}_1,\tilde{x}_2}
+ [2;6]_{\tilde{x}_1,\tilde{x}_2}
) t^{16}
   +\dots~.
   \eea
In terms of the fugacities $x_1$ and $x_2$ the above plethystic logarithm exhibits the moduli space generators with their mesonic charges, where the flavour charges as powers of $x_1$ and $x_2$ take integer values.  They are summarized in \tref{t15agen}. The generators can be presented on a charge lattice. The generators form a convex polygon on the charge lattice which is the dual of the toric diagram of Model 15a. 

As indicated in \eref{esm15A_3}, the generators fall into an irreducible representation of $SU(2)\times SU(2)$ with the character 
\beal{esm15a_xx5}
[2;2]_{\tilde{x}_1,\tilde{x}_2}t^4 
=
\left(\tilde{x}_1^2 + 1 + \frac{1}{\tilde{x}_1^2}\right)
\left(\tilde{x}_2^2 + 1 + \frac{1}{\tilde{x}_2^2}\right)
~~.
\eea
The generators in terms of quiver fields are shown in \tref{t15agen2}.
\\

\begin{table}[H]
\centering
\resizebox{\hsize}{!}{
\begin{minipage}[!b]{0.5\textwidth}
\begin{tabular}{|l|c|c|}
\hline
Generator & $SU(2)_{x_1}$ & $SU(2)_{x_2}$ 
\\
\hline
\hline
$p_{1}^2 p_{3}^2 ~ s$
& 1 & 1
\nn\\
$p_{1} p_{2} p_{3}^2 ~ s$
& 0 & 1
\nn\\
$p_{2}^2 p_{3}^2 ~ s$
& -1 & 1
\nn\\
$p_{1}^2 p_{3} p_{4} ~ s$
& 1 & 0
\nn\\
$p_{1} p_{2} p_{3} p_{4} ~ s$
& 0 & 0
\nn\\
$p_{2}^2 p_{3} p_{4} ~ s$
& -1 & 0
\nn\\
$p_{1}^2 p_{4}^2 ~ s$
& 1 & -1
\nn\\
$p_{1} p_{2} p_{4}^2 ~ s$
& 0 & -1
\nn\\
$p_{2}^2 p_{4}^2 ~ s$
& -1 & -1
\nn\\
   \hline
\end{tabular}
\end{minipage}
\hspace{1cm}
\begin{minipage}[!b]{0.3\textwidth}
\includegraphics[width=4 cm]{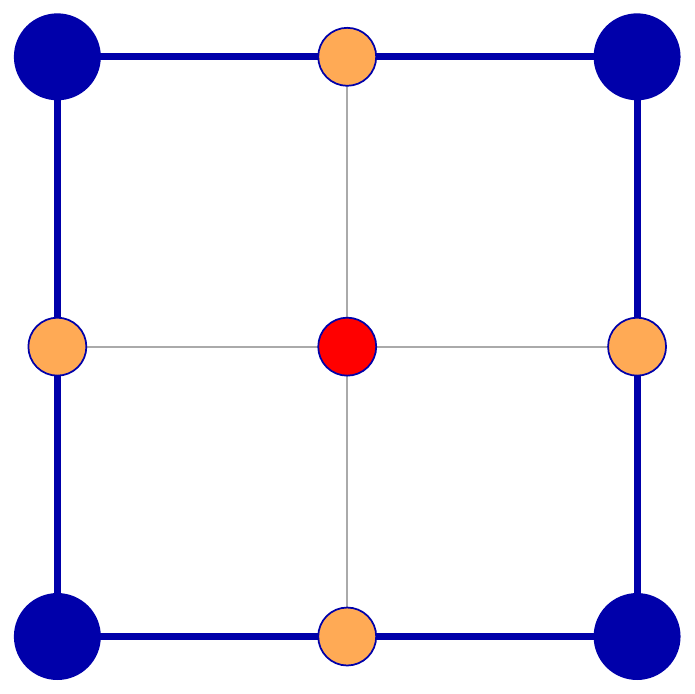}
\end{minipage}
}
\caption{The generators and lattice of generators of the mesonic moduli space of Model 15a in terms of GLSM fields with the corresponding flavor charges. \label{t15agen}\label{f15agen}} 
\end{table}

\begin{table}[H]
\centering
\resizebox{\hsize}{!}{
\begin{tabular}{|l|c|c|}
\hline
Generator & $SU(2)_{x_1}$ & $SU(2)_{x_2}$ 
\\
\hline
\hline
$ X_{12}^{1} X_{23}^{1} X_{34}^{1} X_{41}^{1}$
& 1 & 1
\nn\\
$X_ {12}^{1} X_ {23}^{1} X_ {34}^{2} X_ {41}^{1}=  X_ {12}^{2} X_{23}^{1} X_ {34}^{1} X_ {41}^{1}$
& 0 & 1
\nn\\
$X_ {12}^{2} X_ {23}^{1} X_ {34}^{2} X_ {41}^{1}$
& -1 & 1
\nn\\
$X_ {12}^{1} X_ {23}^{1} X_ {34}^{1} X_ {41}^{2}=  X_ {12}^{1} X_{23}^{2} X_ {34}^{1} X_ {41}^{1}$
& 1 & 0
\nn\\
$X_ {12}^{1} X_ {23}^{1} X_ {34}^{2} X_ {41}^{2}=  X_ {12}^{1} X_{23}^{2} X_ {34}^{2} X_ {41}^{1}=  X_ {12}^{2} X_ {23}^{1} X_{34}^{1} X_ {41}^{2}=  X_ {12}^{2} X_ {23}^{2} X_ {34}^{1} X_ {41}^{1}$
& 0 & 0
\nn\\
$X_ {12}^{2} X_ {23}^{1} X_ {34}^{2} X_ {41}^{2}=  X_ {12}^{2} X_{23}^{2} X_ {34}^{2} X_ {41}^{1}$
& -1 & 0
\nn\\
$ X_{12}^{1} X_{23}^{2} X_{34}^{1} X_{41}^{2}$
& 1 & -1
\nn\\
$X_ {12}^{1} X_ {23}^{2} X_ {34}^{2} X_ {41}^{2}=  X_ {12}^{2} X_{23}^{2} X_ {34}^{1} X_ {41}^{2}$
& 0 & -1
\nn\\
$X_{12}^{2} X_{23}^{2} X_{34}^{2} X_{41}^{2}$
& -1 & -1
\nn\\
   \hline
\end{tabular}
}
\caption{The generators in terms of bifundamental fields (Model 15a).\label{t15agen2}\label{f15agen2}} 
\end{table}

By introducing the fugacity map
\beal{esm15a_x1}
T_1 =\frac{t^4}{x_1 x_2} = y_{s} ~ t_2^2 t_4^2~,~
T_2 = x_1 = \frac{t_1}{t_2}~,~
T_3 = x_2 = \frac{t_3}{t_4}~,
\eea
the mesonic Hilbert series can be expressed as
\beal{esm15a_x2}
&&
g_1(T_1,T_2,T_3;\mathcal{M}_{15a}^{mes})=
\big(
1
+ T_1 T_2 T_3
+ T_1 T_3
+ T_1 T_2^2 T_3
+ T_1 T_2
+ T_1 T_2 T_3^2
\nn\\
&&
\hspace{0.5cm}
- (
T_1^2 T_2^2 T_3^2
+ T_1^2 T_2 T_3^2
+ T_1^2 T_2^3 T_3^2
+ T_1^2 T_2^2 T_3
+ T_1^2 T_2^2 T_3^3
)
- T_1^3 T_2^3 T_3^3
\big) \times
\nn\\
&&
\hspace{0.5cm}
\frac{
1
}{
(1-T_1)
(1-T_1 T_2^2)
(1-T_1 T_3^2)
(1-T_1 T_2^2 T_3^2)
}~~.
\eea
The corresponding plethystic logarithm has the form
\beal{esm15_x3}
&&
PL[g_1(T_1,T_2,T_3;\mathcal{M}_{15a}^{mes})]=
T_1 T_2^2 T_3^2
+ T_1 T_2 T_3^2
+ T_1 T_3^2
+ T_1 T_2^2 T_3
+ T_1 T_2 T_3 
+ T_1 T_3
\nn\\
&&
\hspace{1cm}
+ T_1 T_2^2
+ T_1 T_2
+ T_1
- T_1^2 T_2^2 
- T_1^2 T_2^3 T_3^3
+\dots ~~.
\eea
The above Hilbert series and plethystic logarithm are in terms of three fugacities which carry only positive powers. This illustrates the conical structure of the toric Calabi-Yau 3-fold.
\\

\subsection{Model 15 Phase b}

\begin{figure}[H]
\begin{center}
\includegraphics[trim=0cm 0cm 0cm 0cm,width=4.5 cm]{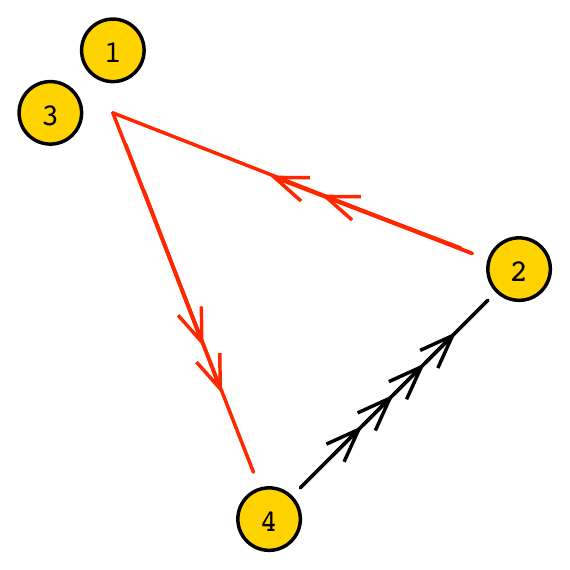}
\includegraphics[width=5 cm]{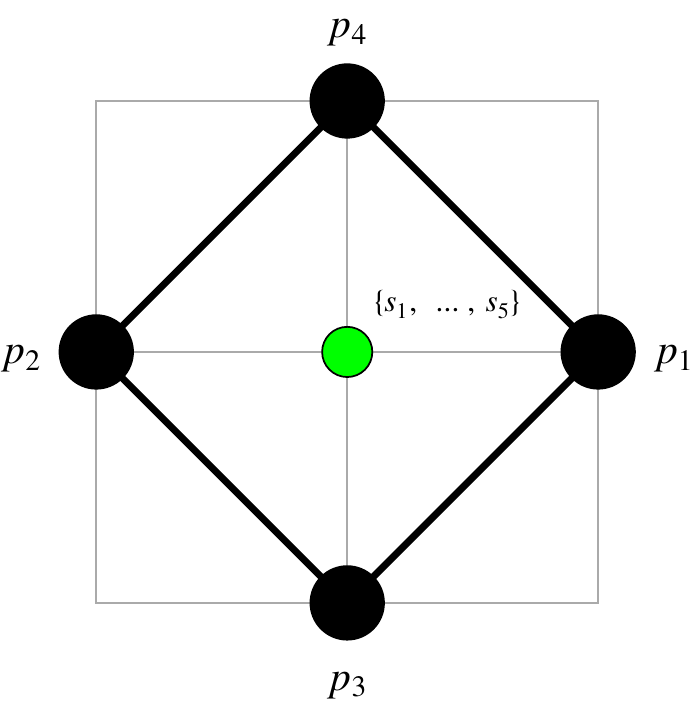}
\includegraphics[width=5 cm]{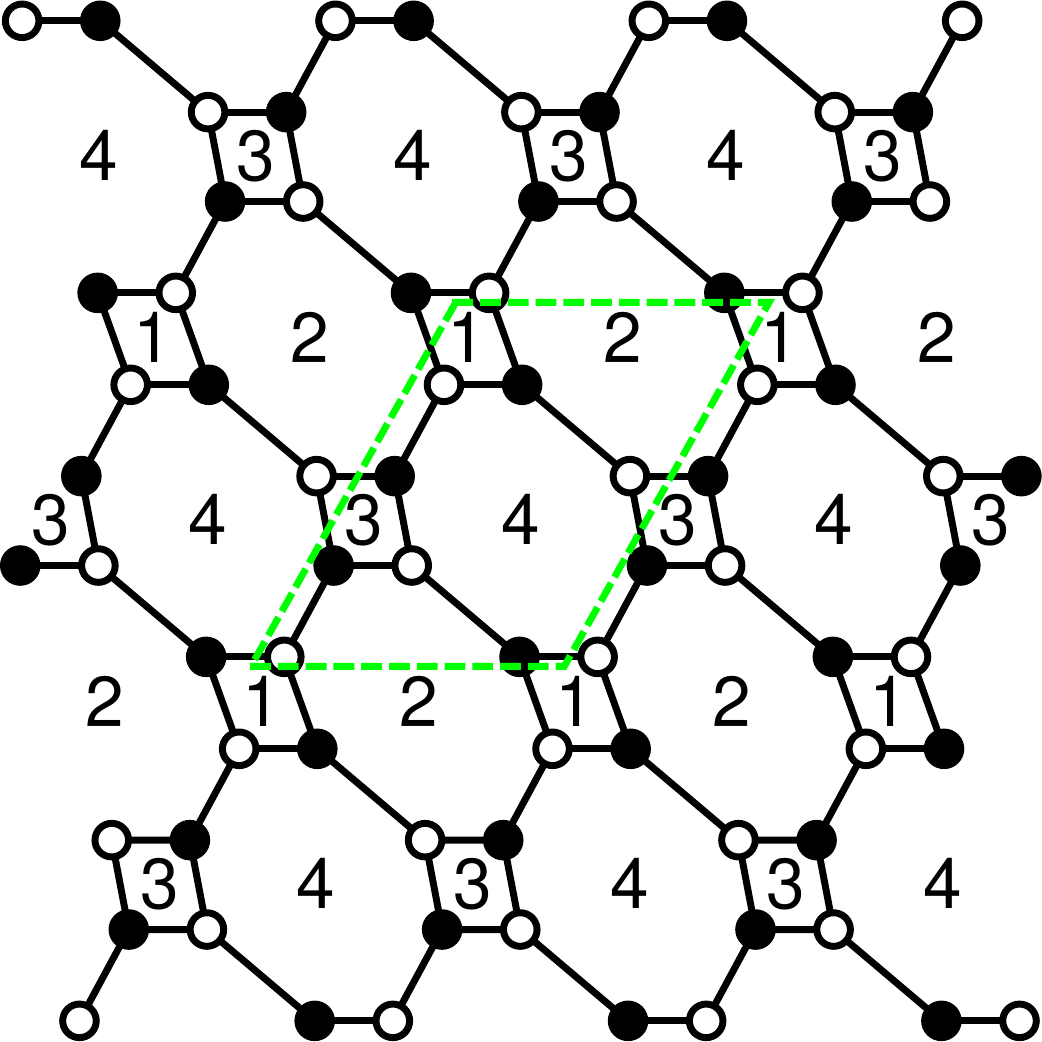}
\caption{The quiver, toric diagram, and brane tiling of Model 15b. The red arrows in the quiver indicate all possible connections between blocks of nodes.}
  \label{f15b}
 \end{center}
 \end{figure}
 
 \noindent The superpotential is 
\beal{esm15b_00}
W&=&
+ X_{21}^{1} X_{14}^{1} X_{42}^{1} 
+ X_{21}^{2} X_{14}^{2} X_{42}^{2} 
+ X_{23}^{1} X_{34}^{2} X_{42}^{3} 
+ X_{23}^{2} X_{34}^{1} X_{42}^{4}
\nn\\
&&
- X_{21}^{1} X_{14}^{2} X_{42}^{3} 
- X_{21}^{2} X_{14}^{1} X_{42}^{4} 
- X_{23}^{1} X_{34}^{1} X_{42}^{2} 
- X_{23}^{2} X_{34}^{2} X_{42}^{1}
  \eea
 
 \noindent The perfect matching matrix is 
 
\noindent\makebox[\textwidth]{%
\footnotesize
$
P=
\left(
\begin{array}{c|cccc|ccccc}
 \; & p_1 & p_2 & p_3 & p_4 & s_1 & s_2 & s_3 & s_4 & s_5 \\
 \hline
 X_{42}^{2} & 1 & 0 & 1 & 0 & 0 & 0 & 1 & 0 & 0 \\
 X_{42}^{3} & 0 & 1 & 1 & 0 & 0 & 0 & 1 & 0 & 0 \\
 X_{42}^{4} & 1 & 0 & 0 & 1 & 0 & 0 & 1 & 0 & 0 \\
 X_{42}^{1} & 0 & 1 & 0 & 1 & 0 & 0 & 1 & 0 & 0 \\
 X_{21}^{1} & 1 & 0 & 0 & 0 & 1 & 0 & 0 & 1 & 0 \\
 X_{21}^{2} & 0 & 1 & 0 & 0 & 1 & 0 & 0 & 1 & 0 \\
 X_{34}^{2} & 1 & 0 & 0 & 0 & 0 & 1 & 0 & 1 & 0 \\
 X_{34}^{1} & 0 & 1 & 0 & 0 & 0 & 1 & 0 & 1 & 0 \\
 X_{23}^{2} & 0 & 0 & 1 & 0 & 1 & 0 & 0 & 0 & 1 \\
 X_{23}^{1} & 0 & 0 & 0 & 1 & 1 & 0 & 0 & 0 & 1 \\
 X_{14}^{1} & 0 & 0 & 1 & 0 & 0 & 1 & 0 & 0 & 1 \\
 X_{14}^{2} & 0 & 0 & 0 & 1 & 0 & 1 & 0 & 0 & 1
\end{array}
\right)
$
}
\vspace{0.5cm}

 \noindent The F-term charge matrix $Q_F=\ker{(P)}$ is

\noindent\makebox[\textwidth]{%
\footnotesize
$
Q_F=
\left(
\begin{array}{cccc|ccccc}
p_1 & p_2 & p_3 & p_4 & s_1 & s_2 & s_3 & s_4 & s_5 \\
\hline
 1 & 1 & 0 & 0 & 0 & 0 & -1 & -1 & 0 \\
 0 & 0 & 1 & 1 & 0 & 0 & -1 & 0 & -1 \\
 0 & 0 & 0 & 0 & 1 & 1 & 0 & -1 & -1
\end{array}
\right)
$
}
\vspace{0.5cm}

\noindent The D-term charge matrix is

\noindent\makebox[\textwidth]{%
\footnotesize
$
Q_D=
\left(
\begin{array}{cccc|ccccc}
p_1 & p_2 & p_3 & p_4 & s_1 & s_2 & s_3 & s_4 & s_5 \\
\hline
 0 & 0 & 0 & 0 & 0 & 1 & -1 & 0 & 0 \\
 0 & 0 & 0 & 0 & 0 & 0 & 1 & -1 & 0 \\
 0 & 0 & 0 & 0 & 0 & 0 & 0 & 1 & -1
\end{array}
\right)
$
}
\vspace{0.5cm}

The total charge matrix $Q_t$ exhibits two pairs of identical columns. Accordingly, the global symmetry is enhanced to $SU(2)_{x_1}\times SU(2)_{x_2} \times U(1)_R$. The mesonic charges on extremal perfect matchings are found following the discussion in \sref{s1_3}. They are identical to the ones for Model 15a and are presented in \tref{t15a}.

The product of all internal perfect matchings is expressed as
\beal{esm15b_x1}
s = \prod_{m=1}^{5} s_m ~.
\eea
The fugacity which counts the above product is $y_{s}$. The fugacity which counts the remaining extremal perfect matchings $p_\alpha$ is $t_\alpha$.

The mesonic Hilbert series for Model 15b is found using the Molien integral formula in \eref{es12_2}. The mesonic Hilbert series of Model 15b is identical to the one for Model 15a in \eref{esm15a_1}. 

The moduli space generators in terms of perfect matchings of Model 15b are shown in \tref{t15agen}. In terms of quiver fields of Model 15b, they are presented in \tref{t15bgen2}. The lattice of generators is a reflexive polygon and the dual of the toric diagram.

\comment{
\begin{table}[H]
\centering
\resizebox{\hsize}{!}{
\begin{minipage}[!b]{0.5\textwidth}
\begin{tabular}{|l|c|c|}
\hline
Generator & $SU(2)_{x_1}$ & $SU(2)_{x_2}$ 
\\
\hline
\hline
$p_{1}^2 p_{3}^2 ~ \prod_{m=1}^{5} s_m$
& 1 & 1
\nn\\
$p_{1} p_{2} p_{3}^2 ~ \prod_{m=1}^{5} s_m$
& 0 & 1
\nn\\
$p_{2}^2 p_{3}^2 ~ \prod_{m=1}^{5} s_m$
& -1 & 1
\nn\\
$p_{1}^2 p_{3} p_{4} ~ \prod_{m=1}^{5} s_m$
& 1 & 0
\nn\\
$p_{1} p_{2} p_{3} p_{4} ~ \prod_{m=1}^{5} s_m$
& 0 & 0
\nn\\
$p_{2}^2 p_{3} p_{4} ~ \prod_{m=1}^{5} s_m$
& -1 & 0
\nn\\
$p_{1}^2 p_{4}^2 ~ \prod_{m=1}^{5} s_m$
& 1 & -1
\nn\\
$p_{1} p_{2} p_{4}^2 ~ \prod_{m=1}^{5} s_m$
& 0 & -1
\nn\\
$p_{2}^2 p_{4}^2 ~ \prod_{m=1}^{5} s_m$
& -1 & -1
\nn\\
   \hline
\end{tabular}
\end{minipage}
\hspace{1cm}
\begin{minipage}[!b]{0.3\textwidth}
\includegraphics[width=4 cm]{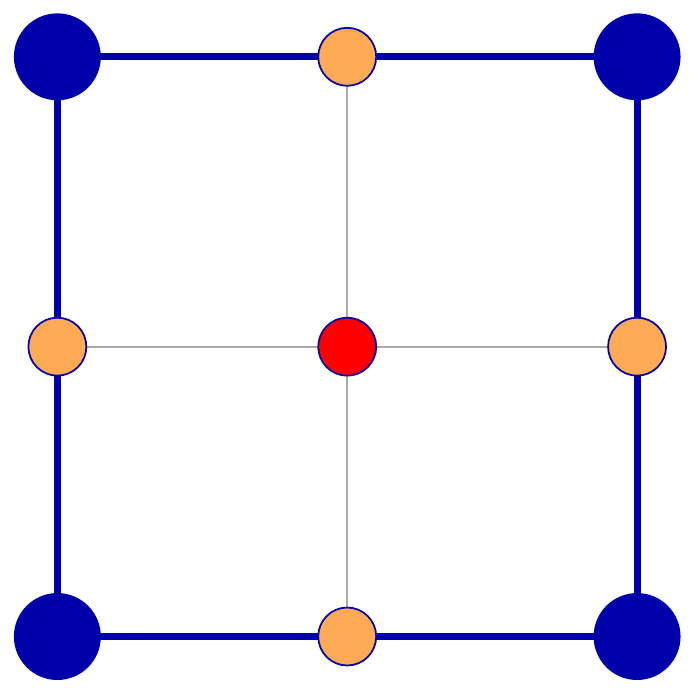}
\end{minipage}
}
\caption{The generators and lattice of generators of the mesonic moduli space of Model 15b in terms of GLSM fields with the corresponding flavor charges. The lattice of generators is the toric diagram of Model 4.\label{t15bgen}\label{f15bgen}} 
\end{table}
}

\begin{table}[H]
\centering
\resizebox{\hsize}{!}{
\begin{tabular}{|l|c|c|}
\hline
Generator & $SU(2)_{x_1}$ & $SU(2)_{x_2}$ 
\\
\hline
\hline
$X_ {14}^{1} X_ {42}^{2} X_ {21}^{1}=  X_ {23}^{2} X_ {34}^{2} X_{42}^{2}$
& 1 & 1
\nn\\
$ X_{14}^{1} X_{42}^{3} X_{21}^{1}=  X_{14}^{1} X_{42}^{2} X_{21}^{2}= X_{23}^{2} X_{34}^{1} X_{42}^{2}=  X_{23}^{2} X_{34}^{2} X_{42}^{3}$
& 0 & 1
\nn\\
$X_{14}^{1} X_{42}^{3} X_{21}^{2}=  X_{23}^{2} X_{34}^{1} X_{42}^{3}$
& -1 & 1
\nn\\
$X_ {14}^{1} X_ {42}^{4} X_ {21}^{1}=  X_ {14}^{2} X_ {42}^{2} X_{21}^{1}=  X_ {23}^{1} X_ {34}^{2} X_ {42}^{2}=  X_ {23}^{2} X_{34}^{2} X_ {42}^{4}$
& 1 & 0
\nn\\
$ X_{14}^{1} X_{42}^{1} X_{21}^{1}=  X_{14}^{1} X_{42}^{4} X_{21}^{2}=  X_{14}^{2} X_{42}^{3} X_{21}^{1}=  X_{14}^{2} X_{42}^{2} X_{21}^{2}=  X_{23}^{1} X_{34}^{1} X_{42}^{2}=  X_{23}^{1} X_{34}^{2} X_{42}^{3}=  X_{23}^{2} X_{34}^{1} X_{42}^{4}=  X_{23}^{2} X_{34}^{2} X_{42}^{1}$
& 0 & 0
\nn\\
$X_ {14}^{1} X_ {42}^{1} X_ {21}^{2}=  X_ {14}^{2} X_ {42}^{3} X_{21}^{2}=  X_ {23}^{1} X_ {34}^{1} X_ {42}^{3}=  X_ {23}^{2} X_{34}^{1} X_ {42}^{1}$
& -1 & 0
\nn\\
$X_ {14}^{2} X_ {42}^{4} X_ {21}^{1}=  X_ {23}^{1} X_ {34}^{2} X_{42}^{4}$
& 1 & -1
\nn\\
$X_ {14}^{2} X_ {42}^{1} X_ {21}^{1}=  X_ {14}^{2} X_ {42}^{4} X_ {21}^{2}=  X_ {23}^{1} X_ {34}^{1} X_ {42}^{4}=  X_ {23}^{1} X_{34}^{2} X_ {42}^{1}$
& 0 & -1
\nn\\
$X_{14}^{2} X_{42}^{1} X_{21}^{2}=  X_{23}^{1} X_{34}^{1} X_{42}^{1}$
& -1 & -1
\nn\\
   \hline
\end{tabular}
}
\caption{The generators in terms of bifundamental fields (Model 15b).\label{t15bgen2}\label{f15bgen2}} 
\end{table}

\section{Model 16: $\mathbb{C}^{3}/\mathbb{Z}_{3}~(1,1,1),~\text{dP}_0$ \label{sm16}}

\begin{figure}[H]
\begin{center}
\includegraphics[trim=0cm 0cm 0cm 0cm,width=4.5 cm]{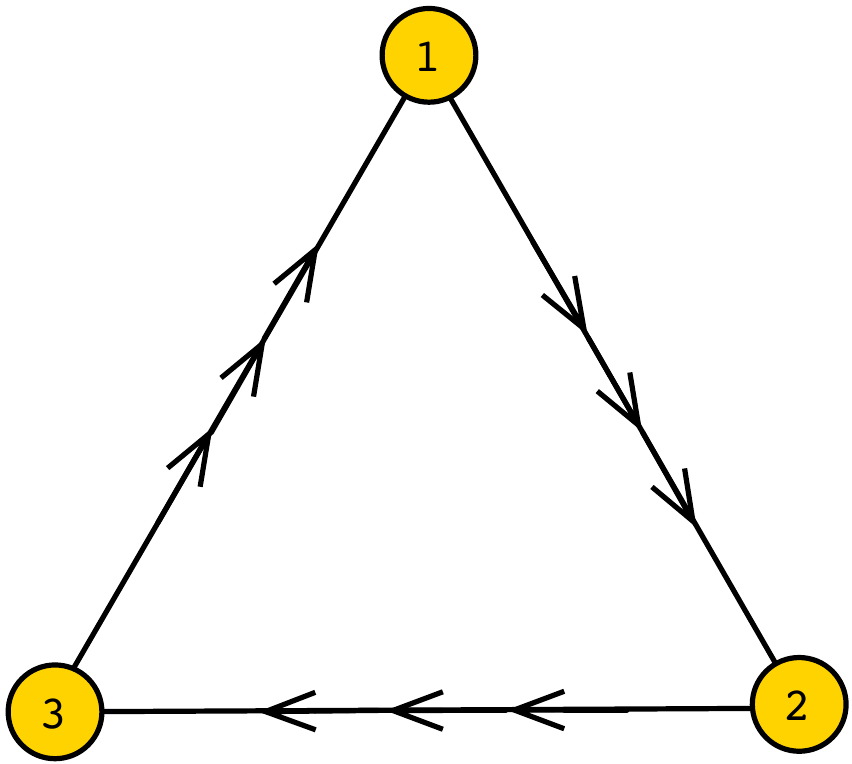}
\includegraphics[width=5 cm]{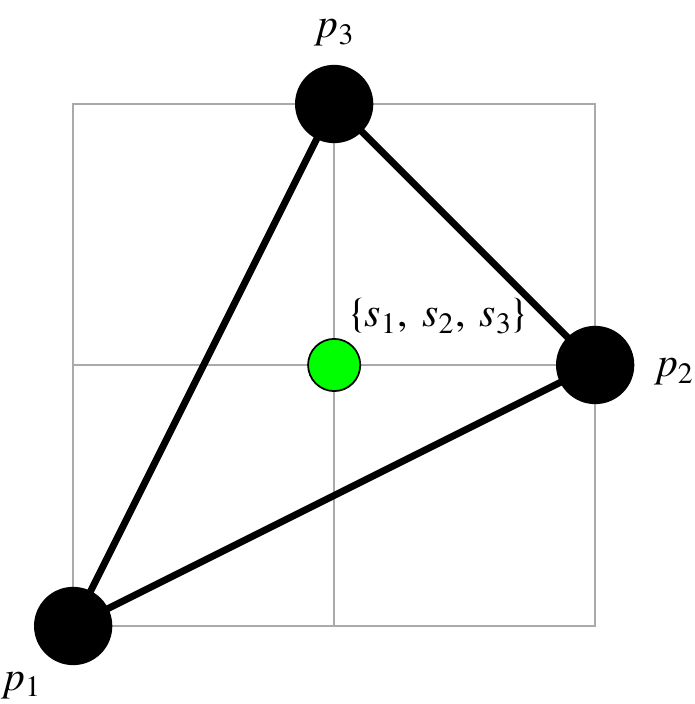}
\includegraphics[width=5 cm]{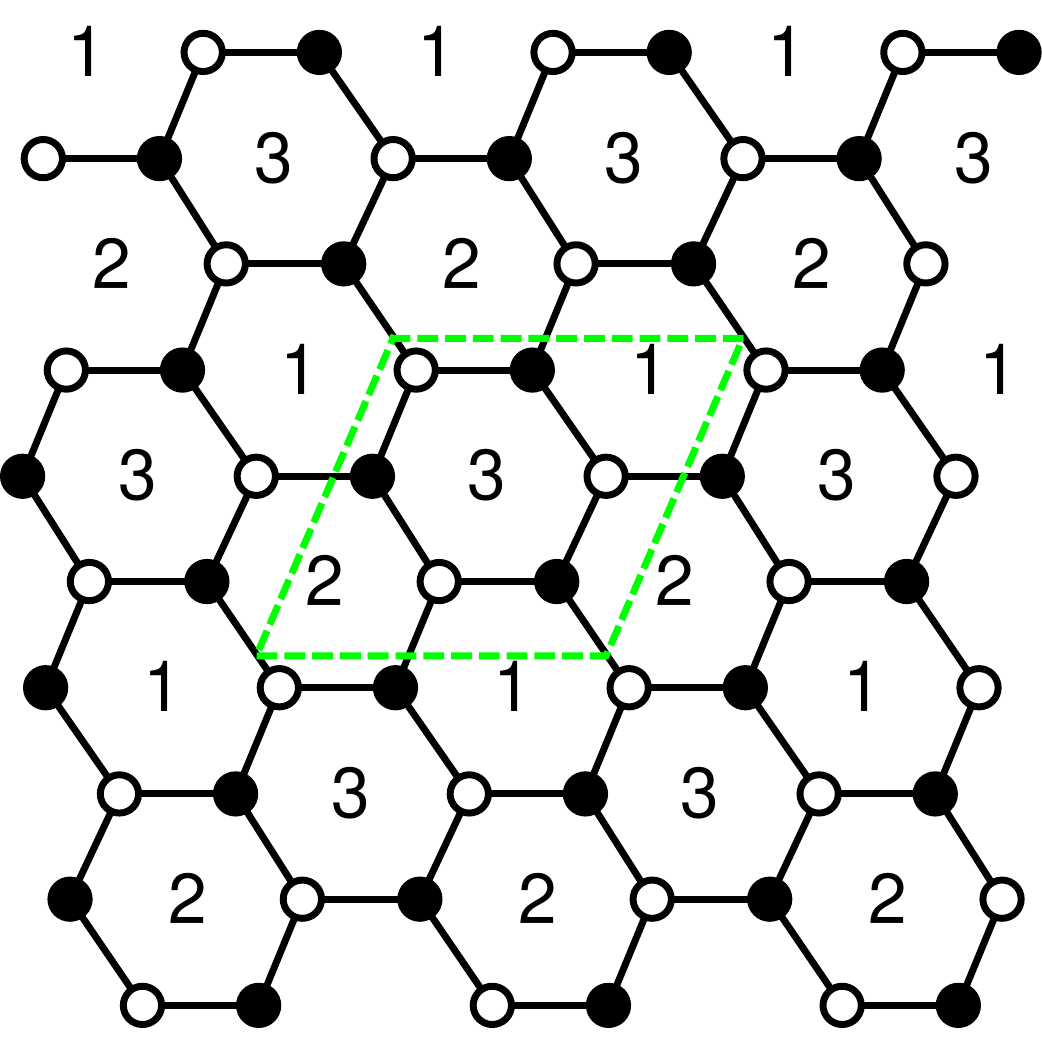}
\caption{The quiver, toric diagram, and brane tiling of Model 16.}
  \label{f16}
 \end{center}
 \end{figure}
 
 \noindent The superpotential is 
\beal{esm14_00}
W&=&
+ X_{12}^{1}X_{23}^{3}X_{31}^{2} 
+ X_{12}^{2}X_{23}^{1}X_{31}^{3}
+ X_{12}^{3}X_{23}^{2}X_{31}^{1}
\nn\\
&&
- X_{12}^{1}X_{23}^{1}X_{31}^{1} 
- X_{12}^{3}X_{23}^{3}X_{31}^{3} 
- X_{12}^{2}X_{23}^{2}X_{31}^{2}
\eea
 
 \noindent The perfect matching matrix is 
 
\noindent\makebox[\textwidth]{%
\footnotesize
$
P=
\left(
\begin{array}{c|ccc|ccc}
 \; & p_1 & p_2 & p_3 & s_1 & s_2 & s_3 \\
 \hline
 X_{12}^{3} & 1 & 0 & 0 & 1 & 0 & 0 \\
 X_{31}^{2} & 1 & 0 & 0 & 0 & 1 & 0 \\
 X_{23}^{1} & 1 & 0 & 0 & 0 & 0 & 1 \\
 X_{12}^{1} & 0 & 1 & 0 & 1 & 0 & 0 \\
 X_{31}^{3} & 0 & 1 & 0 & 0 & 1 & 0 \\
 X_{23}^{2} & 0 & 1 & 0 & 0 & 0 & 1 \\
 X_{12}^{2} & 0 & 0 & 1 & 1 & 0 & 0 \\
 X_{31}^{1} & 0 & 0 & 1 & 0 & 1 & 0 \\
 X_{23}^{3} & 0 & 0 & 1 & 0 & 0 & 1
\end{array}
\right)
$
}
\vspace{0.5cm}

 \noindent The F-term charge matrix $Q_F=\ker{(P)}$ is

\noindent\makebox[\textwidth]{%
\footnotesize
$
Q_F=
\left(
\begin{array}{ccc|ccc}
p_1 & p_2 & p_3 & s_1 & s_2 & s_3 \\
\hline
 1 & 1 & 1 & -1 & -1 & -1
\end{array}
\right)
$
}
\vspace{0.5cm}

\noindent The D-term charge matrix is

\noindent\makebox[\textwidth]{%
\footnotesize
$
Q_D=
\left(
\begin{array}{ccc|ccc}
p_1 & p_2 & p_3 & s_1 & s_2 & s_3 \\
\hline
 0 & 0 & 0 & 1 & -1 & 0 \\
 0 & 0 & 0 & 0 & 1 & -1
\end{array}
\right)
$
}
\vspace{0.5cm}

One observes that the GLSM fields corresponding to the extremal points of the toric diagram in \fref{f16} are equally charged under the F- and D-term constraints. This is shown by three identical columns of the total charge matrix $Q_t$. This leads to the enhancement of the global symmetry from $U(1)^3$ to $SU(3)_{(x_1,x_2)} \times U(1)_R$. Accordingly, the mesonic charges on the GLSM fields corresponding to extremal points in the toric diagram in \fref{f16} can be found following the discussion in \sref{s1_3}. They are presented in \tref{t16}.

\begin{table}[H]
\centering
\begin{tabular}{|c||rl|c||c|} 
\hline
\; & \multicolumn{2}{|c|}{$SU(3)_{(x_1,x_2)}$} & $U(1)_R$ & fugacity \\
\hline
\hline
$p_1$ & (-1/3,&-1/3)& $2/3$ &  $t_1$\\
$p_2$ & (+2/3,&-1/3)& $2/3$ &  $t_2$\\
$p_3$ & (-1/3,&+2/3)& $2/3$ &  $t_3$\\
\hline
\end{tabular}
\caption{The GLSM fields corresponding to extremal points of the toric diagram with their mesonic charges (Model 16).\label{t16}}
\end{table}

The product of all internal perfect matchings expressed as
\beal{esm16_x1}
s = \prod_{m=1}^{3} s_m~.
\eea
The above product is counted by the fugacity $y_{s}$. The remaining extremal perfect matchings $p_\alpha$ are counted by $t_\alpha$.

The mesonic Hilbert series of Model 16 is calculated using the Molien integral formula in \eref{es12_2}. It is
 \beal{esm16_1}
 &&
g_{1}(t_\alpha,y_{s}; \mathcal{M}^{mes}_{16})= 
\nn\\
&&
\hspace{0.5cm}
\frac{
1 
+ y_{s}~ t_1^2 t_2 
+ y_{s}~ t_1 t_2^2 
+ y_{s}~ t_1^2 t_3 
+ y_{s}~ t_1 t_2 t_3 
+ y_{s}~ t_2^2 t_3 
+ y_{s}~ t_1 t_3^2 
+ y_{s}~ t_2 t_3^2 
+ y_{s}^2~ t_1^2 t_2^2 t_3^2
}{
(1 - y_{s}~t_1^3) (1 - y_{s}~t_2^3) (1 - y_{s}~t_3^3)
}
~~.
\nn\\
 \eea
 The plethystic logarithm of the mesonic Hilbert series is
\beal{esm16_3}
&&
PL[g_1(t_\alpha,y_{s};\mathcal{M}_{16}^{mes})]=
y_{s}~t_1^3 
+ y_{s}~t_1^2 t_2 
+ y_{s}~t_1 t_2^2 
+ y_{s}~t_2^3 
+ y_{s}~t_1^2 t_3 
+ y_{s}~t_1 t_2 t_3 
+ y_{s}~t_2^2 t_3 
\nn\\
&&
\hspace{0.5cm}
+ y_{s}~t_1 t_3^2 
+ y_{s}~t_2 t_3^2 
+ y_{s}~t_3^3
- y_{s}^2~t_1^4 t_2^2 
- y_{s}^2~t_1^3 t_2^3 
- y_{s}^2~t_1^2 t_2^4 
- y_{s}^2~t_1^4 t_2 t_3 
- 2~y_{s}^2~ t_1^3 t_2^2 t_3 
\nn\\
&&
\hspace{0.5cm}
- 2~y_{s}^2~ t_1^2 t_2^3 t_3 
- y_{s}^2~t_1 t_2^4 t_3 
- y_{s}^2~t_1^4 t_3^2 
- 2~y_{s}^2~ t_1^3 t_2 t_3^2 
- 3~y_{s}^2~ t_1^2 t_2^2 t_3^2
- 2~y_{s}^2~ t_1 t_2^3 t_3^2 
- y_{s}^2~t_2^4 t_3^2 
\nn\\
&&
\hspace{0.5cm}
- y_{s}^2~t_1^3 t_3^3 
- 2~y_{s}^2~ t_1^2 t_2 t_3^3 
- 2~y_{s}^2~ t_1 t_2^2 t_3^3 
- y_{s}^2~t_2^3 t_3^3 
- y_{s}^2~t_1^2 t_3^4 
- y_{s}^2~t_1 t_2 t_3^4 
- y_{s}^2~t_2^2 t_3^4 
+ \dots~.
\nn\\
\eea

Consider the following fugacity map
\beal{esm16_y1}
x_1 = 
\frac{t_2}{t_1}
~,~
x_2 =
\frac{t_3}{t_1}
~,~
t =
y_s^{1/3} ~ t_1^{1/3} t_2^{1/3} t_3^{1/3}
~,~
\eea
where $x_1$, $x_2$ and $t$ count the mesonic charges. The fugacities $x_1$ and $x_2$ with their powers being integers count integer flavour charges. With a further redefinition of fugacities,
\beal{esm16_y2}
\tilde{x}_1=\frac{1}{x_1^{1/3} x_2^{1/3}}
~,~
\tilde{x}_2=\frac{x_1^{1/3}}{x_2^{2/3}}
\eea
the Hilbert series and plethystic logarithm can be expressed in terms of characters of irreducible representations of $SU(3)$. The expansion of the Hilbert series takes the form
 \beal{esm16b_1}
g_{1}(t,\tilde{x}_1,\tilde{x}_2; \mathcal{M}^{mes}_{16})= 
\sum_{n=0}^{\infty}~
[3n,0]_{(\tilde{x}_1,\tilde{x}_2)}~t^{3n}
~~.
 \eea
The plethystic logarithm is
\beal{esm16_3}
&&
PL[g_1(t,\tilde{x}_1,\tilde{x}_2;\mathcal{M}_{16}^{mes})]=
[3,0]_{(\tilde{x}_1,\tilde{x}_2)} t^3
- [2,2]_{(\tilde{x}_1,\tilde{x}_2)} t^6
+ ([1,1]_{(\tilde{x}_1,\tilde{x}_2)}+[1,4]_{(\tilde{x}_1,\tilde{x}_2)}
\nn\\
&&
\hspace{0.5cm}
+[2,2]_{(\tilde{x}_1,\tilde{x}_2)}+[4,1]_{(\tilde{x}_1,\tilde{x}_2)}) t^9
- (
2[0,3]_{(\tilde{x}_1,\tilde{x}_2)}
+ 2[1,1]_{(\tilde{x}_1,\tilde{x}_2)}
+ 2[1,4]_{(\tilde{x}_1,\tilde{x}_2)}
\nn\\
&&
\hspace{0.5cm}
+ 2[2,2]_{(\tilde{x}_1,\tilde{x}_2)}
+ [2,5]_{(\tilde{x}_1,\tilde{x}_2)}
+ 2[3,0]_{(\tilde{x}_1,\tilde{x}_2)}
+ 2[3,3]_{(\tilde{x}_1,\tilde{x}_2)}
+ 2[4,1]_{(\tilde{x}_1,\tilde{x}_2)}
\nn\\
&&
\hspace{0.5cm}
+ [5,2]_{(\tilde{x}_1,\tilde{x}_2)}
) t^{12}
   +\dots~.
   \eea
In terms of fugacities $x_1$ and $x_2$ the above plethystic logarithm exhibits the moduli space generators with their integer flavour charges and R-charges. They are summarized in \tref{t16gen}. The generators can be presented on a charge lattice. The lattice of generators is the dual polygon of the toric diagram. As indicated in \eref{esm16_3}, the generators fall into an irreduciable representation of $SU(3)$ with the character being
\beal{es16_xx5}
[3,0]_{(\tilde{x}_1,\tilde{x}_2)} t^3
=
\left(
\tilde{x}_1^3
+ \tilde{x}_1 \tilde{x}_2
+ \frac{\tilde{x}_1^2}{\tilde{x}_2}
+ \frac{\tilde{x}_2^2}{\tilde{x}_1}
+ 1
+ \frac{\tilde{x}_2^3}{\tilde{x}_1^3}
+ \frac{\tilde{x}_1}{\tilde{x}_2^2}
+ \frac{\tilde{x}_2}{\tilde{x}_1^2}
+ \frac{1}{\tilde{x}_1 \tilde{x}_2}
+ \frac{1}{\tilde{x}_2^3}
\right) t^3~~.
\nn\\
\eea
The generators of the mesonic moduli space in terms of quiver fields of Model 16 are shown in \tref{f16gen2}.
\\

\begin{table}[H]
\centering
\resizebox{\hsize}{!}{
\begin{minipage}[!b]{0.6\textwidth}
\begin{tabular}{|l|rl|}
\hline
Generator & \multicolumn{2}{|c|}{$SU(3)_{(x_1,x_2)}$} 
\\
\hline
\hline
$p_{1}^3 ~ 
s$
& (-1,&-1)
\nn\\
$p_{1}^2 p_{2} ~ 
s$
& (0,&-1)
\nn\\
$p_{1} p_{2}^2 ~ 
s$
& (1,&-1)
\nn\\
$p_{2}^3 ~ 
s$
& (2,&-1)
\nn\\
$p_{1}^2 p_{3} ~ 
s$
& (-1,&0)
\nn\\
$p_{1} p_{2} p_{3} ~ 
s$
& (0,&0)
\nn\\
$p_{2}^2 p_{3} ~
s$
& (1,&0)
\nn\\
$p_{1} p_{3}^2 ~
s$
& (-1,&1)
\nn\\
$p_{2} p_{3}^2 ~
s$
& (0,&1)
\nn\\
$p_{3}^3 ~
s$
& (-1,&2)
\nn\\
   \hline
\end{tabular}
\end{minipage}
\hspace{1cm}
\begin{minipage}[!b]{0.3\textwidth}
\includegraphics[width=4 cm]{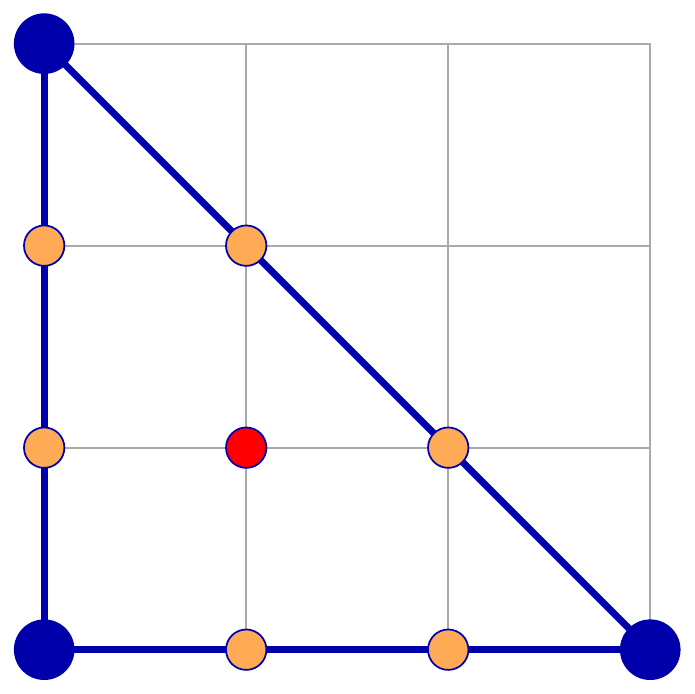}
\end{minipage}
}
\caption{The generators and lattice of generators of the mesonic moduli space of Model 16 in terms of GLSM fields with the corresponding flavor charges. \label{t16gen}\label{f16gen}}
\end{table}

\begin{table}[H]
\centering
\resizebox{\hsize}{!}{
\begin{tabular}{|l|rl|}
\hline
Generator & \multicolumn{2}{|c|}{$SU(3)_{(x_1,x_2)}$} 
\\
\hline
\hline
$ X_{12}^{3} X_{23}^{1} X_{31}^{2}$
& (-1,&-1)
\nn\\
$X_ {12}^{1} X_ {23}^{1} X_ {31}^{2}=  X_ {12}^{3} X_ {23}^{1} X_{31}^{3}=  X_ {12}^{3} X_ {23}^{2} X_ {31}^{2}$
& (0,&-1)
\nn\\
$X_ {12}^{1} X_ {23}^{1} X_ {31}^{3}=  X_ {12}^{1} X_ {23}^{2} X_{31}^{2}=  X_ {12}^{3} X_ {23}^{2} X_ {31}^{3}$
& (1,&-1)
\nn\\
$X_{12}^{1} X_{23}^{2} X_{31}^{3}$
& (2,&-1)
\nn\\
$X_ {12}^{2} X_ {23}^{1} X_ {31}^{2}=  X_ {12}^{3} X_ {23}^{1} X_{31}^{1}=  X_ {12}^{3} X_ {23}^{3} X_ {31}^{2}$
& (-1,&0)
\nn\\
$X_ {12}^{1} X_ {23}^{1} X_ {31}^{1}=  X_ {12}^{1} X_ {23}^{3} X_{31}^{2}=  X_ {12}^{2} X_ {23}^{1} X_ {31}^{3}=  X_ {12}^{2} X_{23}^{2} X_ {31}^{2}=  X_ {12}^{3} X_ {23}^{2} X_ {31}^{1}=  X_{12}^{3} X_ {23}^{3} X_ {31}^{3}$
& (0,&0)
\nn\\
$ X_{12}^{1} X_{23}^{2} X_{31}^{1}=  X_{12}^{1} X_{23}^{3} X_{31}^{3}=  X_{12}^{2} X_{23}^{2} X_{31}^{3}$
& (1,&0)
\nn\\
$X_ {12}^{2} X_ {23}^{1} X_ {31}^{1}=  X_ {12}^{2} X_ {23}^{3} X_{31}^{2}=  X_ {12}^{3} X_ {23}^{3} X_ {31}^{1}$
& (-1,&1)
\nn\\
$X_ {12}^{1} X_ {23}^{3} X_ {31}^{1}=  X_ {12}^{2} X_ {23}^{2} X_{31}^{1}=  X_ {12}^{2} X_ {23}^{3} X_ {31}^{3}$
& (0,&1)
\nn\\
$X_{12}^{2} X_{23}^{3} X_{31}^{1}$
& (-1,&2)
\nn\\
   \hline
\end{tabular}
}
\caption{The generators in terms of bifundamental fields (Model 16).\label{t16gen2}\label{f16gen2}}
\end{table}
\hspace{0.1cm}

With the fugacity map
\beal{esm16_xx1}
T_1 = \frac{t}{x_1^{1/3} x_2^{1/3}} = y_{s}^{1/3} t_1~,~
T_2 = \frac{x_1^{2/3} t}{x_2^{1/3}} = y_{s}^{1/3} t_2~,~
T_3 = \frac{x_2^{2/3} t}{x_1^{1/3}} = y_{s}^{1/3} t_3~,~
\eea
the mesonic Hilbert series becomes
\beal{esm16_xx2}
&&
g_{1}(T_1,T_2,T_3;\mathcal{M}^{mes}_{16})=
\nn\\
&&
\hspace{0.5cm}
\frac{
1 + T_1^2 T_2
+ T_1 T_2^2
+ T_1^2 T_3
+ T_1 T_2 T_3 
+ T_2^2 T_3
+ T_1 T_3^2
+ T_2 T_3^2
+ T_1^2 T_2^2 T_3^2
}{
(1-T_1^3)
(1-T_2^3)
(1-T_3^3)
}~,
\nn\\
\eea
with the plethystic logarithm becoming
\beal{esm16_xx3}
&&
PL[g_{1}(T_1,T_2,T_3;\mathcal{M}^{mes}_{16})]
=
T_1^3 
+ T_1^2 T_2 
+ T_1 T_2^2 
+ T_2^3 
+ T_1^2 T_3 
+ T_1 T_2 T_3 
+ T_2^2 T_3 
+ T_1 T_3^2 
\nn\\
&&
\hspace{0.5cm}
+ T_2 T_3^2 
+ T_3^3
- T_1^4 T_2^2 
- T_1^3 T_2^3 
- T_1^2 T_2^4 
- T_1^4 T_2 T_3 
- 2~ T_1^3 T_2^2 T_3 
- 2~ T_1^2 T_2^3 T_3 
\nn\\
&&
\hspace{0.5cm}
- T_1 T_2^4 T_3 
- T_1^4 T_3^2 
- 2~ T_1^3 T_2 T_3^2 
- 3~ T_1^2 T_2^2 T_3^2
- 2~ T_1 T_2^3 T_3^2 
- T_2^4 T_3^2 
- T_1^3 T_3^3 
\nn\\
&&
\hspace{0.5cm}
- 2~ T_1^2 T_2 T_3^3 
- 2~ T_1 T_2^2 T_3^3 
- T_2^3 T_3^3 
- T_1^2 T_3^4 
- T_1 T_2 T_3^4 
- T_2^2 T_3^4
+ \dots~.
\eea
The above Hilbert series and plethystic logarithm are in terms of three fugacities with positive powers. This illustrates the conical structure of the toric Calabi-Yau 3-fold.
\\

\section{Seiberg Duality Trees \label{strees}}

\begin{figure}[H]
\begin{center}
\resizebox{0.76\hsize}{!}{
\includegraphics[trim=0cm 0cm 0cm 0cm,totalheight=18 cm]{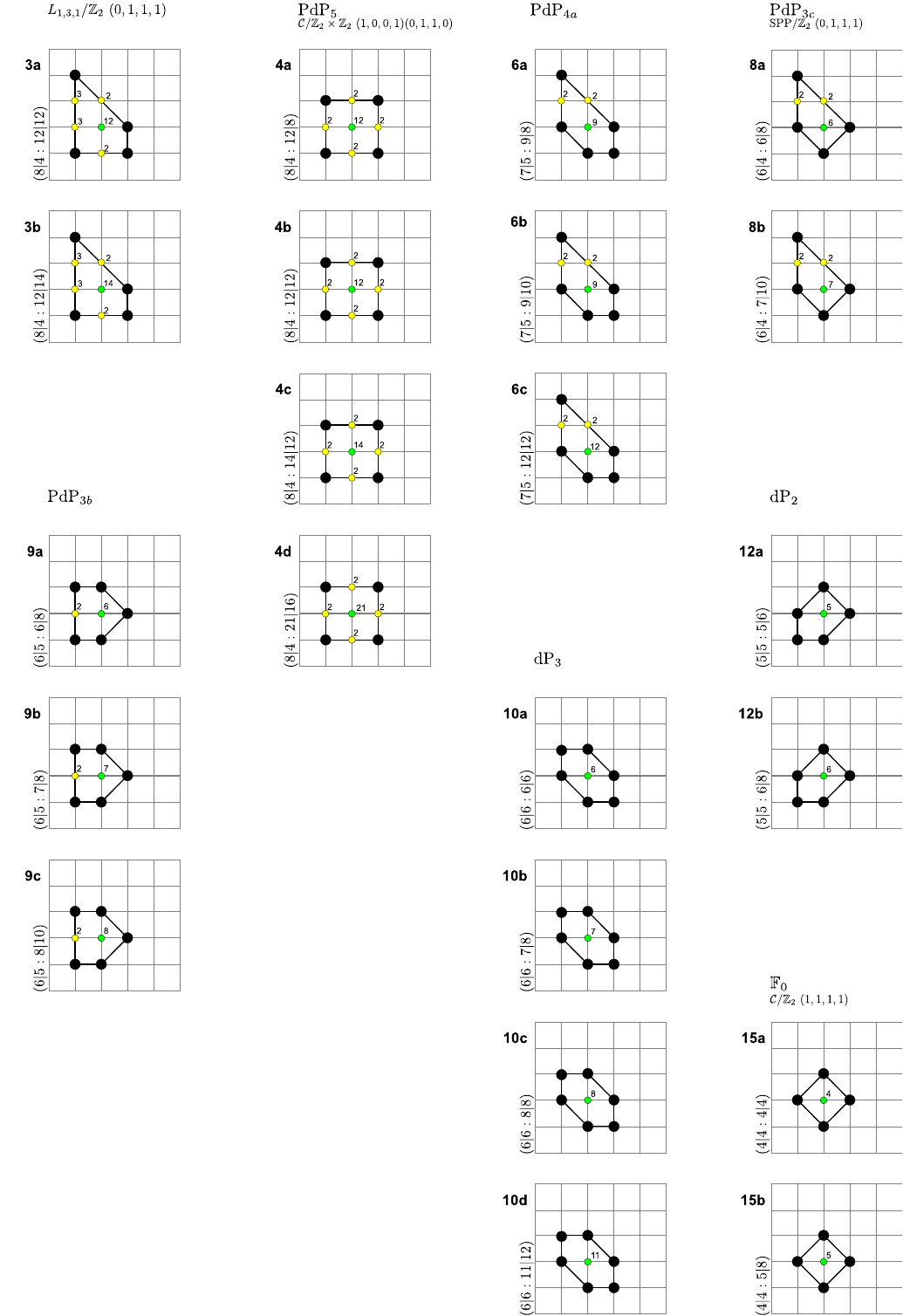}
}
  \caption{
  Toric Diagrams of toric (Seiberg) dual phases of quiver gauge theories with brane tilings. The label $(G|n_p:n_i|n_w)$ is used, where $G$, $n_p$, $n_i$ and $n_w$ are the number of $U(n)$ gauge groups, GLSM fields with non-zero R-charge, internal toric points and superpotential terms respectively.
  \label{f_seibergmodels}}
 \end{center}
 \end{figure}

The above sections have identified all $30$ supersymmetric gauge theories with brane tilings corresponding to the $16$ reflexive polygons. $8$ reflexive polygons are associated to multiple quiver gauge theories as summarized in \fref{f_seibergmodels}. These are called phases of the corresponding toric variety. For a given toric variety, the phases are so called \textit{toric (Seiberg) dual} and are related under toric (Seiberg) duality as discussed in appendix \sref{sapp_seiberg}. Multiple toric duality actions on various $U(n)$ gauge groups corresponding to $4$-sided faces in the brane tiling create closed orbits among the phases. 

\begin{figure}[H]
\begin{center}
\includegraphics[trim=0cm 0cm 0cm 0cm,width=14 cm]{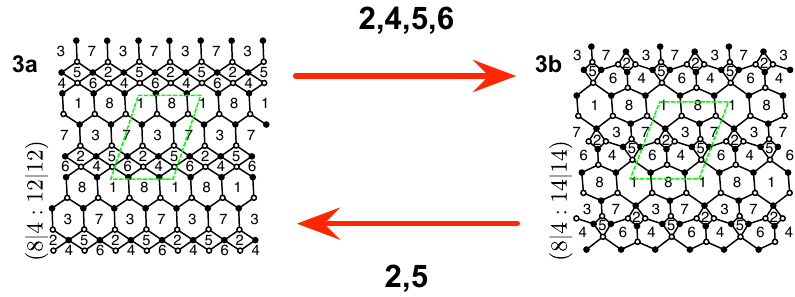}
\caption{The duality tree for $L_{131}/\mathbb{Z}_2$ with orbifold action $(0,1,1,1)$ [Model $3$].}
  \label{fduality3}
 \end{center}
 \end{figure}

\begin{figure}[H]
\begin{center}
\includegraphics[trim=1cm 0cm 0cm 0cm,width=17 cm]{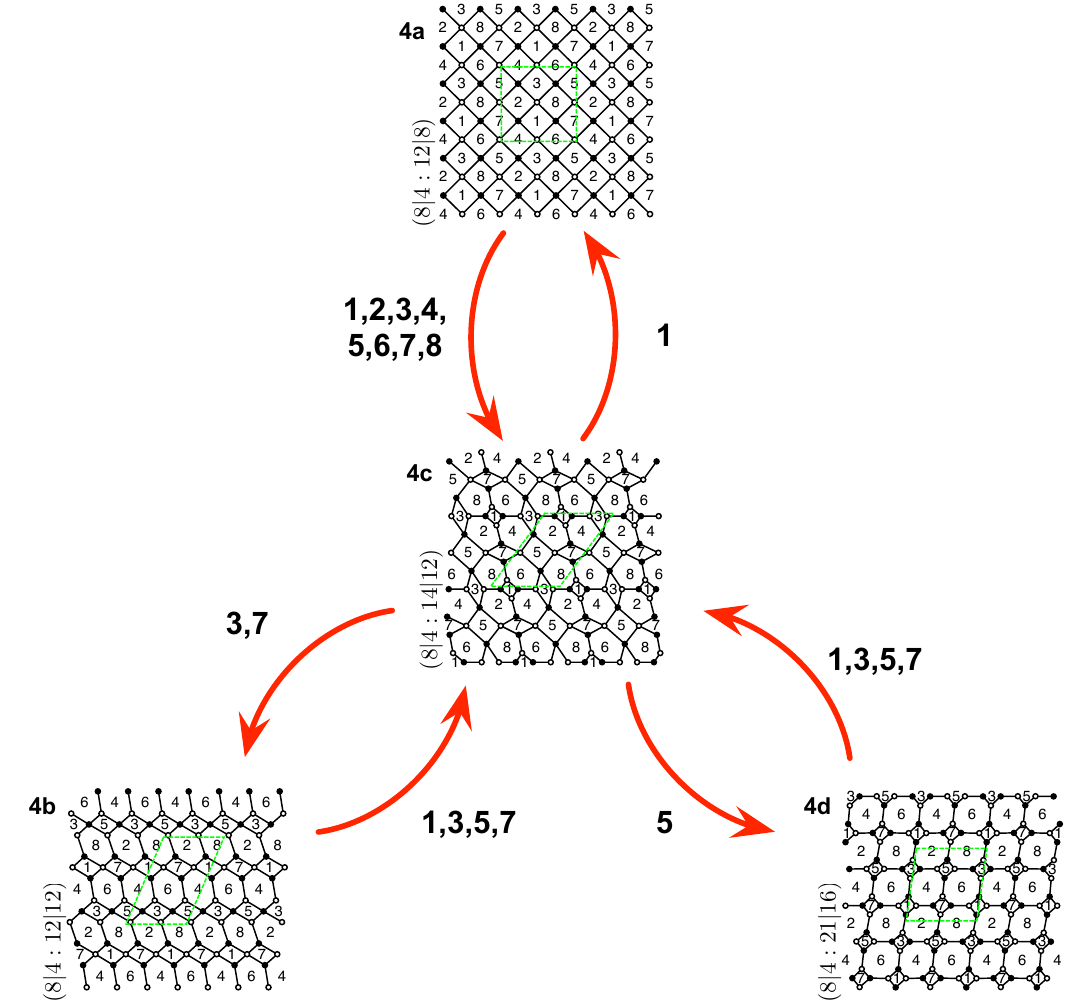}
\caption{The duality tree for $\mathcal{C}/\mathbb{Z}_2\times\mathbb{Z}_2$ with orbifold action $(0,1,1,0)(1,0,0,1)$ [Model $4$].}
  \label{fduality4}
 \end{center}
 \end{figure}

\begin{figure}[H]
\begin{center}
\includegraphics[trim=1.5cm 0cm 0cm 0cm,width=16.5 cm]{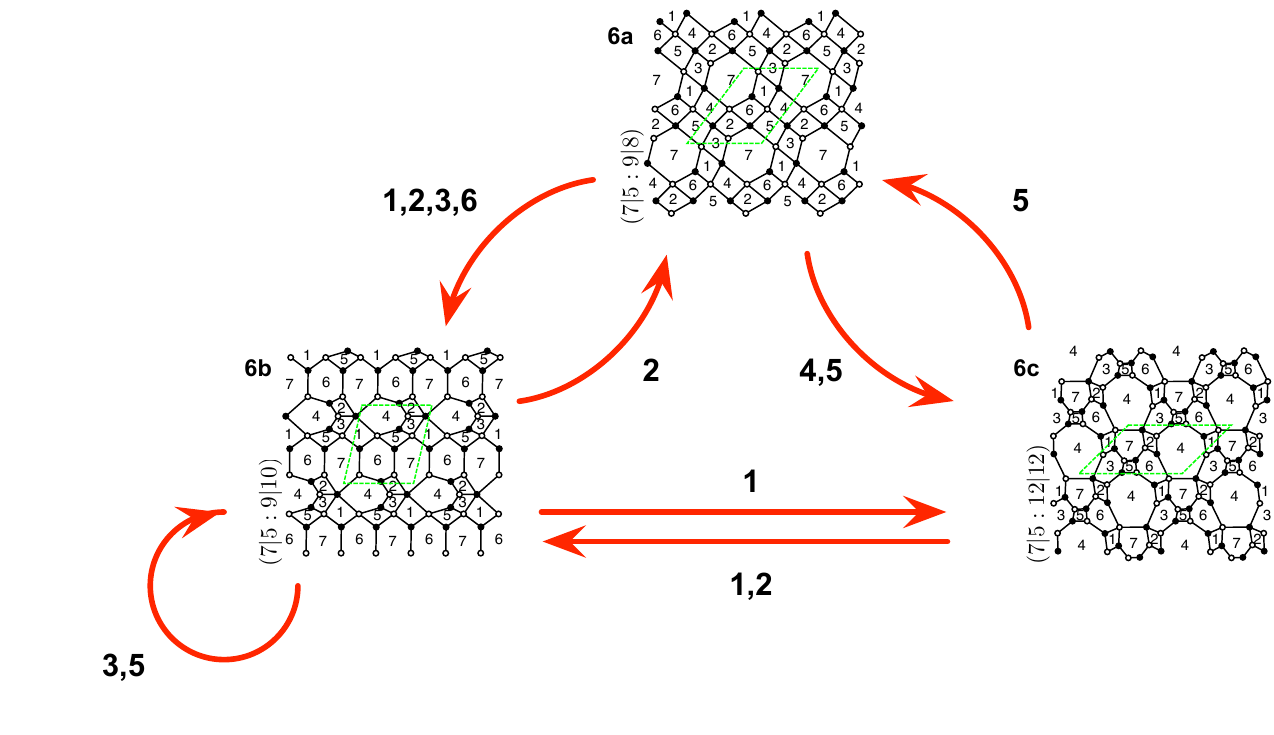}
\caption{The duality tree for $\text{PdP}_{4a}$ [Model $6$].}
  \label{fduality6}
 \end{center}
 \end{figure}

\begin{figure}[H]
\begin{center}
\includegraphics[trim=0cm 0cm 0cm 0cm,width=14 cm]{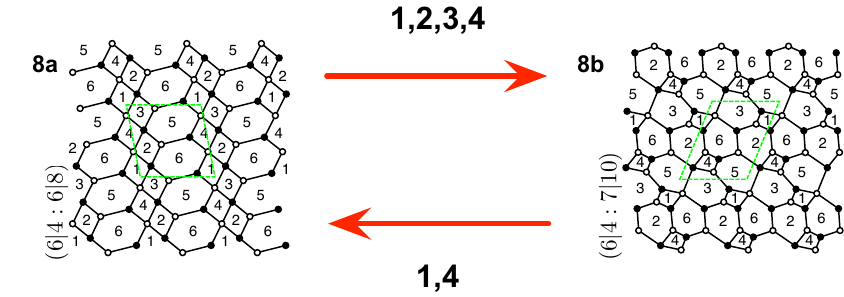}
\caption{The duality tree for $\text{SPP}/\mathbb{Z}_2$ with orbifold action $(0,1,1,1)$ [Model $8$].}
  \label{fduality8}
 \end{center}
 \end{figure}

\begin{figure}[H]
\begin{center}
\includegraphics[trim=1.5cm 0cm 0cm 0cm,width=19 cm]{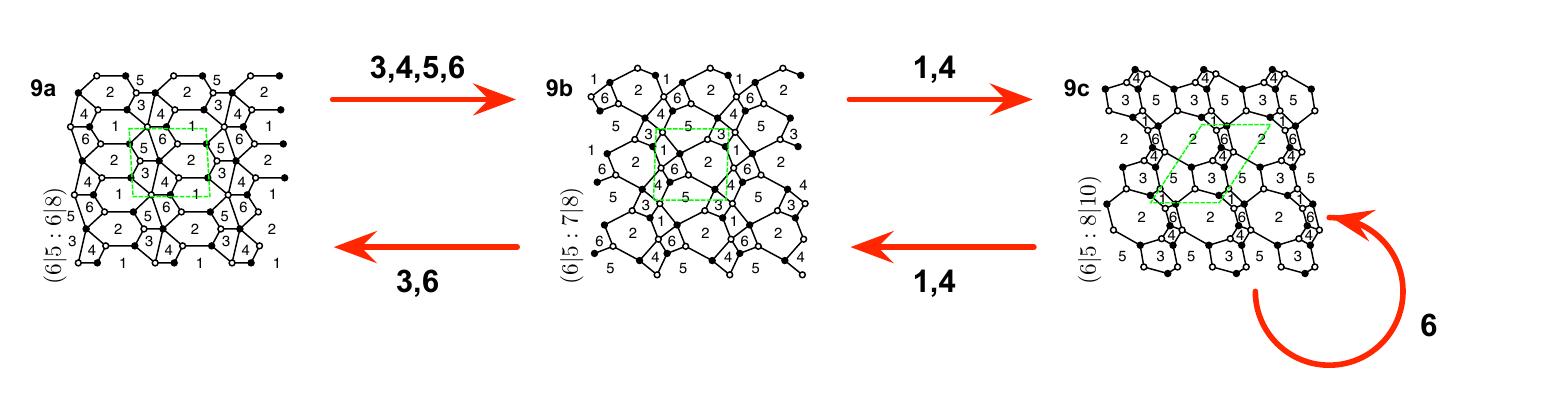}
\caption{The duality tree for $\text{PdP}_{3(b)}$ [Model $9$].}
  \label{fduality9}
 \end{center}
 \end{figure}

\begin{figure}[H]
\begin{center}
\includegraphics[trim=0cm 0cm 0cm 0cm,width=15.5 cm]{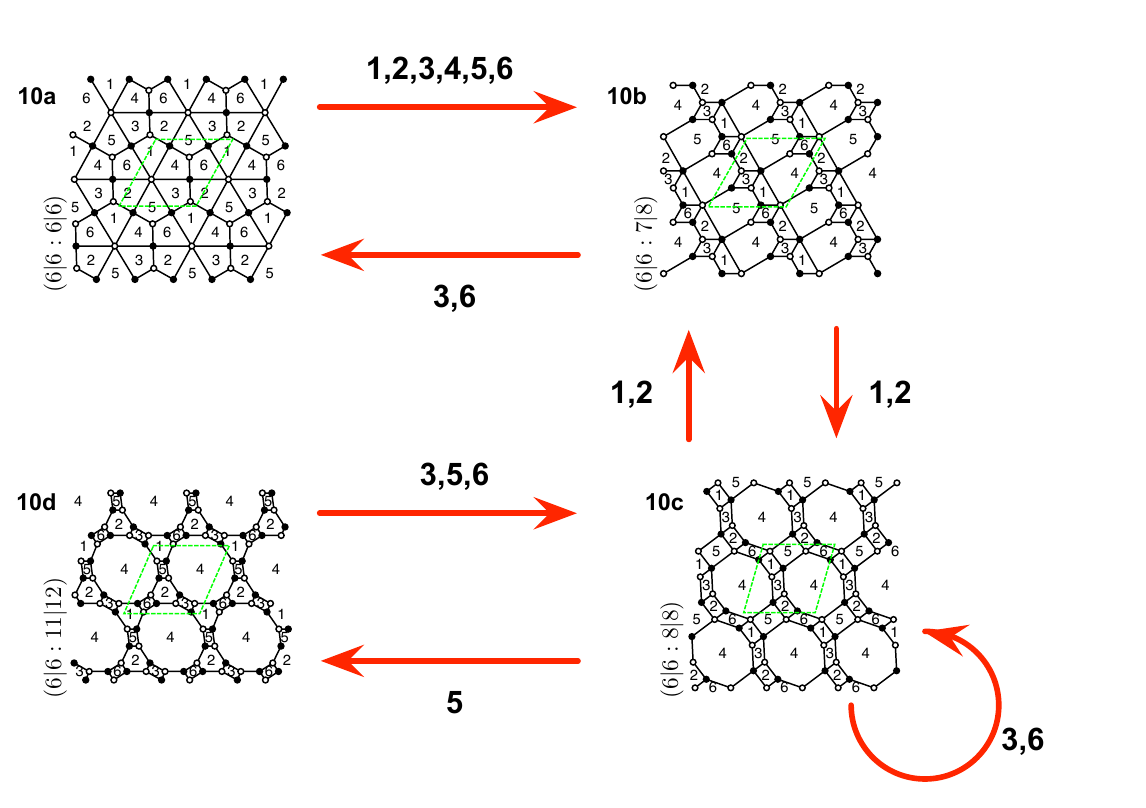}
\caption{The duality tree for $\text{dP}_{3}$ [Model $10$].}
  \label{fduality10}
 \end{center}
 \end{figure}

\begin{figure}[H]
\begin{center}
\includegraphics[trim=2cm 0cm 0cm 0cm,width=18.5 cm]{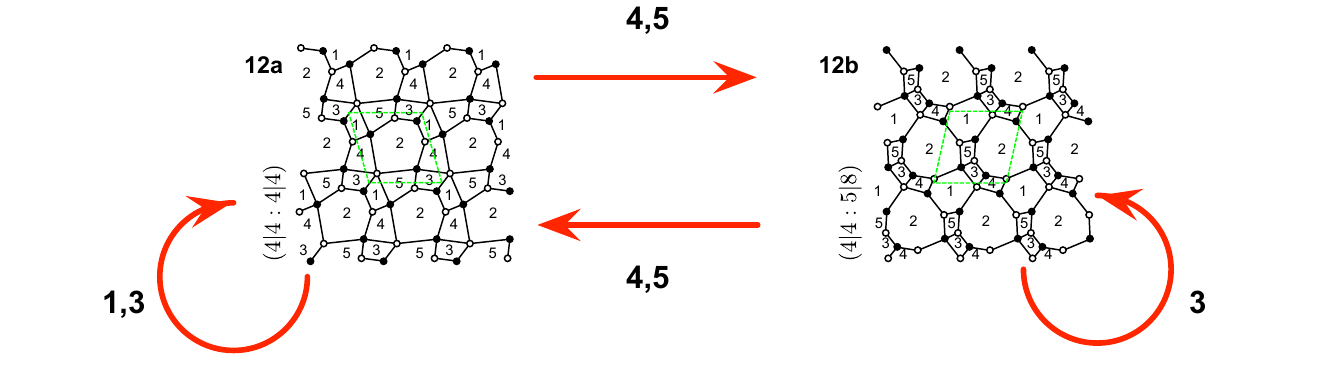}
\caption{The duality tree for $\text{dP}_{2}$ [Model $12$].}
  \label{fduality12}
 \end{center}
 \end{figure}

\begin{figure}[H]
\begin{center}
\includegraphics[trim=0cm 0cm 0cm 0cm,width=15 cm]{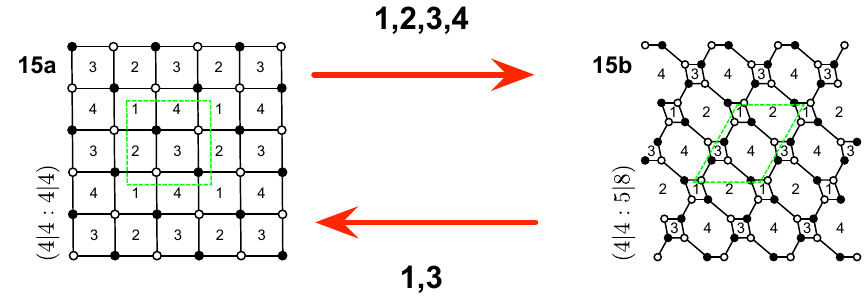}
\caption{The duality tree for $\mathcal{C}/\mathbb{Z}_2$ with orbifold action $(1,1,1,1)$ or the cone over $F_0$ [Model $15$].}
  \label{fduality15}
 \end{center}
 \end{figure}

In \fref{fduality3} to \fref{fduality15}, a summary of the orbits presented as \textit{duality trees} is shown, where nodes represent the brane tiling of the phase, and arrows are labelled with the index of the gauge group on which one acts under toric (Seiberg) duality to obtain the phase at the head of the arrow.

\clearpage
\section{Specular Duality and Conclusions}

The work above uses the $16$ reflexive polygons in \fref{f_sumtoric} as toric diagrams of Calabi-Yau moduli spaces of $3+1$ dimensional $\mathcal{N}=1$ supersymmetric gauge theories. These quiver gauge theories are represented by brane tilings. A natural question to ask from this setup is to identify all brane tilings corresponding to the $16$ reflexive polygons. Motivated by this line of thought, the following comprehensive results have been presented in this paper:
\begin{itemize}
\item There are exactly $30$ brane tilings encoding supersymmetric quiver gauge theories whose mesonic moduli spaces are represented by reflexive polygons. All gauge theories are related by a cascade of Higgs mechanisms. In addition, toric (Seiberg) duality maps multiple gauge theories to the same reflexive polygon.

\item The generating function of mesonic gauge invariant operators known as the mesonic Hilbert series is computed using the Molien integral formula for each of the $30$ quiver theories. Fugacities of the Hilbert series are related both to perfect matchings and hence points in the toric diagram as well as charges under the global symmetry of the gauge theory. Hilbert series of toric dual phases have been shown to be identical.

\item The generators of the mesonic moduli space of all $30$ quiver gauge theories have been found both in terms of chiral fields of the gauge theory as well as the perfect matchings of the brane tiling.

\item The mesonic charges on the moduli space generators have been found such that they form for each generator a point on $\mathbb{Z}^2$. The convex hull of all such points is a reflexive polygon. For all $30$ quiver gauge theories, these reflexive polygons known as lattice of generators are exactly the polar duals to the toric diagrams.
\end{itemize}

The above observations made by classifying all brane tilings corresponding to reflexive polygons lead to a comprehensive overview of a special set of quiver gauge theories. This overview is the precursor to a discovery of a new duality of quiver gauge theories. This \textbf{specular duality} is best observed in the context of toric diagrams with points labelled by perfect matchings of the brane tiling. Recall that extremal perfect matchings correspond to the corner points coloured black in the toric diagrams in \fref{f_sumtoric2}, whereas internal perfect matchings are points lying strictly within the perimeter of the polygon. External perfect matchings are all points on the perimeter of the polygon including the extremal ones. All except extremal perfect matchings correspond to GLSM fields with zero R-charge.

\begin{figure}[H]
\begin{center}
\resizebox{1\hsize}{!}{
\includegraphics[trim=0cm 0cm 0cm 0cm,totalheight=19 cm]{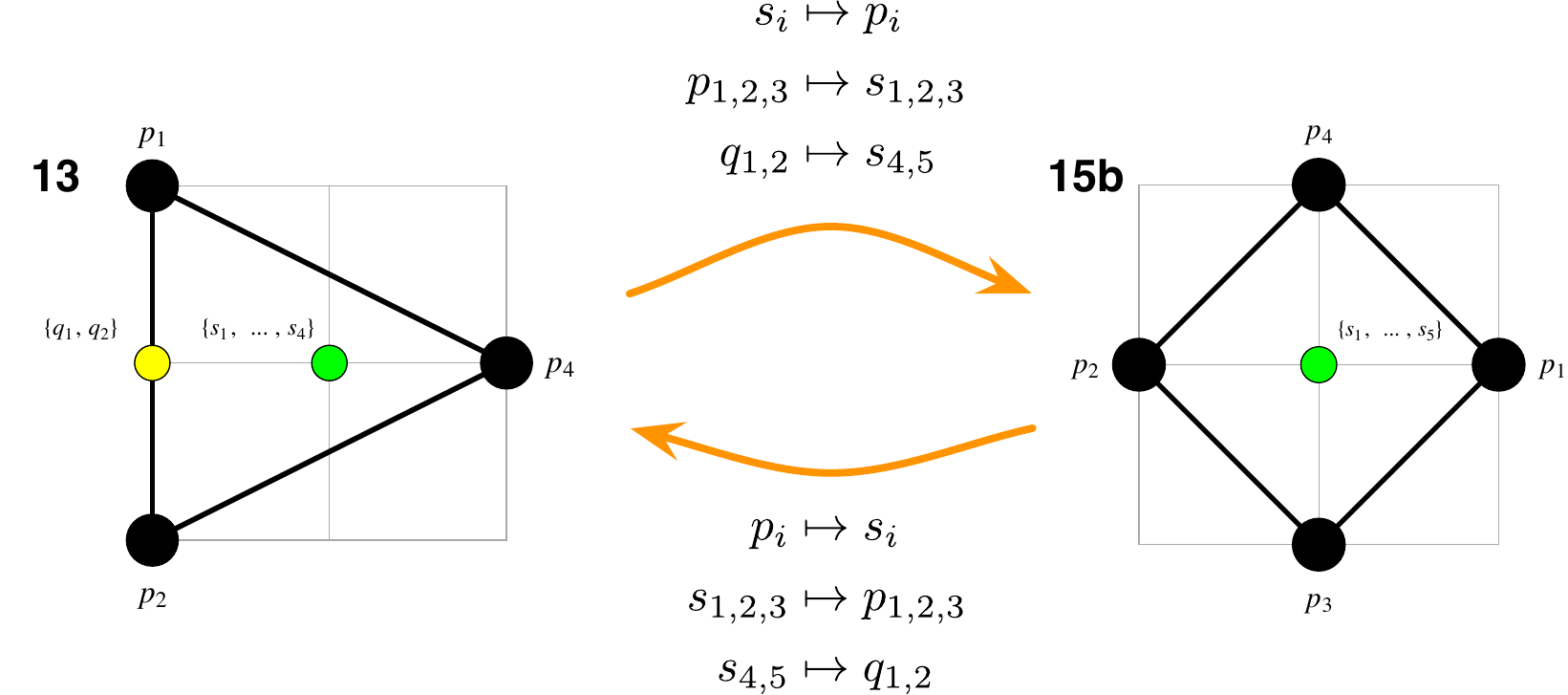}
}
\caption{Specular duality between Model 13 ($\mathbb{C}^3/\mathbb{Z}_{4} (1,1,2)$) and Model 15b  ($\mathbb{F}_{0}$, phase b). The exchange of internal and external perfect matchings map between the two models.
\label{f_dualexample}}
 \end{center}
 \end{figure}

The new duality we propose exchanges the internal perfect matchings with the external perfect matchings. For the set of brane tilings corresponding to reflexive polygons, the duality map is unique by forming duality pairs between models as follows
\beal{esco1}
&
1 \leftrightarrow 1
&
\nn\\
&
2 \leftrightarrow 4d
~,~
3a \leftrightarrow 4c
~,~
3b \leftrightarrow 3b
~,~
4a \leftrightarrow 4a
~,~
4b \leftrightarrow 4b
&
\nn\\
&
5 \leftrightarrow 6c
~,~
6a \leftrightarrow 6a
~,~
6b \leftrightarrow 6b
&
\nn\\
&
7 \leftrightarrow 10d
~,~
8a \leftrightarrow 10c
~,~
8b \leftrightarrow 9c
~,~
9a \leftrightarrow 10b
~,~
9b \leftrightarrow 9b
~,~
10a \leftrightarrow 10a
&
\nn\\
&
11 \leftrightarrow 12b
~,~
12a \leftrightarrow 12a
&
\nn\\
&
13 \leftrightarrow 15b
~,~
14 \leftrightarrow 14
~,~
15a \leftrightarrow 15a
&
\nn\\
&
16 \leftrightarrow 16
&
~.
\eea
For instance, the dual pair $13\leftrightarrow 15b$ in \fref{f_dualexample} is exact under the indicated swap between external and internal perfect matchings.

Accordingly, specular duality maps between brane tilings whose corresponding quiver gauge theories have different mesonic moduli spaces. In \cite{HananySeongSpecular}, it is illustrated how specular duality maps not the mesonic moduli spaces but the master spaces \cite{Forcella:2008bb,Forcella:2008ng,Forcella:2009bv,Hanany:2010zz,Forcella:2008eh,Zaffaroni:2008zz} of the dual pairs in \eref{esco1}. The master space is the complete moduli space including both the mesonic and baryonic branches. It is shown that the master spaces of the dual pairs in \eref{esco1} are \textit{identical} under a translation of fields given by the mapping of perfect matchings of the corresponding brane tilings. Further study of this duality is of great interest and some interpretations are reported in \cite{HananySeongSpecular}.


\section*{Acknowledgements}
We would like to thank Alastair D. King for very interesting discussions that eventually led to the creation of this project. R.-K. S. likes to thank the Yukawa Institute of Theoretical Physics at Kyoto University, the Simons Center for Geometry and Physics at Stony Brook University and the Hebrew University of Jerusalem for hospitality during various stages of this work. He also likes to thank Tohru Eguchi and Kazuo Hosomichi for hospitality in Kyoto, as well as Stefano Cremonesi, Masato Taki and Giuseppe Torri for valuable discussions. He is also grateful to his parents.

\appendix
\section{The theory for $\mathbb{C}^3/\mathbb{Z}_4\times \mathbb{Z}_4~~(1,0,3)(0,1,3)$ \label{s_parent}}

\begin{figure}[H]
\begin{center}
\includegraphics[trim=0cm 0cm 0cm 0cm,height=5cm]{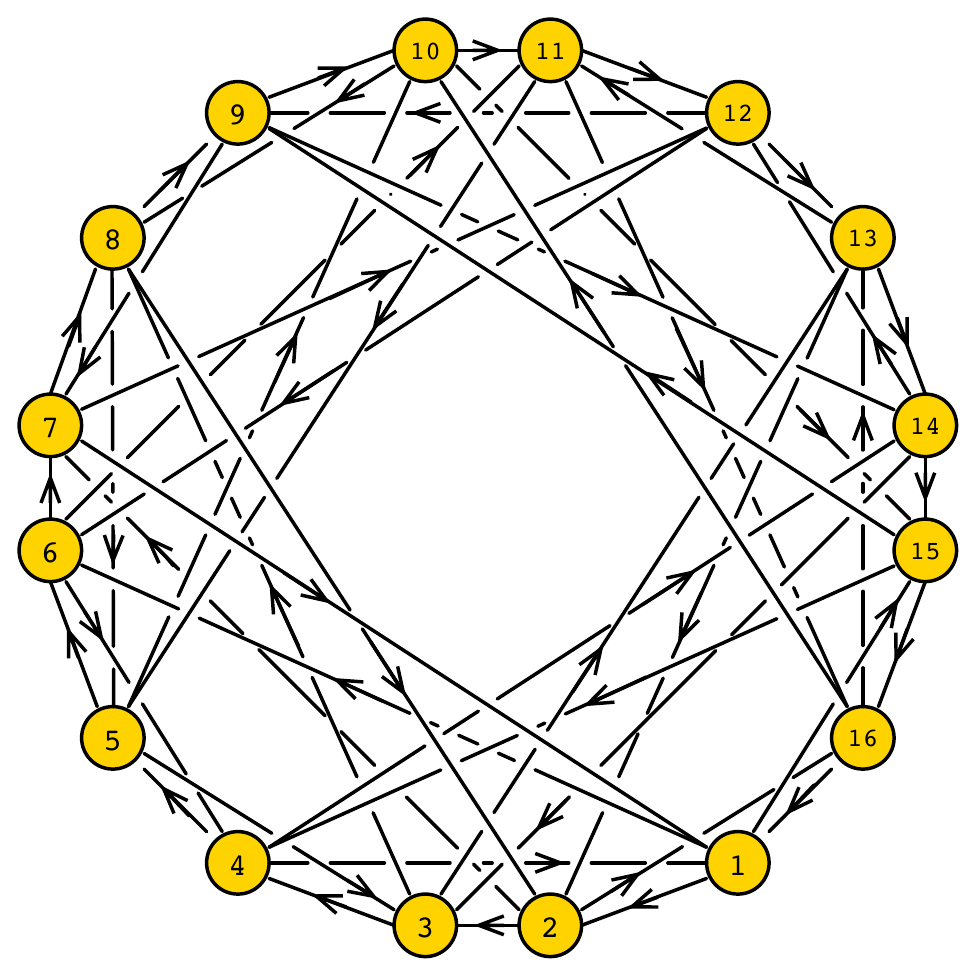}
\includegraphics[trim=0cm 4cm 13cm 0cm,height=5cm]{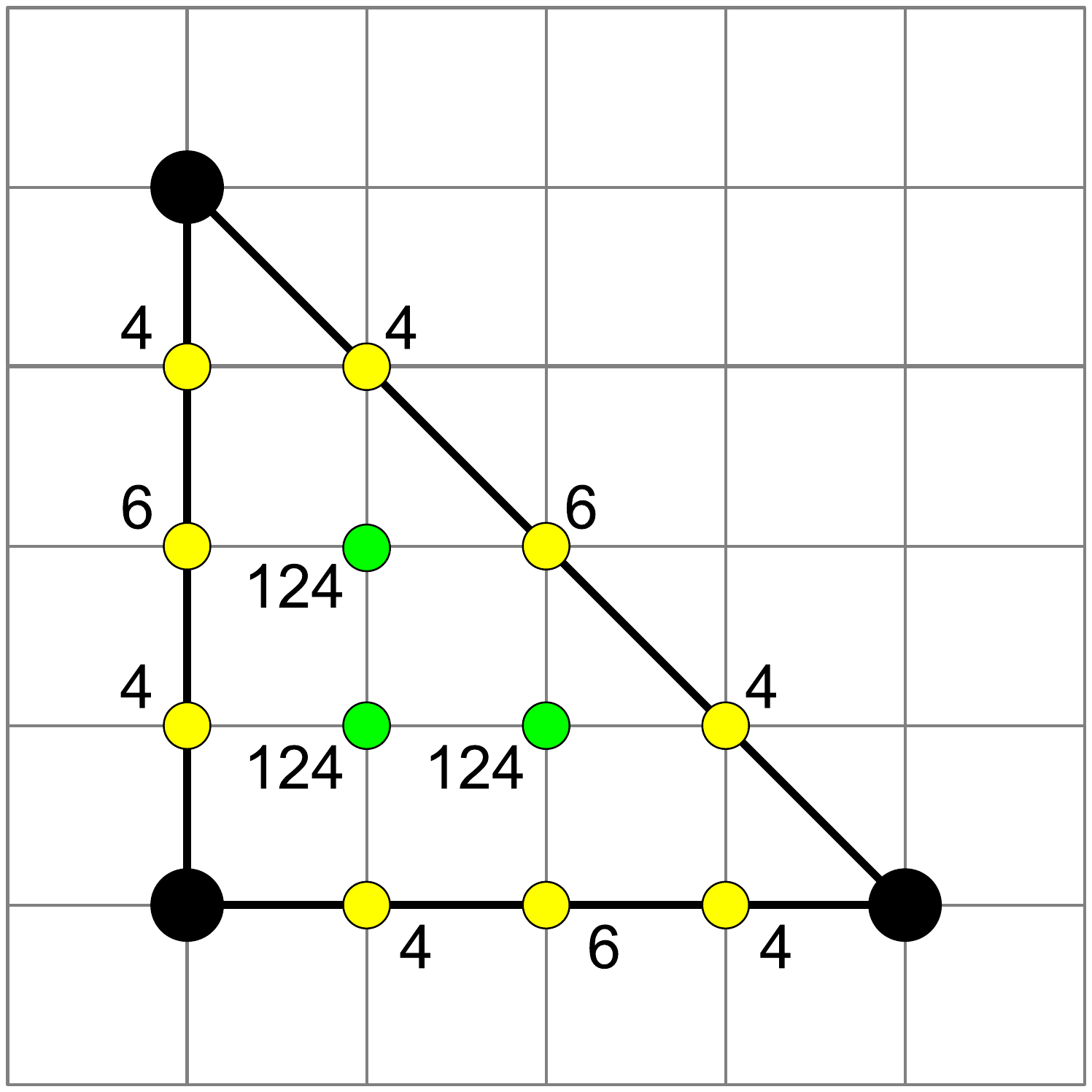}
\includegraphics[trim=0cm 0cm 0cm 0cm,height=5cm]{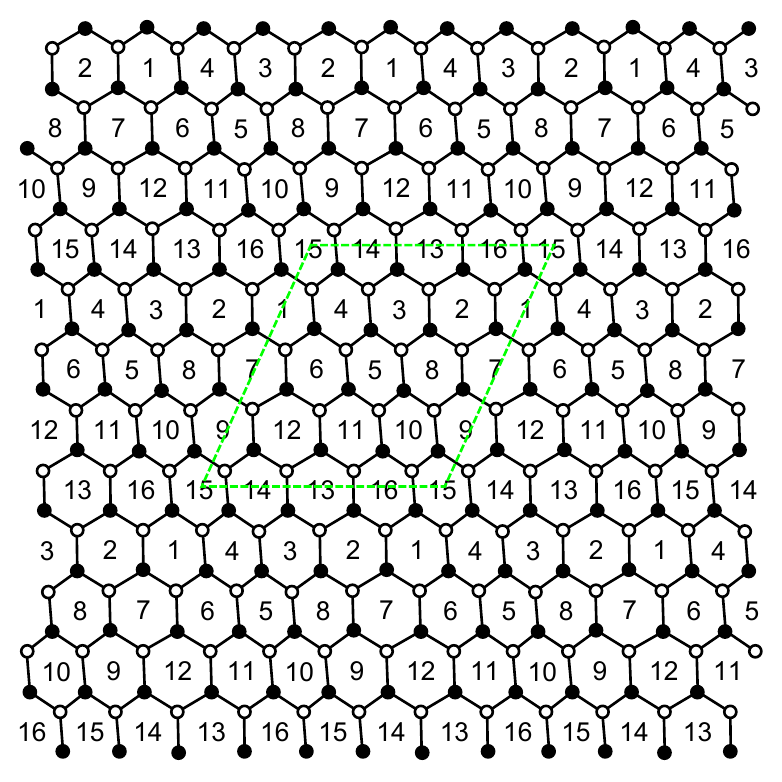}
\caption{The quiver, toric diagram, and brane tiling of the abelian orbifold of the form $\mathbb{C}^3/\mathbb{Z}_4\times\mathbb{Z}_4$ with orbifold action $(1,0,3)(0,1,3)$.\label{forigin}}
 \end{center}
 \end{figure}

The quiver, toric diagram and brane tiling of $\mathbb{C}^3/\mathbb{Z}_4 \times \mathbb{Z}_4~(1,0,3)(0,1,3)$ theory are shown in \fref{forigin} with the superpotential\footnote{Note: The superpotential features an overall trace which is not explicitly written down in the following discussion.} having the form
\beal{es2_9}
W&=&
+ X_{7\hspace{0.9mm}8}~X_{8\hspace{0.9mm}2}~X_{2\hspace{0.9mm}7} +X_{12\hspace{0.9mm}9}~X_{9\hspace{0.9mm}7}~X_{7\hspace{0.9mm}12} +X_{13\hspace{0.9mm}14}~X_{14\hspace{0.9mm}12}~X_{12\hspace{0.9mm}13} +X_{2\hspace{0.9mm}3}~X_{3\hspace{0.9mm}13}~X_{13\hspace{0.9mm}2} \nn\\
&&+X_{8\hspace{0.9mm}5}~X_{5\hspace{0.9mm}3}~X_{3\hspace{0.9mm}8} +X_{9\hspace{0.9mm}10}~X_{10\hspace{0.9mm}8}~X_{8\hspace{0.9mm}9} +X_{14\hspace{0.9mm}15}~X_{15\hspace{0.9mm}9}~X_{9\hspace{0.9mm}14}+X_{3\hspace{0.9mm}4}~X_{4\hspace{0.9mm}14}~X_{14\hspace{0.9mm}3} \nn\\
&&+X_{5\hspace{0.9mm}6}~X_{6\hspace{0.9mm}4}~X_{4\hspace{0.9mm}5} +X_{10\hspace{0.9mm}11}~X_{11\hspace{0.9mm}5}~X_{5\hspace{0.9mm}10} +X_{15\hspace{0.9mm}16}~X_{16\hspace{0.9mm}10}~X_{10\hspace{0.9mm}15} +X_{4\hspace{0.9mm}1}~X_{1\hspace{0.9mm}15}~X_{15\hspace{0.9mm}4} \nn\\
&&+X_{6\hspace{0.9mm}7}~X_{7\hspace{0.9mm}1}~X_{1\hspace{0.9mm}6} +X_{11\hspace{0.9mm}12}~X_{12\hspace{0.9mm}6}~X_{6\hspace{0.9mm}11} +X_{16\hspace{0.9mm}13}~X_{13\hspace{0.9mm}11}~X_{11\hspace{0.9mm}16} +X_{1\hspace{0.9mm}2}~X_{2\hspace{0.9mm}16}~X_{16\hspace{0.9mm}1} \nn\\
&&-X_{7\hspace{0.9mm}8}~X_{8\hspace{0.9mm}9}~X_{9\hspace{0.9mm}7} -X_{12\hspace{0.9mm}9}~X_{9\hspace{0.9mm}14}~X_{14\hspace{0.9mm}12} -X_{13\hspace{0.9mm}14}~X_{14\hspace{0.9mm}3}~X_{3\hspace{0.9mm}13} -X_{2\hspace{0.9mm}3}~X_{3\hspace{0.9mm}8}~X_{8\hspace{0.9mm}2} \nn\\
&&-X_{8\hspace{0.9mm}5}~X_{5\hspace{0.9mm}10}~X_{10\hspace{0.9mm}8} -X_{9\hspace{0.9mm}10}~X_{10\hspace{0.9mm}15}~X_{15\hspace{0.9mm}9} -X_{14\hspace{0.9mm}15}~X_{15\hspace{0.9mm}4}~X_{4\hspace{0.9mm}14} -X_{3\hspace{0.9mm}4}~X_{4\hspace{0.9mm}5}~X_{5\hspace{0.9mm}3} \nn\\
&&-X_{5\hspace{0.9mm}6}~X_{6\hspace{0.9mm}11}~X_{11\hspace{0.9mm}5} -X_{10\hspace{0.9mm}11}~X_{11\hspace{0.9mm}16}~X_{16\hspace{0.9mm}10} -X_{15\hspace{0.9mm}16}~X_{16\hspace{0.9mm}1}~X_{1\hspace{0.9mm}15} -X_{4\hspace{0.9mm}1}~X_{1\hspace{0.9mm}6}~X_{6\hspace{0.9mm}4} \nn\\
&&-X_{6\hspace{0.9mm}7}~X_{7\hspace{0.9mm}12}~X_{12\hspace{0.9mm}6} -X_{11\hspace{0.9mm}12}~X_{12\hspace{0.9mm}13}~X_{13\hspace{0.9mm}11} -X_{16\hspace{0.9mm}13}~X_{13\hspace{0.9mm}2}~X_{2\hspace{0.9mm}16} -X_{1\hspace{0.9mm}2}~X_{2\hspace{0.9mm}7}~X_{7\hspace{0.9mm}1} ~~.
\nn\\
\eea
\\

\section{Review: Seiberg Duality, Integrating out Mass Terms, and the Higgs Mechanism \label{sapp1}}

\subsection{Seiberg Duality \label{sapp_seiberg}}

Two $3+1$ dimensional worldvolume theories are called \textbf{toric (Seiberg) dual} if in the UV they have different Lagrangians with a different field content and superpotential, but flow to the same universality class in the IR. The mesonic moduli spaces of toric (Seiberg) dual theories are toric Calabi-Yau $3$-folds which are identical. The corresponding toric diagrams are $GL(2,\mathbb{Z})$ equivalent, however multiplicities of internal toric points and hence GLSM fields with zero R-charge can differ. 

\begin{figure}[th!]
\begin{center}
\includegraphics[trim=0cm 0cm 0cm 0cm,width=16 cm]{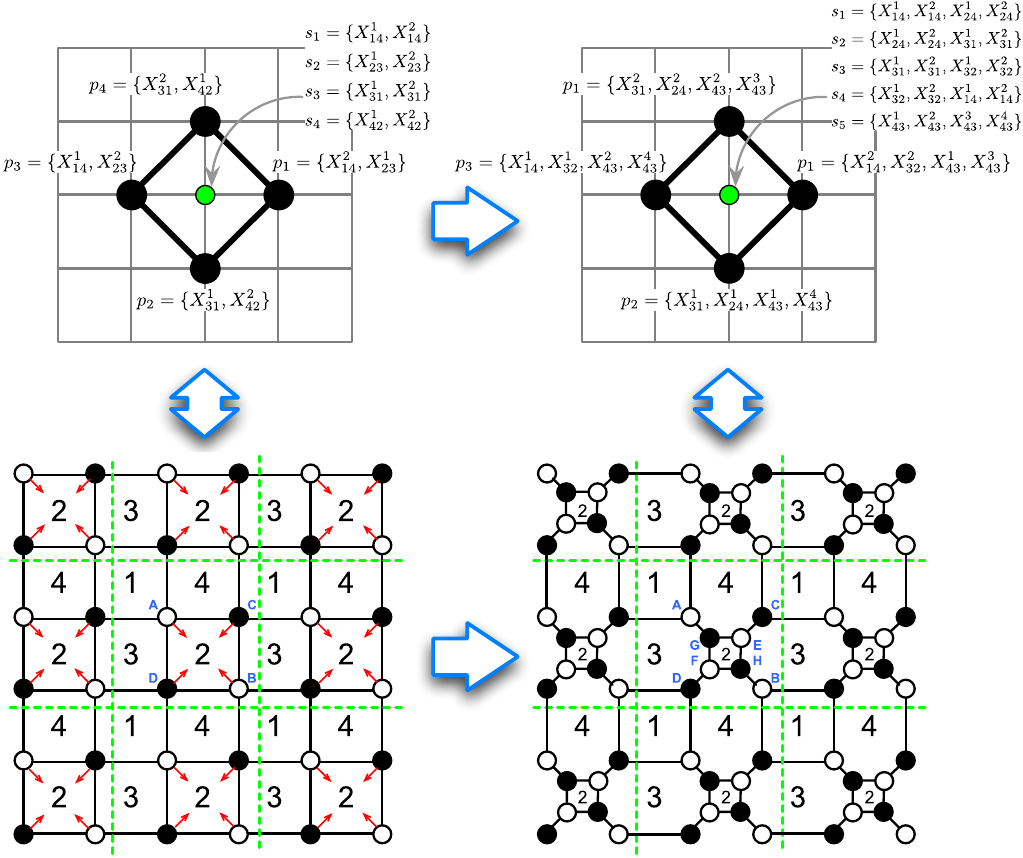}
\caption{The toric (Seiberg) duality action on the brane tiling of the zeroth Hirzebruch surface $F_0$ model with corresponding toric diagrams. The points in the toric diagram correspond to GLSM fields which are presented as perfect matchings or sets of bifundamental fields in the brane tiling picture.}
  \label{fseiberg}
 \end{center}
 \end{figure}

The relationship between two toric (Seiberg) dual theories is best illustrated with an example using brane tilings. Dualizing on a given gauge group $U(n)$ has a natural interpretation in the brane tiling picture. Let us consider the Hirzebruch $\mathbb{F}_0$ model. The corresponding gauge theory has a superpotential of the form 
\beal{es01_1}
W_{I} =
  \underbracket[0.1mm]{X_{14}^{1} X_{42}^{1} X_{23}^{1} X_{31}^{1}}_{A}
+ \underbracket[0.1mm]{X_{14}^{2} X_{42}^{2} X_{23}^{2} X_{31}^{2}}_{B}
- \underbracket[0.1mm]{X_{14}^{2} X_{42}^{1} X_{23}^{2} X_{31}^{1}}_{C}
- \underbracket[0.1mm]{X_{14}^{1} X_{42}^{2} X_{23}^{1} X_{31}^{2}}_{D}
~~,\nn\\
\eea
whose corresponding brane tiling and toric diagram are shown in the first column of \fref{fseiberg}. The terms are labelled $A$ to $D$ and the corresponding brane tiling nodes are indicated in \fref{fseiberg}. By dualizing on the gauge group $U(n_2)$, the superpotential becomes
\beal{es01_2}
W_{II} &=&
  \underbracket[0.1mm]{X_{14}^{1}X_{43}^{1}X_{31}^{1}}_{A}
+ \underbracket[0.1mm]{X_{14}^{2}X_{43}^{2}X_{31}^{2}}_{B}
- \underbracket[0.1mm]{X_{14}^{2}X_{43}^{3}X_{31}^{1}}_{C}
- \underbracket[0.1mm]{X_{14}^{1}X_{43}^{4}X_{31}^{2}}_{D}
\nn\\
&&
+ \underbracket[0.1mm]{X_{14}^{1}X_{43}^{3}X_{31}^{2}}_{E}
+ \underbracket[0.1mm]{X_{14}^{2}X_{43}^{4}X_{31}^{1}}_{F}
- \underbracket[0.1mm]{X_{14}^{1}X_{43}^{1}X_{31}^{1}}_{G}
- \underbracket[0.1mm]{X_{14}^{2}X_{43}^{2}X_{31}^{2}}_{H}
\eea
and the corresponding new brane tiling and quiver are shown in the second column of \fref{fseiberg}. One observes that under toric (Seiberg) duality, the number of gauge groups $G$ remains constant, the number of bifundamental fields $E$ and the number of superpotential terms both increase each by $4$.

The change in the number of bifundamental fields and superpotential terms corresponds to the change in the number of GLSM fields corresponding to internal points of the corresponding toric diagram. The area of the toric diagram corresponding to the number of gauge groups $G$ remains constant. The two toric diagrams and brane tilings in \fref{fseiberg} with the corresponding superpotentials given in \eref{es01_1} and \eref{es01_2} are called \textbf{phases} of the $F_0$ model.

The duality action often leads to superpotentials with quadratic mass terms. Quadratic mass terms relate to massive fields which become non-dynamical in the IR. The removal of quadratic mass terms and the corresponding deformation of the brane tiling are discussed in the following section.
\\

\subsection{Integrating out mass terms}

Quadratic terms in the superpotential relate to massive fields which are non-dynamical in the IR \cite{Franco:2005rj}. We are interested in the IR regime of the quiver gauge theories above, and therefore need to integrate out the quadratic terms in the superpotential.

\begin{figure}[H]
\begin{center}
\includegraphics[trim=0cm 0cm 0cm 0cm,width=12 cm]{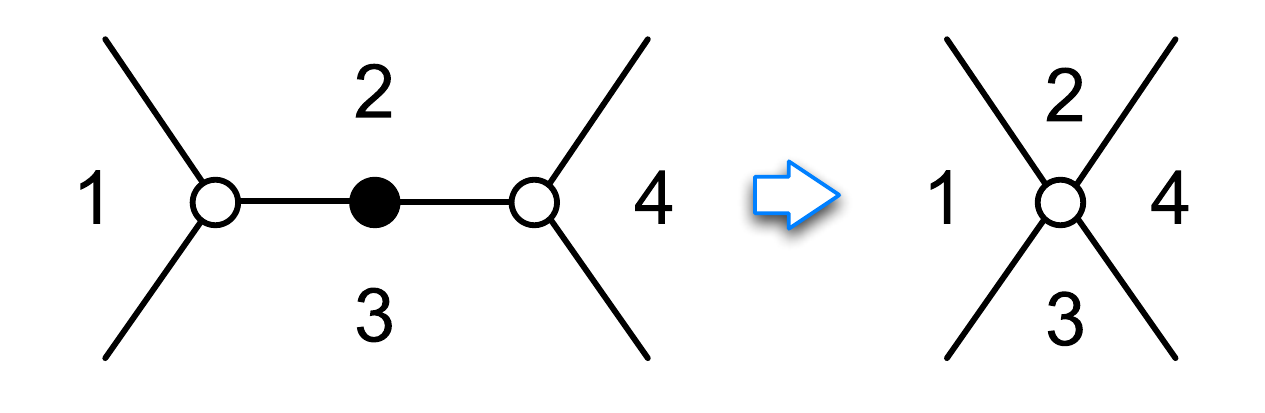}
\caption{The removal of quadratic mass terms in the superpotential corresponds to the removal of $2$-valent vertices in the brane tiling.}
  \label{fintegrate}
 \end{center}
 \end{figure}

The procedure of integrating out quadratic mass terms in the superpotential has a natural interpretation in the brane tiling context as illustrated in \fref{fintegrate}. Let us consider the superpotential corresponding to the case shown in \fref{fintegrate}, 
\beal{es01_5}
W_{I} = \dots + X_{31} X_{12} \underline{X_{23}} + \underline{X_{32}} X_{24} X_{43} - \underline{X_{23} X_{32}} + \dots
~~,
\eea
where the quadratic mass term and matter fields involved have been underlined. The removal of the quadratic mass term in \eref{es01_5} leads to the new superpotential of the form
\beal{es01_6}
W_{II} = \dots + X_{31} X_{12} X_{24} X_{43} + \dots
~~.
\eea
One observes that the process of integrating out mass terms preserves the toric condition discussed in section \sref{s1_1}.
\\

\subsection{Higgs Mechanism \label{sapp_higgs}}

The Higgs Mechanism has a natural interpretation in the brane tiling picture. By giving a non-zero vacuum expectation value (VEV) to a gauge field in gauge theory I, and integrating out resulting quadratic mass terms in the superpotential as explained above, one obtains a new theory II whose mesonic moduli space is a different toric Calabi-Yau $3$-fold to the one of theory I. Giving a VEV to a bifundamental field $X_{ij}$ results in the removal of the corresponding edge in the brane tiling picture. This results in an effective merger between two adjacent faces, analogous of combining two gauge groups $U(n)_i$ and $U(n)_j$ into one.

\begin{figure}[H]
\begin{center}
\includegraphics[trim=0cm 0cm 0cm 0cm,width=15 cm]{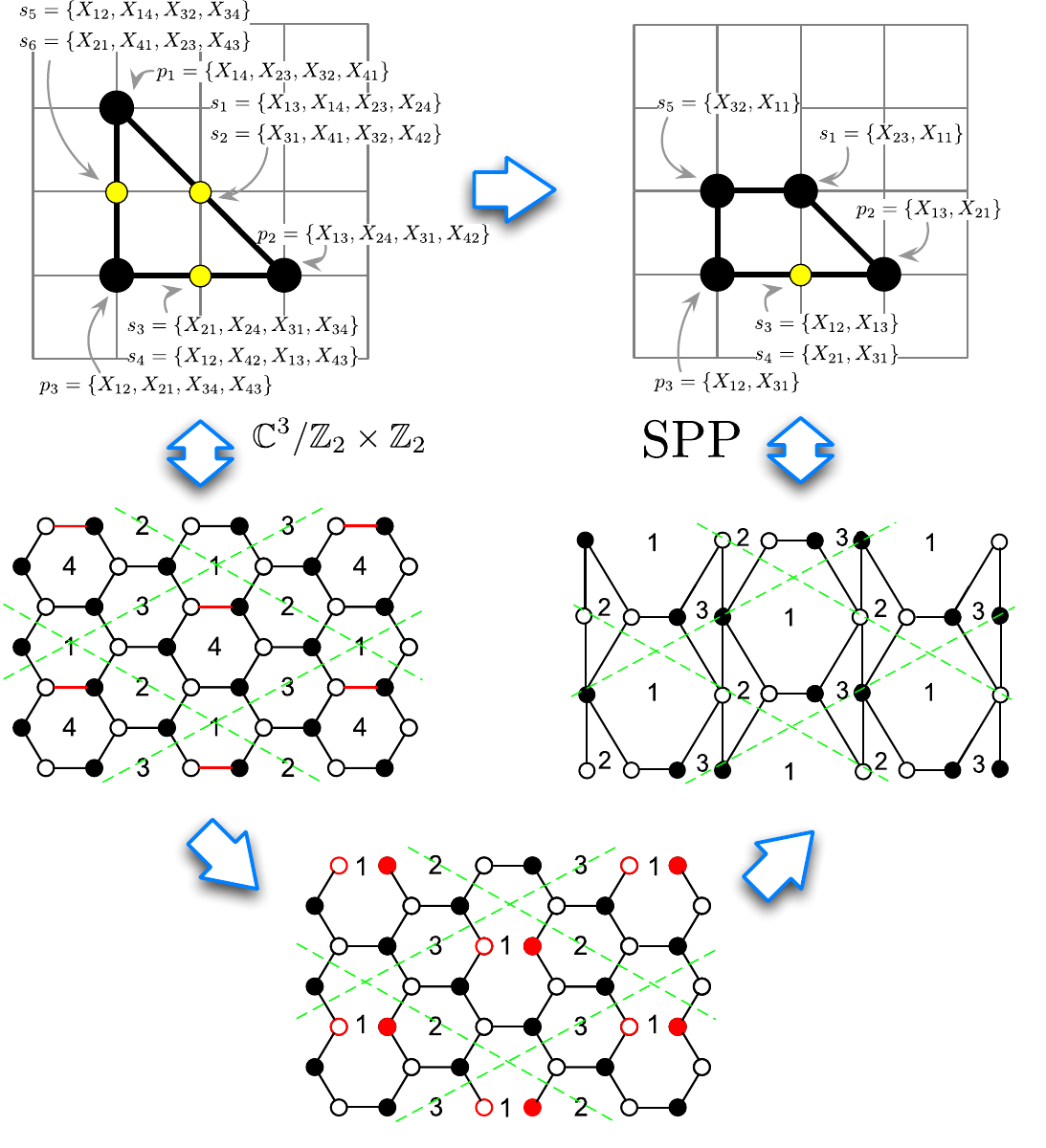}
\caption{By giving a non-zero vacuum expectation value to the bifundamental field $X_{14}$ of the $\mathbb{C}^3/\mathbb{Z}_2 \times \mathbb{Z}_2$ orbifold theory, one obtains the Suspended Pinch Point theory (SPP). The bifundamental field $X_{14}$ is represented by a red edge in the brane tiling. By setting $\langle X_{14}\rangle=1$, one obtains quadratic mass terms represented by red nodes in the second brane tiling, which are integrated out to give the third SPP tiling. The nodes of the corresponding toric diagrams are labelled with perfect matching variables and the corresponding sets of bifundamental fields. The Higgsing procedure corresponds to a blow down from $\mathbb{C}^3/\mathbb{Z}_2\times\mathbb{Z}_2$ to the cone over the Suspended Pinch Point.}
  \label{fhiggs}
 \end{center}
 \end{figure}

Let us consider the example of the $\mathbb{C}^3/\mathbb{Z}_2 \times \mathbb{Z}_2$ orbifold theory with orbifold action $(0,1,1)(1,0,1)$. The corresponding brane tiling and toric diagram is shown in \fref{fhiggs}, and the superpotential is 
\beal{es01_10}
W_{I}&=&
X_{42} X_{23} X_{34} +X_{31} X_{14} X_{43} +X_{24} X_{41} X_{12} 
+X_{13} X_{32} X_{21} \nn\\
&&
-X_{42} X_{21} X_{14} -X_{31} X_{12} X_{23} 
-X_{24} X_{43} X_{32} -X_{13} X_{34} X_{41} 
~~.
\eea
By giving the bifundamental field $X_{14}$ a VEV, such that $\langle X_{14} \rangle=1$, the superpotential becomes,
\beal{es01_11}
W_{I}\prime &=&
X_{42} X_{23} X_{34} +\underline{X_{31} X_{43}} +X_{24} X_{41} X_{12} 
+X_{13} X_{32} X_{21} \nn\\
&&
-\underline{X_{42} X_{21}}-X_{31} X_{12} X_{23} 
-X_{24} X_{43} X_{32} -X_{13} X_{34} X_{41} 
~~,
\eea
which in turn, by integrating out the above underlined quadratic mass terms, becomes
\beal{es01_12}
W_{II} = 
X_{13} X_{32} X_{23} X_{31} +X_{12} X_{21} X_{11} -X_{12} X_{23}
X_{32} X_{21} -X_{13} X_{31} X_{11}~~.
\eea

Theory II with the above superpotential and brane tiling shown in \fref{fhiggs} corresponds to the suspended pinch point (SPP) theory. Thus one has, by giving a VEV to a field in theory I, blown down a toric point in $\mathbb{C}^3/\mathbb{Z}_2\times\mathbb{Z}_2$ to give the SPP model. \fref{fhiggs} shows the perfect matchings and their field content for each toric point of the toric diagrams of $\mathbb{C}^3/\mathbb{Z}_2\times\mathbb{Z}_2$ and SPP.

The claim is that the combination of toric duality procedures, integrating out mass terms, and higgs mechanisms on the $\mathbb{C}^3/\mathbb{Z}_4 \times \mathbb{Z}_4$ orbifold theory with orbifold action $(1,0,3)(0,1,3)$ results in all possible quiver gauge theories whose mesonic moduli space is toric Calabi-Yau and has a toric diagram which is a reflexive polygon on $\mathbb{Z}^2$.
\\

\nocite{Acharya:1998db}
\nocite{Kehagias:1998gn}

\bibliographystyle{JHEP}
\bibliography{mybib}


\end{document}

%% file: pref.tex
\newcommand{\be}{\begin{equation}}
\newcommand{\ee}{\end{equation}}
\newcommand{\beq}{\begin{equation}}
\newcommand{\beql}[1]{\begin{equation}\label{#1}}
\newcommand{\eeq}{\end{equation}}
\newcommand{\ba}{\begin{array}}
\newcommand{\ea}{\end{array}}
\newcommand{\bea}{\begin{eqnarray}}
\newcommand{\beal}[1]{\begin{eqnarray}\label{#1}}
\newcommand{\eea}{\end{eqnarray}}
\newcommand{\ben}{\begin{enumerate}}
\newcommand{\een}{\end{enumerate}}
\newcommand{\bean}{\begin{eqnarray*}}
\newcommand{\eean}{\end{eqnarray*}}
\newcommand{\eref}[1]{(\ref{#1})}
\newcommand{\sref}[1]{\S\ref{#1}}
\newcommand{\tref}[1]{Table~\ref{#1}}
\newcommand{\nn}{\nonumber}

\newcommand{\fref}[1]{Figure \ref{#1}}
\newcommand{\btab}[1]{\begin{tabular}{#1}}
\newcommand{\etab}{\end{tabular}}

\def \Black {\pdfsetcolor{0 0 0 1}}

\newcommand{\comment}[1]{}

\newcommand{\ud}{\mathrm{d}}

\newcommand{\qed}{\nobreak \ifvmode \relax \else
      \ifdim\lastskip<1.5em \hskip-\lastskip
      \hskip1.5em plus0em minus0.5em \fi \nobreak
      \vrule height0.75em width0.5em depth0.25em\fi}